\documentclass{ucbthesis}

\usepackage{adjustbox}

\usepackage{chngcntr}
\counterwithin{equation}{section}

\let\Oldpart\part
\newcommand{\parttitle}{}
\renewcommand{\part}[1]{\renewcommand{\parttitle}{#1}\Oldpart{#1}}

\makeoddhead{myheadings}{\textit{Part \thepart. ~\parttitle}}{}{\protect\thepage}

\newenvironment{coolsubappendices}{
\clearpage
\chapter*{Chapter \thechapter. ~ Appendices}
\addcontentsline{toc}{section}{\emph{Appendices for Chapter \thechapter}}
\counterwithin{figure}{section} %
\counterwithin{table}{section} %
\namedsubappendices %
\begin{subappendices}
}{
\end{subappendices}
\counterwithout{figure}{section} %
\counterwithin{figure}{chapter} %
\counterwithout{table}{section} %
}

\newenvironment{coolcontents}{
\setcounter{tocdepth}{3} 
\startcontents[chapters]
\printcontents[chapters]{}{1}{\subsection*{Chapter \thechapter\ Contents}}
}{
\stopcontents[chapters]
}

\usepackage{caption}

\usepackage{subcaption}

\usepackage{enumitem}

\usepackage{titletoc}

\usepackage{multibib}
\newcommand{\appendixbibname}{Additional References for Appendices}
\newcites{Appendix}{\appendixbibname}

\usepackage{tocloft}
\cftsetindents{section}{0em}{3em}
\cftsetindents{chapter}{0em}{3em}
\cftsetindents{part}{0em}{3em}

\usepackage{mathrsfs}
\usepackage{dsfont}
\DeclareMathAlphabet\mathbfcal{OMS}{cmsy}{b}{n}
\usepackage{graphicx}
\usepackage{float}
\usepackage{relsize}
\usepackage{amsmath}
\usepackage{amssymb}
\usepackage{amsfonts}
\usepackage{amsthm}

\usepackage{multirow}

\usepackage{hyperref}
\hypersetup{
    colorlinks,
    citecolor=black,
    filecolor=black,
    linkcolor=black,
    urlcolor=blue
}

\newcommand{\vectorize}{\operatorname{vect}}

\newcommand{\rvec}[1]{\vec{\mathbf{#1}}}

\newcommand{\rate}{\lambda}
\newcommand{\freq}{f}
\newcommand{\vfreq}{\vec{\freq}}
\newcommand{\efreq}{\hat{\freq}}
\newcommand{\specie}[1][\relax]{\ifx\relax#1 s \else \sigma\fi}
\newcommand{\species}[1][\relax]{\specie[#1]}
\newcommand{\strain}[1][\relax]{\specie[#1]}
\newcommand{\Species}{S}
\newcommand{\Strains}{\Species}
\renewcommand{\time}{t}
\newcommand{\batch}[1][\relax]{\ifx\relax#1 b \else \beta\fi}
\newcommand{\Batches}{B}

\newcommand{\N}{\mathbb{N}_{\ge 0}}
\newcommand{\R}{\mathbb{R}_{\ge 0}}

\newcommand{\countbasesymbol}{M}
\newcommand{\expectedcount}[1][\relax]{\ifx \relax#1 \countbasesymbol \else \countbasesymbol^{(#1)}\fi}
\newcommand{\observedcount}[1][\relax]{\ifx \relax#1 \hat{\countbasesymbol} \else \hat{\countbasesymbol}^{(#1)}\fi}

\newcommand{\countprobbasesymbol}{p}
\newcommand{\countprob}[1][\relax]{\ifx \relax#1 \countprobbasesymbol \else \countprobbasesymbol^{(#1)}\fi}
 
\newcommand{\Population}[1][\relax]{\ifx\relax#1 K \else K\strut^{(#1)}\fi}
\newcommand{\population}[1][\relax]{\ifx\relax#1 n \else n\strut^{(#1)}\fi}
\newcommand{\vPopulation}{\rvec{\Population}}
\newcommand{\vpopulation}{\rvec{\population}}
\newcommand{\prevdropletbase}{\droplet\!-\!1}
\newcommand{\prevdroplet}[1][\relax]{\ifx\relax#1 \prevdropletbase \else \scriptscriptstyle \prevdropletbase\fi}
\newcommand{\popfreq}[1][\relax]{\ifx\relax#1 \kappa\strut \else \kappa\strut^{(#1)}\fi}

\newcommand{\mysteryseq}[1][]{\mathscr{R}_{\mathrm{#1}}}
\newcommand{\uRobbins}{\overline{\mysteryseq}}
\newcommand{\lRobbins}{\underline{\mysteryseq}}
\newcommand{\fudgefunction}{\Phi}

\newcommand{\clustering}{\zeta}
\newcommand{\concentration}{\clustering}
\newcommand{\density}{D}
\newcommand{\dconcentration}{\concentration_{\density}}
\newcommand{\compositional}{C}
\newcommand{\cconcentration}{\concentration_{\compositional}}

\newcommand{\norm}[1]{\left\lVert#1\right\rVert}

\usepackage{euscript}
\newcommand{\skeleton}{\EuScript{S}}
\newcommand{\uskel}{\skeleton}
\newcommand{\sskel}{\skeleton^{\pm}}
\newcommand{\magskel}{\EuScript{M}}

\newcommand{\abundance}[1][\relax]{\ifx\relax#1 N \else N^{(#1)}\fi}
\newcommand{\labundance}[3]{\abundance[#3]_{#1,#2}}
\newcommand{\vabundance}{\rvec{\abundance}}
\newcommand{\lvabundance}[2]{\vabundance_{#1,#2}}
\newcommand{\counts}[1][\relax]{\ifx\relax#1 n \else n^{(#1)}\fi}
\newcommand{\lcounts}[3]{\counts[#3]_{#1,#2}}
\newcommand{\vcounts}{\vec{\counts}}
\newcommand{\lvcounts}[2]{\vcounts_{#1,#2}}
\newcommand{\dcounts}[1][\relax]{\ifx\relax#1 m \else m^{(#1)}\fi}

\newcommand{\vdcounts}{\vec{\dcounts}}
\newcommand{\lvdcounts}[2]{\vdcounts_{#1,#2}}

\newcommand{\strainbasesymbol}{\mathbb{S}}
\newcommand{\straincount}[1][\relax]{\ifx\relax#1 \strainbasesymbol \else \strainbasesymbol^{(#1)}\fi}
\newcommand{\nstraincount}[1][\relax]{\ifx\relax#1 \overline{\strainbasesymbol} \else \overline{\strainbasesymbol}^{(#1)}\fi}
\newcommand{\vstraincount}{\rvec{\straincount}}

\newcommand{\Z}[1][\relax]{\ifx\relax#1 Z \else Z^{(#1)}\fi}

\newcommand{\globe}{\mathcal{G}}

\newcommand{\basedropletsymbol}{d}
\newcommand{\droplet}[1][\relax]{\ifx \relax#1 \basedropletsymbol \else \delta \fi}
\newcommand{\Droplets}{\MakeUppercase{\basedropletsymbol}}
\newcommand{\poisson}{\operatorname{Pois}}
\newcommand{\negbinom}{\operatorname{NB}}
\newcommand{\multinomial}{\operatorname{Mult}}
\newcommand{\mult}{\multinomial}
\newcommand{\dirmult}{\operatorname{DirMult}}
\newcommand{\gaussian}{\operatorname{Gaussian}}
\newcommand{\support}{\operatorname{supp}}
\newcommand{\observed}[1][\relax]{\ifx\relax#1 \support(\vabundance (0))  \else \support(\vabundance (#1))  \fi}

\newcommand{\indicator}[1]{\mathbb{I}\!\left\{#1 \right\}}

\newcommand{\baserate}[1][\specie]{r_{#1}}
\newcommand{\interaction}[2]{\alpha_{#1 #2}}
\newcommand{\noise}{\varepsilon}
\newcommand{\noisescale}{\eta}

\newcommand{\glvindices}[1][\specie]{\mathscr{B}_{#1}}
\newcommand{\fixsigns}{\varphi} 

\newcommand{\sign}{\operatorname{sign}}
\newcommand{\abs}{\operatorname{abs}}

\newcommand{\rateterm}[1][\specie]{G_{#1}}
\newcommand{\carryingcapacity}[1][\specie]{K_{#1}}
\newcommand{\limitingfactor}[1][\specie]{H_{#1}}

\newcommand{\copynumbersbias}{\vec{\mathbf{B}}}
\newcommand{\reads}[1][\relax]{\ifx\relax#1 R \else R^{(#1)}\fi}
\newcommand{\vreads}{\rvec{\reads}}

\newcommand{\merging}{m}

\newcommand{\treatmentsymbol}{\mathcal{T}}
\newcommand{\treatment}[1][\relax]{\ifx\relax#1 \treatmentsymbol \else \treatmentsymbol^{(#1)}\fi}
\newcommand{\controlsymbol}{\mathcal{C}}
\newcommand{\control}[1][\relax]{\ifx\relax#1 \controlsymbol \else \controlsymbol^{(#1)}\fi}

\newcommand{\powersymbol}{\mathscr{P}}
\newcommand{\power}[1][\relax]{\ifx\relax #1 \powersymbol \else \powersymbol^{(#1)}\fi}

\newcommand{\treatmentsizesymbol}{\mathbf{T}}
\newcommand{\treatmentsize}[1][\relax]{\ifx\relax#1 \treatmentsizesymbol \else \treatmentsizesymbol^{(#1)}\fi}
\newcommand{\controlsizesymbol}{\mathbf{C}}
\newcommand{\controlsize}[1][\relax]{\ifx\relax#1 \controlsizesymbol \else \controlsizesymbol^{(#1)}\fi}

\newcommand{\rank}{\mathcal{R}}
\newcommand{\sparrank}[1]{\hspace{-1em}\underset{\scriptscriptstyle\hspace{1em}\graph_1\!,\!\graph_2}{\rank}\hspace{-1em}\left( #1 \right)}

\newcommand{\lrestimatorsymbol}{\hat{\boldsymbol{\Psi}}}
\newcommand{\lrestimatormatrix}[1][\relax]{\ifx\relax#1 \lrestimatorsymbol \else \lrestimatorsymbol^{(#1)}\fi}
\newcommand{\lrestimator}[3]{\lrestimatormatrix[#1]_{#2/#3}}

\newcommand{\lrestimandsymbol}{\Psi}
\newcommand{\lrestimandmatrix}[1][\relax]{\ifx\relax#1 \lrestimandsymbol \else \lrestimandsymbol^{(#1)}\fi}
\newcommand{\lrestimand}[3]{\lrestimandmatrix[#1]_{#2/#3}}

\newcommand{\lrestimandsymbolfull}{\Psi^F}
\newcommand{\lrestimandmatrixfull}[1][\relax]{\ifx\relax#1 \lrestimandsymbolfull \else {\lrestimandsymbolfull}^{(#1)}\fi}
\newcommand{\lrestimandfull}[3]{\lrestimandmatrixfull[#1]_{#2/#3}}

\newcommand{\cvxcoeff}[1]{c_{#1}}
\newcommand{\powerset}[1]{\mathcal{P}\left( #1 \right)}

\usepackage{mathtools}

\newcommand{\expectationsymbol}{\mathbb{E}}
\newcommand{\expectation}[1][]{\expectationsymbol_{#1}\expectarg}
\newcommand{\mean}{\boldsymbol{\hat{\mu}}\expectarg}
\newcommand{\var}{\operatorname{Var}\expectarg}

\newcommand{\evar}{\hat{\operatorname{var}}\expectarg}
\newcommand{\cov}{\operatorname{Cov}\expectarg}
\newcommand{\ecov}{\hat{\operatorname{cov}}\expectarg}
\newcommand{\probability}[1][]{\mathbb{P}_{#1}\probarg}
\newcommand{\likelihoodsymbol}{\mathcal{L}}
\newcommand{\likelihood}{\likelihoodsymbol\probarg}
\newcommand{\loglikelihoodsymbol}{\ell}
\newcommand{\loglikelihood}{\loglikelihoodsymbol\probarg}
\DeclareMathOperator*{\argmin}{argmin}
\DeclareMathOperator*{\argmax}{argmax}

\DeclarePairedDelimiterX{\expectarg}[1]{[}{]}{%
  \ifnum\currentgrouptype=16 \else\begingroup\fi
  \activatebar#1
  \ifnum\currentgrouptype=16 \else\endgroup\fi
}

\DeclarePairedDelimiterX{\probarg}[1]{(}{)}{%
  \ifnum\currentgrouptype=16 \else\begingroup\fi
  \activatebar#1
  \ifnum\currentgrouptype=16 \else\endgroup\fi
}

\newcommand{\innermid}{\nonscript\;\delimsize\vert\nonscript\;}
\newcommand{\activatebar}{%
  \begingroup\lccode`\~=`\|
  \lowercase{\endgroup\let~}\innermid 
  \mathcode`|=\string"8000
}

\newcommand{\exponint}[1]{\operatorname{Ein} \left( #1 \right)}

\newtheorem{corollary}{Corollary}

\newcommand{\adjacencybasesymbol}{A}
\newcommand{\adjacency}{\mathbf{\adjacencybasesymbol}}
\newcommand{\adjacencytruth}{\adjacency_*}
\newcommand{\adjacencyest}{\mathbf{\hat{\adjacencybasesymbol}}}
\newcommand{\graph}{\mathcal{G}}
\newcommand{\edgefunc}{\epsilon}
\newcommand{\jaccard}{\mathcal{S}_J}
\newcommand{\edgeset}{\mathcal{E}}
\newcommand{\nodeset}{\mathcal{N}}

\newcommand{\graphspace}{\mathscr{G}}
\newcommand{\dissimilarity}{\mathscr{D}}
\newcommand{\similarity}{\mathscr{S}}

\newcommand{\spearman}{\rho_S}

\newcommand{\deltacon}{d_{DC}}

\newtheorem{lemma}{Lemma}[chapter]
\newtheorem{definition}{Definition}[subsection]

\title{Statistics of High-Throughput Characterization of Microbial Interactions}
\author{William Krinsman}
\date{}

\newcommand{\dropletsgitrepourl}{https://gitlab.com/krinsman/droplets}
\newcommand{\dropletsgitrepo}{\href{\dropletsgitrepourl}{the relevant GitLab repository}}

\newcommand{\networksgitrepourl}{https://gitlab.com/krinsman/mixed-sign-networks}
\newcommand{\networksgitrepo}{\href{\networksgitrepourl}{the relevant GitLab repository}}

\usepackage[singlelinecheck=false]{caption}

\begin{document}

\title{Statistics of High-Throughput Characterization of Microbial Interactions}
\author{William Edward Krinsman}
\degreesemester{Spring}
\degreeyear{2022}
\degree{Doctor of Philosophy}
\chair{Professor Mark van der Laan}
\othermembers{Professor Adam Arkin \\
Professor Lexin Li}
\numberofmembers{3}
\field{Biostatistics}
\emphasis{Computational and Data Science and Engineering}
\secondemphasis{Computational and Genomic Biology}
\campus{Berkeley}

\maketitle
\copyrightpage

\begin{abstract}

An active area of research interest is the inference of ecological models of complex microbial communities. Inferring such ecological models entails understanding the interactions between microbes and how they affect each other's growth. This dissertation employs a statistical perspective to contribute further to the knowledge currently addressing this problem.

  \paragraph{Part \ref{part:introduction}}
  Part \ref{part:introduction} explains how high-throughput droplet-based microfluidics technology can be used to screen for microbial interactions. An explicit, statistical framework is motivated and developed that can guide the analysis of data from such experiments.
  Chapter \ref{cha:motivation} investigates the specific questions that need to be answered to study microbial interactions. Chapter \ref{cha:motivation} explains why high-throughput droplet-based microfluidics technology overcomes previous limitations to answering these questions. It is shown how answering these questions can be recast as the statistical problem of estimating a network with mixed-sign edge weights.
  Chapter \ref{cha:introduction} investigates how to approach these questions using statistical models. Chapter \ref{cha:introduction} explains that the data from the noisy dynamical systems corresponding to each droplet can be understood as censored observations of a multivariate Markov process. The statistical understanding of a droplet's initial state is identified as crucial to overcoming the main limitation of these experiments, the uncontrolled assignment of microbes to droplets.

  \paragraph{Part \ref{part:modell-init-form}}
  Part \ref{part:modell-init-form} explains how it might be possible to predict, based on the experimental setup, how much data will be produced to infer given microbial interactions. Running the experiment once without incubating the droplets turns out to be necessary to make such predictions.
  Chapter \ref{cha:model-init-form} investigates which statistical (working) models can be used to describe a droplet's initial state.
  Chapter \ref{cha:model-init-form} shows that specific assumptions justify a default working model. New working models are derived by relaxing each of these assumptions.
  Chapter \ref{chap:model_comparison} investigates whether the failure of any of these assumptions leads to substantially new behavior. Chapter \ref{chap:model_comparison} demonstrates that log likelihood ratios can be used to answer this question. Failure of the sampling without replacement assumption turns out to have negligible effects in practice, but failures of the other assumptions could be important.
  Chapter \ref{chap:data_throughput} investigates how failures of the relevant assumptions affect the targeted estimands that enable the prediction of how much data will be produced to infer given microbial interactions.
  Chapter \ref{chap:data_throughput} confirms that more severe failures of these assumptions lead to more severe discrepancies with the predictions derived from the default working model. The nature of the effect depends on the chosen grouping of droplets defining the targeted estimands.
  Chapter \ref{chap:hetero_estimator_performance} investigates how to estimate failures of the relevant assumptions from the data produced by unincubated droplets.
  Chapter \ref{chap:hetero_estimator_performance} presents both plugin and maximum likelihood estimators for doing so. Failures of these assumptions are shown to be understandable non-parametrically.

  \paragraph{Part \ref{part:aver-treatm-effects}}
Part \ref{part:aver-treatm-effects} demonstrates the feasibility of inferring microbial interactions from the data produced by these experiments. Relevant ideas from the microbiological and ecological literature are recast into an explicit, statistical framework.
Chapter \ref{chap:log-ratio-coeff} investigates how a particular measure of relative fitness can be recast into the statistical framework of average treatment effects.
Chapter \ref{chap:log-ratio-coeff} explains how violations of positivity assumptions are inevitable for this problem, making controlling for confounding difficult.
Explicit assumptions are given under which the estimands are identifiable from the observed data produced by incubated droplets, even though initial states of the droplets are not directly observed.
Chapter \ref{chap:network_comparison} investigates how comparisons of ecological interactions can be recast into the statistical framework of loss functions for signed networks.
Chapter \ref{chap:network_comparison} explains how avoiding unexpected behavior requires loss functions for signed networks to satisfy what is called herein ``the double penalization principle''.
Starting from loss functions of unsigned networks, several examples of loss functions for signed networks are derived that satisfy this property.

\paragraph{Conclusion} Future work will explore further choices that can be made when modelling this problem, how to connect these ideas to more sophisticated statistical methodologies, and the biological interpretation of the results of applying these ideas to real experimental datasets.
This work demonstrates the plausibility of characterizing microbial interactions using high-throughput droplet-based microfluidics technologies, and hopefully will guide the analysis of data produced by such experiments in the future.
  
\end{abstract}

\begin{frontmatter}
 
\begin{dedication}
\null\vfil
\begin{center}
For Falcon and Ruby. We will always miss you.
\end{center}
\vfil\null
\end{dedication}

\tableofcontents
\clearpage

{\begingroup
  \let\clearpage\relax
  \chapter*{Code}
  \addcontentsline{toc}{chapter}{Code}
\endgroup}

\noindent Relevant source code and Jupyter notebooks can be found in several public archives.
\\~\\
\textbf{GitLab:} \url{https://gitlab.com/krinsman/dissertation}
\\~\\
\textbf{GitHub:} \url{https://github.com/krinsman/dissertation}
\\~\\
\textbf{Zenodo:} \url{https://doi.org/10.5281/zenodo.6539845} 
\\~\\
\textbf{FigShare:} \url{https://doi.org/10.6084/m9.figshare.19747609}
\\~\\
\noindent Excerpted source code and Jupyter notebooks relevant for Part \ref{part:modell-init-form} can be found in \dropletsgitrepo ~at \url{\dropletsgitrepourl}.
\\~\\
\noindent Excerpted source code and Jupyter notebooks relevant for chapter \ref{chap:network_comparison} can be found in \networksgitrepo ~at \url{\networksgitrepourl}.

\listoffigures
\clearpage
\listoftables

{\begingroup
 \let\clearpage\relax
\chapter*{Notational Conventions}
\addcontentsline{toc}{chapter}{Notational Conventions}
\endgroup}

In most cases bound variables will be denoted by lower-case Latin letters, and their corresponding ``maximum possible value'' (which is a free variable) will be denoted by the corresponding Latin letter. For example, $\strain \in [\Strains]$, or $\droplet \in [\Droplets]$. In cases where the free variable will later be used to define the bound variable for another expression, the free variable will usually be denoted with a lower-case Latin letter (e.g. $\strain$), and the bound variable will be denoted with the corresponding lower-case Greek letter (e.g. $\strain[]$).

Deterministic scalars are generally denoted with lower-case, regular-font variable names. Random scalars are generally denoted with upper-case, regular-font variable names. Deterministic vectors are generally denoted with lower-case, bold-face, and arrow-topped variable names. Random vectors are generally denoted with upper-case, bold-face, and arrow-topped variable names. (Deterministic) matrices are generally denoted with upper-case, bold-face variable names (\textit{without} arrows). Random matrices do not have a notational convention.

$\mathbb{N}$ is the natural numbers $\{1, 2, 3, \dots\}$, $\N$ is the whole numbers $\{0, 1, 2, 3, \dots\}$, $\mathbb{R}$ is the real numbers, and $\R$ is the non-negative real numbers $[0, \infty)$.

Given a positive integer $N \in \mathbb{N}$, define $[N]:= \{1, \dots, n , \dots, N\}$. In general, the entries of a (non-negative) vector $\rvec{v} \in \R^S$ are denoted $v^{(\strain)}$, i.e. 
  \begin{equation*}
    \rvec{v} \overset{def}{=} (v^{(1)}, \dots, v^{(\strain)}, \dots, v^{(\Strains)}) \,.
  \end{equation*}
The shorthand ${v := \sum_{\strain=1}^{\Strains} v^{(\strain)}}$ is used for the sum of the entries of $\rvec{v}$. 

\begin{acknowledgements}
Dr. Fangchao Song for suggesting the topic of the dissertation and generous consultations, feedback, and advice throughout.

Dr. Lauren M. Lui for invaluable recommendations of strategies and resources for scientific writing.
  
Professor Perry de Valpine for the useful suggestion of investigating the Dirichlet-Multinomial and related models for weakening the 'well-mixed' droplets assumption.
\end{acknowledgements}

\end{frontmatter}

\clearpage
\nouppercaseheads

\setcounter{footnote}{0}
\pagestyle{myheadings}
\part{Introduction}
\label{part:introduction}

The ideas explored in Part \ref{part:introduction} belong to the general fields of both:
\begin{enumerate}[label=(\arabic*)]
\item using droplet-based microfluidics technology for high-throughput biological assays, and
\item quantitatively comparing the growth of microbes under different conditions.
\end{enumerate}

Droplet-based microfluidics technology has already been used for several different kinds of high-throughput biological assays. See e.g.  \cite{Kintses2010} or \cite{Guo2012} for reviews of such studies. However, droplet-based microfluidics technology has been particularly successful for the analysis of single cells. See e.g. \cite{single_cell_1} or \cite{single_cell_2} for reviews of such studies.

Experiments quantitatively comparing the growth of microbes under different conditions are often called ``competition assays'' \cite{competition_assay_defn}. At a high level, one can think of there being roughly two different kinds of competition assays. I describe both kinds below.

Competition assays where the treatment conditions correspond to mutant strains of a microbe, and the control condition is the ``wild type'' of the microbe, are very popular. Often the experiment is augmented by comparing the growth of mutant strains and wild types not just under one set of external conditions, but repeatedly under multiple sets of external conditions. The studies \cite{rb_tnseq}, \cite{GE_competition_1}, and \cite{GE_competition_2} are some (possibly non-representative) examples. This kind of competition assay can be thought of as a ``bipartite interaction network'' \cite{Calderer2021} \cite{Weighill2021} \cite{Torrisi2020}, where there is a clear distinction between ``predictor variable nodes'' (the sets of external conditions, ``environmental variable nodes'') and ``response variable nodes'' (the numbers of individuals for the various mutant strains). This kind of competition assay corresponds to a ``gene $\times$ environment interactions problem'' in the framework described in item \ref{item:gene_gene_interactions} of section \ref{broader-field-2}.

In contrast, there are also competition assays where the treatment conditions correspond to co-cultures of different types of microbes, and the control conditions are monocultures of the types of microbes. (See \cite{coculture} for a review of methodologies for co-culturing microbes.) Studies such as \cite{Venturelli} that are based on this kind of competition assay allow us to study how different microbes affect each other's growth. The studies \cite{GG_competition_1} and \cite{GG_competition_2} are other (possibly non-representative) examples. This kind of competition assay can be thought as an ``unpartitioned interaction network'' \cite{Angulo2017}, where there is no distinction between ``predictor variable nodes'' and ``response variable nodes'' (both being the numbers of individuals for the types of microbes). This kind of competition assay corresponds to a ``gene $\times$ gene interactions problem'' in the framework described in item \ref{item:gene_gene_interactions} of section \ref{broader-field-2}.

The dichotomy described above may not always be entirely clear-cut in practice. For example, the studies \cite{ambiguous_competition_1} and \cite{ambiguous_competition_2} describe competition assays where the external conditions are the presence of microbes different from those whose growth is being measured.

Inferring ecological models of microbial communities entails understanding the interactions between microbes and how they affect each other's growth. Therefore high-throughput competition assays using droplet-based microfluidics could potentially allow us to infer ecological models of complex microbial communities consisting of numerous types of microbes.
Hence the specific focus of Part \ref{part:introduction} is on the new subfield of using high-throughput droplet-based microfluidics assays to study how different microbes affect each other's growth. Previous studies in this subfield, e.g. \cite{kChip} and \cite{Hsu2019}, had important limitations. (Cf. the discussion in sections \ref{sec:how-micr-ecol} and \ref{sec:prep-sequ}.) The new MOREI platform (cf. section \ref{sec:using-morei-char}) aims to overcome most of these limitations. In doing so, the form of the resulting data is substantially different from that produced by previous experimental platforms. Thus the specific problem of this dissertation is to identify and address the data analysis challenges for the MOREI platform. This includes those data analysis challenges that are also applicable to more general experimental platforms but that are insufficiently addressed by previous work.

\paragraph{Chapter \ref{cha:motivation}}

The broader field that chapter \ref{cha:motivation} belongs to is the study of ecological models of microbial interactions.
In particular, chapter \ref{cha:motivation} explores the subfield of how droplet-based microfluidics technology can be used to infer microbial interactions.
This chapter begins its exploration by investigating the specific questions that need to be answered to study microbial interactions when using any technology.
Then the chapter follows this by explaining why high-throughput droplet-based microfluidics technology may overcome previous limitations to answering these questions.
The chapter proceeds further by describing how the problem of inferring ecological models of microbial interactions can be recast as a special case of the statistical problem of estimating a network with mixed-sign edge weights.
Finally, the chapter overviews of the details of the particular experiment considered by this dissertation.

\paragraph{Chapter \ref{cha:introduction}}

Hypothetically, the ideas of chapter \ref{cha:introduction} belong to a broader field of ``\textit{statistical inference of (time-evolving) interactions between variables of a (noisy) dynamical system for which the temporal evolution may be incompletely observed}''. However, such a concrete field of research appears not to exist\footnote{
If it does exist, hopefully it has a name that is pithier and catchier than the phrase written above.
} in the literature. In practice, the relevant ideas are spread between numerous disparate fields that are currently unconnected to each other. See section \ref{sec:backgr-sign-2} for a further elaboration.
Hence any of those disparate broader fields may be chosen, according to one's preferences, as the general context within which the ideas of chapter \ref{cha:introduction} may be understood.
Starting from any such general context, chapter \ref{cha:introduction} then focuses on the important issue of statistical inference of (time-dependent) microbial interactions based on pseudo-longitudinal data.
In particular, chapter \ref{cha:introduction} investigates plausible statistical models that could be used to describe the data produced by this experiment.
This chapter explains that the data from the noisy dynamical systems corresponding to each droplet can be understood as censored observations of a multivariate Markov process.
The statistical understanding of a droplet's initial state is identified as crucial to overcoming the main limitation of these experiments, the uncontrolled assignment of microbes to droplets.

\clearpage
\pagestyle{headings}

\chapter{Biological Context of Experiment}
\label{cha:motivation}

I show how inferring an ecological model of a microbial community can be understood as a special case of the statistical problem of estimating a network with mixed-sign edge weights.
See section \ref{sec:form-as-stat}.
Herein I explain why high-throughput droplet-based microfluidics technology overcomes previous limitations to studying microbial interactions.
See section \ref{sec:how-micr-ecol}.
Finally, I explain why and how the data-generating process for the experiment motivating this dissertation can be divided into (at least) three phases.
See section \ref{sec:using-morei-char}.

Section \ref{sec:what-why-microbial} clarifies some issues motivating the scientific questions this dissertation helps to address. Section \ref{sec:form-as-stat} explains how these scientific questions are simplified to a statistical questions for the purposes of this work. Section \ref{sec:how-micr-ecol} discusses the kinds of data necessary for answering such questions in general. Finally, section \ref{sec:using-morei-char} describes the particular experiment being used to answer these questions and specific features of the data it generates.

\begin{coolcontents}

\section{Why and What of Microbial Ecology}
\label{sec:what-why-microbial}

Microbial ecology is the study of microbial interactions with other microbes, with non-microbial organisms, and with the environment. 

A central assumption of this thesis is that microbial interactions are interesting and therefore that characterizing microbial interactions is valuable. While there are many reasons why that is the case, many more than I can list here, I want to share with the unfamiliar reader some such reasons why one might be interested in microbial ecology. There are also reviews discussing this, cf. e.g. \cite{Faust2012}, \cite{coculture}, or \cite{Antwis2017}. Readers already familiar with or interested in microbial ecology might consider skipping the rest of this section.

\paragraph{Why} 

One general reason to study microbial interactions is advancing fundamental biological knowledge. Microbial interactions have greatly shaped the formation and development of (macroscopic) ecosystems throughout deep time. Microbial interactions crucially influence biogeochemical nutrient cycles and thus Earth's climate\cite{biogeochemical}, for example via the marine carbon pump or the oxygenation of Earth's atmosphere and oceans billions of years ago. 

Microbial interactions also greatly shaped selective pressures that favored many key evolutionary innovations. Examples include multicellularity, endosymbiosis (including eukaryogenesis), pathogenicity, and antibiotic resistance. 

The above two categories are not mutually exclusive. Microbial interactions impact the fitness of macroscopic organisms \cite{Biswas2015} and can lead to key evolutionary innovations that greatly shape macroscopic ecosystems. For example, microbial associations are also important for plants' being ``ecosystem engineers'' and for animals to exploit new food sources. 

Another general reason to study microbial interactions is the potential to develop new technological applications that benefit humanity \cite{coculture} \cite{Angulo2019}. Via their influence on non-microscopic organisms and biogeochemical nutrient cycles, understanding microbial interactions helps to develop new techniques in agriculture \cite{Mahmood2016}, medicine and health \cite{Porter2017}, climate modelling and protection of the environment \cite{Wagner2002} \cite{Graham2011} \cite{Yoon2019}, and astrobiology and human space colonization \cite{Lopez2019}, among other areas. 

From a purely anthropocentric point of view, understanding the role of microbial interactions in the evolution of life on Earth also means better understanding our origins.

Microbial interactions with other microbes have historically been the most difficult to study because all participants in the interactions are microscopic. The great diversity and number of microbes leads to a great diversity and complexity of microbial interactions with other microbes. It is arguably more important to understand microbial interactions with other microbes than to understand microbial interactions with non-microbial organisms and the environment. The former interactions mediate the latter, while also being more diverse, more complex, and more understudied. Herein, unless specified otherwise, ``microbial interactions'' refers to microbial interactions with other microbes.

\paragraph{What}

The general field that chapter \ref{cha:motivation} belongs to is the study of ecological models of microbial interactions.
The review \cite{Faust2012} describes how such models are often formulated as pairwise interaction networks, an approach also espoused elsewhere, cf. e.g. \cite{Angulo2019} or \cite{Angulo2017}.
  (Although see e.g. \cite{Momeni2017} or \cite{higher_order} regarding potential limitations of models that consider only pairwise interactions.)
Ecological interactions have been classified at least since \cite{Lidicker1979} in terms of signs $(\pm)$ describing how organisms affect each other's growth.
The review \cite{coculture} describes several technologies that can be used for studying how various microbes affect the growth of other microbes with their presence. See also \cite{Venturelli} for an example.
The notion of a network describing ecological interactions, and ecological interactions being described by $(\pm)$ signs, leads naturally to the notion of a signed network, something that has been discussed (in other contexts) at least since \cite{Harary}.
The signed values of the edges can also be interpreted as the coefficients of some (parametric) dynamical system, see e.g. \cite{Gonze2018} for a review or either of \cite{signs_longitudinal_compositional} and \cite{Xiao2017} for discussion.
Cf. the introduction to Part \ref{part:introduction}.
Chapter \ref{cha:motivation} explores the subfield of how droplet-based microfluidics technology can be used to infer microbial interactions.
Published studies in this subfield include \cite{kChip}, \cite{Hsu2019}, and \cite{positive_kChip}. In this chapter I investigate the specific questions that need to be answered to study microbial interactions.

\section{Modelling Microbial Interactions}
\label{sec:form-as-stat}

We need both (1) an estimand to estimate and (2) data from which to make estimates. Then a (parametric) statistical model posits how the estimand specifies the data generating distribution. In this section I loosely describe the estimand to estimate for characterizing microbial interactions. Later sections discuss the data used for making estimates.

We can use a network to characterize the pairwise interactions of a microbial community (cf. \cite{Faust2012}). The approach to do this used herein is as follows. Each of the $\Strains$ nodes of the network corresponds uniquely to one of the $\Strains$ strains\footnote{Herein I use ``strains'' to refer equally to strains belonging to the same species(/genus/family/etc.) as well as to strains belonging to different species(/genera/families/etc.), because the distinction is irrelevant for setting up the abstract problem. It may matter for the implementation of a specific experiment.}, where $\Strains$ denotes the number of strains in the microbial community. An edge is directed from (the node for) strain $\strain_1$ to (the node for) strain $\strain_2$ if strain $\strain_1$ affects the growth of strain $\strain_2$. If strain $\strain_1$ promotes the growth of strain $\strain_2$ then the weight of the edge is positive, whereas if strain $\strain_1$ suppresses the growth of strain $\strain_2$ then the weight of the edge is negative. The estimand for characterizing microbial interactions is a directed network with mixed-sign edge weights \cite{Harary} \cite{Lidicker1979} \cite{signs_longitudinal_compositional}, whose number of nodes is the same as the number of strains. A ``microbial interaction'' is operationally defined as one strain affecting the growth of another strain. Cf. figure \ref{fig:example_microbial_interaction_network}.

\begin{figure}[H]
  \centering
  \includegraphics[width=\textwidth,height=0.4\textheight,keepaspectratio]{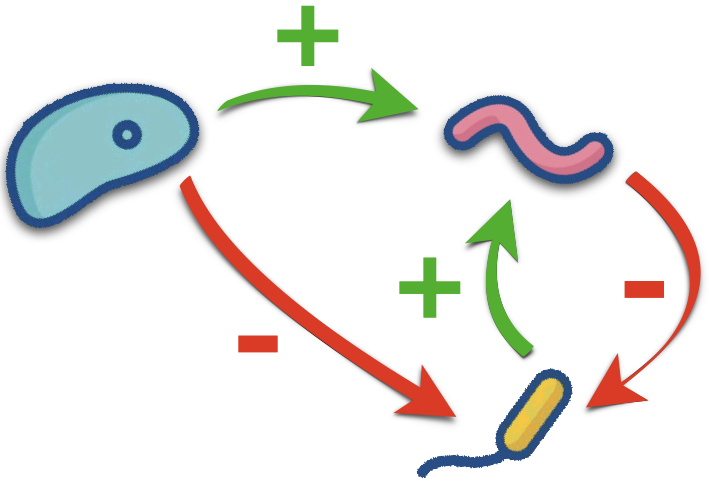}
  \caption[Signed, weighted, directed network representing microbial community interactions.]{A directed network with mixed-sign edge weights that represents the interactions within a given microbial community containing three strains. Arrows on the edges indicate the directions of the interactions. Green indicates positive edge weights (interaction promotes growth). Red indicates negative edge weights (interaction suppresses growth).}
  \label{fig:example_microbial_interaction_network}
\end{figure}

Herein I only consider how to characterize pairwise microbial interactions. There are a panoply of formalisms generalizing networks to objects that describe ``higher order interactions''. (Hypergraphs, simplicial complexes, multilayered networks, etc.) Choosing among them requires us to be much more precise about the notion of ``higher order'' microbial interaction in which we are interested. Moreover, compared to pairwise interactions, there is usually less data available to infer higher order interactions\footnote{
Cf. the discussions in sections \ref{sec:why-empir-distr} and \ref{sec:why-expect-posit}, which both make arguments to this effect.
}. Thus both aspects of formulating ``higher order'' microbial interactions as a statistical estimation problem are substantially more difficult. Hence these tasks are left to future work. Nevertheless, failing to model or account for the effects of ``higher order interactions'' might limit the usefulness or validity of any ecological interpretations of our results; cf. \cite{higher_order} or \cite{Momeni2017} for arguments along these lines. For a contrasting viewpoint, see \cite{Angulo2019} which (using simplifying assumptions) argues that it is sufficient to characterize only the pairwise interactions of a microbial community to be able to manipulate it (e.g. for biotechnological applications, cf. \cite{coculture}).

In chapter \ref{chap:network_comparison} I discuss methods for comparing two directed networks with mixed-sign edge weights. This allows us to quantify how similar an estimated microbial interaction network is to the true microbial interaction network.

\section{How of Microbial Ecology}
\label{sec:how-micr-ecol}

In section \ref{sec:form-as-stat} above I described the estimand to estimate for characterizing microbial interactions. Now I will discuss the data used for making estimates.

As described in section \ref{sec:form-as-stat}, herein ``microbial interactions'' are operationally defined as one strain affecting the growth of another strain, and we only characterize pairwise microbial interactions. Therefore to characterize microbial interactions, we want to quantify for all (ordered) pairs of microbial strains the effect that one strain has on the other strain's growth\footnote{
Note that inferring self-interactions is a problem. The ``competition assay'' framework described below provides no obvious way to do that. A priori it's unclear whether self-interactions are actually identifiable from such observations without more information (or assumptions). Previous work has either assumed values for self-interactions (e.g. \cite{Xiao2017}) or used models that require longitudinal data (e.g. \cite{Venturelli} or \cite{eco_model_time_series}). See \cite{Gonze2018} for an overview. Preliminary results (data not shown) suggested that attempting to modify methodology similar to that used in \cite{Venturelli} \cite{eco_model_time_series} \cite{Marino2014} \cite{signs_longitudinal_compositional} \cite{Fisher2014} \cite{Mounier2008} to not require longitudinal data, and apply to pseudo-longitudinal data, leads to very poor estimates of all of the interactions, including the self-interactions. (The attempted modification is described in detail in appendix \ref{sec:param-estim-inter}.) Other approaches that are unable to estimate self-interactions led to much better results (data not shown) for the remaining interactions. Addressing this shortcoming is left to future work.
}.

While there are many approaches that have been used for this problem, herein I focus exclusively on controlled laboratory studies. Biology is immensely complicated with many ``unknown unknowns'', and ecology even more so. Study designs for this problem based on observational data are exposed to many more potentially uncontrolled confounders. Therefore the conclusions from study designs based on observational data are much less reliable\footnote{
  One argument for observational data is that conclusions based on the synthetic microbial communities (``SynComs'') used in controlled laboratory studies might not transfer to naturally occurring systems. Thus it is useful for SynComs to be as realistic as possible, e.g. vis a vis the number of strains in the community.
}.

Another reason to not use observational data (e.g. metagenomic studies) for \textit{attempting} to infer microbial interactions is that these studies are problematic \cite{Mounier2008} \cite{Barlow2020} inasmuch as they usually only produce ``compositional data'' consisting of relative abundances, see section \ref{sec:prep-sequ}. Cf. e.g. \cite{microbiome_time_series}, \cite{eco_model_time_series}, \cite{signs_longitudinal_compositional}, or \cite{Marino2014} for counterpoints.

To infer the effect of strain $\strain_1$ on the growth of strain $\strain_2$, we can compare the growth of strain $\strain_2$ when strains $\strain_1$ and $\strain_2$ co-occur with the growth of strain $\strain_2$ when strains $\strain_1$ and $\strain_2$ do \textit{not} co-occur. If growth under both conditions occurs in a controlled laboratory setting, we can conclude that any difference in growth of strain $\strain_2$ between the two conditions is most likely the effect of strain $\strain_1$. The condition where strain $\strain_2$ co-occurs with strain $\strain_1$ can be thought of as the ``treatment'', and the condition where strain $\strain_2$ does \textit{not} co-occur with strain $\strain_1$ can be thought of as the ``control''. Cf. figure \ref{fig:experiment_example} or the discussion of ``competition assays'' from the introduction to Part \ref{part:introduction}. See also \cite{coculture} for a (perhaps now somewhat dated) review of such ``co-culture'' systems for studying cellular interactions. 

\begin{figure}[H]
  \centering
  \includegraphics[width=\textwidth,height=0.58\textheight,keepaspectratio]{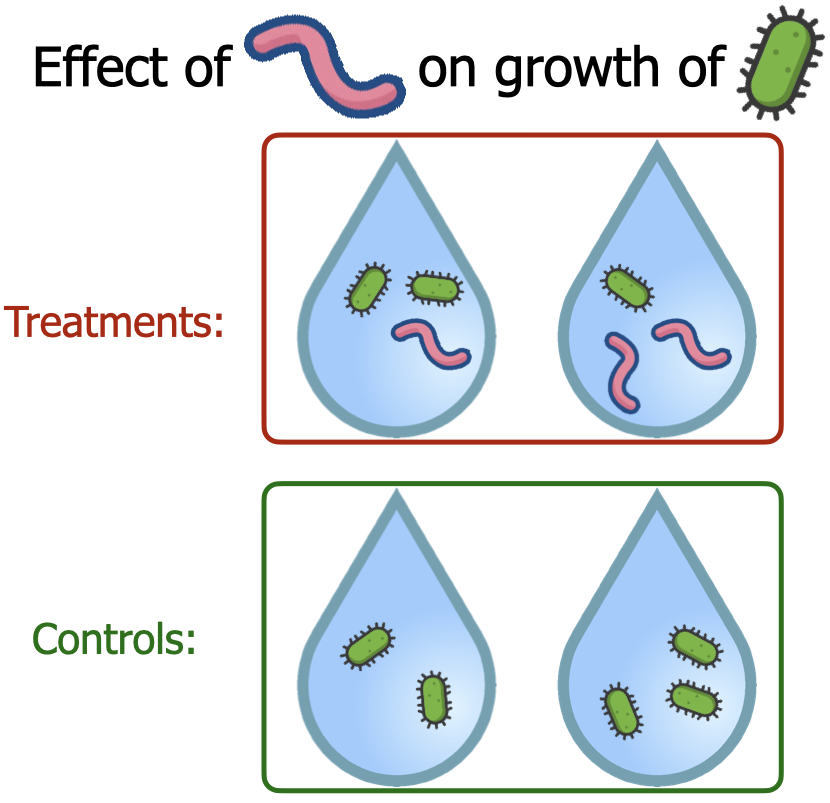}
  \caption[Inferring the effect of one strain on the growth of another strain.]{Inferring the effect of one strain on the growth of another strain. Cf. the discussion of ``competition assays'' from the introduction to Part \ref{part:introduction}.}
  \label{fig:experiment_example}
\end{figure}

As before, let $\Strains$ denote the total number of strains in the microbial community. Then to implement this framework for all (ordered) pairs of strains, we need at least $\binom{\Strains}{2} + \Strains$ replicates. $\binom{\Strains}{2}$ of the replicates correspond to ``treatments'' where two distinct strains co-occur, while $\Strains$ of the replicates correspond to ``controls'' where a single strain does not co-occur with other strains. Therefore, to characterize all pairwise microbial interactions, a lower bound on the number of required replicates grows quadratically with the total number of strains\footnote{
Conceivably the only way we could escape quadratic scaling for this problem is with a ``sparsity assumption''. Otherwise this appears to be an ``inherently quadratic'' problem. More specifically, to escape quadratic scaling it seems we would need some a priori biological knowledge that only a subquadratic number of microbial interactions can actually exist. (The other interactions would be ``zero'', i.e. correspond to missing edges in the microbial interaction network.) That would allow us to get away with having a subquadratic number of replicates. One conceivable way this could occur is if the microbial community was structured into ``subcommunities'' with minimal overlaps, such that pairwise interactions could only occur for two strains within the same subcommunity. Taking the number of pairwise interactions within the largest subcommunity as a constant, it follows that the scaling would be roughly linear in the number of subcommunities. The scaling following from this ``divide and conquer'' approach could in practice be much more favorable than the quadratic scaling in the total number of strains. That being said, we would need biological justification for why either (1) no pairwise microbial interactions occur between strains from different subcommunities, or (2) why we don't care about pairwise microbial interactions between strains from different subcommunities and are content to consider the microbial interaction \textit{sub}network where all of the edges corresponding to these interactions are missing.
}.

The most controlled way to generate these replicates is to manually plate the cells from the aforementioned $\binom{\Strains}{2} + \Strains$ combinations of strains. However such an approach is fairly labor intensive and does not scale well with the number of plates. Therefore the approach becomes practically infeasible for even relatively small numbers of strains. For example, the current state of the art as performed in e.g. \cite{Venturelli} is not much larger than $10$ strains.

Therefore to characterize all pairwise microbial interactions in a community with a reasonably large number of strains, we need to substantially increase the throughput of replicates. One way to do this is using droplet-based microfluidics technology. Using droplet-based microfluidics, the number of replicates (which corresponds to the number of microfluidic droplets containing cells) can be in the hundreds of thousands or even millions. See any of \cite{Teh2008} \cite{Kintses2010} \cite{Guo2012} \cite{Zhu2017} \cite{Shang2017} \cite{Sohrabi2020}, sorted in roughly chronological order, for a review of droplet-based microfluidics technology\footnote{
Droplet-based microfluidics is a special case of a wider variety of microfluidics technologies developed in recent decades. See any of \cite{Whitesides2006} \cite{Tian2008} \cite{Sackmann2014} \cite{Chiu2017} \cite{Convery2019}, sorted in roughly chronological order, for a review of general microfluidics technology.
}.

Droplet-based microfluidics has already proven useful for single-cell, single-type experiments. See, for example, \cite{single_cell_1} or \cite{single_cell_2}, for a review of single-cell studies. New experimental platforms, such as kChip \cite{kChip} \cite{positive_kChip} or MINI-Drop \cite{Hsu2019}, use droplet microfluidics to probe the interactions of multiple types of cells. For kChip and MINI-Drop specifically the motivation is to enable more systematic explorations of microbial interactions. Microbial ecology would benefit from experimental, rather than merely observational, data characterizing the interactions of large numbers of microbial strains. The current state of the art for such experiments, by manually culturing and plating combinations of cell types, probes the interactions of fewer than 20 microbial strains simultaneously \cite{Venturelli}. Because droplet microfluidics devices produce large numbers of microfluidic droplets, multi-cell, multi-type droplet microfluidics experimental platforms potentially could drastically increase the number of microbial strains whose interactions can be probed simultaneously. This has not yet happened with current droplet-based microfluidics studies of microbial interactions, which were severely limited in the number of strains considered \cite{Hsu2019} \cite{kChip} \cite{positive_kChip} due (in part) to requiring fluorescently labelled strains. Cf. section \ref{sec:prep-sequ}.

A major tradeoff associated with this approach, compared to manually culturing and plating combinations of cell types, is that droplet microfluidics chips form microfluidic droplets randomly. Thus compared to manually plating cells we lose some control over each individual replicate. Cf. figure \ref{fig:control_throughput_tradeoff}. Therefore, even starting with known microbial communities, it is a priori unclear approximately how many droplets will be produced containing any given combination of cell types. Filling this gap will increase the relative advantage of multi-cell, multi-type droplet microfluidics experimental platforms over manually culturing and plating combinations of cell types. This will help ensure that the drastic increase in throughput in the number of replicates more than compensates for any loss of control. Cf. chapters \ref{cha:model-init-form} and \ref{chap:data_throughput}. Hopefully this can abet future dramatic increases in the state of the art for the number of microbial strains whose interactions can be probed simultaneously. Being able to characterize all of the interactions in more realistically sized microbial communities under controlled laboratory settings could substantially advance the field of microbial ecology.

\section{Description of Experiment}
\label{sec:using-morei-char}

MOREI (pronounced ``more-ray'') stands for ``More Interactions'' or (treating MORE itself as a backronym) ``Microfluidics Offers Replicate Experiments for Interactions''. Starting from a microbial community sample, MOREI creates on the order of $10^5-10^6$ microfluidic droplets for characterizing microbial interactions.

Multiple batches of droplets are created using droplet-based microfluidics chips\footnote{
If you are unfamiliar with this technology, cf. \cite{microfluidic_chip_review} for a recent helpful review.
}. Droplets within each batch are incubated for the same amount of time, while different batches are incubated for different amounts of time. Cf. figure \ref{fig:droplets_batches}.

\begin{figure}
  \centering
  \includegraphics[width=\textwidth,height=0.5\textheight,keepaspectratio]{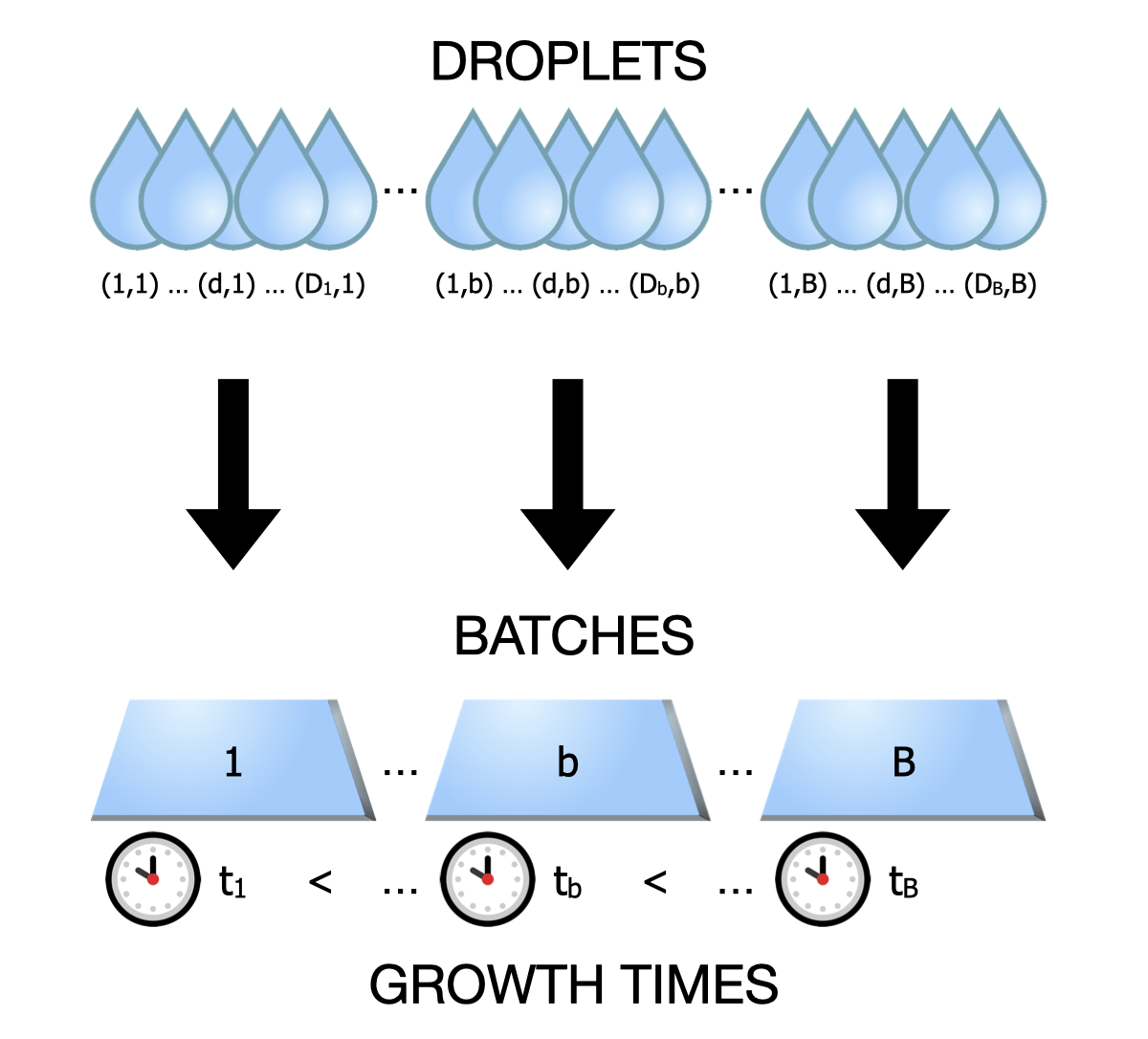}
  \caption{Notation; Organization of Droplets into Batches}
  \label{fig:droplets_batches}
\end{figure}

The ``life cycle'' of each droplet can be divided into roughly three phases. 
\begin{enumerate}
\item The first phase is ``\textbf{initial droplet formation}''. In the first phase, fluid from the microbial community sample is pumped from a syringe into the microfluidics chip. The microfluidics chip uses oil to encapsulate the fluid, as well as any cells that may have been inside the fluid, into a droplet.
\item The second phase is ``\textbf{growth of cells inside of the droplet}''. In the second phase, the droplet is incubated for the amount of time prescribed for its batch, during which any cells inside are able to divide and grow.
\item Finally the third phase is ``\textbf{preparation for sequencing}''. In the third phase, the droplet is fed into another microfluidics chip, where it is merged with another droplet containing materials that enable the sequencing and unique identification of the contents of the original droplet.
\end{enumerate}

After all three phases are completed, all of the (merged) droplets are pooled together to undergo PCR\footnote{polymerase chain reaction} amplification and finally sequencing. Cf. figure \ref{fig:droplet_life_cycle} below. I will also describe all three phases in more detail below. Readers uninterested in the technical details of the experiment or of the resulting data might consider skipping the rest of this section.

\begin{figure}
  \centering
  \includegraphics[width=\textwidth,height=0.95\textheight,keepaspectratio]{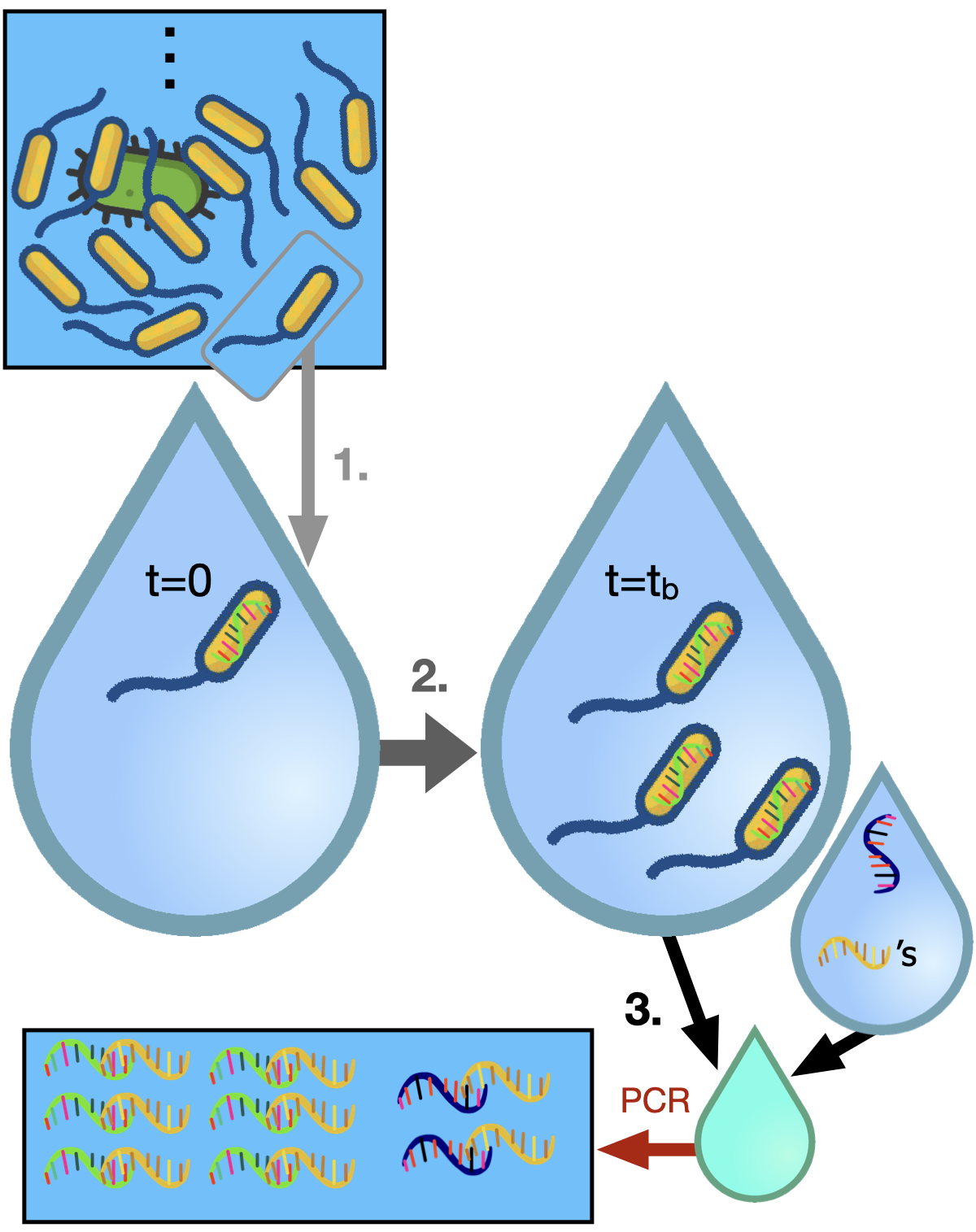}
  \caption{Schematic of the ``life cycle'' of a droplet. See text for details.}
  \label{fig:droplet_life_cycle}
\end{figure}

\subsection{Initial Droplet Formation}
\label{sec:init-dropl-form}

The microfluidics chip\footnote{
If you are unfamiliar with this technology, cf. \cite{microfluidic_chip_review} for a recent helpful review.
} encapsulates cells into droplets randomly. This makes each droplet's initial condition, both the number of cells in the droplet, as well as which the strains the cells belong to, random. The growth and division of cells during the second phase means that the droplets' random initial conditions are obscured in the final data. Therefore the best we can do is to make predictions of the droplets' unobserved random initial conditions based on average behavior. 

We can control the concentration/dilution of the microbial community sample to guarantee a specified average number of cells per droplet. The average number of cells per droplet is usually chosen to be around two. This is intended to guarantee that, when there are two (or more) strains present in a droplet, the most likely scenario is that there is at most one cell of each strain.

The small average number of cells per droplet has other effects. The data is incredibly sparse. Even in a typical non-empty droplet, almost all strains will be absent. This is very much unlike, for example, typical single-cell RNAseq data. Another effect is that for any given interaction, the number of droplets that serve as controls will usually vastly exceed the number of droplets that serve as treatments.

\subsection{Growth of Cells Inside of the Droplet}
\label{sec:growth-cells-inside}

The contents of the droplets are only observed once, after the end of the second phase. In particular, the details of what occurs during the second phase inside of any given droplet are completely opaque to us. The resulting data is therefore not longitudinal. Thus many standard methods for inferring microbial interactions, which assume time series data\footnote{
See e.g. \cite{eco_model_time_series} or \cite{Marino2014}. An introduction to time series for microbial ecologists is \cite{microbiome_time_series}.
}, are not applicable. Currently there appears to be no scalable method that can measure the abundances of microbial strains within the same microfluidic droplet at multiple time points \cite{Dressler2017}, so the limitation posed by the absence of longitudinal data appears unavoidable.

The droplets are all small, and thus nutrient-limited. Models of microbial interactions like the generalized Lotka-Volterra equations cannot make accurate predictions for the second phase, because they allow for indefinite growth. Cf. appendix \ref{sec:nutr-limit-indir}. The maximum possible number of cells that could exist inside of a droplet after incubation is \textit{not} unlimited. 

How long the droplets are incubated will determine whether cells are still actively dividing at the end of the second phase. It could be more difficult to infer the strengths of interactions from cells that exhausted the nutrients in their droplet and are undergoing logistic-like growth. This makes it important to incubate distinct batches for distinct amounts of time, to get ``pseudo-longitudinal'' data. Future work needs to determine how to best combine data from different batches, for example how to best balance averaging results within batches and averaging results between batches, while also accounting for how batches that were incubated longer are more likely to correspond to logistic-like growth.

\subsection{Preparation for Sequencing}
\label{sec:prep-sequ}

During the third phase, a second microfluidics chip\footnote{
If you are unfamiliar with this technology, cf. \cite{microfluidic_chip_review} for a recent helpful review.
} merges each droplet with a ``PCR mix'' droplet, whose contents prepare the contents of the original droplet for sequencing. 

First there is a lysis buffer, which lyses the cells in the droplet. This kills any cells which are still alive, and makes the genetic material inside all cells (living or dead) accessible. The third phase is necessary for observing the contents of the droplet, but the lysis buffer also ensures that it is destructive and cannot be repeated. Therefore the third phase is what forces the data to not be longitudinal. Because dead cells are also lysed and have their genetic material available for sequencing, this also introduces potential biases from counting such ``relic'' genetic material the same as if it had come from living cells. Cf. figure \ref{fig:relic_dna_intro}.

\begin{figure}[H]
  \centering
  \includegraphics[width=0.95\textwidth,height=\textheight,keepaspectratio]{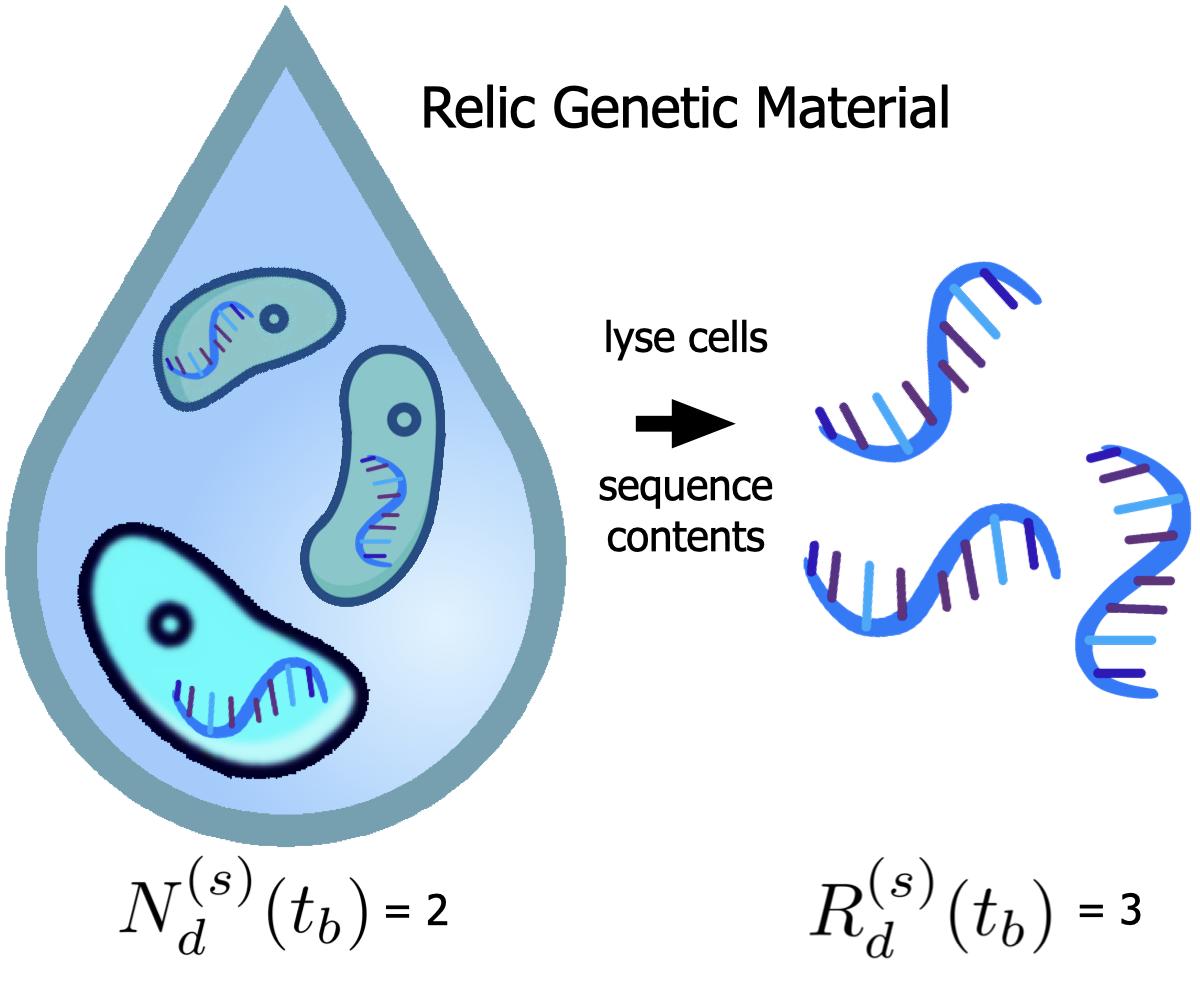}
  \caption[Relic genetic material (schematic).]{Relic genetic material can inflate inferred count of cells, potentially introducing biases. This droplet had two living cells and one dead cell after incubation. However, all three cells (living and dead) are lysed and thus also have their contents sequenced.}
  \label{fig:relic_dna_intro}
\end{figure}

Second there are the barcodes, represented by the yellow nucleotides in figure \ref{fig:droplet_life_cycle}. (For more details about barcode methodology, and examples of applications, see for example any of the papers \cite{barcode_1}, \cite{barcode_2}, \cite{barcode_3}, \cite{Roodgar2021}, or \cite{barcode_4}.) More precisely, the barcodes consist of (1) an Illumina Miseq adapter, (2) a barcode sequence, and (3) a 515F primer. The adapter ensures that the sequence and any other nucleotides that bind to it will be sequenced. The barcode sequences are (approximately\footnote{Choosing a barcode sequence of $L$ nucleotides, there are $4^{L}$ possible barcode sequences of length $L$. Therefore, choosing a long enough length $L$ for the barcode sequences, and generating them randomly, in principle one will have with high probability that the barcode sequence in each ``PCR mix'' droplet is unique, or that almost all are. This is much easier said than done, of course, but still feasible with today's technology.}) unique for each ``PCR mix'' droplet. Therefore the barcode sequences identify everything they bind to as having come from the same droplet. Finally the 515F primers bind to 16S rRNA genes from the lysed microbial cells. This ensures that primarily the 16S rRNA genes from the cells will be amplified in the PCR step that occurs following the third phase. Pooled replicate reactions may be unnecessary to avoid ``jackpot effects and chimera formation'' during PCR, at least for 16S data \cite{Marotz2019}. Hence, we should be able to use the PCR-amplified 16S rRNA gene sequences to distinguish different strains and (approximately) count reads within droplets.

Third there are spikein genes, represented by the blue nucleotides in figure \ref{fig:droplet_life_cycle}. A random, but approximately constant\footnote{\label{footnote:spikein}
Assuming a roughly Poisson distribution\cite{Collins2015} (for which the mean equals the variance), the larger the mean, the smaller the ratio of the standard deviation to the mean. For example, $\frac{\pm 10}{100}$ is a better relative uncertainty than $\frac{\pm 5}{25}$, even though $\pm 10$ is clearly a worse absolute uncertainty than $\pm 5$. Relative uncertainty is all that matters in this context because we use the PCR amplified count of spikein genes as a normalization factor. Thus, as long as the average number of copies of spikein gene per ``PCR mix'' droplet is large enough, we may treat it as ``approximately constant''. However, we don't want to add too many copies of the spikein gene on average, because then after PCR amplification the spikein genes will likely ``swamp out'' any signal from strains that are (relatively) less abundant inside of the droplet.
}, number of these are incorporated into each ``PCR mix'' droplet. Like the 16S genes from inside of the lysed cells, these will also bind to the barcodes, and therefore will also be amplified in the PCR step that occurs following the third phase. Assuming that all sequences from the same droplet that bind to the barcodes will be amplified by roughly the same amount during the PCR step, for each droplet we can divide all of the PCR amplified counts by the ratio of the PCR amplified spikein count to the average number of spikein gene copies originally in each ``PCR mix'' droplet.

While details of the spikein genes may seem unimportant from a purely statistical perspective, they actually imply a very important feature of the data.

Specifically, the spikein gene protocol enables us to normalize by an (approximate) PCR amplification factor, which means that we get \textbf{absolute abundance} values from the sequencing results. Cf. \cite{Barlow2020} for details of a related spikein protocol. Without estimates for the PCR amplification factor, normally one can only infer \textit{relative abundance} values. However relative abundance data may be problematic for inferring microbial interactions \cite{Mounier2008} or \cite[Figure 1]{Barlow2020}. This also implies that \textit{none} of the vast, pre-existing literature on methods for studying the so-called ``compositional data'' coming from relative abundances is (directly) relevant to this problem. (Cf. e.g. \cite{Silverman2017}, \cite{microbiome_time_series}, \cite{McDavid2019}, \cite{Harrison2020}, or the introduction of \cite{Prost2021} for examples of methods for dealing with compositional data.) This is true although previous work for inferring microbial (or genetic) interactions uses working models (superficially) similar\footnote{
E.g. working models related to the Poisson or Dirichlet-Multinomial distributions, or working models using generalized Lotka-Volterra models, cf. \cite{Marino2014}, \cite{signs_longitudinal_compositional}, and \cite{eco_model_time_series}.
} to those discussed herein.

A previous attempt, MINI-drop \cite{Hsu2019}, to study microbial interactions using microfluidic droplets while getting absolute abundance measurements was limited to only $3$ distinct strains. Studies using kChip \cite{kChip} \cite{positive_kChip} were able to investigate larger number of strains (e.g. $20$ in the case of \cite{positive_kChip}), but still nothing vastly surpassing the state of art using cell plating \cite{Venturelli}. The reason for this is the reliance of these methods on fluorescent labeling of cells to get absolute abundance measurements. That means that the total number of strains that can possibly be distinguished, and thus included in the experiment, is limited by the number of distinct possible fluorescent proteins available whose fluorescence can be distinguished by automated imaging techniques. Moreover, it also requires either pre-existing transgenic strains of interest that express (compatible) fluorescent proteins, or genetically modifying de novo the organisms one wants to study with the hope of being able to get them to express fluorescent proteins. (This seems to be why \cite{positive_kChip} only studied strains of \textit{Enterobacterales} and \textit{Pseudomonadales} $\gamma$-Proteobacteria. Such bacteria are closely related to intensively studied ``lab rats''of microbiology, namely \textit{E. Coli} and \textit{Pseudomonas} respectively.) In contrast, MOREI is in principle capable of studying microbial communities with many more strains, while still producing absolute abundance measurements. This is because MOREI does not use automated fluorescence imaging to get absolute abundance measurements\footnote{
Cf. \cite{Taylor2022} for one fluorescence imaging protocol for counting bacteria in microfluidic droplets.
}, and instead uses a spikein gene protocol (cf. \cite{Barlow2020}). Therefore MOREI is not limited to studying transgenic strains, nor does the available number of distinguishable fluorescent proteins limit the number of strains that MOREI can study.

\section{Conclusion}
\label{sec:outline-rest-thesis}

\paragraph{Findings and Contributions}

I explained why high-throughput droplet-based microfluidics technology overcomes previous limitations to answering these questions.

I showed how answering these questions can be recast as the statistical problem of estimating a network with mixed-sign edge weights.

\paragraph{Practical Implications} MOREI needs to increase the number of strains whose interactions can be simultaneously characterized to advance the state of the art (cf. section \ref{sec:how-micr-ecol}). Being able to characterize the interactions of more strains requires characterizing the interactions of strains with low relative abundance (a.k.a. frequency) in the microbial community. For example, if all strains have the same frequency, then simultaneously characterizing the interactions of 20 strains requires characterizing the interactions of strains with $5\%$ relative abundance, while simultaneously characterizing the interactions of $2,000$ strains requires characterizing the interactions of strains with $.05\%$ relative abundance. Thus understanding how to use MOREI to advance the state of the art in microbial ecology requires understanding how to analyze the data produced by MOREI in a way that characterizes the interactions of the least abundant strains as accurately as possible.
 
\paragraph{Next Steps and Open Questions} Given highly accurate simulations of the data-generating process for MOREI, we could compare the performance of different estimators to see which most accurately characterizes the interactions of the least abundant strains. This would determine the methods that are best for analyzing the data produced by MOREI. That in turn would determine how far we can realistically expect to advance the state of the art in microbial ecology using MOREI.

\end{coolcontents}

\chapter{Formulation as a Statistical Problem}
\label{cha:introduction}

Herein I explain that the data from the noisy dynamical systems corresponding to each droplet can be understood as censored observations of a multivariate Markov process. 
See section \ref{sec:likelihoods}.
I identify the statistical understanding of a droplet's initial state as crucial to overcoming the main limitation of these experiments, the uncontrolled assignment of microbes to droplets.
See section \ref{sec:motiv-targ-estim}.
I also give the precise definitions of the targeted parameters that will be the focus of part \ref{part:modell-init-form}.
See sections \ref{sec:defin-thro-intro}, \ref{sec:proj-targ-param-hpomu}, and \ref{sec:proj-targ-param-appendix}.

Section \ref{sec:likelihoods} discusses the statistical model and questions for the entire experiment, when droplets are incubated. This corresponds to all three phases of the data generating process from section \ref{sec:using-morei-char}. Section \ref{sec:t=0-experiment} discusses the statistical model and questions for the important ``$t=0$'' experiment special case, when droplets are \textit{not} incubated. This corresponds only to the first and third phases of the data generating process from section \ref{sec:using-morei-char}.

In this chapter I make no attempt to explicitly model the third phase of the droplet life cycle described in section \ref{sec:using-morei-char} (although see appendix \ref{chap:simul-impl} for some very preliminary thoughts). The problem of choosing good statistical models of PCR amplification is applicable to an extremely wide range of biological experiments. It should be its own project and fully addressing it is outside of the scope of this work. Previous work that has more substantially considered similar problems includes \cite{PCR1} \cite{PCR2} \cite{PCR3} \cite{PCR4}\cite{bias_framework}\cite{Marotz2019}.

\begin{coolcontents}
\section{Background}
\label{sec:backgr-sign-2}

Section \ref{broader-field-2} overviews the general context of related problems across many fields. Section \ref{sec:specific-problem-2} narrows this down to only issues that are directly relevant to analysis of the data produced by MOREI. Finally, section \ref{sec:particular-approach-2} overviews the ideas behind the concrete proposals that will be made further in this chapter and throughout the dissertation.

\subsection{Broader field}
\label{broader-field-2}

The ideas of this chapter belong to the hypothetical broader field of ``\textit{statistical inference of (time-evolving) interactions between variables of a (noisy) dynamical system for which the temporal evolution may be incompletely observed}''.
Cf. the discussion earlier from the introduction to Part \ref{part:introduction}.
  \cite{Golightly2011} appears to describe something very similar as ``\textit{inference for the parameters of complex nonlinear multivariate stochastic process models}''. The ideas in this chapter can be understood as belonging to any of the following general fields:
  \begin{enumerate}[label=(\arabic*)]
  \item The statistical inference of dynamical systems. See for example \cite{Ellner2016}, \cite{Ramsay2017}, and \cite{Wilkinson2018} for introductory monographs. The papers \cite{Ionides2006}, \cite{Golightly2011}, \cite{Sugihara}, \cite{Estrela2022}, \cite{Angulo2017}, and \cite{Lu2011} are examples of important contributions to the field. This field apparently overlaps with both functional data analysis, see \cite{Hooker_Ramsay2007} and \cite{Dubey2021} for examples, and with ecological modelling, see \cite{eco_dynamics1}, \cite{eco_dynamics2}, \cite{eco_dynamics3}, \cite{eco_dynamics_time_series} for examples.
Notice that this field appears to be unrelated\footnote{
\cite{dynamical_system_approach} explains how many problems in multivariate data analysis can be formulated as matrix fitting constrained optimization problems, which in turn can be reformulated into initial value problems for (ordinary) differential equations on the matrix manifolds defined by the constraints of the optimization problem. To clarify, ``matrix manifold'' refers to an abstract manifold whose points can be thought of as elements of a(n abstract) space of matrices. Then the idea is that dynamical systems theory can often be used to better understand the corresponding initial value problem. However, given the connection of the theory of statistical inference of dynamical systems with functional data analysis mentioned above, and the connection between functional data analysis and multivariate data analysis described in e.g. \cite{Ramsay2005}, this may still suggest an \textit{indirect} (and non-trivial) connection between statistical inference of dynamical systems and the ``dynamical system approach to multivariate data analysis''. Of course this still remains to be confirmed.
    } to the ``dynamical system approach to multivariate data analysis'' described in \cite{dynamical_system_approach}. 
    
  \item The study and mathematical theory of random dynamical systems. See \cite{Humphries2002} for a relatively accessible introduction, or \cite{ludwig} for a comprehensive reference. This appears to include much of the theory of stochastic differential equations (SDEs) as a special case, cf. \cite[section 2.3]{ludwig} or \cite{Golightly2011}. Thus I do not discuss modelling and inference with SDEs as its own field below, although one could reasonably do so.
  \item The field of functional data analysis. See any of \cite{fda_review_1}, \cite{fda_review_2}, \cite{Ramsay2009}, \cite{Kokoszka2021}, \cite{Ramsay2005}, \cite{Horvath2012}, \cite{Hsing2015}, or \cite{Grenander1981} for introductions of varying levels of difficulty. As mentioned above, this apparently overlaps with the statistical inference of dynamical systems, with \cite{Hooker_Ramsay2007} and \cite{Dubey2021} as examples. The overlap of this field with manifold learning may be particularly important for multivariate data like this problem, perhaps because the phase space of the dynamical system is a (random) manifold. See \cite{Chiou2013}, \cite{Chen2012}, and \cite{Chen2018} for examples.
  \item \label{item:gene_gene_interactions} The study of variable interactions, or the inference of interaction networks. See \cite{ruan_interactions}, \cite{Angulo2017}, or even \cite{Mani2008} for examples of the kinds of studies that I intend to describe. In regression terminology, in this context all of the response variables are also predictor variables for each other, so given $n$ (response=predictor) variables, there are potentially $O(n^k)$ $k$-th order interactions. In the terminology of genetics studies, this corresponds only to ``gene $\times$ gene'' interactions (pairwise interactions grow \textit{quadratically} with the number of genes), but \textit{not} to e.g. ``gene $\times$ environment'' interactions (pairwise interactions grow only \textit{linearly} with the number of genes). The former, ``response=predictor'' or ``gene $\times$ gene'', interactions problem is much more difficult than the latter problem, even if only for combinatorial reasons\footnote{ \label{footnote:n_instances}
Let $n$ be the number of genes. Then a ``gene $\times$ gene'' interactions problem amounts to $n$ instances of ``(one) gene $\times$ (other genes)'', or ``gene $\times$ environment'', interactions problems. For each of the $n$ genes, construe the $(n-1)$ ``other genes'' as the ``environment'' to get a ``gene $\times$ environment'' interactions problem. Cf. \cite{Torrisi2020} for other ideas about relating ``gene $\times$ environment'' and ``gene $\times$ gene'' problems.
}. Unfortunately, this more difficult ``gene $\times$ gene'' interactions problem also seems to be less frequently studied than the easier ``gene $\time$ environment'' interactions problem. In the terminology from the Part \ref{part:introduction} introduction, ``gene $\times$ gene'' interactions are the same as ``unpartitioned interactions'', whereas ``gene $\times$ environment'' interactions are the same as ``bipartite interactions''. The variables of the ``environment'' can also be described as ``features'', such that the ``gene $\times$ environment'' interactions problem also corresponds to the ``feature enrichment'' or ``differential analysis'' problem commonly studied with high-throughput biological assays, see \cite{Ge2021} for a discussion.

Gene regulatory networks \cite{Weighill2021} \cite{Calderer2021} \cite{MacKay2020} \cite{Torrisi2020} \cite{Lu2011} are a common example of the ``gene $\times$ environment'' interactions problem, with transcription factors corresponding to the ``environmental variable nodes''. The preprint \cite{ruan_interactions} suggests \cite{interaction_regression_1} or \cite{interaction_regression_2} as recommendations for papers on regression with large numbers of (interaction) covariates, and any of \cite{gene_gene_interactions_1}, \cite{gene_gene_interactions_2}, or \cite{gene_gene_interactions_3} as recommendations for papers studying the inference of ``gene $\times$ gene'' interactions. The paper \cite{Lu2011}, which takes a dynamical systems approach, also appears to be related. See also \cite{McDavid2019} vis a vis the ``gene $\times$ gene'' interactions problem.
 
  \item The study of pseudotemporal, or ``pseudotemporally ordered'', cross-sectional data\footnote{
Described in \cite{phenopath} as datasets where ``individuals in the cross-sectional cohort behave asynchronously and each is at a different stage of progression''. This should (in some sense) apply to the data produced by this experiment due to the use of batches incubated for different amounts of time. Cf. section \ref{sec:using-morei-char}.
    }. While no textbook account thereof seems to exist, the references \cite{pseudotime1}, \cite{pseudotime2}, \cite{phenopath}, and \cite{pseudotime3} are all important contributions. Current studies in this field appear to all be conceived within the context of analyzing data produced by high-throughput droplet-based microfluidics experiments.
  \end{enumerate}
In particular, valuable future work would further clarify the connections that these fields have not only with each other, but also with the fields of
  \begin{enumerate}[label=(\roman*)]
  \item the statistical analysis of longitudinal (also known as
    ``time series'') data (for longitudinal studies of similar biological problems, cf. e.g. \cite{eco_dynamics_time_series}, \cite{eco_model_time_series}, \cite{signs_longitudinal_compositional}, \cite{Roodgar2021}, or cf. \cite{microbiome_time_series} for an introduction geared towards microbial ecologists),
  \item  the mathematical theory of Markov (or specifically Feller) stochastic processes (cf. e.g. \cite{Ionides2006}, \cite[chapter 6]{Grenander1981}, \cite[appendix A.4]{ludwig}, or \cite[section 1.6]{Kloeden} for discussion of how this mathematical theory is relevant to several of the aforementioned fields),
  \item  (macro)ecological modelling studies (for relevant examples compare e.g. any of the following papers: \cite{higher_order}, \cite{Sugihara}, \cite{monte_carlo_chisquare_stat}, \cite{eco_model_time_series}, \cite{Gonze2018}, \cite{Momeni2017}, \cite{Erez2020}, \cite{Goldford2018}, \cite{Estrela2021}, \cite{Estrela2022}, \cite{Goyal2021},
    \cite{Marino2014}, \cite{Lewis2010}, \cite{eco_dynamics1}, \cite{eco_dynamics2}, \cite{eco_dynamics3}, \cite{eco_dynamics_time_series}, \cite{Golightly2011}, \cite{Browning2021}, \cite{MacKay2020}), and
  \item ``repeated cross-sectional'' or ``pseudo-longitudinal'' epidemiological studies.
\end{enumerate}
Some amount of ``synthesis'' or ``unification'' of the relevant ideas from \textit{all} of these fields would probably lead to a substantial increase in the breadth and sophistication of the techniques available for many problems. Nevertheless, it would be difficult to accomplish.

From this more general context, chapter \ref{cha:introduction} focuses on the important issue of statistical inference of (time-dependent) microbial interactions, using either longitudinal or pseudo-longitudinal data.
The interacting variables are the populations of microbes.
See \cite{umibato} for one example of this type of problem, and \cite{Gonze2018} for a helpful review of related work.

\subsection{Specific problem}
\label{sec:specific-problem-2}

Herein I investigate how to approach the data produced by MOREI using statistical models.
We want an explicitly specified statistical model to be the basis of our statistical analyses.
This allows us to keep track of which assumptions we are making and how realistic they are.
Previous work has implicitly (but not explicitly) specified longitudinal statistical models for inferring microbial interactions that amount to discrete-time Markov processes. See any of 
\cite{Fisher2014} \cite{Mounier2008} \cite{Venturelli} \cite{Marino2014}
  for examples. However, explicitly specifying something like this would not be enough for MOREI because in this case the data is only pseudo-longitudinal.
As mentioned in the review \cite{Gonze2018}, most previous work using dynamical system models does not claim the ability to infer microbial interactions without access to (strictly) longitudinal data. The paper \cite{Xiao2017} is a noteworthy exception. The paper begins by making strong differentiability assumptions about the population dynamics. Based on these assumptions, it then claims that it should be possible to infer ecological interactions from entirely cross-sectional data, not even necessarily pseudo-longitudinal, provided the dynamical system is always observed in a steady state or local equilibrium. Similarly, the pseudotime method from \cite{phenopath} also claims to be able to make similar inferences (cf. the discussion from \ref{item:gene_gene_interactions} of \ref{broader-field-2} above) from entirely cross-sectional data.

  The assumptions required by \cite{Xiao2017} are questionable within the context of the data produced by MOREI. For example, the use of a pseudo-longitudinal data structure (from making multiple batches) already implies that we do not generally believe the equilibrium assumptions to be valid for most droplets. Likewise the method from \cite{phenopath} also fails to exploit the pseudo-longitudinal structure of the MOREI data. Moreover, the proposed algorithms from \cite{Xiao2017} and \cite{phenopath} both do not exploit the sparsity of the MOREI data, and both fail to scale well with the number of observations (which for MOREI is massive). These issues lead to prohibitively severe performance problems when applied to MOREI data.
Given these problems, I decided to not attempt identifying and describing explicit models used implicitly in \cite{Xiao2017} and \cite{phenopath}. The models would be unrealistic, and the inference methods they would suggest would not actually be useful for MOREI data.

 Interestingly, \cite{Sikorski2017} claims to have developed a method based on the generalized Lotka-Volterra equations, analogous to e.g. \cite{Fisher2014}, yet which also claims to work on cross-sectional data. \textit{However}, the claim does not seem to be that they can make inferences about microbial interactions based on pseudo-longitudinal data. Instead, \cite{Sikorski2017} seems to essentially require cross-sectional data that is analogous to longitudinal data but with some other quantity as the ``independent variable'' besides time:
  \begin{quote}
 series of cross-sectional data which originate from samples along gradients in environmental parameters but do not have a temporal dimension
  \end{quote}
  This does not characterize the data produced by MOREI any more than longitudinality does. Hence the methodology of \cite{Sikorski2017} also seems to not be applicable to this problem. Nevertheless it does seem to warrant closer comparison in future work.

\subsection{Particular approach}
\label{sec:particular-approach-2}

This chapter characterizes the droplet life cycle using a discrete-time Markov process whose initial state is in $\N^{\Strains}$ but which thereafter can take values in $\R^{\Strains}$ (cf. equation (\ref{eq:simulation_update}) from appendix \ref{chap:simul-impl}). This choice of model can be thought of as ``implicitly'' discretizing a continuous-time Markov process that is the ``true model''. 

The idea of modelling a system similar to this as a Markov process observed at discrete time points has precedent in \cite{Ionides2006}, as well as in references [24, 25, 27, 30, 32] of \cite{Golightly2011}.
Cf. also \cite[Part IV]{Kloeden}, which explains and discusses procedures for discrete-time approximations of stochastic differential equation solutions.
Similarly, precedent for applying models that are essentially stochastic differential equations (SDEs) to Lotka-Volterra-like systems and to microbial (prokaryotic) population dynamics can be found in \cite{Golightly2011} itself.
Cf. also \cite[section 7.1]{Kloeden}, which discusses applying SDEs to model population dynamics.
The paper \cite{Browning2021} borrows tools from mathematical finance to investigate SDE models of bacterial growth that include ``hedging'' via persisters (slow-growing but resilient cells).

However, much previous work, such as \cite{Venturelli} \cite{Fisher2014} \cite{Marino2014} \cite{eco_model_time_series} \cite{Mounier2008}, implicitly\footnote{The references do not explicitly state the underlying statistical models, which would be needed to precisely define or justify the regression procedures that they propose.} used discrete-time Markov processes. So it is left to future work to fully consider e.g. \cite[chapter 6]{Kloeden}, which discusses the applications of stochastic differential equations\footnote{Which correspond to special kinds of continuous-time Markov processes.} to modelling problems, in particular \cite[section 6.2]{Kloeden} that discusses the continuous-time ``diffusion limits'' of discrete-time Markov processes. For example, it is doubtful that equation (\ref{eq:simulation_update_SDE_version}) from appendix \ref{chap:simul-impl} is actually the appropriate diffusion limit of the discrete-time Markov processes used in previous work. A starting point might be \cite[section 2.1]{ludwig}, which discusses how discrete-time random dynamical systems can be derived from processes like the discrete-time Markov process models from previous work, and vice versa.

The assumption of a discrete-time Markov process is made not only for the purpose of facilitating comparisons with prior work, but also for the purpose of model simplicity. In particular, there is the issue of the distinction between (using the terminology of \cite{ludwig}) ``random differential equations'' \cite[section 2.2]{ludwig} and ``stochastic differential equations'' \cite[section 2.3]{ludwig}. Specifically, even if there might only be one standard way, via the diffusion limit, to go from a discrete-time Markov process to a continuous-time Markov process, there appears to be at least two standard ways to go from a (deterministic) differential equation to a ``non-deterministic'' differential equation, corresponding to ``random differential equations'' and ``stochastic differential equations'' respectively. The discrete-time Markov process models implicit in \cite{Venturelli} \cite{Fisher2014} \cite{Marino2014} \cite{eco_model_time_series}\cite{Mounier2008} were all derived by discretizing and then ``stochasticizing'' a (deterministic) differential equation\footnote{
Cf. the procedure used for deriving equation (\ref{eq:simulation_update}) in the appendix section \ref{sec:modell-cell-counts}.
}. Hence we can ask whether, when extending these models to continuous-time, it might be better to ``skip the intermediary'' and seek to instead directly ``stochasticize'' the (deterministic) differential equation. If we believe it is better to do so, then it becomes incumbent upon us to clearly articulate a reason for a preference of a ``random differential equation'' model versus a ``stochastic differential equation'' model. To ensure that all of this is done correctly and well requires considerable mathematical sophistication and hence is left to future work.

At the same time, some choice of a continuous-time Markov process model (cf. equation (\ref{eq:simulation_update_SDE_version}) from appendix \ref{chap:simul-impl}) would still probably be more accurate or realistic than the current discrete-time model. Another benefit of using a continuous-time model would be that time periods of unequal length between the batches' incubation times would not complicate the interpretation of such a continuous-time model, cf. the corresponding argument made in \cite{umibato}. The references \cite{Ionides2006} and \cite{Golightly2011} also seem relevant to this question.

There is precedent for using continuous-time stochastic processes to model the growth of bacterial populations in microfluidics droplets. See \cite{Taylor2022} which claims to find empirical support for such a model using automated fluorescence imaging, elaborating on prior work \cite{Barizien2019} that generalized the ``classical'' Bellman-Harris branching process to apply to the growth of bacteria inside of microfluidic droplets. The Bellman-Harris process (as well as its generalizations in \cite{Barizien2019} and \cite{Taylor2022}) is a continuous-time stochastic process with \textit{discrete} state space, i.e. taking values in $\mathbb{N}$. Contrast this with an SDE model, which is continuous-time with continuous state space, and the model proposed below in section \ref{sec:likelihoods}, where initial states are required to be discrete but subsequent states are allowed to be ``pseudocounts'' i.e. continuous. Note that both the models in \cite{Taylor2022} and \cite{Barizien2019} are explicitly designed for the count of cells of a single, non-interacting strain, i.e. they are single-type branching processes. Hence future work trying to apply similar models to MOREI would need to begin by identifying the generalization of these models as multi-type branching processes. Cf. \cite{Athreya1972} for a reference about multi-type branching processes.

While this model was intended to describe MOREI specifically, in principle it should not be too difficult to modify in a way that makes it applicable to other ``co-culture'' systems as well. (See \cite{coculture} for a perhaps somewhat dated review of examples of ``co-culture'' systems for studying cellular interactions.) These modified models might even describe their corresponding systems more accurately than this model describes MOREI. For example, the assumption that different experimental replicates (in the case of MOREI, droplets) are statistically independent is probably more realistic for other systems.

\section{``Full'' Experiment}
\label{sec:likelihoods}

Below I propose a general form for likelihoods characterizing the first and second phases of the droplet life cycle that were described in section \ref{sec:using-morei-char}. This corresponds to a discrete time Markov process whose initial state is in $\N^{\Strains}$ but which thereafter can take values in $\R^{\Strains}$.

\subsection{Preliminaries}
\label{sec:descr-likel}

Let ``$\Strains$'' for ``\textbf{s}trains'' denote the total number of microbial strains\footnote{Herein I use ``strains'' to refer equally to strains belonging to the same species(/genus/family/etc.) as well as to strains belonging to different species(/genera/families/etc.), because the distinction is irrelevant for setting up the abstract problem. It may matter for the implementation of a specific experiment.} in the microbial community sample. For each strain $\strain$, let $\freq^{(\strain)}$ (``$\freq$'' for ``\textbf{f}requency'') denote the relative abundance of strain $\strain$ in the sampling pool. In particular, we necessarily have that $\sum_{\strain=1}^{\Strains} \freq^{(\strain)} = 1$. 

Given a positive integer $N \in \mathbb{N}$, define $[N]:= \{1, \dots, n , \dots, N\}$. In general, the entries of a (non-negative) vector $\rvec{v} \in \R^S$ are denoted $v^{(\strain)}$, i.e. 
  \begin{equation}
    \label{eq:vector_notation_likelihoods}
    \rvec{v} \overset{def}{=} (v^{(1)}, \dots, v^{(\strain)}, \dots, v^{(\Strains)}) \,.
  \end{equation}
The shorthand ${v := \sum_{\strain=1}^{\Strains} v^{(\strain)}}$ is used for the sum of the entries of $\rvec{v}$. 

For a given droplet $\droplet$ (``$\droplet$'' for ``\textbf{d}roplet''), the random variable $\abundance_{\droplet}(0)$ is the number of cells from the population encapsulated in the droplet $\droplet$ during its formation (``$\abundance$'' for ``\textbf{n}umber''). Equivalently, $\abundance_{\droplet}(0)$ is the number of cells in droplet $\droplet$ at time $0$. This is the sum of the entries of the random vector:
  \begin{equation}
    \label{eq:total_number_cells_is_sum_likelihoods}
    \vabundance_{\droplet}(0) := \left( \abundance[1]_{\droplet}(0), \dots, \abundance[\specie]_{\droplet}(0), \dots, \abundance[\Species]_{\droplet}(0) \right) \,,
  \end{equation}
where for each strain $\specie$ (``$\strain$'' for ``\textbf{s}train'') of the $\Species$ total strains, the random variable $\abundance[\specie]_{\droplet}(0)$ is the number of cells from strain $\specie$ in droplet $\droplet$ at time $0$. 

For the sake of brevity, often the subscript $\droplet$ is omitted, i.e. $\abundance(0)$, $\vabundance(0)$, and $\abundance[\strain](0)$ are used in place of $\abundance_{\droplet}(0)$, $\vabundance_{\droplet}(0)$, and $\abundance[\strain]_{\droplet}(0)$, respectively. This highlights how the random variables corresponding to the droplets are assumed (for the sake of simplicity) to be identically distributed. Therefore the corresponding statements should not depend on the specific droplet $\droplet$, making it unnecessary or even misleading to specify $\droplet$. Cf. figure \ref{fig:count_categorical}.

For simplicity, the $\vabundance_{\droplet}(0)$ are assumed to not only be identically distributed, but also mutually statistically independent. This becomes relevant in sections \ref{sec:likelihood-batch} and \ref{sec:likelihood-experiment} below, and is implicit throughout everything that follows, e.g. the derivations in chapter \ref{chap:hetero_estimator_performance}. In practice this is probably unrealistic.

Let ``$\Batches$'' for ``\textbf{b}atches'' denote the total number of batches. For any given batch ${\batch \in [ \Batches]}$, let $\Droplets_{\batch} \in \mathbb{N}$ for ``\textbf{d}roplets'' denote the total number of microfluidic droplets belonging to the batch. With current droplet microfluidics technology, $D_{\batch}$ can be on the order of $10^6$ or greater. Whether one assumes for simplicity that ${\Droplets_1 = \cdots = \Droplets_B}$ does not substantially affect the analysis. Droplets in different batches are measured at different times, and for the entire experiment the total number of droplets, denoted $\Droplets$, is
\begin{equation}
  \label{total-droplets-notation}
  \Droplets := \sum_{\batch = 1}^{\Batches} {\Droplets}_{\batch} = \Batches \bar{\Droplets} \,, 
\end{equation}
where $\bar{\Droplets}$ denotes the arithmetic mean of the $\Droplets_{\batch}$. Compare Figure \ref{fig:droplets_batches}.

 During the second phase that was described in section \ref{sec:using-morei-char}, the cells of all $\Droplets_{\batch}$ droplets in a given batch $\batch$, for any $\batch \in [\Batches]$, are incubated until time $\time_{\batch}$ has elapsed (``$\time$'' for ``\textbf{t}ime''), until the droplets are measured (which begins the third phase). The incubation time $\time_{\batch}$ for any given batch $\batch$ can be subdivided into $\batch$ time periods, corresponding to ${0=:\time_0 < \time_1 < \cdots < \time_{\batch[]} < \cdots < \time_{\batch}}$, the incubation times for all batches that are not incubated longer than batch $\batch$. Again, compare Figure \ref{fig:droplets_batches}. Each time period does not necessarily need to have the same length, although that could simplify interpretation of the analysis. 

\subsubsection{Observables and Format of Data}
\label{sec:observ-form-data}

After the end of the experiment, for each batch $\batch \in [\Batches]$ what we observe are (roughly\footnote{Cf. appendix \ref{chap:simul-impl} for some distortions that the third phase from section \ref{sec:using-morei-char} causes.}) the $\Droplets_{\batch}$ vectors $\vabundance_{\droplet}(\time_{\batch}) \in \R^{\Strains}$, one for each $\droplet \in [\Droplets_{\batch}]$. These data of course can be formatted into a $\Strains \times \Droplets_{\batch}$ or $\Droplets_{\batch} \times \Strains$ matrix.

\subsection{Likelihood of a Single Droplet}
\label{sec:likelihood-droplet}

For a given droplet, at time $\time = 0$ the unobserved likelihood $\N^{\Strains} \to [0,1]$ for the initial strain distribution vector, i.e. the random vector $\vabundance(0) \in \N^{\Strains}$ giving the initial counts of the number of living cells in the droplet from each of the $\Strains$ strains in the population, is for any $\vcounts \in \N^{\Strains}$ denoted
\begin{equation}
  \label{eq:droplet_initial}
  \probability*{\vabundance(0) = \vcounts} \,,
\end{equation}
where $\vcounts =: (\counts[1], \dots, \counts[\Strains])$ (such that for all $\strain \in [\Strains]$ one has $\counts[\strain] \in \N$). This corresponds to the first phase that was described in section \ref{sec:using-morei-char}.

Various possible likelihood models specifying ${\probability*{\vabundance(0) = \vcounts} }$ are derived and described in chapter \ref{cha:model-init-form}. Chapter \ref{chap:model_comparison} compares these various models. Chapter \ref{chap:data_throughput} discusses the different ``downstream'' impact that each of these models has on the amount of data available for inferring interactions. Finally, chapter \ref{chap:hetero_estimator_performance} discusses how we might go about choosing which of these models is most realistic based on observations of $\vabundance(0)$ (which can be accomplished by not incubating the droplets and thereby skipping the second phase of the droplet life cycle that was discussed above in section \ref{sec:growth-cells-inside}).

The unobserved full likelihood ${\N^{\Strains} \bigtimes \left( \prod_{\batch[]=1}^{\batch} \R^{\Strains} \right) \to [0,1]}$ for a given droplet in batch $\batch$ over the entire duration of the growth of its cells is assumed to correspond to a discrete time Markov process. In other words the unobserved full likelihood for a single droplet of a given batch $\batch$ is assumed to be of the form:
\begin{equation}
\label{eq:droplet_full}
  \begin{split}
    & \likelihood*{\vabundance(0) = \vcounts(0), \vabundance(\time_1)=\vcounts(\time_1), \dots, \vabundance(\time_{\batch}) = \vcounts(\time_{\batch})} \\
    =& \left[ \prod_{\batch[]=1}^{\batch} \likelihood*{
        \vabundance(\time_{\batch[]}) = \vcounts(\time_{\batch[]}) |
        \vabundance(\time_{\batch[]\!-\!1}) =
        \vcounts(\time_{\batch[]\!-\!1}) } \right] \cdot \left[
      \probability*{\vabundance(0) = \vcounts(0) } \right]\,,
  \end{split}
\end{equation}
where $\vcounts(\time_{\batch[]}) \in \R^{\Strains}$ for all $\batch[] \in [\batch]$, $\vcounts(0) \in \N^{\Strains}$, and $\vcounts(0) =:(\counts[1](0), \dots, \counts[\Strains](0))$. The transition likelihoods $\likelihoodsymbol$ from (\ref{eq:droplet_full}) correspond to the second phase that was described in section \ref{sec:using-morei-char}. Non-integer values of the {(pseudo-)counts} ${\counts[\strain](\batch[])\in\R}$ can be interpreted to correspond to e.g. incomplete cell division.

Since the contents of the droplets in a given batch $\batch$ are only observed at time $\time_{\batch}$ (when the cells inside are lysed and their genetic material sequenced), starting from (\ref{eq:droplet_full}) it follows that the \textit{observed} likelihood for a single droplet of batch $\batch$ is of the form:
\begin{equation}
  \label{eq:droplet_observed}
  \begin{split}
    &  \likelihood*{\vabundance(\time_{\batch}) = \vcounts(\time_{\batch})  } \\
    =& \int_{\R^{\Strains}} \cdots \int_{\R^{\Strains}} \left[
      \sum_{\vdcounts(0) \in \N^{\Strains}}
      \likelihood*{\vabundance(\time_{\batch}) =
        \vcounts(\time_{\batch}) | \vabundance(\time_{\batch\! -\! 1})
        = \vdcounts(\time_{\batch\! -\! 1}) } \cdot
    \right.\\
    &\left[ \prod_{\batch[]=1}^{\batch - 1} \likelihood*{
        \vabundance(\time_{\batch[]}) = \vdcounts(\time_{\batch[]}) |
        \vabundance(\time_{\batch[]\!-\!1}) =
        \vdcounts(\time_{\batch[]\! -\! 1}) }
    \right] \cdot 
\left. \vphantom{\sum_{\vdcounts(0) \in \N^{\Strains}}} \left[
        \probability*{\vabundance(0) = \vdcounts(0)} \right] \right]
    \mathrm{d}\vdcounts(\time_1) \cdots
    \mathrm{d}\vdcounts(\time_{\batch\! -\! 1}) \,,
  \end{split}
\end{equation}
where $\vdcounts (0) \in \N^{\Strains}$ and $\vdcounts(\time_{\batch[]}) =: (\dcounts[1](\time_{\batch[]}), \dots, \dcounts[\strain](\time_{\batch[]}), \dots, \dcounts[\Strains](\time_{\batch[]}) )\in \R^{\Strains} $ for all ${\batch[] \in [\batch\!-\!1]}$ denote dummy variables, corresponding to the values from the full likelihood (\ref{eq:droplet_full}) that are unobserved and which therefore need to be marginalized out. Also, $\int_{\R^{\Strains}} \cdots \mathrm{d}\vdcounts(\time_{\batch[]})$ is used as shorthand for the multiple integral $\int_{\R} \cdots \int_{\R} \cdots \mathrm{d}\dcounts[1](\time_{\batch[]}) \cdots \mathrm{d}\dcounts[\Strains](\time_{\batch[]})$ for all $\batch[] \in [\batch-1]$.

\subsection{Likelihood of a Single Batch}
\label{sec:likelihood-batch}

Letting $\lvabundance{\droplet}{\batch}(\time)$ denote the strain distribution vector of the $\droplet$'th droplet of batch $\batch$ at time $\time$, and $\labundance{\droplet}{\batch}{\strain}(\time)$ denote the number of living cells of strain $\strain$ in the $\droplet$'th droplet of batch $\batch$ at time $\time$, i.e.
\begin{equation}
  \label{eq:droplet_abundance_definition}
  \lvabundance{\droplet}{\batch}(\time) := (\labundance{\droplet}{\batch}{1}(\time), \dots, \labundance{\droplet}{\batch}{\strain}(\time), \dots, \labundance{\droplet}{\batch}{\Strains}(\time)) \in \R^{\Strains} \,,
\end{equation}
for all $\droplet \in [\Droplets_{\batch}]$, $\strain \in [\Strains]$, and $\time \ge 0$, then the unobserved full likelihood for (all of the droplets of) a given batch $\batch$ is of the form:
\begin{equation}
  \label{eq:batch_full}
  \begin{split}
    &  \likelihoodsymbol  \left(\lvabundance{1}{\batch}(0) = \lvcounts{1}{\batch}(0), \dots, \lvabundance{1}{\batch}(\time_{\batch}) = \lvcounts{1}{\batch}(\time_{\batch}), \lvabundance{2}{\batch}(0)=\lvcounts{2}{\batch}(0), \dots, 
\right. \\
    & \left.  
\hphantom{\likelihoodsymbol\likelihoodsymbol}
\lvabundance{\Droplets_{\batch}}{\batch}(0) = \lvcounts{\Droplets_{\batch}}{\batch}(0), \dots, \lvabundance{\Droplets_{\batch}}{\batch}(\time_{\batch}) = \lvcounts{\Droplets_{\batch}}{\batch}(\time_{\batch}) \right)\\
    =& \prod_{\droplet=1}^{\Droplets_{\batch}} \left( \left[
        \prod_{\batch[]=1}^{\batch}
        \likelihood*{\lvabundance{\droplet}{\batch}(\time_{\batch[]})
          = \lvcounts{\droplet}{\batch}(\time_{\batch[]}) |
          \lvabundance{\droplet}{\batch}(\time_{\batch[]\!-\!1}) =
          \lvcounts{\droplet}{\batch}(\time_{\batch[]\! -\! 1}) }
      \right] \cdot
      \left[ \probability*{\lvabundance{\droplet}{\batch}(0) = \lvcounts{\droplet}{\batch}(0)  } \right] \right)
    \,,
  \end{split}
\end{equation}
where $\lvcounts{\droplet}{\batch}(\time_{\batch[]}) =: (\lcounts{\droplet}{\batch}{1}(\time_{\batch[]}), \dots, \lcounts{\droplet}{\batch}{\strain}(\time_{\batch[]}), \dots, \lcounts{\droplet}{\batch}{\Strains}(\time_{\batch[]})) \in \R^{\Strains}$ for all $\batch[] \in \{0\} \cup [\batch]$ and all $\droplet \in [ \Droplets_{\batch}]$. Extending equation (\ref{eq:droplet_observed}) analogously to how equation (\ref{eq:droplet_full}) was extended to equation (\ref{eq:batch_full}), the observed likelihood for (all of the droplets of) a given batch $\batch$ is assumed to have the following form:
\begin{equation}
  \label{eq:batch_observed}
  \begin{split}
    &  \likelihood*{\lvabundance{1}{\batch}(\time_{\batch})=\lvcounts{1}{\batch}(\time_{\batch}), \lvabundance{2}{\batch}(\time_{\batch}) = \lvcounts{2}{\batch}(\time_{\batch}), \dots, \lvabundance{\Droplets_{\batch}}{\batch}(\time_{\batch}) = \lvcounts{\Droplets_{\batch}}{\batch}(\time_{\batch}) } \\
     = & \prod_{\droplet=1}^{\Droplets_{\batch}} \left(
      \int_{\R^{\Strains}} \cdots \int_{\R^{\Strains}} \left[
        \sum_{\lvdcounts{\droplet}{\batch}(0) \in \N^{\Strains}}
        \likelihood*{\lvabundance{\droplet}{\batch}(\time_{\batch}) =
          \lvcounts{\droplet}{\batch}(\time_{\batch}) |
          \lvabundance{\droplet}{\batch}(\time_{\batch\! -\! 1}) =
          \lvdcounts{\droplet}{\batch}(\time_{\batch\! -\! 1}) } \cdot \right.
    \right. \\
    & \left[ \prod_{\batch[]=1}^{\batch-1}
      \likelihood*{\lvabundance{\droplet}{\batch}(\time_{\batch[]}) =
        \lvdcounts{\droplet}{\batch}(\time_{\batch[]}) |
        \lvabundance{\droplet}{\batch}(\time_{\batch[]\!-\!1}) =
        \lvdcounts{\droplet}{\batch}(\time_{\batch[]\! -\! 1}) } \right] 
\! \cdot \!
\left[ \probability*{\lvabundance{\droplet}{\batch}(0) = \lvdcounts{\droplet}{\batch}(0)} \right] 
\left.
      \vphantom{\left[ \sum_{\lvdcounts{\droplet}{\batch}(0) \in \N^{\Strains}}
          \likelihood*{\lvabundance{\droplet}{\batch}} \right]}
    \right]
\\ &
\hphantom{
\prod_{\droplet=1}^{\Droplets_{\batch}} \left( \right.
\int_{\R^{\Strains}} \cdots \int_{\R^{\Strains}} 
}
\mathrm{d}\lvdcounts{\droplet}{\batch}(\time_1) \cdots \mathrm{d}
    \lvdcounts{\droplet}{\batch}(\time_{\batch\! -\! 1}) \left.  \vphantom{
        \prod_{\droplet=1}^{\Droplets} \left[ \sum_{\lvdcounts{\droplet}{\batch}(0) \in
            \N^{\Strains}}
          \likelihood*{\lvabundance{\droplet}{\batch}} \right] }
    \right) \,.
  \end{split}
\end{equation}

I claim it is valid, going from (\ref{eq:batch_full}) to (\ref{eq:batch_observed}), to put the multiple integrals (and sum\footnote{Which from a measure-theoretic point of view is a special case of an integral.}) inside the product $\prod_{\droplet=1}^{\Droplets}$ (thereby changing the order of integration) because the integrands (the likelihood factors) are non-negative, and because the values of the multiple integral $\int_{\R^{\Strains}} \cdots \int_{\R^{\Strains}}  \sum_{\lvdcounts{\droplet_1}{\batch}(0) \in \N^{\Strains}} \cdots \mathrm{d}\lvdcounts{\droplet_1}{\batch}(\time_1) \cdots \mathrm{d}  \lvdcounts{\droplet_1}{\batch}(\time_{\batch\! -\! 1}) $ does not depend on the values of the dummy variables $\lvdcounts{\droplet_2}{\batch}(0) , \lvdcounts{\droplet_2}{\batch}(\time_1), \dots, \lvdcounts{\droplet_2}{\batch}(\time_{\batch\! - \! 1})$ whenever $\droplet_1 \not= \droplet_2$. (The integrations ``can be performed separately for each droplet and therefore distributed across the product''.)

\subsection{Likelihood of Entire Experiment}
\label{sec:likelihood-experiment}

For each given $\batch \in [\Batches]$ and  each $\droplet \in [\Droplets_{\batch}]$, the $\droplet$'th droplet of batch $\batch$ and its strain distribution vector at time $\time$:
\begin{equation}
  \label{experiment-droplet-vector-notation}
  \lvabundance{\droplet}{\batch}(\time) := \vabundance_{\droplet + \sum_{\batch[]=1}^{\batch - 1} \Droplets_{\batch[]} } (\time) \,,
\end{equation}
are the same as the $\displaystyle{\left[\droplet + \sum_{\batch[]=1}^{\batch - 1} \Droplets_{\batch[]} \right]}$'th droplet of the entire experiment and its strain distribution vector at time $\time$. Consequently, for every $\batch \in [\Batches], \droplet \in [\Droplets_{\batch}]$, and $\strain \in [\Strains]$:
\begin{equation}
  \label{experiment-droplet-scalar-notation}
  \labundance{\droplet}{\batch}{\strain}(\time) := \abundance[\strain]_{\droplet + \sum_{\batch[]=1}^{\batch - 1} \Droplets_{\batch[]} } (\time) \,,
\end{equation}
is the number of living strain $\strain$ cells in the $\displaystyle \left[\droplet + \sum_{\batch[]=1}^{\batch - 1} \Droplets_{\batch[]} \right]$'th droplet at time $\time$. 

Even though it may be unrealistic to assume that the distributions of individual droplets within a batch are statistically independent, it is definitely realistic to assume that the distributions for distinct batches are statistically independent. Assuming the distributions for distinct batches are statistically independent motivates the derivation below.

Because we assume all droplets are sampled independently, and that they have the same distribution at any common given time, it follows that the unobserved full likelihood for the entire experiment must have the form:
\begin{equation}
  \label{eq:experiment_full}
  \begin{adjustbox}{max width=\textwidth,keepaspectratio}
 $ \begin{split}
    & \likelihoodsymbol \left( \lvabundance{1}{1}(0) =
      \lvcounts{1}{1}(0), \dots, \lvabundance{\Droplets_1}{1}(\time_1)
      = \lvcounts{\Droplets_1}{1}(\time_1), \lvabundance{1}{2}(0) =
      \lvcounts{1}{2}(0), \dots,
    \right. \\
    & 
\hphantom{\likelihoodsymbol\likelihoodsymbol}
\lvabundance{1}{\batch}(0)=\lvcounts{1}{\batch}(0), \dots,
    \lvabundance{\Droplets_{\batch}}{\batch}(\time_{\batch}) =
    \lvcounts{\Droplets_{\batch}}{\batch}(\time_{\batch}),
    \lvabundance{1}{\batch+1}(0) = \lvcounts{1}{\batch+1}(0),
    \\
    & \left.  \vphantom{\lvabundance{1}{1}}
\hphantom{\likelihoodsymbol\likelihoodsymbol}
 \dots,
      \lvabundance{1}{\Batches}(0) = \lvcounts{1}{\Batches}(0), \dots,
      \lvabundance{\Droplets_{\Batches}}{\Batches}(\time_{\Batches}) =
      \lvabundance{\Droplets_{\Batches}}{\Batches}(\time_{\Batches})
    \right) \\
    =& \prod_{\batch=1}^{\Batches} \left(
      \prod_{\droplet=1}^{\Droplets_{\batch}} \left( \left[
          \prod_{\batch[]=1}^{\batch} \likelihood*{
            \lvabundance{\droplet}{\batch}(\time_{\batch[]}) =
            \lvcounts{\droplet}{\batch}(\time_{\batch[]}) |
            \lvabundance{\droplet}{\batch}(\time_{\batch[]\!-\!1}) =
            \lvcounts{\droplet}{\batch}(\time_{\batch[]\!-\!1}) }
        \right] \cdot
        \left[ \probability*{\lvabundance{\droplet}{\batch}(0) = \lvcounts{\droplet}{\batch}(0)}\right]
      \right) \right) \,.
  \end{split} $
\end{adjustbox}
\end{equation}
Thus the observed likelihood for the entire experiment has the form:
\begin{equation}
  \label{eq:experiment_observed}
  \begin{split}
    & \likelihoodsymbol \left( \lvabundance{1}{1}(\time_1) =
      \lvcounts{1}{1}(\time_1), \dots,
      \lvabundance{\Droplets_1}{1}(\time_1) =
      \lvcounts{\Droplets_1}{1}(\time_1), \lvabundance{1}{2}(\time_2)
      = \lvcounts{1}{2}(\time_2), \dots,
    \right. \\
    &
\hphantom{\likelihoodsymbol\likelihoodsymbol}
    \lvabundance{1}{\batch}(\time_{\batch})=\lvcounts{1}{\batch}(\time_{\batch}),
    \dots, \lvabundance{\Droplets_{\batch}}{\batch}(\time_{\batch}) =
    \lvcounts{\Droplets_{\batch}}{\batch}(\time_{\batch}),
    \lvabundance{1}{\batch+1}(\time_{\batch\!+\!1}) =
    \lvcounts{1}{\batch+1}(\time_{\batch\!+\!1}),
    \\
    & \left.  \vphantom{\lvabundance{1}{1}} 
\hphantom{\likelihoodsymbol\likelihoodsymbol}
\dots,
      \lvabundance{1}{\Batches}(\time_{\Batches}) =
      \lvcounts{1}{\Batches}(\time_{\Batches}), \dots,
      \lvabundance{\Droplets_{\Batches}}{\Batches}(\time_{\Batches}) =
      \lvcounts{\Droplets_{\Batches}}{\Batches}(\time_{\Batches})
    \right) \\
    =& \prod_{\batch=1}^{\Batches} \left(
      \prod_{\droplet=1}^{\Droplets_{\batch}} \left(
        \int_{\R^{\Strains}} \cdots \int_{\R^{\Strains}} \left[
          \sum_{\lvdcounts{\droplet}{\batch}(0) \in \N^{\Strains}}
          \likelihood*{\lvabundance{\droplet}{\batch}(\time_{\batch})
            = \lvcounts{\droplet}{\batch}(\time_{\batch}) |
            \lvabundance{\droplet}{\batch}(\time_{\batch\! -\! 1}) =
            \lvdcounts{\droplet}{\batch}(\time_{\batch\! -\! 1})} \cdot \right.  \right.
    \right.
    \\
    & \left[ \prod_{\batch[]=1}^{\batch-1}
      \likelihood*{\lvabundance{\droplet}{\batch}(\time_{\batch[]}) =
        \lvdcounts{\droplet}{\batch}(\time_{\batch[]}) |
        \lvabundance{\droplet}{\batch}(\time_{\batch[]\!-\!1}) =
        \lvdcounts{\droplet}{\batch}(\time_{\batch[]\!-\!1})} \right] \cdot
    \left[ \probability*{\lvabundance{\droplet}{\batch}(0) = \lvdcounts{\droplet}{\batch}(0) } \right] \left.
 \vphantom{\sum_{\lvdcounts{\droplet}{\batch}(0) \in
              \N^{\Strains}}\likelihood*{\lvabundance{\droplet}{\batch}}
          } \right]
\\
&
\left.\left.
 \vphantom{\sum_{\lvdcounts{\droplet}{\batch}(0) \in
              \N^{\Strains}}\likelihood*{\lvabundance{\droplet}{\batch}}
          }
\hphantom{
\prod_{\batch=1}^{\Batches} \left( \right.
      \prod_{\droplet=1}^{\Droplets_{\batch}} \left( \right.
\int_{\R^{\Strains}} \cdots \int_{\R^{\Strains}} 
}
 \mathrm{d}\lvdcounts{\droplet}{\batch}(\time_1) \cdots
        \mathrm{d}\lvdcounts{\droplet}{\batch}(\time_{\batch\! -\! 1})
        \vphantom{\int_{\R^{\Strains}}} \right)
      \vphantom{\prod_{\droplet=1}^{\Droplets}} \right) \,.
  \end{split} 
\end{equation}
Note that, as implied by the above equations, each droplet has its contents measured exactly once, regardless of the batch to which it belongs. In particular the data is not longitudinal in the sense that no droplet is tracked across more than one time point. The batch a droplet belongs to only determines how long its contents are allowed to grow before they are destroyed (killed via cell lysis) in order to measure them. Cf. again section \ref{sec:growth-cells-inside} above.

Going from (\ref{eq:experiment_full}) to (\ref{eq:experiment_observed}), the multiple integrals and sum can be distributed across the product $\prod_{\batch=1}^{\Batches} \prod_{\droplet=1}^{\Droplets_{\batch}}$ for reasons analogous to those for going from (\ref{eq:batch_full}) to (\ref{eq:batch_observed}).

\subsection{Possible Form of the Transition Likelihoods $\likelihoodsymbol$}
\label{sec:form-trans-likel}

Cells within a microfluidic droplet have the opportunity to interact only while the droplet is being incubated, corresponding to the second phase of the experiment described in section \ref{sec:using-morei-char}. Because the transition likelihoods $\likelihoodsymbol$ correspond to the second phase, it is clear that the interactions between microbes must be encoded within the form of the transition likelihoods $\likelihoodsymbol$, regardless of whether that corresponds to a discrete time Markov process or an SDE. The challenge lies in (1) choosing a non-parametric estimand that sensibly quantifies the notion of ``interaction'' regardless of the underlying model\footnote{Allowing us to escape the need to posit any explicit form for the transition likelihoods.}, or (2) choosing a parametric model that is both biologically realistic and that has parameters directly quantifying the interactions. Neither strategy is easy to implement well. All comparable prior work that I know of corresponds to strategy (2), although my opinion is that in the long-term only strategy (1) is likely to have indefinite ``marginal returns''. Cf. \cite{vanderLaan2011} for a discussion of parametric vs. non-parametric models.

For the purposes of facilitating comparisons with prior work where equations similar to the following were either introduced explicitly \cite{Fisher2014} \cite{eco_model_time_series} or implicitly \cite{Venturelli} \cite{Marino2014} \cite{Mounier2008}, the transition likelihoods in equations (\ref{eq:batch_full})(\ref{eq:batch_observed})(\ref{eq:experiment_full})(\ref{eq:experiment_observed}) would most often be assumed below to be implicitly specified by the following relationships for a given droplet in a given batch $\batch$, applicable for all $\batch[] \in [\batch]$ and $\strain \in [\Strains]$:
\begin{equation}
  \label{eq:transitions_complicated}
  \begin{cases}
    \log(\abundance[\strain](\time_{\batch[]})) - \log(\abundance[\strain](\time_{\batch[]\!-\!1})) =
 \baserate + \displaystyle\sum_{\strain[*]=1}^{\Strains} \interaction{\strain[*]}{\strain} \abundance[{\strain[]}] (\time_{\batch[]\! -\! 1}) + \noise & \abundance[\strain](\time_{\batch[]\!-\!1}) > 0 \\
\abundance[\strain](\time_{\batch[]}) = 0 & \abundance[\strain](\time_{\batch[]\!-\!1}) = 0
  \end{cases}
\end{equation}
where the $\noise$ denote i.i.d. random variables with a $\gaussian(0,\noisescale^2)$ distribution ($\noisescale > 0$ being some ``noise scale'' parameter), and the $\baserate$ and $\interaction{\strain[]}{\strain} \in \mathbb{R}$ are parameters to be estimated from the data. Using induction, we can see that the above relationships (\ref{eq:transitions_complicated}) are equivalent to
\begin{equation}
\label{eq:transitions}
  \begin{cases}
    \log(\abundance[\strain](\time_{\batch[]})) - \log(\abundance[\strain](\time_{\batch[]\!-\!1})) =
 \baserate + \displaystyle\sum_{\strain[*]=1}^{\Strains} \interaction{\strain[*]}{\strain} \abundance[{\strain[]}] (\time_{\batch[]\! -\! 1}) + \noise & \abundance[\strain](0) > 0 \\
\abundance[\strain](\time_{\batch[]}) = 0 & \abundance[\strain](0) = 0
  \end{cases} \,, 
\end{equation}
valid for all $\batch[] \in [\batch]$ and all $\strain \in [ \Strains]$. This implies that observations of $\abundance[\strain](\time_{\batch})$ alone should in principle be sufficient to determine whether $\abundance[\strain](0)$ was nonzero, except in cases where one of the strains in the droplet goes ``extinct'' during the course of incubation (but cf. also appendix \ref{chap:simul-impl} as well as the extensive discussion of this in section \ref{sec:ident-observed}).

It is worth noting explicitly that for this model the coefficients $\interaction{\strain_1}{\strain_2}$ encode the microbial interactions. Cf. appendix \ref{chap:relev-diff-equat} for more detail.

The equations (\ref{eq:transitions}) imply the biologically plausible assumption that cells from a given strain cannot spontaneously appear in a droplet if they were not previously present. They are motivated by the goal of creating a discretized and stochastic version of the generalized Lotka-Volterra equations (\ref{eq:glv_equations}) from appendix \ref{chap:relev-diff-equat} in such a way that the coefficients can be estimated from time series\footnote{
See \cite{microbiome_time_series} for an introduction to time series geared towards microbial ecologists.
}/longitudinal data using some form of regression\cite{Venturelli} \cite{Fisher2014} \cite{Marino2014} \cite{eco_model_time_series} \cite{signs_longitudinal_compositional} \cite{Mounier2008}. However, like the generalized Lotka-Volterra equations they also allow indefinite growth under some conditions, which as discussed in section \ref{sec:growth-cells-inside} is unrealistic. Cf. also the discussion in appendices \ref{chap:relev-diff-equat} and \ref{chap:simul-impl}.

\section{``$\time=0$'' experiment}
\label{sec:t=0-experiment}

The ``$\time=0$'' (``sub-'')experiment differs from the ``full'' experiment described in section \ref{sec:likelihoods} in that droplets are not incubated. Cells are not allowed to grow or divide. The droplets' contents are sequenced immediately after formation, skipping the second phase from \ref{sec:using-morei-char}. Section \ref{sec:notation-intro} reviews the relevant notation and observed data. Section \ref{sec:non-parametric-statistical-model} clarifies the statistical model. Section \ref{sec:defin-thro-intro} discusses the original target parameter definitions. Section \ref{sec:why-empir-distr} explains why in practice one most likely has to use working models for this problem. Section \ref{sec:hpomu_definition_main} gives the default working model that an experimentalist would be expected to use, as well as corresponding projected target parameters. Section \ref{sec:failures-hpomu-model-intro} outlines the reasons why we might expect this working model to be inadequate in practice. Section \ref{sec:model-arbitr-comb_intro} defines a generalized working model which mitigates some of these problems, as well as corresponding projected target parameters.

\subsection{Notation and Observed Data}
\label{sec:notation-intro}

Let ``$\Droplets$'' for ``\textbf{d}roplets'' denote the total number of microfluidic droplets formed from a sampling pool of cells, and 
$\Strains$ 
denote
the total number of cell types in the sampling pool. In the motivating microbial ecology example, ``$\Strains$'' for ``\textbf{s}trains'' is the total number of microbial strains in the microbial community sample.

Given a positive integer $N \in \mathbb{N}$, define $[N]:= \{1, \dots, n , \dots, N\}$. Unless specified otherwise, given a vector $\rvec{v} \in \R^S$, its entries are denoted
  \begin{equation}
    \label{eq:vector_notation_intro_initial_state}
    \rvec{v} \overset{def}{=} (v^{(1)}, \dots, v^{(\strain)}, \dots, v^{(\Strains)}) \,.
  \end{equation}
The shorthand ${v := \sum_{\strain=1}^{\Strains} v^{(\strain)}}$ is used for the sum of the entries of $\rvec{v}$. 
Another useful notion is the \textbf{support} of a (non-negative) vector $\rvec{v} \in \R^S$:
  \begin{equation}
    \label{eq:support_definition_intro}
    \support(\rvec{v}) := \{ \strain \in [\Strains] : v^{(\strain)}> 0  \} \subseteq [\Strains] \,.
  \end{equation}

For a given droplet
 $\droplet$ (``$\droplet$'' for ``\textbf{d}roplet'')
, the random variable 
$\abundance_{\droplet}(0)$ 
is the number of cells from the population encapsulated in the droplet
 $\droplet$
 during its formation (``$\abundance$'' for ``\textbf{n}umber''). Equivalently, 
$\abundance_{\droplet}(0)$ 
is the number of cells 
in droplet $\droplet$ 
at ``time $0$''. This is the sum of the entries of the random vector:
  \begin{equation}
    \label{eq:total_number_cells_is_sum}
    \vabundance_{\droplet}(0) := \left( \abundance[1]_{\droplet}(0), \dots, \abundance[\specie]_{\droplet}(0), \dots, \abundance[\Species]_{\droplet}(0) \right) \,,
  \end{equation}
where for each type $\specie$ (``$\strain$'' for ``\textbf{s}train'') of the $\Species$ total types, 
$\abundance[\specie]_{\droplet}(0)$
 is the random variable equalling the number of cells from type $\specie$ 
in droplet $\droplet$
 at time $0$. \textbf{These are the observed data.}
 For the sake of brevity, often the subscript $\droplet$ is omitted, i.e. $\abundance(0)$, $\vabundance(0)$, and $\abundance[\strain](0)$ are used instead. This highlights how the random variables corresponding to the droplets are assumed (for the sake of simplicity) to have \textbf{independent and identical probability distributions}, and thus how most statements should not depend on the specific droplet $\droplet$.

For each cell type $\strain$, let $\freq^{(\strain)}$ (``$\freq$'' for ``\textbf{f}requency'') denote the relative abundance of cell type $\strain$ in the sampling pool. In particular, $\sum_{\strain=1}^{\Strains} \freq^{(\strain)} = 1$.

\subsection{Statistical Motivation for Targeted Estimands}
\label{sec:motiv-targ-estim}

The treatment group $\treatment[\strain_1, \strain_2]$ for the effect of strain $\strain_1$ on strain $\strain_2$ consists of droplets containing both strains. Likewise, the control group $\control[\strain_1, \strain_2]$ for the effect of strain $\strain_1$ on strain $\strain_2$ consists of droplets containing strain $\strain_2$ but not strain $\strain_1$. With these groups, we can quantify the effect of strain $\strain_1$ on strain $\strain_2$ by targeting an average treatment effect (ATE) or other similar causal estimand.

However, even when we can implement estimators for such causal estimands, sometimes their results may still not be scientifically reliable. For example, consider a situation where our treatment group $\treatment[\strain_1, \strain_2]$ consists of only $\treatmentsize[\strain_1, \strain_2] = 5$ droplets and our control group $\control[\strain_1, \strain_2]$ consists of $\controlsize[\strain_1, \strain_2] \approx \! 50,000$ droplets. This example may seem unrealistic, but it is actually optimistic compared to some of the size imbalances between the treatment and control group sizes $\treatmentsize[\strain_1, \strain_2], \controlsize[\strain_1, \strain_2]$ that one may expect to see in practice. Cf. section \ref{sec:toy-model}. Regardless of the causal estimand we choose to quantify such interaction effects, in order for our estimates to be scientifically useful we need to estimate their reliability. We need to estimate their statistical power $\power(\treatmentsize[\strain_1, \strain_2], \controlsize[\strain_1, \strain_2] )$.

Of course (even for a fixed causal estimand) the statistical power $\power$ function will depend on the particular estimator, so at this level of generality we cannot complete such a calculation. However, a priori we do also know that the statistical power $\power(\treatmentsize[\strain_1, \strain_2], \controlsize[\strain_1, \strain_2] )$ will always depend on the sizes $\treatmentsize[\strain_1, \strain_2]$, $\controlsize[\strain_1, \strain_2]$ of the treatment group and control group, respectively. Thus, to estimate statistical power $\power(\treatmentsize[\strain_1, \strain_2], \controlsize[\strain_1, \strain_2] )$, we always need to know the treatment group and control group sizes $\treatmentsize[\strain_1, \strain_2], \controlsize[\strain_1, \strain_2] $, regardless of the particular estimator.

Unfortunately, due to the nature of the experiment, the scientist can never directly control the treatment group and control group sizes $\treatmentsize[\strain_1, \strain_2], \controlsize[\strain_1, \strain_2] $. The treatment group and control group sizes $\treatmentsize[\strain_1, \strain_2], \controlsize[\strain_1, \strain_2] $ are random variables. Therefore, for any estimator of any causal estimand, the statistical power $\power(\treatmentsize[\strain_1, \strain_2], \controlsize[\strain_1, \strain_2] )$ will itself be a random variable. This is not inherently a ``deal-breaker'', but we do still need to constrain the possibilities for the distribution of the statistical power $\power(\treatmentsize[\strain_1, \strain_2], \controlsize[\strain_1, \strain_2] )$. At the very least, we would like to be able to compute the expected statistical power $\expectation{\power(\treatmentsize[\strain_1, \strain_2], \controlsize[\strain_1, \strain_2] ) }$. 

Specifically, for a given estimator of a given causal estimand, the statistical power $\power(\treatmentsize[\strain_1, \strain_2], \controlsize[\strain_1, \strain_2] )$ will be a deterministic function $\power$ of the treatment and control group sizes $\treatmentsize[\strain_1, \strain_2], \controlsize[\strain_1, \strain_2] $ (and possibly other factors which for simplicity we will assume to be known and fixed). If we know specific values $\treatmentsize[\strain_1, \strain_2] = T^{(\strain_1, \strain_2)}$, $\controlsize[\strain_1, \strain_2]  = C^{(\strain_1, \strain_2)}$ for the treatment and control group sizes $\treatmentsize[\strain_1, \strain_2], \controlsize[\strain_1, \strain_2] $, then we can evaluate this statistical power function $\power$ and get a deterministic value $\power(T^{(\strain_1, \strain_2)}, C^{(\strain_1, \strain_2)})$. If instead the treatment and control group sizes $\treatmentsize[\strain_1, \strain_2], \controlsize[\strain_1, \strain_2] $ are random variables, then when evaluating this statistical power function we get a new random variable $\power(\treatmentsize[\strain_1, \strain_2], \controlsize[\strain_1, \strain_2] )$ whose distribution is determined by the pushforward measure.

Obviously, without specifying the particular causal estimand and its particular estimator, we still cannot complete the calculation of the expectation $\expectation*{\power(\treatmentsize[\strain_1, \strain_2], \controlsize[\strain_1, \strain_2] )} $ of a random statistical power $\power(\treatmentsize[\strain_1, \strain_2], \controlsize[\strain_1, \strain_2] )$. Nevertheless, for any given estimator of any given causal estimand, the distribution of the statistical power $\power(\treatmentsize[\strain_1, \strain_2], \controlsize[\strain_1, \strain_2] )$ will be determined by the distribution of the treatment group and control group sizes $\treatmentsize[\strain_1, \strain_2], \controlsize[\strain_1, \strain_2]$. Thus, for any estimator of any causal estimand, as a prerequisite for being able to estimate expected statistical power $\expectation*{\power(\treatmentsize[\strain_1, \strain_2], \controlsize[\strain_1, \strain_2] )} $, we must first understand the distribution of the treatment group and control group sizes $\treatmentsize[\strain_1, \strain_2], \controlsize[\strain_1, \strain_2] $.

Note that the treatment group and control group sizes $\treatmentsize[\strain_1, \strain_2], \controlsize[\strain_1, \strain_2]  $ can be understood as sums of indicator random variables:
\begin{equation}
  \label{eq:group_sizes_sums_of_indicators}
  \treatmentsize[\strain_1, \strain_2] = \sum_{\droplet \in [\Droplets]} \indicator{ \droplet\text{'th droplet in } \treatment[\strain_1, \strain_2] }  \,, \quad \controlsize[\strain_1, \strain_2] = \sum_{\droplet \in [\Droplets]} \indicator{\droplet\text{'th droplet in } \control[\strain_1, \strain_2]} \,.
\end{equation}
To completely characterize the distribution of indicator random variables it suffices to know their expectations, i.e. the probabilities:
\begin{equation}
  \label{eq:probabilities_group_sizes}
  \probability*{ \droplet\text{'th droplet in }\treatment[\strain_1, \strain_2]  } \,, \quad \quad \probability*{ \droplet\text{'th droplet in }\control[\strain_1, \strain_2] } \,.
\end{equation}
If we assume the droplets are identically distributed, then linearity of expectation allows us to compute the expected treatment and control group sizes:
\begin{equation}
  \label{eq:expected_group_sizes}
  \expectation*{ \treatmentsize[\strain_1, \strain_2]  } = \Droplets \cdot   \probability*{ \droplet\text{'th droplet in }\treatment[\strain_1, \strain_2]  } \,, \quad
  \expectation*{ \controlsize[\strain_1, \strain_2] } = \Droplets \cdot \probability*{ \droplet\text{'th droplet in }\control[\strain_1, \strain_2] } \,.
\end{equation}
This is already enough to compute $\power \left( \expectation*{\treatmentsize[\strain_1, \strain_2]}, \expectation*{\controlsize[\strain_1, \strain_2]}  \right)$, a quantity that we can use as a (potentially suboptimal\footnote{
Cf. the ``arithmetic'' estimands from section \ref{sec:append-relat-less}.
}) stand-in for the expected statistical power $\expectation*{\power (\treatmentsize[\strain_1, \strain_2], \controlsize[\strain_1, \strain_2] ) }$ in situations where we don't want to make additional assumptions about the distributions of the droplets.

However, if we additionally assume that the droplets are mutually independent, then we can compute the distributions of the treatment and control group sizes as marginals of a Multinomial distribution\footnote{Cf. sections \ref{sec:glob-picky-goodn} and \ref{sec:pairw-glutt-goodn}.}. Given the statistical power function $\power$, i.e. choice of a particular causal estimand and estimator thereof, this would then in principle\footnote{
In practice, given realistic problem sizes, evaluating this exact formula would most likely be intractable, such that one would probably also need to use additional heuristics and approximations. The choice of those heuristics and approximations would depend on the particular causal estimand and estimator thereof, and thus is outside the scope of this discussion.
} be enough information to compute the distribution of the statistical power $\power (\treatmentsize[\strain_1, \strain_2], \controlsize[\strain_1, \strain_2] )$ or any summary statistic thereof. 

One might object to treating the statistical power $\power (\treatmentsize[\strain_1, \strain_2], \controlsize[\strain_1, \strain_2] )$ as a random variable. Certainly, after the experiment has been run, the treatment group and control group sizes $\treatmentsize[\strain_1, \strain_2], \controlsize[\strain_1, \strain_2] $ will be known, allowing us to condition on their observed values $T^{(\strain_1, \strain_2)}, C^{(\strain_1, \strain_2)}$ and ``restore determinism'' by evaluating $\power(T^{(\strain_1, \strain_2)}, C^{(\strain_1, \strain_2)})$. However, this is insufficient to help the scientist before running the experiment. The scientist has \textit{no} direct control over what the treatment group and control group sizes $\treatmentsize[\strain_1, \strain_2], \controlsize[\strain_1, \strain_2] $ will be.

However, the distribution of the treatment group and control group sizes $\treatmentsize[\strain_1, \strain_2], \controlsize[\strain_1, \strain_2] $ should depend on deterministic factors that the scientist can directly control, like the total number of droplets $\Droplets$, the expected number of cells per droplet $\rate$, and the relative abundances $\freq^{(\strain)}$ of the various strains $\strain$ in the sampling population. The scientist would like to plan the experiment (\textit{before} running it) in a way that allows them to (most likely) achieve certain thresholds of statistical power $\power (\treatmentsize[\strain_1, \strain_2], \controlsize[\strain_1, \strain_2] )$ for certain interactions of strains. Therefore the best way to help the scientist is to predict which values of statistical power $\power (\treatmentsize[\strain_1, \strain_2], \controlsize[\strain_1, \strain_2] )$ they will most likely be able to achieve as a function of the deterministic factors that the scientist can directly control. From that perspective it is inescapable that
\begin{enumerate}[label=(\arabic*)]
\item the statistical power $\power (\treatmentsize[\strain_1, \strain_2], \controlsize[\strain_1, \strain_2] )$ must be treated as random,
\item the most basic prerequisite for constraining the distribution of the statistical power $\power (\treatmentsize[\strain_1, \strain_2], \controlsize[\strain_1, \strain_2] )$ is to characterize the distributions of the treatment group and control group sizes $\treatmentsize[\strain_1, \strain_2], \controlsize[\strain_1, \strain_2] $, and
\item (if possible) the distributions of the treatment group and control group sizes $\treatmentsize[\strain_1, \strain_2]$, $\controlsize[\strain_1, \strain_2] $ should be described as functions of the deterministic factors that the scientist can directly control.
\end{enumerate}

There is no need to incubate the droplets in order to estimate the probabilities (\ref{eq:probabilities_group_sizes}). Indeed, it should be possible to estimate the probabilities using a much smaller number of total droplets $\Droplets$ than what might be required in the final experiment to achieve certain (expected) statistical power thresholds. Therefore I propose that the scientist, to be able to design the experiment in a way that allows them achieve certain (expected) statistical power thresholds, first run a ``$\time=0$'' version of the experiment that skips incubating the droplets and uses a smaller number of total droplets $\Droplets$. This both (i) maximizes the usefulness of the data for estimating the distributions of the treatment group and control group sizes (because it should mitigate artifacts resulting from microbial growth and competition, e.g. censoring) and (ii) decreases the cost (in money and time) of this ``pre-experiment'' to the scientist.

\subsection{Non-Parametric Statistical Model}
\label{sec:non-parametric-statistical-model}

The non-parametric statistical model $\mathcal{M}$ containing the distribution of $\vabundance(0)$ for any of the $D$ independent and identically distributed droplets is
\begin{equation}
  \label{eq:model_definition}
  \mathcal{M} = \left\{ \mathcal{P}: \mathcal{P} \text{ is supported on }\N^{\Strains}  \right\} \,.
\end{equation}

\subsection{Target Parameter Definitions}
\label{sec:defin-thro-intro}

The below sections define two different classes of target parameters. Cf. section \ref{sec:defin-thro} for more details.

\subsubsection{``Gluttonous'' Target Parameters}
\label{sec:glutt-defin-data-intro}

The treatment group for inferring the effect of type $\strain_1$ on the growth of type $\strain_2$ corresponds to droplets where $\strain_1$ and $\strain_2$ co-occur. 
The set of all droplets satisfying this condition is
  \begin{equation}
    \label{eq:formal_gluttonous_treatment_definition_intro}
    \begin{split}
      \treatment[\strain_1, \strain_2]_g := & \left\{ \droplet \in
        [\Droplets]: \support(\vabundance_{\droplet}(0)) \supseteq
        \{\strain_1, \strain_2\} \right\} \,.
    \end{split}
  \end{equation}
This is called the ``\textbf{gluttonous}'' \textbf{treatment} group because \textit{all} droplets where $\strain_1$ and $\strain_2$ co-occur are included. 
Cf. figure \ref{fig:gluttonous_vs_picky}. Because of the i.i.d. assumption, this leads to the statistical queries\footnote{Cf. section \ref{sec:append-funct-notat}, in particular section \ref{sec:function-notation} on the notation for functions used herein.} for all $\strain_1, \strain_2 \in [\Strains]$, $\strain_1 \not= \strain_2$:
  \begin{equation}
    \label{eq: formal_gluttonous_treatment_probability_definition}
    \begin{split}
  \Psi^{(\strain_1, \strain_2)}_{\treatment, g} :& \mathcal{M} \to [0,1] \,, \\    
\Psi^{(\strain_1, \strain_2)}_{\treatment, g} : &\mathcal{P} \mapsto    \probability[\mathcal{P}]*{ \support(\vabundance(0)) \supseteq \{\strain_1, \strain_2\}  } \,.
\end{split}
  \end{equation}
The control group for inferring the effect of type $\strain_1$ on the growth of type $\strain_2$ corresponds to droplets where $\strain_2$ occurs but $\strain_1$ does not. 
The set of all droplets satisfying this condition is
  \begin{equation}
    \label{eq:formal_gluttonous_control_definition_intro}
    \begin{split}
      \control[\strain_1, \strain_2]_g := & \left\{ \droplet \in
        [\Droplets] : \support(\vabundance_{\droplet}(0)) \supseteq
        \{\strain_2\} \,,\, \strain_1 \not\in
        \support(\vabundance_{\droplet}(0)) \right\} \,.
    \end{split}
  \end{equation}
This is called the ``\textbf{gluttonous}'' \textbf{control} group because \textit{all} droplets where $\strain_2$ occurs but $\strain_1$ does not are included. Cf. figure \ref{fig:gluttonous_vs_picky}. Because of the i.i.d. assumption, this leads to the statistical queries for all $\strain_1, \strain_2 \in [\Strains]$, $\strain_1 \not= \strain_2$:
  \begin{equation}
    \label{eq: formal_gluttonous_control_probability_definition_intro}
    \begin{split}
    \Psi^{(\strain_1, \strain_2)}_{\control,g} :& \mathcal{M} \to [0,1] \,, \\   
 \Psi^{(\strain_1, \strain_2)}_{\control,g} : &\mathcal{P}  \mapsto  \probability[\mathcal{P}]*{ \support(\vabundance(0)) \supseteq \{ \strain_2\} \,, \, \strain_1 \not\in \support(\vabundance(0))  } \,.
\end{split}
  \end{equation}
These definitions place no constraints on the presence or absence of types $\strain \in [\Strains]$ besides $\strain_1$ and $\strain_2$. This can be favorable for example when studying very rare types, for which there may be no or very few droplets containing $\strain_1$ and $\strain_2$ only, but some containing $\strain_1$ and $\strain_2$ along with other types. These definitions give us the best possible chance of avoiding ``data starvation''.

\subsubsection{``Picky'' Target Parameters}
\label{sec:picky-defin-data-intro}

The above ``gluttonous'' definitions correspond to more possible combinations of cell types than we probably would have included when studying the effect of type $\strain_1$ on the growth of type $\strain_2$ by manually plating combinations of cell types. This motivates the following definitions.

The set of droplets where $\strain_1$ and $\strain_2$ co-occur in a way corresponding to the combination of cell types we would use if we were manually plating combinations of cell types is
  \begin{equation}
    \label{eq:formal_picky_treatment_definition_intro}
    \begin{split}
      \treatment[\strain_1, \strain_2]_p := & \left\{ \droplet \in
        [\Droplets] : \support(\vabundance_{\droplet}(0)) = \{
        \strain_1, \strain_2 \} \right\} \,.
    \end{split}
  \end{equation}
This is called the ``\textbf{picky}'' \textbf{treatment} group because droplets where $\strain_1$ and $\strain_2$ co-occur are included \textit{only when all other types are absent}. Cf. figure \ref{fig:gluttonous_vs_picky}. Because of the i.i.d. assumption, this leads to the statistical queries\footnote{Cf. section \ref{sec:append-funct-notat}, in particular section \ref{sec:function-notation} on the notation for functions used herein.} for all $\strain_1, \strain_2 \in [\Strains]$, $\strain_1 \not= \strain_2$:
  \begin{equation}
    \label{eq: formal_picky_treatment_probability_definition}
    \begin{split}
      \Psi^{(\strain_1, \strain_2)}_{\treatment, p} : & \mathcal{M} \to [0,1] \,, \\  
  \Psi^{(\strain_1, \strain_2)}_{\treatment, p} :&  \mathcal{P} \mapsto  \probability[\mathcal{P}]*{ \support(\vabundance(0)) = \{\strain_1, \strain_2\}  }  \,.
\end{split}
  \end{equation}
The set of droplets where \textit{only} $\strain_2$ occurs, 
and all other types are absent,
corresponding to the combination of cell types we would use if we were manually plating combinations of cell types, is
  \begin{equation}
    \label{eq:formal_picky_control_definition_intro}
    \begin{split}
      \control[\strain_1, \strain_2]_p := & \left\{ \droplet \in
        [\Droplets] : \support(\vabundance_{\droplet}(0)) = \{
        \strain_2 \} \right\} \,.
    \end{split}
  \end{equation}
This 
is called the ``\textbf{picky}'' \textbf{control} group
  because droplets where $\strain_2$ occurs but $\strain_1$ does not occur are included \textit{only when all other types are absent}. It is not enough for $\strain_1$ alone to be absent. 
Cf. figure \ref{fig:gluttonous_vs_picky}. Because of the i.i.d. assumption, this leads to the statistical queries for all $\strain_1, \strain_2 \in [\Strains]$, $\strain_1 \not= \strain_2$:
  \begin{equation}
    \label{eq: formal_picky_control_probability_definition}
    \begin{split}
      \Psi^{(\strain_1, \strain_2)}_{\control, p}: & \mathcal{M} \to [0,1] \,, \\
 \Psi^{(\strain_1, \strain_2)}_{\control, p}:& \mathcal{P} \mapsto   \probability[\mathcal{P}]*{ \support(\vabundance(0)) = \{\strain_2\}  } \,.
\end{split}
  \end{equation}
Note that unlike any of the other group definitions, this one is shared between all questions investigating effects on the growth of strain $\strain_2$.

Unlike the gluttonous groups, the picky groups \textit{do} place constraints on the presence or absence of types $\strain \in [\Strains]$ besides $\strain_1$ and $\strain_2$. For both the treatments and the controls, for ``picky'' groups all types besides $\strain_1$ or $\strain_2$ must have zero counts. Every picky group is a subset of its gluttonous counterpart.

The picky definitions can be favorable when we have plenty of droplets from which to make estimates. They reduce the possibility of confounding effects on the growth of type $\strain_2$ that could be caused by types that are not type $\strain_1$. On the other hand, in instances where there are only very few or no droplets without ``extra'' types, for example for very rare types for which there are few droplets containing the type at all, the picky definitions could lead to ``data starvation''.

\subsection{Why Empirical Estimates are not Useful}
\label{sec:why-empir-distr}

The empirical estimates for these probabilities are inefficient for several reasons. What connects these reasons is the fact that the empirical estimates do not provide any way to connect these estimands with each other. Thus information about one estimand, which should ideally be applicable for learning about related estimands, is wasted. More specifically, note that we have a combinatorial explosion in the number of estimands as a function of the number of strains $\Strains$. This is at least quadratic if we only consider the subsets of $\Strains$ corresponding to the groups defined above, and exponential if we want to consider the probabilities associated with arbitrary subsets of $[\Strains]$ (e.g. for the purpose of investigating higher-order interactions). Cf. section \ref{sec:why-expect-posit}, which discusses a related issue. This is in contrast to the number of parameters in the working models which (if including the relative abundances $\freq^{(\strain)}$) only grow linearly with the number of strains $\Strains$. Thus using the empirical estimates leads to not only exponential increases in ``description length'' or ``information'' required to characterize the initial distribution of droplets, but also to exponential increases in computational runtime or algorithmic complexity for computing estimates of all estimands. The exponential number of parameters has practical implications. Cf. the problem motivating \cite{dirichlet_process}.

Related to this, especially for the treatment groups as well as types $\strain$ with small relative abundances $\freq^{(\strain)}$, even when $\Droplets$ is large, the expected counts can still be very small. In practice this leads to positivity violations, where the empirical estimates force us to state that certain unobserved events happen with zero probability when in principle information from related events should allow us make some small but non-zero estimate for their probabilities. This, combined with the combinatorial explosion, means that we would have to use very large values of $\Droplets$ to make useful (i.e. non-zero and low-variance) estimates for all probabilities. Again, cf. the related discussion in section \ref{sec:why-expect-posit}. This is very inefficient because we would prefer to use a smaller or medium sized $\Droplets$ for the initial trial run (the ``$\time = 0$ pre-experiment'', section \ref{sec:t=0-experiment}). Fitting a model for predicting data throughput is simply a prerequisite for estimating the power we will have to answer certain questions as a function of $\Droplets$, rather than the main question itself. Cf. again section \ref{sec:motiv-targ-estim}.

Similarly, the empirical estimates do not allow us to describe the probabilities in terms of parameters such as the relative abundances $\freq^{(\strain)}$ or the mean number of cells per droplet. That means that, e.g. if we slightly modify our experiment by swapping out two strains with five different strains (which will obviously have different relative abundances), then when using empirical estimates we will have to re-estimate everything from scratch. In contrast, to the extent that parameters like the density concentration $\dconcentration$ or compositional concentration $\cconcentration$ are believed to correspond to ``intrinsic features'' of the experimental apparatus itself, we could reasonably believe that we could re-use estimates for them even if we change the number or relative abundances of strains or the mean number of cells per droplet.

\subsection{Default Working Model: hPoMu}
\label{sec:hpomu_definition_main}

The ``hierarchical Poisson Multinomial'' (hPoMu) distributions $\mathcal{M}_{\operatorname{hPoMu}} \subset \mathcal{M}$ are hierarchical distributions. The total number of cells in the droplet has a Poisson distribution, while (\textit{conditional upon the total number of cells}) the numbers of cells belonging to the multiple types have a joint Multinomial distribution. Cf. figure \ref{fig:count_categorical}. Both choices of distribution correspond to assuming uniform and perfectly homogeneous sampling for each droplet. Cf. section \ref{sec:derivations} for two detailed arguments. The likelihood for the hPoMu distributions equals
  \begin{equation}
    \label{eq:hpomu_defn_1_main}
    \begin{split}
      \probability[\operatorname{hPoMu}(\rate, \vfreq)]*{\vabundance(0) = \vpopulation} = & \probability[\operatorname{hPoMu}(\rate, \vfreq)]*{
        \abundance(0) = \population } \!\cdot\!
      \probability[\operatorname{hPoMu}(\rate, \vfreq)]*{\vabundance(0) = \vpopulation | \abundance(0) =
        \population}
      \\
      =& 
\hphantom{\probability[\operatorname{hPoMu}(\rate, \vfreq)]*{}}
\frac{ e^{-\rate} \rate^{\population} }{ \population !  }
      \cdot \binom{\population}{\population[1] \! \cdots \!
        \population[\Species]} \prod_{\specie=1}^{\Species}
      (\freq^{(\strain)})^{\population[\specie]} \,.
    \end{split}
  \end{equation}
An alternative form of the likelihood of hPoMu reveals that its marginal distributions are mutually independent Poisson distributions:
  \begin{equation}
    \label{eq:hpomu_defn_2_main}
    \begin{split}
      \probability[\operatorname{hPoMu}(\rate, \vfreq)]*{\vabundance(0) = \vpopulation} =
      \prod_{\strain=1}^{\Strains} \frac{ e^{-\freq^{(\strain)}\rate}
        (\freq^{(\strain)}\rate)^{\population[\specie]} }{
        (\population[\specie]) !  } \,.
    \end{split}
  \end{equation}
Heuristically speaking, equation (\ref{eq:hpomu_defn_2_main}) says that during the formation of each droplet the expected number of cells, or ``sampling rate'', $\rate$ is evenly spread out among the multiple types according to their frequencies in the population.

\subsubsection{Projected Target Parameters Under hPoMu Working Model}
\label{sec:proj-targ-param-hpomu}

Let $\iota_{\operatorname{hPoMu}}: \mathcal{M}_{hPoMu} \to \mathcal{M}$, $\iota_{\operatorname{hPoMu}}: \mathcal{P} \mapsto \mathcal{P}$ denote the inclusion map\footnote{Cf. section \ref{sec:append-funct-notat}, in particular the last part of section \ref{sec:conv-funct-comp}.} ${\mathcal{M}_{\operatorname{hPoMu}} \xhookrightarrow{\iota_{\operatorname{hPoMu}}} \mathcal{M}}$. Suppose $\Pi_{\operatorname{hPoMu}} : \mathcal{M} \to \mathcal{M}_{\operatorname{hPoMu}}$ is a projection\footnote{Cf. section \ref{sec:append-funct-notat}, in particular section \ref{sec:function-notation} on the notation for functions used herein.}:
\begin{equation}
  \label{eq:hPoMu_projection}
  \Pi_{\operatorname{hPoMu}} : \mathcal{P} \mapsto \argmin_{\tilde{\mathcal{P}} \in \mathcal{M}_{\operatorname{hPoMu}} }  \int_{\N^{\Strains}} L(\tilde{\mathcal{P}}(\vcounts)) d\mathcal{P}(\vcounts) \,,
\end{equation}
for a given loss function $L$, e.g. $L: p \mapsto - \log(p)$, onto the working parametric submodel $\mathcal{M}_{\operatorname{hPoMu}}$ from the nonparametric model $\mathcal{M}$. In particular one should have the relationships that ${\Pi_{\operatorname{hPoMu}} \circ \iota_{\operatorname{hPoMu}} = \operatorname{Id}_{\operatorname{hPoMu}} : \mathcal{M}_{\operatorname{hPoMu}} \to \mathcal{M}_{\operatorname{hPoMu}}}$ and that ${ \iota_{\operatorname{hPoMu}} \circ \Pi_{\operatorname{hPoMu}}  : \mathcal{M} \to \mathcal{M}}$ is idempotent.

Given the projection of a distribution $\mathcal{P}$ from the non-parametric model $\mathcal{M}$ onto the working parametric submodel $\mathcal{M}_{\operatorname{hPoMu}}$, we can now evaluate any of the non-parametrically defined parameters (\ref{eq: formal_gluttonous_treatment_probability_definition}), (\ref{eq: formal_gluttonous_control_probability_definition_intro}), (\ref{eq: formal_picky_treatment_probability_definition}), (\ref{eq: formal_picky_control_probability_definition}) from section \ref{sec:defin-thro-intro} at the projection, defining a new set of target parameters. This new set of target parameters corresponding to $\mathcal{M}_{\operatorname{hPoMu}}$ is defined below.

For a given $\mathcal{P} \in \mathcal{M}$, let $\iota_{\operatorname{hPoMu}}(\Pi_{\operatorname{hPoMu}}(\mathcal{P})) =: \operatorname{hPoMu}(\rate_{\mathcal{P}}, \vfreq_{\mathcal{P}}) \in \mathcal{M}_{\operatorname{hPoMu}}$. The projected target parameters for the gluttonous treatment groups $\Psi^{(\strain_1, \strain_2)}_{\treatment, g, \operatorname{hPoMu}} 
:= \Psi^{(\strain_1, \strain_2)}_{\treatment, g} \circ  \iota_{\operatorname{hPoMu}} \circ \Pi_{\operatorname{hPoMu}} $ are defined as (cf. section \ref{sec:pairw-glutt-goodn})
\begin{equation}
  \label{eq:hPoMu_projected_parameters_gluttonous_treatment}
  \begin{array}{rccl}
    \Psi^{(\strain_1, \strain_2)}_{\treatment, g, \operatorname{hPoMu}}:
    & \mathcal{M}
    &
      \to
    &
      [0,1] \\
    &&&
    \\
\Psi^{(\strain_1, \strain_2)}_{\treatment, g, \operatorname{hPoMu}}:
& \mathcal{P} &
\mapsto &
\probability[\iota_{\operatorname{hPoMu}}(\Pi_{\operatorname{hPoMu}}(\mathcal{P}))]*{ \support(\vabundance(0)) \supseteq \{\strain_1, \strain_2\}  }
\\
&&=&
\probability[\operatorname{hPoMu}(\rate_{\mathcal{P}}, \vfreq_{\mathcal{P}})]*{ \support(\vabundance(0)) \supseteq \{\strain_1, \strain_2\}  } \\
&&=&
(1 - e^{-\freq^{(\strain_1)}_{\mathcal{P}} \rate_{\mathcal{P}}}) \cdot  (1 - e^{-\freq^{(\strain_2)}_{\mathcal{P}} \rate_{\mathcal{P}}}) \,,
  \end{array}
\end{equation}
the projected target parameters for the gluttonous control groups $ \Psi^{(\strain_1, \strain_2)}_{\control,g, \operatorname{hPoMu}} :=  \Psi^{(\strain_1, \strain_2)}_{\control,g} \circ \iota_{\operatorname{hPoMu}} \circ \Pi_{\operatorname{hPoMu}}$ are defined as (cf. section \ref{sec:pairw-glutt-goodn})
\begin{equation}
  \label{eq:hPoMu_projected_parameters_gluttonous_control}
  \begin{array}{rccl}
    \Psi^{(\strain_1, \strain_2)}_{\control,g, \operatorname{hPoMu}}: & \mathcal{M} & \to & [0,1] \\
    &&&\\
    \Psi^{(\strain_1, \strain_2)}_{\control,g, \operatorname{hPoMu}}:
&\mathcal{P}&
\mapsto &
\probability[\iota_{\operatorname{hPoMu}}(\Pi_{\operatorname{hPoMu}}(\mathcal{P}))]*{ \support(\vabundance(0)) \supseteq \{ \strain_2\} \,, \, \strain_1 \not\in \support(\vabundance(0))  }\\
&&=&
\probability[\operatorname{hPoMu}(\rate_{\mathcal{P}}, \vfreq_{\mathcal{P}})]*{ \support(\vabundance(0)) \supseteq \{ \strain_2\} \,, \, \strain_1 \not\in  \support(\vabundance(0))  } \\
&& = & 
e^{-\freq^{(\strain_1)}_{\mathcal{P}} \rate_{\mathcal{P}}} \cdot (1 - e^{-\freq^{(\strain_2)}_{\mathcal{P}} \rate_{\mathcal{P}}}) \,,
  \end{array}
\end{equation}
the projected target parameters for the picky treatment groups $\Psi^{(\strain_1, \strain_2)}_{\treatment, p, \operatorname{hPoMu}} := \Psi^{(\strain_1, \strain_2)}_{\treatment, p} \circ \iota_{\operatorname{hPoMu}} \circ \Pi_{\operatorname{hPoMu}}$ are defined as (cf. section \ref{sec:glob-picky-goodn})
\begin{equation}
\label{eq:hPoMu_projected_parameters_picky_treatment}
\begin{array}{rccl}
  \Psi^{(\strain_1, \strain_2)}_{\treatment, p, \operatorname{hPoMu}} : & \mathcal{M} & \to & [0,1] \\
  &&& \\
\Psi^{(\strain_1, \strain_2)}_{\treatment, p, \operatorname{hPoMu}} :  &
\mathcal{P} & \mapsto &
\probability[\iota_{\operatorname{hPoMu}}(\Pi_{\operatorname{hPoMu}}(\mathcal{P}))]*{ \support(\vabundance(0)) = \{\strain_1, \strain_2\}  } \\
&&=&
\probability[\operatorname{hPoMu}(\rate_{\mathcal{P}}, \vfreq_{\mathcal{P}})]*{ \support(\vabundance(0)) = \{\strain_1, \strain_2\}  } \\
&&=&
(1 - e^{-\freq^{(\strain_1)}_{\mathcal{P}}\rate_{\mathcal{P}}})(1 - e^{-\freq^{(\strain_2)}_{\mathcal{P}}\rate_{\mathcal{P}}}) \cdot e^{-(1-(\freq^{(\strain_1)}_{\mathcal{P}} + \freq^{(\strain_2)}_{\mathcal{P}}))\rate_{\mathcal{P}}} \,,
\end{array}
\end{equation}
and the projected target parameters for the picky control groups $ \Psi^{(\strain_1, \strain_2)}_{\control, p, \operatorname{hPoMu}} :=  \Psi^{(\strain_1, \strain_2)}_{\control, p} \circ \iota_{\operatorname{hPoMu}} \circ \Pi_{\operatorname{hPoMu}}$ are defined as (cf. section \ref{sec:glob-picky-goodn})
\begin{equation}
  \label{eq:hPoMu_projected_parameters_picky_control}
  \begin{array}{rccl}
    \Psi^{(\strain_1, \strain_2)}_{\control, p, \operatorname{hPoMu}}:  & \mathcal{M} & \to & [0,1] \\
    &&& \\
    \Psi^{(\strain_1, \strain_2)}_{\control, p, \operatorname{hPoMu}}: &
\mathcal{P}& \mapsto &
\probability[\iota_{\operatorname{hPoMu}}(\Pi_{\operatorname{hPoMu}}(\mathcal{P}))]*{ \support(\vabundance(0)) = \{\strain_2\}  } \\
&& = &
\probability[\operatorname{hPoMu}(\rate_{\mathcal{P}}, \vfreq_{\mathcal{P}})]*{ \support(\vabundance(0)) = \{\strain_2\}  } \\
&& = &
(1 - e^{-\freq^{(\strain_2)}_{\mathcal{P}}\rate_{\mathcal{P}}}) \cdot e^{-(1 - \freq^{(\strain_2)}_{\mathcal{P}})\rate_{\mathcal{P}}} \,.
  \end{array}
\end{equation}
To estimate these, we just need to estimate $\rate_{\mathcal{P}}$ and $\vfreq_{\mathcal{P}}$ and then use the corresponding plugin estimators. Estimating $\rate_{\mathcal{P}}$ and $\vfreq_{\mathcal{P}}$ also gives us plugin estimators for the probabilities associated with \textit{all} $2^{\Species}$ subsets of $[\Species]$. Thus ``information is shared'' between estimates for the probabilities of related events. Moreover, as long as all estimates of the $\freq^{(\strain)}_{\mathcal{P}}$ are nonzero, all estimated probabilities will also all be nonzero. Therefore these estimates should both be ``smoother'' and more efficient than the empirical estimates, especially when $\Species$ is large (corresponding to some of the $\freq^{(\strain)}_{\mathcal{P}}$ being small). The large $\Strains$ regime would substantially advance the state of the art in microbial ecology, and thus is of particular interest.

\subsubsection{Assumptions Justifying Default Working Model}
\label{sec:impl-model-assumpt-intro}

Both of the two derivations justifying hPoMu as the default working model, cf. section \ref{sec:derivations}, make the following implicit assumptions (among others):

\begin{itemize}
\item For \textit{every} type $\strain$, the number of cells in the population which belong to type $\strain$ is very large.
\item Each cell (regardless of its type) has the same small probability of ending up in the droplet as any other cell.
\item Whether a given cell (regardless of its type) ends up in the droplet is completely independent of what happens to any other cell.
\end{itemize}

The first can be thought of as stating that the population size for each type is effectively infinite. The second and third can together be thought of as stating that the sampling pool is perfectly homogeneous.

\subsection{Failures of Default Working Model Assumptions}
\label{sec:failures-hpomu-model-intro}

While hPoMu is a sensible working model to start with, its implicit assumptions failing to be realistic could undermine its usefulness in practice.

\subsubsection{Populations are Finite}
\label{sec:sampling-from-finite-intro}

We assumed that \textit{for each individual type $\strain$} the number of cells is ``effectively infinite''. For ``rare'' types it is a priori unclear whether this assumption is reasonable. However in practice, even for types with relative abundance as low as $0.01\%$, taking the finiteness of the populations into account does not lead to substantially different predictions than the default hPoMu working model. (Cf. sections \ref{sec:sampl-without-repl} and \ref{sec:distr-again-fall} where this claim is substantiated.)

\subsubsection{Density Heterogeneity}
\label{sec:dens-heter-intro}

Different sections of the sampling pool could have average numbers of cells per unit volume that differ from the average number of cells per unit volume for the entire sampling pool (i.e. the total number of cells divided by the volume of the sampling pool). Herein I call this phenomenon ``density heterogeneity''. Higher density heterogeneity implies more variance in the numbers of cells per unit volume throughout the sampling pool. Cf. figure \ref{fig:density_heterogeneity}.

The default Poisson working model for the total number of cells per droplet can not account for additional variance that density heterogeneity might cause. Its variance is the smallest possible value that could hypothetically exist in practice, which seems unrealistic.

Because this describes extra variance in the total number of cells, regardless of their types, density heterogeneity is relevant even when there is only one type. Thus density heterogeneity is relevant even for the single-cell\footnote{The goal for these experiments is to have one cell per droplet, whence the name ``single-cell''. However, because the droplet formation process is random, in practice some droplets in ``single-cell'' experiments may still end up with more than one cell.}, single type experiments which have already been intensively developed.

\subsubsection{Compositional Heterogeneity}
\label{sec:comp-heter-intro}

The relative abundance of any given cell type $\strain$ across the entire sampling pool is a fixed value $\freq^{(\strain)}$. Different sections of the sampling pool could have relative abundances of the cell types that differ from those for the entire sampling pool. Herein I call this phenomenon ``compositional heterogeneity''. Higher compositional heterogeneity implies more variance in the relative abundances for the cell types throughout the sampling pool. Cf. figure \ref{fig:compositional_heterogeneity}.

The default Multinomial working model for the relative abundances of the cell types per droplet (\textit{conditional upon the total number of cells}) does not account for additional variance that would be introduced by compositional heterogeneity. Its variance is the smallest possible value that could hypothetically exist in practice, which is seems unrealistic.

Compositional heterogeneity is relevant only for multi-cell, multi-type experiments. Such experiments are newer and still less common than single-cell, single type experiments. It seems that compositional heterogeneity may not have been identified before as a potential issue affecting the data throughput of droplet microfluidics experiments.

\subsection{Working Model to Address Assumption Failures: ghNBDM}
\label{sec:model-arbitr-comb_intro}

Any of the three assumptions from section \ref{sec:impl-model-assumpt-intro} can be unrealistic in practice. As mentioned before, even for extremely rare strains, taking the finiteness of the populations into account (section \ref{sec:sampling-from-finite-intro}) does not lead to predictions substantially different than those made by the default hPoMu model. On the other hand, models for which the perfect homogeneity assumption is false \textit{can} lead to substantially different predictions. This section introduces a generalization of the hPoMu working model that makes fewer homogeneity assumptions than hPoMu does. In particular, both nonzero density heterogeneity (section \ref{sec:dens-heter-intro}) and nonzero compositional heterogeneity (section \ref{sec:comp-heter-intro}) are allowed.

The ``generalized hierarchical Negative Binomial Dirichlet-Multinomial''  (ghNBDM) working model $\mathcal{M}_{\operatorname{ghNBDM}} \subset \mathcal{M}$ is a family of hierarchical distributions. The number of cells in the droplet has a Negative Binomial distribution with ``density concentration'' parameter $\dconcentration\Strains$, while (\textit{conditional upon the total number of cells}) the numbers of cells belonging to the multiple types have a joint Dirichlet-Multinomial distribution with ``compositional concentration'' parameter $\cconcentration \Strains$. The likelihood for this parametric submodel $\mathcal{M}_{\operatorname{ghNBDM}} $ is
  \begin{equation}
    \label{eq:ghnbdm_definition_intro}
    \begin{split}
      &    \probability[\operatorname{ghNBDM}(\rate,\vfreq, \dconcentration, \cconcentration)]*{\vabundance(0) = \vpopulation} \\
      = & \quad \probability[\operatorname{ghNBDM}(\rate,\vfreq, \dconcentration, \cconcentration)]*{ \abundance(0) = \population } 
\hphantom{\frac{\rate^{\counts}}{\counts!}}
\!\cdot
      \probability[\operatorname{ghNBDM}(\rate,\vfreq, \dconcentration, \cconcentration)]*{\vabundance(0) = \vpopulation |
        \abundance(0) = \population}
      \\
      = & \quad \frac{ \Gamma(\dconcentration \Species + \population)
      }{ \Gamma (\dconcentration \Species) } \!\cdot \!  \frac{
        (\dconcentration \Species)^{\dconcentration \Species} }{
        (\dconcentration \Species + \lambda)^{\dconcentration \Species
          + \population} } \!\cdot\!  \frac{ \rate^n }{ n!  } \! \cdot
      \!  \frac{\Gamma( \cconcentration \Species
        )}{\Gamma(\cconcentration \Species + \population )} \!\cdot \!
      \left[\prod_{\specie=1}^{\Species} \frac{ \Gamma
          (\cconcentration\Species \freq^{(\strain)} +
          \population[\specie] ) }{ \Gamma \left(\cconcentration
            \Species \freq^{(\strain)} \right) } \right] \! \cdot \!
      \binom{\population}{\population[1] \! \cdots \!
        \population[\Species]} \,.
    \end{split}
  \end{equation}
In general the marginal distributions of members of the ghNBDM family $\mathcal{M}_{\operatorname{ghNBDM}} \subset \mathcal{M}$ have non-zero cross-covariance. Thus the entries $\abundance[\specie](0)$ of the random vector $\vabundance(0)$ will in general \textit{not} be mutually independent. This is unlike the default working model (hPoMu) $\mathcal{M}_{\operatorname{hPoMu}} \subset \mathcal{M}_{\operatorname{ghNBDM}} \subset \mathcal{M}$, for which the $\abundance[\specie](0)$ are mutually independent, cf. again section \ref{sec:hpomu_definition_main}, particularly equation (\ref{eq:hpomu_defn_2_main}).

Notice how this working model has two additional parameters compared to the hPoMu working model, namely the density concentration $\dconcentration$ parameter and the compositional concentration $\cconcentration$ parameter. Higher values of the density concentration $\dconcentration$ parameter correspond to lower density heterogeneity (section \ref{sec:dens-heter-intro}). As $\dconcentration \to \infty$ the total number of cells in the droplet approaches a Poisson distribution. Cf. section \ref{sec:proof-that-nb}. Similarly, higher values of the compositional concentration $\cconcentration$ parameter correspond to lower compositional heterogeneity (section \ref{sec:comp-heter-intro}). As $\cconcentration \to \infty$ the joint distribution of the numbers of cells belonging to the multiple types approaches a Multinomial distribution. Cf. section \ref{sec:proof-that-hpodm}. The hPoMu model is approached in the limit as both $\dconcentration \to \infty$ and $\cconcentration \to \infty$ jointly. Cf. figure \ref{fig:ghNBDM} and section \ref{sec:ghnbdm-family-defin}.
For details about the calculations demonstrating how the somewhat complicated expressions involving the $\Gamma$ function do indeed generalize the likelihoods of the Poisson and Multinomial distributions, cf. sections \ref{sec:hnbdm-gener-hpomu}, \ref{sec:hpodm-gener-hpomu}, and \ref{sec:append-results-bound}. At a high level, the idea is to use Stirling's approximation, in particular the formulations given by \cite{Robbins} and \cite{Gordon1994}.

\subsubsection*{Projected Target Parameters Under ghNBDM Working Model}
\label{sec:proj-targ-param-ghnbdm}

This is similar to what was done in section \ref{sec:proj-targ-param-hpomu}. See section \ref{sec:proj-targ-param-ghnbdm-appendix} for details.

\subsection{Intermediate Choice: hNBDM Working Model}
\label{sec:proj-targ-param-hNBDM}

Given estimates of $\rate_{\mathcal{P}}$, $\vfreq_{\mathcal{P}}$, ${\dconcentration}_{,\mathcal{P}}$, and ${\cconcentration}_{,\mathcal{P}}$, plugin estimators for the above target parameters (\ref{eq:ghNBDM_projected_parameters_gluttonous_treatment}), (\ref{eq:ghNBDM_projected_parameters_gluttonous_control}), (\ref{eq:ghNBDM_projected_parameters_picky_treatment}), (\ref{eq:ghNBDM_projected_parameters_picky_control}) can be computed to arbitrary accuracy, e.g. by truncating terms of the series expansions. Nevertheless, it is at best tedious to derive explicit bounds on the accuracy as a function of the number of terms in the truncated series expansions. An alternative is to use the ``hierarchical Negative Binomial Dirichlet-Multinomial'' (hNBDM) working model $\mathcal{M}_{\operatorname{hNBDM}} \subset  \mathcal{M}_{\operatorname{ghNBDM}} \subset \mathcal{M}$.

The hNBDM working model is intermediate in complexity and flexibility between the ghNBDM and hPoMu working models, indeed $\mathcal{M}_{\operatorname{hPoMu}} \subset \mathcal{M}_{\operatorname{hNBDM}} \subset \mathcal{M}_{\operatorname{ghNBDM}}$. Distributions in the hNBDM working model are distributions in the ghNBDM working model such that the density concentration $\dconcentration$ and compositional concentration $\cconcentration$ parameters are equal, $\dconcentration = \cconcentration =: \concentration$. Hence the hNBDM working model has one more parameter than the hPoMu working model, but also one less parameter than the ghNBDM working model.

The real ``magic'' of the hNBDM working model happens in its likelihood: when $\dconcentration = \cconcentration$, the resulting cancellations and simplifications in the ghNBDM likelihood cause it to factorize over its marginal distributions:
  \begin{equation}
    \label{eq:hnbdm_definition_intro}
    \begin{split}
      &    \probability[\operatorname{hNBDM}(\rate,\vfreq, \concentration)]*{\vabundance(0) = \vpopulation} \\
      = & \quad
\prod_{\strain=1}^{\Strains}
\frac{
\Gamma(\concentration\Strains\freq^{(\strain)} + \counts[\strain] )
}{
\Gamma(\concentration\Strains\freq^{(\strain)})
}
\frac{
(\concentration\Strains\freq^{(\strain)})^{\concentration\Strains\freq^{(\strain)}}
}{
(\concentration\Strains\freq^{(\strain)} + \freq^{(\strain)}\rate )^{\concentration\Strains\freq^{(\strain)} + \counts[\strain]}
}
\frac{
(\freq^{(\strain)} \rate)^{\counts[\strain]}
}{
(\counts[\strain])!
}
\,.
    \end{split}
  \end{equation}
In other words, the marginal distributions are mutually independent Negative Binomial distributions. The ``sampling rate'' $\rate$ is proportionately ``spread out'' across the marginals according to the frequencies $\freq^{(\strain)}$, exactly analogous to what occurs for the hPoMu working model. Indeed, the hPoMu distributions can be considered the ``boundary'' of the hNBDM model reached by the limit $\concentration \to \infty$.

\subsubsection*{Projected Target Parameters Under hNBDM Working Model}
\label{sec:proj-targ-param-hnbdm}

This is similar to what was done in section \ref{sec:proj-targ-param-hpomu}. See section \ref{sec:proj-targ-param-hnbdm-appendix} for details.

\section{Conclusion}
\label{sec:conclusion-2}

\paragraph{Findings and Contributions}

The data from the noisy dynamical systems corresponding to each droplet can be understood as censored observations of a multivariate Markov process.
The statistical understanding of a droplet's initial state is crucial to overcoming the main limitation of these experiments, the uncontrolled assignment of microbes to droplets.

\paragraph{Practical Implications}

Giving explicit descriptions of the intended underlying statistical model has several benefits. By formalizing intuitive insights, like the division of the experiment into three phases, it allows us use those insights as the basis for data analysis. It helps to clarify the similarities and differences between this problem and other problem, as well the similarities and differences between methods used for this problem and methods used for other problems. It makes explicit which assumptions we are making when analyzing the data, so that we might later question those assumptions if necessary. It guides us to understanding clearly what the most challenging aspects of the data analysis are. In short, it gives us a starting point that anchors everything else we might do.

\paragraph{Next Steps and Open Questions}

There are several different choices we could have made when positing the explicit statistical models given in this chapter. It remains to be fully investigated what the consequences of such different choices would be. Reasonable possible choices, and their consequences, are explained in the next part. Later, methods for inferring interactions that would apply to observations produced by these models are considered.

\begin{coolsubappendices}
\section{Conventions and Notation for Functions}
\label{sec:append-funct-notat}

Section \ref{sec:function-notation} discusses the notation used for functions. Section \ref{sec:conv-funct-comp} discusses the convention used for function composition.

\subsection{Function Notation}
\label{sec:function-notation}

Herein I use a definition of function that requires not only (i) a rule of assignment, but also (ii) a specified domain (source) and codomain (target). A notion of function requiring only (i) corresponds more to ``classical'' mathematics, and in some sources is called an ``intensional'' definition of function\cite[section 1.1]{selinger}. In contrast, a notion of function requiring (i) and (ii) corresponds more to ``modern'' mathematics, and in some sources is called an ``extensional'' definition of function\cite[section 1.1]{selinger}. Herein I only use the ``extensional'' notion\footnote{
In programming language terminology, the ``intensional'' notion is ``untyped'', whereas the ``extensional'' notion is ``typed''. Thus one motivation for using the latter notion more often in ``modern'' mathematics is that the resulting statements are more ``type safe''.
}.

Given a function $f$ whose domain is a set $X$ and whose codomain is a set $Y$, thus sending every element $x \in X$ to a unique element $f(x) \in Y$, I write
\begin{equation}
  \label{eq:function_notation}
  \begin{split}
    f: & X \to Y \,, \\
    f: & x \mapsto f(x) \,.
  \end{split}
\end{equation}
The first line, $f: X \to Y$, indicates the domain and codomain of the function. The second line (using ``anonymous function'' notation from lambda calculus), $f: x \mapsto f(x)$, indicates the rule of assignment of the function.

In particular, I use the convention that for, two functions $f_1 : X_1 \to Y_1$ and $f_2: X_2 \to Y_2$ to be equal, not only must they have the same rule of assignment, but they also must have the same domain and codomain. For example, the function $f_1: x \mapsto x$ and $f_2: x \mapsto x$ are considered different when $f_1: \mathbb{N} \to \mathbb{R}$ and $f_2:\mathbb{N} \to \mathbb{N}$, even though they both have the same rule of assignment and same domain, because their codomains are different. With this convention, a rule of assignment is insufficient to specify a function.

\subsection{Conventions for Function Composition}
\label{sec:conv-funct-comp}

There are at least two different possible conventions for when, given two functions
\[f_1: X_1 \to Y_1  \quad \quad \text{and} \quad \quad  f_2: X_2 \to Y_2 \,,\]
  we say that the composition of the two functions, ${f_2 \circ f_1: X_1 \to Y_2}$, is defined:
\begin{enumerate}[label=(\roman*)]
\item whenever $Y_1 \subseteq X_2$,
\item whenever $Y_1 = X_2$.
\end{enumerate}
Using either convention, the rule of assignment of $f_2 \circ f_1$ is $x_1 \mapsto f_2(f_1(x_1))$. However, with the first convention, there are potentially infinitely many functions $f_1$ such that $f_2 \circ f_1: X_1 \to Y_2$ is defined and has the same rule of assignment, each corresponding to a distinct $Y_1 \subseteq X_2$. In particular, the first convention corresponds to defining a function only in terms of a rule of assignment, and not requiring that a specific domain and codomain be specified.

Hence the first convention for composition of functions is ill-suited for the definition of function explained in section \ref{sec:function-notation} above. It is incompatible with the demand that not only the rule of assignment of a function must be specified but also its domain and codomain.

The second convention for composition of functions is compatible with the definition of function explained in section \ref{sec:function-notation} above. The second convention requires a specific codomain for $f_1$ (namely exactly the domain of $f_2$) is specified in order for the composition $f_2 \circ f_1$ to even be defined. This convention is less ambiguous and is the convention I will use.

The choice between the two conventions can have practical consequences. For example, when using the first convention, the composition of two surjective (a.k.a. onto) functions need not again be surjective, and the composition of two bijections need not again be a bijection. This can be subtle and confusing and lead to hard to detect mistakes. On the other hand, when using the second convention, the composition of two surjective functions is always surjective, and the composition of two bijections is always a bijection.

For example, say that $f_1: X_1 \to Y_1$ and $f_2: X_2 \to Y_2$ are both surjective, with $Y_1 \subsetneq X_2$. If the restriction of $f_2$ to $Y_1$ is not surjective (onto) $Y_2$, then, when using the first convention, the composition $f_2 \circ f_1$ will not be surjective, even though $f_1$ and $f_2$ are both surjective. However, when using the second convention, the composition $f_2 \circ f_1$ is not even defined. Instead, we first need to define the inclusion function $\iota: Y_1 \xhookrightarrow{} X_2$, $\iota: y_1 \mapsto y_1$, which is clearly injective but not surjective because $Y_1 \subsetneq X_2$ by assumption. Then the function $X_1 \to Y_2$ we are interested in is $f_2 \circ \iota \circ f_1$. It becomes clear then why the composite function is not surjective. Although $f_1$ and $f_2$ are both surjective, $\iota$ is not.

\section{Additional Projected Target Parameters}
\label{sec:proj-targ-param-appendix}

Section \ref{sec:proj-targ-param-ghnbdm-appendix} discusses the analogue of section \ref{sec:proj-targ-param-hpomu} but for the ghNBDM working model. Section \ref{sec:proj-targ-param-hnbdm-appendix} does likewise but for the hNBDM working model.

\subsection{Projected Target Parameters Under ghNBDM Working Model}
\label{sec:proj-targ-param-ghnbdm-appendix}

Let $\iota_{\operatorname{ghNBDM}}: \mathcal{M}_{ghNBDM} \to \mathcal{M}$, $\iota_{\operatorname{ghNBDM}}: \mathcal{P} \mapsto \mathcal{P}$ denote the inclusion map\footnote{Cf. section \ref{sec:append-funct-notat}, in particular the last part of section \ref{sec:conv-funct-comp}.}
\[{\mathcal{M}_{\operatorname{ghNBDM}} \xhookrightarrow{\iota_{\operatorname{ghNBDM}}} \mathcal{M}} \,.\]
Suppose $\Pi_{\operatorname{ghNBDM}} : \mathcal{M} \to \mathcal{M}_{\operatorname{ghNBDM}}$ is a projection\footnote{Cf. section \ref{sec:append-funct-notat}, in particular section \ref{sec:function-notation} on the notation for functions used herein.}:
\begin{equation}
  \label{eq:ghNBDM_projection}
  \Pi_{\operatorname{ghNBDM}} : \mathcal{P} \mapsto \argmin_{\tilde{\mathcal{P}} \in \mathcal{M}_{\operatorname{ghNBDM}} }  \int_{\N^{\Strains}} L(\tilde{\mathcal{P}}(\vcounts)) d\mathcal{P}(\vcounts) \,,
\end{equation}
for a given loss function $L$, e.g. $L: p \mapsto - \log(p)$, onto the working parametric submodel $\mathcal{M}_{\operatorname{ghNBDM}}$ from the nonparametric model $\mathcal{M}$. One should have the relationships:
\begin{itemize}
\item ${\Pi_{\operatorname{ghNBDM}} \circ \iota_{\operatorname{ghNBDM}} = \operatorname{Id}_{\operatorname{ghNBDM}} : \mathcal{M}_{\operatorname{ghNBDM}} \to \mathcal{M}_{\operatorname{ghNBDM}}}$, and
\item ${ \iota_{\operatorname{ghNBDM}} \circ \Pi_{\operatorname{ghNBDM}}  : \mathcal{M} \to \mathcal{M}}$
  is idempotent.
\end{itemize}

Given the projection of a distribution $\mathcal{P}$ from the non-parametric model $\mathcal{M}$ onto the working parametric submodel $\mathcal{M}_{\operatorname{ghNBDM}}$, we can now evaluate any of the non-parametrically defined parameters (\ref{eq: formal_gluttonous_treatment_probability_definition}), (\ref{eq: formal_gluttonous_control_probability_definition_intro}), (\ref{eq: formal_picky_treatment_probability_definition}), (\ref{eq: formal_picky_control_probability_definition}) from section \ref{sec:defin-thro-intro} at the projection, defining a new set of target parameters. This new set of target parameters corresponding to $\mathcal{M}_{\operatorname{ghNBDM}}$ is defined below.

I use the notation $(x)^{\overline{n}}$ ostensibly popularized by Knuth for the \textit{rising factorial},
\begin{equation}
  \label{eq:rising_factorial_defn}
  {(x)^{\overline{n}} := \prod_{\nu=0}^{n-1} (x+i) = \frac{\Gamma(x+n)}{\Gamma(x)} } \,,
\end{equation}
assuming $x >0$, real, and $n \in \N$. Cf. footnote \ref{footnote:gamma_interpolate} of section \ref{sec:mle-dens}.

For a given $\mathcal{P} \in \mathcal{M}$, let $\iota_{\operatorname{ghNBDM}}(\Pi_{\operatorname{ghNBDM}}(\mathcal{P})) =: \operatorname{ghNBDM}(\rate_{\mathcal{P}}, \vfreq_{\mathcal{P}}, {\dconcentration}_{,\mathcal{P}}, {\cconcentration}_{,\mathcal{P}}) \in \mathcal{M}_{\operatorname{ghNBDM}}$. The projected target parameters for the gluttonous treatment groups 
\[\Psi^{(\strain_1, \strain_2)}_{\treatment, g, \operatorname{ghNBDM}} 
  := \Psi^{(\strain_1, \strain_2)}_{\treatment, g} \circ  \iota_{\operatorname{ghNBDM}} \circ \Pi_{\operatorname{ghNBDM}} \]
  are defined as
  \begin{equation}
  \label{eq:ghNBDM_projected_parameters_gluttonous_treatment}
  \begin{adjustbox}{max width=\textwidth,keepaspectratio}
$  \begin{array}{rccl}
    \Psi^{(\strain_1, \strain_2)}_{\treatment, g, \operatorname{ghNBDM}}: & \mathcal{M} & \to & [0,1] \\
    &&& \\
\Psi^{(\strain_1, \strain_2)}_{\treatment, g, \operatorname{ghNBDM}}:
& \mathcal{P} &
\mapsto &
\probability[\iota_{\operatorname{ghNBDM}}(\Pi_{\operatorname{ghNBDM}}(\mathcal{P}))]*{ \support(\vabundance(0)) \supseteq \{\strain_1, \strain_2\}  }
\\
&&=&
\probability[\operatorname{ghNBDM}(\rate_{\mathcal{P}}, \vfreq_{\mathcal{P}}, {\dconcentration}_{,\mathcal{P}}, {\cconcentration}_{,\mathcal{P}})]*{ \support(\vabundance(0)) \supseteq \{\strain_1, \strain_2\}  } \\[10pt]
&&=&
\displaystyle
\sum_{\counts=2}^{\infty}
\!
\frac{
({\dconcentration}_{,\mathcal{P}}\Strains)^{{\dconcentration}_{,\mathcal{P}}\Strains}
}{
({\dconcentration}_{,\mathcal{P}}\Strains \!+\! \rate_{\mathcal{P}})^{{\dconcentration}_{,\mathcal{P}}\Strains \!+\! \counts}
}
\frac{({\dconcentration}_{,\mathcal{P}}\Strains)^{\overline{\counts}}}{({\cconcentration}_{,\mathcal{P}}\Strains)^{\overline{\counts}}}
\frac{
\rate_{\mathcal{P}}^{\counts}
}{\counts!}
\left[
({\cconcentration}_{,\mathcal{P}}\Strains)^{\overline{\counts}}
\!-\!
({\cconcentration}_{,\mathcal{P}}\Strains(1 \!-\! \freq^{(\strain_1)}_{\mathcal{P}}))^{\overline{\counts}}
\right.
\\
&&&
\displaystyle
\left.
\hphantom{
\sum_{\counts=2}^{\infty}
\frac{
(\dconcentration\Strains)^{\dconcentration\Strains}
}{
(\dconcentration\Strains \!+\! \rate)^{\dconcentration\Strains \!+\! \counts}
}
\frac{(\dconcentration\Strains)^{\overline{\counts}}}{(\cconcentration\Strains)^{\overline{\counts}}}
\frac{
\rate^{\counts}
}{\counts!}
}
\!-\!
({\cconcentration}_{,\mathcal{P}}\Strains(1 \!-\! \freq^{(\strain_2)}_{\mathcal{P}}))^{\overline{\counts}}
\!+\!
({\cconcentration}_{,\mathcal{P}}\Strains(1 \!-\! \freq^{(\strain_1)}_{\mathcal{P}} \!-\! \freq^{(\strain_2)}_{\mathcal{P}} ) )^{\overline{\counts}}
\right]
 \,,
  \end{array} $
\end{adjustbox}
\end{equation}
the projected target parameters for the gluttonous control groups
\[\Psi^{(\strain_1, \strain_2)}_{\control,g, \operatorname{ghNBDM}} :=  \Psi^{(\strain_1, \strain_2)}_{\control,g} \circ \iota_{\operatorname{ghNBDM}} \circ \Pi_{\operatorname{ghNBDM}}\]
  are defined as
\begin{equation}
  \label{eq:ghNBDM_projected_parameters_gluttonous_control}
  \begin{array}{rccl}
    \Psi^{(\strain_1, \strain_2)}_{\control,g, \operatorname{ghNBDM}}: & \mathcal{M} & \to & [0,1] \\
    &&& \\
    \Psi^{(\strain_1, \strain_2)}_{\control,g, \operatorname{ghNBDM}}:
&\mathcal{P}&
\mapsto &
\probability[\iota_{\operatorname{ghNBDM}}(\Pi_{\operatorname{ghNBDM}}(\mathcal{P}))]*{ \support(\vabundance(0)) \supseteq \{ \strain_2\} \,, \, \strain_1 \not\in \support(\vabundance(0))  }\\
&&=&
\probability[\operatorname{ghNBDM}(\rate_{\mathcal{P}}, \vfreq_{\mathcal{P}}, {\dconcentration}_{,\mathcal{P}}, {\cconcentration}_{,\mathcal{P}})]*{ \support(\vabundance(0)) \supseteq \{ \strain_2\} \,, \, \strain_1 \not\in  \support(\vabundance(0))  } \\[10pt]
&&=&
\displaystyle
\sum_{\counts=1}^{\infty}
\!
\frac{
({\dconcentration}_{,\mathcal{P}}\Strains)^{{\dconcentration}_{,\mathcal{P}}\Strains}
}{
({\dconcentration}_{,\mathcal{P}}\Strains \!+\! \rate_{\mathcal{P}})^{{\dconcentration}_{,\mathcal{P}}\Strains \!+\! \counts}
}
\frac{({\dconcentration}_{,\mathcal{P}}\Strains)^{\overline{\counts}}}{({\cconcentration}_{,\mathcal{P}}\Strains)^{\overline{\counts}}}
\frac{
\rate_{\mathcal{P}}^{\counts}
}{\counts!} 
\,\, \cdot 
\\
&&&
\displaystyle
\hphantom{
\sum_{\counts=1}^{\infty}
\sum_{\counts=1}^{\infty}
}
\left[
({\cconcentration}_{,\mathcal{P}}\Strains(1 \!-\! \freq^{(\strain_1)}_{\mathcal{P}} ))^{\overline{\counts}}
\!-\!
({\cconcentration}_{,\mathcal{P}}\Strains(1 \!-\! \freq^{(\strain_1)}_{\mathcal{P}} \!-\! \freq^{(\strain_2)}_{\mathcal{P}} ) )^{\overline{\counts}}
\right]
 \,,
  \end{array}
\end{equation}
the projected target parameters for the picky treatment groups
\[\Psi^{(\strain_1, \strain_2)}_{\treatment, p, \operatorname{ghNBDM}} := \Psi^{(\strain_1, \strain_2)}_{\treatment, p} \circ \iota_{\operatorname{ghNBDM}} \circ \Pi_{\operatorname{ghNBDM}}\]
are defined as
\begin{equation}
\label{eq:ghNBDM_projected_parameters_picky_treatment}
\begin{adjustbox}{max width=\textwidth,keepaspectratio}
$\begin{array}{rccl}
\Psi^{(\strain_1, \strain_2)}_{\treatment, p, \operatorname{ghNBDM}} : & \mathcal{M} & \to & [0,1] \\
  &&& \\
\Psi^{(\strain_1, \strain_2)}_{\treatment, p, \operatorname{ghNBDM}} :  &
\mathcal{P} & \mapsto &
\probability[\iota_{\operatorname{ghNBDM}}(\Pi_{\operatorname{ghNBDM}}(\mathcal{P}))]*{ \support(\vabundance(0)) = \{\strain_1, \strain_2\}  } \\
&&=&
\probability[\operatorname{ghNBDM}(\rate_{\mathcal{P}}, \vfreq_{\mathcal{P}}, {\dconcentration}_{,\mathcal{P}}, {\cconcentration}_{,\mathcal{P}})]*{ \support(\vabundance(0)) = \{\strain_1, \strain_2\}  } \\[10pt]
&&=&
\displaystyle
\sum_{\counts=2}^{\infty}
\frac{
({\dconcentration}_{,\mathcal{P}} \Strains)^{{\dconcentration}_{,\mathcal{P}} \Strains}
}{
({\dconcentration}_{,\mathcal{P}} \Strains \!+\! \rate_{\mathcal{P}})^{{\dconcentration}_{,\mathcal{P}} \Strains + \counts}
}
\frac{
({\dconcentration}_{,\mathcal{P}} \Strains)^{\overline{\counts}}
}{
({\cconcentration}_{,\mathcal{P}} \Strains)^{\overline{\counts}}
}
\frac{\rate_{\mathcal{P}}^{\counts}}{\counts!}
\left[
({\cconcentration}_{,\mathcal{P}} \Strains (\freq^{(\strain_1)}_{\mathcal{P}} \!+\! \freq^{(\strain_2)}_{\mathcal{P}}))^{\overline{\counts}}
\right.
\\
&&&
\displaystyle
\hphantom{
\sum_{\counts=2}^{\infty}
\frac{
({\dconcentration}_{,\mathcal{P}} \Strains)^{{\dconcentration}_{,\mathcal{P}} \Strains}
}{
({\dconcentration}_{,\mathcal{P}} \Strains \!+\! \rate_{\mathcal{P}})^{{\dconcentration}_{,\mathcal{P}} \Strains + \counts}
}
\frac{
({\dconcentration}_{,\mathcal{P}} \Strains)^{\overline{\counts}}
}{
({\cconcentration}_{,\mathcal{P}} \Strains)^{\overline{\counts}}
}
\frac{\rate_{\mathcal{P}}^{\counts}}{\counts!}
}
\left.
- ({\cconcentration}_{,\mathcal{P}} \Strains \freq^{(\strain_1)}_{\mathcal{P}})^{\overline{\counts}}
- ({\cconcentration}_{,\mathcal{P}} \Strains \freq^{(\strain_2)}_{\mathcal{P}})^{\overline{\counts}}
\right]
\,,
\end{array} $
\end{adjustbox}
\end{equation}
and the projected target parameters for the picky control groups
\[ \Psi^{(\strain_1, \strain_2)}_{\control, p, \operatorname{ghNBDM}} :=  \Psi^{(\strain_1, \strain_2)}_{\control, p} \circ \iota_{\operatorname{ghNBDM}} \circ \Pi_{\operatorname{ghNBDM}}\]
are defined as
\begin{equation}
  \label{eq:ghNBDM_projected_parameters_picky_control}
  \begin{array}{rccl}
    \Psi^{(\strain_1, \strain_2)}_{\control, p, \operatorname{ghNBDM}}: & \mathcal{M} & \to & [0,1] \\
    &&& \\
    \Psi^{(\strain_1, \strain_2)}_{\control, p, \operatorname{ghNBDM}}: &
\mathcal{P}& \mapsto &
\probability[\iota_{\operatorname{ghNBDM}}(\Pi_{\operatorname{ghNBDM}}(\mathcal{P}))]*{ \support(\vabundance(0)) = \{\strain_2\}  } \\
&& = &
\probability[\operatorname{ghNBDM}(\rate_{\mathcal{P}}, \vfreq_{\mathcal{P}}, {\dconcentration}_{,\mathcal{P}}, {\cconcentration}_{,\mathcal{P}})]*{ \support(\vabundance(0)) = \{\strain_2\}  } \\[10pt]
&& = &
\displaystyle
\sum_{\counts=1}^{\infty}
\frac{
({\dconcentration}_{,\mathcal{P}} \Strains)^{{\dconcentration}_{,\mathcal{P}} \Strains}
}{
({\dconcentration}_{,\mathcal{P}} \Strains + \rate_{\mathcal{P}})^{{\dconcentration}_{,\mathcal{P}} \Strains + \counts}
}
\frac{
({\dconcentration}_{,\mathcal{P}} \Strains)^{\overline{\counts}}
}{
({\cconcentration}_{,\mathcal{P}} \Strains)^{\overline{\counts}}
}
\frac{\rate_{\mathcal{P}}^{\counts}}{\counts!}
({\cconcentration}_{,\mathcal{P}} \Strains \freq^{(\strain_2)}_{\mathcal{P}})^{\overline{\counts}}
 \,.
  \end{array}
\end{equation}
To estimate these, we just need to estimate $\rate_{\mathcal{P}}$, $\vfreq_{\mathcal{P}}$, ${\dconcentration}_{,\mathcal{P}}$, and ${\cconcentration}_{,\mathcal{P}}$ and then use the corresponding plugin estimators. Estimating $\rate_{\mathcal{P}}$, $\vfreq_{\mathcal{P}}$, ${\dconcentration}_{,\mathcal{P}}$, and ${\cconcentration}_{,\mathcal{P}}$ also gives us plugin estimators for the probabilities associated with \textit{all} $2^{\Species}$ subsets of $[\Species]$. So again, like for the hPoMu working model, ``information is shared'' between estimates for the probabilities of related events. Again, like for the hPoMu working model, as long as all estimates of the $\freq^{(\strain)}_{\mathcal{P}}$ are nonzero, all estimated probabilities will also all be nonzero. Thus these estimates should again be useful particularly for the important large $\Strains$ regime.

\subsection{Projected Target Parameters Under hNBDM Working Model}
\label{sec:proj-targ-param-hnbdm-appendix}

Let $\iota_{\operatorname{hNBDM}}: \mathcal{M}_{hNBDM} \to \mathcal{M}$, $\iota_{\operatorname{hNBDM}}: \mathcal{P} \mapsto \mathcal{P}$ denote the inclusion map\footnote{Cf. section \ref{sec:append-funct-notat}, in particular the last part of section \ref{sec:conv-funct-comp}.}
\[{\mathcal{M}_{\operatorname{hNBDM}} \xhookrightarrow{\iota_{\operatorname{hNBDM}}} \mathcal{M}} \,. \]
Suppose $\Pi_{\operatorname{hNBDM}} : \mathcal{M} \to \mathcal{M}_{\operatorname{hNBDM}}$ is a projection\footnote{Cf. section \ref{sec:append-funct-notat}, in particular section \ref{sec:function-notation} on the notation for functions used herein.}:
\begin{equation}
  \label{eq:hNBDM_projection}
  \Pi_{\operatorname{hNBDM}} : \mathcal{P} \mapsto \argmin_{\tilde{\mathcal{P}} \in \mathcal{M}_{\operatorname{hNBDM}} }  \int_{\N^{\Strains}} L(\tilde{\mathcal{P}}(\vcounts)) d\mathcal{P}(\vcounts) \,,
\end{equation}
for a given loss function $L$, e.g. $L: p \mapsto - \log(p)$, onto the working parametric submodel $\mathcal{M}_{\operatorname{hNBDM}}$ from the nonparametric model $\mathcal{M}$. One should have the relationships:
\begin{itemize}
\item ${\Pi_{\operatorname{hNBDM}} \circ \iota_{\operatorname{hNBDM}} = \operatorname{Id}_{\operatorname{hNBDM}} : \mathcal{M}_{\operatorname{hNBDM}} \to \mathcal{M}_{\operatorname{hNBDM}}}$, and
\item  ${ \iota_{\operatorname{hNBDM}} \circ \Pi_{\operatorname{hNBDM}}  : \mathcal{M} \to \mathcal{M}}$ is idempotent.
\end{itemize}

Given the projection of a distribution $\mathcal{P}$ from the non-parametric model $\mathcal{M}$ onto the working parametric submodel $\mathcal{M}_{\operatorname{hNBDM}}$, we can now evaluate any of the non-parametrically defined parameters (\ref{eq: formal_gluttonous_treatment_probability_definition}), (\ref{eq: formal_gluttonous_control_probability_definition_intro}), (\ref{eq: formal_picky_treatment_probability_definition}), (\ref{eq: formal_picky_control_probability_definition}) from section \ref{sec:defin-thro-intro} at the projection, defining a new set of target parameters. This new set of target parameters corresponding to $\mathcal{M}_{\operatorname{hNBDM}}$ is defined below.

For a given $\mathcal{P} \in \mathcal{M}$, let $\iota_{\operatorname{hNBDM}}(\Pi_{\operatorname{hNBDM}}(\mathcal{P})) =: \operatorname{hNBDM}(\rate_{\mathcal{P}}, \vfreq_{\mathcal{P}}, {\concentration}_{\mathcal{P}}) \in \mathcal{M}_{\operatorname{hNBDM}}$. The projected target parameters for the gluttonous treatment groups
\[\Psi^{(\strain_1, \strain_2)}_{\treatment, g, \operatorname{hNBDM}} 
:= \Psi^{(\strain_1, \strain_2)}_{\treatment, g} \circ  \iota_{\operatorname{hNBDM}} \circ \Pi_{\operatorname{hNBDM}} \]
are defined as
\begin{equation}
  \label{eq:hNBDM_projected_parameters_gluttonous_treatment}
  \begin{adjustbox}{max width=\textwidth,keepaspectratio}
$  \begin{array}{rccl}
    \Psi^{(\strain_1, \strain_2)}_{\treatment, g, \operatorname{hNBDM}}: & \mathcal{M} & \to & [0,1] \\
    &&& \\
\Psi^{(\strain_1, \strain_2)}_{\treatment, g, \operatorname{hNBDM}}:
& \mathcal{P} &
\mapsto &
\probability[\iota_{\operatorname{hNBDM}}(\Pi_{\operatorname{hNBDM}}(\mathcal{P}))]*{ \support(\vabundance(0)) \supseteq \{\strain_1, \strain_2\}  }
\\
&&=&
\probability[\operatorname{hNBDM}(\rate_{\mathcal{P}}, \vfreq_{\mathcal{P}}, {\concentration}_{\mathcal{P}})]*{ \support(\vabundance(0)) \supseteq \{\strain_1, \strain_2\}  } \\[10pt]
&&=&
\displaystyle
\left(
1
-
\left(
\frac{
\concentration_{\mathcal{P}} \Strains\freq^{(\strain_1)}_{\mathcal{P}}
}{
\concentration_{\mathcal{P}} \Strains\freq^{(\strain_1)}_{\mathcal{P}} 
+
\freq^{(\strain_1)}_{\mathcal{P}} \rate_{\mathcal{P}}
}
\right)^{\concentration_{\mathcal{P}} \Strains\freq^{(\strain_1)}_{\mathcal{P}}}
\right)
\cdot
\left(
1
-
\left(
\frac{
\concentration_{\mathcal{P}} \Strains\freq^{(\strain_2)}_{\mathcal{P}}
}{
\concentration_{\mathcal{P}} \Strains\freq^{(\strain_2)}_{\mathcal{P}} 
+
\freq^{(\strain_2)}_{\mathcal{P}} \rate_{\mathcal{P}}
}
\right)^{\concentration_{\mathcal{P}} \Strains\freq^{(\strain_2)}_{\mathcal{P}}}
\right)
 \,,
  \end{array}$
\end{adjustbox}
\end{equation}
the projected target parameters for the gluttonous control groups
\[ \Psi^{(\strain_1, \strain_2)}_{\control,g, \operatorname{hNBDM}} :=  \Psi^{(\strain_1, \strain_2)}_{\control,g} \circ \iota_{\operatorname{hNBDM}} \circ \Pi_{\operatorname{hNBDM}}\]
are defined as
\begin{equation}
  \label{eq:hNBDM_projected_parameters_gluttonous_control}
  \begin{adjustbox}{max width=\textwidth,keepaspectratio}
 $ \begin{array}{rccl}
    \Psi^{(\strain_1, \strain_2)}_{\control,g, \operatorname{hNBDM}}: & \mathcal{M} & \to & [0,1] \\
    &&& \\
    \Psi^{(\strain_1, \strain_2)}_{\control,g, \operatorname{hNBDM}}:
&\mathcal{P}&
\mapsto &
\probability[\iota_{\operatorname{hNBDM}}(\Pi_{\operatorname{hNBDM}}(\mathcal{P}))]*{ \support(\vabundance(0)) \supseteq \{ \strain_2\} \,, \, \strain_1 \not\in \support(\vabundance(0))  }\\
&&=&
\probability[\operatorname{hNBDM}(\rate_{\mathcal{P}}, \vfreq_{\mathcal{P}}, {\concentration}_{\mathcal{P}})]*{ \support(\vabundance(0)) \supseteq \{ \strain_2\} \,, \, \strain_1 \not\in  \support(\vabundance(0))  } \\[10pt]
&&=&
\displaystyle
\left(
\frac{
\concentration_{\mathcal{P}} \Strains\freq^{(\strain_1)}_{\mathcal{P}}
}{
\concentration_{\mathcal{P}} \Strains\freq^{(\strain_1)}_{\mathcal{P}} 
+
\freq^{(\strain_1)}_{\mathcal{P}} \rate_{\mathcal{P}}
}
\right)^{\concentration_{\mathcal{P}} \Strains\freq^{(\strain_1)}_{\mathcal{P}}}
\cdot
\left(
1
-
\left(
\frac{
\concentration_{\mathcal{P}} \Strains\freq^{(\strain_2)}_{\mathcal{P}}
}{
\concentration_{\mathcal{P}} \Strains\freq^{(\strain_2)}_{\mathcal{P}} 
+
\freq^{(\strain_2)}_{\mathcal{P}} \rate_{\mathcal{P}}
}
\right)^{\concentration_{\mathcal{P}} \Strains\freq^{(\strain_2)}_{\mathcal{P}}}
\right)
 \,,
  \end{array}$
\end{adjustbox}
\end{equation}
the projected target parameters for the picky treatment groups
\[\Psi^{(\strain_1, \strain_2)}_{\treatment, p, \operatorname{hNBDM}} := \Psi^{(\strain_1, \strain_2)}_{\treatment, p} \circ \iota_{\operatorname{hNBDM}} \circ \Pi_{\operatorname{hNBDM}}\]
are defined as
\begin{equation}
  \label{eq:hNBDM_projected_parameters_picky_treatment}
  \begin{adjustbox}{max width=\textwidth,keepaspectratio}
$\begin{array}{rccl}
\Psi^{(\strain_1, \strain_2)}_{\treatment, p, \operatorname{hNBDM}} : & \mathcal{M} & \to & [0,1] \\
  &&& \\
\Psi^{(\strain_1, \strain_2)}_{\treatment, p, \operatorname{hNBDM}} :  &
\mathcal{P} & \mapsto &
\probability[\iota_{\operatorname{hNBDM}}(\Pi_{\operatorname{hNBDM}}(\mathcal{P}))]*{ \support(\vabundance(0)) = \{\strain_1, \strain_2\}  } \\
&&=&
\probability[\operatorname{hNBDM}(\rate_{\mathcal{P}}, \vfreq_{\mathcal{P}}, {\concentration}_{\mathcal{P}})]*{ \support(\vabundance(0)) = \{\strain_1, \strain_2\}  } \\[10pt]
&&=&
\displaystyle
\left(
1
-
\left(
\frac{
\concentration_{\mathcal{P}} \Strains\freq^{(\strain_1)}_{\mathcal{P}}
}{
\concentration_{\mathcal{P}} \Strains\freq^{(\strain_1)}_{\mathcal{P}} 
+
\freq^{(\strain_1)}_{\mathcal{P}} \rate_{\mathcal{P}}
}
\right)^{\concentration_{\mathcal{P}} \Strains\freq^{(\strain_1)}_{\mathcal{P}}}
\right)
\cdot
\left(
1
-
\left(
\frac{
\concentration_{\mathcal{P}} \Strains\freq^{(\strain_2)}_{\mathcal{P}}
}{
\concentration_{\mathcal{P}} \Strains\freq^{(\strain_2)}_{\mathcal{P}} 
+
\freq^{(\strain_2)}_{\mathcal{P}} \rate_{\mathcal{P}}
}
\right)^{\concentration_{\mathcal{P}} \Strains\freq^{(\strain_2)}_{\mathcal{P}}}
\right)
\cdot
\\
&&&
\displaystyle
\left(
\frac{
\concentration_{\mathcal{P}} \Strains (1 \!-\! \freq^{(\strain_1)}_{\mathcal{P}} \!-\! \freq^{(\strain_2)}_{\mathcal{P}})
}{
\concentration_{\mathcal{P}} \Strains (1 \!-\! \freq^{(\strain_1)}_{\mathcal{P}} \!-\! \freq^{(\strain_2)}_{\mathcal{P}})
+
(1 \!-\! \freq^{(\strain_1)}_{\mathcal{P}} \!-\! \freq^{(\strain_2)}_{\mathcal{P}}) \rate_{\mathcal{P}}
}
\right)^{\concentration_{\mathcal{P}} \Strains (1 - \freq^{(\strain_1)}_{\mathcal{P}} - \freq^{(\strain_2)}_{\mathcal{P}})}
\,,
\end{array}$
\end{adjustbox}
\end{equation}
and the projected target parameters for the picky control groups
\[ \Psi^{(\strain_1, \strain_2)}_{\control, p, \operatorname{hNBDM}} :=  \Psi^{(\strain_1, \strain_2)}_{\control, p} \circ \iota_{\operatorname{hNBDM}} \circ \Pi_{\operatorname{hNBDM}}\]
are defined as
\begin{equation}
  \label{eq:hNBDM_projected_parameters_picky_control}
  \begin{array}{rccl}
    \Psi^{(\strain_1, \strain_2)}_{\control, p, \operatorname{hNBDM}}:  & \mathcal{M} & \to & [0,1] \\
    &&& \\
    \Psi^{(\strain_1, \strain_2)}_{\control, p, \operatorname{hNBDM}}: &
\mathcal{P}& \mapsto &
\probability[\iota_{\operatorname{hNBDM}}(\Pi_{\operatorname{hNBDM}}(\mathcal{P}))]*{ \support(\vabundance(0)) = \{\strain_2\}  } \\
&& = &
\probability[\operatorname{hNBDM}(\rate_{\mathcal{P}}, \vfreq_{\mathcal{P}}, {\concentration}_{\mathcal{P}})]*{ \support(\vabundance(0)) = \{\strain_2\}  } \\[10pt]
&& = &
\displaystyle
\left(
1
-
\left(
\frac{
\concentration_{\mathcal{P}} \Strains\freq^{(\strain_2)}_{\mathcal{P}}
}{
\concentration_{\mathcal{P}} \Strains\freq^{(\strain_2)}_{\mathcal{P}} 
+
\freq^{(\strain_2)}_{\mathcal{P}} \rate_{\mathcal{P}}
}
\right)^{\concentration_{\mathcal{P}} \Strains\freq^{(\strain_2)}_{\mathcal{P}}}
\right)
\cdot
\\
&&&
\displaystyle
\left(
\frac{
\concentration_{\mathcal{P}} \Strains (1 \!-\! \freq^{(\strain_2)}_{\mathcal{P}})
}{
\concentration_{\mathcal{P}} \Strains (1 \!-\! \freq^{(\strain_2)}_{\mathcal{P}})
+
(1 \!-\! \freq^{(\strain_2)}_{\mathcal{P}}) \rate_{\mathcal{P}}
}
\right)^{\concentration_{\mathcal{P}} \Strains (1 - \freq^{(\strain_2)}_{\mathcal{P}})}
 \,.
  \end{array}
\end{equation}
Note that (cf. Lemma \ref{lem:exponential_ratios} from section \ref{sec:append-results-bound}) as $\concentration \to \infty$:
\begin{equation}
  \label{eq:NB_probs_approach_poisson_probs}
  \lim_{\concentration \to \infty} \left(  \frac{ \concentration \Strains \freq^{(\strain)}  }{ \concentration\Strains \freq^{(\strain)} + \freq^{(\strain)} \rate  } \right)^{\concentration\Strains\freq^{(\strain)}} = e^{-\freq^{(\strain)} \rate} \,,
\end{equation}
and therefore equations (\ref{eq:hNBDM_projected_parameters_gluttonous_treatment}), (\ref{eq:hNBDM_projected_parameters_gluttonous_control}), (\ref{eq:hNBDM_projected_parameters_picky_treatment}), (\ref{eq:hNBDM_projected_parameters_picky_control}) \textit{do} approach equations (\ref{eq:hPoMu_projected_parameters_gluttonous_treatment}), (\ref{eq:hPoMu_projected_parameters_gluttonous_control}), (\ref{eq:hPoMu_projected_parameters_picky_treatment}), (\ref{eq:hPoMu_projected_parameters_picky_control}) respectively in the limit as $\concentration_{\mathcal{P}} \to \infty$, as we would have expected a priori. To estimate  (\ref{eq:hNBDM_projected_parameters_gluttonous_treatment}), (\ref{eq:hNBDM_projected_parameters_gluttonous_control}), (\ref{eq:hNBDM_projected_parameters_picky_treatment}), (\ref{eq:hNBDM_projected_parameters_picky_control}), we just need to estimate $\rate_{\mathcal{P}}$, $\vfreq_{\mathcal{P}}$, and ${\concentration}_{\mathcal{P}}$ and then use the corresponding plugin estimators. Again, estimating $\rate_{\mathcal{P}}$, $\vfreq_{\mathcal{P}}$, and ${\concentration}_{\mathcal{P}}$ also gives us plugin estimators for the probabilities associated with \textit{all} $2^{\Species}$ subsets of $[\Species]$. Thus like for the hPoMu and ghNBDM working models, again ``information is shared'' between estimates for the probabilities of related events and all estimated probabilities will be nonzero as long as all estimates of the $\freq^{(\strain)}_{\mathcal{P}}$ are nonzero. While the $\dconcentration = \cconcentration$ constraint provides less flexibility than the full ghNBDM working model, it still has both more flexibility than the hPoMu working model and plugin estimators with the same low level of complexity as those of the hPoMu working model.

\section{Toy Model}
\label{sec:toy-model}

Consider the following toy model. Droplets have one cell with $50\%$ probability, two cells with two different strains with $50\%$ probability. In both cases the distribution is assumed Multinomial with parameters corresponding to population relative abundances. Now consider the effect of strain $\strain_1$, with relative abundance $0.01\%$, on strain $\strain_2$, with relative abundance $20\%$. If the total number of droplets is $\Droplets = 500,000$, then we expect $50,000$ droplets containing strain $\strain_2$ only, whereas we would expect only $(50\% \times 20\% \times 0.01\%) \times 500,000 = 5$ droplets containing both strains. Cf. section \ref{sec:why-expect-posit} which expresses a similar idea.

Keep in mind that the reality is actually almost always \textit{worse} than this toy model. For example, there will be a large and non-negligible number of empty droplets with zero cells. Moreover, when droplets are non-empty, all of the simplifying assumptions implicitly made above can and often will fail to be true. When any given strain is present, it's possible for there to be more than one cell belonging to that strain in the droplet (the ``multiple representatives problem'', cf. section \ref{sec:append-mult-repr}). When the droplet is non-empty, there can be more than two strains present in the droplet. Finally, even when conditioning only on droplets with either one or two strains present, there usually will not be a $1:1$ ratio of droplets with one strain versus droplets with two strains.

Even the $1:1$ ratio from the toy model isn't ideal, because there are $O(\Strains^2)$ possible two-strain treatment groups, and $O(\Strains)$ possible one-strain control groups, so actually in an ideal world we would like a $ \approx 1: \Strains$ ratio of droplets with one strain versus droplets with two strains. So even for the toy model, the sizes of the treatment groups compared to the sizes of the control groups would become increasingly imbalanced as the number of strains becomes larger. This is problematic to the extent that we want $\Strains$ as large as possible.

\end{coolsubappendices}

\end{coolcontents}

\clearpage
\setcounter{footnote}{0}
\pagestyle{myheadings}
\part{Modelling Initial Formation of Droplets}
\label{part:modell-init-form}

Inferring ecological models of complex microbial communities entails understanding the interactions between microbes and how they affect each other's growth. The initial counts of cells from each strain within a droplet are the baseline for comparing the growth of microbes under different conditions. Being better able to better characterize this baseline for growth means being better able to characterize effects on growth. Thus modelling and understanding these initial counts will help us to understand microbial interactions.

The broader field for part \ref{part:modell-init-form} is the study of statistical models for multivariate count data. One common approach for defining and categorizing these involves generalized linear models. Cf. \cite{multivar_count_regression_topic_model} or \cite{multivar_count_regression_rna_seq}. Hierarchical models, often based in some way on the Poisson distribution (which is important for univariate count data), are also commonly used. Cf. \cite{Inouye2017} for a review. Models for multivariate count data are important to many problems, such as topic modelling. See \cite{Zhou2018} or \cite{multivar_count_regression_topic_model} for examples.

Part \ref{part:modell-init-form} concerns more particularly statistical models for multivariate count data arising from two sources: high-throughput biological assays and ecological studies.

For multivariate count data arising from high-throughput biological assays, such models have been considered for (single-cell) droplet-based microfluidics experiments, and for RNA-seq experiments. Generally the Poisson distribution is used as the default statistical model for single-cell droplet-based microfluidics experiments \cite{Collins2015}. (Although this is univariate count data, not multivariate. See any of \cite{single_cell_1}, \cite{Kintses2010}, \cite{single_cell_2}, or \cite{Guo2012} for reviews of such experiments.) Examples of work considering statistical models for characterizing count data from RNA-seq experiments include \cite{Yu2013} and \cite{multivar_count_regression_rna_seq}.

The literature for multivariate count data arising from ecological studies is likely more developed. Relevant studies are often described as investigating ``joint species abundances'' or ``joint species distribution''. See \cite{dirichlet_process}, \cite{Ovaskainen2017}, and \cite{Chiquet2021}. Work such as \cite{Billheimer2001} has considered the possible effects of preferential (dis)association of certain organisms (cf. section \ref{sec:clar-comp-heter}). Such work is not limited only to macroecology. Studies such as \cite{Biswas2015}, \cite{Yu2014}, \cite{Harrison2020}, or \cite{bias_framework} investigate count data from microbial ecology.

\paragraph{Specific Problem}
The next four chapters of this thesis discuss the ``$t=0$ experiment''. This means focusing on candidate statistical working models for the first phase of the overall experiment described in section \ref{sec:using-morei-char}. Chapter \ref{cha:model-init-form} defines and discusses the candidate statistical working models. Chapter \ref{chap:model_comparison} qualitatively compares the candidate statistical working models. We want to determine whether any behave substantially differently from the default working model, which is the simplest. Motivated by the results of chapter \ref{chap:model_comparison}, chapter \ref{chap:data_throughput} investigates and compares the target estimands for each of the candidate statistical working models. We want to determine whether substantially different behavior between the candidate working models translates into substantially different values of the target estimands in practice. Chapter \ref{chap:hetero_estimator_performance} constructs estimators and obtains inference for all of the candidate statistical working models. Even under model misspecification, these estimators return sensible results.

\paragraph{Chapter \ref{cha:model-init-form}}

Chapter \ref{cha:model-init-form} belongs to the general field of describing and categorizing statistical models for multivariate count data.
For any particular problem with multivariate count data, some thought is usually needed to find reasonable models.
Chapter \ref{cha:model-init-form} investigates which statistical (working) models are reasonable choices for describing the initial state of a droplet.
Chapter \ref{cha:model-init-form} shows that specific assumptions justify a default working model for this problem.
New working models are derived by relaxing each of these assumptions.

\paragraph{Chapter \ref{chap:model_comparison}}

The problem considered in chapter \ref{chap:model_comparison} falls under the general field of ``model selection''.
Herein we only need to consider the subfield of so-called ``goodness of fit'' tests, in particular those using the classic likelihood ratio statistic.
Concretely, chapter \ref{chap:model_comparison} investigates whether the failure of any of the assumptions identified in chapter \ref{cha:model-init-form} leads to substantially new behavior, and
demonstrates that log likelihood ratio statistics can be used to address this question.
Failure of the sampling without replacement assumption turns out to have negligible effects in practice, but failures of the other assumptions could be important.

\paragraph{Chapter \ref{chap:data_throughput}}

Like chapter \ref{chap:model_comparison}, the problem of chapter \ref{chap:data_throughput} also belongs both to the general field of model selection and to the subfield of goodness of fit tests in particular.
However, the focus herein is on goodness of fit tests for contingency tables using the $\chi^2$-divergence test statistic. (Cf. section \ref{sec:pears-categ-diverg} for terminology.)
Chapter \ref{chap:data_throughput} investigates how failures of the assumptions identified in chapter \ref{cha:model-init-form} affect the targeted estimands from section \ref{sec:defin-thro-intro}.
The results of this chapter confirm that more severe failures of these assumptions lead to more severe discrepancies with the predictions derived from the default working model.
The nature of the effect depends on the chosen grouping of droplets defining the targeted estimands.

\paragraph{Chapter \ref{chap:hetero_estimator_performance}}

The problem of chapter \ref{chap:hetero_estimator_performance} belongs to the general field of point estimation.
Chapter \ref{chap:hetero_estimator_performance} is concerned in particular with the plug-in (also known as ``method of moments'') and maximum likelihood approaches to point estimation.
The chapter investigates how to measure, from the data produced by unincubated droplets, failures of the assumptions identified in chapter \ref{cha:model-init-form}. This is framed as a point estimation problem.
The chapter presents both plugin and maximum likelihood estimators for solving this problem.
In doing so, failures of the violations are also shown to be understandable non-parametrically.

\paragraph{Significance} We have to investigate specific methods (cf. part \ref{part:aver-treatm-effects}) to predict how much data is needed to infer microbial interactions. To configure the experiment in accordance with such predictions, we first need to predict how much usable data will be produced. (Cf. again the discussion from section \ref{sec:motiv-targ-estim}.) Part \ref{part:modell-init-form} helps with the latter kind of prediction. Running the experiment once without incubating the droplets allows us to predict how much usable data will be produced by understanding the initial distribution of cells within droplets. This includes assumptions for the distribution of the total numbers of cells within droplets, and for the distribution (given the total number of cells) of proportions of each strain.

\clearpage
\pagestyle{headings}

\chapter[Motivation for and Definitions of Candidate Statistical Working Models for the Initial Formation of Droplets][Candidate Statistical Working Models for Initial Formation of Droplets]{Motivation for and Definitions of Candidate Statistical Working Models for the Initial Formation of Droplets}
\label{cha:model-init-form}

Herein I show that specific assumptions justify a default working model for the initial state of droplets.  See section \ref{sec:impl-model-assumpt}. I derive new working models by relaxing each of these assumptions. See sections \ref{sec:model-sampl-with}, \ref{sec:model-comp-heter}, \ref{sec:model-dens-heter}, and \ref{sec:model-arbitr-comb}. I discuss concrete ways these assumptions could fail to be true in practice, so the relaxations are not purely hypothetical. See section \ref{sec:failures-hpomu-model}.

Section \ref{sec:backgr-sign-3} provides context for what follows. Section \ref{sec:hier-count-categ} proposes a general framework for defining any statistical model of the initial formation of droplets. Section \ref{sec:hpom-compl-homog} proposes a ``default'' working model of the initial formation of droplets. Section \ref{sec:failures-hpomu-model} explains three ways the assumptions underlying this default working model could fail to be true in practice.

The rest of chapter explains how to model these assumption failures. Section \ref{sec:model-sampl-with} proposes a working model accounting for the finiteness of the microbial population. Section \ref{sec:model-comp-heter} proposes a working model accounting for heterogeneity in the relative abundances of microbial strains throughout the microbial population. Section \ref{sec:model-dens-heter} proposes a working model accounting for heterogeneity in the average number of cells throughout the microbial population. Section \ref{sec:model-arbitr-comb} describes a family of distributions subsuming both of the previous two working models as well as the default working model. Section \ref{sec:conclusion} summarizes the chapter.

\begin{coolcontents}
\section{Background and Significance}
\label{sec:backgr-sign-3}

\paragraph{Broader field}

This chapter belongs to the general field of describing and categorizing statistical models for multivariate count data. The references \cite{Zhou2018} and \cite{multivar_count_regression_topic_model} discuss some examples of these in the field of topic modelling. The article \cite{Inouye2017} is a helpful review of some other relevant statistical models. Approaches related to generalized linear models are often used to define statistical models for multivariate count data, cf. \cite{multivar_count_regression_rna_seq} and \cite{multivar_count_regression_topic_model} for examples. See also the discussion earlier from the introduction to Part \ref{part:modell-init-form}.

In general, some thought is required to find models that are reasonable for a given, specific problem involving multivariate count data. The references \cite{Billheimer2001}, \cite{Biswas2015}, \cite{Ovaskainen2017}, and \cite{Chiquet2021} are examples of statistical models tailored to problems from ecological studies. Similarly, the references \cite{Yu2013}, \cite{Collins2015}, and \cite{multivar_count_regression_rna_seq} are examples of statistical models tailored to the analysis of data from high-throughput biological assays. (See any of \cite{single_cell_1}, \cite{Kintses2010}, \cite{single_cell_2}, or \cite{Guo2012} for reviews of some examples of such experiments.)

\paragraph{Specific problem}

Herein I investigate which statistical (working) models can be used to describe a droplet's initial state. A statistical model of the initial formation of droplets helps us to characterize microbial interactions. One of the major challenges of the MOREI data is the randomness of the initial droplet formation process. Unlike manually plating everything, with MOREI the scientist has no \textit{direct} control over which microbial interactions are observed. This means a loss of control over which questions that can be answered from the data. The premise of MOREI is that the vastly increased throughput makes the tradeoff of decreased control worthwhile. The more control we can retain for the scientist when using MOREI, compared to direct plating, the stronger the case for its superiority over existing methods, and the greater its ability to advance the state of the art. Cf. figure \ref{fig:control_throughput_tradeoff}. 

\begin{figure}[H]
\centering
\includegraphics[width=\textwidth,height=\textheight,keepaspectratio]{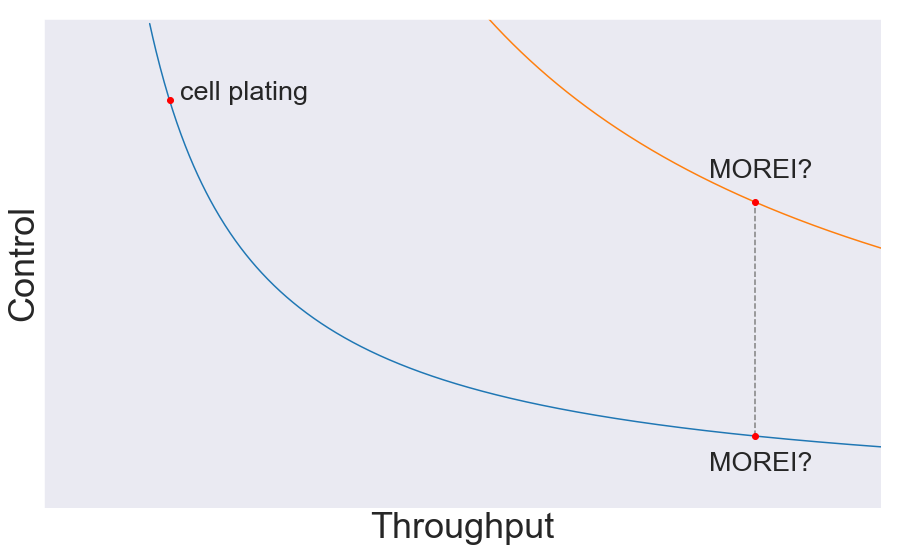}
  \caption[Control-Throughput Tradeoff Curve (heuristic).]{Control-Throughput Tradeoff Curve. Having a predictive statistical model of initial droplet formation would allow us to increase control without sacrificing throughput, moving us to the higher tradeoff curve.}
  \label{fig:control_throughput_tradeoff}
\end{figure}

To restore the scientist's control over which questions they can answer using MOREI data, we want to account for the effects which the randomness of the initial droplet formation process has on the data. To account for these effects, we need to make accurate predictions about what these effects will be. Cf. again section \ref{sec:motiv-targ-estim}. To make such predictions, we need to understand the randomness of the initial droplet formation process. To understand its randomness, we need to have adequate statistical models for describing the initial droplet formation process. Similar previous work involving the selection of statistical models for count data has helped to improve the design of RNA-seq experiments, see e.g. \cite{Yu2013} or \cite{multivar_count_regression_rna_seq}.
I hope that this work is an initial step towards analogous progress in the design of MOREI and related future experiments.

The ultimate goal of making accurate predictions about the droplet formation process may sound overly ambitious. Yet the amount of throughput that we can potentially achieve with a MOREI experiment is enormous. With such enormous sample sizes, the effects of asymptotic concentration of measure phenomena (such as the law of large numbers or the central limit theorem) almost certainly come into play. To phrase it very crudely, ``randomness + enormous sample sizes = quasi-determinism''. To exploit the ``quasi-determinism'' resulting from enormous sample sizes, all we need is to understand and characterize the underlying probability distribution well enough.

\paragraph{Particular approach}

I review how we might model the initial droplet formation process. I begin with a general framework for how we might describe special multivariate count distributions as hierarchical models. Then, within this general framework, I derive specific distributions that follow from given assumptions about the initial droplet formation process. 

An approach similar to that used in \cite{Billheimer2001} might be useful to explore for future work. However, such an approach requires us to infer a correlation pattern based on highly sparse observations\footnote{In the sense that observations are vectors such that most of the entries are zero.}. This would most likely require substantial work to identify a suitable regularization technique to avoid over-fitting spurious correlations. Moreover, even assuming that issue was addressed, there would be the issue of the scientific interpretation. It would be tempting to infer interactions based on possible preferential grouping or distancing of strains in the sampling pool. Yet it is unclear how much such inferred parameters would actually reveal about the spatial preferences of strains, much less whether such spatial preferences actually correspond to interactions. Cf. section \ref{sec:clar-comp-heter}.

Other work, such as \cite{Ovaskainen2017}, \cite{Chiquet2021}, \cite{Biswas2015}, or \cite{Inouye2017}, has also presented possibly relevant working models for multivariate count data. However, like the situation with \cite{Billheimer2001}, all of these fairly sophisticated approaches could possibly cause subtle complications when trying to apply them to the data produced by MOREI.

In any case, the working models considered herein are relatively simple and thus suitable as starting points for generalization, e.g. in the directions of \cite{Billheimer2001} or \cite{Inouye2017}, and so should be considered complementary with those other approaches.

Providing this general framework, and detailed descriptions of simple working models within it, is our first step towards understanding the form of randomness that controls the droplet formation process, and thus ultimately towards allowing the scientist to control which questions they can answer using MOREI data. Cf. again section \ref{sec:motiv-targ-estim}.

\section{Hierarchical Count-Categorical Distributions}
\label{sec:hier-count-categ}

By \textbf{count distribution}, I mean a probability distribution $\mathcal{P}$ with values in the non-negative integers $\N$. $\abundance(0)$ corresponds to a count distribution, as does $\abundance[\strain](0)$ for all $\strain \in [\Strains]$.

Similarly, I define a \textbf{multivariate count distribution} to be a probability distribution taking values in $\N^{\Species} := \bigtimes_{\strain=1}^{\Species} \N$. $\vabundance(0)$ corresponds to a multivariate count distribution. All marginal distributions of a multivariate count distribution are themselves count distributions.

A typical way to specify a multivariate count distribution is to specify the $\Strains$ count distributions that are its marginal distributions along with the statistical dependence structure amongst the marginals. There are infinitely many ways to define such a statistical dependence structure. Arguably the simplest of such ways is to declare the marginal distributions to be mutually independent, but making such a restriction drastically reduces the variety of multivariate count distributions that can be described. In general, directly specifying the statistical dependence structure amongst the marginal distributions can be quite complicated and difficult. Specifying $\Strains$ count distributions can also be cumbersome.

In this thesis I will use another framework for specifying multivariate count distributions. This framework decomposes the multivariate count distribution into a hierarchical\footnote{
Some sources call such distributions ``compound''. I use ``hierarchical'' here instead to avoid confusion. A ``compound Poisson distribution'' often refers to the distribution of a sum of $N$ i.i.d. random variables where $N$ itself is Poisson distributed. Using the latter notion, hPoMu is ``compound Poisson Multinoulli'', but using the former notion (``compound = hierarchical''), hPoMu is ``compound Poisson Multinomial''.
} distribution. One of the distributions in the hierarchical distribution is always easy to describe. If the other distribution (more properly/technically speaking a family of distributions) in the hierarchical distribution is also easy to describe, then this hierarchical framework allows us to describe a multivariate count distribution easily using two distributions.

Consider a random vector $\rvec{X} \in \N^{\Species}$ corresponding to a multivariate count distribution. As mentioned before, this always corresponds to $\Species$ count distributions, one for each of its $\Species$ marginal distributions $X^{(\strain)}$. However, there is also always a way to associate with $\rvec{X}$ a single count distribution, namely the sum of the marginal distributions $X := \sum_{\specie \in [\Species]} X^{(\specie)}$. To define a hierarchical distribution from this, we need to specify a conditional distribution for every possible value of $X$. In other words, instead of considering $X^{(\strain)}$ for every $\strain \in [\Strains]$ to specify $\rvec{X}$, I propose instead considering the pair $X$ and ${\rvec{X} | X}$.

For any $\counts \in \N$, ${\rvec{X}|X=\counts}$ takes values in $\{0, \dots, \counts \}^{\Species}$ such that the sum of the marginals equals $\counts$. I define a \textbf{categorical distribution} to be any \footnote{
This definition is much more general than what is more commonly referred to as ``\textit{the} categorical distribution'' (sensu stricto), for which $\counts=1$, and which is also called a Multinoulli distribution. Actually all categorical distributions (sensu lato) with $\counts=1$ have to reduce to a Multinoulli distribution. This is analogous to how they must reduce to the Dirac Delta at $0$ in the case where $\counts=0$. Note also that while for $\counts > 1$ the sum of $\counts$ independent Multinoulli random variables does correspond to a categorical distribution (sensu lato), specifically the Multinomial distribution, not every categorical distribution (sensu lato) for $\counts >1$ is Multinomial. So one could reasonably say that categorical distributions (sensu lato) strictly generalize categorical distributions (sensu stricto) only for $\counts > 1$.
} probability distribution $\mathcal{P}_{\counts}$ that takes values in $\{0, \dots, \counts \}^{\Species}$ with the restriction that the sum of its marginals must equal $\counts$, or which equivalently takes as values partitions of $\counts$ into $\Species$ (possibly empty) distinct categories.

Two observations are worth making here. First, technically speaking this framework can be used to characterize any multivariate count distribution. No restrictions\footnote{
Even the restriction that $\rvec{X}$ is a multivariate count distribution may be immaterial, although it does imply that $X$ is a discrete distribution. That makes rigorous mathematical proofs easier to achieve.
} on $\rvec{X}$ are required to define $X$ and $\rvec{X}|X$. Second is that $\rvec{X}|X$ is not a proper random variable. At best $\rvec{X}|X$ is a family of random variables or perhaps more accurately a ``random random variable''. For every $\counts \in \N$, $\rvec{X}|X=\counts$ is a distinct (proper) random variable with its own distinct probability distribution (a categorical distribution). Thus this framework can only provide a simplification in the case that the infinite family of probability distributions for the $\rvec{X}|X=\counts$ for all ${\counts \in \N}$ can itself be described simply (ideally as a straightforward function of $\counts$). Otherwise this framework clearly complicates\footnote{
In the worst case this framework requires specifying one count distribution and infinitely many completely unrelated categorical distributions. Compare this to specifying $\Species$ count distributions and one description of their statistical dependence structure for the other aforementioned framework.
} the description of $\rvec{X}$.

Given a single count distribution $\mathcal{P}$ and a family of categorical distributions $\mathcal{P}_{\counts}$ for all $\counts \in \N$, we can construct a random vector $\rvec{X}$ following what I call a \textbf{hierarchical count-categorical distribution} as follows: given a random variable $X \sim \mathcal{P}$, draw from the categorical distribution $\mathcal{P}_{\counts}$ whenever the value of $X$ is $\counts$. Observe how, by construction, the sum of the marginals of $\rvec{X}$ is $X$, and that ${(\rvec{X}|X=\counts)} \sim \mathcal{P}_{\counts}$ for all $\counts \in \N$. Again, any multivariate count distribution can be constructed this way, but whether this is a simple description of the multivariate count distribution depends entirely on whether the infinite family of categorical distributions $\mathcal{P}_{\counts}$ for all ${\counts \in \N}$ can be simply described.

In what follows, I indulge in an abuse of terminology by referring to the infinite family of categorical distributions corresponding to the $\rvec{X}|X=\counts$ for all ${\counts \in \N}$ as ``the categorical distribution''. This is to emphasize the henceforth implicit assumption that this infinite family of categorical distributions admits a simple parameterization in terms of $\counts$, simple enough to \textit{not} require much (or any) more effort to describe than a single categorical distribution. When\footnote{
This assumption may not be possible to satisfy for every multivariate count distribution. It is impossible to falsify either way without making the assumption more precise, of course.
} this assumption holds, describing $X$ and $\rvec{X}|X$ is not much more difficult than specifying a single count distribution and a single categorical distribution.

For MOREI, the count distribution part of a hierarchical count-categorical distribution specifies how many cells $\abundance(0)$ are in the droplet. The categorical distribution part specifies which proportions of cells $\vabundance(0)|\abundance(0)=\counts$ belong to each of the $\Strains$ possible strains\footnote{Herein I use ``strains'' to refer equally to strains belonging to the same species(/genus/family/etc.) as well as to strains belonging to different species(/genera/families/etc.), because the distinction is irrelevant for setting up the abstract problem. It may matter for the implementation of a specific experiment.}. Hence using the hierarchical count-categorical distribution framework allows us to split the problem of selecting a statistical model for initial droplet formation into two (smaller and easier) parts: (1) selecting which statistical model describes how many cells are in the droplet via the count distribution and (2) selecting which statistical model describes which strains any cells in the droplet belong to via the categorical distribution. Cf. figure \ref{fig:count_categorical}.

\begin{figure}[H]
  \centering
  \includegraphics[width=\textwidth,height=\textheight,keepaspectratio]{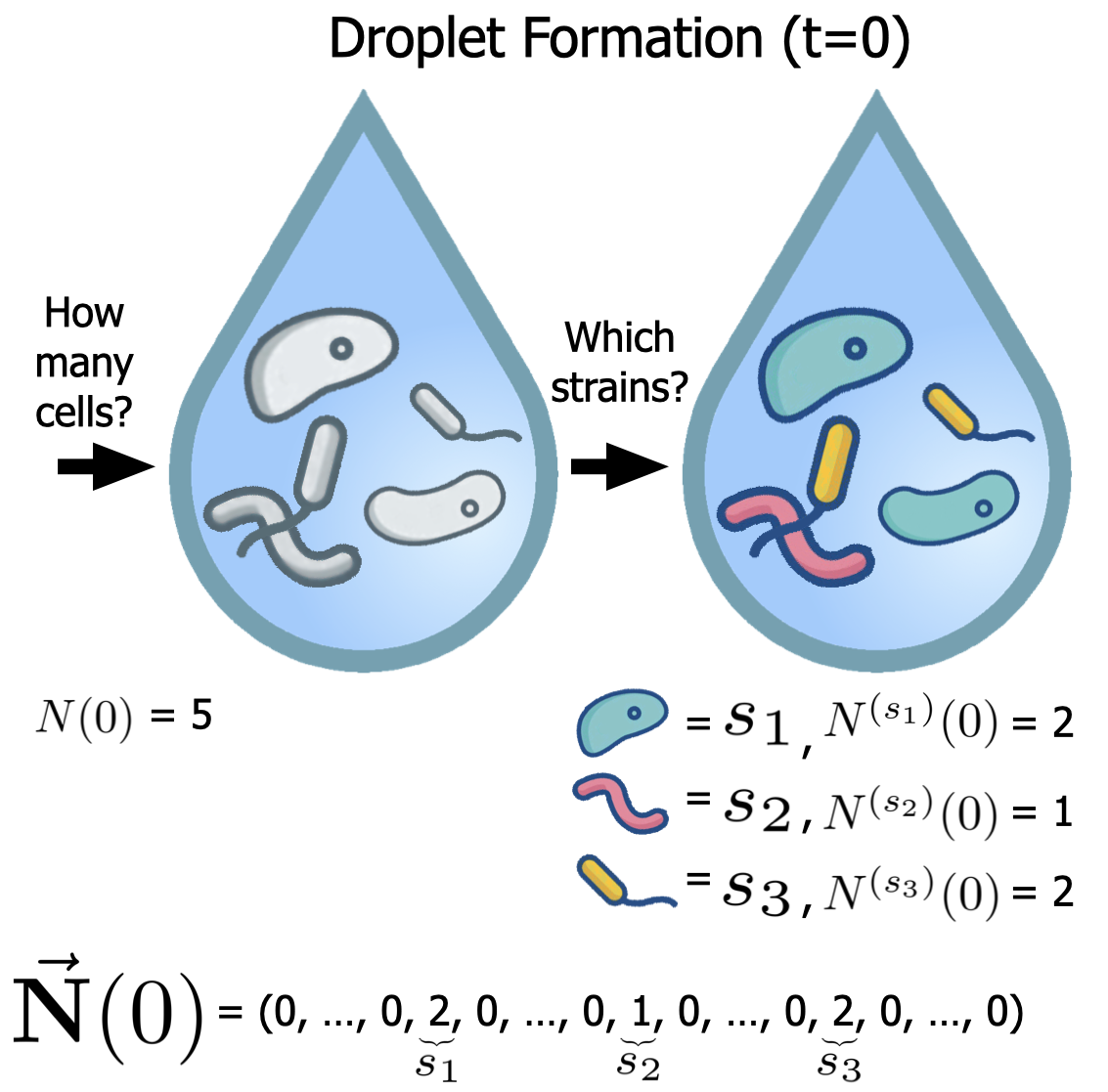}
  \caption{Schematic of count-categorical distributions and notation.}
  \label{fig:count_categorical}
\end{figure}

Splitting the problem up this way not only simplifies it, but also makes it easier to connect the model with details of the experimental setup. For example, because we expect the dilution of cells in the microbial community sample to only affect the number of cells that end up in each droplet, we can connect it with a parameter for the count distribution (without affecting the categorical distribution). Similarly, because we expect the relative abundances of the strains in the microbial community sample to only affect the proportions of the cells in each droplet belonging to each strain, we can connect them to parameters for the categorical distribution (without affecting the count distribution).

\subsection{Naming Convention}
\label{sec:naming-convention}

Hierarchical count-categorical distributions will be named (or abbreviated) as follows.
\begin{itemize}
\item First, a prefix ``h'' for ``hierarchical''.
\item Two letters corresponding to the count distribution.
\item Two letters corresponding to the categorical distribution.
\end{itemize}
For both the count and categorical distributions, the pairs of letters are assigned as follows.
\begin{itemize}
\item If the name of the distribution is one word, then the two letters are the first two letters of the word, the first upper case and the second lower case.
\item If the name of the distribution is more than one word, then the two letters are the first two initials of the name, both upper case.
\end{itemize}
For example the ``hierarchical geometric multivariate Wallenius' noncentral hypergeometric distribution'' would be ``hGeMW''.

\subsection{Connection with Topic Modelling}
\label{sec:conn-with-docum}

Imagine a document as a droplet, and the cells within the droplet as words. We can count the number of the words in the document the same way we can count the number of cells in a droplet. Similarly, the way cells can be assigned to strains, words within a document can be assigned to topics. Imagining the strains of the cells as topics of words, we can also count the number of words corresponding to each given topic. Cf. figure \ref{fig:droplet_document_analogy}. Statistical questions related to modelling documents and the topics of words within them is already a well-studied field, given the name ``topic modelling''.

\begin{figure}[H]
  \centering
  \includegraphics[width=0.5\textwidth,height=\textheight,keepaspectratio]{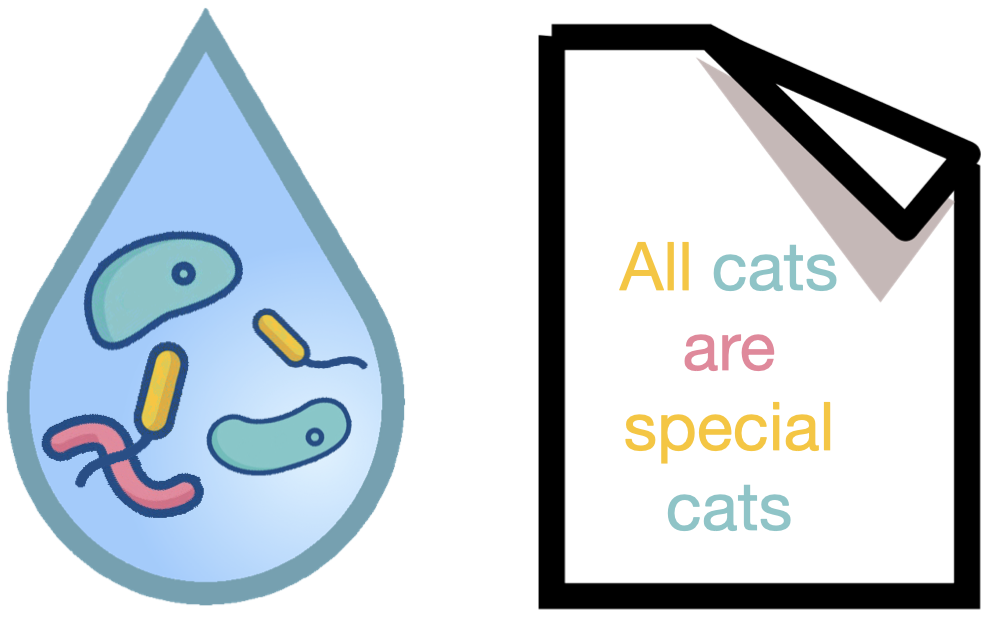}
  \caption[Analogy between droplet capture of microbes and topic modelling of documents.]{Analogy between documents with droplets, words with cells, and topics with strains. Yellow cells $\leftrightarrow$ adjectives; blue cells $\leftrightarrow$ nouns; pink cells $\leftrightarrow$ verbs.}
  \label{fig:droplet_document_analogy}
\end{figure}

Given the above analogy of documents with droplets, words with cells, and topics with strains, a valuable opportunity for future work is to apply what is already known about topic modelling to modelling the initial formation of droplets in MOREI. See e.g. Theorem 1 of \cite{Zhou2018} for one relevant example, which discusses both the hPoMu and hNBDM distributions (using different names). Cf. also \cite{multivar_count_regression_topic_model}. The analogy plausibly suggests that mathematical abstractions relevant for one problem can be relevant for the other problem.

Of course this comes with caveats. One obvious difference is that parameter values for these models which are realistic for topic modelling are unlikely to be realistic for modelling the initial formation of droplets in MOREI, and vice versa. For example, the mean number of words in a document should generally be much larger than the mean number of cells in each droplet. A second caveat is that models like those in \cite{Zhou2018} allow different documents to follow different distributions, whereas we might not want different droplets to follow different distributions (although doing so might be useful for e.g. modelling batch effects).

\section{hPoMu: A Default Working Model}
\label{sec:hpom-compl-homog}

Section \ref{sec:hpomu_definition} gives the explicit definition of the hPoMu working model, mostly for future reference. Section \ref{sec:derivations} gives two heuristic arguments which establish hPoMu as the ``baseline'' or ``default'' (working) model for the initial formation of droplets against which other models are compared. Finally section \ref{sec:impl-model-assumpt} outlines some of the assumptions implicit in the derivation of the hPoMu working model, hinting at some of its potential weaknesses for accurately modelling the initial formation of droplets in practice.

\subsection{hPoMu Definition}
\label{sec:hpomu_definition}

The ``hierarchical Poisson Multinomial'' (abbreviated as hPoMu) distribution is a hierarchical count-categorical distribution whose count distribution is Poisson and whose categorical distribution is Multinomial. Hence the probability mass function (PMF) for droplet $\droplet$ equals
\begin{equation}
  \label{eq:hpomu_defn_1}
  \begin{split}
    \probability*{\vabundance_{\droplet}(0) = \vpopulation} 
 = & 
 \probability*{ \abundance_{\droplet}(0) = \population  } 
\!\cdot\!
\probability*{\vabundance_{\droplet}(0) = \vpopulation | \abundance_{\droplet}(0) = \population} 
\\
=& 
\frac{
e^{-\rate} \rate^{\population}
}{
\population !
}
\cdot
\binom{\population}{\population[1] \! \cdots \! \population[\Species]} \prod_{\specie=1}^{\Species} (\freq^{(\specie)})^{\population[\specie]}
\,.
  \end{split}
\end{equation}
An alternative form of expressing the PMF of hPoMu, which reveals that its marginals are mutually independent, is
\begin{equation}
  \label{eq:hpomu_defn_2}
  \begin{split}
  \probability*{\vabundance_{\droplet}(0) = \vpopulation} 
=  
\prod_{\strain=1}^{\Strains}  
\frac{
e^{-\freq^{(\strain)}\rate} (\freq^{(\strain)}\rate)^{\population[\specie]}
}{
(\population[\specie]) !
}
\,.
  \end{split}
\end{equation}
Via tedious algebra, one can show that equation (\ref{eq:hpomu_defn_2}) really does equal equation (\ref{eq:hpomu_defn_1}). The $\population!$ from the multinomial coefficient cancels with that from the Poisson PMF. The remaining factorials from the multinomial coefficient can be distributed across the product, and $e^{-\rate} \rate^n = \prod_{\specie=1}^{\Species} e^{-\freq^{(\strain)} \rate} \rate^{\population[\specie]}$, which can also be distributed across the product. So despite looking different, the first and second definitions are actually consistent with one another.

Heuristically, equation (\ref{eq:hpomu_defn_2}) says that, during the formation of a droplet, the sampling rate $\rate$ is spread out among the strains according to their frequencies in the population.

\subsection{Derivations}
\label{sec:derivations}

The first derivation determines the count distribution and then the marginals. The second derivation determines the marginals and then the count distribution. 

Assume we are sampling from a population of $\Population \in \mathbb{N}$ cells. Therefore we are sampling from a population of $\freq^{(1)} \Population \in \mathbb{N}$ cells of strain $1$, $\dots$, $\freq^{(\strain)} \Population \in \mathbb{N}$ cells of strain $\strain$, $\dots$, and $\freq^{(\Strains)}\Population \in \mathbb{N}$ cells of strain $\Strains$.

When forming a droplet $\droplet$, we assume each cell has the same small probability $p$ of ending up in the droplet. Thus the indicator random variable of ending up in the droplet is a $\operatorname{Bernoulli}(p)$ random variable for every cell.

We also assume that whether a given cell ends up in the droplet $\droplet$ is completely independent of what happens to any of the other cells.

\subsubsection{First Derivation}
\label{sec:first-derivation}

The number of cells that end up in the droplet $\droplet$ is the sum of all $\Population$ of the indicator random variables. Because the sum of i.i.d. Bernoullis is binomial, the distribution of the number of cells that end up in the droplet is $\operatorname{Binomial}(p,K)$.

Due to the way the experiment was calibrated, we know a priori that the expected number of cells in the droplet $\droplet$ is (some fixed constant) $\rate$. Because the number of cells follows a binomial distribution, the expected number of cells in the droplet $\droplet$ is also $p\Population$. (Thus $p = \frac{\rate}{\Population}$.)

Because $\Population$ is very large, we may as well use the approximating distribution as $\Population \to \infty$. By the ``Law of Small Numbers'', a.k.a. the Poisson Limit Theorem, this distribution is Poisson. In other words:
\begin{equation}
  \label{eq:first_derivation_num_prob}
  \probability*{ \abundance_{\droplet}(0) = \population  } = 
\frac{
e^{-\rate} \rate^{\population}
}{
\population !
} \,.
\end{equation}

Now that we know the number of cells in droplet $\droplet$, we need to determine which strain each cell belongs to. Because any cell was equally likely to have been sampled, the strains follow a multivariate hypergeometric distribution with categories of sizes $\freq^{(1)} \Population$, $\dots$, $\freq^{(\Strains)} \Population$. Because the $\Population$ is very large, it is very unlikely\footnote{Even more unlikely than being sampled once, which is already very unlikely.} that any given cell would be sampled twice. Thus we might as well use the approximating distribution as $\Population \to \infty$, multinomial with category probabilities $\freq^{(1)}, \dots, \freq^{(\Strains)}$, which corresponds to sampling with replacement. So
\begin{equation}
  \label{eq:first_derivation_categorical_prob}
  \probability*{\vabundance_{\droplet}(0) = \vpopulation | \abundance_{\droplet}(0) = \population} = 
\binom{\population}{\population[1] \! \cdots \! \population[\Species]} \prod_{\specie=1}^{\Species} (\freq^{(\specie)})^{\population[\specie]} \,.
\end{equation}

Combining the count distribution and categorical distribution derived above, it follows that the joint distribution must approximately equal
\begin{equation}
  \label{eq:first_derivation_hpomu}
  \begin{split}
    \probability*{\vabundance_{\droplet}(0) = \vpopulation} = & \probability*{\vabundance_{\droplet}(0) = \vpopulation | \abundance_{\droplet}(0) = \population} \cdot \probability*{ \abundance_{\droplet}(0) = \population  } \\
= &  
\binom{\population}{\population[1] \! \cdots \! \population[\Species]} \prod_{\specie=1}^{\Species} (\freq^{(\specie)})^{\population[\specie]}
\cdot
\frac{
e^{-\rate} \rate^{\population}
}{
\population !
} \,.
  \end{split}
\end{equation}

\subsubsection{Second Derivation}
\label{sec:second-derivation}

For any given strain $\strain$, the number of cells \textit{of that strain} that end up in the droplet $\droplet$ is the sum of all $\freq^{(\strain)} \Population$ of the indicator random variables belonging to cells \textit{of that strain}. Because the sum of i.i.d. Bernoullis is binomial, the marginal distribution of the number of cells of strain $\strain$ that end up in the droplet $\droplet$ is again binomially distributed, $\operatorname{Binomial}(p, \freq^{(\strain)}\Population)$. 

Because\footnote{Due to the argument found in the first derivation in section \ref{sec:first-derivation}.} $p = \frac{\rate}{\Population}$, the expected number of cells of strain $\strain$ that end up in the droplet $\droplet$ is $p(\freq^{(\strain)}\Population) = \frac{\rate}{\Population} (\freq^{(\strain)}\Population) = \freq^{(\strain)} \rate$. If we assume that $\Population$ is sufficiently large such that $\freq^{(\strain)}\Population$ is also very large, then as before we may as well use the approximating distribution as\footnote{Equivalently, as $\Population \to \infty$ along a subsequence such that $\freq^{(\strain)}\Population \in \mathbb{N}$ (i.e. remains an integer).} $(\freq^{(\strain)}\Population) \to \infty$. As in the first derivation, this distribution is Poisson as a result of the Poisson limit theorem, so for the marginal distributions:
\begin{equation}
  \label{eq:second_derivation_marginal}
    \probability*{ \abundance[\specie]_{\droplet}(0) = \population[\specie]  } = 
\frac{
e^{-\freq^{(\strain)}\rate} (\freq^{(\strain)}\rate)^{\population[\specie]}
}{
(\population[\specie]) !
} \,.
\end{equation}

Because what happens to any given cell is completely independent of what happens to any other cells, the marginal distributions for each strain are sums of mutually disjoint sets of i.i.d. Bernoulli random variables. Therefore the marginal distributions must be mutually independent Binomial random variables, and thus in the limit approach mutually independent Poisson random variables. Thus the joint distribution must approximately equal
\begin{equation}
  \label{eq:second_derivation_hpomu}  
 \probability*{ \vabundance_{\droplet}(0) = \vpopulation  } = 
\prod_{\strain=1}^{\Strains}  
\frac{
e^{-\freq^{(\strain)}\rate} (\freq^{(\strain)}\rate)^{\population[\specie]}
}{
(\population[\specie]) !
} \,.
\end{equation}

\subsection{Implicit Model Assumptions}
\label{sec:impl-model-assumpt}

Both of the above two derivations share the following implicit assumptions:

\begin{itemize}
\item The experiment can be calibrated to a priori guarantee a certain \textit{average} number of cells per droplet.
\item The total number of cells in the sampling pool, $\Population$, is very large.
\item For \textit{every} strain $\strain$, the number of cells $\freq^{(\strain)} \Population$ in the sampling pool which belong to strain $\strain$ is very large.
\item Each cell has the same small probability (regardless of its strain) of ending up in the droplet as any other cell.
\item (Regardless of their strains), whether a given cell ends up in the droplet is completely independent of what happens to any other cell.
\end{itemize}

The second and third can be thought of as stating that the population size is effectively infinite. The fourth and fifth can be thought of as stating that the sampling pool is completely homogeneous, or perfectly ``well-mixed''. 

My collaborator tells me that, using commercially available microfluidic devices, one can calibrate the experiment to a priori guarantee a certain average number of cells per droplet, and that the total number of cells in the sampling pool will be $\sim 10^{10}$. However, the remaining three assumptions are more questionable. In the next section I explain three ways these assumptions could fail to be true in practice.

\subsection{Similar Work in Ecology}
\label{sec:similar-work-ecology}

Similar (frameworks for) hierarchical working models for multivariate count data have been described for joint species distributions in (macro)ecology and observational metagenomic studies in microbial ecology \cite{Billheimer2001}\cite{Biswas2015} \cite{Inouye2017} \cite{Ovaskainen2017} \cite{Chiquet2021}. However hPoMu appears to be far simpler than the other working models proposed in those studies. Moreover, hPoMu is also a much more direct generalization of the univariate Poisson distribution that is used as the default working model for single-cell (single-type) droplet microfluidics \cite{Collins2015}. Hence in my opinion it makes much more sense to propose hPoMu as the ``default'' working model for multivariate count data than any of these alternatives.

Nevertheless, the independent convergence of this and similar work onto the approach of hierarchical models (cf. also section \ref{sec:conn-with-docum}) seems to support the claim from section \ref{sec:hier-count-categ} that hierarchical models are a good strategy for characterizing multivariate count data.

\section{Failures of hPoMu Model Assumptions}
\label{sec:failures-hpomu-model}

While hPoMu is definitely a sensible starting point, in practice its implicit assumptions failing to be realistic could undermine its usefulness. Sections \ref{sec:sampling-from-finite}, \ref{sec:dens-heter}, and \ref{sec:comp-heter} define three plausible ways that these assumptions could fail to be true in practice. Section \ref{sec:model-heter} outlines the framework used to describe the latter failures of the hPoMu assumptions.

\subsection{Sampling is from a Finite Population without Replacement}
\label{sec:sampling-from-finite}

In both of the above derivations from section \ref{sec:derivations}, we assumed that the number of cells being sampled from is ``effectively infinite''. This implied that no meaningful loss of accuracy was occurred when invoking the Poisson limit theorem for the number of cells in each droplet. As mentioned before, with $\sim10^{10}$ cells, this is a reasonable assumption.

However, we assumed more than this. We also assumed that \textit{for each individual strain $\strain$} the number of cells is ``effectively infinite''. This assumption is reasonable and uncontroversial in the case of relatively abundant strains. However, in the case of ``rare'' strains, whose relative abundance is e.g. $0.1\%$, $0.01\%$, or even lower, whether this assumption remains reasonable is a priori unclear. Admittedly $0.01\%$ of $\sim 10^{10}$ cells is still many, but perhaps not enough for the asymptotically approximate distributions to remain accurate.

\subsection{Density Heterogeneity}
\label{sec:dens-heter}

In both of the above derivations we assumed that each cell everywhere in the sampling pool had the same probability of being sampled for every droplet. We also assumed that the probability of ending up in the droplet for any given cell was statistically independent of what happened to any other cell. This would make sense in a theoretical world where each cell corresponded to a scalar field uniformly spread across the entire sampling pool. In reality cells have finite volume and are limited at any given moment in time to a single location in the sampling pool. Only cells reasonably close to where the droplet is being formed have any nonzero probability of ending up in the droplet. Cells that are closer to one another are more likely to end up in the same droplet than those that are further apart.

These details hypothetically might not matter in the case where the average number of cells per unit volume\footnote{For volumes ``large enough''. This is obviously false for ``infinitesimal volumes'' as discussed above.} was constant throughout the entire sampling pool. However in practice we have no guarantee that this would be the case, and conceivably different sections of the sampling pool could have average numbers of cells per unit volume which differ from the average number of cells per unit volume for the entire sampling pool (i.e. the total number of cells divided by the volume of the sampling pool). 

Herein I call this phenomenon ``density heterogeneity''. Compare figures \ref{fig:low_density_heterogeneity} and \ref{fig:high_density_heterogeneity}. In those hypothetical examples, the average number of cells per unit volume for the entire sampling pool is $2$, but the average number of cells per unit volume differs between the specific sections. In figure \ref{fig:low_density_heterogeneity} the differences are small, introducing not much more variance compared to what would be expected under the uniform/homogeneous density assumed by the default hPoMu model. In figure \ref{fig:high_density_heterogeneity} the differences are larger, introducing much more variance compared to what would be expected under uniform/homogeneous density assumed by the default hPoMu model.

\begin{figure}[p]
  \centering
  
  \begin{subfigure}{\textwidth}
  \centering \includegraphics[width=\textwidth,height=0.47\textheight,keepaspectratio]{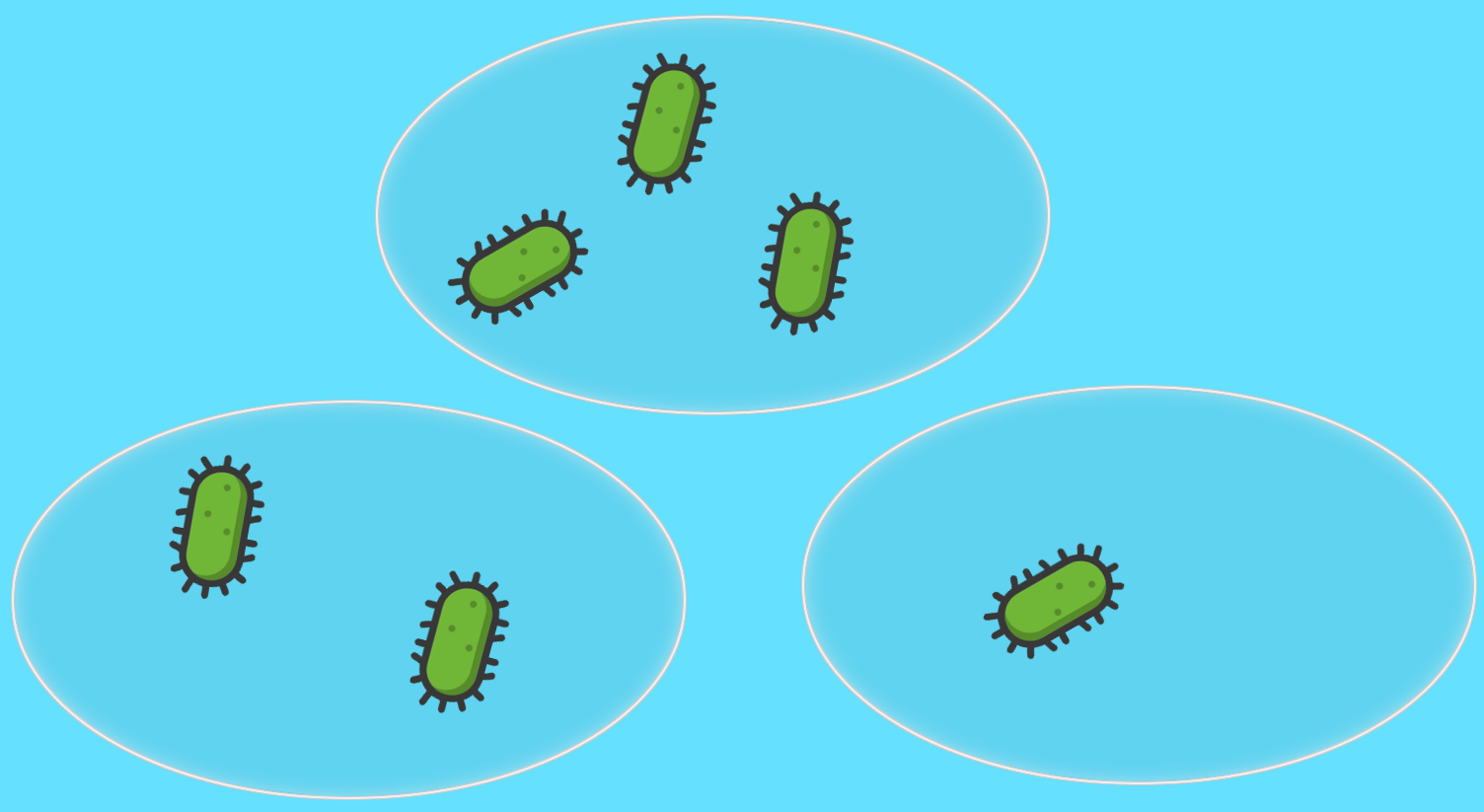}
  \caption{Low Density Heterogeneity}
  \label{fig:low_density_heterogeneity}
\end{subfigure}

\begin{subfigure}{\textwidth}
  \centering \includegraphics[width=\textwidth,height=0.47\textheight,keepaspectratio]{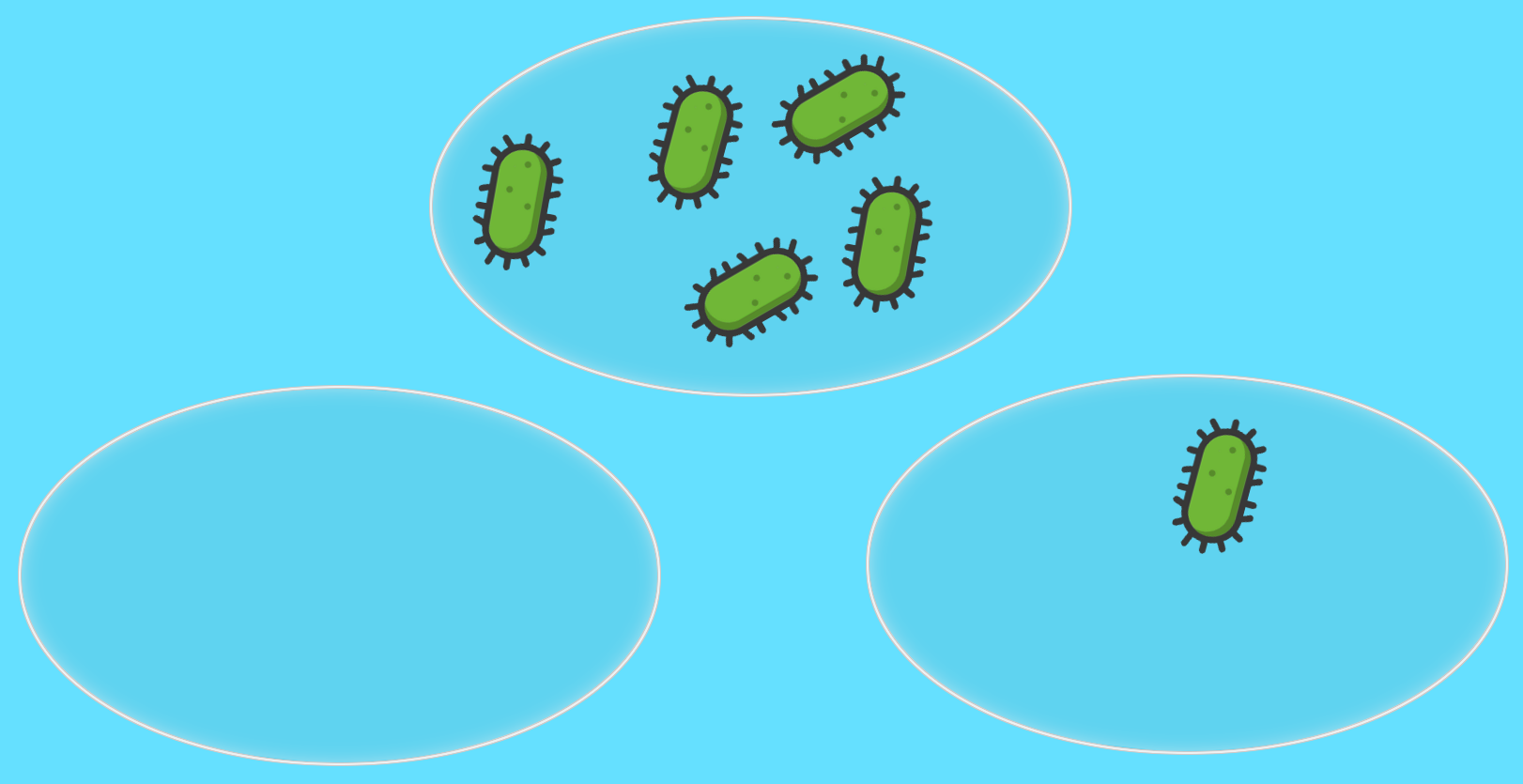}
  \caption{High Density Heterogeneity}
  \label{fig:high_density_heterogeneity}
\end{subfigure}

\caption{Schematic representation of varying levels of density heterogeneity.}
\label{fig:density_heterogeneity}
\end{figure}

In the hierarchical count-categorical distribution framework, density heterogeneity is an issue which affects the count distribution. The default Poisson model for the count distribution will not account for additional variance which might be introduced in practice by density heterogeneity. See \cite{Yu2013} for modelling of density heterogeneity for RNA-seq.

\subsection{Compositional Heterogeneity}
\label{sec:comp-heter}

In making the homogeneous assumptions that all cells have equal probability of ending up in the droplet, and that what happens to a given cell is completely independent of what happens to any other cells, we implicitly ignore the fact that cells belonging to different strains may tend to behave very differently. 

In particular, although the relative abundance of strain $\strain$ across the entire sampling pool may be $\freq^{(\strain)}$, conceivably cells of this strain could be slightly more concentrated or slightly less concentrated than this in different sections of the sampling pool. This means different sections of the sampling pool could have average relative abundances which differ from the average relative abundances for the entire sampling pool. Thus the fact that each droplet is formed only from cells nearby a certain location in the sampling pool could affect not only the number of cells likely to end up in the droplet, but also which strains are most likely to end up in the droplet. This is probably guaranteed to happen to some extent due to random chance given that cells occupy finite, not infinitesimal volume. The probabilities of belonging to given strains cannot be perfectly uniform scalar fields. For an example of one mechanism that could aggravate this, cells of a certain strain could preferentially aggregate with, or avoid, cells of another strain due to their interactions with one another. 

Herein I call this phenomenon ``compositional heterogeneity''. Compare figures \ref{fig:low_compositional_heterogeneity} and \ref{fig:high_compositional_heterogeneity}. In those hypothetical examples, the average relative abundances of each strain for the entire sampling pool is $\frac{1}{2}$ blue, $\frac{1}{4}$ pink, $\frac{1}{4}$ yellow, but the average relative abundances of each strain differs between the specific sections. In figure \ref{fig:low_compositional_heterogeneity} the differences are small, introducing not much more variance compared to what would be expected under the uniform/homogeneous density assumed by the default hPoMu model. In figure \ref{fig:high_compositional_heterogeneity} the differences are larger, introducing much more variance compared to what would be expected under uniform/homogeneous density assumed by the default hPoMu model.

\begin{figure}[p]
\centering

  \begin{subfigure}{\textwidth}
  \centering \includegraphics[width=\textwidth,height=0.47\textheight,keepaspectratio]{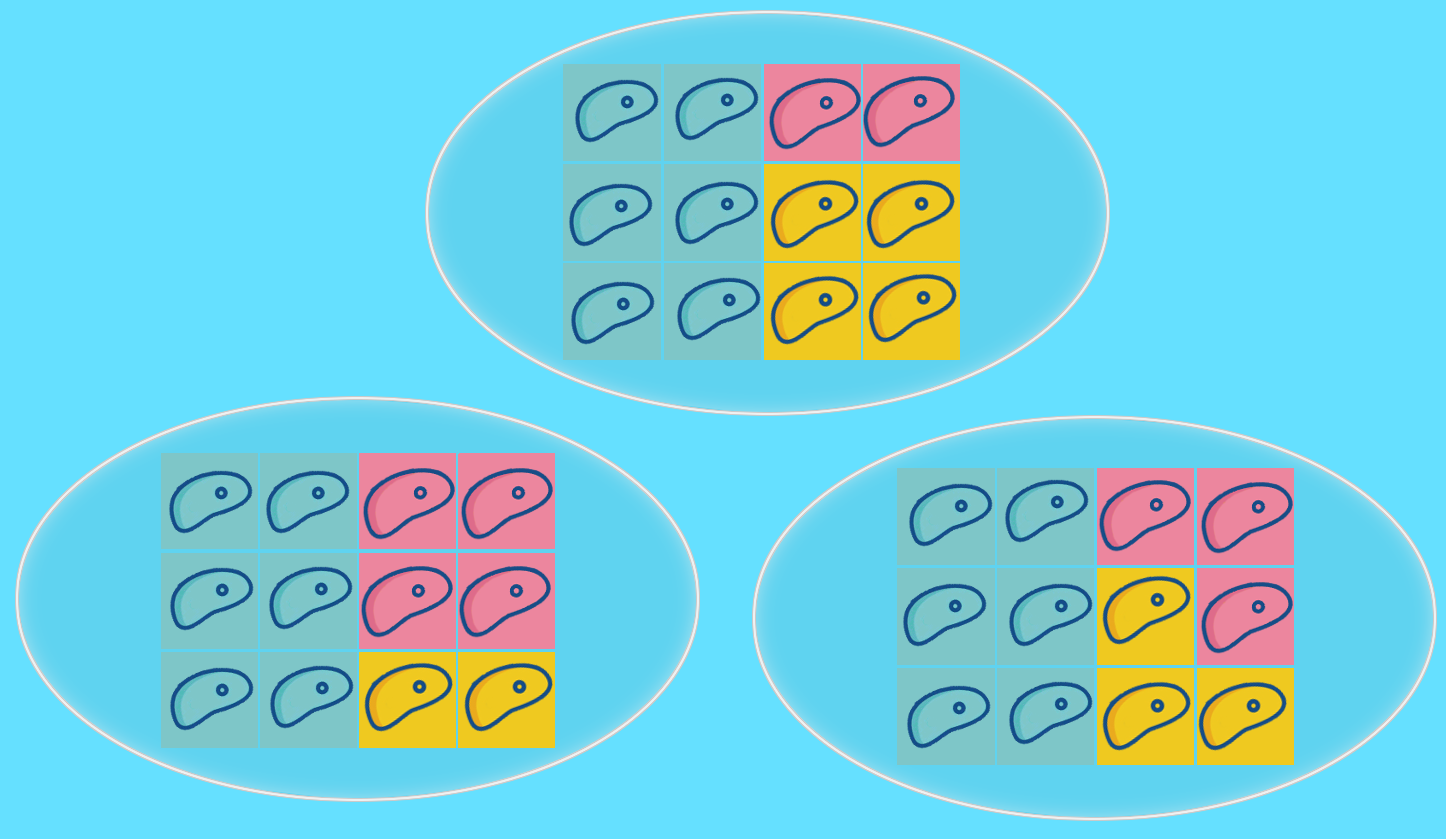}
  \caption{Low Compositional Heterogeneity}
  \label{fig:low_compositional_heterogeneity}
\end{subfigure}

\begin{subfigure}{\textwidth}
  \centering \includegraphics[width=\textwidth,height=0.47\textheight,keepaspectratio]{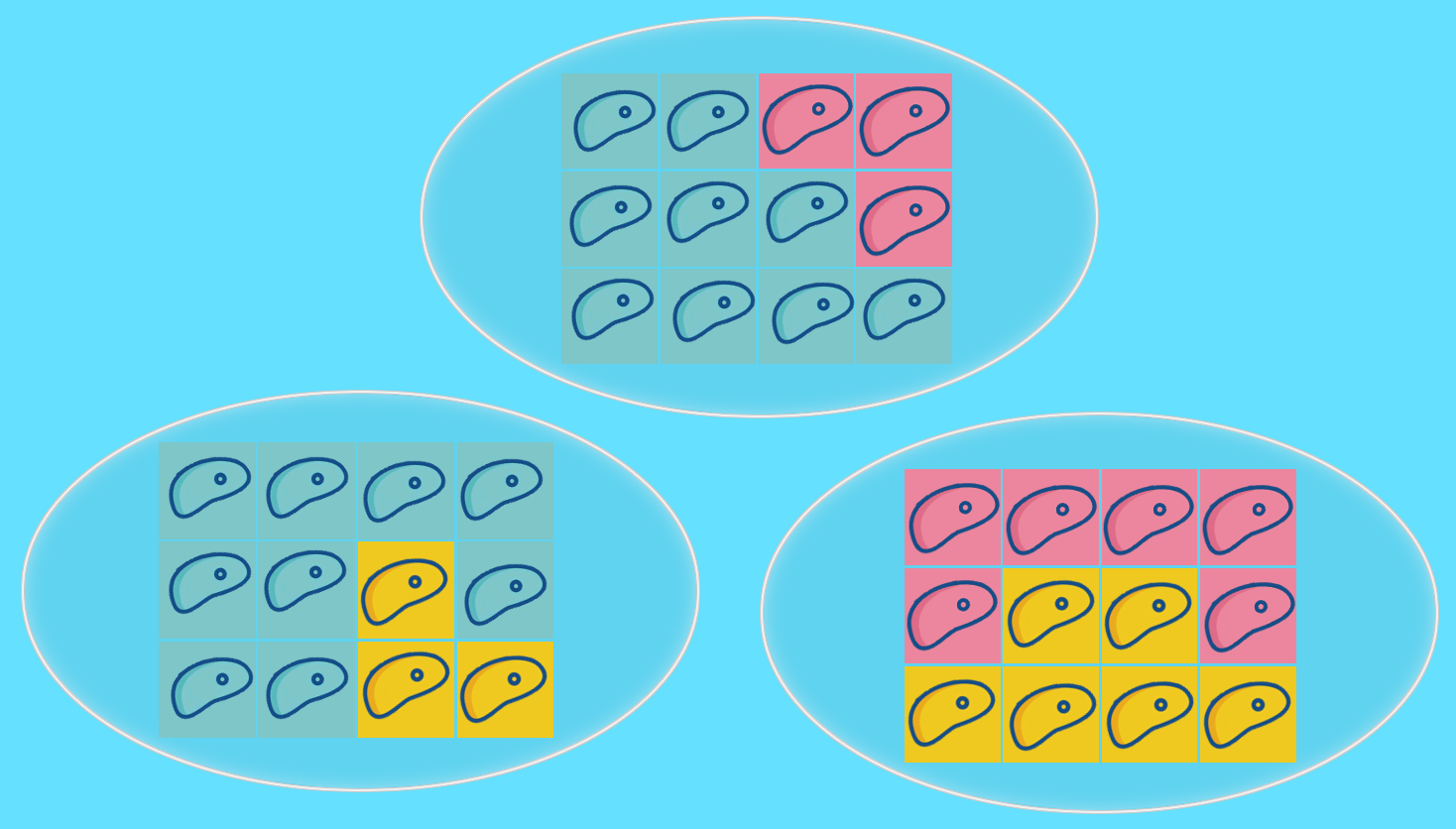}
  \caption{High Compositional Heterogeneity}
  \label{fig:high_compositional_heterogeneity}
\end{subfigure}

\caption{Schematic representation of varying levels of compositional heterogeneity.}
\label{fig:compositional_heterogeneity}
\end{figure}

In the hierarchical count-categorical distribution framework, compositional heterogeneity is an issue which affects the categorical distribution. The default multinomial model for the categorical distribution (found in hPoMu) is a ``null model'' that does not account for additional variance that might be introduced in practice by compositional heterogeneity.

See \ref{sec:clar-comp-heter} for a clarification of the intended notion of ``compositional heterogeneity''.

\subsection{Modelling Heterogeneities}
\label{sec:model-heter}

For both density heterogeneity and compositional heterogeneity, extra variance is introduced that is not accounted for by the default hPoMu model. In trying to predict and model the probability distribution of droplet formation, instead of trying to develop complicated mechanistic models describing these spatial heterogeneities whose impacts on the initial droplet formation distribution we might then infer, we can instead seek the humbler goal of modelling this extra variance. In other words, instead of directly modelling the causes of density and compositional heterogeneity, we can restrict ourselves to modelling their ``downstream'' effect on variance. This should be enough by itself to make our models for the distribution of initial droplet formation more accurate.

I quantify this extra variance via the over-dispersion (defined below in section \ref{sec:over-dispersion}), and describe later in sections \ref{sec:over-disp-dirichl} and \ref{sec:over-disp-negat} how the parameters of the proposed models relate to their over-dispersion, and therefore can describe the increased variance caused by either density heterogeneity or compositional heterogeneity.

\subsubsection{Over-Dispersion}
\label{sec:over-dispersion}

Herein we define the over-dispersion of $Y$ with respect to $X$, where $Y$ and $X$ are scalar random variables, as
  \begin{equation}
    \label{eq:scalar_overdispersion}
    \frac{
\var{Y} - \var{X}
}{
\var{X}
} \,.
  \end{equation}

If $Y$ and $X$ are random vectors, and if there exists a constant $C$ such that
\begin{equation}
  \label{eq:vector_overdispersion}
  \frac{
\cov{Y_i, Y_j} - \cov{X_i, X_j}
}{
\cov{X_i, X_j}
}
= C
\end{equation}
for all $i,j$ (including $i=j$), then $C$ is herein defined to be the over-dispersion of $Y$ with respect to $X$.

The over-dispersion of $\mathcal{P}_2$ with respect to $\mathcal{P}_1$, where $\mathcal{P}_2$ and $\mathcal{P}_1$ are probability distributions, is defined to be the over-dispersion of $Y$ with respect to $X$, where the random variables $Y \sim \mathcal{P}_2$ and $X \sim \mathcal{P}_1$ but are otherwise arbitrary.

\section{Working Model for Sampling Without Replacement}
\label{sec:model-sampl-with}

The hPoMu working model assumes that the number of cells of any given strain $\strain$ is effectively infinite, and that therefore sampling can effectively be modeled as without replacement. However, for very rare strains the adequacy of this asymptotic approximation becomes questionable. Therefore what we need is a distribution whose definition explicitly acknowledges the finiteness of the populations of cells being sampled. In section \ref{sec:introduction-htpmh} I introduce a working model which fills this gap, and then in section \ref{sec:htpmh-gener-hpomu} I show that this working model extends what can be described using hPoMu alone.

\subsection{Introduction to hTPMH}
\label{sec:introduction-htpmh}

As suggested by its name, the ``hierarchical truncated Poisson multivariate hypergeometric'' (hTPMH) distribution uses a multivariate hypergeometric model of sampling without replacement for the categorical distribution. The multivariate hypergeometric distribution is introduced in section \ref{sec:mult-hyperg-mh}. The truncated Poisson distribution is introduced in section \ref{sec:truncated-poisson-tp}, and is a slight modification of the Poisson distribution that explicitly acknowledges that the number of cells to be sampled from is finite. Finally I combine the new information learned to define the hTPMH model in section \ref{sec:htpmh-definition}.

The total number of cells in the sampling pool before any droplets are formed is $\Population =: \Population_0$ (cf. the notation from section \ref{sec:derivations}), and the total number that belong to strain $\strain$ before any droplets have formed is $\Population[\strain] = \freq^{(\strain)} \Population =: \Population[\strain]_0$. This leads us to define $\vPopulation =: \vPopulation_0$ as
\begin{equation}
  \label{eq:population_vector_defn}
  \vPopulation := (\Population[1], \dots, \Population[\strain], \dots, \Population[\Strains]) \,, \quad \sum_{\strain=1}^{\Strains} \Population[\strain] = \Population \,.
\end{equation}

Similarly, $\Population_{\droplet}$ is the total number of cells in the sampling pool \textbf{after} $\droplet$ droplets have been formed, whereas $\Population[\strain]_{\droplet}$ is the total number of cells of strain $\strain$ in the sampling pool \textbf{after} $\droplet$ droplets have been formed. Thus \textbf{after} $\droplet$ droplets have been formed we also have a corresponding vector $\vPopulation_{\droplet}$ such that
\begin{equation}
  \label{eq:population_decremented_vector_defn}
  \vPopulation_{\droplet} := (\Population[1]_{\droplet}, \dots, \Population[\strain]_{\droplet}, \dots, \Population[\Strains]_{\droplet}) \,, \quad \sum_{\strain=1}^{\Strains} \Population[\strain]_{\droplet} = \Population_{\droplet} \,.
\end{equation}
In particular, for all $\droplet \in [\Droplets]$ we have (by definition) that $\Population_{\droplet}  = \Population_0 - \sum_{\droplet[]=1}^{\droplet} \abundance_{\droplet[*]}(0)$ and $\Population[\strain]_{\droplet} = \Population[\strain]_0 - \sum_{\droplet[]=1}^{\droplet} \abundance[\strain]_{\droplet[*]}(0)$.

\subsubsection{Multivariate Hypergeometric (MH) Distribution}
\label{sec:mult-hyperg-mh}
The PMF of the multivariate hypergeometric distribution can be written as
\begin{equation}
  \label{eq:mhg_pmf}
  \frac{(\Population_{\prevdroplet[*]} - \population)!}{(\Population_{\prevdroplet[*]})!}
\!\cdot\!
\left[\prod_{\specie=1}^{\Species} 
 \frac{(\Population[\specie]_{\prevdroplet[*]})!}{(\Population[\specie]_{\prevdroplet[*]} - \population[\specie])!}
\right]
\!\cdot\!
\binom{\population}{\population[1] \! \cdots \! \population[\Species]} \,.
\end{equation}
As a reminder from section \ref{sec:descr-likel}, $\population := \sum\limits_{\specie=1}^{\Species} \population[\specie]$ (the sum of the entries of $\vpopulation$).

\subsubsection{Truncated Poisson (TP) Distribution}
\label{sec:truncated-poisson-tp}

The mass function of the truncated Poisson distribution is
\begin{equation}
  \label{eq:truncated_poisson_pmf}
  \frac{1}{\probability*{\poisson (\rate) \le \Population_{\prevdroplet[*]} }  }
\cdot
\frac{
e^{-\rate} \rate^{\population}
}{
\population !
} \,.
\end{equation}
Unless $\Population_{\prevdroplet[*]}$ is very small, this is effectively almost identical to the corresponding probabilities from the Poisson distribution.

\subsubsection{hTPMH Definition}
\label{sec:htpmh-definition}
Under the hTPMH distribution, for the $\droplet$'th droplet: the probability of the strain distribution vector given the number of cells
\[\probability*{ \vabundance_{\droplet}(0) = \vpopulation | \abundance_{\droplet}(0) = \population, \vPopulation_{\prevdroplet[*]} }\]
is determined by a multivariate hypergeometric distribution, while the probability of the number of cells ${\probability*{\abundance_{\droplet}(0) = \population | \vPopulation_{\prevdroplet[*]}}}$ is determined by a truncated Poisson distribution.

Thus for the unconditional probability:
  \begin{equation}
    \label{eq:catpaw_mahjong}
    \begin{split}
&    \probability*{\vabundance_{\droplet}(0) = \vpopulation | \vPopulation_{\prevdroplet[*]}} \\
 = & \quad
 \probability*{ \abundance_{\droplet}(0) = \population | \vPopulation_{\prevdroplet[*]} } 
\!\cdot\!
\probability*{\vabundance_{\droplet}(0) = \vpopulation | \abundance_{\droplet}(0) = \population, \vPopulation_{\prevdroplet[*]}} 
\\
= & \quad
\probability*{\poisson (\rate) \le \Population_{\prevdroplet[*]}}^{-1} 
\!\cdot\!
\frac{
e^{-\rate} \rate^{\population}
}{
\population !
}
  \frac{(\Population_{\prevdroplet[*]} - \population)!}{(\Population_{\prevdroplet[*]})!}
\!\cdot\!
\left[\prod_{\specie=1}^{\Species} 
 \frac{(\Population[\specie]_{\prevdroplet[*]})!}{(\Population[\specie]_{\prevdroplet[*]} - \population[\specie])!}
\right]
\!\cdot\!
\binom{\population}{\population[1] \! \cdots \! \population[\Species]} \,.
    \end{split}
  \end{equation}

\subsection{hTPMH Generalizes hPoMu}
\label{sec:htpmh-gener-hpomu}

One can show that hTPMH converges to hPoMu under appropriate limits. Hence the hPoMu working model can be considered a submodel of the hTPMH working model.

Sections \ref{sec:appr-form-likel}, \ref{sec:appr-form-likel-1}, and \ref{sec:proof-that-htpmh} give the needed context and then show that hTPMH really is an extension of hPoMu. Because hTPMH and hPoMu have nearly the same count distribution, at its core this reduces to showing how the multivariate hypergeometric distribution extends the multinomial distribution, which section \ref{sec:appr-form-likel} discusses.

\subsubsection{Approximate Form of Likelihood Ratio of Multivariate Hypergeometric with respect to Multinomial and Bounds}
\label{sec:appr-form-likel}
Applying Lemma \ref{lem:factorial_stirling} and then algebraic manipulations (e.g. multiplying numerator and denominator by the same factor) gives
\begin{equation}
  \label{eq:MHg_stirling}
  \begin{split}
&  \frac{(\Population_{\prevdroplet[*]} - \population)!}{(\Population_{\prevdroplet[*]})!}
\!\cdot\!
\left[\prod_{\specie=1}^{\Species} 
 \frac{(\Population[\specie]_{\prevdroplet[*]})!}{(\Population[\specie]_{\prevdroplet[*]} - \population[\specie])!}
\right] \\
=& \exp \left( \mysteryseq[MHg](\vpopulation, \vPopulation_{\prevdroplet[*]}) \right)
\!\cdot\!
\left(
\frac{
\Population_{\prevdroplet[*]} - \population
}{
\Population_{\prevdroplet[*]}
}
\right)^{
\Population_{\prevdroplet[*]} - \population + \frac{1}{2}
} 
\\
&
\!\cdot\!
\prod_{\specie=1}^{\Species} \left(
 \frac{\Population[\specie]_{\prevdroplet[*]}}{\Population[\specie]_{\prevdroplet[*]} - \population[\specie]}
\right)^{
\Population[\specie]_{\prevdroplet[*]} - \population[\specie] + \frac{1}{2}
}
\!\cdot\!
\prod_{\specie=1}^{\Species}
\left(
\frac{
\Population[\specie]_{\prevdroplet[*]}
}{
\freq^{(\specie)} \Population_{\prevdroplet[*]}
}
\right)^{\population[\specie]}
\!\cdot\!
\prod_{\specie=1}^{\Species} (\freq^{(\specie)})^{\population[\specie]} \,.
  \end{split}
\end{equation}

 In equation \ref{eq:MHg_stirling}, based on \cite{Robbins}, Lemma 1 implies that
 \begin{equation}
   \label{eq:MHg_Robbins}
   \lRobbins\left( \vPopulation_{\prevdroplet[*]} - \vpopulation, \vPopulation_{\prevdroplet[*]} \right) < 
\mysteryseq[MHg] \left(\vpopulation, \vPopulation_{\prevdroplet[*]} \right) <
\uRobbins \left( \vPopulation_{\prevdroplet[*]} - \vpopulation, \vPopulation_{\prevdroplet[*]} \right) \,,
 \end{equation}
with the ``lower Robbins function''  $\lRobbins$ defined for convenience as
\begin{equation}
  \label{eq:lRobbins}
  \lRobbins(\vec{N}, \vec{M}) := 
\frac{1}{12N+1} - \sum_{\specie=1}^{\Species} \frac{1}{12N_{\specie} }
- \frac{1}{12M} + \sum_{\specie=1}^{\Species} \frac{1}{12M_{\specie} + 1} = -\uRobbins(\vec{M}, \vec{N}) \,,
\end{equation}
and the ``upper Robbins function'' $\uRobbins$ defined for convenience as
\begin{equation}
  \label{eq:uRobbins}
  \uRobbins(\vec{N}, \vec{M}) := \frac{1}{12N} - \sum_{\specie=1}^{\Species} \frac{1}{12N_{\specie}+1} - \frac{1}{12M+1} + \sum_{\specie=1}^{\Species}\frac{1}{12M_{\specie}} = -\lRobbins(\vec{M}, \vec{N}) \,. 
\end{equation}
Above we again used the notational the convention that, given a vector $\vec{X}=(X_1, \dots, X_{\Species})$, $X := \sum_{\specie=1}^{\Species} X_{\specie}$.

Equation \ref{eq:MHg_stirling} implies that the likelihood ratio of Multivariate Hypergeometric distribution
  with respect to the Multinomial distribution may be written
  \begin{equation}
    \label{eq:MHg_fudge_factor}
    \begin{adjustbox}{max width=\textwidth,keepaspectratio}
$\displaystyle    \begin{split}
& \exp \left( \mysteryseq[MHg](\vpopulation, \vPopulation_{\prevdroplet[*]}) \right)
\!\cdot\!
\left(
\frac{
\Population_{\prevdroplet[*]} - \population
}{
\Population_{\prevdroplet[*]}
}
\right)^{
\Population_{\prevdroplet[*]} - \population + \frac{1}{2}
} 
\!\cdot\!
\prod_{\specie=1}^{\Species} \left(
 \frac{\Population[\specie]_{\prevdroplet[*]}}{\Population[\specie]_{\prevdroplet[*]} - \population[\specie]}
\right)^{
\Population[\specie]_{\prevdroplet[*]} - \population[\specie] + \frac{1}{2}
}
\!\cdot\!
\prod_{\specie=1}^{\Species}
\left(
\frac{
\Population[\specie]_{\prevdroplet[*]}
}{
\freq^{(\specie)} \Population_{\prevdroplet[*]}
}
\right)^{\population[\specie]} \,.
\end{split} $
\end{adjustbox}
  \end{equation}
In other words, multiplying the PMF of the Multinomial distribution by the quantity in \ref{eq:MHg_fudge_factor} gives the PMF of the Multivariate Hypergeometric Distribution.

\subsubsection{Approximate Form of Likelihood Ratio of hTPMH with respect to hPoMu and Bounds}
\label{sec:appr-form-likel-1}
From equation \ref{eq:MHg_fudge_factor} it follows readily that the likelihood ratio of the hTPMH distribution with respect to the hPoMu distribution is
  \begin{equation}
    \label{eq:hTPMH_fudge_factor}
    \begin{split}
&\probability*{ \poisson (\rate) \le \Population_{\prevdroplet[*]} }^{-1} \exp \left( \mysteryseq[MHg](\vpopulation, \vPopulation_{\prevdroplet[*]}) \right)
\!\cdot\!
\left(
\frac{
\Population_{\prevdroplet[*]} - \population
}{
\Population_{\prevdroplet[*]}
}
\right)^{
\Population_{\prevdroplet[*]} - \population + \frac{1}{2}
} \\
&
\!\cdot\!
\prod_{\specie=1}^{\Species} \left(
 \frac{\Population[\specie]_{\prevdroplet[*]}}{\Population[\specie]_{\prevdroplet[*]} - \population[\specie]}
\right)^{
\Population[\specie]_{\prevdroplet[*]} - \population[\specie] + \frac{1}{2}
}
\!\cdot\!
\prod_{\specie=1}^{\Species}
\left(
\frac{
\Population[\specie]_{\prevdroplet[*]}
}{
\freq^{(\specie)} \Population_{\prevdroplet[*]}
}
\right)^{\population[\specie]} \,.
\end{split} 
  \end{equation}
The factor $\probability*{ \poisson (\rate) \le \Population_{\prevdroplet[*]} }^{-1}$ reflects how the number of cells in droplet $\droplet$ will be truncated Poisson distributed, not Poisson distributed. Otherwise this is the same as (\ref{eq:MHg_fudge_factor}).

\subsubsection{Sketch of Proof that hTPMH Converges in Distribution to hPoMu}
\label{sec:proof-that-htpmh}

Because the hTPMH and hierarchical Poisson Multinomial (hPoMu) distributions are discrete, to show that the former converges in distribution to the latter, it suffices to show that the probability mass function of the former converges to the latter.

Equation \ref{eq:hTPMH_fudge_factor} gives a quantity such that, when it is multiplied with the PMF of the hPoMu distribution, the result is the PMF of the corresponding hTPMH distribution.

Fixing the values of $\vabundance_{\droplet[*]}(0)$ for all $\droplet[*] \in [\prevdroplet]$, from the above it follows that the quantity in \ref{eq:hTPMH_fudge_factor} converging to $1$ as $\Population \to \infty$ implies that the hTPMH distribution converges to the hPoMu distribution as $\Population \to \infty$.

The desired convergence follows from the Squeeze Theorem. Note that both
\[\lRobbins \left( \vPopulation_{\prevdroplet[*]} - \vpopulation, \vPopulation_{\prevdroplet[*]} \right) \quad \text{and} \quad \uRobbins \left( \vPopulation_{\prevdroplet[*]} - \vpopulation, \vPopulation_{\prevdroplet[*]} \right)\]
converge to $0$ as $\Population \to \infty$ because of additivity of limits and that ${\pm \lim\limits_{N \to \infty} \frac{1}{a_N} = 0} $ for any sequence $a_N$ such that $\lim\limits_{N \to \infty} a_N = \infty$ (due to the Archimedean property of $\mathbb{R}$).

The factor $\probability*{ \poisson (\rate) \le \Population_{\prevdroplet[*]} }^{-1}$ converges to $1$ because of the defintion of $\Population_{\prevdroplet[*]}$ and that any valid cumulative distribution function must converge to $1$ as its input goes to $\infty$ (or otherwise there would be nonzero probability mass ``at infinity'').

The last expression in (\ref{eq:hTPMH_fudge_factor}) converges to $1$ because of L'H\^{o}spital's Rule, and the remaining factors converge to $1$ because of Lemma \ref{lem:exponential_ratios}.

We may assume without loss of generality that, when taking the limit as $\Population \to \infty$, we do so along a subsequence such that all ${\Population[\specie] = \freq^{(\specie)} \Population}$ are in $\mathbb{N}$. This of course also guarantees that all of the ${\Population[\specie]_{\prevdroplet[*]} := \Population[\specie] - \sum_{\droplet[*]=1}^{\prevdroplet} \abundance[\specie]_{\droplet[*]}(0)}$ are also integer-valued.

Instead of showing that the likelihood ratio converges to $1$, one can also show (e.g. according to one's personal preference) that the logarithm of the likelihood ratio converges to $0$, using Lemma \ref{lem:fudge_function} (the ``logarithmic version'' of Lemma \ref{lem:exponential_ratios}). The ``fudge function'' $\fudgefunction$ is defined for convenience in equation (\ref{eq:fudge_function}). Starting from equation (\ref{eq:hTPMH_fudge_factor}), tedious algebra shows that the logarithm of the likelihood ratio of hTPMH with respect to hPoMu is
\begin{equation}
  \label{eq:hTPMH_log_fudge}
  \begin{split}
 &   -\log \left( \probability*{ \poisson (\rate) \le \Population_{\prevdroplet[*]} } \right) 
\!+\! 
\sum_{\specie=1}^{\Species} \fudgefunction (\Population[\specie]_{\prevdroplet[*]} - \population[\specie], \population[\specie], \frac{1}{2}  )
 \\
& 
\!-\! \fudgefunction (\Population_{\prevdroplet[*]} - \population, \population, \frac{1}{2})  
\!+\!
\sum_{\specie=1}^{\Species} \population[\specie] \left(  \log(\Population[\specie]_{\prevdroplet[*]}) - \log(\freq^{(\specie)} \Population_{\prevdroplet[*]}) \right) 
\!+\! 
\mysteryseq[MHg](\vpopulation, \vPopulation_{\prevdroplet[*]}) \,.
  \end{split}
\end{equation}
The only subtlety involves recalling the definitions of $\Population[\specie]_{\prevdroplet[*]}$ and $\Population_{\prevdroplet[*]}$ so as to correctly substitute them into (\ref{eq:hTPMH_log_fudge}):
\begin{equation}
  \label{eq:plug_in_definitions}
  \log(\Population[\specie]_{\prevdroplet[*]}) - \log(\freq^{(\specie)} \Population_{\prevdroplet[*]})
=
\log \left(  \frac{
\freq^{(\specie)} \Population - \sum_{\droplet[*]=1}^{\prevdroplet} \abundance[\specie]_{\droplet[*]}(0)
}{
\freq^{(\specie)} \Population - \freq^{(\specie)} \left( \sum_{\droplet[*]=1}^{\prevdroplet} \abundance_{\droplet[*]}(0) \right)
}  \right) \,.
\end{equation}

\section{Working Model for Compositional Heterogeneity}
\label{sec:model-comp-heter}

The hTPMH distribution uses a categorical distribution (the multivariate hypergeometric) which does not capture over-dispersion relative to the multinomial. In fact, its categorical distribution can also be derived using homogeneity assumptions, and only relaxes the assumptions regarding effectively infinite population size. So both hPoMu and hTPMH are unable to model over-dispersion of the categorical distribution caused by compositional heterogeneity. Thus the main change we need to make is to use a categorical distribution over-dispersed relative to the multinomial distribution.

In section \ref{sec:introduction-hpodm} I introduce a working model which fills this gap, and then in section \ref{sec:hpodm-gener-hpomu} I show that this working model extends what can be described using hPoMu alone.

\subsection{Introduction to hPoDM}
\label{sec:introduction-hpodm}

As suggested by its name, the ``hierarchical Poisson Dirichlet-Multinomial'' (hPoDM) distribution fills this gap by using a new distribution for the categorical distribution, the Dirichlet-Multinomial.

In section \ref{sec:dirichl-mult-dm} I introduce the Dirichlet-Multinomial distribution. In section \ref{sec:over-disp-dirichl} I explain how the Dirichlet-Multinomial distribution models over-dispersion relative to the multinomial distribution. In section \ref{sec:hpodm-definition} I use the new information learned to define the hPoDM family of distributions.

\subsubsection{Dirichlet-Multinomial (DM) Distribution}
\label{sec:dirichl-mult-dm}
The PMF for the Dirichlet-Multinomial distribution may be written
  \begin{equation}
    \label{eq:DM_pmf}
    \frac{\Gamma( \cconcentration \Species )}{\Gamma(\cconcentration \Species +  \population )} 
\!\cdot \!
\left[\prod_{\specie=1}^{\Species}  \frac{ \Gamma (\cconcentration\Species \freq^{(\specie)} + \population[\specie] )
}{
\Gamma \left(\cconcentration \Species \freq^{(\specie)} \right) 
}  \right]
\! \cdot \!
\binom{\population}{\population[1] \! \cdots \! \population[\Species]} \,.
  \end{equation}
More typically this is parameterized in terms of $\cconcentration \Species$. I parameterize in terms of $\cconcentration$ instead to facilitate interpretation. Parameterized this way, $\cconcentration = 1$ always corresponds to the same concentration as the uniform distribution on the $(\Species -1)$-dimensional simplex, regardless of the number of strains $\Strains$. 

\subsubsection{Clarification of Compositional Heterogeneity}
\label{sec:clar-comp-heter}

The intended notion of ``compositional heterogeneity'' herein does not include irregularities corresponding to e.g. strains preferentially grouping with or distancing from each other.

First, note that the Dirichlet-Multinomial can be thought of as a hierarchical distribution, where the $\freq^{(\strain)}$ parameters of the Multinomial distribution (living on the unit simplex due to the $\sum_{\strain=1}^{\Strains} \freq^{(\strain)} = 1$ constraint) are Dirichlet distributed. (Whence the name ``Dirichlet-Multinomial''.) Second, while the marginal distributions of the Dirichlet distribution are anti-correlated with one another, this is in some sense due only to the normalization condition, $\sum_{\strain=1}^{\Strains} \freq^{(\strain)} = 1$. It can be shown that the joint distribution of mutually independent Gamma random variables, when normalized by their sum, is Dirichlet. In particular, the Dirichlet distribution can not model a situation where e.g. particular strains $\strain_1$ and $\strain_2$ tend to attach to one another within the sampling pool.

One way to get distributions on the unit simplex able to model such preferential grouping or distancing of strains is to use correlated random variables with distributions derived from Gamma distributions and again normalize by their sum. Such an approach would most closely resemble the intuitive idea of a ``correlated Dirichlet distribution''. Another approach is to use the logistic-normal distribution. See \cite{Billheimer2001} for a study applying methodology based on the latter approach to the study of multivariate count data from ecology.

\subsubsection{Over-Disperson of Dirichlet-Multinomial with respect to Multinomial}
\label{sec:over-disp-dirichl}

The over-dispersion of the Dirichlet-Multinomial distribution with concentration $\cconcentration \Species$, total number $\population$, and frequencies $\vfreq$ with respect to the corresponding Multinomial distribution with total number $\population$ and frequencies $\vfreq$ is
\begin{equation}
  \label{eq:DM_overdispersion}
  \frac{
\population - 1
}{
1 + \cconcentration \Species 
} \,.
\end{equation}
Thus the concentration is approximately proportional to the reciprocal of the over-dispersion. (They are asymptotically equivalent up to a constant, i.e. the limit of their ratio as ${\cconcentration \to \infty}$ is the constant $n-1$.)

\subsubsection{hPoDM Definition}
\label{sec:hpodm-definition}

Under the hPoDM distribution, for the $\droplet$'th droplet: the probability of the strain distribution vector given the number of cells
\[\probability*{ \vabundance_{\droplet}(0) = \vpopulation | \abundance_{\droplet}(0) = \population }\]
is determined by a Dirichlet-Multinomial distribution, while the probability of the number of cells ${\probability*{\abundance_{\droplet}(0) = \population}}$ is determined by a Poisson distribution.

Thus for the unconditional probability:
  \begin{equation}
    \label{eq:hpodm_definition}
    \begin{split}
&    \probability*{\vabundance_{\droplet}(0) = \vpopulation} \\
 = & \quad
 \probability*{ \abundance_{\droplet}(0) = \population  } 
\!\cdot\!
\probability*{\vabundance_{\droplet}(0) = \vpopulation | \abundance_{\droplet}(0) = \population} 
\\
= & \quad
\frac{
e^{-\rate} \rate^{\population}
}{
\population !
}
\! \cdot \!
    \frac{\Gamma( \cconcentration \Species )}{\Gamma(\cconcentration \Species +  \population )} 
\!\cdot \!
\left[\prod_{\specie=1}^{\Species}  \frac{ \Gamma (\cconcentration\Species \freq^{(\specie)} + \population[\specie] )
}{
\Gamma \left(\cconcentration \Species \freq^{(\specie)} \right) 
}  \right]
\! \cdot \!
\binom{\population}{\population[1] \! \cdots \! \population[\Species]}  \,.
    \end{split}
  \end{equation}

\subsection{hPoDM Generalizes hPoMu}
\label{sec:hpodm-gener-hpomu}

Just like for hTPMH, hPoDM converges to the default hPoMu model under appropriate limits. Therefore when using the hPoDM family of distributions to model the distribution of the initial formation of droplets, we ``retain the same language'' used by the hPoMu distribution for that task.

In sections \ref{sec:appr-form-likel-2} and \ref{sec:proof-that-hpodm} I give the context needed and then show that the hNBDM family really is an extension of hPoMu. Because hPoDM and hPoMu have the same count distribution, this reduces to showing how the Dirichlet-Multinomial distribution extends the multinomial distribution.

\subsubsection{Approximate Form of Likelihood Ratio of DM with respect to Multinomial and of hPoDM with respect to hPoMu and Bounds}
\label{sec:appr-form-likel-2}
Using Lemma \ref{lem:gamma_stirling} we get that
\begin{equation}
  \label{eq:DM_stirling}
  \begin{split}
    & \frac{\Gamma(\cconcentration \Species)}{\Gamma(\cconcentration \Species + \population)}
\!\cdot \!
\left[\prod_{\specie=1}^{\Species} \frac{
\Gamma(\cconcentration \Species \freq^{(\specie)} + \population[\specie])
}{
\Gamma(\cconcentration \Species \freq^{(\specie)})
}\right] \\
= & \exp \left( \mysteryseq[DM](\vpopulation, \cconcentration \Species \vfreq)  \right)
\!\cdot\!
\left(
\frac{
\cconcentration \Species
}{
\cconcentration \Species + \population
}
\right)^{\cconcentration \Species + \population - \frac{1}{2}}\\
&\! \cdot \!
\left[  \prod_{\specie=1}^{\Species} \left(
\frac{
\cconcentration \Species \freq^{(\specie)} + \population[\specie]
}{
\cconcentration \Species \freq^{(\specie)}
}
\right)^{\cconcentration \Species \freq^{(\specie)} + \population[\specie] - \frac{1}{2}}  
\right]
\!\cdot\!
\prod_{\specie=1}^{\Species} (\freq^{(\specie)})^{\population[\specie]} \,,
  \end{split}
\end{equation}
where (with the upper $\uRobbins$ and lower $\lRobbins$ functions as defined before)
\begin{equation}
  \label{eq:DM_Robbins}
\lRobbins(\cconcentration \Species \vfreq, \cconcentration \Species \vfreq + \vpopulation) 
< \mysteryseq[DM](\vpopulation, \cconcentration \Species \vfreq) 
< \uRobbins(\cconcentration \Species \vfreq, \cconcentration \Species \vfreq + \vpopulation) \,.
\end{equation}
Unfortunately the bounds in (\ref{eq:DM_Robbins}) appear to be tight in general only for larger values of $\cconcentration$. (They become tighter as $\cconcentration \to \infty$; cf. section \ref{sec:proof-that-hpodm}.)

Thus the likelihood ratio of the Dirichlet-Multinomial distribution with respect to the Multinomial distribution, which is the same as the likelihood ratio of the hPoDM distribution with respect to the hPoMu distribution, based on equation \ref{eq:DM_stirling} above equals 
\begin{equation}
  \label{eq:DM_fudge_factor}
\exp \left( \mysteryseq[DM](\vpopulation, \cconcentration \Species \vfreq)  \right)
\!\cdot\!
\left(
\frac{
\cconcentration \Species
}{
\cconcentration \Species + \population
}
\right)^{\cconcentration \Species + \population - \frac{1}{2}}
\! \cdot \!
\left[  \prod_{\specie=1}^{\Species} \left(
\frac{
\cconcentration \Species \freq^{(\specie)} + \population[\specie]
}{
\cconcentration \Species \freq^{(\specie)}
}
\right)^{\cconcentration \Species \freq^{(\specie)} + \population[\specie] - \frac{1}{2}}  
\right]  
\,.
\end{equation}

\subsubsection{Sketch of Proof that hPoDM Converges in Distribution to hPoMu}
\label{sec:proof-that-hpodm}

Again, it suffices to show that the probability mass function of hPoDM converges to that of hPoMu. Cf. section \ref{sec:proof-that-htpmh}.

That the expression in (\ref{eq:DM_fudge_factor}) approaches $1$ as $\cconcentration \to \infty$ again follows from the Squeeze Theorem and Lemma \ref{lem:exponential_ratios}. (As before, that the expressions involving the upper and lower Robbins functions approach $0$ is an elementary consequence of the Archimedean property of the real numbers.) Thus the hPoDM distribution can be seen to converge to the hPoMu distribution as $\cconcentration \to \infty$. 

Because the likelihood ratio of the Dirichlet-Multinomial distribution with respect to the Multinomial distribution is the same as the likelihood ratio of the hPoDM distribution with respect to the hPoMu distribution, it follows that the above also shows how the Dirichlet-Multinomial distribution converges to the Multinomial distribution as $\cconcentration \to \infty$.

Instead of showing that the likelihood ratio converges to $1$, one can also show (e.g. according to one's personal preference) that the logarithm of the likelihood ratio converges to $0$, using Lemma \ref{lem:fudge_function} (the ``logarithmic version'' of Lemma \ref{lem:exponential_ratios}). The ``fudge function'' $\fudgefunction$ is defined for convenience in equation (\ref{eq:fudge_function}). Then, starting from (\ref{eq:DM_fudge_factor}), tedious algebra shows that the logarithm of the likelihood ratio of (hPo)DM with respect to (hPo)Mu is
  \begin{equation}
    \label{eq:DM_log_fudge}
    \sum_{\specie=1}^{\Species} \fudgefunction (\cconcentration \Species \freq^{(\specie)}, \population[\specie], \population[\specie] - \frac{1}{2})  - \fudgefunction (\cconcentration \Species, \population, \population - \frac{1}{2}) +  \mysteryseq[DM](\vpopulation, \cconcentration \Species \vfreq) \,.
  \end{equation}

\section{Working Model for Density Heterogeneity}
\label{sec:model-dens-heter}

The hPoDM working model has the same count distribution as hPoMu, and hence is unable to model over-dispersion of the count distribution caused by density heterogeneity. Thus the main change we need to make is from using the Poisson distribution as the count distribution to using a distribution over-dispersed relative to the Poisson distribution. Section \ref{sec:introduction-hnbdm} introduces a working model that fills this gap, and then section \ref{sec:hnbdm-gener-hpomu} shows that this working model can also be considered an extension of hPoMu.

\subsection{Introduction to hNBDM}
\label{sec:introduction-hnbdm}

As suggested by its name, the ``hierarchical Negative Binomial Dirichlet-Multinomial'' (abbreviated as hNBDM) working model fills this gap by using a new distribution for the count distribution, the negative binomial. Section \ref{sec:negative-binomial-nb} introduces the negative binomial distribution. Section \ref{sec:over-disp-negat} explains how the negative binomial distribution models over-dispersion relative to the Poisson distribution. Section \ref{sec:hnbdm-definition} defines the hNBDM working model.

\subsubsection{Negative Binomial (NB) Distribution}
\label{sec:negative-binomial-nb}

One may write the PMF of the negative binomial distribution as
  \begin{equation}
    \label{eq:NB_pmf}
    \frac{
\Gamma(\dconcentration \Species + \population)
}{
\Gamma (\dconcentration \Species)
}
\!\cdot \!
\frac{
(\dconcentration \Species)^{\dconcentration \Species}
}{
(\dconcentration \Species + \lambda)^{\dconcentration \Species + \population}
}
\!\cdot\!
\frac{
\rate^n
}{
n!
}
  \end{equation}
where $\rate$ denotes the expected value of the distribution.

More typically this is parameterized in terms of $\dconcentration\Species$ instead of $\dconcentration$. (Indeed in principle it should not even be necessary to specify a number of strains $\Species$ in order to define this distribution.) I parameterize in terms of $\dconcentration$ to be compatible with the parameterization of the Dirichlet-Multinomial I use in section \ref{sec:dirichl-mult-dm}. See section \ref{sec:hnbdm-definition} for how this pays off later by simplifying the math.

The parameter $\dconcentration$ is herein called the ``density concentration''. Using the negative binomial distribution for modelling density heterogeneity has precedent in the literature. For macroecological studies, the parameter $\rate$ ``has been defined as the density of organisms in the area of interest'' \cite{Willson1984}. Herein $\rate$ will be interpreted as the \textit{average} cell density in the sampling pool. Previous work has also claimed that $\dconcentration$ should be able to capture variability in the density of organisms: ``The practical use of [$\dconcentration\Species$] ... in ecological studies of aggregation requires caution because [$\dconcentration\Species$] is usually density-dependent'' \cite{ClarkPerry1989}. 

If a Poisson distribution with rate $\rate$ corresponds to a homogeneous cell density $\rate$ throughout the sampling pool, a negative binomial distribution with rate $\rate$ is interpreted to have the same \textit{average} cell density. Larger values of density concentration $\dconcentration$ correspond to \textit{smaller} heterogeneity around the average density, and thus more closely resemble the Poisson case (which is the limit as $\dconcentration \to \infty$), cf. figure \ref{fig:low_density_heterogeneity}. Similarly, smaller values of density concentration $\dconcentration$ correspond to \textit{larger} heterogeneity around the average density, cf. figure \ref{fig:high_density_heterogeneity}.

\subsubsection{Over-Disperson of Negative Binomial with respect to Poisson}
\label{sec:over-disp-negat}

The over-dispersion of the negative binomial distribution with concentration $\dconcentration \Species$ and mean $\rate$ with respect to the Poisson distribution with mean $\rate$ is
  \begin{equation}
    \label{eq:NB_overdispersion}
    \frac{\rate}{\dconcentration \Species} \,.
  \end{equation}
Thus the concentration is proportional to the reciprocal of the over-dispersion.

\subsubsection{hNBDM Definition}
\label{sec:hnbdm-definition}
For hNBDM we set $\dconcentration = \cconcentration =: \concentration$. So for the $\droplet$'th droplet, the strain distribution vector $\probability*{ \vabundance_{\droplet}(0) = \vpopulation | \abundance_{\droplet}(0) = \population }$ given the number of cells is Dirichlet-Multinomial distributed with parameter $\concentration$, and the number of cells ${\probability*{\abundance_{\droplet}(0) = \population}}$ is Negative Binomial distributed with parameter $\concentration$. Thus the PMF defining the unconditional probability is
  \begin{equation}
    \label{eq:hNBDM_pmf_hierarchical}
    \begin{split}
&    \probability*{\vabundance_{\droplet}(0) = \vpopulation} \\
 = & \quad
 \probability*{ \abundance_{\droplet}(0) = \population  } 
\!\cdot\!
\probability*{\vabundance_{\droplet}(0) = \vpopulation | \abundance_{\droplet}(0) = \population} 
\\
&    \frac{
\Gamma(\clustering \Species + \population)
}{
\Gamma (\clustering \Species)
}
\!\cdot \!
\frac{
(\clustering \Species)^{\clustering \Species}
}{
(\clustering \Species + \lambda)^{\clustering \Species + \population}
}
\!\cdot\!
\frac{
\rate^n
}{
n!
} 
    \frac{\Gamma( \clustering \Species )}{\Gamma(\clustering \Species +  \population )} 
\!\cdot \!
\left[\prod_{\specie=1}^{\Species}  \frac{ \Gamma (\clustering\Species \freq^{(\specie)} + \population[\specie] )
}{
\Gamma \left(\clustering \Species \freq^{(\specie)} \right) 
}  \right]
\! \cdot \!
\binom{\population}{\population[1] \! \cdots \! \population[\Species]}     
    \end{split}
  \end{equation}
which via tedious algebraic manipulations can be seen to equal
\begin{equation}
  \label{eq:hNBDM_pmf}
  \prod_{\specie=1}^{\Species} 
\left[
\frac{
\Gamma(\clustering \Species \freq^{(\specie)} + \population[\specie])
}{
\Gamma(\clustering \Species \freq^{(\specie)})
}
\!\cdot\!
\frac{
(\clustering\Species \freq^{(\specie)})^{\clustering \Species \freq^{(\specie)}}
}{
  (\clustering \Species \freq^{(\specie)} + \freq^{(\specie)} \rate)^{\clustering \Species \freq^{(\specie)} + \population[\specie]}
} 
\!\cdot\!
\frac{
(\freq^{(\specie)}\rate)^{\population[\specie]}
}{
(\population[\specie])!
}
\right] \,.
\end{equation}
The marginal distributions of the hNBDM distribution correspond to mutually independent Negative Binomial random variables whose parameters have been rescaled by the $\freq^{(\specie)}$. 

This makes the relationship of the Negative Binomial distribution to the Dirichlet-Multinomial a direct analogue of the relationship of the Poisson distribution to the Multinomial Distribution. Unsurprisingly these observations have precedent in the literature. Cf. for example Theorem 1 of \cite{Zhou2018}.

\subsection{hNBDM Generalizes hPoMu}
\label{sec:hnbdm-gener-hpomu}

Just like for hTPMH and hPoDM, hNBDM converges to the default hPoMu model under appropriate limits. Therefore when using the hNBDM family of distributions to model the distribution of the initial formation of droplets, we ``retain the same language'' used by the hPoMu distribution for that task.

In sections \ref{sec:appr-form-likel-3} and \ref{sec:proof-that-nb} I give the context needed and then show that, as a count distribution, the negative binomial distribution truly is a generalization of the Poisson. In sections \ref{sec:appr-form-likel-4} and \ref{sec:proof-that-hnbdm} I give the context needed and then show that the hNBDM family really is an extension of hPoMu.

\subsubsection{Approximate Form of Likelihood Ratio of Negative Binomial with respect to Poisson and Bounds}
\label{sec:appr-form-likel-3}
Applying Lemma \ref{lem:gamma_stirling}, one has
  \begin{equation}
    \label{eq:NB_stirling}
      \frac{
\Gamma(\dconcentration \Species + \population)
}{
\Gamma (\dconcentration \Species)
}
= 
e^{-n} \!\cdot \!
\left(
\frac{
\dconcentration \Species + \population
}{
\dconcentration \Species
}
\right)^{\dconcentration \Species + \population - \frac{1}{2}}
\!\cdot \!
 (\dconcentration \Species)^{\population}
\! \cdot \!
\exp \left( \mysteryseq[NB] (\dconcentration \Species, \population)  \right) \,,
  \end{equation}
where $\mysteryseq[NB](\dconcentration \Species, \population)$ satisfies
\begin{equation}
  \label{eq:NB_Robbins}
  \frac{1}{12(\dconcentration \Species + \population) + 1} - \frac{1}{12\dconcentration \Species}
< \mysteryseq[NB](\dconcentration \Species, \population)
< \frac{1}{12(\dconcentration \Species + \population)} - \frac{1}{12\dconcentration \Species + 1} \,.
\end{equation}
It follows from equations (\ref{eq:NB_pmf}) and (\ref{eq:NB_stirling}) that the likelihood ratio of the negative binomial distribution (with mean $\rate$) with respect to the Poisson distribution (with mean $\rate$) is
  \begin{equation}
    \label{eq:NB_fudge_factor}
    e^{-n} \!\cdot \!
\left(
\frac{
\dconcentration \Species + \population
}{
\dconcentration \Species
}
\right)^{\dconcentration \Species + \population - \frac{1}{2}}
\!\cdot \!
\left(
\frac{
\dconcentration \Species
}{
\dconcentration \Species + \rate
}
\right)^{\dconcentration \Species + \population}
\!\cdot \!
e^{\rate}
\!\cdot \!
\exp \left( \mysteryseq[NB](\dconcentration\Species, \population) \right) \,.
  \end{equation}

\subsubsection{Sketch of Proof that NB Converges in Distribution to Poisson}
\label{sec:proof-that-nb}
It follows nearly immediately from Lemma \ref{lem:exponential_ratios} that the above expression (\ref{eq:NB_fudge_factor}) converges to $1$ as $\dconcentration \to \infty$. (Again, both of the bounds for $\mysteryseq[NB](\dconcentration \Species, \population)$ from (\ref{eq:NB_Robbins}) approach $0$ for trivial reasons, allowing us to apply the Squeeze Theorem.) In particular, it follows that the negative binomial distribution converges to the Poisson distribution as $\dconcentration \to \infty$.

\subsubsection{Approximate Form of Likelihood Ratio of hNBDM with respect to hPoMu and Bounds}
\label{sec:appr-form-likel-4}

Using either of the expressions (\ref{eq:hNBDM_pmf_hierarchical}) or (\ref{eq:hNBDM_pmf}) for the PMF of the hNBDM distribution given above, more tedious algebraic manipulations (including application of Lemma \ref{lem:gamma_stirling}) allow one to show that the likelihood ratio of the hNBDM distribution relative to the corresponding hPoMu distribution is
\begin{equation}
  \label{eq:hNBDM_fudge_factor}
  \begin{adjustbox}{max width=\textwidth,keepaspectratio}
$\displaystyle  \begin{split}
& 
\left[
  \prod_{\specie=1}^{\Species} 
e^{-\population[\specie]} 
\!\cdot \!
\left(
\frac{
\clustering \Species \freq^{(\specie)} + \population[\specie]
}{
\clustering \Species \freq^{(\specie)}
}
\right)^{\clustering \Species \freq^{(\specie)} + \population[\specie] - \frac{1}{2}}
\!\cdot\!
\left(
\frac{
\clustering \Species \freq^{(\specie)}
}{
\clustering\Species \freq^{(\specie)} + \freq^{(\specie)} \rate
}
\right)^{\clustering\Species \freq^{(\specie)} + \population[\specie]}
\! \cdot \!
e^{\freq^{(\specie)} \rate}
\right] 
\! \cdot\!
\mysteryseq[hNBDM](\clustering \Species \vfreq, \vpopulation) \,,
  \end{split}$
\end{adjustbox}
\end{equation}
where
\begin{equation}
  \label{eq:hNBDM_Robbins}
\mysteryseq[hNBDM](\clustering \Species \vfreq, \vpopulation) = \mysteryseq[NB](\clustering\Species, \population) + \mysteryseq[DM](\clustering\Species\vfreq, \vpopulation) 
= \sum_{\specie=1}^{\Species}  \mysteryseq[NB](\clustering\Species \freq^{(\specie)}, \population[\specie]) \,.
\end{equation}

Observe that adding the bounds from (\ref{eq:DM_Robbins}) and (\ref{eq:NB_Robbins}) gives the following bounds for $\mysteryseq[hNBDM](\clustering\Species\vfreq, \vpopulation)$:
  \begin{equation}
    \label{eq:hNBDM_Robbins_hierarchical}
    \begin{split}
     & \scalebox{0.75}{ $
\displaystyle
\sum_{\specie=1}^{\Species} \left[
\frac{1}{12(\clustering\Species\freq^{(\specie)} \!+\! \population[\specie])\!+\!1}
\!-\!
\frac{1}{12\clustering \Species\freq^{(\specie)}}
\right]
\!+\!
\left(
\frac{1}{12\clustering\Species \!+\!1}
\!-\!
\frac{1}{12\clustering\Species}
\right)
\!+\!
\left(
\frac{1}{12(\clustering\Species \!+\! \population)\!+\! 1}
\!-\!
\frac{1}{12(\clustering\Species\!+\!\population)}
\right)
$ } \\
= & \quad
\lRobbins(\clustering\Species\vfreq, \clustering \Species \vfreq + \vpopulation) + \frac{1}{12(\clustering\Species + \population) + 1} - \frac{1}{12\clustering\Species} \\
< & \quad \mysteryseq[DM](\clustering\Species\vfreq, \vpopulation) + \mysteryseq[NB](\clustering\Species, \population) = \mysteryseq[hNBDM](\clustering\Species\vfreq, \vpopulation) \\
< & \quad\uRobbins(\clustering\Species\vfreq, \clustering\Species\vfreq + \vpopulation) + \frac{1}{12(\clustering\Species + \population)} - \frac{1}{12\clustering\Species + 1} \\
= & \scalebox{0.75}{$
\displaystyle
\sum_{\specie=1}^{\Species} \left[
\frac{1}{12(\clustering \Species \freq^{(\specie)}\!+\! \population[\specie])}
\!-\!
\frac{1}{12\clustering\Species \freq^{(\specie)} \!+\! 1}
\right]
\!+\!
\left(
\frac{1}{12\clustering \Species}
\!-\!
\frac{1}{12\clustering\Species\!+\!1}
\right)
\!+\!
\left(
\frac{1}{12 (\clustering \Species \!+\! \population)} 
\!-\!
\frac{1}{12(\clustering\Species \!+\! \population) \!+\! 1}
\right)
$}  \,.
    \end{split}
  \end{equation}

Compared to the above, observe that summing (over $\strain$) the bounds from (\ref{eq:NB_Robbins}) based on  (\ref{eq:NB_pmf}) actually gives strictly tighter bounds:
\begin{equation}
  \label{eq:hNBDM_Robbins_tight}
  \begin{split}
    & \sum_{\specie=1}^{\Species} \left[
\frac{1}{12(\clustering\Species\freq^{(\specie)} \!+\! \population[\specie])\!+\!1}
\!-\!
\frac{1}{12\clustering \Species\freq^{(\specie)}}
\right] \\
< & \sum_{\specie=1}^{\Species} \mysteryseq[NB](\clustering\Species\freq^{(\specie)}, \population[\specie]) = \mysteryseq[hNBDM](\clustering\Species\vfreq, \vpopulation) \\
< & \sum_{\specie=1}^{\Species} \left[
\frac{1}{12(\clustering \Species \freq^{(\specie)}\!+\! \population[\specie])}
\!-\!
\frac{1}{12\clustering\Species \freq^{(\specie)} \!+\! 1}
\right] \,.
  \end{split}
\end{equation}
The additional terms from the bounds in (\ref{eq:hNBDM_Robbins_hierarchical}) not present in the bounds from (\ref{eq:hNBDM_Robbins_tight}) are the result of twice applying Lemma \ref{lem:gamma_stirling} unnecessarily, once to $\frac{\Gamma(\clustering\Species+\population)}{\Gamma(\clustering\Species)}$ and once to $\frac{\Gamma(\clustering\Species)}{\Gamma(\clustering\Species +\population)}$. Because they cancel each other out, they obviously do not need to be bounded. Therefore the bounds do not contradict one another.

\subsubsection{Sketch of Proof that hNBDM Converges in Distribution to hPoMu}
\label{sec:proof-that-hnbdm}

Again, it suffices to show that the probability mass function of hNBDM converges to that of hPoMu. Cf. section \ref{sec:proof-that-htpmh}.

 That (\ref{eq:hNBDM_fudge_factor}) converges to $1$ as $\clustering \to \infty$ quickly follows from Lemma \ref{lem:exponential_ratios}. This convergence implies that the hNBDM distribution converges to the hPoMu distribution as $\clustering \to \infty$.

Equivalently, the convergence of the hNBDM distribution to the hPoMu distribution as $\clustering \to \infty$ follows from the convergence of (\ref{eq:NB_fudge_factor}) to $1$ as $\clustering \to \infty$ (replacing $\clustering$, $\population$, and $\rate$ with $\clustering \freq^{(\specie)}$, $\population[\specie]$, and $\freq^{(\specie)} \rate$ respectively). Because (\ref{eq:hNBDM_fudge_factor}) is the product of (\ref{eq:NB_fudge_factor}) and (\ref{eq:DM_fudge_factor}), it also equivalently follows from the convergence of both (\ref{eq:NB_fudge_factor}) and (\ref{eq:DM_fudge_factor}) to $1$ as $\clustering \to \infty$. 

Instead of showing that the likelihood ratio converges to $1$, one can also show (e.g. according to one's personal preference) that the logarithm of the likelihood ratio converges to $0$, using Lemma \ref{lem:fudge_function} (the ``logarithmic version'' of Lemma \ref{lem:exponential_ratios}). The ``fudge function'' $\fudgefunction$ is defined for convenience in equation (\ref{eq:fudge_function}). Then, starting from (\ref{eq:hNBDM_fudge_factor}), tedious algebra shows that the logarithm of the likelihood ratio of hNBDM with respect to hPoMu is
  \begin{equation}
    \label{eq:hNBDM_log_fudge}
    \rate - \population
+
\sum_{\specie=1}^{\Species} \fudgefunction (\concentration \Species \freq^{(\specie)}, \population[\specie], \population[\specie] \!-\! \frac{1}{2}) 
-
\sum_{\specie=1}^{\Species} \fudgefunction (\concentration \Species \freq^{(\species)}, \freq^{(\specie)} \rate, \population[\specie]) 
+\!
 \mysteryseq[hNBDM](\clustering\Species\vfreq, \vpopulation) \,.
  \end{equation}

\section{Working Model for Arbitrary Combinations of Heterogeneities}
\label{sec:model-arbitr-comb}

While the hNBDM family of distributions is useful for modelling density heterogeneity, it does have the limitation that the value of density heterogeneity is assumed to be coupled with corresponding value of compositional heterogeneity. By relaxing this restriction we can generalize the hNBDM family to the ``generalized hNBDM'' or ``ghNBDM'' family of distributions. This makes it possible to model arbitrary combinations of density and compositional heterogeneities.

\subsection{ghNBDM Family Definition}
\label{sec:ghnbdm-family-defin}

Under a distribution in the ghNBDM family, for the $\droplet$'th droplet: the probability for the strain distribution vector given the number of cells ${\probability*{ \vabundance_{\droplet}(0) = \vpopulation | \abundance_{\droplet}(0) = \population }}$ is described by a Dirichlet-Multinomial distribution with concentration parameter $\cconcentration$, while the probability for the number of cells ${\probability*{\abundance_{\droplet}(0) = \population}}$ is described by a Negative binomial distribution with concentration parameter $\dconcentration$. Thus for the unconditional probability:
  \begin{equation}
    \label{eq:ghnbdm_definition}
    \begin{split}
&    \probability*{\vabundance_{\droplet}(0) = \vpopulation} \\
 = & \quad
 \probability*{ \abundance_{\droplet}(0) = \population  } 
\!\cdot\!
\probability*{\vabundance_{\droplet}(0) = \vpopulation | \abundance_{\droplet}(0) = \population} 
\\
= & \quad
    \frac{
\Gamma(\dconcentration \Species + \population)
}{
\Gamma (\dconcentration \Species)
}
\!\cdot \!
\frac{
(\dconcentration \Species)^{\dconcentration \Species}
}{
(\dconcentration \Species + \lambda)^{\dconcentration \Species + \population}
}
\!\cdot\!
\frac{
\rate^n
}{
n!
}
\! \cdot \!
    \frac{\Gamma( \cconcentration \Species )}{\Gamma(\cconcentration \Species +  \population )} 
\!\cdot \!
\left[\prod_{\specie=1}^{\Species}  \frac{ \Gamma (\cconcentration\Species \freq^{(\specie)} + \population[\specie] )
}{
\Gamma \left(\cconcentration \Species \freq^{(\specie)} \right) 
}  \right]
\! \cdot \!
\binom{\population}{\population[1] \! \cdots \! \population[\Species]}  \,.
    \end{split}
  \end{equation}
Unlike for the hNBDM family, for arbitrary members of the ghNBDM family one may have that $\cconcentration \not= \dconcentration$. In general the marginal distributions of members of the ghNBDM family are \textit{not} mutually independent (and in fact have non-zero cross-covariance) and need not be negative binomial distributed.

The hNBDM family is the special case of the ghNBDM family when $\cconcentration = \dconcentration$. The hPoDM and hPoMu distributions emerge as limiting cases of the ghNBDM family, or are bona fide members if we allow ``$\infty$'' as a valid concentration parameter value. Cf. figure \ref{fig:ghNBDM}. The case where $\cconcentration$ is allowed to approach infinity while $\dconcentration$ remains finite corresponds to the hierarchical Negative Binomial Multinomial (hNBMu) distribution.

\begin{figure}[H]
  \centering
  \includegraphics[width=\textwidth,height=\textheight,keepaspectratio]{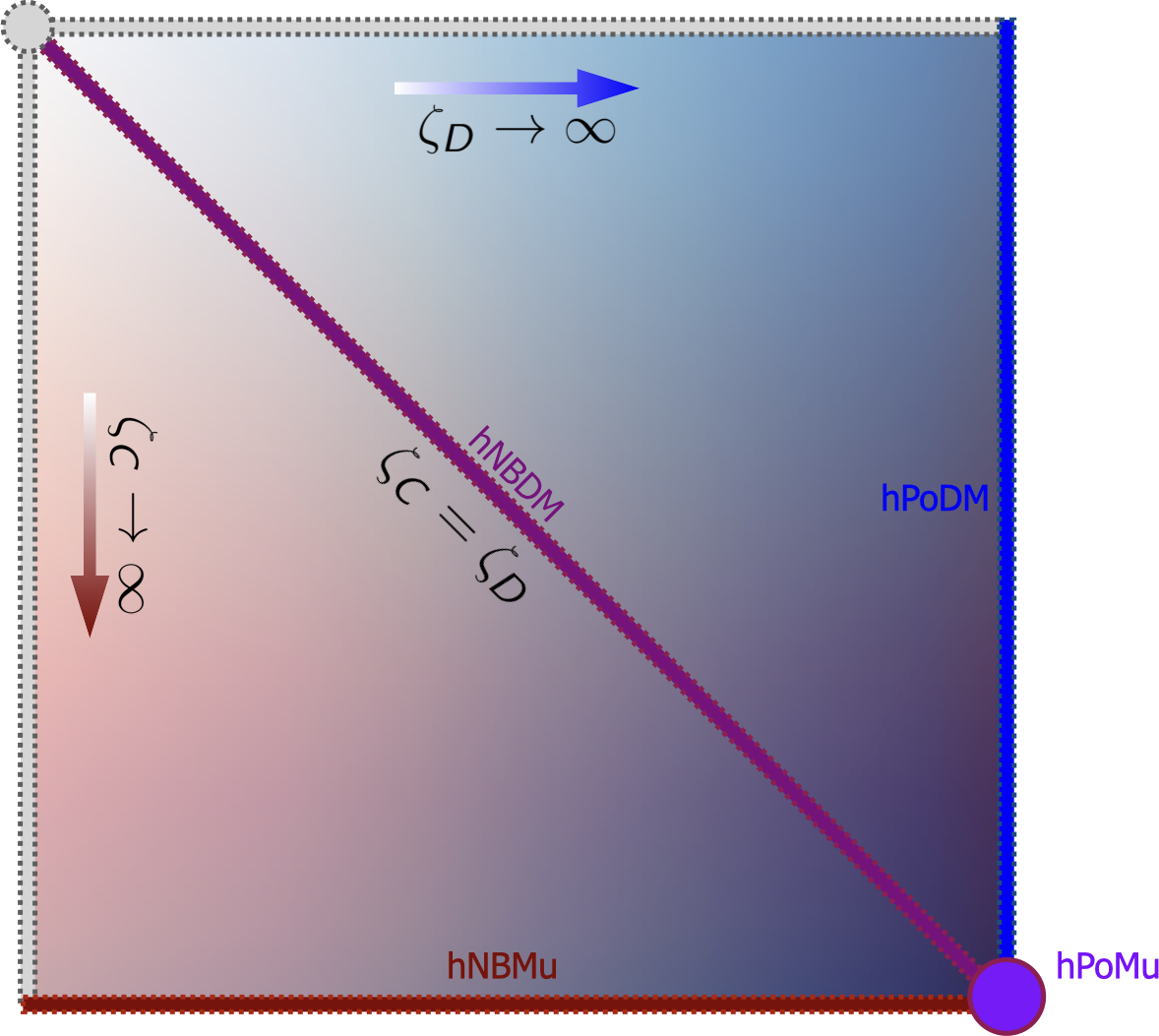}
  \caption[Heuristic schematic of ghNBDM working model.]{Schematic of the ghNBDM family of distributions, assuming  a fixed $\rate$ and vector of population strain frequencies $\vfreq$.}
  \label{fig:ghNBDM}
\end{figure}

\subsection{Over-Dispersions for these Distributions}
\label{sec:over-disp-these}

Since the non-diagonal covariances of the hPoMu distribution are $0$, it of course does not make sense to speak of over-dispersion relative to them using the definition (\ref{eq:vector_overdispersion}). However, we may still sensibly speak of the over-dispersion of their marginals with respect to the corresponding marginals of the hPoMu distribution (which unsurprisingly we will now do).

The over-dispersion of the $\strain$'th marginal of the ghNBDM distribution with respect to the $\strain$'th marginal of the hPoMu distribution is
  \begin{equation}
    \label{eq:GHNBDM_overdispersion}
   \frac{[(1+ \dconcentration\Species) + \freq^{(\strain)}(\cconcentration \Species - \dconcentration \Species)]}{(1+\cconcentration \Species)(\dconcentration \Species)} \rate \,.
\end{equation}
Observe how this value depends on the particular $\strain$. Moreover, the expression (\ref{eq:GHNBDM_overdispersion}) above: 
\begin{itemize}
\item reduces to (\ref{eq:HNBDM_overdispersion}) when $\dconcentration = \cconcentration$, 
\item approaches (\ref{eq:HPoDM_overdispersion}) when $\dconcentration \to \infty$,
\item approaches (\ref{eq:HNBMu_overdispersion}) when $\cconcentration \to \infty$, 
\item approaches $0$ when both $\dconcentration \to \infty$ and $\cconcentration \to \infty$.
\end{itemize}
The last corresponds to hPoMu, of course. Details about the remaining special cases (hNBDM, hPoDM, hNBMu) are given below.

\subsubsection{hNBDM Over-dispersion}
\label{sec:hnbdm-overdispersion}
  The over-dispersion of the $\strain$'th marginal of the hNBDM distribution with respect to the $\strain$'th marginal of the hPoMu distribution is
\begin{equation}
  \label{eq:HNBDM_overdispersion}
  \frac{\rate}{\concentration\Species} \,,
\end{equation}
which interestingly does \textit{not} depend on the particular strain $\strain$, and moreover equals the over-dispersion of a Negative Binomial distribution with respect to a Poisson distribution, cf. (\ref{eq:NB_overdispersion}).

The latter is to be expected given that the marginals of the hPoMu distribution are $\poisson(\freq^{(\specie)} \rate)$ distributed and the marginals of the hNBDM distribution are $\negbinom(\concentration\Species \freq^{(\specie)}, \freq^{(\specie)}\rate)$ distributed. Note that (\ref{eq:HNBDM_overdispersion}) also tends to $0$ as $\concentration \to \infty$ as it should. Finally observe how the over-dispersion (\ref{eq:HNBDM_overdispersion}) is inversely proportional to the concentration parameter $\clustering$, just like (\ref{eq:NB_overdispersion}) which it equals.

\subsubsection{hPoDM Over-dispersion}
\label{sec:hpodm-over-disp}

The over-dispersion of the $\strain$'th marginal of the hPoDM distribution with respect to the $\strain$'th marginal of the hPoMu distribution is
\begin{equation}
  \label{eq:HPoDM_overdispersion}
  \frac{1 - \freq^{(\strain)}}{1+\cconcentration\Species} \rate \,.
\end{equation}
Observe how this value depends on the particular $\strain$. And of course (\ref{eq:HPoDM_overdispersion}) tends to $0$ as $\cconcentration \to \infty$. Further observe how the over-dispersion (\ref{eq:HPoDM_overdispersion}) is (asymptotically) proportional to the inverse of the concentration parameter $\cconcentration$, just like (\ref{eq:DM_overdispersion}).

\subsubsection{hNBMu Over-dispersion}
\label{sec:hnbmu-over-disp}

The over-dispersion of the $\strain$'th marginal of the hNBMu distribution with respect to the $\strain$'th marginal of the hPoMu distribution is
  \begin{equation}
    \label{eq:HNBMu_overdispersion}
    \frac{\freq^{(\strain)}\rate}{\dconcentration\Species} \,.
  \end{equation}
Observe how this value depends on the particular $\strain$. And of course (\ref{eq:HNBMu_overdispersion}) tends to $0$ as $\dconcentration \to \infty$. Further observe how the over-dispersion (\ref{eq:HNBMu_overdispersion}) is proportional to the inverse of the concentration parameter $\dconcentration$, just like (\ref{eq:DM_overdispersion}).

\section{Conclusion}
\label{sec:conclusion}

\paragraph{Findings and Contributions} We have identified
\begin{itemize}
\item a default working model we can use to for the initial droplet formation,
\item key implicit assumptions made by that default working model,
\item concrete ways those assumptions could fail to be true in practice, and
\item working models generalizing the default working model which can account for any of those failed assumptions.
\end{itemize}

\paragraph{Practical Implications}
We can now start to ask how realistic the default working model is and what changes might need to be made to use a more realistic working model instead. Being able to use a more realistic working model means that the data-generating process for MOREI can be simulated more accurately. Being able to simulate the data-generating process for MOREI more accurately will ultimately help to identify best-performing methods for characterizing microbial interactions based on MOREI data, cf. part \ref{part:aver-treatm-effects}.

\paragraph{Next Steps and Open Questions} In chapter \ref{chap:model_comparison} we compare the default model with several of these alternative models to see whether there are noticeable differences. In chapter \ref{chap:data_throughput} we look at the effects of these alternative models on the throughput of data useful for characterizing microbial interactions. Finally in chapter \ref{chap:hetero_estimator_performance} we propose and investigate the performance of estimators which will allow us to ascertain which of these models is most realistic in practice. We want to know how much the default working model differs from other, potentially more realistic working models and how to quantify any such differences.

\begin{coolsubappendices}
\section{Results for Bounding Approximate Likelihood Ratios}
\label{sec:append-results-bound}

\begin{lemma}
\label{lem:factorial_stirling}
 Using Stirling's Approximation as given in \cite{Robbins},
 \begin{equation}
   \label{eq:factorial_stirling}
   \begin{split}
&     \frac{1}{12(n+m) + 1} - \frac{1}{12n}  \\
< &\log \left(   \left(  \frac{(n+m)!}{n!} \right)    \middle/ 
\left(  e^{-m} \left( \frac{n+m}{n} \right)^{n+m+\frac{1}{2}} n^m  \right)
   \right)  \\
< & \frac{1}{12(n+m)} - \frac{1}{12n + 1} \,.
   \end{split}
 \end{equation}
  \end{lemma}
\textbf{Proof:} This follows basically immediately from the bounds in \cite{Robbins} via simple but tedious algebraic manipulations. $\square$

Heuristically, this says that $\frac{(n+m)!}{n!} \approx n^m$ when $n$ is large. This is because the expression ${\left( \frac{n+m}{n} \right)^{n+m+\frac{1}{2}} }$ approaches $e^m$ as $n \to \infty$ by Lemma \ref{lem:exponential_ratios}.

 \begin{lemma}
\label{lem:simple_lhospital}
    Given $\alpha, c, y_1, y_2 \in \mathbb{R}$ (in particular $c$ is constant w.r.t. $x$):
\[ \lim_{x \to \infty} \left(  \frac{\alpha x + y_1 }{\alpha x + y_2} \right)^c = 1 \,.  \]
  \end{lemma}
\noindent \textbf{Proof:} By properties of limits, since $f(x) = x^c$ is continuous, this equals
\[ \left[ \lim_{x \to \infty} \left(  \frac{\alpha x + y_1 }{\alpha x + y_2} \right) \right]^c \,,  \]
which by L'H\^{o}spital's Rule equals $1^c = 1$. $\square$

\begin{lemma}
\label{lem:exponential_ratios}
    Given $\alpha, c, y_1, y_2 \in \mathbb{R}$, $\alpha > 0$, one has 
\[ \lim_{x \to \infty} \left( \frac{\alpha x + y_1}{ \alpha x + y_2} \right)^{\alpha x + c} = e^{y_1} e^{-y_2} \,.  \]
  \end{lemma}
\noindent \textbf{Proof:} By properties of limits, the LHS of the above equals 
\[ \left[ \lim_{x \to \infty} \left( \frac{\alpha x + y_1}{\alpha x + y_2}  \right)^{\alpha x} \right] \left[ \lim_{x \to \infty} \left( \frac{\alpha x + y_1}{\alpha x + y_2}  \right)^c \right] = \lim_{x \to \infty}  \left( \frac{\alpha x + y_1}{\alpha x + y_2}  \right)^{\alpha x}  \,,    \]
where the second equality uses Lemma \ref{lem:simple_lhospital}. Rearranging further:
\[  = \lim_{x \to \infty}\left[ \left(\left( \frac{ x + \frac{y_1}{\alpha}}{x}  \right)^{ x} \right)^{\alpha} \left( \left( \frac{x + \frac{y_2}{\alpha} }{x}  \right)^{ x} \right)^{-\alpha} \right]
\,, \]
from which the result more or less immediately follows from the properties of limits and the definition of the exponential function.\textit{ (In particular because $f_1(x) = x^{\alpha}$ and $f_2(x) = x^{-\alpha}$ are continuous.)} $\square$

\begin{lemma}
\label{lem:gamma_stirling}
      Using the generalization of the bounds on Stirling's approximation to the entire $\Gamma$ function as given in Theorem 5 of \cite{Gordon1994}
      \begin{equation}
        \label{eq:gamma_stirling}
        \begin{split}
          & \frac{1}{12(x+\beta) + 1} - \frac{1}{12x} \\
< &  \log \left(
\left(  \frac{\Gamma(x+\beta)}{\Gamma(x)}  \right)
\middle/
\left(  e^{-\beta} \left( \frac{x+\beta}{x}  \right)^{x+\beta - \frac{1}{2}} x^{\beta}  \right)
\right)
\\
< & \frac{1}{12(x+\beta)} - \frac{1}{12x + 1} \,.
        \end{split}
      \end{equation}
  \end{lemma}
\noindent \textbf{Proof:} Lemma \ref{lem:gamma_stirling} follows almost immediately from Theorem 5 of \cite{Gordon1994} via tedious algebraic manipulations. $\square$

 Heuristically, Lemma \ref{lem:gamma_stirling} can be interpreted as saying that $\frac{\Gamma(x + \beta)}{\Gamma(x)} \approx x^{\beta}$ for large $x$. Again, Lemma \ref{lem:exponential_ratios} implies that ${\left(\frac{x+\beta}{x}\right)^{x + \beta - \frac{1}{2}}}$ approaches $e^{\beta}$ as $x \to \infty$.

The ``fudge function'' $\fudgefunction$ is defined as
\begin{equation}
  \label{eq:fudge_function}
  \fudgefunction (x,y,z) := (x+z) \left[ \log(x+y)  - \log(x) \right] \,.
\end{equation}

\begin{lemma}
    \label{lem:fudge_function}
Given $y,z \in \mathbb{R}$ which are constant with respect to $x\in \mathbb{R}$,
\[ \lim_{x \to \infty} \fudgefunction(x,y,z) = y \,.   \]
  \end{lemma}
\noindent \textbf{Proof:} This is a corollary of Lemma \ref{lem:exponential_ratios}. Specifically, because the logarithm is continuous and continuous functions commute with limits,
\begin{equation*}
  \begin{split}
&\lim_{x \to \infty} \fudgefunction(x,y,z)  = \log \left( \lim_{x \to \infty}\exp( \fudgefunction(x,y,z) )  \right)  = \log \left( \lim_{x \to \infty} \left(  \frac{x+y}{x} \right)^{x+z} \right) \\
& = \log(e^y) = y\,,
  \end{split}
\end{equation*}
where the penultimate equality was an application of \ref{lem:exponential_ratios}. $\square$

\subsection{Comments about Robbins-like Bounds}
\label{sec:comm-about-robb}

The bounds on $n!$ given in \cite{Robbins} follow from those for the $\Gamma$ function given in Theorem 5 of \cite{Gordon1994} using the identity $n! = n \Gamma(n) = n (n-1)!$ (for all $n \in \mathbb{N}$). They are still slightly different though, because for $n \in \mathbb{N}$ the expression from \cite{Gordon1994} gives ${\Gamma(n) = \frac{n!}{n} =  (n-1)! \sim \sqrt{2\pi} n^{n-\frac{1}{2}}e^{-n}}$, whereas \textit{directly}\footnote{As opposed to indirectly, by dividing the expression for $n!$ by $n$.} applying the expression from \cite{Robbins} to $(n-1)$ gives ${(n-1)! \sim \sqrt{2\pi} (n-1)^{n-\frac{1}{2}}e^{-(n-1)}}$. Although these expressions are technically distinct, it follows from Lemma \ref{lem:exponential_ratios} that they are asymptotically equivalent ($\sim$, the limit of their ratio approaches $1$ as $n \to \infty$). Therefore they are consistent with one another.

\section{Cumulants of Hierarchical Count-Categorical Distributions}
\label{sec:append-cumul-hier}

The following are consequences of the law of total covariance (of which the law of total variance is a special case) and in turn from the law of total expectation.

  \begin{lemma}
    If $\rvec{X}$ is such that $\rvec{X}|N \sim \mult (N, \vfreq)$ or $\rvec{X} | N \sim \dirmult(\cconcentration\Species, N, \vfreq)$, then for all $\specie \in [\Species]$:
\[  \expectation{X_{\specie}} = \freq^{(\specie)} \expectation{N} \,. \]
  \end{lemma}

  \begin{lemma}
      \label{lem:HMult_variance}
    If $\rvec{X}$ is such that $\rvec{X}|N \sim \mult(N,\vfreq)$, then for all ${\strain \in [\Species]}$:
    \begin{equation}
      \label{eq:HMult_variance}
      \var{X_{\strain}} = (\freq^{(\strain)})^2 \var{N} + \freq^{(\strain)} (1- \freq^{(\strain)}) \expectation{N} \,.
    \end{equation}
  \end{lemma}

  \begin{lemma}
      \label{lem:HDirMult_variance}
    If $\rvec{X}$ is such that $\rvec{X}|N \sim \dirmult(\cconcentration\Species, N, \vfreq)$, then for all ${\strain \in [\Species]}$:
    \begin{equation}
      \label{eq:HDirMult_variance}
      \var{X_{\strain}} = \left[ \frac{\freq^{(\specie)}(1-\freq^{(\strain)})}{1 + \cconcentration \Species} + (\freq^{(\specie)})^2  \right] \var{N}
+
\frac{\freq{\strain}(1-\freq^{(\strain)})}{1 + \cconcentration\Species} \expectation{N} (\expectation{N} + \cconcentration \Species) \,.
    \end{equation}
  \end{lemma}
\noindent Observe how (using L'H\^{o}spital's rule) equation (\ref{eq:HDirMult_variance}) approaches (\ref{eq:HMult_variance}) as ${\cconcentration \to \infty}$.

  \begin{lemma}
\label{lem:HMult_covariance}
    If $\rvec{X} $ is such that $\rvec{X}|N \sim \mult(N, \vfreq)$, then for all ${\strain_1, \strain_2 \in [\Species]}$ ($\strain_1 \not=\strain_2$):
    \begin{equation}
\label{eq:HMult_covariance}
      \cov{X_{\strain_1}, X_{\strain_2}} = \freq^{(\strain_1)} \freq^{(\strain_2)} \left( \var{N} - \expectation{N} \right) \,.
    \end{equation}
  \end{lemma}

  \begin{lemma}
      \label{lem:HDirMult_covariance}
    If $\rvec{X}$ is such that $\rvec{X} | N \sim \dirmult(\cconcentration\Species, N, \vfreq)$, then for all ${\strain_1, \strain_2 \in [\Species]}$ (${\strain_1 \not= \strain_2}$):
    \begin{equation}
      \label{eq:HDirMult_covariance}
      \cov{X_{\strain_1}, X_{\strain_2}} = \freq^{(\strain_1)}\freq^{(\strain_2)} \left(  \var{N} - \frac{1}{1+\cconcentration \Species} \left[
\cconcentration \Species \expectation{N} + \expectation{N^2}
\right] \right) \,.
    \end{equation}
  \end{lemma}
\noindent Observe how (using L'H\^{o}spital's rule) equation (\ref{eq:HDirMult_covariance}) approaches (\ref{eq:HMult_covariance}) as ${\cconcentration \to \infty}$.

\subsection{Variance and Covariance of ghNBDM}
\label{sec:vari-covar-ghnbdm}
From Lemma \ref{lem:HDirMult_variance}, for the ghNBDM distributions, for each ${\strain \in [\Species]}$:
\begin{equation}
  \label{eq:GHNBDM_variance}
\var{\abundance[\strain](0)} =  \freq^{(\strain)} \rate + \frac{[\freq^{(\strain)}(1+ \dconcentration\Species) + (\freq^{(\strain)})^2(\cconcentration \Species - \dconcentration \Species)]}{(1+\cconcentration \Species)(\dconcentration \Species)} \rate^2  \,.
\end{equation}
Note how (\ref{eq:GHNBDM_variance}) reduces to (\ref{eq:HNBDM_variance}) when $\dconcentration = \cconcentration$. Similarly, (\ref{eq:GHNBDM_variance}) approaches (\ref{eq:HPoDM_variance}) as $\dconcentration \to \infty$, and (\ref{eq:GHNBDM_variance}) approaches (\ref{eq:HNBMu_variance}) as $\cconcentration \to \infty$, both conclusions following from L'H\^{o}spital's rule.

From Lemma \ref{lem:HDirMult_covariance}, for the ghNBDM distributions, for each ${\strain_1, \strain_2 \in [\Species]}$ (${\strain_1 \not= \strain_2}$):
\begin{equation}
  \label{eq:GHNBDM_covariance}
  \cov{\abundance[\strain_1](0), \abundance[\strain_2](0)} = \freq^{(\strain_1)} \freq^{(\strain_2)} \left[ \frac{\cconcentration \Species - \dconcentration \Species}{(\dconcentration \Species)(1 + \cconcentration \Species)} \right] \rate^2 \,.
\end{equation}
Note how (\ref{eq:GHNBDM_covariance}) reduces to (\ref{eq:HNBDM_covariance}) when $\dconcentration = \cconcentration$. Similarly, (\ref{eq:GHNBDM_covariance}) approaches (\ref{eq:HPoDM_covariance}) as $\dconcentration \to \infty$, and (\ref{eq:GHNBDM_covariance}) approaches (\ref{eq:HNBMu_covariance}) as $\cconcentration \to \infty$, both conclusions again following from L'H\^{o}spital's rule. Information about special cases (hPoMu, hNBDM, hPoDM, hNBMu) is given below.

\subsubsection{Variance and Covariance of hPoMu}
\label{sec:vari-covar-hpomu}
For the hPoMu distribution, for each ${\strain \in [\Species]}$:
\begin{equation}
  \label{eq:HPoMu_variance}
  \var{\abundance[\strain](0)} = \freq^{(\strain)} \rate \,,
\end{equation}
and for all ${\strain_1, \strain_2 \in [\Species]}$ (${\strain_1 \not= \strain_2}$):
\begin{equation}
  \label{eq:HPoMu_covariance}
 \cov{\abundance[\strain_1](0), \abundance[\strain_2](0)} = 0 \,.
\end{equation}
This follows either from applying Lemmas \ref{lem:HMult_variance} and \ref{lem:HMult_covariance} directly, or recalling that the marginals of the hPoMu distribution are mutually independent $\poisson (\freq^{(\strain)} \rate)$ random variables.

\subsubsection{Variance and Covariance of hNBDM}
\label{sec:vari-covar-hnbdm}
For the hNBDM distribution, for each ${\strain \in [\Species]}$:
\begin{equation}
  \label{eq:HNBDM_variance}
  \var{\abundance[\specie](0)} = \freq^{(\strain)} \rate + \frac{\freq^{(\strain)} }{\concentration \Species}\rate^2 \,,
\end{equation}
and for all ${\strain_1, \strain_2 \in [\Species]}$ (${\strain_1 \not= \strain_2}$):
\begin{equation}
  \label{eq:HNBDM_covariance}
  \cov{\abundance[\strain_1](0), \abundance[\strain_2](0)} = 0.
\end{equation}
These follow from applying Lemmas \ref{lem:HDirMult_variance} and \ref{lem:HDirMult_covariance} directly, or recalling that the marginals of the hNBDM distribution are mutually independent $\negbinom(\concentration\Species \freq^{(\strain)}, \freq^{(\strain)}\rate)$ random variables. Notice also how (\ref{eq:HNBDM_variance}) approaches (\ref{eq:HPoMu_variance}), as $\concentration \to \infty$, as it should. Of course (\ref{eq:HNBDM_covariance}) already equals (\ref{eq:HPoMu_covariance}).

\subsubsection{Variance and Covariance of hPoDM}
\label{sec:vari-covar-hpodm}
 For the hPoDM distribution, for each ${\strain \in [\Species]}$:
\begin{equation}
  \label{eq:HPoDM_variance}
  \var{\abundance[\specie](0)} =\freq^{(\strain)}\rate +  \frac{\freq^{(\strain)}(1-\freq^{(\strain)})}{1+\cconcentration \Species} \rate^2  \,,
\end{equation}
and for all ${\strain_1, \strain_2 \in [\Species]}$ (${\strain_1 \not= \strain_2}$):
\begin{equation}
  \label{eq:HPoDM_covariance}
  \cov{\abundance[\strain_1](0), \abundance[\strain_2](0)} = - \frac{\freq^{(\strain_1)}\freq^{(\strain_2)}}{1 + \cconcentration \Species} \rate^2 \,.
\end{equation}
These follow from applying Lemmas \ref{lem:HDirMult_variance} and \ref{lem:HDirMult_covariance}. Notice also how (\ref{eq:HPoDM_variance}) approaches (\ref{eq:HPoMu_variance}), and (\ref{eq:HPoDM_covariance}) approaches (\ref{eq:HPoMu_covariance}), as $\cconcentration \to \infty$, as it should.

\subsubsection{Variance and Covariance of hNBMu}
\label{sec:vari-covar-hnbmu}
From Lemma \ref{lem:HMult_variance}, for the hNBMu distribution, for each ${\strain \in [\Species]}$:
  \begin{equation}
    \label{eq:HNBMu_variance}
    \var{\abundance[\specie](0)} = \freq^{(\strain)} \rate + \frac{(\freq^{(\strain)})^2 }{\dconcentration \Species}\rate^2   \,.
  \end{equation}
As expected, (\ref{eq:HNBMu_variance}) approaches (\ref{eq:HPoMu_variance}) as $\dconcentration \to \infty$.

From Lemma \ref{lem:HMult_covariance}, for the hNBMu distribution, for each ${\strain_1, \strain_2 \in [\Species]}$ (${\strain_1 \not= \strain_2}$):
\begin{equation}
  \label{eq:HNBMu_covariance}
  \cov{\abundance[\specie_1](0), \abundance[\strain_2](0)} = \frac{\freq^{(\strain_1)}\freq^{(\strain_2)} }{\dconcentration \Species}\rate^2 \,.
\end{equation}
Again, as expected, (\ref{eq:HNBMu_covariance}) approaches (\ref{eq:HPoMu_covariance}) as $\dconcentration \to \infty$.

\end{coolsubappendices}
\end{coolcontents}

\chapter{Log-Likelihood Ratio Statistics for Goodness of Fit}
\label{chap:model_comparison}

Herein I show that log likelihood ratios can be used to determine whether failure of assumptions from section \ref{sec:impl-model-assumpt} causes substantially new behavior.
See sections \ref{sec:preliminaries}, \ref{sec:log-likel-rati}, and \ref{sec:model_comparison_results}.
Failure of the sampling without replacement assumption has negligible effects in practice, but failures of the other assumptions could be important.
See sections \ref{sec:model_comparison_results} and \ref{sec:model_comparison_discussion}.

Section \ref{sec:backgr-sign-4} provides context for the chapter. Section \ref{sec:preliminaries} explains some of the technical details of the comparisons. Section \ref{sec:model_comparison_methods} explains the implementation details of the comparisons. Section \ref{sec:model_comparison_results} describes and explains the comparisons. Section \ref{sec:model_comparison_discussion} interprets the results from section \ref{sec:model_comparison_results}, explaining what they mean for modelling the initial formation of droplets. Section \ref{sec:conclusion-4} summarizes the findings and concludes with implications and next steps.

\begin{coolcontents}

\section{Background and Significance}
\label{sec:backgr-sign-4}

The overall scientific problem remains how to design the experiment so that enough data is collected from a single experimental run to adequately power statistical inference, without additional costly and time-consuming experimental runs being required. Cf. section \ref{sec:motiv-targ-estim}. As argued in section \ref{sec:t=0-experiment}, this amounts to knowing the joint distribution of the strain counts well enough to predict the probabilities associated with any possible combination of strains. In the previous chapter \ref{cha:model-init-form}, I identified a simple working model (hPoMu) for the initial formation of droplets which practitioners are likely to assume is unquestionably accurate. I explained the assumptions required to justify such a simple working model, and presented working models resulting from relaxing some of the most unreasonable of those assumptions.

\paragraph{Broader field}
The problem considered in this chapter falls under the general field of ``model selection''. See \cite[chapter 7]{Hastie2009}  for a more widely accessible introduction to the general theory of model selection, or \cite{Barron1999} for a highly sophisticated discussion thereof.
Cf. the discussion from the introduction to Part \ref{part:modell-init-form}.
Herein we only need to consider the subfield of model selection corresponding to so-called ``goodness of fit'' tests, in particular those using the classic likelihood ratio statistic. For a basic introduction, tailored for non-statisticians, on the use of the likelihood ratio statistic for model selection with ecological count data, see \cite{Lewis2010}.
For a more sophisticated discussion of the likelihood ratio statistic, see \cite[chapter 17]{Keener2010} or \cite[chapter 16]{vaart_1998} for information such as a general definition, or the statement and proof of the result about its asymptotic $\chi^2$ distribution.

\paragraph{Specific problem}
Herein I investigate whether the failure of any of the implicit hPoMu assumptions from section \ref{sec:impl-model-assumpt} causes substantially new behavior. As explained previously, failure of any of these assumptions can be associated with a new working model that generalizes hPoMu.
Therefore, if data generated by a given such generalized working model fits hPoMu well, it seems reasonable to say that the generalized working model, and thus also the corresponding assumption failure, is not causing substantially new behavior.

On the other hand, given two working models which are both adequate for describing the initial droplet formation process, the simpler working model will be easier to understand. Therefore, given two working models with the same descriptive power, we should choose the simpler working model. It follows that justifying the use of a more complicated working model over a simpler working model requires showing that the more complicated working model has substantially more descriptive power than the simpler working model.

Therefore, to justify potentially using any of the other, more complicated working models in place of hPoMu, we must first ascertain whether any of the other working models have substantially more descriptive power than hPoMu. Thus we want to ascertain how well hPoMu can describe data generated by the other working models. Only those working models which do not always produce data that can be described well by hPoMu have substantially more descriptive power than hPoMu. Any other working models are not as useful.

In statistical terms, we know that a working model as simple as hPoMu will inevitably lead to biased estimators. So we want to investigate whether the failure of any of these assumptions can cause estimates derived from hPoMu to become biased enough that they are no longer useful. Thinking of the estimates produced by each of these working models as an algorithm, in the targeted learning roadmap \cite{vanderLaan2011} we want to investigate whether we can rely on a single ``default'' algorithm based on questionable assumptions.

\paragraph{Particular approach} In this chapter I describe the information needed to implement goodness of fit tests based on the log-likelihood ratio statistic that compare any one of these more complicated working models against the ``default'' working model (hPoMu). After generating data from any of these working models whose assumptions are more relaxed than those of hPoMu, if these tests demonstrate that the resulting data are significantly different than what would be expected under the null model (hPoMu), then this suggests that the predictions made by hPoMu are inadequate. We could conclude that we may need to pay attention to whether the corresponding assumption of hPoMu is violated in practice. In contrast, if the tests demonstrate no significant difference between the resulting data and what would be expected under the null model (hPoMu), then this suggests we do not need to care whether the corresponding assumption of hPoMu is violated in practice.

Comparing the fit of two models for a given dataset by comparing the values of their likelihoods is a straightforward idea. In turn the (arguably) simplest way to make that comparison is to consider (the logarithm of) the ratio of those likelihoods. Because this is exploratory and preliminary work, there does not seem to be any compelling reason to use approaches that are more sophisticated than the simplest possible.

For problems similar to these, either the Wald statistic or the Rao score statistic is often used in place of the likelihood ratio statistic. However, in addition to being less straightforward, these statistics are also ``equivalent'' (in a sense that can be made precise) to the likelihood ratio statistic. See \cite[section 16.2]{vaart_1998} or \cite[section 17.4]{Keener2010} for discussion of the relationship of the Wald, Rao score, and likelihood ratio statistics.

Similarly, the AIC (Akaike Information Criterion) and BIC (Bayesian Information Criterion) methods can both (roughly) be thought of as more sophisticated versions of the likelihood ratio statistic. See, for example, \cite{multimodel}, in particular \cite[section 6.9.3]{multimodel}. Thus the same preference for simplicity described above is also a reason for not using AIC or BIC.

The ``null model'' in this case, hPoMu, is (effectively by design) nested within all of these generalized working models against which it is being compared.
Nevertheless, as explained in \cite{Lewis2010}, the assumption that the null model is nested within the alternative model is not necessary in general to make comparisons using the likelihood ratio statistic. So failure of nesting in other contexts is not a reason to avoid using the likelihood ratio statistic.

In comparing the performances of ``Negative Binomial Factor Analysis'' (related to the hNBDM model) and ``Poisson Factor Analysis'' (related to the hPoMu model), it can be argued that \cite[section 6]{Zhou2018} effectively addresses a problem similar to the one considered in this chapter. However, the underlying questions being asked therein are substantially different from those questions being asked herein. Hence, there seems to be no reason to expect that the comparison from \cite[section 6]{Zhou2018} would map (well) to this problem.

\section{Preliminaries}
\label{sec:preliminaries}

Below I give the formulae for the log likelihood ratios comparing more complex working models to hPoMu. These exact formulae for the log likelihood ratios do not invoke any version of Stirling's approximation. I used these formulae as described in section \ref{sec:log-likel-rati}.

\subsection{hTPMH}
\label{sec:exact_log_lr_htpmh}

The likelihood ratio of the hTPMH distribution with respect to the corresponding hPoMu distribution is
\begin{equation}
  \label{eq:exact_lr_htpmh}
  \probability*{\poisson (\rate) \le \Population_{\prevdroplet[*]}}^{-1} 
\!\cdot\!
  \frac{(\Population_{\prevdroplet[*]} - \population)!}{(\Population_{\prevdroplet[*]})!}
\!\cdot\!
\left[\prod_{\specie=1}^{\Species} 
 \frac{(\Population[\specie]_{\prevdroplet[*]})!}{(\Population[\specie]_{\prevdroplet[*]} - \population[\specie])!}
\right]
\!\cdot\!
\left[
\prod_{\specie=1}^{\Species}
(\freq^{(\specie)})^{\population[\specie]}
\right]^{-1} \,.
\end{equation}
Therefore the log likelihood ratio is
\begin{equation}
  \label{eq:exact_log_lr_htpmh}
  \begin{split}
     -\log\left(\probability*{\poisson (\rate) \le \Population_{\prevdroplet[*]}}\right) 
\!+\!
\log\left( (\Population_{\prevdroplet[*]} - \population)! \right) 
\!-\!
\log\left( (\Population_{\prevdroplet[*]})!  \right) 
\!+\!
\\
\sum_{\specie=1}^{\Species}
\left[
\log \left( (\Population[\specie]_{\prevdroplet[*]})! \right)
\!-\!
\log \left( (\Population[\specie]_{\prevdroplet[*]} - \population[\specie])! \right)
\right]
\!-\!
\sum_{\specie=1}^{\Species}
\population[\specie] \log(\freq^{(\specie)}) \,.
  \end{split}
\end{equation}
Using the identity $\Gamma(n+1) = n!$, the above can be rewritten using the log-gamma function\footnote{Which is most likely more numerically stable and computationally tractable than directly computing the logarithm of the factorial.} (which is directly implemented in SciPy \cite{SciPy}):
\begin{equation}
  \label{eq:exact_log_lr_htpmh_loggamma}
  \begin{split}
     -\log\left(\probability*{\poisson (\rate) \le \Population_{\prevdroplet[*]}}\right) 
\!+\!
\log\left( \Gamma(\Population_{\prevdroplet[*]} - \population + 1) \right) 
\!-\!
\log\left( \Gamma(\Population_{\prevdroplet[*]} + 1)  \right) 
\!+\!
\\
\sum_{\specie=1}^{\Species}
\left[
\log \left( \Gamma (\Population[\specie]_{\prevdroplet[*]} + 1) \right)
\!-\!
\log \left( \Gamma(\Population[\specie]_{\prevdroplet[*]} - \population[\specie] + 1) \right)
\right]
\!-\!
\sum_{\specie=1}^{\Species}
\population[\specie] \log(\freq^{(\specie)}) \,.
  \end{split}
\end{equation}
As an aside, because the above expression equals (\ref{eq:hTPMH_log_fudge}) from section \ref{sec:proof-that-htpmh}, one can derive an explicit formula for $\mysteryseq[MHg](\vpopulation, \vPopulation_{\prevdroplet[*]})$. It is not very elegant.

For all observed values of $\Population_{\prevdroplet[*]}$, $\probability*{\poisson (\rate) \le \Population_{\prevdroplet[*]}}^{-1}$ was indistinguishable from $1$ up to numerical precision. (Indeed for $\rate=2$ that probability appears to be indistinguishable from $1$ up to numerical precision already for $\Population_{\prevdroplet[*]} \ge 20$, values many orders of magnitude smaller than any observed value.) Therefore the contribution of the logarithm of this term was treated as if it was exactly $0$ in the code implementation and not explicitly included.

\subsection{hPoDM}
\label{sec:exact_log_lr_hpodm}

The likelihood ratio of the hPoDM distribution with respect to the corresponding hPoMu distribution is
\begin{equation}
  \label{eq:exact_lr_hpodm}
  \frac{
\Gamma(\cconcentration \Species)
}{
\Gamma(\cconcentration\Species + \population)
}
\!\cdot\!
\left[\prod_{\specie=1}^{\Species}  \frac{ \Gamma (\cconcentration\Species \freq^{(\specie)} + \population[\specie] )
}{
\Gamma \left(\cconcentration \Species \freq^{(\specie)} \right) 
}  \right]
\! \cdot \!
\left[
\prod_{\specie=1}^{\Species}
(\freq^{(\specie)})^{\population[\specie]}
\right]^{-1} \,.
\end{equation}
Therefore the log likelihood ratio is
\begin{equation}
  \label{eq:exact_log_lr_hpodm}
  \begin{adjustbox}{max width=\textwidth,keepaspectratio}
$\displaystyle  \log \left(  \Gamma(\cconcentration \Species) \right)
-
\log \left( \Gamma(\cconcentration\Species\! +\! \population)  \right)
+
\sum_{\specie=1}^{\Species}
\left[
\log \left( \Gamma (\cconcentration\Species \freq^{(\specie)}\! +\! \population[\specie] )  \right)
\!-\!
\log \left(  \Gamma \left(\cconcentration \Species \freq^{(\specie)} \right)  \right)
\right]
\!-\!
\sum_{\specie=1}^{\Species}
\population[\specie] \log(\freq^{(\specie)}) \,. $
\end{adjustbox}
\end{equation}
As an aside, because the above expression equals (\ref{eq:DM_log_fudge}) from section \ref{sec:proof-that-hpodm}, one can derive an explicit formula for $\mysteryseq[DM](\vpopulation, \cconcentration \Species \vfreq)$. Again, it is not elegant.

\subsection{hNBDM}
\label{sec:exact_log_lr_hnbdm}

The likelihood ratio of the hNBDM distribution with respect to the corresponding hPoMu distribution is
\begin{equation}
  \label{eq:exact_lr_hnbdm}
  \prod_{\specie=1}^{\Species}
\left[
\frac{
\Gamma(\concentration\Species\freq^{(\specie)} + \population[\specie])
}{
\Gamma(\concentration\Species\freq^{(\specie)})
}
\!\cdot \!
\frac{
(\concentration\Species\freq^{(\specie)})^{\concentration\Species \freq^{(\specie)}}
}{
(\concentration\Species\freq^{(\specie)} + \freq^{(\specie)}\rate)^{\concentration\Species\freq^{(\specie)} + \population[\specie]}
}
\right]
\! \cdot \!
\left[
\prod_{\specie=1}^{\Species}
e^{-\freq^{(\specie)}\rate}
\right]^{-1} \,.
\end{equation}
Therefore the log likelihood ratio is
\begin{equation}
  \label{eq:exact_log_lr_hnbdm}
  \begin{split}
    \sum_{\specie=1}^{\Species}
\left[
\log\left( \Gamma(\concentration\Species\freq^{(\specie)} \!+\! \population[\specie]) \right)
-
\log \left( \Gamma(\concentration\Species\freq^{(\specie)}) \right)
+
\concentration \Species \freq^{(\specie)} \log\left( \concentration\Species\freq^{(\specie)} \right)
\right.
\\
\left. -
(\concentration\Species\freq^{(\specie)} + \population[\specie]) \log \left( \concentration\Species\freq^{(\specie)} + \freq^{(\specie)}\rate  \right)
\right]
+
\sum_{\specie=1}^{\Species}
\freq^{(\specie)} \rate
\,.
  \end{split}
\end{equation}
As an aside, because the above expression equals (\ref{eq:hNBDM_log_fudge}) from section \ref{sec:proof-that-hnbdm}, one can derive an explicit formula for $\mysteryseq[hNBDM](\clustering\Species\vfreq, \vpopulation)$. Again, it is not elegant.

\section{Methods}
\label{sec:model_comparison_methods}

Section \ref{sec:simul-distr} explains which distributions were simulated, why, and how. Section \ref{sec:log-likel-rati} explains how logarithms of the likelihood ratios were computed for five of the distributions. I used NumPy \cite{NumPy} version 1.20.2 and SciPy \cite{SciPy} version 1.6.2 for computations, and Matplotlib \cite{Matplotlib} version 3.4.1 and/or Seaborn \cite{Seaborn} version 0.11.1 for plots. Complete implementation details can be found in the code at \dropletsgitrepo. See \url{\dropletsgitrepourl}.

\subsection{Simulated Distributions}
\label{sec:simul-distr}

To test how much the assumptions of hPoMu may be violated with hPoMu remaining an adequate model, I simulated five distributions. 
The first distribution, the hierarchical truncated Poisson Multivariate Hypergeometric (hTPMH) accounts for sampling without replacement (section \ref{sec:simul-sampl-without}), removing the sampling with replacement approximation of hPoMu. The remaining four distributions removed the homogeneity assumption of hPoMu by modelling density and compositional heterogeneities via over-dispersion (sections \ref{sec:simul-comp-het} and \ref{sec:simul-comp-dens}). Details of the simulations common to all distributions are explained in section \ref{sec:simul-impl-deta}.

\subsubsection{Simulating Sampling without Replacement}
\label{sec:simul-sampl-without}

I used an initial total population size of $500,000,000$ ($500$ million) cells for defining the parameter $\Population$ of hTPMH. This is substantially less than the number of cells ($\sim 10^{10}$) which would be sampled from in practice. The NumPy version of the multivariate hypergeometric distribution does not scale beyond a total population size of $10^9$.

\subsubsection{Simulating Compositional Heterogeneity Only}
\label{sec:simul-comp-het}

The first two distributions model only compositional heterogeneity and have no density heterogeneity. hPoDM (cf. section \ref{sec:model-comp-heter}) with compositional concentration $\cconcentration = 100$ has the lowest compositional heterogeneity, followed by hPoDM with $\cconcentration = 1$.

\subsubsection{Simulating Compositional and Density Heterogeneities}
\label{sec:simul-comp-dens}

The final two distributions model both compositional and density heterogeneity, with both assumed to correspond to equal over-dispersion. hNBDM (cf. section \ref{sec:model-dens-heter}) with concentration $\concentration:= \cconcentration = \dconcentration = 100$ has the lowest heterogeneities, followed by hNBDM with $\concentration=1$.

\subsubsection{Simulation Implementation Details}
\label{sec:simul-impl-deta}

For all simulations and distributions, the same ``simulated community'' (or ``SimCom'') of $91$ strains\footnote{Herein I use ``strains'' to refer equally to strains belonging to the same species(/genus/family/etc.) as well as to strains belonging to different species(/genera/families/etc.), because the distinction is irrelevant for setting up the abstract problem. It may matter for the implementation of a specific experiment.} was used. These correspond to $90$ strains distributed across $9$ distinct relative abundances or population frequencies: $10$ strains for each of $.01\%$, $.02\%$, $.05\%$, $.1\%$, $.2\%$, $.5\%$, $1\%$, $2\%$, and $5\%$. The $91$st strain was a ``remainder'' strain with abundance $\approx 12\%$. All distributions also used the same value of the rate parameter, $\rate =2$. This equals the expected number of cells for each droplet for all distributions except\footnote{The expectation of a truncated Poisson distribution will be less than that of the corresponding Poisson distribution. In this context any deficit compared to $\rate =2$ is negligible in practice, because the support was always truncated to a value no smaller than $\approx 400$ million.} technically for hTPMH.

Each of the distributions was statistically independently simulated $500$ times, with each simulation having $15$ million statistically independently simulated droplets, except for hTPMH, for which the statistical dependence structure of successive droplets is inherently Markov. (This corresponds to $5$ batches, each with $3$ million droplets.)

\subsection{Log Likelihood Ratios}
\label{sec:log-likel-rati}

I derived explicit formulae for the logarithms of the likelihood ratios relative to hPoMu for the hTPMH, hPoDM, and hNBDM families of distributions (cf. section \ref{sec:preliminaries}). I did not attempt to derive analogous formulae for the hExhPoDM and hExhNBDM families of distributions to avoid making mistakes when attempting to compute anti-derivatives of expressions involving the (logarithm of the) Gamma function. (Cf. section \ref{sec:simul-distr-redux}.)

Then, for each of the five distributions (hTPMH, hPoDM $\cconcentration=100$, hNBDM $\concentration=100$, hPoDM $\cconcentration=1$, hNBDM $\concentration=1$) with explicit formulae for the logarithm of their likelihood ratios with respect to hPoMu, of the $500$ simulations I arbitrarily chose one (the $18$th) and evaluated the respective formula on all $15$ million droplets. For each of the five distributions, I also collected the results of doing so into gluttonous groups\footnote{See section \ref{sec:gluttonous-groups} for clarification of the specific meaning of the term ``gluttonous groups''. See also footnote \ref{footnote:gluttonous_faux_controls} of section \ref{sec:gluttonous-groups} regarding the groups defining the diagonals of the heatmaps.} (which by definition overlap) and computed the geometric mean of the observed likelihood ratios for all groups.

\section{Results}
\label{sec:model_comparison_results}

Section \ref{sec:sampl-without-repl} discusses evidence indicating that sampling with replacement is the most innocuous of the assumptions behind hPoMu. Section \ref{sec:very-simil-distr} discusses evidence indicating that ghNBDM distributions with low (but nonzero) heterogeneity may be adequately modeled by hPoMu. Finally, section \ref{sec:slightly-diff-distr} discusses evidence indicating that ghNBDM distributions with only moderate heterogeneity can be easily distinguished from hPoMu.

\subsection{Sampling without Replacement}
\label{sec:sampl-without-repl}

All available evidence suggests that, even after $15,000,000$ droplets have been formed, the hTPMH distribution for the chosen value of $\Population$ is extremely similar to the corresponding hPoMu distribution.

\begin{figure}[p]
  \centering
  \begin{subfigure}{\textwidth}
  \centering
\includegraphics[width=\textwidth,height=0.45\textheight,keepaspectratio]{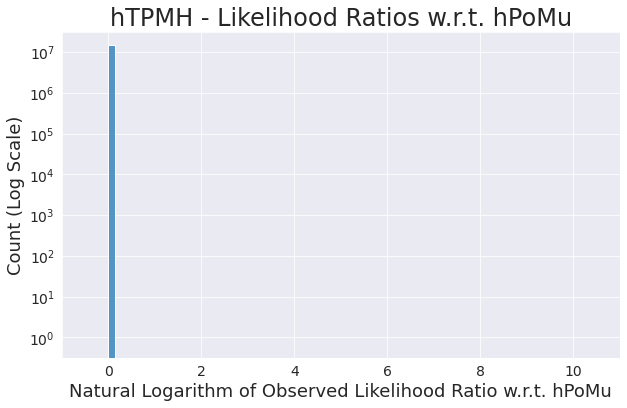}  
  \caption[]{Likelihood ratio histogram for hTPMH.}
  \label{fig:hTPMH_LRs_histogram}
\end{subfigure}

\begin{subfigure}{\textwidth}
  \centering
\includegraphics[width=\textwidth,height=0.45\textheight,keepaspectratio]{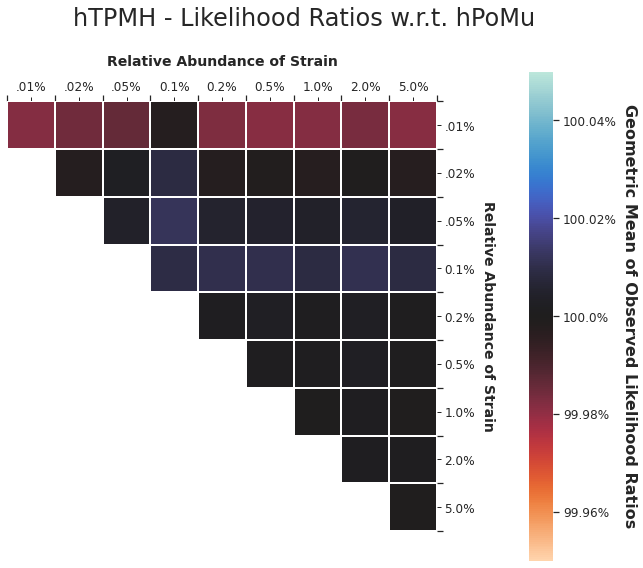}  
  \caption[]{Grouped likelihood ratios for hTPMH.}
  \label{fig:hTPMH_LRs_heatmap}
\end{subfigure}

\caption[Likelihood ratios for hTPMH, histogram and grouped.]{The likelihood ratio with respect to hPoMu never substantially differs from $1$.}
\label{fig:hTPMH_LRs}
\end{figure}

\subsubsection{Distribution of Likelihood Ratios w.r.t. hPoMu}
\label{sec:distr-likel-rati}

 When plotted on the same scale as that used for the other distributions, the histogram for hTPMH of observed likelihood ratios relative to hPoMu seen in figure \ref{fig:hTPMH_LRs_histogram} has no right tail whatsoever. This indicates few, if any, of the droplets were substantially ``better explained'' using the hTPMH distribution compared to using hPoMu. 

\subsubsection{Typical Likelihood Ratios as a Function of Strain Frequency}
\label{sec:typic-likel-rati}

The heat map of typical values for hTPMH of the likelihood ratio across gluttonous groups showin in figure \ref{fig:hTPMH_LRs_heatmap} also shows, even at worst, negligible differences from hPoMu. The values are slightly smaller for the least abundant strains, reflecting how the sampling with approximation is weakest for those strains. Even so, the values are still very close to $1$ and well within the range observed for other distributions.

\subsection{Very Similar Distributions}
\label{sec:very-simil-distr}

All available evidence suggests that the hPoDM with $\cconcentration =100$ and hNBDM with $\concentration =100$ distributions are very similar to the corresponding hPoMu distribution. Nevertheless, they also appear to differ slightly more than the hTPMH distribution.

\subsubsection{Distributions of Likelihood Ratios w.r.t. hPoMu}
\label{sec:distr-likel-rati-1}

Unlike what occurs for hTPMH, figures \ref{fig:hPoDM_100_LRs_histogram} and \ref{fig:hNBDM_100_LRs_histogram} respectively do show the distributions of observed likelihood ratios for hPoDM with $\cconcentration=100$ and hNBDM with $\concentration=100$ having right tails. However, these right tails are small. 

\begin{figure}[p]
  \centering
  
  \begin{subfigure}{\textwidth}
  \centering
\includegraphics[width=\textwidth,height=0.45\textheight,keepaspectratio]{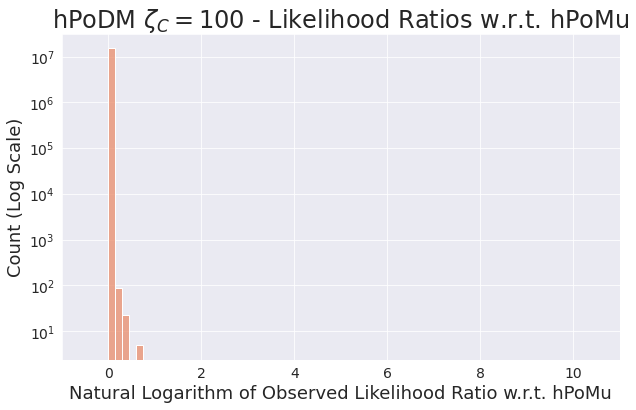}
  \caption[]{Compare with figure \ref{fig:hTPMH_LRs_histogram}.}
  \label{fig:hPoDM_100_LRs_histogram}
\end{subfigure}

\begin{subfigure}{\textwidth}
  \centering
\includegraphics[width=\textwidth,height=0.45\textheight,keepaspectratio]{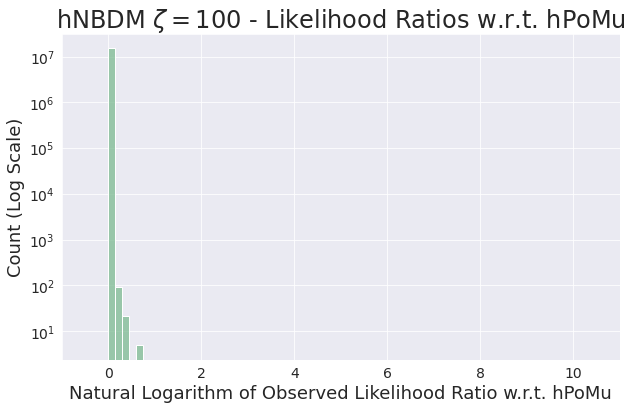}
  \caption[]{Compare with figure \ref{fig:hTPMH_LRs_histogram}.}
  \label{fig:hNBDM_100_LRs_histogram}
\end{subfigure}

\caption{Likelihood ratio histograms for very similar distributions.}
\label{fig:very_similar_LRs_histogram}
\end{figure}

\subsubsection{Typical Likelihood Ratios as a Function of Strain Frequency}
\label{sec:typic-likel-rati-1}

Similarly, while figures \ref{fig:hPoDM_100_LRs_heatmap} and \ref{fig:hNBDM_100_LRs_heatmap} also show greater differences relative to hPoMu compared to those observed for the hTPMH distribution, the differences still trend very small. (That the geometric means across the gluttonous groups tend to be slightly \textit{less} than $1$, rather than slightly greater, most likely reflects, in addition to the overlapping nature of the gluttonous groups, the depletion of droplets with $3$ or more strains relative to hPoMu that will be observed in these distributions in chapter \ref{chap:data_throughput}.)

\begin{figure}[p]
  \centering

  \begin{subfigure}{\textwidth}
  \centering
\includegraphics[width=\textwidth,height=0.45\textheight,keepaspectratio]{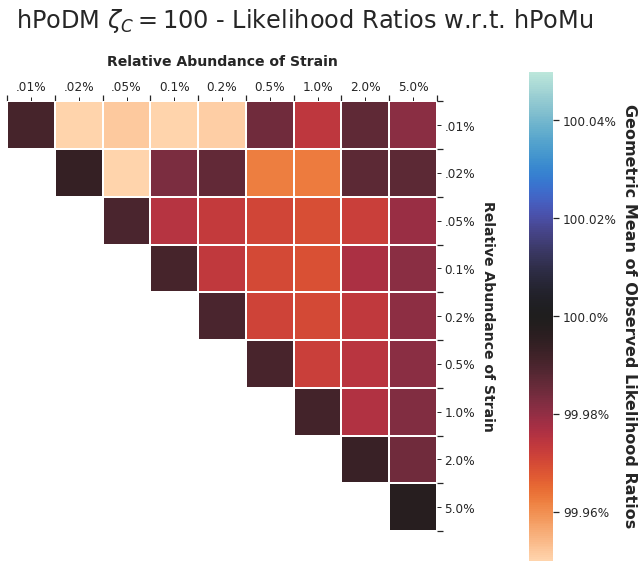}  
  \caption[]{Compare with figure \ref{fig:hTPMH_LRs_heatmap}.}
  \label{fig:hPoDM_100_LRs_heatmap}
\end{subfigure}

\begin{subfigure}{\textwidth}
  \centering
\includegraphics[width=\textwidth,height=0.45\textheight,keepaspectratio]{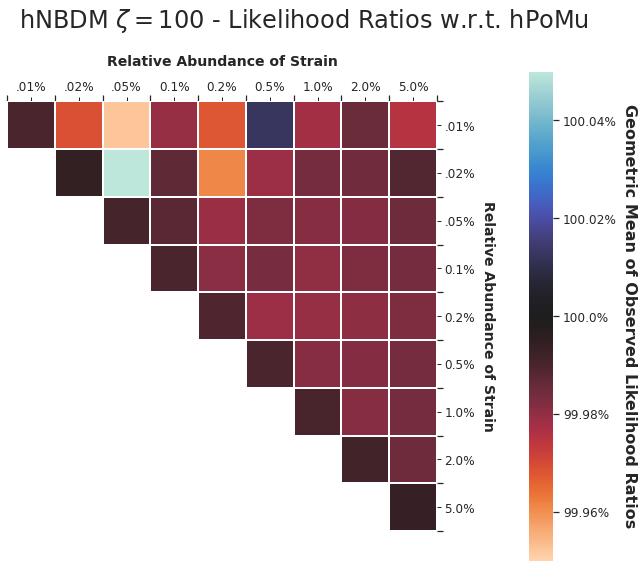}
  \caption[]{Compare with figure \ref{fig:hTPMH_LRs_heatmap}.}
  \label{fig:hNBDM_100_LRs_heatmap}
\end{subfigure}

\caption{Grouped likelihood ratios for very similar distributions.}
\label{fig:very_similar_LRs_heatmap}
\end{figure}

\subsection{Slightly Different Distributions}
\label{sec:slightly-diff-distr}

All available evidence suggests that the hPoDM with $\cconcentration =1$ and hNBDM with $\concentration =1$ distributions are (at least) slightly different from the corresponding hPoMu distribution.

\subsubsection{Distributions of Likelihood Ratios w.r.t. hPoMu}
\label{sec:distr-likel-rati-2}

Figures \ref{fig:hPoDM_1_LRs_histogram} and \ref{fig:hNBDM_1_LRs_histogram} show very large right tails for the distributions of observed likelihood ratios for both the hPoDM ($\cconcentration=1$) and the hNBDM ($\concentration=1$). This shows that many droplets were substantially ``better explained'' using the true simulation distribution rather than hPoMu, something which did not occur for any of the less heterogeneous distributions. It is unclear whether either large right tail is ``larger'' than the other.

\begin{figure}[p]
  \centering
  
  \begin{subfigure}{\textwidth}
  \centering
\includegraphics[width=\textwidth,height=0.45\textheight,keepaspectratio]{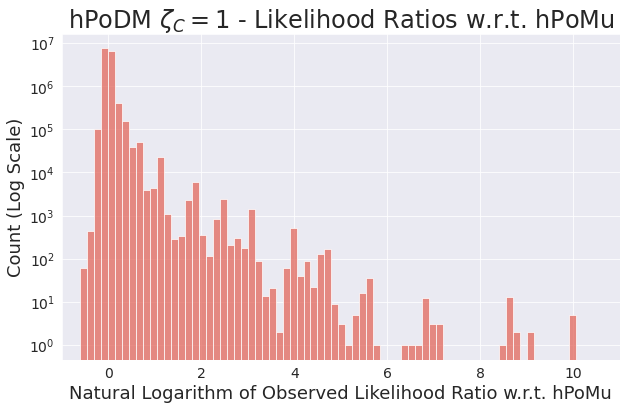}
  \caption[]{Compare with figure \ref{fig:hTPMH_LRs_histogram} as well as with figure \ref{fig:hPoDM_100_LRs_histogram}.}
  \label{fig:hPoDM_1_LRs_histogram}
\end{subfigure}

\begin{subfigure}{\textwidth}
  \centering
\includegraphics[width=\textwidth,height=0.45\textheight,keepaspectratio]{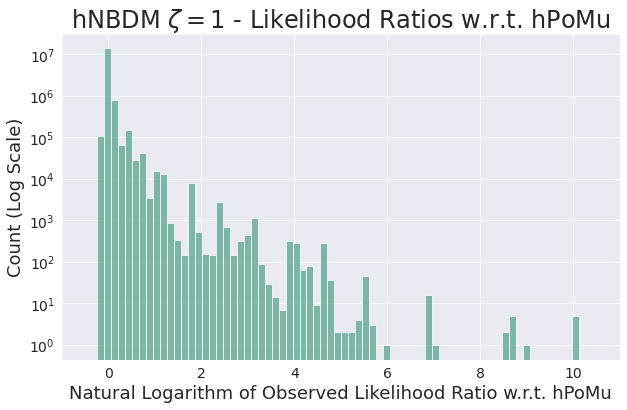}
  \caption[]{Compare with figure \ref{fig:hTPMH_LRs_histogram} as well as with figure \ref{fig:hNBDM_100_LRs_histogram}.}
  \label{fig:hNBDM_1_LRs_histogram}
\end{subfigure}

\caption{Likelihood ratio histograms for slightly similar distributions.}
\label{fig:slightly_similar_LRs_histogram}
\end{figure}

\subsubsection{Typical Likelihood Ratios as a Function of Strain Frequency}
\label{sec:typic-likel-rati-2}

Finally, figures \ref{fig:hPoDM_1_LRs_heatmap} and \ref{fig:hNBDM_1_LRs_heatmap} show typical values for the likelihood ratios in each gluttonous group for hPoDM $\cconcentration=1$ and hNBDM $\concentration=1$ that consistently ``float above'' $1$, in clear contrast to the ``hovering around $1$'' trend observed in figures \ref{fig:hTPMH_LRs_heatmap}, \ref{fig:hPoDM_100_LRs_heatmap}, and \ref{fig:hNBDM_100_LRs_heatmap} that correspond to less heterogeneous distributions. One also sees that the values in figure \ref{fig:hNBDM_1_LRs_heatmap} trend slightly higher than those in figure \ref{fig:hPoDM_1_LRs_heatmap}, reflecting again how the hNBDM family of distributions incorporates both kind of heterogeneities whereas the hPoDM family does not.

\begin{figure}[p]
  \centering
  
  \begin{subfigure}{\textwidth}
  \centering
\includegraphics[width=\textwidth,height=0.45\textheight,keepaspectratio]{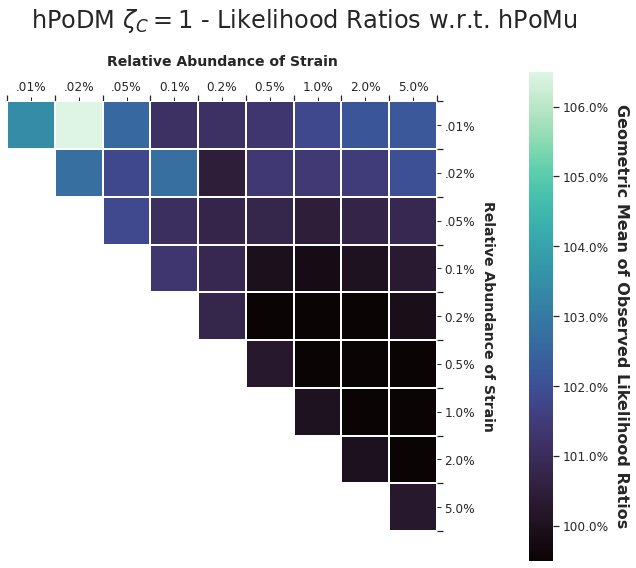}
  \caption[]{Compare with figure \ref{fig:hTPMH_LRs_heatmap} as well as with figure \ref{fig:hPoDM_100_LRs_heatmap}.}
  \label{fig:hPoDM_1_LRs_heatmap}
\end{subfigure}

\begin{subfigure}{\textwidth}
  \centering
\includegraphics[width=\textwidth,height=0.45\textheight,keepaspectratio]{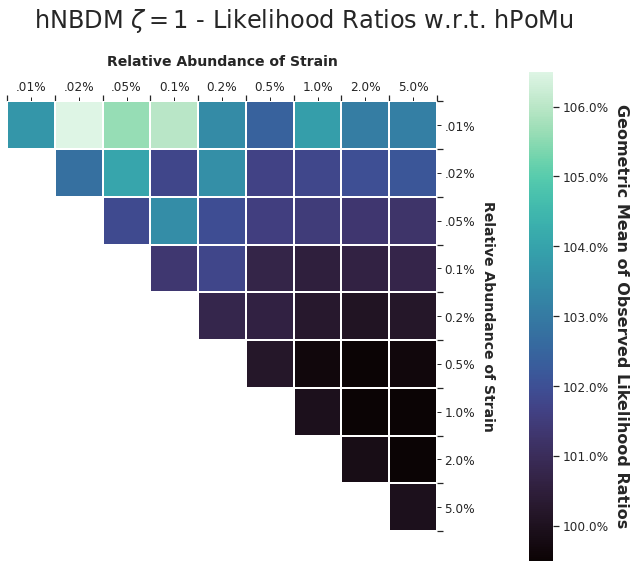}
  \caption[]{Compare with figure \ref{fig:hTPMH_LRs_heatmap} as well as with figure \ref{fig:hNBDM_100_LRs_heatmap}.}
  \label{fig:hNBDM_1_LRs_heatmap}
\end{subfigure}

\caption{Grouped likelihood ratios for slightly similar distributions.}
\label{fig:slightly_similar_LRs_heatmap}
\end{figure}

\section{Discussion}
\label{sec:model_comparison_discussion}

The results show the following. The sampling with replacement assumption of hPoMu is completely harmless for adequately modelling the data. Depending on our target parameters, and which values of heterogeneities are realistic, the homogeneity assumptions of hPoMu could potentially inadequately model the data. Density and compositional heterogeneities are the only failure of the hPoMu assumptions with the potential to be important in practice.

Section \ref{sec:overview-groups-distr} discusses the taxonomy of distributions identified in section \ref{sec:model_comparison_results}. Section \ref{sec:very-simil-distr-1} discusses the two most similar groups, which are either very difficult or impossible to distinguish from hPoMu. Section \ref{sec:slightly-diff-distr-1} discusses the distributions which can be easily distinguished from hPoMu, but whose differences from hPoMu may not be \textit{practically} important. Section \ref{sec:next-steps} reviews the next steps to ascertain which differences from hPoMu might affect our practical goal of being able to infer microbial interactions.

\subsection{Overview of Groups of Distributions}
\label{sec:overview-groups-distr}

The simulated distributions can be grouped into three levels of assumption violation. 

For the first two levels, the first consisting of hTPMH and the second consisting of hPoDM $\cconcentration=100$ and hNBDM $\concentration=100$, the data produced is difficult to distinguish from data produced by the hPoMu distribution. Therefore hPoMu is most likely an adequate model and approximation for data generated by distributions from these two levels. 

The third level consists of hPoDM $\cconcentration=1$ and hNBDM $\concentration=1$, which produce data that can be relatively easily distinguished from data produced by hPoMu.

\subsection{Very Similar Distributions}
\label{sec:very-simil-distr-1}

While hTPMH, hPoDM $\cconcentration=100$, and hNBDM $\concentration=100$ are all very similar to hPoMu, the above results suggest that hTPMH is even more similar to hPoMu than the other two.

Moreover, the simulations actually \textit{overstate} how different data generated by hTPMH should be in practice from data generated by hPoMu. I showed in section \ref{sec:proof-that-htpmh} that hTPMH converges in distribution to hPoMu as the size of the total population of cells being sampled from increases. While the simulation assumed a population size of $500$ \textbf{m}illion, or $5 \times 10^8$, cells, a real MOREI experiment is likely to sample from a population containing upwards of $10$ \textbf{b}illion, or $1 \times 10^{10}$, cells, more than twenty times as many. 

Thus, to the extent we consider that data generated by hPoDM $\cconcentration=100$ and hNBDM $\concentration=100$ are difficult to distinguish from data generated by hPoMu, data that would be generated by hTPMH in practice should be considered effectively indistinguishable from data generated by hPoMu. In particular, even for extremely low abundance strains, these results show that the sampling with replacement approximation is completely irrelevant in practice to accurate modelling of the data generated by MOREI.

\subsection{Slightly Different Distributions}
\label{sec:slightly-diff-distr-1}

The results show that there are evident differences between data generated by distributions in the third level, hPoDM $\cconcentration=1$ and hNBDM $\concentration=1$, and data generated by hPoMu. Yet this still is not enough to be certain that there are \textit{practically} significant differences. (Setting aside how to be more precise about what is a ``practically significant'' difference.) 

For example, the long right tails from figures \ref{fig:hPoDM_1_LRs_histogram} and \ref{fig:hNBDM_1_LRs_histogram} are only easily visible when plotting the number of counts on a log scale. The total fraction of droplets represented by those long right tails is actually relatively small. It's unclear whether the existence of a small faction of droplets which are \textit{much} better explained by the true models than by hPoMu is an important demerit against hPoMu when most droplets are nearly equally well explained by both. Focusing on ``typical'' values of the likelihood ratios as depicted in figures \ref{fig:hPoDM_1_LRs_heatmap} and \ref{fig:hNBDM_1_LRs_heatmap}, the advantage of the true models over hPoMu is usually \textit{at most} around $6\%$. 

On the other hand, the ``typical'' likelihood ratio values in figures figures \ref{fig:hPoDM_1_LRs_heatmap} and \ref{fig:hNBDM_1_LRs_heatmap} also show a distinct upwards trend as the abundance of one of the involved strains decreases. As explained in section \ref{sec:outline-rest-thesis}, quantifying the extent of MOREI's ability to advance the state of the art requires focusing on the \textit{least} abundant strains. Therefore, inasmuch as the discrepancies of these models compared to hPoMu may be particularly important for the least abundant strains, the discrepancies may also be important overall.

Thus the results in this chapter seem to point both ways about whether hPoMu is an adequate working model for data generated by these distributions.

\section{Conclusion}
\label{sec:conclusion-4}

\paragraph{Findings and Contributions}

I demonstrated that log likelihood ratios can be used to determine whether the failure of assumptions from section \ref{sec:impl-model-assumpt} causes substantially new behavior.
Failure of the sampling without replacement assumption turns out to have negligible effects in practice, but failures of the other assumptions could be important.

\paragraph{Practical Implications}

The results from this chapter clarify that we can safely ignore concerns regarding the sampling without replacement assumption of hPoMu when choosing a (working) model for the initial formation of droplets.
The practical implications of the results from this chapter are otherwise limited to the extent that they do not unambiguously indicate whether or not we can ignore concerns regarding the homogeneity assumptions of hPoMu.
These results are useful, however, for at least raising the possibility that violations of the homogeneity assumptions could have important effects on our ability to predict the numbers of droplets available for making inferences about given microbial interactions. 

\paragraph{Next Steps and Open Questions}
\label{sec:next-steps}

Although these results do suggest that data generated by certain distributions is unlikely to be adequately modeled by hPoMu, we need to be more precise to substantiate such a claim. We need to decide precisely which aspects of the data we want to be model accurately, to target, or otherwise we have no way to measure whether hPoMu is an adequate model for the data. Only experiments measuring those targeted aspects of the data will give results which could definitively determine whether or when hPoMu is an adequate model. This is chapter \ref{chap:data_throughput}. 

Moreover, even if hPoMu is inaccurate for those targeted aspects of the data, it could still be an adequate working model as long as its predictions are conservative compared to the truth, rather than overly optimistic. To make that assessment, we need to know what values of heterogeneity might be realistic in practice. That in turn requires having estimators for the heterogeneities to use on empirical data. We fill that gap later in chapter \ref{chap:hetero_estimator_performance}.

\end{coolcontents}

\chapter[Targeted Estimands and Goodness of Fit of Contingency Tables][Goodness of Fit Using Contingency Tables of the Targeted Estimands]{Goodness of Fit Using Contingency Tables of the Targeted Estimands\\ {\large Effects of Heterogeneities on Data Throughput}}
\label{chap:data_throughput}

Herein I confirm that more severe failures of the assumptions from section \ref{sec:impl-model-assumpt} cause more severe discrepancies with the predictions derived from hPoMu. 
See for example sections \ref{sec:slightly-diff-distr-conting}, \ref{sec:very-diff-distr-conting}, \ref{sec:distr-again-fall}.
The nature of the effect depends on the chosen grouping (picky or gluttonous) of droplets defining the targeted estimands.
See sections \ref{sec:effect-incr-heter} and \ref{sec:effect-heter-data}.
Effects of the failure of the sampling with replacement assumption are further confirmed to be negligible in practice.
See for example sections \ref{sec:sampl-without-repl-conting} or \ref{sec:distr-again-fall}.

Section \ref{sec:preliminaries-1} explains some of the technical details of the analyses. Section \ref{sec:data_throughput_methods} explains implementation details of the analyses. Section \ref{sec:data_throughput_results} explains the results of the analyses. Section \ref{sec:data_throughput_discussion} explains how the results are relevant to modelling the initial formation of droplets.

\begin{coolcontents}

\section{Background and Significance}
\label{sec:backgr-sign-5}

When planning the previous chapter \ref{chap:model_comparison}, I had anticipated the effect of the homogeneity assumptions to be negligible. This was true of the effect of the sampling with replacement assumption, which was investigated earlier. Similarity of underlying distributions $\mathcal{P}_1$ and $\mathcal{P}_2$ should be enough to imply similarity of target parameters $\boldsymbol{\Psi}(\mathcal{P}_1)$ and $\boldsymbol{\Psi}(\mathcal{P}_2)$. The log-likelihood ratio statistics provide evidence of similarity of the underlying distributions $\mathcal{P}_1$ and $\mathcal{P}_2$. Hence, if the anticipated findings of the previous chapter \ref{chap:model_comparison} had been valid, then the log-likelihood ratio statistics alone would have been sufficient to demonstrate them.

The questions here are the same as those of the previous chapter \ref{chap:model_comparison}. In practice we only care about the \textit{relevant} portion $\mathcal{Q}_O$ of the probability distribution of the observed data that affects our target parameters $\boldsymbol{\Psi}(\mathcal{P}_O)$. (Cf. the overview of these ideas in \cite[section 1.5]{vanderLaan2011}.) Again, in the previous chapter \ref{chap:model_comparison} we found evidence that violations of the homogeneity assumptions could cause the \textit{entire} underlying probability distribution of the observed data $\mathcal{P}_O$ to substantially differ from that under hPoMu. Yet even if two probability distributions $\mathcal{P}_1$ and $\mathcal{P}_2$ substantially differ, that does not necessarily mean that the \textit{relevant} portions $\mathcal{Q}_1$ and $\mathcal{Q}_2$ also substantially differ. Hence, as mentioned before in section \ref{sec:conclusion-4}, the evidence from chapter \ref{chap:model_comparison} is insufficient to say whether violations of the homogeneity assumptions lead to practically significant differences, ones that affect  $\mathcal{Q}_O$ / $\boldsymbol{\Psi}(\mathcal{P}_O)$. Thus the goal for this chapter is to look for and implement goodness of fit procedures that speak more directly to comparisons of target parameters $\boldsymbol{\Psi}(\mathcal{P})$ and relevant portions $\mathcal{Q}$ of probability distributions.

\paragraph{Broader field}

As was the case for the previous chapter \ref{chap:model_comparison}, the problem of this chapter also belongs to the general field of model selection, and to the subfield of goodness of fit tests in particular.
Again, the more accessible \cite[chapter 7]{Hastie2009}  
or the highly sophisticated \cite{Barron1999} are recommended as references for the general theory of model selection.
Cf. the discussion earlier from the introduction to Part \ref{part:modell-init-form} or from section \ref{sec:backgr-sign-4}.
While the previous chapter \ref{chap:model_comparison} discusses goodness of fit tests using the likelihood ratio statistic, this chapter focuses on goodness of fit tests for contingency tables that use the $\chi^2$-divergence \cite[p. 57]{amari} statistic. (Cf. section \ref{sec:pears-categ-diverg} for terminology.)
See the first two chapters of \cite{Huber_Carol}, or the introduction of \cite{Johnson2004}, for reviews of the classical procedure for using the $\chi^2$-divergence statistic (``Pearson'' $\chi^2$ statistic) in goodness of fit tests for contingency tables. The article \cite{monte_carlo_chisquare_stat} discusses ``non-classical'' use of the $\chi^2$-divergence statistic for goodness of fit via Monte Carlo approximations of null distributions, and applications thereof to ecology. The review \cite{Fienberg2011} discusses the use of classical $\chi^2$ procedure as well as other, more sophisticated methods for analyzing contingency tables (for goodness of fit and for other reasons). 

\paragraph{Specific problem}

Herein ``practically significant'' differences correspond to the target parameters defined in section \ref{sec:defin-thro-intro}, the ``data throughput'' of the experiment. These target parameters allow us to predict how much data (measured in numbers of droplets) will be produced that can be used to infer given microbial interactions, cf. again section \ref{sec:motiv-targ-estim}. Thus, in this chapter I investigate how failures of the implicit hPoMu assumptions, in particular failures of the homogeneity assumptions, affect these target parameters. 

As long as the average number of cells per droplet is not too large, the number of droplets available to serve as controls will usually be much larger than the number of droplets available to serve as treatments (i.e. droplets where strains co-occur).  Cf. the heuristic argument in section \ref{sec:toy-model}. Hence, if an estimation method is ever too starved of data to make inferences about a given microbial interaction, most likely the cause will be too few treatment droplets. In terms of ``data throughput'', bottlenecks that occur will usually be caused by insufficiently many treatment droplets. Therefore the \textit{main} goal of this chapter is to understand effects on those target parameters corresponding specifically to the treatment droplets.

These target parameters, and estimators thereof, can be grouped meaningfully into contingency tables. (Cf. sections \ref{sec:glob-picky-goodn} and \ref{sec:pairw-glutt-goodn}.)
Therefore we can use goodness of fit procedures designed for contingency tables to accomplish our goal, to find and implement goodness of fit procedures that speak more directly to comparisons of these target parameters.
  
The paper
\cite{monte_carlo_chisquare_stat}
discusses goodness of fit procedures for ecological simulations. This includes ecological models that produce count data that can be grouped into a contingency table, but the authors also show how the same methodology can be applied more generally. The approach is particularly useful because it can be applied in situations that violate the assumptions of the ``classical'' Pearson goodness of fit procedure for contingency tables.
  
The paper \cite{Zhou2018} studies mixed-membership models involving distributions similar to those in this work. The analysis of mixed-membership models is known to be related to the analysis of contingency tables, see for example the review article \cite{Fienberg2011}. The key idea relevant here appears to be that multivariate count data is closely related to multivariate categorical data. For finite samples, observed multivariate count data can always be converted into multivariate categorical data via truncation. In fact, the procedures in e.g. section \ref{sec:categories} can be thought of as ``collapsing'' or ``marginalizing'' a sparse ``multi-dimensional'' contingency table with $2^{\Strains}$ cells (``exclusive and nonempty subsets'' \cite{Zhou2018}) into a ``one-dimensional'' contingency table whose number of cells is at most $O(\Strains^2)$. So although the connection with the problems and ideas in \cite{Zhou2018} is certainly currently somewhat hazy, it also seems reasonable to expect that the connection could be clarified in future work.

\paragraph{Particular approach}

Under the assumptions of the classical goodness of fit test for contingency tables developed by Pearson \cite{Huber_Carol}, the asymptotic distribution of the $\chi^2$-divergence \cite[p. 57]{amari} statistic is the Pearson $\chi^2$ distribution \cite[p. 53]{monte_carlo_chisquare_stat}. Because this is only an asymptotic result, for finite samples the sampling distribution of the $\chi^2$-divergence statistic may actually substantially differ from the $\chi^2$-distribution, even under the null hypothesis. Cf. section \ref{sec:computing-p-values}. Therefore I computed a Monte Carlo approximation of the sampling distribution of the $\chi^2$-divergence statistic and used that to compute $p$-values. Cf. \cite[section 3]{monte_carlo_chisquare_stat} for a fairly detailed explanation of Monte Carlo hypothesis testing.
  
Details of the classical approach, which assumes that the asymptotic distribution is a valid approximation, again can be found in \cite{Huber_Carol}. That assumption seemed unrealistic here, cf. again section \ref{sec:computing-p-values}.
Bayesian versions of this procedure also exist, cf. \cite{Johnson2004}. However such an approach seemed unnecessarily complicated for exploratory work like this.

\section{Preliminaries}
\label{sec:preliminaries-1}

In this section I clarify details needed to analyze the data throughput predicted by each model of initial droplet formation. Section \ref{sec:defin-thro} gives precise definitions of the statistics that are relevant for studying data throughput (following the argument given in section \ref{sec:append-mult-repr}). Section \ref{sec:pears-categ-diverg} defines the divergence used to compare distributions. Section \ref{sec:glob-picky-goodn} clarifies the details of the hypothesis test comparing distributions to hPoMu using this divergence. In addition to the ``picky'' and ``global'' way of grouping of droplets by strain\footnote{Herein I use ``strains'' to refer equally to strains belonging to the same species(/genus/family/etc.) as well as to strains belonging to different species(/genera/families/etc.), because the distinction is irrelevant for setting up the abstract problem. It may matter for the implementation of a specific experiment.} for hypothesis testing defined already in section \ref{sec:glob-picky-goodn}, section \ref{sec:pairw-glutt-goodn} defines a ``gluttonous'', ``local'', or ``pairwise'' way of grouping droplets by strain for hypothesis testing.

\subsection{Definition of Throughput}
\label{sec:defin-thro}

Previously we have been looking at distributions of numbers of cells in a given droplet (of any strain or of a given strain). As argued in section \ref{sec:append-mult-repr}, I believe that understanding the distributions of numbers of \textit{strains} in a given droplet is the most relevant for predicting data throughput. In section \ref{sec:dist-distr-numb} I clarify how to define these as statistics computed from the numbers of cells. Then sections \ref{sec:glutt-defin-data} and \ref{sec:picky-defin-data} give two different possible definitions of throughput. This section is a more detailed version of section \ref{sec:defin-thro-intro}.

\subsubsection{Strain Count Distribution}
\label{sec:dist-distr-numb}

The notation $\indicator{A}$ denotes the indicator function for the event $A$.

For a given strain $\strain$, the \textbf{strain presence} RV (random variable):
\begin{equation}
  \label{eq:strain_presence}
  \straincount[\strain](0) := \indicator{\abundance[\strain](0) \ge 1} = \indicator{\abundance[\strain](0) > 0} \,.
\end{equation}
indicates whether any cells of strain $\strain$ belong to the droplet. It takes values in $\{0,1\}$ like any indicator random variable. Specifically, it equals $1$ if strain $\strain$ is present in the droplet and $0$ if strain $\strain$ is absent.

Similarly, for a given strain $\strain$, the \textbf{strain absence} RV:
\begin{equation}
  \label{eq:strain_absence}
  \nstraincount[\strain](0) := 1 - \straincount[\strain](0) = \indicator{\abundance[\strain](0) = 0 } \,,
\end{equation}
takes values in $\{0,1\}$ like any indicator random variable, equalling $1$ if strain $\strain$ is absent in the droplet and $0$ if strain $\strain$ is present.

For a given droplet, the joint distribution of the strain presence RVs for all strains corresponds to a binary random vector:
\begin{equation}
  \label{eq:vector_strain_presence}
  \vstraincount(0) := (\straincount[1](0), \dots, \straincount[\Strains](0)) \,.
\end{equation}
Its values are in $\bigtimes_{\strain=1}^{\Strains} \{0,1\}$ (the $\Strains$-fold Cartesian product of $\{0,1\}$ with itself).

For a given droplet, its \textbf{strain count} $\straincount(0)$ is the number of strains present in the droplet. The strain count equals the sum of the strain presence RVs:
\begin{equation}
  \label{eq:strain_count}
  \straincount(0) := \sum_{\strain \in [\Strains]} \straincount[\strain](0) = \sum_{\strain \in [\Strains]} \indicator{\abundance[\strain](0) \ge 1} \,.
\end{equation}
It is an RV taking values in $\{0\} \cup [\Species]$.

Another useful notion is the \textbf{support} of a (non-negative) vector $\rvec{v} \in \R^S$:
\begin{equation}
  \label{eq:support_definition}
  \support(\rvec{v}) := \{ \strain \in [\Strains] : [\rvec{v}]_{\strain} > 0  \} \subseteq [\Strains] \,.
\end{equation}
It follows directly from the definitions that $\support(\vabundance(0)) = \support(\vstraincount(0))$ always, corresponding to the set of $\strain \in [\Strains]$ such that $\straincount[\strain](0)=1$. Moreover we also always have\footnote{
The \textbf{\textit{cardinality}} $|\mathcal{S}|$ of a finite set $\mathcal{S}$ is the number of elements of the set. E.g. ${|[\Strains]| = \Strains}$.
} that $\straincount(0) = |\support(\vabundance(0))| = |\support(\vstraincount(0))|$.

As argued in section \ref{sec:append-mult-repr}, I believe that the notion of ``throughput'' that is important for this problem corresponds to $\vstraincount(0)$ and not (directly\footnote{Only indirectly as mediated via $\vstraincount(0)$, which of course can be computed from $\vabundance(0)$.}) to $\vabundance(0)$. While choosing $\rate \approx 2$ is intended to make $\vabundance(0)$ as similar to $\vstraincount(0)$ as possible, they are definitely not the same, something which e.g. the multiple representatives problem from section \ref{sec:append-mult-repr} makes clear.

Even when the first moments of $\vabundance(0)$ remain the same, changes to the second moments of $\vabundance(0)$ can still cause changes in the expected numbers of droplets serving as treatments or controls. A heuristic way to understand this is to observe that the expected data throughput corresponds to the second (or higher order) moments of $\vstraincount(0)$; see sections \ref{sec:glutt-defin-data} and \ref{sec:picky-defin-data}. Although for any distribution belonging to the ghNBDM working model (see section \ref{sec:ghnbdm-family-defin}) the expected cell counts for each strain are the same as those under hPoMu, the choice of distribution from the ghNBDM working model still affects the predicted data throughput.

\begin{figure}
  \centering
  \includegraphics[width=\textwidth,height=\textheight,keepaspectratio]{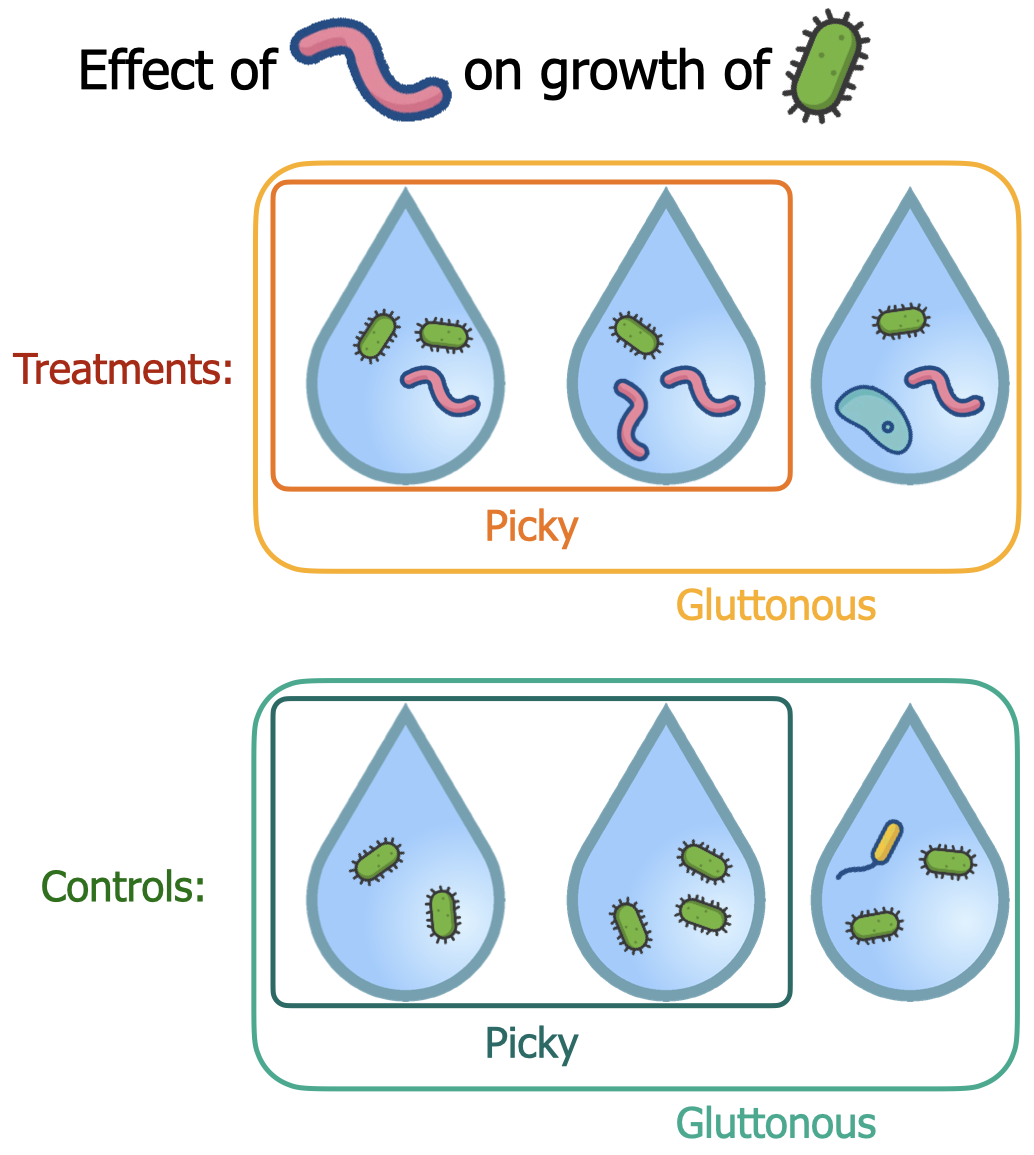}
  \caption{Gluttonous vs. Picky Definitions of Data Throughput. Cf. text.}
  \label{fig:gluttonous_vs_picky}
\end{figure}

\subsubsection{Gluttonous Definition of Data Throughput}
\label{sec:glutt-defin-data}

Recall from section \ref{sec:how-micr-ecol}, or the introduction to Part \ref{part:introduction}, that the treatment group for inferring the effect of strain $\strain_1$ on the growth of strain $\strain_2$ corresponds to droplets where $\strain_1$ and $\strain_2$ co-occur. The set of all droplets satisfying this condition is
\begin{equation}
  \label{eq:formal_gluttonous_treatment_definition}
  \begin{split}
\treatment[\strain_1, \strain_2]_g
 := & \left\{ \droplet \in [\Droplets] :  \straincount[\strain_1]_{\droplet}(0) \cdot \straincount[\strain_2]_{\droplet}(0) = 1  \right\} \\
=& \left\{ \droplet \in [\Droplets] : \indicator{ \abundance[\specie_1]_{\droplet}(0) \ge 1 \text{ \& } \abundance[\specie_2]_{\droplet}(0) \ge 1 } = 1    \right\} \\
=&
\left\{ \droplet \in [\Droplets]: \support(\vstraincount_{\droplet}(0)) \supseteq \{\strain_1, \strain_2\} \right\}
\,.    
  \end{split}
\end{equation}
This is called the ``\textbf{gluttonous}'' \textbf{treatment} group because \textit{all} droplets where $\strain_1$ and $\strain_2$ co-occur are included. Cf. figure \ref{fig:gluttonous_vs_picky}.

Recall also from section \ref{sec:how-micr-ecol}, or the introduction to Part \ref{part:introduction}, that the control group for inferring the effect of strain $\strain_1$ on the growth of strain $\strain_2$ corresponds to droplets where $\strain_2$ occurs but $\strain_1$ does not. The set of all droplets satisfying this condition is
\begin{equation}
  \label{eq:formal_gluttonous_control_definition}
  \begin{split}
\control[\strain_1, \strain_2]_g := &
 \left\{ \droplet \in [\Droplets] :  \nstraincount[\strain_1]_{\droplet}(0) \cdot \straincount[\strain_2]_{\droplet}(0) = 1  \right\} \\
=&
 \left\{ \droplet \in [\Droplets] : \indicator{ \abundance[\specie_1]_{\droplet}(0) = 0 \text{ \& } \abundance[\specie_2]_{\droplet}(0) \ge 1 } = 1    \right\} \\
=&
\left\{ \droplet \in [\Droplets] :  \support(\vstraincount_{\droplet}(0)) \supseteq \{\strain_2\} \,,\, \strain_1 \not\in \support(\vstraincount_{\droplet}(0)) \right\}
 \,.    
  \end{split}
\end{equation}
This is called the ``\textbf{gluttonous}'' \textbf{control} group because \textit{all} droplets where $\strain_2$ occurs but $\strain_1$ does not are included. Cf. figure \ref{fig:gluttonous_vs_picky}.

These definitions place no constraints on the presence or absence of strains $\strain \in [\Strains]$ besides $\strain_1$ and $\strain_2$. This can be favorable e.g. when studying very rare strains, for which there may be no or very few droplets containing $\strain_1$ and $\strain_2$ only, but some containing $\strain_1$ and $\strain_2$ along with other strains. These definitions give the best chance of avoiding ``data starvation''.

From equation (\ref{eq:formal_gluttonous_treatment_definition}) we get (assuming the droplets are identically distributed) that the expected number of droplets for the gluttonous treatment group for inferring the effect of strain $\strain_2$ on the growth of $\strain_1$ is
\begin{equation}
  \label{eq:expected_number_gluttonous_treatments}
  \expectation*{\left|  \treatment[\strain_1, \strain_2]_g \right|} = \Droplets \cdot \expectation*{  \straincount[\strain_1](0) \  \straincount[\strain_2](0) } \,.
\end{equation}
Similarly, from equation (\ref{eq:formal_gluttonous_control_definition}) we get (assuming the droplets are identically distributed) that the expected number of droplets for the corresponding gluttonous control group for inferring the effect of strain $\strain_2$ on the growth of $\strain_1$ is
\begin{equation}
  \label{eq:expected_number_gluttonous_controls}
  \expectation*{\left|  \control[\strain_1, \strain_2]_g \right|}  = \Droplets \cdot \expectation*{\nstraincount[\strain_1](0) \cdot \straincount[\strain_2](0)} = 
\Droplets \cdot \expectation*{\straincount[\strain_2](0) - \straincount[\strain_1](0) \cdot \straincount[\strain_2](0)} \,.
\end{equation}
As claimed in section \ref{sec:dist-distr-numb}, these expressions for the expected gluttonous data throughput involve the second moments of $\vstraincount(0)$.

\subsubsection{Picky Definition of Data Throughput}
\label{sec:picky-defin-data}

Comparing figures \ref{fig:gluttonous_vs_picky} and \ref{fig:experiment_example}, we see that the above ``gluttonous'' definitions correspond to more possible combinations of strains than we probably would have included when studying the effect of strain $\strain_1$ on the growth of strain $\strain_2$ by manually plating cells. This motivates the following definitions.

The set of droplets where $\strain_1$ and $\strain_2$ co-occur in a way corresponding to the combination of strains we would use if we were manually plating cells is
\begin{equation}
  \label{eq:formal_picky_treatment_definition}
  \begin{split}
\treatment[\strain_1, \strain_2]_p
 := &    
 \left\{ \droplet \in [\Droplets] :  \straincount[\strain_1]_{\droplet}(0) \cdot \straincount[\strain_2]_{\droplet}(0) \prod_{\strain[] \not\in \{ \strain_1, \strain_2 \}} \nstraincount[{\strain[]}]_{\droplet}(0) = 1  \right\}
\\
= &
\left\{ \droplet \in [\Droplets] : \indicator{ \abundance[\specie_1]_{\droplet}(0) \ge 1 \text{ \& } \abundance[\specie_2]_{\droplet}(0) \ge 1 \text{ \& } \forall \strain[] \not\in \{\strain_1, \strain_2\}, \  
\abundance[{\strain[]}]_{\droplet}(0) =0 } = 1     \right\} 
\\
= &
\left\{ \droplet \in [\Droplets] :  \support(\vstraincount_{\droplet}(0)) = \{ \strain_1, \strain_2 \} \right\}
\,.
  \end{split}
\end{equation}
This is called the ``\textbf{picky}'' \textbf{treatment} group because droplets where $\strain_1$ and $\strain_2$ co-occur are included \textit{only when all other strains are absent}. Cf. figure \ref{fig:gluttonous_vs_picky}.

The set of droplets where \textit{only} $\strain_2$ occurs, corresponding to the combination of strains we would use if we were manually plating cells, is
\begin{equation}
  \label{eq:formal_picky_control_definition}
  \begin{split}
\control[\strain_1, \strain_2]_p
 := &    
 \left\{ \droplet \in [\Droplets] :  \nstraincount[\strain_1]_{\droplet}(0) \cdot \straincount[\strain_2]_{\droplet}(0) \prod_{\strain[] \not\in \{ \strain_1, \strain_2 \}} \nstraincount[{\strain[]}]_{\droplet}(0) = 1  \right\}
\\
= &
\left\{ \droplet \in [\Droplets] : \indicator{ \abundance[\specie_1]_{\droplet}(0) = 0 \text{ \& } \abundance[\specie_2]_{\droplet}(0) \ge 1 \text{ \& } \forall \strain[] \not\in \{\strain_1, \strain_2\}, \  
\abundance[{\strain[]}]_{\droplet}(0) =0 } = 1     \right\} 
\\
=&
\left\{ \droplet \in [\Droplets] : \support(\vstraincount_{\droplet}(0)) = \{ \strain_2 \}  \right\}
\,.
  \end{split}
\end{equation}
This is called the ``\textbf{picky}'' \textbf{control} group because droplets where $\strain_2$ occurs but $\strain_1$ does not occur belong \textit{only when all other strains are absent}. It is not enough for $\strain_1$ alone to be absent. Cf. figure \ref{fig:gluttonous_vs_picky}.

Unlike the gluttonous groups, the picky groups \textit{do} place constraints on the presence or absence of strains $\strain \in [\Strains]$ besides $\strain_1$ and $\strain_2$. For both the treatments and the controls, for ``picky'' groups all strains besides $\strain_1$ or $\strain_2$ must have zero counts. Every picky group is by definition a subset of its gluttonous counterpart.

The picky definitions can be favorable when we have plenty of droplets from which to make estimates. They reduce the possibility of confounding effects on the growth of strain $\strain_2$ that could be caused by strains that are not strain $\strain_1$. On the other hand, in instances where there are only very few or no droplets without ``extra'' strains, the picky definitions could lead to ``data starvation''.

From equation (\ref{eq:formal_picky_treatment_definition}) we get (assuming the droplets are identically distributed) that the expected number of droplets for the picky treatment group for inferring the effect of strain $\strain_2$ on the growth of $\strain_1$ is
\begin{equation}
  \label{eq:expected_number_picky_treatments}
  \expectation*{\left|  \treatment[\strain_1, \strain_2]_p \right|} = \Droplets \cdot \expectation*{ \straincount[\strain_1](0) \cdot \straincount[\strain_2](0) \prod_{\strain[] \not\in \{ \strain_1, \strain_2 \}} \left(1 - \straincount[{\strain[]}](0) \right) } \,.
\end{equation}
Similarly, from equation (\ref{eq:formal_picky_control_definition}) we get (assuming the droplets are identically distributed) that the expected number of droplets for the corresponding picky control group for inferring the effect of strain $\strain_2$ on the growth of $\strain_1$ is
\begin{equation}
  \label{eq:expected_number_picky_controls}
  \expectation*{\left|  \control[\strain_1, \strain_2]_p \right|}  = \Droplets \cdot \expectation*{ \left(1 - \straincount[\strain_1](0) \right) \cdot \straincount[\strain_2](0) \prod_{\strain[] \not\in \{ \strain_1, \strain_2 \}} \left(1 - \straincount[{\strain[]}](0) \right) } \,.
\end{equation}
As claimed in section \ref{sec:dist-distr-numb}, these expressions for the expected picky data throughput involve the higher order moments of $\vstraincount(0)$.

\subsection{Pearson Categorical Divergence}
\label{sec:pears-categ-diverg}

The Pearson categorical divergence (more often referred to as the ``Pearson $\chi^2$ divergence'' or ``Pearson $\chi^2$ statistic'') of $\rvec{v} = (v^{(1)}, \dots, v^{(I)}) \in \R$ relative to $\rvec{w} = (w^{(1)}, \dots, w^{(I)}) \in \R$ is defined\cite[p. 57]{amari} as
  \begin{equation}
    \label{eq:pearson_divergence_definition}
    \sum_{i=1}^I  \frac{(v^{(i)} - w^{(i)})^2}{w^{(i)}} \,,
  \end{equation}
where both $\rvec{v}$ and $\rvec{w}$ belong to the unit simplex (entries sum to $1$ and are all $\ge 0$), and $\rvec{w}$ belongs to the interior of the unit simplex (entries are all $> 0$). This definition is not symmetric. In general the divergence of $\rvec{w}$ relative to $\rvec{v}$ differs from that of $\rvec{v}$ relative to $\rvec{w}$.

Herein I call this the ``Pearson categorical divergence'' to avoid confusion with either
\begin{enumerate}[label=(\alph*)]
\item the ``Pearson divergence'', ${1 - \rho_{\vec{x}\vec{y}}}$, where $\rho_{\vec{x}\vec{y}}$ is the Pearson correlation of $\vec{x}$ and $\vec{y}$, or
\item the ``Pearson $\chi^2$ distribution''\footnote{Technically a parameterized family of distributions and not a single distribution.}, the asymptotic sampling distribution of this statistic under the null distribution of ``Pearson-style'' goodness of fit tests.
\end{enumerate}
In particular, it would sound confusing to say (in section \ref{sec:computing-p-values}) that the sampling distribution of the ``$\chi^2$ divergences'' is not a ``$\chi^2$ distribution''.

\subsection{Global Picky Goodness of Fit Test}
\label{sec:glob-picky-goodn}

The strains present in a droplet determine the microbial interactions it can help characterize. Therefore we want to group the droplets according to the strains that are present in each droplet. Different models for the initial formation of droplets make different predictions about which such groups of droplets will be most common. A hypothesis test whose null distribution is a sampling distribution derived from the hPoMu working model compares the predictions made by other (working) models with those made by hPoMu. Rejecting the null hypothesis corresponds to asserting that hPoMu is unable to describe the observed data, so the true model that generated the data has more ``descriptive power'', cf. section \ref{sec:backgr-sign-4}. 

The definition of the hypothesis test is split into several parts. Section \ref{sec:categories} defines how the droplets are grouped according to strains for the hypothesis test. Section \ref{sec:probabilities} gives probabilities for each group under the null hypothesis (hPoMu), while section \ref{sec:expected-counts} clarifies the expected counts. Section \ref{sec:observed-counts} defines the observed counts used in defining the goodness of fit test. Section \ref{sec:symmetry-definitions} explains why some groups are redundant. Finally, section \ref{sec:test-statistic} uses the conclusions of the previous sections to define the test statistic.

\subsubsection{Categories}
\label{sec:categories}

There are technically exponentially many, $2^\Species$, potential combinations of strains that could occur in a given droplet. All of these could be used in defining mutually exclusive categories for a goodness of fit test. However, for both reasons of computational feasibility and data availability\footnote{
Most combinations of three or more strains are highly unlikely and will be represented by few or no data points. Cf. the related discussion in sections \ref{sec:form-as-stat}, \ref{sec:why-empir-distr}, and \ref{sec:why-expect-posit}.
}, herein only combinations with two or fewer strains are given distinct categories, with all combinations of three or more strains lumped into a category.

Thus the number of categories for the ``global'' goodness of fit test is
\begin{equation}
  \label{eq:global_num_categories}
  1 + \Species + \binom{\Species}{2} + 1 \,.
\end{equation}
This is one category for all droplets with zero strains (empty droplets), $S$ categories for each of the picky control groups (droplets with exactly one strain), $\binom{\Species}{2}$ categories for each of the picky treatment groups (droplets with exactly two strains), and one category for all droplets with three or more (``multiple'') strains. The groups being ``picky'' and not ``gluttonous'' ensures that categories do not overlap. This makes it feasible to define a single ``global'' test.

Cf. section \ref{sec:defin-thro} for more details about ``picky'' and ``gluttonous'' groups.

\subsubsection{Probabilities}
\label{sec:probabilities}

The probabilities associated with each category are
\begin{equation}
  \label{eq:global_test_probs}
  \begin{split}
    \countprob[\emptyset]_{\globe}
& := 
\probability{\abundance(0) = 0 } = e^{-\rate} \,,
\\
\countprob[\specie]_{\globe} 
& :=
\probability{\abundance[\specie](0) \ge 1, \forall \specie[]\not=\specie: \abundance[{\specie[]}](0) = 0} 
\\
& \, =
\probability{\abundance[\specie](0) \ge 1} \cdot \prod_{\specie[] \not= \specie} \probability{\abundance[{\specie[]}](0) = 0}
\\
& \, =
(1 - e^{-\freq^{(\specie)}\rate}) \cdot e^{-(1 - \freq^{(\specie)})\rate} \,,
\\
\countprob[\specie_1, \specie_2]_{\globe}
& :=
\probability{\abundance[\specie_1](0) \ge 1, \abundance[\specie_2](0) \ge 1, \forall \specie[]\not\in \{ \specie_1, \specie_2 \}: \abundance[{\specie[]}](0) = 0} 
\\
& \, =
\probability{\abundance[\specie_1](0) \ge 1} \cdot \probability{\abundance[\specie_2](0) \ge 1} \cdot \prod_{\specie[] \not\in  \{ \specie_1, \specie_2 \} } \probability{\abundance[{\specie[]}](0) = 0}
\\
& \, =
(1 - e^{-\freq^{(\strain_1)}\rate})(1 - e^{-\freq^{(\strain_2)}\rate}) \cdot e^{-(1-(\freq^{(\strain_1)} + \freq^{(\strain_2)}))\rate} \,,
\\
\countprob[3+]_{\globe} 
& := 
1 - \countprob[\emptyset]_{\globe} - \sum_{{\strain[]}=1}^{\Species} \countprob[{\strain[]}]_{\globe} - \sum_{{\strain[]_1}=1}^{\Species} \sum_{{\strain[]_2} > {\strain[]_1}} \countprob[{\strain[]_1}, {\strain[]_2}]_{\globe} \, .
  \end{split}
\end{equation}  

\subsubsection{Expected Counts}
\label{sec:expected-counts}

The expected counts equal the total number of droplets $\Droplets$ multiplied by the corresponding probabilities.

\begin{equation}
  \label{eq:global_test_expectations}
  \begin{split}
\expectedcount[\emptyset]_{\globe}
& := 
\Droplets \cdot     \countprob[\emptyset]_{\globe} \,,
\\
\expectedcount[\specie]_{\globe}
& :=
\Droplets \cdot \countprob[\specie]_{\globe}  \,,
\\
\expectedcount[\specie_1, \specie_2]_{\globe}
& :=
\Droplets \cdot \countprob[\specie_1, \specie_2]_{\globe} \,,
\\
\expectedcount[3+]_{\globe}
& := 
\Droplets \cdot \countprob[3+]_{\globe}  \,.
  \end{split}
\end{equation}

\subsubsection{Observed Counts}
\label{sec:observed-counts}

The observed counts for the global goodness of fit test are
\begin{equation}
  \label{eq:global_test_observed}
  \begin{split}
\observedcount[\emptyset]_{\globe}
& := 
\sum_{\droplet \in [\Droplets]} \indicator{\abundance_{\droplet}(0) = 0} \,,
\\
\observedcount[\specie]_{\globe}
& :=
\sum_{\droplet \in [\Droplets]} \indicator{\abundance[\specie]_{\droplet}(0) \ge 1} \cdot \prod_{{\specie[]}\not=\specie} \indicator{ \abundance[{\specie[]}]_{\droplet}(0) = 0} \,,
\\
\observedcount[\specie_1, \specie_2]_{\globe}
& :=
\sum_{\droplet \in [\Droplets]}  \left[ \vphantom{\prod_{{\specie[]}\not\in \{ \specie_1, \specie_2 \}  }}  
\indicator{\abundance[\specie_1]_{\droplet}(0) \ge 1}\cdot \indicator{\abundance[\specie_2]_{\droplet}(0) \ge 1} \right.
\\
& \phantom{:= \sum_{\droplet \in [\Droplets]}} \left.
\cdot \prod_{{\specie[]}\not\in \{ \specie_1, \specie_2 \}  } \indicator{\abundance[{\specie[]}]_{\droplet}(0) = 0}  \right] \,,
\\
\observedcount[3+]_{\globe}
& := 
\Droplets - \observedcount[\emptyset]_{\globe}  - \sum_{{\specie[]}=1}^{\Species} \observedcount[{\specie[]}]_{\globe} - \sum_{{\specie[]_1}=1}^{\Species} \sum_{{\specie[]_2 > \specie[]_1}} \observedcount[{\specie[]_1, \specie[]_2}]_{\globe} \,.
  \end{split}
\end{equation}

\subsubsection{Symmetry of Definitions}
\label{sec:symmetry-definitions}

Observe how, for any given pair of strains $\strain_1, \strain_2$:
\begin{equation}
  \label{eq:comment_global_test}
  \begin{array}{rcl}
\countprob[\strain_1, \strain_2]_{\globe} & = & \countprob[\strain_2, \strain_1]_{\globe} \,, \\
\expectedcount[\strain_1, \strain_2]_{\globe} & = & \expectedcount[\strain_2, \strain_1]_{\globe} \,, \\
\observedcount[\strain_1, \strain_2]_{\globe} & = & \observedcount[\strain_2, \strain_1]_{\globe} \,.
  \end{array}
\end{equation}
Thus for any given pair of strains $\strain_1, \strain_2$, only the left hand sides of (\ref{eq:comment_global_test}) are considered, avoiding redundancy and overlapping categories.

\subsubsection{Test Statistic}
\label{sec:test-statistic}
The test statistic is the Pearson categorical divergence of
\begin{equation}
  \label{eq:global_pearson_divergence_vectors}
  \begin{split}
 &   \Droplets^{-1}
\!\cdot \!
\left( \observedcount[\emptyset]_{\globe}, \observedcount[1]_{\globe}, \dots, \observedcount[\Species]_{\globe}, \observedcount[1,2]_{\globe}, \dots, \observedcount[\Species \! - \! 1, \Species]_{\globe}, \observedcount[3+]_{\globe} \right)
\\
&\text{relative to}   \\
&
\Droplets^{-1}
\!\cdot \!
\left( \expectedcount[\emptyset]_{\globe}, \expectedcount[1]_{\globe}, \dots, \expectedcount[\Species]_{\globe}, \expectedcount[1,2]_{\globe}, \dots, \expectedcount[\Species \! - \! 1, \Species]_{\globe}, \expectedcount[3+]_{\globe} \right) \,.
  \end{split}
\end{equation}
We can get an asymptotic $p$-value in the standard way, via the survival function of the $\chi^2$ distribution with $\left(1 + \Species + \binom{\Species}{2} + 1\right) - 1$ degrees of freedom.

\subsection{Pairwise Gluttonous Goodness of Fit Tests}
\label{sec:pairw-glutt-goodn}

In section \ref{sec:glob-picky-goodn} droplets were grouped according to strain in a non-overlapping way. This made it possible to compare other distributions to hPoMu using a single hypothesis test, but for practical reasons it also required lumping together all droplets with three or more strains. However, if we are willing to perform multiple hypothesis tests, we can better utilize the information contained in those droplets. By performing a separate hypothesis test for each pair of strains, we can include more droplets in both our ``treatment'' and ``control'' groups.

The definition of the hypothesis test corresponding to each pair of strains is split into several parts. Section \ref{sec:categories-1} defines how the droplets are grouped. Section \ref{sec:probabilities-1} gives probabilities for each group under the null hypothesis (hPoMu), while section \ref{sec:expected-counts-1} clarifies the expected counts. Section \ref{sec:observed-counts-1} defines the observed counts used in defining the goodness of fit test. Section \ref{sec:symmetry-definitions-1} explains why some of the groups are redundant. Finally, section \ref{sec:test-statistic-1} combines the information from the previous sections to define the test statistic.

\subsubsection{Categories}
\label{sec:categories-1}

Note that the above definitions correspond to ``gluttonous'' groups, unlike the ``global'' test defined before which corresponds to ``picky'' groups. The ``gluttony'' means that the categories for distinct combinations of strains can overlap. This is what necessitates separate tests for each combination of strains, i.e. why they are ``pairwise''.

\subsubsection{Probabilities}
\label{sec:probabilities-1}

The probabilities $\countprob[\strain_1,\strain_2]_{0,0}$,$ \countprob[\strain_1,\strain_2]_{1,0}$, $ \countprob[\strain_1,\strain_2]_{1,1}$, $ \countprob[\strain_1,\strain_2]_{0,1}$, for the pairwise tests are
\begin{equation}
\label{eq:unconditional_prob_defns}
\begin{split}
\countprob[\strain_1, \strain_2]_{0,0}
& := \probability[\scalebox{0.75}{$\mathrm{hPoMu}(\rate,\vfreq)$} ]*{  
 \scalebox{0.85}{$\abundance[\strain_1](0) = 0, \abundance[\strain_2](0) = 0$}
} \\
& = e^{-\freq^{(\strain_1)} \rate} \cdot e^{-\freq^{(\strain_2)} \rate}
 \,,
\\
\countprob[\strain_1, \strain_2]_{1,0}
& := \probability[\scalebox{0.75}{$\mathrm{hPoMu}(\rate,\vfreq)$} ]*{  
 \scalebox{0.85}{$\abundance[\strain_1](0) \ge 1, \abundance[\strain_2](0) = 0$}
} \\
& = (1 - e^{-\freq^{(\strain_1)} \rate})   \cdot \cdot e^{-\freq^{(\strain_2)} \rate}
 \,,
\\
\countprob[\strain_1,\strain_2]_{1,1}
& :=  \probability[\scalebox{0.75}{$\mathrm{hPoMu}(\rate,\vfreq)$} ]*{
\scalebox{0.85}{$\abundance[\strain_1](0) \ge 1, \abundance[\strain_2](0) \ge 1 $}
}
\\
& = (1 - e^{-\freq^{(\strain_1)} \rate}) \cdot  (1 - e^{-\freq^{(\strain_2)} \rate})
\,,
\\
\countprob[\strain_1, \strain_2]_{0,1}
& := \probability[\scalebox{0.75}{$\mathrm{hPoMu}(\rate,\vfreq)$} ]*{  
 \scalebox{0.85}{$\abundance[\strain_1](0) = 0, \abundance[\strain_2](0) \ge 1$}
} 
\\
& = e^{-\freq^{(\strain_1)} \rate} \cdot (1 - e^{-\freq^{(\strain_2)} \rate})
 \,.
  \end{split}
\end{equation}
Note the implicit dependence on an assumed $\rate$ in the definitions of $\countprob[\strain_1,\strain_2]_{0,0}$, $ \countprob[\strain_1,\strain_2]_{1,0}$, $ \countprob[\strain_1,\strain_2]_{1,1}$, $ \countprob[\strain_1,\strain_2]_{0,1}$. Thus the scientist needs an estimate, or some a priori knowledge, of the value of $\rate$ to use these tests.

\subsubsection{Expected Counts}
\label{sec:expected-counts-1}

The expected counts for the pairwise tests are
\begin{equation}
\label{eq:unconditional_expected_count_defns}
\begin{split}
  \expectedcount[\strain_1,\strain_2]_{0,0} & := \Droplets \cdot \countprob[\strain_1,\strain_2]_{0,0} \,, \\
  \expectedcount[\strain_1,\strain_2]_{1,0} & :=\Droplets \cdot \countprob[\strain_1,\strain_2]_{1,0} \,, \\
\expectedcount[\strain_1,\strain_2]_{1,1} &:=\Droplets \cdot \countprob[\strain_1, \strain_2]_{1,1} \,, \\
\expectedcount[\strain_1,\strain_2]_{0,1} &:=\Droplets \cdot \countprob[\strain_1, \strain_2]_{0,1} \,.
\end{split}
\end{equation}

\subsubsection{Observed Counts}
\label{sec:observed-counts-1}

The observed counts are defined as
\begin{equation}
  \label{eq:observed_counts_defns}
  \begin{split}
    \observedcount[\strain_1,\strain_2]_{0,0} &
:= \left|  \left\{
\droplet:
\abundance[\strain_1]_{\droplet}(0) = 0 , \abundance[\strain_2]_{\droplet}(0) = 0
\right\}  \right| \,, 
\\
    \observedcount[\strain_1,\strain_2]_{1,0} &
:= \left|  \left\{
\droplet:
\abundance[\strain_1]_{\droplet}(0) \ge 1 , \abundance[\strain_2]_{\droplet}(0) = 0
\right\}  \right| \,, 
\\
\observedcount[\strain_1, \strain_2]_{1,1} &
:= \left| \left\{
\droplet:
\abundance[\strain_1]_{\droplet}(0) \ge 1, \abundance[\strain_2]_{\droplet} (0) \ge 1
\right\} \right| \,,
\\
\observedcount[\strain_1, \strain_2]_{0,1} &
:= \left|\left\{
\droplet:
\abundance[\strain_1]_{\droplet}(0) = 0, \abundance[\strain_2]_{\droplet} (0) \ge 1
\right\}  \right| \,.
  \end{split}
\end{equation}

\subsubsection{Symmetry of Definitions}
\label{sec:symmetry-definitions-1}

Of course, 
\begin{equation}
  \label{eq:pairwise_defns_symmetry}
  \begin{array}{rcl}
    \observedcount[\strain_1, \strain_2]_{1,0} = \observedcount[\strain_2, \strain_1]_{0,1} \,,
& \quad &
\countprob[\strain_1, \strain_2]_{1,0} = \countprob[\strain_2, \strain_1]_{0,1} \,,
\\
\observedcount[\strain_1, \strain_2]_{0,0} = \observedcount[\strain_2, \strain_1]_{0,0} \,,
& \quad &
\countprob[\strain_1, \strain_2]_{0,0} = \countprob[\strain_2, \strain_1]_{0,0} \,,
\\
\observedcount[\strain_1, \strain_2]_{1,1} = \observedcount[\strain_2, \strain_1]_{1,1} \,,
& \quad &
\countprob[\strain_1, \strain_2]_{1,1} = \countprob[\strain_2, \strain_1]_{1,1} \,.
  \end{array}
\end{equation}
Therefore the results of the test for $(\strain_2, \strain_1)$ will always be the same as those for $(\strain_1, \strain_2)$, and so at most one of the two tests needs to be performed.

\subsubsection{Test Statistic}
\label{sec:test-statistic-1}

The test statistic is the Pearson categorical divergence of
\begin{equation}
  \label{eq:unconditional_pearson_divergence_vectors}
  \begin{split}
 &   \Droplets^{-1}
\!\cdot \!
\left( \observedcount[\specie_1, \specie_2]_{0,0}, \observedcount[\specie_1, \specie_2]_{1,0}, \observedcount[\specie_1, \specie_2]_{1,1}, \observedcount[\specie_1, \specie_2]_{0,1}   \right)
\\
\text{relative to} \quad &
\Droplets^{-1}
\!\cdot \!
\left(  \expectedcount[\specie_1, \specie_2]_{0,0},  \expectedcount[\strain_1, \strain_2]_{1,0} , \expectedcount[\strain_1, \strain_2]_{1,1} ,  \expectedcount[\strain_1, \strain_2]_{0,1}  \right) \,,
  \end{split}
\end{equation}
{namely}
\begin{equation}
  \label{eq:unconditional_pairwise_pearson_divergence}
  \sum_{i =0}^1 \sum_{j=0}^1   \frac{
\left(  \observedcount[\strain_1, \strain_2]_{i,j}   - \expectedcount[\strain_1, \strain_2]_{i,j}  \right)^2
}{
\expectedcount[\strain_1, \strain_2]_{i,j}
} \,.
\end{equation}
One can then get an asymptotic $p$-value for this test in the standard way, via the survival function of the $\chi^2$ distribution with $3$ degrees of freedom.

\section{Methods}
\label{sec:data_throughput_methods}

Section \ref{sec:simul-distr-redux} explains which distributions were simulated, why, and how. Section \ref{sec:computing-p-values} explains how categorical divergences and corresponding approximate $p$-values were computed. Section \ref{sec:numb-dropl-with} explains how percentages and percent changes for droplets with a given number of strains were computed. Section \ref{sec:gluttonous-groups} explains how I computed the percent changes for gluttonous groups. Section \ref{sec:picky-groups} explains how I computed the percent changes for picky groups. I plotted results using Matplotlib \cite{Matplotlib} version 3.4.1 and Seaborn \cite{Seaborn} version 0.11.1. The bar plot was made using the \textit{microbiome} \cite{microbiome_r}, \textit{phyloseq} \cite{phyloseq}, and \textit{ggplot2} \cite{ggplot2} packages for the \textit{R} programming language. Complete implementation details can be found in the code at \dropletsgitrepo. See \url{\dropletsgitrepourl}.

\subsection{Simulated Distributions}
\label{sec:simul-distr-redux}

To test how much the assumptions of hPoMu may be violated with hPoMu remaining an adequate working model, I simulated seven distributions. Below, I describe the two new distributions beyond the five distributions that were already described in section \ref{sec:simul-distr}. Details of the simulations common to all seven distributions were given already in section \ref{sec:simul-impl-deta}.

\subsubsection{Simulating High Compositional Heterogeneity Only}
\label{sec:simul-comp-het-high}

This distribution was a hierarchical Exponential hPoDM model (hExhPoDM), with $\cconcentration$ distributed as an $\operatorname{Exponential}(1)$ random variable.

\subsubsection{Simulating High Compositional and Density Heterogeneities}
\label{sec:simul-comp-dens-high}

This distribution was a hierarchical Exponential hNBDM model (hExhNBDM), and $\concentration$ was distributed as an $\operatorname{Exponential}(1)$ random variable.

\subsection{Computing $p$-values and $\chi^2$ Divergences}
\label{sec:computing-p-values}

Using the categories and expected counts defined for the ``global'' goodness of fit hypothesis test of the null hPoMu using ``picky'' groups defined in section \ref{sec:picky-defin-data}, I computed Pearson categorical (``$\chi^2$'', cf. section \ref{sec:pears-categ-diverg}) divergence values and $\chi^2$-approximated $p$-values for all $500$ simulations of all $7$ distributions. 

Despite that each simulation corresponded to an extremely large number of droplets, the expected counts for some categories were very small ($<5$), making the asymptotic approximation provided by the $\chi^2$ assumption possibly inadequate. To account for this, I generated $10^9$ (one \textbf{b}illion) independent replicates of the Multinomial distribution corresponding to the number of droplets per simulation and the probabilities for each category in the goodness of fit test. From this, I generated $10^9$ replicates from the sampling distribution under the null of the Pearson categorical divergence statistic. I then used this Monte Carlo distribution of Pearson categorical divergence statistics to compute ``Monte Carlo $p$-values'' for all $500$ simulations from all $7$ distributions. Given the very large number of replicates, I expect these $p$-values to be more accurate than those from the $\chi^2$ asymptotic approximation. Even so the results using the $\chi^2$ approximation were similar, cf. section \ref{sec:p-values-from}.

\subsection{Numbers of Droplets with $n$ Strains}
\label{sec:numb-dropl-with}

Using the relationships:
\begin{equation}
  \label{eq:strain_count_strata_general}
  \begin{split}
 \expectation*{\left| \left\{ \droplet \in [\Droplets]: \straincount_{\droplet}(0) = \counts  \right\}  \right|  }   
 = &
\Droplets \cdot \probability{ \straincount(0) = \counts } \\
= &
\Droplets \cdot
\left[
\sum_{\mathcal{S} \subseteq [\Strains]: |\mathcal{S}| = \counts}
\probability{ \support(\vstraincount(0)) = \mathcal{S}  }
\right]
\,,
  \end{split}
\end{equation}
I computed via brute force these quantities for $\counts \in [5]$ for hPoMu:
 \begin{equation}
   \label{eq:strain_count_strata_hPoMu}
   \begin{split}
&
\Droplets \cdot
\left[
\sum_{\mathcal{S} \subseteq [\Strains]: |\mathcal{S}| = \counts}
e^{-\sum_{\strain \not \in \mathcal{S}} \freq^{(\strain)} \rate }
\cdot
\prod_{\strain \in \mathcal{S}} \left( 1 - e^{-\freq^{(\strain)} \rate}  \right)
\right] \\
= &
\Droplets \cdot
\left[
\sum_{\mathcal{S} \subseteq [\Strains]: |\mathcal{S}| = \counts}
e^{-(1 - \sum_{\strain  \in \mathcal{S}} \freq^{(\strain)}) \rate }
\cdot
\prod_{\strain \in \mathcal{S}} \left( 1 - e^{-\freq^{(\strain)} \rate}  \right)
\right]
 \,.
   \end{split}
 \end{equation}
The value for $6+$ strains is just $\Droplets$ minus the sum of the values for all $\counts \in [5]$.

For each of the seven distributions, I stratified and then counted the droplets from each of the 500 simulations according to how many strains were present. I then computed the average values for each distribution as the arithmetic mean over the 500 simulations. (By the law of large numbers the arithmetic mean is consistent for the true expectation and the only reasonable choice of measure of ``central tendency'' in this context.) I then computed the average percent change for each distribution by subtracting the value expected under hPoMu from the mean observed value, and then dividing by the expected value.

\subsection{Gluttonous Groups}
\label{sec:gluttonous-groups}

Using the formulae (\ref{eq:unconditional_prob_defns}) and (\ref{eq:unconditional_expected_count_defns}), I computed the expected count for each treatment gluttonous group\footnote{\label{footnote:gluttonous_faux_controls}
For the gluttonous ``controls'', i.e. corresponding to the diagonals of the heatmaps, I actually used the sets ${ \{ \droplet \in [\Droplets]: \abundance[\strain](0) \ge 1  \} }$, which is a superset of all gluttonous controls for experiments measuring effects on the growth of strain $\strain$. The expected size of these sets is ${\Droplets \cdot \expectation{\straincount[\strain](0)} =\Droplets\cdot \probability{\abundance[\strain](0) \ge 1}}$, which under hPoMu equals $\Droplets \cdot (1 - e^{-\freq^{(\strain)} \rate})$. 
} under hPoMu. For each of the seven distributions, for each of the 500 simulations using NumPy \cite{NumPy} version 1.20.2 I computed via brute force the number of occurrences of each of the gluttonous groups, and then reported the arithmetic means over all 500 simulations. I again computed the average percent change as the observed value minus the value expected under hPoMu divided by the value expected under hPoMu.

Initially I considered ``filtering'' differences depending on whether they were ``significant'' according to the (asymptotic) $p$-values of the hypothesis tests defined in section \ref{sec:pairw-glutt-goodn}. However, I found that doing so made the results more difficult to interpret by obscuring overall trends. The inferred overall trends were qualitatively the same in either case. Moreover, because each gluttonous pair corresponds to a distinct hypothesis test, and the gluttonous groups overlap, this leads to a difficult multiple comparisons problem. One has to not only decide how to account for multiple testing between simulations but also ``within'' simulations. Each of the multiple testing correction schemes I tried seemed objectionable in some way. Given all of the above, I decided to report the ``raw'' average percent change without ``filtering'' for ``significance''.

\subsection{Picky Groups}
\label{sec:picky-groups}

Using the formulae (\ref{eq:global_test_probs}) and (\ref{eq:global_test_expectations}), cf. also the equations (\ref{eq:expected_number_picky_treatments}) and (\ref{eq:expected_number_picky_controls}), I computed the expected count for each picky group under hPoMu. Note that, unlike for gluttonous treatment groups, picky treatment groups do not overlap. Moreover, unlike for gluttonous control groups, the picky control group for investigating the effect of strain $\strain_1$ on the rate of growth of strain $\strain_2$ is the same regardless of what $\strain_1$ is. So the values shown on the diagonals of the heat maps really correspond to the actual picky control groups. The procedure was then otherwise exactly the same as for the gluttonous groups. I did not ``filter'' according to ``significance''.

\section{Results}
\label{sec:data_throughput_results}

Section \ref{sec:results-from-global} examines the results of the global $\chi^2$ goodness of fit tests computed from the picky contingency tables. Section \ref{sec:results-contingency-tables} directly examines both the picky and gluttonous contingency tables.

\subsection{Global $\chi^2$ Tests}
\label{sec:results-from-global}

Section \ref{sec:sampl-without-repl-conting} discusses evidence indicating that sampling with replacement is the most innocuous of the assumptions behind hPoMu. Section \ref{sec:very-simil-distr-conting} discusses evidence indicating that ghNBDM distributions with low (but nonzero) heterogeneity may be adequately modeled by hPoMu. Section \ref{sec:slightly-diff-distr-conting} discusses evidence indicating that ghNBDM distributions with only moderate heterogeneity can be easily distinguished from hPoMu. Finally, section \ref{sec:very-diff-distr-conting} discusses evidence indicating that even further heterogeneity can make differences from hPoMu extremely obvious.

\begin{figure}
  \centering
\includegraphics[width=\textwidth,height=0.47\textheight,keepaspectratio]{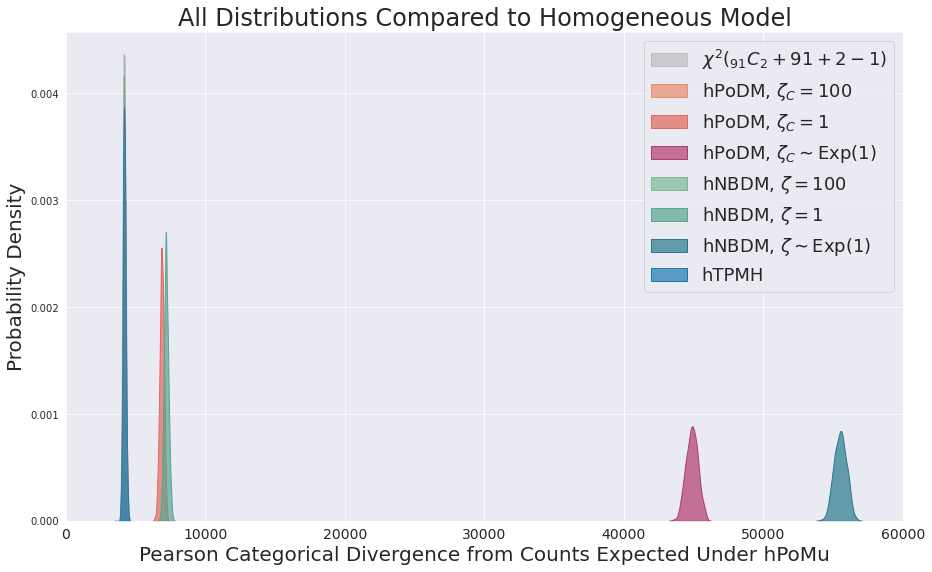}
  \caption[]{}
  \label{fig:divergences_all}
\end{figure}

\subsubsection{Sampling without Replacement}
\label{sec:sampl-without-repl-conting}

All available evidence suggests that, even after $15,000,000$ droplets have been formed, the hTPMH distribution for the chosen value of $\Population$ is extremely similar to the corresponding hPoMu distribution.

\paragraph{Goodness of Fit $p$-Values Are Approximately Uniformly Distributed}
\label{sec:goodness-fit-p}

For the hTPMH distribution, figure \ref{fig:hTPMH_pvals} shows how the distribution of (Monte Carlo) $p$-values over $500$ simulations is approximately uniform, as it would be if we had actually sampled from the true hPoMu null distribution instead.

\begin{figure}
  \centering
  \includegraphics[width=\textwidth,height=0.47\textheight,keepaspectratio]{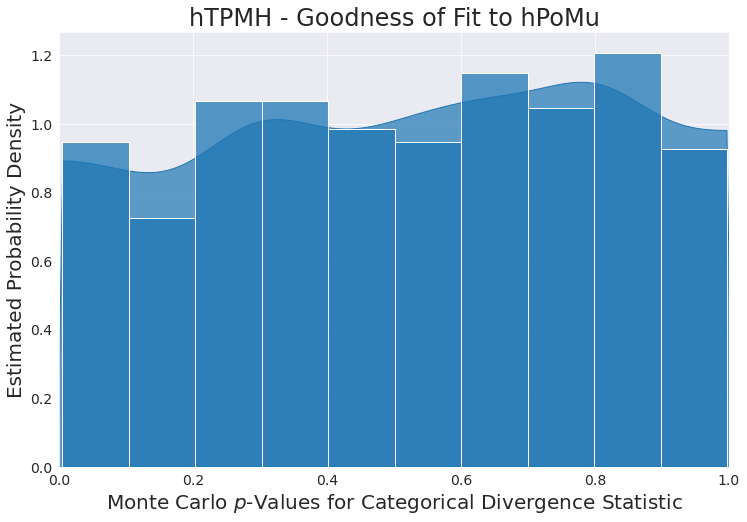}
  \caption{Monte Carlo $p$-value distribution for sampling without replacement.}
  \label{fig:hTPMH_pvals}
\end{figure}

\paragraph{Distribution of Pearson Categorical Divergences for Sampling without Replacement}
\label{sec:distr-diverg-sampl}

Moreover, as seen clearly in figure \ref{fig:divergences_similar} (and less so in figure \ref{fig:divergences_all}), the observed distribution of Pearson categorical divergences from the hTPMH distribution overlaps strongly with approximately what would have been expected under the hPoMu null distribution.

\begin{figure}
  \centering
  \includegraphics[width=\textwidth,height=0.47\textheight,keepaspectratio]{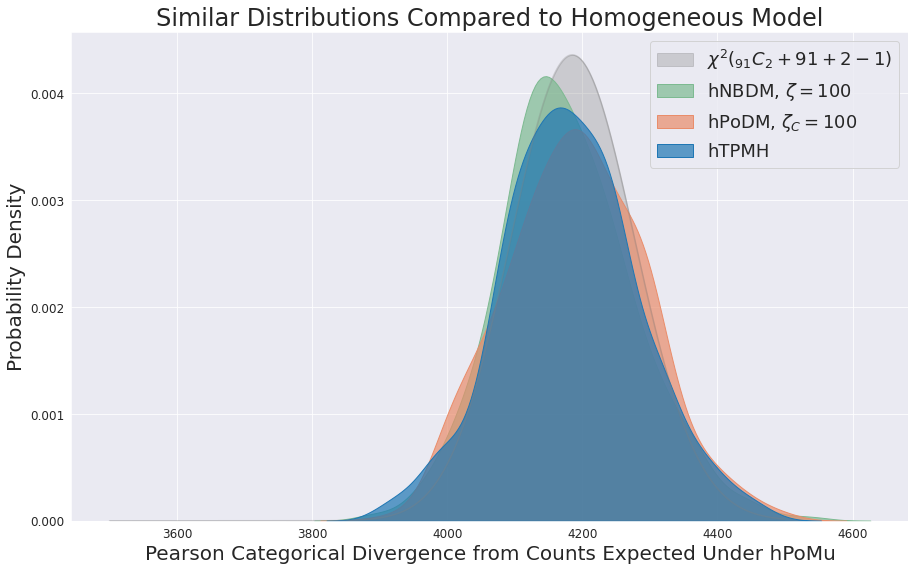}
  \caption[]{}
  \label{fig:divergences_similar}
\end{figure}

\subsubsection{Very Similar Distributions}
\label{sec:very-simil-distr-conting}

All available evidence suggests that the hPoDM with $\cconcentration =100$ and hNBDM with $\concentration =100$ distributions are very similar to the corresponding hPoMu distribution. Nevertheless, they also appear to differ slightly more than the hTPMH distribution.

\paragraph{Goodness of Fit $p$-Values Are Roughly Uniformly Distributed}
\label{sec:goodness-fit-p-1}

For the hPoDM distribution with $\cconcentration=100$ and the hNBDM distribution with $\concentration=100$, figures \ref{fig:hPoDM_100_pvals} and \ref{fig:hNBDM_100_pvals} show respectively how their distributions of (Monte Carlo) $p$-values over $500$ simulations is roughly similar to the uniform distribution. 

\begin{figure}[p]

\begin{subfigure}{\textwidth}
  \centering
  \includegraphics[width=\textwidth,height=0.45\textheight,keepaspectratio]{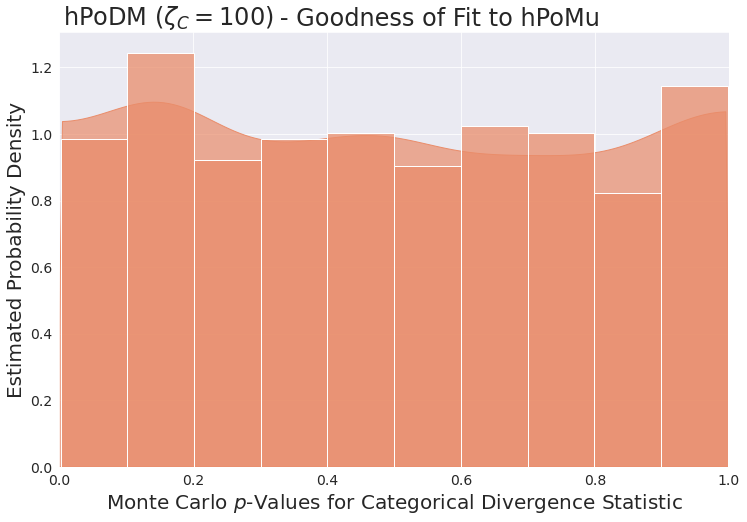}
  \caption[]{}
  \label{fig:hPoDM_100_pvals}
\end{subfigure}

\begin{subfigure}{\textwidth}
  \centering
\includegraphics[width=\textwidth,height=0.45\textheight,keepaspectratio]{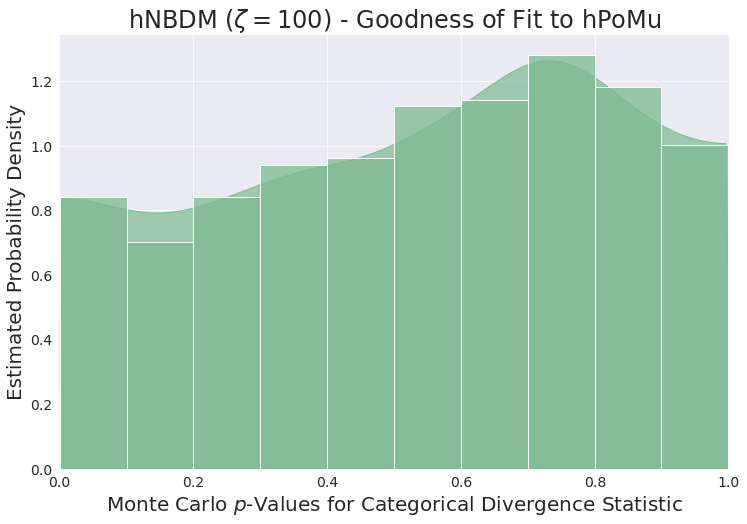}
  \caption[]{}
  \label{fig:hNBDM_100_pvals}
\end{subfigure}

\caption{Monte Carlo goodness of fit $p$-value distributions for very similar distributions.}
\label{fig:very_similar_pvals}
\end{figure}

The $p$-value distribution for hNBDM with $\concentration=100$ in figure \ref{fig:hNBDM_100_pvals} appears to fit the uniform distribution worse than the corresponding distribution for hPoDM with $\cconcentration=100$ in figure \ref{fig:hPoDM_100_pvals}. This could possibly be a reflection of how the hNBDM family incorporates both kind of heterogeneities whereas the hPoDM family does not. However, it could also possibly be a reflection merely of the relatively small ($n=500$) sample size. 

\paragraph{Distributions of Pearson Categorical Divergences for Very Similar Distributions}
\label{sec:distr-diverg-very-1}

Furthermore, figure \ref{fig:divergences_similar} indicates how the distribution of their Pearson categorical divergence statistics also overlaps well with roughly what would be expected under the null distribution. 

The overlap for hNBDM $\concentration=100$ seen in figure \ref{fig:divergences_similar} does not seem to be substantially worse (nor better) than that observed for hPoDM $\cconcentration=100$, in contrast to the extent of difference that might have been anticipated after comparing figures \ref{fig:hPoDM_100_pvals} and \ref{fig:hNBDM_100_pvals}.

\subsubsection{Slightly Different Distributions}
\label{sec:slightly-diff-distr-conting}

All available evidence suggests that the hPoDM with $\cconcentration =1$ and hNBDM with $\concentration =1$ distributions are (at least) slightly different from the corresponding hPoMu distribution.

\paragraph{Goodness of Fit $p$-Values Are All Effectively Zero}
\label{sec:goodness-fit-p-2}

For the hPoDM distribution with $\cconcentration=1$ and the hNBDM distribution with $\concentration=1$, all of the approximate $p$-values were indistinguishable from $0$ up to numerical precision. Therefore their distributions of approximate $p$-values, being a ``single point spike'' or ``Dirac delta'', were not plotted. 

\paragraph{Distributions of Pearson Categorical Divergences for Slightly Different Distributions}
\label{sec:distr-diverg-slightl}

The reason behind these low $p$-values is fairly obvious from either figure \ref{fig:divergences_all} or \ref{fig:divergences_slightly_different}: their distributions of observed Pearson categorical divergences overlap almost not at all with what would be anticipated under the hPoMu null distribution. Also, the distribution of Pearson categorical divergences for hNBDM with $\concentration=1$ is slightly further to the right than that of hPoDM with $\cconcentration=1$, reflecting how the former incorporates both density and compositional heterogeneity while the latter incorporates only compositional heterogeneity.

\begin{figure}
  \centering
\includegraphics[width=\textwidth,height=0.47\textheight,keepaspectratio]{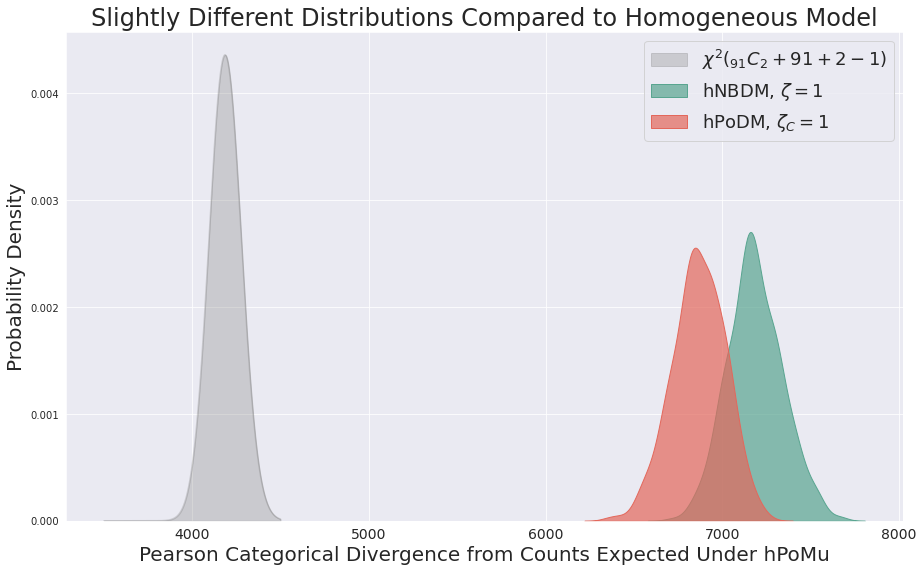}
  \caption[]{}
  \label{fig:divergences_slightly_different}
\end{figure}

\subsubsection{Very Different Distributions}
\label{sec:very-diff-distr-conting}

All available evidence suggests that the hExhPoDM with $\mathbb{E}[\cconcentration] =1$ and hExhNBDM with $\mathbb{E}[\concentration] =1$ distributions are very different from the corresponding hPoMu distribution.

\paragraph{Goodness of Fit $p$-Values Are All Effectively Zero}
\label{sec:goodness-fit-p-3}

For the hExhPoDM with $\mathbb{E}[\cconcentration] =1$ and hExhNBDM with $\mathbb{E}[\concentration] =1$ distributions, all of the (approximate) $p$-values were again indistinguishable from $0$ up to numerical precision, as was also the case for hPoDM $\cconcentration=1$ and hNBDM $\concentration=1$. The approximate $p$-values were again not plotted for the same reasons as for hPoDM $\cconcentration=1$ and hNBDM $\concentration=1$. 

\paragraph{Distributions of Pearson Categorical Divergences for Very Different Distributions}
\label{sec:distr-diverg-very}

The reason for these small $p$-values is also the same as for hPoDM $\cconcentration=1$ and hNBDM $\concentration=1$: as can be seen clearly in either figure \ref{fig:divergences_all} or figure \ref{fig:divergences_very_different}, the peaks of their distributions of Pearson categorical divergences completely fail to overlap with the range of values that would be anticipated under the hPoMu null distribution. Both peaks are also \textit{much} further to the right than anything observed for any of the other five distributions, indicating that the hExhPoDM $\mathbb{E}[\cconcentration] =1$ and hExhNBDM $\mathbb{E}[\concentration] =1$ distributions are (by far) the most heterogeneous of the seven distributions simulated. 

\begin{figure}
  \centering
\includegraphics[width=\textwidth,height=0.47\textheight,keepaspectratio]{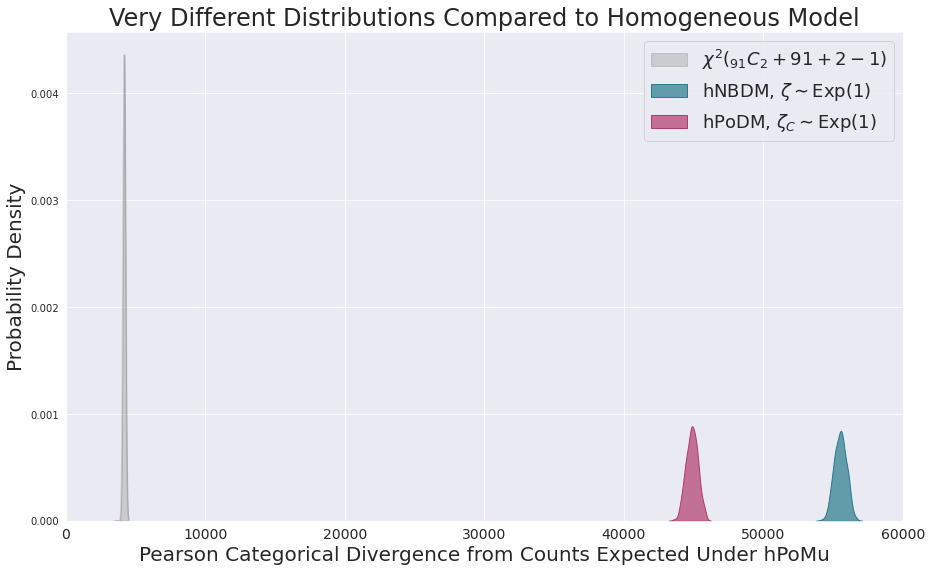}
  \caption[]{}
  \label{fig:divergences_very_different}
\end{figure}

Analogous to what occurs for hPoDM $\cconcentration=1$ and hNBDM $\concentration=1$, the peak for hExhNBDM $\mathbb{E}[\concentration] =1$ is further to the right than the peak for hExhPoDM $\mathbb{E}[\cconcentration] =1$, reflecting how the former incorporates both kind of heterogeneities whereas the latter does not. (Thus hExhNBDM $\mathbb{E}[\concentration] =1$ can be considered the single most heterogeneous distribution of the seven that were simulated.) However, in contrast to the situation observed for hPoDM $\cconcentration=1$ and hNBDM $\concentration=1$, the peaks for hExhNBDM $\mathbb{E}[\concentration] =1$ and hExhPoDM $\mathbb{E}[\cconcentration] =1$ do not overlap at all, with the peak for hExhNBDM $\mathbb{E}[\concentration] =1$ being greatly further to the right than that for hExhPoDM $\mathbb{E}[\cconcentration] =1$.

\subsection{Direct Examination of Contingency Tables}
\label{sec:results-contingency-tables}

Section \ref{sec:distr-again-fall} explains how the observed effects on data throughput support the ``taxonomy'' of distributions previously proposed in section \ref{sec:model_comparison_results}. Section \ref{sec:incr-heter-decr} explains the evidence indicating that increased heterogeneity decreases the average number of droplets with 3 or more strains. Section \ref{sec:effect-incr-heter} explains the effect of increased heterogeneity on data throughput for droplets with 2 or fewer strains, with section \ref{sec:incr-comp-heter} explaining what happens for compositional heterogeneity without density heterogeneity, and section \ref{sec:incr-comp-dens} explaining what happens when both types of heterogeneity are present.

\begin{figure}[p]

  \begin{subfigure}{\textwidth}
        \centering
  \includegraphics[width=\textwidth,height=0.45\textheight,keepaspectratio]{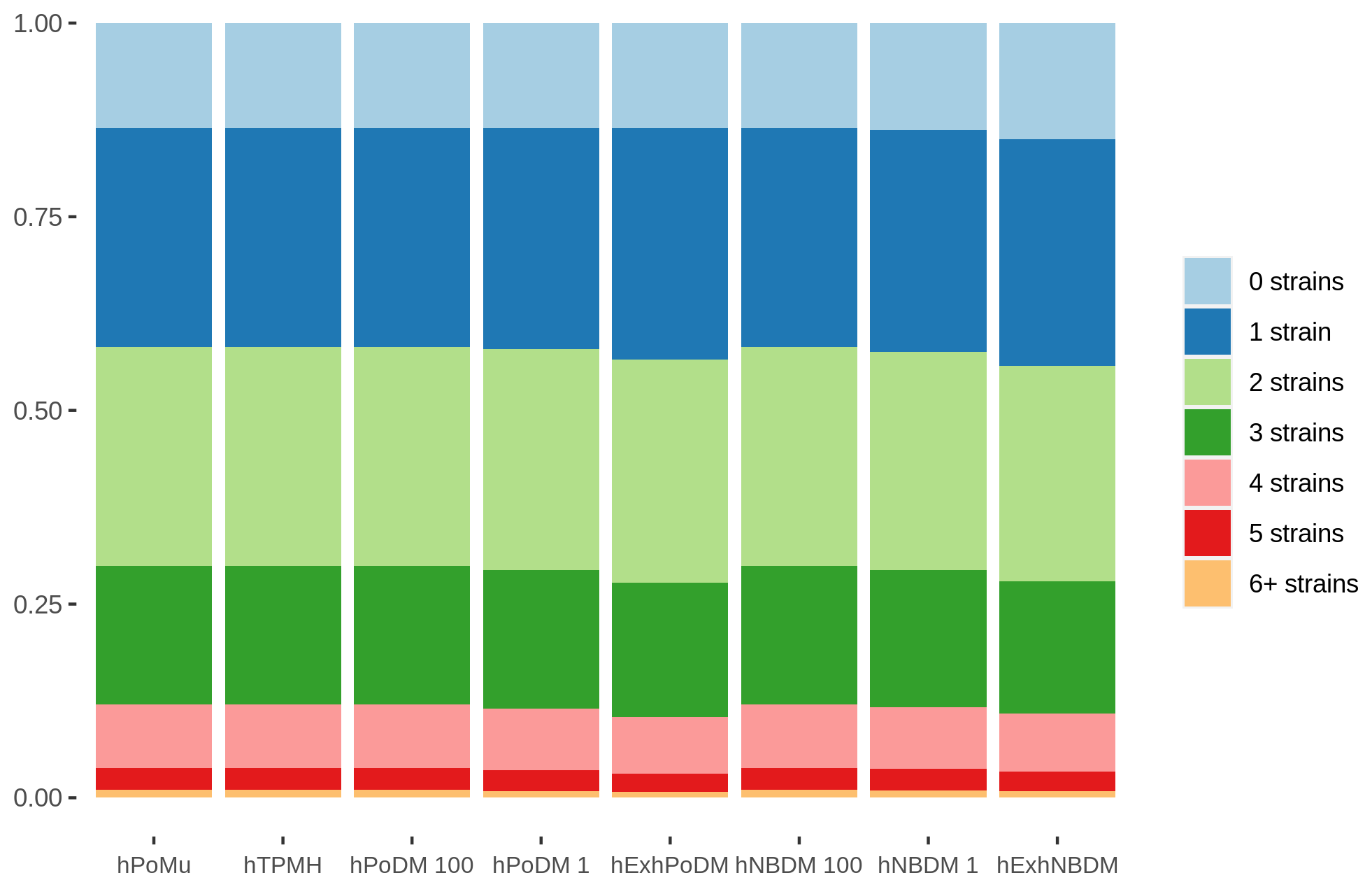}
  \caption[]{}
\label{fig:num_strain_strata_avg_compositions}
\end{subfigure}

\begin{subfigure}{\textwidth}
    \centering
  \includegraphics[width=\textwidth,height=0.45\textheight,keepaspectratio]{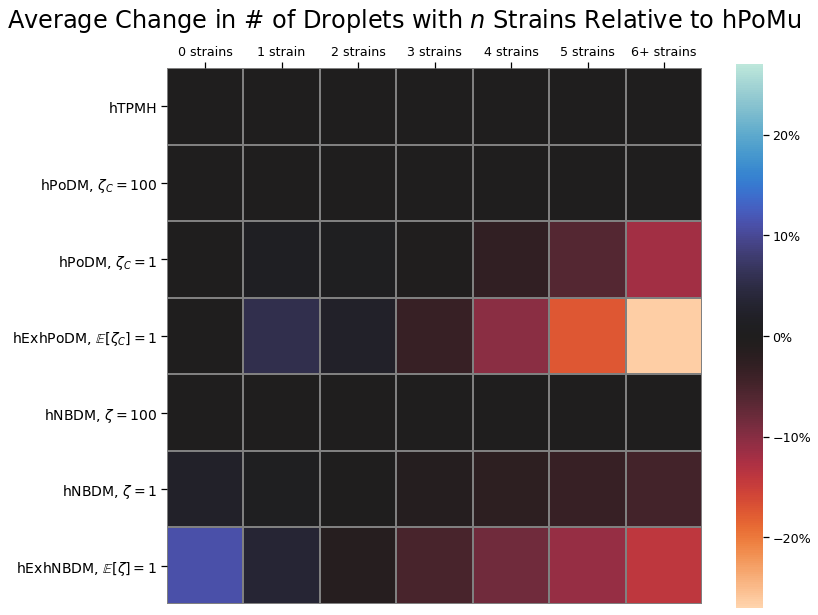}
  \caption[]{}
  \label{fig:num_strain_strata_avg_changes}
\end{subfigure}

\caption{Droplets stratified by numbers of strains found inside droplets.}
\label{fig:num_strain_strata}
\end{figure}

\subsubsection{Distributions Again Fall Into Same Groups}
\label{sec:distr-again-fall}

From figures \ref{fig:num_strain_strata_avg_compositions}, \ref{fig:num_strain_strata_avg_changes}, \ref{fig:hTPMH_gluttonous}, \ref{fig:hPoDM_100_gluttonous}, \ref{fig:hNBDM_100_gluttonous}, \ref{fig:hTPMH_picky}, \ref{fig:hPoDM_100_picky}, and \ref{fig:hNBDM_100_picky} we see that hTPMH, hPoDM with $\zeta_C = 100$, and hNBDM with $\zeta_C = 100$ are again (cf. section \ref{sec:model_comparison_results}) either indistinguishable from or very similar to hPoMu. Any differences in average composition are only visible for medium and high heterogeneities, and easily visible usually only for high heterogeneity.

{
  \newcommand{\figlabel}{fig:low_het_gluttonous}
  \newcommand{\figcaption}{Gluttonous groups for low heterogeneity distributions.}

\begin{figure}
  \begin{subfigure}{\textwidth}
      \centering
  \includegraphics[width=\textwidth,height=0.47\textheight,keepaspectratio]{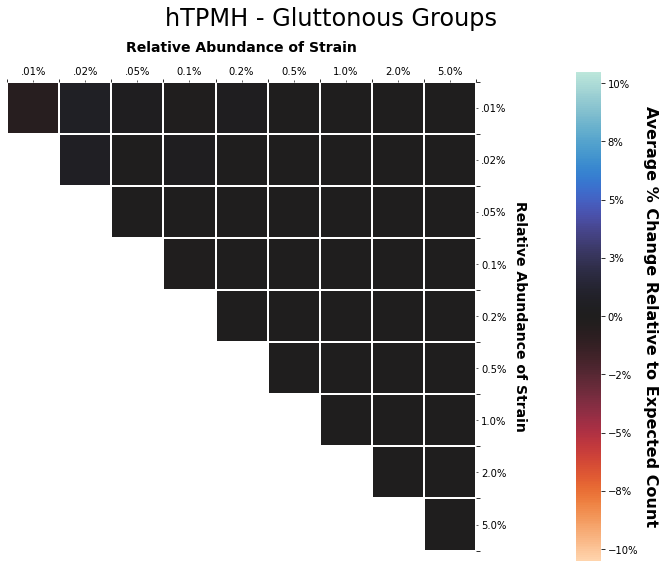}
 \caption[]{}
  \label{fig:hTPMH_gluttonous}
\end{subfigure}

\caption{\figcaption}
\label{\figlabel}
\end{figure}

\begin{figure}[p]
  \ContinuedFloat
  \begin{subfigure}{\textwidth}
      \centering
  \includegraphics[width=\textwidth,height=0.45\textheight,keepaspectratio]{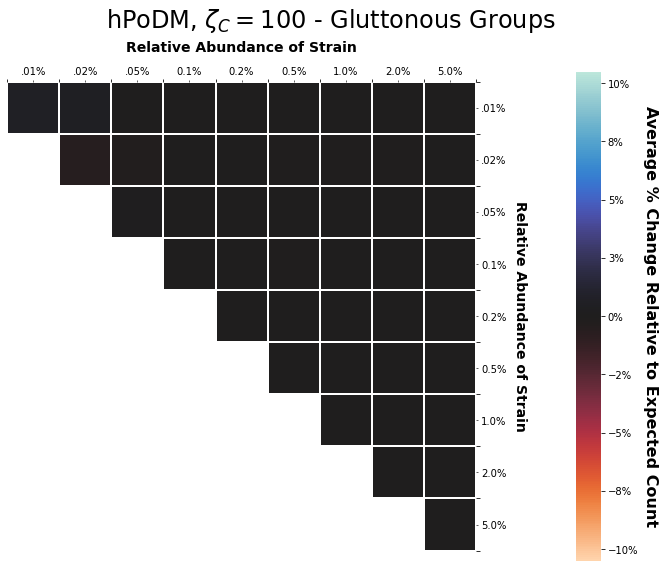}
  \caption[]{}
  \label{fig:hPoDM_100_gluttonous}
\end{subfigure}

\begin{subfigure}{\textwidth}
   \centering
  \includegraphics[width=\textwidth,height=0.45\textheight,keepaspectratio]{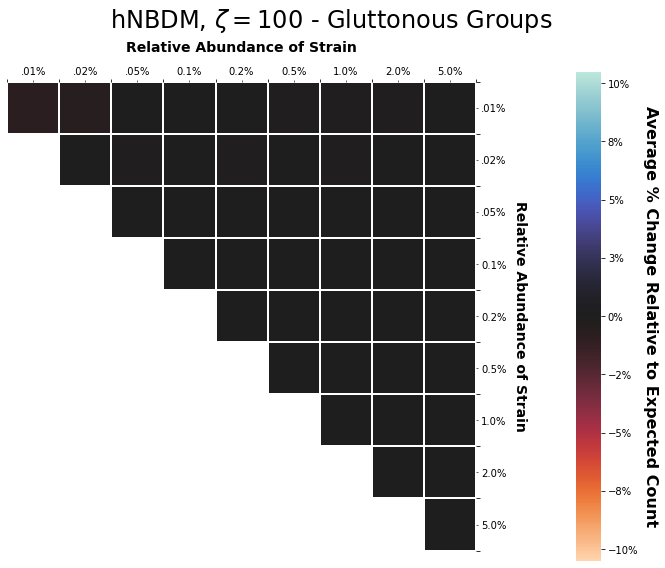}
  \caption[]{}
  \label{fig:hNBDM_100_gluttonous}
\end{subfigure}

\caption[]{\figcaption}
\label{\figlabel}

\end{figure}

}

{
  \newcommand{\figlabel}{fig:low_het_picky}
  \newcommand{\figcaption}{Picky groups for low heterogeneity distributions.}

\begin{figure}
  \centering
  \begin{subfigure}{\textwidth}
  \includegraphics[width=\textwidth,height=0.47\textheight,keepaspectratio]{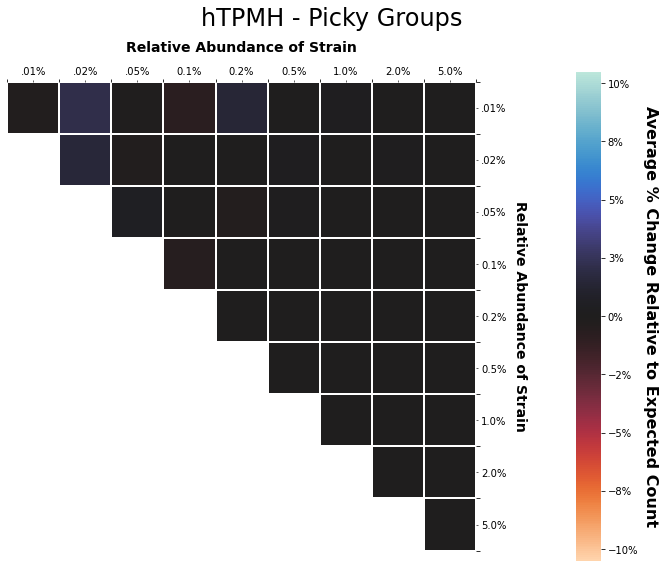}
  \caption[]{}
  \label{fig:hTPMH_picky}
\end{subfigure}

\caption{\figcaption}
\label{\figlabel}

\end{figure}

\begin{figure}[p]
  \ContinuedFloat
  \begin{subfigure}{\textwidth}
     \centering
  \includegraphics[width=\textwidth,height=0.45\textheight,keepaspectratio]{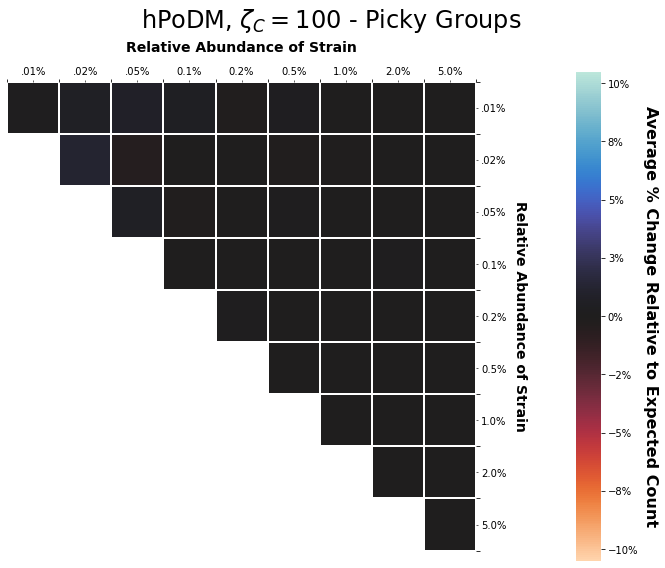}
  \caption[]{}
  \label{fig:hPoDM_100_picky}
\end{subfigure}

\begin{subfigure}{\textwidth}
    \centering
  \includegraphics[width=\textwidth,height=0.45\textheight,keepaspectratio]{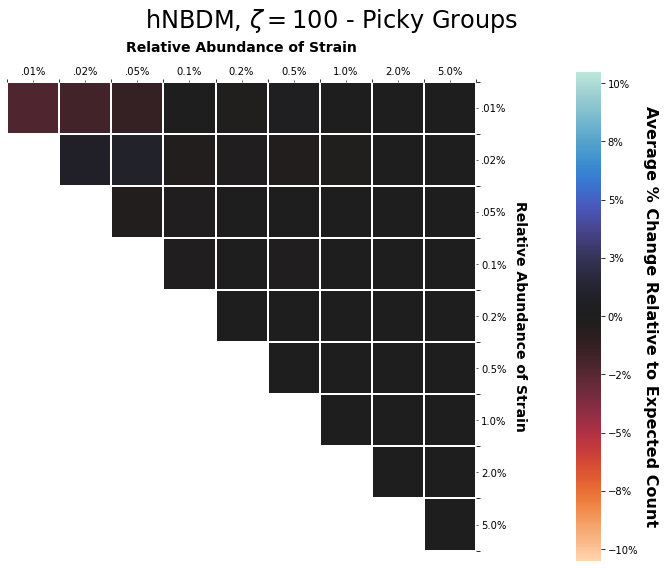}
  \caption[]{}
  \label{fig:hNBDM_100_picky}
\end{subfigure}

\caption[]{\figcaption}
\label{\figlabel}

\end{figure}

}

\subsubsection{Increased Heterogeneity Decreases Number of Droplets with 3 or More Strains}
\label{sec:incr-heter-decr}

From figures \ref{fig:num_strain_strata_avg_compositions} and \ref{fig:num_strain_strata_avg_changes} we see that, both for the hPoDM and hNBDM families, as the compositional heterogeneity increases, the number of droplets with 3 or more strains decreases in both absolute and relative terms. Since droplets with 3 or more strains are a component of the gluttonous groups, their decrease with increased compositional heterogeneity is also reflected in figures \ref{fig:hPoDM_1_gluttonous}, \ref{fig:hNBDM_1_gluttonous}, \ref{fig:hExhPoDM_gluttonous}, and \ref{fig:hExhNBDM_gluttonous}. We can also see from figures \ref{fig:num_strain_strata_avg_changes}, \ref{fig:hPoDM_1_gluttonous}, \ref{fig:hNBDM_1_gluttonous}, \ref{fig:hExhPoDM_gluttonous}, and \ref{fig:hExhNBDM_gluttonous} that this decrease in the number of droplets with 3 or more strains is more pronounced in the absence of density heterogeneity.

{

  \newcommand{\figcaption}{Gluttonous groups for high heterogeneity distributions.}
  \newcommand{\figlabel}{fig:high_het_gluttonous}
  
\begin{figure}[p]
  \begin{subfigure}{\textwidth}
    \centering
  \includegraphics[width=\textwidth,height=0.45\textheight,keepaspectratio]{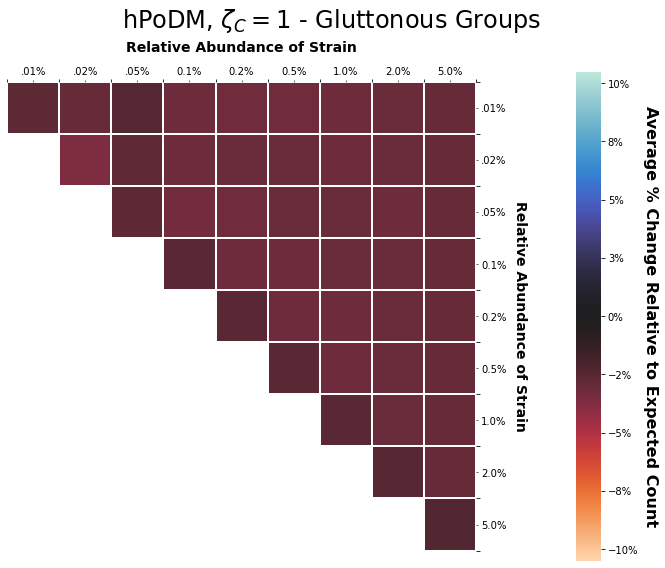}
  \caption[]{}
  \label{fig:hPoDM_1_gluttonous}
\end{subfigure}

\begin{subfigure}{\textwidth}
    \centering
  \includegraphics[width=\textwidth,height=0.45\textheight,keepaspectratio]{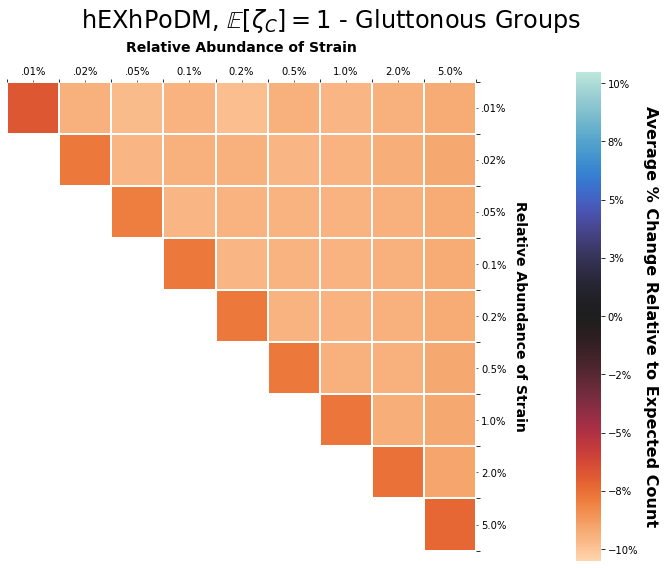}
  \caption[]{}
  \label{fig:hExhPoDM_gluttonous}
\end{subfigure}

\caption{\figcaption}
\label{\figlabel}

\end{figure}

\begin{figure}[p]
  \ContinuedFloat
  \begin{subfigure}{\textwidth}
    \centering
  \includegraphics[width=\textwidth,height=0.45\textheight,keepaspectratio]{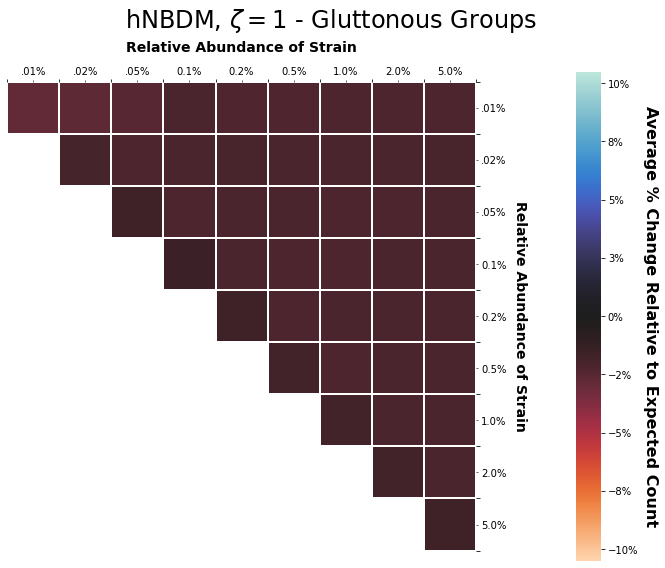}
  \caption[]{}
  \label{fig:hNBDM_1_gluttonous}
\end{subfigure}

\begin{subfigure}{\textwidth}
    \centering
  \includegraphics[width=\textwidth,height=0.45\textheight,keepaspectratio]{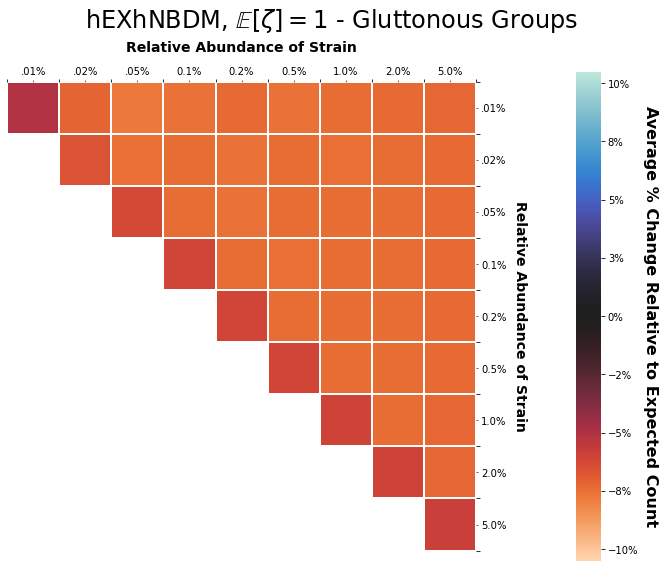}
  \caption[]{}
  \label{fig:hExhNBDM_gluttonous}
\end{subfigure}

\caption[]{\figcaption}
\label{\figlabel}

\end{figure}

} %

\subsubsection{Effect of Increased Heterogeneity on Droplets with Fewer than 3 Strains}
\label{sec:effect-incr-heter}

For droplets with fewer than 3 strains, the effects that occur as compositional heterogeneity increases depend on whether density heterogeneity is present. 

\paragraph{Increased Compositional Heterogeneity Only}
\label{sec:incr-comp-heter}

In the absence of density heterogeneity, we see an increase in the number of droplets with one strain, with a smaller increase in the number of droplets with two strains. This is best evidenced in figure \ref{fig:num_strain_strata_avg_changes}. This slight increase in the number of droplets with two strains causes the slight increase in the number of picky treatments shown in figures \ref{fig:hPoDM_1_picky} and \ref{fig:hExhPoDM_picky}. As is clear also from figure \ref{fig:num_strain_strata_avg_compositions}, there is little change in the number of empty droplets with zero strains regardless of the level of compositional heterogeneity.

{

    \newcommand{\figcaption}{Picky groups for high heterogeneity distributions.}
  \newcommand{\figlabel}{fig:high_het_picky}

\begin{figure}[p]

  \begin{subfigure}{\textwidth}
      \centering
  \includegraphics[width=\textwidth,height=0.45\textheight,keepaspectratio]{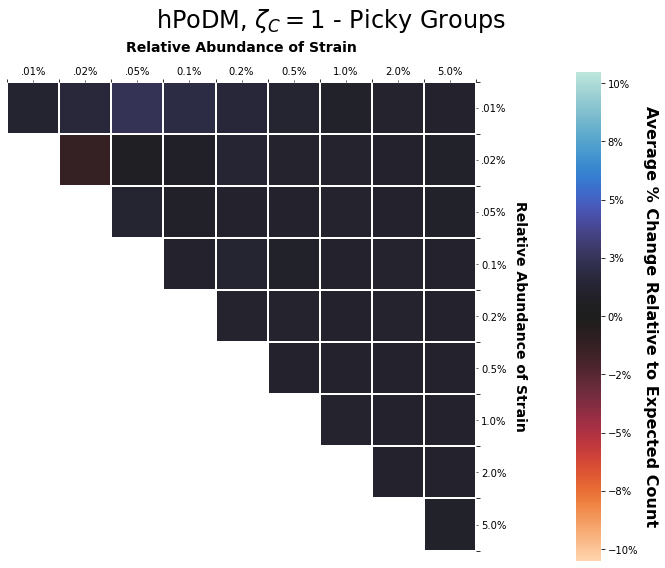}
  \caption[]{}
  \label{fig:hPoDM_1_picky}
\end{subfigure}

\begin{subfigure}{\textwidth}
    \centering
  \includegraphics[width=\textwidth,height=0.45\textheight,keepaspectratio]{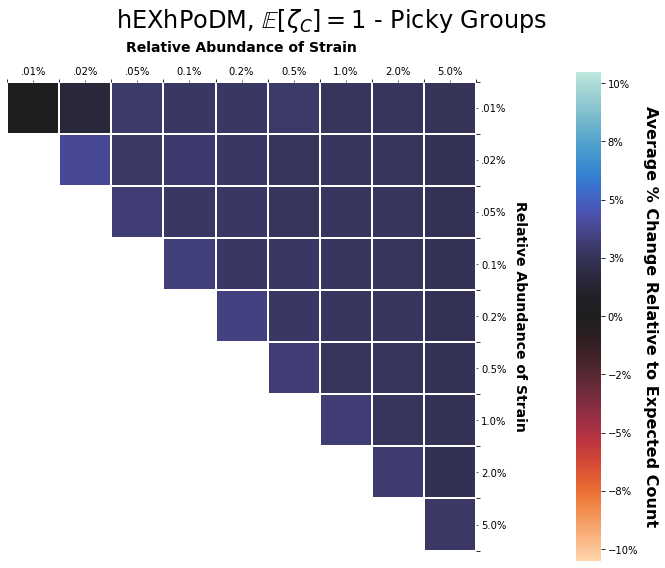}
  \caption[]{}
  \label{fig:hExhPoDM_picky}
\end{subfigure}

\caption{\figcaption}
\label{\figlabel}

\end{figure}

\begin{figure}[p]
  \ContinuedFloat
  
  \begin{subfigure}{\textwidth}
  \centering
    \includegraphics[width=\textwidth,height=0.45\textheight,keepaspectratio]{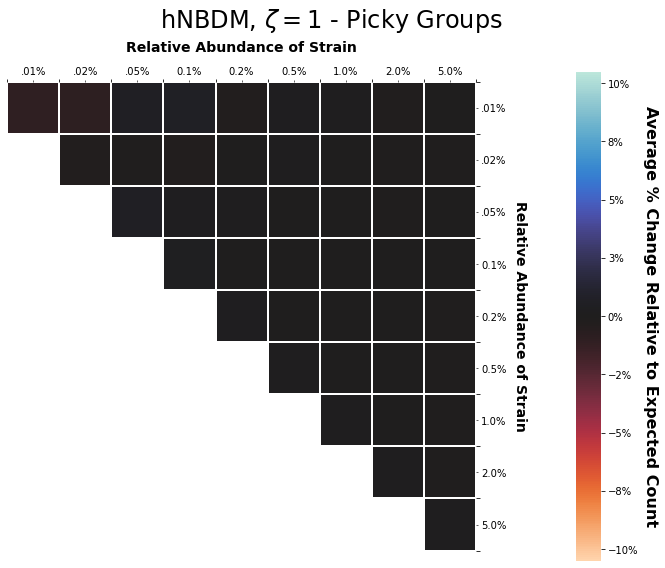}
  \caption[]{}
  \label{fig:hNBDM_1_picky}
\end{subfigure}

  \begin{subfigure}{\textwidth}
  \centering
    \includegraphics[width=\textwidth,height=0.45\textheight,keepaspectratio]{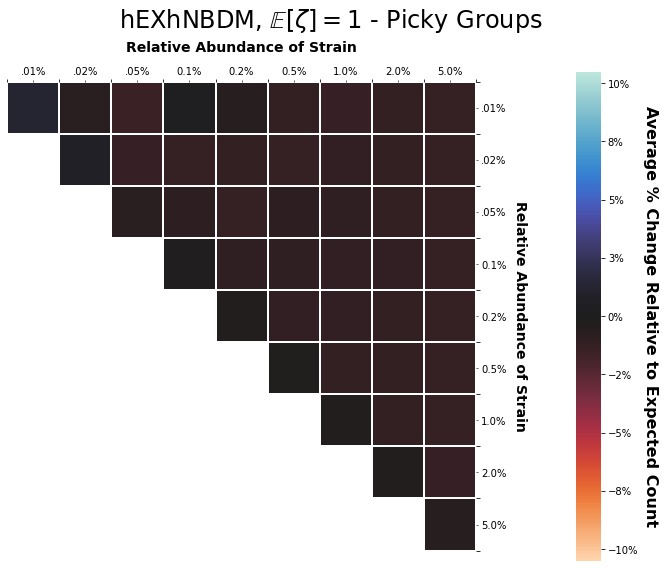}
  \caption[]{}
  \label{fig:hExhNBDM_picky}
\end{subfigure}

\caption[]{\figcaption}
\label{\figlabel}
 
\end{figure}

}  %

\paragraph{Increased Compositional and Density Heterogeneity}
\label{sec:incr-comp-dens}

In the presence of density heterogeneity, as seen in figure \ref{fig:num_strain_strata_avg_changes} the number of droplets with a single strain does still increase slightly. However, in contrast to what happens in the absence of density heterogeneity, in the presence of density heterogeneity there is an observable decrease in the number of droplets with two strains. While this trend is less obvious based only on figure \ref{fig:num_strain_strata_avg_changes}, it is clear based on the observed decrease in the number of picky treatments visible in figures \ref{fig:hNBDM_1_picky} and \ref{fig:hExhNBDM_picky}. Nevertheless it is also clear from figure \ref{fig:num_strain_strata_avg_changes} and from a comparison of figures \ref{fig:hExhNBDM_picky} and \ref{fig:hExhNBDM_gluttonous} (and \ref{fig:hNBDM_1_picky} and \ref{fig:hNBDM_1_gluttonous}) that the decrease in the number of droplets with two strains is not as pronounced (at least in relative terms) as the decrease in the number of droplets with three or more strains. Also in contrast to what happens in the absence of density heterogeneity, in the presence of density heterogeneity figures \ref{fig:num_strain_strata_avg_compositions} and \ref{fig:num_strain_strata_avg_changes} show a marked increase in the number of empty droplets with zero strains as the compositional heterogeneity increases.

\section{Discussion}
\label{sec:data_throughput_discussion}

Section \ref{sec:overview-groups-distr-conting} discusses the taxonomy of four groups of distributions identified in section \ref{sec:data_throughput_results}. Section \ref{sec:very-diff-distr-1-conting} discusses the distributions which are most likely to exhibit differences from hPoMu that are important in practice. Section \ref{sec:effect-heter-data} discusses how the interpretation of the results from section \ref{sec:data_throughput_results} depends fundamentally on our opinion of which droplets constitute ``good data''. Section \ref{sec:next-steps-1} discusses how the results from section \ref{sec:data_throughput_results} indicate what the next step is in order to recover true interaction networks from MOREI data.

\subsection{Overview of Groups of Distributions}
\label{sec:overview-groups-distr-conting}

The seven simulated distributions can be grouped into roughly four levels of assumption violation. 

For the first two levels, the first consisting of hTPMH and the second consisting of hPoDM $\cconcentration=100$ and hNBDM $\concentration=100$, hPoMu is most likely an adequate working model. Cf. again section \ref{sec:very-simil-distr-1} from the previous chapter. 

The third level consists of hPoDM $\cconcentration=1$ and hNBDM $\concentration=1$, which produce data that can be relatively easily distinguished from data produced by hPoMu. Cf. again section \ref{sec:slightly-diff-distr-1}.

In all cases, the results using the contingency tables for the target parameters and resulting $\chi^2$-divergences do nothing to change the qualitative conclusions that were made in the previous chapter. Similarity (or lack thereof) of these working models, as measured by log-likelihood ratios, corresponds to similarity (or lack thereof) of their contingency tables of target parameters.

The fourth level consists of hExhPoDM $\mathbb{E}[\cconcentration] =1$ and hExhNBDM $\mathbb{E}[\concentration] =1$, which produce data so substantially different from that produced by hPoMu that it is almost impossible to \textit{not} distinguish. It is very difficult to believe that hPoMu could possibly serve as an adequate working model for such data.

\subsection{Very Different Distributions}
\label{sec:very-diff-distr-1-conting}

The simulated hExhPoDM and hExhNBDM distributions technically do not belong to the ghNBDM family of distributions, since their concentration ``parameters'' do not have fixed values, and instead are exponentially distributed. This distinction will be particularly relevant in sections \ref{sec:results-are-qual} and \ref{sec:behav-ml-estim}.

Moreover, although the mean of these exponential distributions is $1$, the behavior of the simulated hExhPoDM with $\mathbb{E}[\cconcentration] =1$ and hExhNBDM with $\mathbb{E}[\concentration] =1$ distributions is substantially different from that of the hPoDM $\cconcentration=1$ and hNBDM $\concentration=1$ distributions. This is most likely because concentration parameter values sampled from the $\operatorname{Exponential}(1)$ are $1 - \frac{1}{e} \approx 63.2\%$ of the time smaller than $1$. (The median of the distribution is $\ln (2) \approx .693 < 1$.) 

Thus the simulated hExhPoDM with $\mathbb{E}[\cconcentration] =1$ and hExhNBDM with $\mathbb{E}[\concentration] =1$ can be assumed a priori to resemble more closely hPoDM and hNBDM distributions with values of $\cconcentration$ and $\concentration$ (substantially) smaller than $1$. This corresponds to what we see in practice, because heterogeneities, as measured by over-dispersion, are inversely proportional to the concentration parameters. So for example $\cconcentration = \ln (2)$ would correspond to an hPoDM distribution with $\approx 44\%$ more heterogeneity than the hPoDM $\cconcentration=1$ distribution. 

Considering that the simulated hExhPoDM with $\mathbb{E}[\cconcentration] =1$ and hExhNBDM with $\mathbb{E}[\concentration] =1$ correspond $\approx 63.2\%$ to distributions with heterogeneities no less than those of hPoDM $\cconcentration=1$ and hNBDM $\concentration=1$, and $1 - \frac{1}{\sqrt{e}} \approx 39.3\%$ of the time to distributions with heterogeneities no less than \textit{twice} those of hPoDM $\cconcentration=1$ and hNBDM $\concentration=1$, it is unsurprising that the aggregate/average heterogeneities observed for these distributions is much larger than those observed for either the hPoDM $\cconcentration=1$ or hNBDM $\concentration=1$ distributions. 

Therefore fact that data generated by these two distributions is very unlikely to be adequately modeled by hPoMu further suggests that data generated by hPoDM or hNBDM distributions with $\cconcentration$ or $\concentration$ values not much smaller than $1$ will also be inadequately modeled by hPoMu. It is a priori unclear which, if any, values of the concentration parameters smaller than $1$ are realistic in practice.

\subsection{The Effect of Heterogeneity on Data Throughput is Equivocal}
\label{sec:effect-heter-data}

The general effect that heterogeneities will have on the data throughput of MOREI is not straightforward. The answer depends on the particular combination of density heterogeneity and compositional heterogeneity present. Given both density and compositional heterogeneity, it is clear that the effect on data throughput is negative. Both the numbers of picky treatments and gluttonous treatments decrease. 

In the presence of only compositional heterogeneity but no density heterogeneity, whether the effect on data throughput is positive depends primarily on our opinion of what is useful data. If our primary concern is avoiding ``data starvation'' and being able to say anything about the interactions of the rarest strains, then the strong decrease in the number of gluttonous treatment groups clearly means that the effect of heterogeneity is negative. On the other hand, if we consider only the data from the picky treatments useful, then in the absence of density heterogeneity the effect of compositional heterogeneity is slightly positive.

\section{Conclusion}
\label{sec:conclusion-5}

\paragraph{Findings and Contributions}

I confirmed that more severe failures of the assumptions from section \ref{sec:impl-model-assumpt} cause more severe discrepancies with the predictions derived from hPoMu.
The nature of the effect depends on the chosen grouping (picky or gluttonous) of droplets defining the targeted estimands. This demonstrates that failures of the homogeneity assumptions of hPoMu potentially could be severe enough to require attention in practice.

\paragraph{Practical Implications}

The practical implications of this chapter remain limited by ignorance of which values of density and compositional concentration parameters are realistic in practice. If realistic values correspond to sufficiently small violations of the homogeneity assumptions, then this chapter actually provides evidence that such violations could be ignored in practice. Likewise, if realistic values correspond to sufficiently large violations of the homogeneity assumptions, then this chapter provides evidence of the opposite.

\paragraph{Next Steps and Open Questions}
\label{sec:next-steps-1}

What is also clear is that, if we wish to be able to predict generally what effects heterogeneities may have on the data throughput of MOREI in practice, we must be able to estimate both density heterogeneity and compositional heterogeneity from empirical data. This is the topic of chapter \ref{chap:hetero_estimator_performance}. Given such estimates from real-world data, we can get some sense for the amount of either type of heterogeneity one is likely to encounter in practice. That makes it possible to create more accurate simulations. More accurate simulations in turn better assess which inference methods are most capable of recovering true interaction networks from MOREI data.

\begin{coolsubappendices}
\section{$p$-Values from $\chi^2$ Approximation}
\label{sec:p-values-from}

Even for 15 million droplets, some categories corresponding to the rarest strains had very small expected counts ($<5$). Thus the $\chi^2$ approximation for the sampling distribution of the categorical divergence under the null distribution was unlikely to be accurate. In contrast, the Monte Carlo approximation to the null distribution is from $10^9$ (one \textbf{b}illion) replicates. Therefore the $p$-values from the Monte Carlo approximation are more likely to be representative of the true $p$-values. For completeness, here I report the $p$-value distributions computed from the $\chi^2$ approximation. While similar to the $p$-values from the Monte Carlo approximation, they appear to over-estimate the differences from the null.

{ 

  \newcommand{\figcaption}{~Approximate goodness of fit $p$-value distributions using $\chi^2$ approximation.}
  \newcommand{\figlabel}{fig:pvals_chi2}
  
\begin{figure}[p]
  \begin{subfigure}{\textwidth}
  \centering
\includegraphics[width=\textwidth,height=0.287\textheight,keepaspectratio]{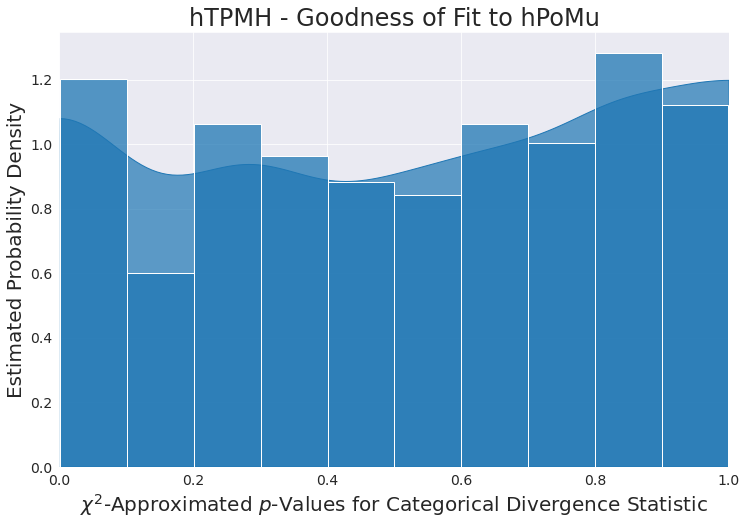}  
  \caption[]{}
  \label{fig:hTPMH_pvals_chi2}
\end{subfigure}

  \begin{subfigure}{\textwidth}
  \centering
\includegraphics[width=\textwidth,height=0.287\textheight,keepaspectratio]{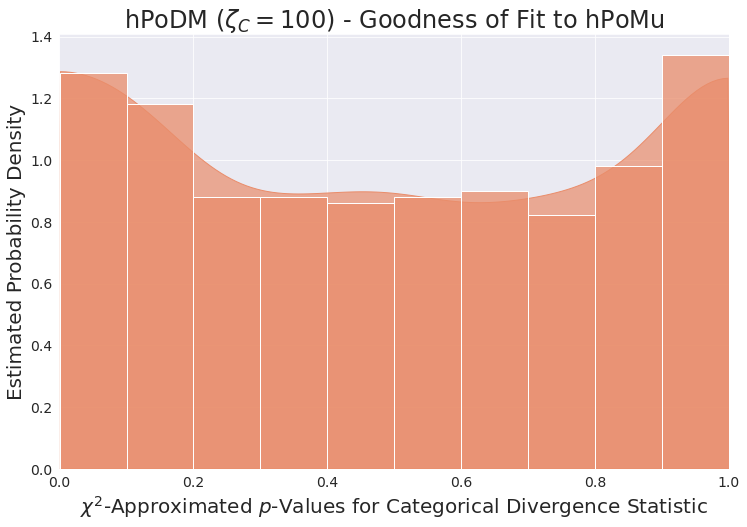}
  \caption[]{}
  \label{fig:hPoDM_100_pvals_chi2}
\end{subfigure}

  \begin{subfigure}{\textwidth}
  \centering
\includegraphics[width=\textwidth,height=0.287\textheight,keepaspectratio]{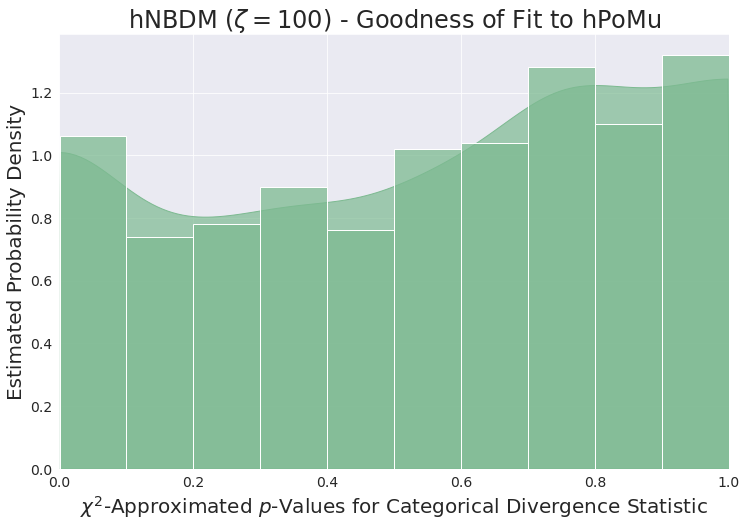}
  \caption[]{}
  \label{fig:hNBDM_100_pvals_chi2}
\end{subfigure}

\caption{\figcaption}
\label{\figlabel}

\end{figure}

} %

\section{Additional Pearson Categorical Divergence Distributions}
\label{sec:categ-diverg-distr}

Here are two additional plots, one focusing on those distributions with compositional heterogeneity only, and the other on those distributions with both heterogeneities.

\begin{figure}[p]
  
  \begin{subfigure}{\textwidth}
  \centering
\includegraphics[width=\textwidth,height=0.47\textheight,keepaspectratio]{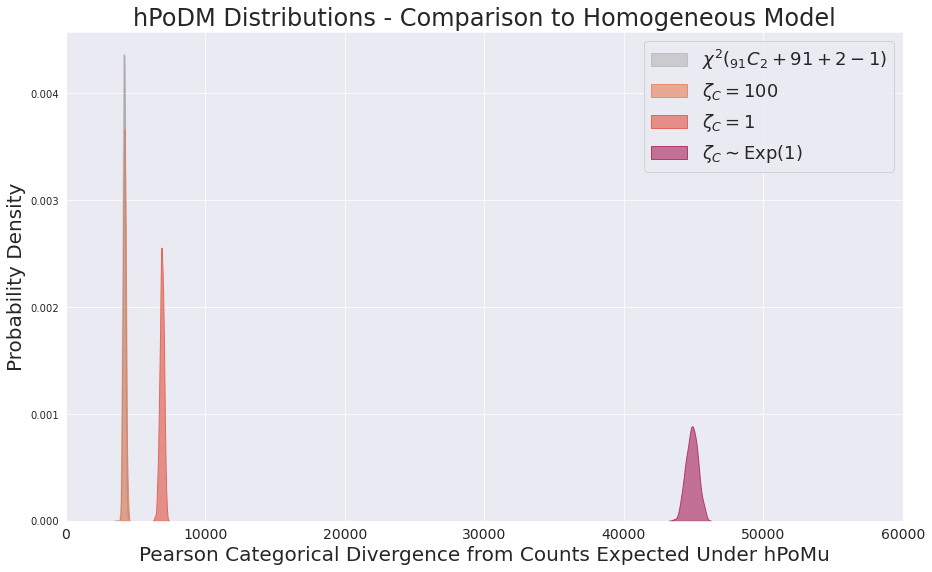}  
  \caption[]{}
  \label{fig:divergences_hPoDM}
\end{subfigure}

  \begin{subfigure}{\textwidth}
  \centering
\includegraphics[width=\textwidth,height=0.47\textheight,keepaspectratio]{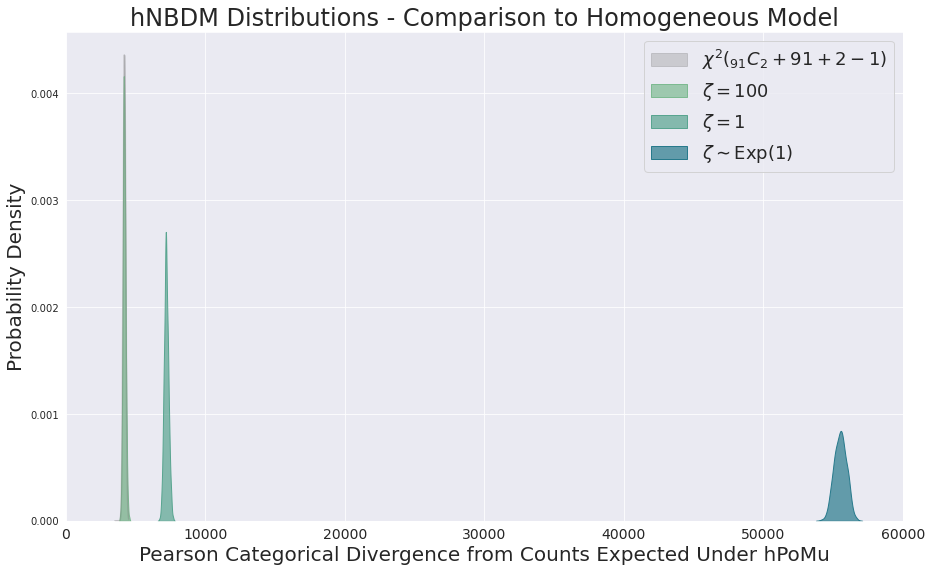}
  \caption[]{}
  \label{fig:divergences_hNBDM}
\end{subfigure}

\end{figure}

\clearpage

\section{Multiple Representatives Problem}
\label{sec:append-mult-repr}

Another potential challenge arising from the randomness of the initial droplet formation process is the ``multiple representatives problem'', where within a given droplet there is initially a strain that is represented by multiple cells. Cf. figure \ref{fig:multiple_representatives}. Even if we were more concerned about the effects randomness of the initial droplet formation process causes via the ``multiple representatives problem'' than the effects it has on data throughput, I argue we should still first focus on the effects it has on data throughput, for two reasons. 

\begin{figure}[p]
  \centering
  \includegraphics[width=\textwidth,height=\textheight,keepaspectratio]{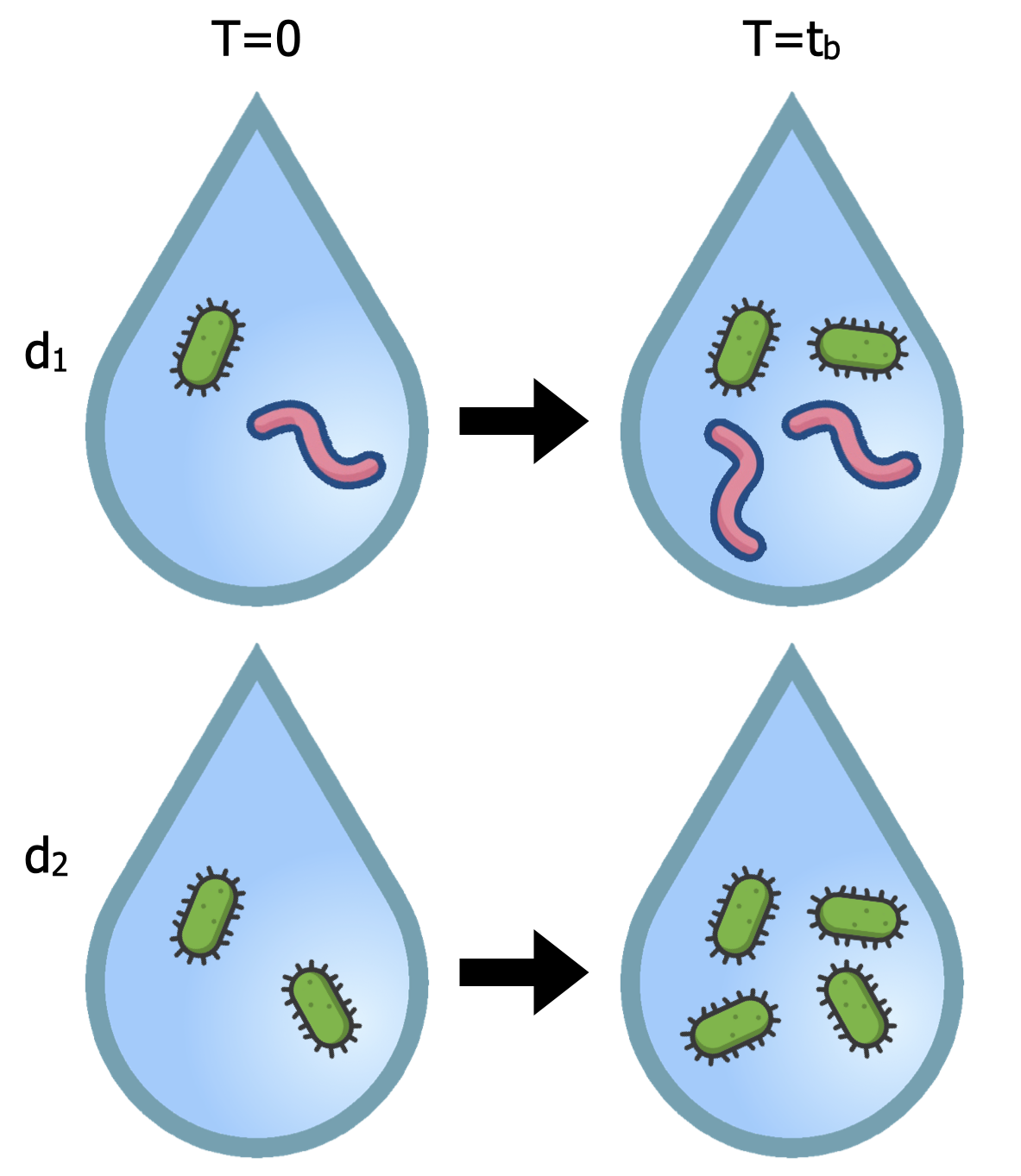}
  \caption[~The multiple representatives problem.]{Multiple Representatives Problem. The droplets are only observed at time $\time_{\batch}$. If we incorrectly assume both droplets began with the same number of green cells, then we incorrectly conclude that the pink cells halve the growth rate of the green cells. Yet the pink cells have no effect on the growth rate of the green cells. Cf. section \ref{sec:pract-limit-ident}.}
  \label{fig:multiple_representatives}
\end{figure}

First, to the extent that the ``multiple representatives problem'' might create ``noise'' or ``fluctuations'' in the MOREI data, these effects will on average ``cancel out'' if the data throughput is great enough to ensure large sample sizes for both the treatment and control groups. Cf. section \ref{sec:proofs} (in particular section \ref{sec:ident-fitness}) for an example of when this is true. Second, even to the extent the ``multiple representatives problem'' might create systematic errors which do not ``cancel out'' given large enough sample sizes, these systematic errors should become easier to identify, and therefore also easier to account for,  with greater data throughput. In either case, figuring out where we can and need to (or where we can't) account for effects caused by the ``multiple representatives problem'' begins with identifying the bottlenecks in data throughput that the particular form of randomness of the initial droplet formation process causes.

\end{coolsubappendices}
\end{coolcontents}

\chapter{Estimators for Density and Compositional Heterogeneities}
\label{chap:hetero_estimator_performance}

Herein I present both plugin and maximum likelihood estimators for failures of the assumptions from section \ref{sec:impl-model-assumpt}. (The failures correspond to the so-called ``concentration'' parameters defined in sections \ref{sec:model-comp-heter}, \ref{sec:model-dens-heter}, and \ref{sec:model-arbitr-comb}.)
See section \ref{sec:inference-methods}.
I show that failures of the assumptions from section \ref{sec:impl-model-assumpt} are understandable non-parametrically, despite the parametric definitions implicit from sections \ref{sec:model-comp-heter}, \ref{sec:model-dens-heter}, and \ref{sec:model-arbitr-comb}.
See sections \ref{sec:effect-dens-conc}, \ref{sec:effect-comp-conc}, \ref{sec:effect-comp-conc-1}, and \ref{sec:heter-might-corr} for arguments to this effect.
In particular, the estimators manage to still behave reasonably even under (mild) model misspecification.
See sections \ref{sec:results-are-qual} and \ref{sec:behav-ml-estim}.

Section \ref{sec:backgr-sign-6} provides context. Section \ref{sec:inference-methods} goes over technical details related to defining the estimators. Section \ref{sec:hetero_estimator_performance_methods} describes some implementation details of the analyses. Section \ref{sec:hetero_estimator_performance_results} explains the results of the analyses. Section \ref{sec:hetero_estimator_performance_discussion} interprets the results and explains how they are relevant to modelling the initial formation of droplets. Finally section \ref{sec:usef-heter-estim} concludes.

\begin{coolcontents}
\section{Background and Significance}
\label{sec:backgr-sign-6}

\paragraph{Broader field}

The problem for this chapter belongs to the broader field of point estimation.
See for example any of \cite{lehmann_cassella}, \cite{Keener2010}, \cite{serfling_1980}, or \cite{cramer_1946a} for general introductions to this field.
Cf. also the discussion earlier from the introduction to Part \ref{part:modell-init-form}.
This chapter considers the plug-in (also known as ``method of moments'') and maximum likelihood approaches to point estimation in particular.
See \cite[Example 4.7, p.456]{lehmann_cassella} for the definition and basic theory of plug-in estimators, which recommends \cite[Section 33.1]{cramer_1946a} and \cite[Section 4.3.1]{serfling_1980} for further reading.
Likewise, see \cite[Section 8.4]{Keener2010} or \cite[Section 6.3, p. 444]{lehmann_cassella} for the definition and basic theory of maximum likelihood estimators, the latter of which also recommends \cite[Sections 64.4 and 84.2]{strasser_1985} and \cite{efron_1982a} for further reading.

\paragraph{Specific problem}

To account for effects of the randomness of the initial droplet formation process we need to make realistic predictions about what these effects will be.
Previous chapters showed that different statistical working models for the initial droplet formation process can make substantially different predictions.
Thus, to make realistic predictions, we need to evaluate which of these statistical working models is most realistic.
However, the more general statistical working models have additional parameters compared to hPoMu, and unlike the parameters in hPoMu these additional parameters are not obviously connected to the configuration of the experiment.
Hence, to evaluate which of these statistical working models is most realistic, we need to estimate these additional parameters from real data.

Therefore the specific problem for this chapter is to demonstrate that these additional parameters are identifiable from the observed data generated by unincubated droplets.

Using the hierarchical count-categorical framework (cf. section \ref{sec:hier-count-categ}), we can separate the estimation problems for the density concentration parameters and the compositional concentration parameters.
  Doing so, estimating the density concentration parameter for the ghNBDM family (cf. again section \ref{sec:model-arbitr-comb}) is the same as estimating the density concentration parameter for the negative binomial distribution, and estimating the compositional concentration parameter for the ghNBDM family is the same as estimating the compositional concentration parameter the Dirichlet-Multinomial.

Because of the use of spikein genes (cf. again section \ref{sec:prep-sequ}), we can assume that the observed data generated by unincubated droplets is (roughly) integer count data.
Plentiful previous work already exists about how to estimate the density concentration parameter (or ``equivalent''  parameters) of the negative binomial distribution from (univariate) count data.
See for example any of
  \cite{Anscombe1950}, \cite{Shenton1962}, \cite{Willson1984}, \cite{Bowman1984}, \cite{RossPreece1985}, \cite{Willson1986},
  \cite{ClarkPerry1989}, \cite{Piegorsch1990}, \cite{Aragon1992}, \cite{Nakashima1997}, \cite{LloydSmith2007}, or \cite{Yu2013}.
Less previous work seems to exist regarding estimation of the compositional concentration parameter of the Dirichlet-Multinomial distribution from multivariate count data. See e.g. 
\cite{Sklar}, \cite{Yu2014}, \cite{Bouneffouf}, or \cite{Harrison2020}.

\paragraph{Particular approach}

I derive plugin and maximum likelihood estimators for these additional parameters and demonstrate their effectiveness in practice. Plugin and maximum likelihood estimation are probably the most widely employed approaches for point estimation. Therefore employing them is a useful starting point for this problem.

The maximum likelihood estimators are more efficient when the models are correctly specified. (This is unsurprising in light of general results about the asymptotic optimality of maximum likelihood estimation; cf. \cite[Chapter 6]{lehmann_cassella}.) Nevertheless the difference in efficiency is small for the density concentration parameter, confirming previous reports \cite{Bowman1984}\cite{Nakashima1997} of the somewhat pathological nature of that particular estimation problem.

Previous work
  \cite{ClarkPerry1989}  \cite{Nakashima1997}
  has used maximum pseudo-likelihood or maximum quasi-likelihood approaches for the density concentration of the negative binomial distribution. However, such approaches are not as standard as plugin (``method of moments'') or maximum likelihood estimation. Moreover, there does not appear to be any consensus in the literature that those approaches actually perform better.
Similar reasoning also applies for choosing not to use the methods of neither \cite{Willson1984} (density concentration of negative binomial) nor \cite{Yu2014} (compositional concentration of Dirichlet-Multinomial).

 This chapter also attempts to isolate non-parametric definitions which correspond to the concentration parameters in the special cases of these parametric working models. Being able to characterize these as non-parametric estimands (more precisely pathwise differentiable functionals definable on non-parametric statistical models) will facilitate future work applying targeted minimum-loss based estimation (TMLE) \cite{vanderLaan2011} to this problem.

\section{Inference of Concentration Parameters: ML and Plugin Approaches}
\label{sec:inference-methods}

Using the hierarchical count-categorical framework defined in section \ref{sec:hier-count-categ}, I treat the estimation of the density concentration $\dconcentration$ and compositional concentration $\cconcentration$ parameters of the ghNBDM family (cf. section \ref{sec:model-arbitr-comb}) as two separate estimation problems. The first seeks to estimate the density concentration $\dconcentration$ from the empirical count distribution. The second seeks to estimate the compositional concentration $\cconcentration$ from the empirical categorical distributions\footnote{Distribution\textit{\textbf{s}} (plural) because again technically speaking there is a different categorical distribution for each possible value of the count distribution.}.

For both estimation problems, I employ two strategies: (1) plugin (``method of moments'') estimation, and (2) maximum likelihood (``ML'') estimation. Both approaches give statistically consistent and asymptotically normal estimators. The maximum likelihood strategy has the advantage of being more efficient and asymptotically optimal (in the sense of satisfying the Cramer-Rao lower bound). The plugin strategy has the advantage of being computationally simpler (and thus also easier for non-specialists to implement). I argue in sections \ref{sec:effect-dens-conc}, \ref{sec:effect-comp-conc}, and \ref{sec:heter-might-corr} that the plugin strategy also has the advantage of being ``semi-parametric''\footnote{\label{footnote:plugin_robustness}
Section 4 of \cite{Nakashima1997} argues (at least for density concentration) that the plugin estimator is also ``semi-parametric'' when compared to the MLE due to having less bias and standard error. This is because the plugin estimator retains consistency (and supposedly also efficiency) for any distribution with the same overdispersion relative to the Poisson while the ML estimator does not. Cf. sections \ref{sec:over-dispersion} and \ref{sec:over-disp-negat} of this work with formula (2.3) and section 4 of \cite{Nakashima1997}. This work does not investigate the estimators' behavior under model misspecification enough to evaluate this robustness claim. The true values of the effective density concentration (cf. section \ref{sec:effect-dens-conc}) and the effective compositional concentration (cf. section \ref{sec:effect-comp-conc}) are never computed for the misspecifed models. Moreover the misspecified models (hExhPoDM and hExhNBDM) are only slight modifications of the ghNBDM family. An investigation of the estimators' behavior under model misspecification thorough enough to make a legitimate and informed judgment about this issue is a worthwhile subject for future work.
}, at least in the sense that its estimand is still straightforward to interpret even under model misspecification, although this is debatable.

Sections \ref{sec:plug-estim-dens} and \ref{sec:mle-dens} discuss the estimation problem corresponding to the count distributions and the density concentration parameter $\dconcentration$. This is formulated as a parametric estimation problem involving the negative binomial distribution. Section \ref{sec:plug-estim-dens} provides details of the plugin strategy, while section \ref{sec:mle-dens} provides details of the ML strategy.

Sections \ref{sec:plug-estim-comp} and \ref{sec:mle-comp} discuss the estimation problem corresponding to the categorical distributions and the compositional concentration parameter $\cconcentration$. This is formulated as a univariate parametric estimation problem involving the Dirichlet-Multinomial distribution. The remaining nuisance parameters (the strain\footnote{Herein I use ``strains'' to refer equally to strains belonging to the same species(/genus/family/etc.) as well as to strains belonging to different species(/genera/families/etc.), because the distinction is irrelevant for setting up the abstract problem. It may matter for the implementation of a specific experiment.} frequencies) can be assumed to be either already known or estimated separately (cf. section \ref{sec:append-plug-estim}). Section \ref{sec:plug-estim-comp} provides details of the plugin strategy, while section \ref{sec:mle-comp} provides details of the ML strategy.

\subsection*{Notation}
\label{sec:notation_estimators}

The notation $\indicator{A}$ denotes the indicator function for the event $A$.

Given any estimand which is a conditional expectation, i.e. of the form $  \expectation{ X | A}$ for some random variable $X$ and some event $A$ in the underlying sigma-algebra, the corresponding empirical (arithmetic) mean is denoted
\begin{equation}
  \label{eq:conditional_empirical_mean}
  \mean{X | A} :=  \frac{\sum_{\droplet \in [\Droplets]} X_{\droplet} \indicator{A} }{  \sum_{\droplet \in [\Droplets]} \indicator{A} } \,.
\end{equation}
In particular the unconditional empirical mean is
\begin{equation}
  \label{eq:empirical_mean}
  \mean{X} := \frac{1}{\Droplets} \sum_{\droplet \in [\Droplets]} X_{\droplet} \,.
\end{equation}
Analogously a conditional variance estimand $\var{X|A}$ has a corresponding empirical conditional variance:
\begin{equation}
  \label{eq:conditional_empirical_variance}
  \evar{X | A} :=  \frac{\sum_{\droplet \in [\Droplets]} (X_{\droplet} - \mean{X|A})^2 \indicator{A} }{  \sum_{\droplet \in [\Droplets]} \indicator{A} } \,,
\end{equation}
and an unconditional variance estimand $\var{X}$ has the empirical counterpart:
\begin{equation}
  \label{eq:empirical_variance}
    \evar{X} := \frac{1}{\Droplets} \sum_{\droplet \in [\Droplets]} (X_{\droplet} - \mean{X})^2 \,.
\end{equation}
Generalizing the above, for a conditional \textit{co}variance estimand $\cov{X, Y | A}$ its empirical counterpart is denoted and defined as
\begin{equation}
  \label{eq:conditional_empirical_covariance}
    \ecov{X, Y | A} :=  \frac{\sum_{\droplet \in [\Droplets]} (X_{\droplet} - \mean{X|A})(Y_{\droplet} - \mean{Y|A}) \indicator{A} }{  \sum_{\droplet \in [\Droplets]} \indicator{A} } \,,
\end{equation}
and likewise for an unconditional covariance estimand $\cov{X,Y}$ its empirical counterpart is denoted and defined as
\begin{equation}
  \label{eq:empirical_covariance}
      \ecov{X, Y} := \frac{1}{\Droplets} \sum_{\droplet \in [\Droplets]} (X_{\droplet} - \mean{X})(Y_{\droplet} - \mean{Y}) \,.
\end{equation}

\subsection{Plugin Estimator for Density Concentration}
\label{sec:plug-estim-dens}

Section \ref{sec:deriv-plug-estim} derives the formula for the plugin estimator. Section \ref{sec:effect-dens-conc} discusses how the plugin strategy implies an estimand that is defined for count distributions more general than the negative binomial distribution alone.

\subsubsection{Derivation of Plugin Estimator for Density Concentration}
\label{sec:deriv-plug-estim}

Recall that when $\abundance(0)$ is negative binomial distributed\footnote{Using the parametrization for the negative binomial described in section \ref{sec:negative-binomial-nb}.}, the variance equals
\begin{equation}
  \label{eq:NB_variance_review}
  \var{\abundance(0)} = \rate \left(  1 + \frac{\rate}{\dconcentration \Species} \right) \,,
\end{equation}
while the variance that would have been expected under the Poisson distribution equals
\begin{equation}
  \label{eq:Po_variance_for_NB}
  \expectation{\abundance (0)} = \rate \,.
\end{equation}
Thus when $\abundance(0)$ is negative binomial distributed its over-dispersion relative to the Poisson distribution equals
\begin{equation}
  \label{eq:NB_overdispersion_with_NB_moments}
  \frac{\var{\abundance(0)} - \expectation{\abundance(0)}}{\expectation{\abundance(0)}} = \frac{\rate}{\dconcentration \Species} = \frac{\expectation{\abundance(0)} }{\dconcentration \Species} \,.
\end{equation}
Solving for $\dconcentration$ leads to
\begin{equation}
  \label{eq:dconcentration_formula_NB}
  \dconcentration = \frac{1}{\Species} \cdot \frac{ \left(\expectation{\abundance(0)}\right)^2 }{ \var{\abundance(0)}  - \expectation{\abundance(0)}} \,.
\end{equation}
The above relationship (\ref{eq:dconcentration_formula_NB}) motivates the following plugin estimator for $\dconcentration$:
\begin{equation}
  \label{eq:dconcentration_plugin}
\hat{\dconcentration} :=
  \begin{cases}
    \infty & \evar{\abundance(0)} \le \mean{\abundance(0)} 
\\
\frac{1}{\Species} \frac{\left(\mean{\abundance(0)}  \right)^2}{\evar{\abundance(0)} - \mean{\abundance(0)}} & \text{else}
  \end{cases} \,.
\end{equation}
Cf. the derivation of the analogous equation (3.10) in \cite{Anscombe1950}.

\subsubsection{Effective Density Concentration}
\label{sec:effect-dens-conc}

 If one defines for any possible distribution of $\abundance(0)$ with finite variance (e.g. including the Poisson) the ``effective density concentration'' as
\begin{equation}
  \label{eq:effective_dconcentration}
{\dconcentration}_{e\!f\!f} :=
  \begin{cases}
    \infty & \var{\abundance(0)} \le \expectation{\abundance(0)}
\\
\frac{1}{S}\frac{\left( \expectation{\abundance(0)} \right)^2  }{ \var{\abundance(0)} - \expectation{\abundance(0)} } & \text{else}
  \end{cases} \,,
\end{equation}
then the above plugin estimator (\ref{eq:dconcentration_plugin}) is always consistent for the ``effective density concentration'' ${\dconcentration}_{e\!f\!f}$. Consistency follows from the weak law of large numbers and the continuous mapping theorem for convergence in probability.

The effective density concentration ${\dconcentration}_{e\!f\!f}$ being infinite merely flags that the distribution of $\abundance(0)$ is not over-dispersed relative to the Poisson. An obvious limitation of this is that provides no quantification of the extent to which the distribution might be \textit{under}-dispersed relative to the Poisson. Such under-dispersed distributions are not necessarily ``exotic'', consider for example the distribution of $Y = \frac{1}{2}X$ when $X \sim \poisson$. Then $\var{Y} = \left(\frac{1}{2}\right)^2 \var{X} = \frac{1}{4}\expectation{X} < \frac{1}{2}\expectation{X} = \expectation{Y}$. This of course works also for $pX$ for any $0 < p <1$. Considering the possible ramifications of under-dispersion is left to future work.

When the distribution of $\abundance(0)$ \textit{is} over-dispersed relative to the Poisson, the effective density concentration measures the over-dispersion on a normalized and inverted scale. Because the effective density concentration ${\dconcentration}_{e\!f\!f}$ equals the density concentration parameter $\dconcentration$ whenever $\abundance(0)$ is negative binomial distributed, the estimand ${\dconcentration}_{e\!f\!f}$ may be considered a generalization of the parameter $\dconcentration$ to other distributions. This resembles the argument made in \cite[section 4]{Nakashima1997} that the negative binomial plugin estimator\footnote{\cite{Nakashima1997} considers a reparameterized version of the negative binomial distribution such that the corresponding plugin estimator is the reciprocal of that considered herein. This would correspond to an estimand that is the reciprocal of the effective density concentration.} is only ``semi-parametric''.

\subsection{Maximum Likelihood Estimator for Density Concentration}
\label{sec:mle-dens}

This section explains how to use the maximum likelihood strategy to estimate the density concentration $\dconcentration$. Section \ref{sec:negat-binom-log} computes the relevant likelihood and log likelihood for the count distributions, while section \ref{sec:negat-binom-score} computes the score. Section \ref{sec:negat-binom-maxim} uses those results to derive the maximum likelihood estimators for the negative binomial distribution. Sections \ref{sec:exist-uniq-finit} and \ref{sec:estim-diff} survey observations from previous literature about this estimator.

\subsubsection{Negative Binomial (Log) Likelihood}
\label{sec:negat-binom-log}

Starting from equation (\ref{eq:NB_pmf}),
\begin{itemize}
\item if we assume for mathematical convenience that the count distributions of all droplets are mutually independent,
\item using the identity\footnote{\label{footnote:gamma_interpolate}This follows from how $\Gamma$ interpolates the factorials, i.e. $\Gamma(x+1) = x\Gamma(x)$.} $\frac{\Gamma(x+M)}{\Gamma(x)} = \prod_{m=0}^{M-1} (x+m)$ for all positive integers $M$,
\end{itemize}
then the following is the full likelihood for a batch of $\Droplets$ droplets:
\begin{equation}
  \label{eq:NB_full_likelihood}
  \begin{split}
  \likelihood{\rate, \dconcentration}
 = &
\prod_{\droplet \in [\Droplets]} \!
\left[
\left(
\prod_{\nu=0}^{\counts_{\droplet}-1}\! (\dconcentration\Species + \nu)\!
\right)    
\!\cdot\!
\left(\!
1 + \frac{\rate}{\dconcentration\Species}
\!\right)^{
-\dconcentration\Species
} \!\cdot\!
\left(\!
\frac{1}{
 \dconcentration\Species + \rate
}
\!\right)^{\counts_{\droplet}}
\!\cdot\!
\frac{\rate^{\counts_{\droplet}}}{
\counts_{\droplet}!
}
\right] 
\,.
  \end{split}
\end{equation}
(Recall that $[\Droplets]:=\{1,\dots, \Droplets\}$.) Thus this is the log likelihood:
\begin{equation}
  \label{eq:NB_full_log_likelihood}
  \begin{split}
   \loglikelihood{\rate, \dconcentration} := & \\
\log \left(
\likelihood{\rate, \dconcentration}
\right) = &
\sum_{\droplet \in [\Droplets]}
\!
\left[
\sum_{\nu=0}^{\counts_{\droplet}-1}
\log\left(
\dconcentration\Species + \nu
\right)
\right]
\!-\!
\Droplets\dconcentration\Species
\log\left(
1 + \frac{\rate}{\dconcentration\Species}
\right)
+ \\
&
\Droplets \overline{\counts}
\log \left(
\frac{1}{
\dconcentration\Species + \rate
}
\right)
+
\Droplets \overline{\counts} \log (\rate)
-
\sum_{\droplet \in [\Droplets]}
\log\left( \counts_{\droplet} ! \right) \,,
  \end{split}
\end{equation}
where I have defined $\overline{\counts} := \frac{1}{\Droplets} \sum_{\droplet \in [\Droplets]} \counts_{\droplet}$. (Each variable $\counts_{\droplet}$ corresponds to a possible value of the random variable $\abundance_{\droplet}(0)$, the number of cells in droplet $\droplet$.)

\subsubsection{Negative Binomial Score}
\label{sec:negat-binom-score}

It follows that the score with respect to $\rate$ is
\begin{equation}
  \label{eq:NB_score_rate}
\frac{
\partial \loglikelihood{\rate, \dconcentration}
}{
\partial \rate
}
=
  -\frac{
\Droplets\dconcentration\Species
}{
\dconcentration\Species + \rate
} 
+
\frac{\overline{\counts}}{\rate}
\cdot
\left(
\frac{
\Droplets\dconcentration\Species
}{
\dconcentration\Species + \rate
}
\right) \,.
\end{equation}
Similarly, from (very) tedious calculus we get the following expression for the score with respect to $\dconcentration$:
\begin{equation}
  \label{eq:NB_score_concentration}
  \begin{split}
\frac{\partial \loglikelihood{\rate, \dconcentration}}{
\partial \dconcentration
} = &
\sum_{\droplet \in [\Droplets]}
\!
\left[
\sum_{\nu=0}^{\counts_{\droplet}-1}
\frac{\Species}{
\dconcentration\Species + \nu
}    
\right]
+
\frac{
\Droplets\Species\rate
}{
\dconcentration\Species + \rate
}
-
\\
&
\Droplets\Species
\log \left(
1 +
 \frac{\rate}{\dconcentration\Species}
\right)
-
\frac{
\Droplets \Species \overline{\counts}
}{
\dconcentration\Species + \rate
} \,.
  \end{split}
\end{equation}

\subsubsection{Negative Binomial Maximum Likelihood Estimator}
\label{sec:negat-binom-maxim}

Setting $\frac{
\partial \loglikelihood{\rate, \dconcentration}
}{
\partial \rate
}
=0$, equation (\ref{eq:NB_score_rate}) gives us that the maximum likelihood estimator $\hat{\rate}$ for $\rate$ is $\hat{\rate} = \overline{\counts}$. This means that the plugin and maximum likelihood estimators for $\rate$ coincide.

  Setting $\frac{
\partial \loglikelihood{\rate, \dconcentration}
}{
\partial \dconcentration
}
=0$, and substituting into (\ref{eq:NB_score_concentration}) the relationship $\rate = \hat{\rate} = \overline{\counts}$, we get that, if the maximum likelihood estimator $\hat{\dconcentration}$ of $\dconcentration$ exists, then it must be a solution of the following equation:
\begin{equation}
  \label{eq:NB_dconcentration_MLE_eq}
  \Droplets \log \left(
1 + \frac{
\overline{\counts}
}{
\hat{\dconcentration} \Species
}
\right)
=
\sum_{\droplet \in [\Droplets]}
\!
\left[
\sum_{\nu=0}^{\counts_{\droplet}-1}
\frac{1}{
\hat{\dconcentration}\Species + \nu
} 
\right]
\,.
\end{equation}
Rearranging terms for the purposes of computational efficiency, the equation (\ref{eq:NB_dconcentration_MLE_eq}) above can be rewritten
\begin{equation}
  \label{eq:NB_dconcentration_MLE_eq_efficient}
  \Droplets \log \left(
1 + \frac{
\overline{\counts}
}{
\hat{\dconcentration} \Species
}
\right)
=
\sum_{m=1}^M
 \frac{\Droplets_{\ge m}}{
\hat{\dconcentration}\Species + (m-1)
}   \,,
\end{equation}
where $M := \max_{\droplet \in [\Droplets]} \abundance_{\droplet}(0)$ and $\Droplets_{\ge m} := \left| \left\{ \droplet: \abundance_{\droplet}(0) \ge m \right\} \right|$.

\subsubsection{Existence, Uniqueness, and Finiteness Conditions}
\label{sec:exist-uniq-finit}

It is believed\cite{Anscombe1950}\cite{Willson1984}\cite{Willson1986} that a solution to the above equation (\ref{eq:NB_dconcentration_MLE_eq_efficient}) exists if and only if $\evar{\abundance(0)} > \mean{\abundance(0)}$ (i.e. if and only if the empirical distribution is over-dispersed with respect to the Poisson), and that if a solution exists it is always unique. A purported\footnote{The proof is probably correct, but I have inspected it only cursorily and not in detail. Therefore I feel I cannot legitimately attest to its validity firsthand.} proof of these claims is given in \cite{Aragon1992}.

Note how the above purported conditions for the existence and uniqueness of the negative binomial MLE are the same as the conditions for the finiteness of the plugin estimator. When the MLE does not exist, it is because the likelihood is increasing but has no maximum. Therefore in those situations it makes sense to define the MLE to be infinite (cf. \cite{Anscombe1950}). Thus the plugin estimator and MLE agree on when the estimate should be finite or infinite. Just like for the plugin estimator, the MLE being infinite has the interpretation that the Poisson distribution fits the data better than any negative binomial distribution.

\subsubsection{Estimation Difficulties}
\label{sec:estim-diff}

The two parameter (i.e. both $\rate$ and $\dconcentration$ unknown) estimation problem for the negative binomial may be intrinsically difficult due to there being no complete sufficient statistic, corresponding to how the two parameter negative binomial family is not an exponential family \cite{Willson1986}.

At least for (relatively) small samples \cite{Bowman1984}\cite{ClarkPerry1989}\cite{Piegorsch1990}\cite{LloydSmith2007}, previous work has raised the concern that the MLE has unusually large bias \cite{Shenton1962} \cite{Willson1984} \cite{Bowman1984} \cite{Piegorsch1990} \cite{LloydSmith2007}. However at least one source \cite{Piegorsch1990} conjectures that the MLE becomes a more viable option for large sample sizes.

The maximum likelihood estimation of $\dconcentration$ is believed to be particularly difficult for ``small'' values of $\rate$ and large values of $\dconcentration$ \cite{Shenton1962}\cite{Willson1984}\cite{Bowman1984}\cite{LloydSmith2007}. This may be due to the contours of the likelihood function being relatively ``flat'' or slowly increasing with respect to $\dconcentration$\cite{Willson1986}. However, other work \cite{ClarkPerry1989} has found that the small $\rate$ and large $\dconcentration$ regime is also the most difficult for other kinds of estimators of $\dconcentration$. \cite{Bowman1984} found that $\evar{\abundance(0)} \le \mean{\abundance(0)}$ occurred most often in the small $\rate$, large $\dconcentration$ regime, a condition for which both the plugin and ML estimators return infinite estimates (cf. section \ref{sec:exist-uniq-finit}).

Because of this, some authors have suggested that the MLE does not perform much better than the corresponding plugin estimator (despite having lower variance) \cite{Anscombe1950} \cite{Shenton1962} \cite{Willson1984} and thus that the plugin estimator should be preferred due to its greater computational simplicity \cite{Anscombe1950} \cite{Willson1984} \cite{Yu2013}. 

Other authors have suggested instead reparameterizing the negative binomial in terms of $\frac{1}{\dconcentration\Species}$ \cite{RossPreece1985}\cite{ClarkPerry1989}\cite{Piegorsch1990}\cite{Nakashima1997}\cite{LloydSmith2007}. This makes sense because the difficulties occur mostly for large values of $\dconcentration$ for which the resulting distributions are all very similar to the corresponding Poisson\cite{Bowman1984}\cite{LloydSmith2007} and because this removes the need for infinite estimates \cite{ClarkPerry1989}\cite{LloydSmith2007}. \cite{RossPreece1985} argue that reparameterizing the negative binomial using $\frac{1}{\dconcentration\Species}$ allows a continuous transition between the negative binomial, which is over-dispersed relative to the Poisson, and the (''positive'') binomial, which is under-dispersed relative to the Poisson. Nevertheless even the reparameterized MLE was not always necessarily found to behave much better than the reparameterized plugin estimator \cite{Piegorsch1990}. 

Exact formulae for the asymptotic variance of both the plugin and ML estimators were given in \cite{Anscombe1950}, and exact formulae for the asymptotic bias of both the plugin and ML estimators were given in \cite{Shenton1962}. (\cite{Nakashima1997} gives exact formulae for the asymptotic variances of the $\frac{1}{\dconcentration\Species}$ reparameterized versions of the plugin and ML estimators.) Because the asymptotic formulae of the plugin and ML estimators have ``the same general structure''\cite{Shenton1962} for both the bias and variance\footnote{Cf. Fig.3 of \cite{Shenton1962} and equations 3.6 and 3.11 of \cite{Anscombe1950}.}, the larger sample sizes typical for this problem may lead to the performance of the MLE and plugin estimator being comparable.

\subsection{Plugin Estimator for Compositional Concentration}
\label{sec:plug-estim-comp}

Section \ref{sec:deriv-plug-estim-1} outlines the big picture ideas behind the plugin estimator's somewhat complicated derivation. Section \ref{sec:solv-comp-conc} provides the details behind most of the derivation. Section \ref{sec:choice-conv-coeff} discusses a possible ambiguity and choices for resolving it. Section \ref{sec:plug-estim-defin} gives the formal definition of the final plugin estimator used for all later analyses. Section \ref{sec:effect-comp-conc} discusses how the plugin strategy implies an estimand that is defined for categorical distributions more general than the Dirichlet-Multinomial distribution alone.

\subsubsection{Overview of Derivation of Plugin Estimator for Compositional Concentration}
\label{sec:deriv-plug-estim-1}

The general idea is as follows:

\begin{itemize}
\item For each number of cells $\counts \ge 2$  and each tuple of strains $(\strain_1, \strain_2)$ get a plugin estimator for $\frac{1}{1 + \cconcentration}$.
\item Take a weighted average to get a single plugin estimator for $\frac{1}{1 + \cconcentration}$.
\item Solve to get a plugin estimator for $\cconcentration$.
\end{itemize}
The motivation for averaging before solving for $\cconcentration$ in the expression $\frac{1}{1+\cconcentration \Species}$ is because I anticipate the former operation to be ``better behaved'' than the reciprocals involved in the latter. Solving after averaging makes it possible to take reciprocals only once. This hopefully leads to a plugin estimator for $\cconcentration$ ``better behaved'' than plugin estimators derived by solving (multiple times) first and then averaging. Nevertheless the latter in principle is possible too.

Note that when $\counts=1$, the Dirichlet-Multinomial distribution reduces completely to the corresponding Multinomial (``Multinoulli'') distribution and any compositional concentration parameter is completely unidentifiable. (This corresponds to how $n-1=0$ in the over-dispersion formula, but follows ultimately from the probability mass function of the Dirichlet-Multinomial distribution.)

\subsubsection{Solving for Compositional Concentration Parameter}
\label{sec:solv-comp-conc}

Whenever $\expectation{\vabundance(0) | \abundance(0)}$ is Dirichlet-Multinomial distributed, for all fixed $n\ge 2$ one has, it follows from (\ref{eq:HDirMult_variance}) that for any strain $\strain$: 
\begin{equation}
  \label{eq:DirMult_variance_review}
  \frac{1}{1 + \cconcentration \Species} =\frac{\var{\abundance[\strain](0) | \abundance(0) = \counts } - \counts \freq^{(\strain)}(1 - \freq^{(\strain)})}{ \counts(\counts-1) \freq^{(\strain)}(1- \freq^{(\strain)})  } \,.
\end{equation}
Similarly, it follows from (\ref{eq:HDirMult_covariance}) that for any pair of strains $\strain_1$, $\strain_2$:
\begin{equation}
  \label{eq:DirMult_covariance_review}
  \frac{1}{1+ \cconcentration \Species} = \frac{-\cov{ \abundance[\specie_1](0), \abundance[\strain_2](0) | \abundance(0) = \counts }  - \counts \freq^{(\strain_1)}\freq^{(\strain_2)} }{
\counts(\counts-1)\freq^{(\strain_1)}\freq^{(\strain_2)}}  \,.
\end{equation}
Therefore, for any convex combination $\cvxcoeff{\strain[]}$, $\cvxcoeff{\strain[]_1, \strain[]_2}$ of the right hand sides of (\ref{eq:DirMult_variance_review}) and (\ref{eq:DirMult_covariance_review}) one has
\begin{equation}
  \label{eq:cvxcoeff_DirMult_cconcentration}
  \begin{split}
    \frac{1}{1+\cconcentration\Species} =& \frac{1}{\counts(\counts-1)} \left[
  \sum_{\strain[] \in [\Species]} 
\cvxcoeff{\strain[]} \left[ 
\frac{
\var{\abundance[{\strain[]}](0) }  - \counts \freq^{(\strain[])}(1- \freq^{(\strain[])})     
}{
\freq^{(\strain[])}(1-\freq^{(\strain[])})
} 
 \right] \right. \\
&\left. 
\vphantom{ \left[\sum_{\strain[] \in [\Species]} \frac{\var{\abundance[{\strain[]}](0) }  - \counts \freq^{(\strain[])}(1- \freq^{(\strain[])})     }{\freq^{(\strain[])}(1-\freq^{(\strain[])})} \right.}
+ \sum_{\strain[]_1 \in [\Species]} \sum_{\strain[]_2 \not= \strain[]_1}\cvxcoeff{\strain[]_1, \strain[]_2}
\left[
\frac{
-\cov{\abundance[{\strain[]_1}](0),  \abundance[{\strain[]_2}](0)}  - \counts \freq^{(\strain[]_1)}\freq^{(\strain[]_2)} 
}{
\freq^{(\strain[]_1)} \freq^{(\strain[]_2)}
}
\right]
  \right]  \,.
  \end{split}
\end{equation}
Therefore one has further that
\begin{equation}
  \label{eq:averaged_cvxcoeff_HDirMult_cconcentration}
  \begin{split}
    \frac{1}{1+\cconcentration\Species} =& 
\sum_{n \ge 2} \left(
\scalebox{0.75}{
$\displaystyle
\left[ 
\frac{1}{\counts(\counts-1)} 
\left[
  \sum_{\strain[] \in [\Species]} 
\cvxcoeff{\strain[]} \left[ 
\frac{
\var{\abundance[{\strain[]}](0) | \abundance(0) = \counts }  - \counts \freq^{(\strain[])}(1- \freq^{(\strain[])})     
}{
\freq^{(\strain[])}(1-\freq^{(\strain[])})
} 
 \right] 
\vphantom{\left[ \sum_{\strain[]_1 \in [\Species]} \sum_{\strain[]_2 \not= \strain[]_1} \right]}
\right.
\right.
$
} 
\right.
\\
&
\scalebox{0.75}{
$\displaystyle
\left.
\left. 
\vphantom{ \left[\sum_{\strain[] \in [\Species]} \frac{\var{\abundance[{\strain[]}](0) }  - \counts \freq^{(\strain[])}(1- \freq^{(\strain[])})     }{\freq^{(\strain[])}(1-\freq^{(\strain[])})} \right.}
+ \sum_{\strain[]_1 \in [\Species]} \sum_{\strain[]_2 \not= \strain[]_1}\cvxcoeff{\strain[]_1, \strain[]_2}
 \left[
\frac{
-\cov{\abundance[{\strain[]_1}](0),  \abundance[{\strain[]_2}](0) | \abundance(0) =\counts  }  - \counts \freq^{(\strain[]_1)}\freq^{(\strain[]_2)} 
}{
\freq^{(\strain[]_1)} \freq^{(\strain[]_2)}
}
\right]
  \right]
\right]
$
} \\
&
\left.
\vphantom{\left( \sum_{\counts \ge 2}  \right)}
 \cdot \probability{\abundance(0)=\counts|\abundance(0) \ge 2} 
\right)
 \,,
  \end{split}
\end{equation}
allowing one to solve for the compositional concentration parameter
\begin{equation}
  \label{eq:HDirMult_cconcentration}
  \begin{split}
    \cconcentration =& 
\frac{1}{\Species}
\left(
\left( \sum_{n \ge 2} \left(
\scalebox{0.75}{
$\displaystyle
\left[ \frac{1}{\counts(\counts-1)} \left[
  \sum_{\strain[] \in [\Species]} 
\cvxcoeff{\strain[]} \left[ 
\frac{
\var{\abundance[{\strain[]}](0) | \abundance(0) = \counts }  - \counts \freq^{(\strain[])}(1- \freq^{(\strain[])})     
}{
\freq^{(\strain[])}(1-\freq^{(\strain[])})
} 
 \right] 
\vphantom{\left[ \sum_{\strain[]_1 \in [\Species]} \sum_{\strain[]_2 \not= \strain[]_1} \right]}
\right.
\right.
$
} 
\right.
\right.
\right.
\\
&
\scalebox{0.75}{
$\displaystyle
\left.
\left. 
\vphantom{ \left[\sum_{\strain[] \in [\Species]} \frac{\var{\abundance[{\strain[]}](0) }  - n \freq^{(\strain[])}(1- \freq^{(\strain[])})     }{\freq^{(\strain[])}(1-\freq^{(\strain[])})} \right.}
+ \sum_{\strain[]_1 \in [\Species]} \sum_{\strain[]_2 \not= \strain[]_1}\cvxcoeff{\strain[]_1, \strain[]_2}
 \left[
\frac{
-\cov{\abundance[{\strain[]_1}](0),  \abundance[{\strain[]_2}](0) | \abundance(0) =n   }  - n \freq^{(\strain[]_1)}\freq^{(\strain[]_2)} 
}{
\freq^{(\strain[]_1)} \freq^{(\strain[]_2)}
}
\right]
  \right]
\right]
$
} \\
&
\left.
\left.
\left.
\vphantom{\left( \sum_{n \ge 2}  \right)}
 \cdot \probability{\abundance(0)=n|\abundance(0) \ge 2} 
\right) \right)^{-1}
-1
\right)
 \,.
  \end{split}
\end{equation}

\subsubsection{Choice of Convex Coefficients}
\label{sec:choice-conv-coeff}

Choosing $\cvxcoeff{\strain[]}, \cvxcoeff{\strain[]_1, \strain[]_2} = \frac{1}{\Species^2}$ for all strains (and pairs thereof) corresponds to the $L_2$ projection of the covariance matrices onto the vector space of matrices whose entries all equal a common value. Thus the corresponding plugin estimator might initially appear attractive as a ``least-squares'' estimator. 

However, the corresponding plugin estimator can have an extremely slow rate of convergence, e.g. when there are many ``rare'' strains. Preliminary investigations (data not shown) found that the corresponding plugin estimator effectively failed to converge towards the true value at all even after sampling $15$ million droplets. 

A more pragmatic strategy is to choose the weights $\cvxcoeff{\strain[]}, \cvxcoeff{\strain[]_1, \strain[]_2}$ such that estimates with the most data behind them also receive the most weight. This means assigning higher weights to (pairs of) more common strains. Using the choices $\cvxcoeff{\strain[]} = (\freq^{(\strain[])})^2$ and $\cvxcoeff{\strain[]_1,\strain[]_2} = \freq^{(\strain[]_1)} \freq^{(\strain[]_2)}$, the resulting formula for $\cconcentration$ is
\begin{equation}
  \label{eq:HDirMult_cconcentration_freq_weights}
  \begin{split}
    \cconcentration =& 
\frac{1}{\Species}
\left(
\left( \sum_{\counts \ge 2} \left(
\scalebox{0.75}{
$\displaystyle
\left[ \frac{1}{
\counts(\counts-1)} 
\left[
  \sum_{\strain[] \in [\Species]} 
(\freq^{(\strain[])})^2 \left[ 
\frac{
\var{\abundance[{\strain[]}](0) | \abundance(0) = \counts }  - \counts \freq^{(\strain[])}(1- \freq^{(\strain[])})     
}{
\freq^{(\strain[])}(1-\freq^{(\strain[])})
} 
 \right] 
\vphantom{\left[ \sum_{\strain[]_1 \in [\Species]} \sum_{\strain[]_2 \not= \strain[]_1} \right]}
\right.
\right.
$
} 
\right.
\right.
\right.
\\
&
\scalebox{0.75}{
$\displaystyle
\left.
\left. 
\vphantom{ \left[\sum_{\strain[] \in [\Species]} \frac{\var{\abundance[{\strain[]}](0) }  - n \freq^{(\strain[])}(1- \freq^{(\strain[])})     }{\freq^{(\strain[])}(1-\freq^{(\strain[])})} \right.}
+ \sum_{\strain[]_1 \in [\Species]} \sum_{\strain[]_2 \not= \strain[]_1}\freq^{(\strain[]_1)} \freq^{(\strain[]_2)}
 \left[
\frac{
-\cov{\abundance[{\strain[]_1}](0),  \abundance[{\strain[]_2}](0) | \abundance(0) =\counts   }  - \counts \freq^{(\strain[]_1)}\freq^{(\strain[]_2)} 
}{
\freq^{(\strain[]_1)} \freq^{(\strain[]_2)}
}
\right]
  \right]
\right]
$
} \\
&
\left.
\left.
\left.
\vphantom{\left( \sum_{\counts \ge 2}  \right)}
 \cdot \probability{\abundance(0)=\counts|\abundance(0) \ge 2} 
\right) \right)^{-1}
-1
\right)
 \,.
  \end{split}
\end{equation}
The corresponding plugin estimator converges much faster to the true value $\cconcentration$.

\subsubsection{Plugin Estimator Definition}
\label{sec:plug-estim-defin}

The resulting plugin estimator for $\cconcentration$ is
\begin{equation}
  \label{eq:general_cconcentration_plugin_official}
\hat{\cconcentration}_{final} =
  \begin{cases}
\infty
&
\frac{1}{1+\hat{\cconcentration}\Species} \le 0
\\
0
&
\frac{1}{1+\hat{\cconcentration}\Species} \ge 1
\\
\hat{\cconcentration}
&
\text{else}
\end{cases}
\end{equation}
where $\hat{\cconcentration}$ is defined as
\begin{equation}
  \label{eq:general_cconcentration_plugin}
  \begin{split}
\hat{\cconcentration} :=
&\frac{1}{\Species}
\left(
\left( \sum_{\counts \ge 2} \left(
\scalebox{0.7}{
$\displaystyle
\left[ \frac{1}{
\counts(\counts-1)} 
\left[
  \sum_{\strain[] \in [\Species]} 
(\efreq^{(\strain[])})^2
\left[ 
\frac{
\evar{\abundance[{\strain[]}](0) | \abundance(0) = \counts }  - \counts \efreq^{(\strain[])}(1- \efreq^{(\strain[])})     
}{
\efreq^{(\strain[])}(1-\efreq^{(\strain[])})
} 
 \right] 
\vphantom{\left[ \sum_{\strain[]_1 \in [\Species]} \sum_{\strain[]_2 \not= \strain[]_1} \right]}
\right.
\right.
$
} 
\right.
\right.
\right.
\\
&\scalebox{0.7}{
$\displaystyle
\left.
\left. 
\vphantom{ \left[\sum_{\strain[] \in [\Species]} \frac{\evar{\abundance[{\strain[]}](0) }  - \counts \efreq^{(\strain[])}(1- \efreq^{(\strain[])})     }{\efreq^{(\strain[])}(1-\efreq^{(\strain[])})} \right.}
+ \sum_{\strain[]_1 \in [\Species]} \sum_{\strain[]_2 \not= \strain[]_1}
\efreq^{(\strain[]_1)} \efreq^{(\strain[]_2)}
 \left[
\frac{
-\ecov{\abundance[{\strain[]_1}](0),  \abundance[{\strain[]_2}](0) | \abundance(0) =\counts   }  - \counts \efreq^{(\strain[]_1)}\efreq^{(\strain[]_2)} 
}{
\efreq^{(\strain[]_1)} \efreq^{(\strain[]_2)}
}
\right]
  \right]
\right]
$
} \\
&\left.
\left.
\left.
\vphantom{\left( \sum_{n \ge 2}  \right)}
 \cdot \probability{\abundance(0)=n|\abundance(0) \ge 2} 
\right) \right)^{-1}
-1
\right)
 \,.
  \end{split}
\end{equation}
In the above definition, $\efreq^{(\strain[])}$ denote statistically consistent estimators of the frequencies $\freq^{(\strain[])}$. One possibile definition for such estimators (which was used for the implementation seen in section \ref{sec:hetero_estimator_performance_results}) is given in section \ref{sec:append-plug-estim}. I did not want to assume that the scientist had, or wanted to rely on, a priori accurate estimates of the true frequency values. In principle, if such knowledge were available and trusted, then one would use those values for $\efreq^{(\strain[])}$ in equation \ref{eq:general_cconcentration_plugin_official}.

Note that when $\frac{1}{1+\hat{\cconcentration}} > 1$, the result of solving for $\hat{\cconcentration}$ is less than $0$. The interpretation in this case is extremely unclear given that the limit as $\cconcentration \to 0$ corresponds to ``infinite heterogeneity''. This is the reason for thresholding the values of $\hat{\cconcentration}_{final}$ below by $0$.

\subsubsection{Effective Compositional Concentration}
\label{sec:effect-comp-conc}

Using the chosen values of $\cvxcoeff{\strain[]}$ and $\cvxcoeff{\strain[]_1, \strain[]_2}$ gives one possible choice for a definition of ``effective compositional concentration''. This estimand exists for any distribution of $\vabundance(0)$ with finite covariances, not just when $\expectation{\vabundance(0)|\abundance(0)}$ is Dirichlet-Multinomial distributed:
\begin{equation}
  \label{eq:effective_compositional_concentration_official}
( {\cconcentration}_{e\!f\!f})_{final} =
  \begin{cases}
\infty
&
\frac{1}{1+{\cconcentration}_{e\!f\!f}\Species} \le 0
\\
0
&
\frac{1}{1+{\cconcentration}_{e\!f\!f}\Species} \ge 1
\\
{\cconcentration}_{e\!f\!f}
&
\text{else}
\end{cases}
\end{equation}
with ${\cconcentration}_{e\!f\!f}$ defined (when $\frac{1}{1 + {\cconcentration}_{e\!f\!f} \Species}  \not=0$) as
\begin{equation}
  \label{eq:effective_compositional_concentration}
  \begin{split}
    {\cconcentration}_{e\!f\!f} =& 
\frac{1}{\Species}
\left(
\left( \sum_{\counts \ge 2} \left(
\scalebox{0.75}{
$\displaystyle
\left[ 
\frac{1}{\counts(\counts-1)} 
\left[
  \sum_{\strain[] \in [\Species]} 
(\freq^{(\strain[])})^2
\left[ 
\frac{
\var{\abundance[{\strain[]}](0) | \abundance(0) = \counts }  - \counts \freq^{(\strain[])}(1- \freq^{(\strain[])})     
}{
\freq^{(\strain[])}(1-\freq^{(\strain[])})
} 
 \right] 
\vphantom{\left[ \sum_{\strain[]_1 \in [\Species]} \sum_{\strain[]_2 \not= \strain[]_1} \right]}
\right.
\right.
$
} 
\right.
\right.
\right.
\\
&
\scalebox{0.75}{
$\displaystyle
\left.
\left. 
\vphantom{ \left[\sum_{\strain[] \in [\Species]} \frac{\var{\abundance[{\strain[]}](0) }  - n \freq^{(\strain[])}(1- \freq^{(\strain[])})     }{\freq^{(\strain[])}(1-\freq^{(\strain[])})} \right.}
+ \sum_{\strain[]_1 \in [\Species]} \sum_{\strain[]_2 \not= \strain[]_1}
\freq^{(\strain[]_1)} \freq^{(\strain[]_2)}
 \left[
\frac{
-\cov{\abundance[{\strain[]_1}](0),  \abundance[{\strain[]_2}](0) | \abundance(0) = \counts   }  - \counts \freq^{(\strain[]_1)}\freq^{(\strain[]_2)} 
}{
\freq^{(\strain[]_1)} \freq^{(\strain[]_2)}
}
\right]
  \right]
\right]
$
} \\
&
\left.
\left.
\left.
\vphantom{\left( \sum_{n \ge 2}  \right)}
 \cdot \probability{\abundance(0)=n|\abundance(0) \ge 2} 
\right) \right)^{-1}
-1
\right)
 \,,
  \end{split}
\end{equation}
The law of large numbers combined with the continuous mapping theorem for convergence in probability guarantees the consistency of the estimator (\ref{eq:general_cconcentration_plugin_official}) for the effective compositional concentration $( {\cconcentration}_{e\!f\!f})_{final}$ whenever, under the given distribution for $\vabundance(0)$, the chosen estimators $\efreq^{(\strain[])}$ are also consistent for the true frequencies. (Asymptotic consistency is actually guaranteed, due to the same reasons, for the analogous plugin estimator and estimand given any choice of convex coefficients $\cvxcoeff{\strain[]}, \cvxcoeff{\strain[]_1, \strain[]_2}$. The result does \textit{not} depend on the chosen definitions $\cvxcoeff{\strain[]}:= \freq^{(\strain[])}$, $\cvxcoeff{\strain[]_1, \strain[]_2} := \freq^{(\strain[]_1)} \freq^{(\strain[]_2)}$.)

Note also that $( {\cconcentration}_{e\!f\!f})_{final}$ is thresholded below by $0$ for the same reason that $\hat{\cconcentration}_{final}$ is thresholded below by $0$.

 Because the effective compositional concentration, as well as ${\cconcentration}_{e\!f\!f}$, both equal the compositional concentration parameter $\cconcentration$ whenever $\expectation{\vabundance(0)|\abundance(0)}$ is Dirichlet-Multinomial distributed, the estimand $({\cconcentration}_{e\!f\!f})_{final}$ may be considered a generalization of the parameter $\cconcentration$ to other distributions.

\subsubsection{Effective Compositional Concentration without Frequencies}
\label{sec:effect-comp-conc-1}

There could be distributions which are not parameterized in terms of frequencies $\freq^{(\specie)}$ to which we might still want to apply the notion of effective compositional concentration. One strategy we could employ is based on the expectations of the multinomial and Dirichlet-Multinomial distributions. For the marginal distribution corresponding to any strain $\strain$, when $\vabundance(0)$ conditional on $\abundance (0) = \counts$ is multinomial or Dirichlet-Multinomial distributed, then one always has
\begin{equation}
  \label{eq:simple_moment_equation}
  \expectation*{\abundance[\strain](0) | \abundance(0) = \counts} = \counts \freq^{(\strain)} \,.
\end{equation}
Rearranging the relationship from (\ref{eq:simple_moment_equation}) gives us a proxy which we can use in place of the frequency parameters $\freq^{(\strain)}$ (at least when $\counts \ge 1$):
\begin{equation}
  \label{eq:proxy_frequency}
  \freq^{(\strain)} = \frac{
\expectation*{\abundance[\strain](0) | \abundance(0) = \counts}
}{
\counts
} \,,
\end{equation}
related to the so-called ``effective frequency'', see section \ref{sec:append-plug-estim} for a derivation. Note this has the drawback of being dependent on the value of $\counts$.

Using these proxies for the frequencies (\ref{eq:proxy_frequency}), under the above distributional assumptions we can rewrite equation (\ref{eq:DirMult_variance_review}) as
\begin{equation}
  \label{eq:frequency_less_effective_variance_overdispersion}
  \begin{adjustbox}{max width=\textwidth,keepaspectratio}
$\displaystyle \frac{1}{1 + \cconcentration \Species} 
=
\frac{1}{\counts - 1}
\!\cdot \!
\left[
\frac{
\counts\var{\abundance[\strain](0) | \abundance(0) = \counts } 
-
\expectation*{\abundance[\strain](0) | \abundance(0) = \counts  } 
\left(
\counts - \expectation*{\abundance[\strain](0) | \abundance(0) = \counts  } 
\right)
}{ 
\expectation*{\abundance[\strain](0) | \abundance(0) = \counts  } 
\left(
\counts - \expectation*{\abundance[\strain](0) | \abundance(0) = \counts  } 
\right)
} \right]
\,, $
\end{adjustbox}
\end{equation}
and rewrite equation (\ref{eq:DirMult_covariance_review}) as
\begin{equation}
  \label{eq:frequency_less_effective_covariance_overdispersion}
  \begin{adjustbox}{max width=\textwidth,keepaspectratio}
$\displaystyle \frac{1}{1 + \cconcentration \Species} 
=
\frac{1}{\counts - 1}
\!\cdot \!
\left[
\frac{
-\counts\cov{\abundance[\strain_1](0), \abundance[\strain_2](0) \! | \abundance(0) = \counts } 
-
\expectation*{\abundance[\strain_1](0)\! | \abundance(0) = \counts  } 
\expectation*{\abundance[\strain_2](0) \!| \abundance(0) = \counts  } 
}{ 
\expectation*{\abundance[\strain_1](0) | \abundance(0) = \counts  } 
\expectation*{\abundance[\strain_2](0) | \abundance(0) = \counts  } 
} \right]
\,. $
\end{adjustbox}
\end{equation}
Based on this we can easily generalize the formulae from sections \ref{sec:solv-comp-conc}-\ref{sec:effect-comp-conc} to define a notion of effective compositional concentration that does not require the distribution to have frequency parameters $\freq^{(\strain)}$. Again, cf. section \ref{sec:append-plug-estim}. 

I do not claim that this is the only possibly proxy one could use in place of the frequency parameters for more general distributions. However, it does have the benefit of being widely applicable. Any definition of effective compositional concentration will (seemingly) require the distribution to have finite second (conditional) cumulants, which in turn guarantees the finiteness of the first (conditional) cumulants in formulae (\ref{eq:frequency_less_effective_variance_overdispersion}) and (\ref{eq:frequency_less_effective_covariance_overdispersion}) above.

\subsection{Maximum Likelihood Estimator for Compositional Concentration}
\label{sec:mle-comp}

This section explains how to use maximum likelihood to estimate the compositional concentration $\cconcentration$. Section \ref{sec:dirichl-mult-log} computes the likelihood and log likelihood for the categorical distributions, while section \ref{sec:dirichl-mult-score} computes the score. Section \ref{sec:dirichl-mult-maxim} uses those results to derive the maximum likelihood estimators for the Dirichlet-Multinomial distribution.

\subsubsection{Dirichlet-Multinomial (Log) Likelihood}
\label{sec:dirichl-mult-log}

Starting from equation (\ref{eq:DM_pmf}),
\begin{itemize}
\item if we assume for mathematical convenience that the distributions of all droplets are mutually independent,
\item using the identity\footnote{Cf. footnote \ref{footnote:gamma_interpolate}.} $\frac{\Gamma(x+M)}{\Gamma(x)} = \prod_{m=0}^{M-1} (x+m)$ for all positive integers $M$,
\item if we assume that the droplets are Dirichlet-Multinomial distributed conditional on their count distributions,
\end{itemize}
then this is the full likelihood for an entire batch of $\Droplets$ droplets:
\begin{equation}
  \label{eq:DM_full_likelihood}
  \begin{split}
    \likelihood{\vfreq, \cconcentration} = & \prod_{\droplet \in [\Droplets]}
\left[
\probability{\abundance_{\droplet}(0) = \counts_{\droplet}}
\!\cdot\!
\left(
\prod_{\nu=0}^{\counts_{\droplet} - 1}
(\cconcentration\Species + \nu)^{-1}
\right)
\vphantom{
\prod_{\strain=1}^{\Strains}
\left(
\prod_{\nu=0}^{\counts[\strain]_{\droplet}-1}
(\cconcentration\Species\freq^{(\strain)} + \nu)
\right)
\!\cdot\!
\binom{\counts_{\droplet}}{\counts[1]_{\droplet} \cdots \counts[\Strains]_{\droplet}}
}
\!\cdot\!
\right.
\\
&
\left.
\hphantom{
\prod_{\droplet \in [\Droplets]}
}
\vphantom{
\prod_{\droplet \in [\Droplets]}
\probability{\abundance_{\droplet}(0) = \counts_{\droplet}}
\!\cdot\!
\left(
\prod_{\nu=0}^{\counts_{\droplet} - 1}
(\cconcentration\Species + \nu)^{-1}
\right)
}
\prod_{\strain \in [\Strains]}
\left(
\prod_{\nu=0}^{\counts[\strain]_{\droplet}-1}
(\cconcentration\Species\freq^{(\strain)} + \nu)
\right)
\!\cdot\!
\binom{\counts_{\droplet}}{\counts[1]_{\droplet} \cdots \counts[\Strains]_{\droplet}}
\right] \,,
  \end{split}
\end{equation}
where (as a reminder) by definition $\counts_{\droplet} = \sum_{\strain \in [\Strains]} \counts[\strain]_{\droplet}$.

Therefore this is the full log likelihood:
\begin{equation}
  \label{eq:DM_full_log_likelihood}
  \begin{split}
    \loglikelihood{\vfreq, \cconcentration} := &
\\
\log\left(\likelihood{\vfreq, \cconcentration}\right)
= &
\sum_{\droplet \in [\Droplets]}
\left[
\log\left( \probability{\abundance_{\droplet}(0) = \counts_{\droplet}} \right)
-
\sum_{\nu=0}^{\counts_{\droplet}-1}
\log(\cconcentration\Species + \nu)
\vphantom{
\sum_{\strain \in [\Strains]}
\left(
\sum_{\nu=0}^{\counts[\strain]_{\droplet} -1}
\log (\cconcentration\Species\freq^{(\strain)} + \nu)
\right)
+
\log\left(
\binom{\counts_{\droplet}}{\counts[1]_{\droplet} \cdots \counts[\Strains]_{\droplet}}
\right)
}
\right. 
+
\\
&
\hphantom{\sum_{\droplet \in [\Droplets]} [}
\left.
\vphantom{
\log\left( \probability{\abundance_{\droplet}(0) = \counts_{\droplet}} \right)
-
\sum_{\nu=0}^{\counts_{\droplet}-1}
\log(\cconcentration\Species + \nu)
}
\sum_{\strain \in [\Strains]}
\left(
\sum_{\nu=0}^{\counts[\strain]_{\droplet} -1}
\log (\cconcentration\Species\freq^{(\strain)} + \nu)
\right)
+
\log\left(
\binom{\counts_{\droplet}}{\counts[1]_{\droplet} \cdots \counts[\Strains]_{\droplet}}
\right)
\right] \,.
  \end{split}
\end{equation}

\subsubsection{Dirichlet-Multinomial Score}
\label{sec:dirichl-mult-score}

Assuming that the count distribution has no dependence\footnote{Or ignoring such a dependence if it exists, e.g. as in the case of hNBDM, and choosing to consider the maximum \textit{conditional} likelihood estimator instead of the MLE sensu stricto.} on $\cconcentration$, it follows that the score with respect to $\cconcentration$ is
  \begin{equation}
    \label{eq:DM_score_concentration}
    \begin{split}
      \frac{
\partial \loglikelihood{\vfreq, \cconcentration}
}{
\partial \cconcentration
} = & 
\sum_{\droplet \in [\Droplets]}
\left[
-\sum_{\nu=0}^{\counts_{\droplet}-1}
\frac{\Species}{
\cconcentration\Species + \nu
}
+
\sum_{\strain \in [\Strains]}
\left(
\sum_{\nu=0}^{\counts[\strain]_{\droplet}-1}
\frac{\Strains\freq^{(\strain)}
}{
\cconcentration\Strains\freq^{(\strain)} + \nu
}
\right)
\right] \,.
    \end{split}
  \end{equation}
Using the commutativity of addition (i.e. rearranging terms) for purposes of computational efficiency, the above may be rewritten
\begin{equation}
  \label{eq:DM_score_concentration_efficient}
  \begin{split}
      \frac{
\partial \loglikelihood{\vfreq, \cconcentration}
}{
\partial \cconcentration
} = & 
- \sum_{m=1}^M
\frac{
\Droplets_{\ge m}
\Species
}{
\cconcentration \Species + (m-1)
}
+
\sum_{\strain \in [\Strains]}
\left(
\sum_{m=1}^{M^{(\strain)}}
\frac{
\Droplets_{\ge m}^{(\strain)} \Species \freq^{(\strain)}
}{
\cconcentration \Strains \freq^{(\strain)} + (m-1)
}
\right) \,,
  \end{split}
\end{equation}
where I have defined
\begin{equation}
  \label{eq:DM_score_definitions}
  \begin{split}
   M := & \max_{\droplet \in [\Droplets]} \abundance_{\droplet}(0) \,, \\
 \forall \strain \in [\Strains] \quad
M^{(\strain)} := & \max_{\droplet \in [\Droplets]} \abundance[\strain]_{\droplet}(0)\,, \\
\Droplets_{\ge m} := &
\left| \left\{ \droplet:  \abundance_{\droplet}(0) \ge m  \right\}\right| \,, \\
\forall \strain \in [\Strains] \quad
\Droplets_{\ge m}^{(\strain)} := &
\left| \left\{ \droplet: \abundance[\strain]_{\droplet}(0) \ge m  \right\} \right| \,.
  \end{split}
\end{equation}
This is basically the same approach to grouping terms as suggested in \cite{Sklar}.

\subsubsection{Dirichlet-Multinomial Maximum Likelihood Estimator}
\label{sec:dirichl-mult-maxim}

Assuming that the frequencies $\vfreq := (\freq_1, \dots, \freq^{(\Strains)})$ are already known or estimated (and thus don't need to be treated as nuisance parameters), it follows from the above formula (\ref{eq:DM_score_concentration_efficient}) for the score that a maximum likelihood estimator $\hat{\cconcentration}$ for the compositional concentration $\cconcentration$ must be a solution of the following equation:
  \begin{equation}
    \label{eq:DM_cconcentration_MLE_eq}
    \sum_{m=1}^M
\frac{
\Droplets_{\ge m}
}{
\hat{\cconcentration \Species} + (m-1)
}
=
\sum_{\strain \in \Strains}
\left(
\sum_{m=1}^{M^{(\strain)}}
\frac{
\Droplets_{\ge m}^{(\strain)} \freq^{(\strain)}
}{
\hat{\cconcentration} \Species \freq^{(\strain)} + (m-1)
}
\right) \,.
  \end{equation}
I only claim that satisfying the above equation is necessary, not sufficient. In particular, I make no claims regarding the conditions under which a root for the above equation exists, nor regarding the conditions under which a root (if it exists) will be unique. These conditions appear to be unknown.

\section{Methods}
\label{sec:hetero_estimator_performance_methods}

 Section \ref{sec:comp-estim-distr} explains how the empirical distributions of the plugin heterogeneity estimators were computed as a function of batch size. Section \ref{sec:computing-mles} provides analogous details for the ML estimators.  Section \ref{sec:plott-estim-distr} explains how these results were depicted. Complete implementation details can be found in the code at \dropletsgitrepo. See \url{\dropletsgitrepourl}.

\subsection{Computing Plugin Estimator Distributions}
\label{sec:comp-estim-distr}

For each of the seven simulated distributions, for each of the 500 simulations, for each of the three batch sizes (small = $10,000$ droplets, medium = $500,000$ droplets, large = $15,000,000$ droplets), using NumPy \cite{NumPy} version 1.20.2 I partitioned the simulation results into equal parts of the given batch size. Since each simulation contains $15,000,000$ droplets, this corresponds to $1,500$ small batches per simulation, $30$ medium batches per simulation, and $1$ large batch per simulation. Then for each of the batches, I evaluated the plugin density heterogeneity estimator (\ref{eq:dconcentration_plugin}) and the plugin compositional heterogeneity estimator (\ref{eq:general_cconcentration_plugin_official}) and stored the results.

\subsection{Computing MLE Distributions}
\label{sec:computing-mles}

Simulated droplets were partitioned into batches exactly as was done for the plugin estimators. Cf. section \ref{sec:comp-estim-distr} above for details.

I tried to use Brent's method \cite[Ch.~3-4]{Brent} whenever possible to estimate roots of the score functions, because Powell's method \cite{Powell} was often unstable. Preliminary investigations (data not shown) suggested that Powell's method often either failed to converge, or failed to converge to a reasonable value (for example returning a negative answer, or one that is far too large), under conditions where Brent's method converged to a reasonable answer similar to the truth. While it was always possible to supply a reasonable interval of values for Brent's method to restrict its search, the method requires the signs of the function to differ at both ends of the interval. It was difficult to systematically choose (i.e. without manual intervention) interval endpoints which would guarantee opposite signs at both ends of the interval for all possible simulated datasets. Therefore Powell's method often had to be used as an alternative, which may have introduced many inaccurate estimates. While gradient-based methods probably would have been more accurate and converged faster, this level of accuracy seemed sufficient for a proof of principle. Estimates that were less than zero were obviously erroneous and therefore discarded -- this affected fewer than $3\%$ of the estimates in general and usually less (data not shown).

\paragraph{For the Negative Binomial (Density Concentration):}

\begin{itemize}
\item Check that the sample variance is larger than the sample mean, otherwise return $\infty$ and terminate.
\item Set initial guess value for the root-finding algorithm to be the value of the plugin estimator\footnote{This never exceeded 10,000 when the sample variance was larger than the sample mean.}, equation (\ref{eq:dconcentration_plugin}).
\item If the score function, computed using the terms of equation (\ref{eq:NB_dconcentration_MLE_eq_efficient}) rearranged to one side to match (\ref{eq:NB_score_concentration}), had different signs at $10^{-4}$ and $10^4$:
  \begin{itemize}
  \item Use Brent's method \cite[Ch.~3-4]{Brent} implemented in SciPy \cite{SciPy} with brackets $10^{-4}$ and $10^4$ to look for a root of the score function.
  \end{itemize}
\item If the score function, computed using the terms of equation (\ref{eq:NB_dconcentration_MLE_eq_efficient}) rearranged to one side to match (\ref{eq:NB_score_concentration}), had the same sign at $10^{-4}$ and $10^4$:
  \begin{itemize}
  \item Use Powell's hybrid method \cite{Powell} as implemented in the HYBRD routine of MINPACK \cite{MINPACK1}\cite[Ch.~5]{MINPACK2} and made available through SciPy \cite{SciPy} to search for a root of the score function.
  \item If the method fails to converge and the final guess is greater than $10^4$, return $\infty$.
  \item Otherwise (even if the method fails to converge) return the final guess (or discard if the final guess is less than zero).
  \end{itemize}
\end{itemize}

\paragraph{For the Dirichlet-Multinomial (Compositional Concentration):}

\begin{itemize}
\item Set the initial guess value for the root-finding algorithm to be the value of the plugin estimator, equation (\ref{eq:general_cconcentration_plugin_official}).
\item If the plugin estimator was larger than $10^6$ (including $\infty$), set the initial guess value to be $0.5 \times 10^6$.
\item If the score function, computed using equation (\ref{eq:DM_score_concentration_efficient}), had different signs at $10^{-6}$ and $10^6$:
  \begin{itemize}
  \item Use Brent's method \cite[Ch.~3-4]{Brent} implemented in SciPy \cite{SciPy} with brackets $10^{-6}$ and $10^6$ to look for a root of the score function.
  \end{itemize}
\item If the score function, computed using equation (\ref{eq:DM_score_concentration_efficient}), had the same sign at $10^{-6}$ and $10^6$:
  \begin{itemize}
  \item Use Powell's hybrid method \cite{Powell} as implemented in the HYBRD routine of MINPACK \cite{MINPACK1}\cite[Ch.~5]{MINPACK2} and made available through SciPy \cite{SciPy} to search for a root of the score function.
  \item If the method fails to converge and the final guess is greater than $10^6$, return $\infty$.
  \item Otherwise (even if the method fails to converge) return the final guess (or discard if the final guess is less than zero).
  \end{itemize}
\end{itemize}

\subsection{Plotting Estimator Distributions}
\label{sec:plott-estim-distr}

I plotted results using Matplotlib \cite{Matplotlib} version 3.4.1 and Seaborn \cite{Seaborn} version 0.11.1. Kernel density estimates were made using the default Seaborn \cite{Seaborn} settings\footnote{Via SciPy \cite{SciPy}, which chooses bandwidths according to Scott's Rule \cite{ScottsRule}.}. I computed empirical cumulative density functions using StatsModels \cite{statsmodels} version 0.12.2, which uses the binary search algorithm from NumPy \cite{NumPy}. Finally, I computed the empirical survival functions from the empirical cumulative density functions.

\section{Results}
\label{sec:hetero_estimator_performance_results}

Section \ref{sec:estim-cons-infin} explains what indicates that the plugin estimators are consistent in practice when the concentration parameters are infinite. Section \ref{sec:estim-cons-finite} explains the analogous evidence when the concentration parameters are finite. 

\subsection{Estimators are Consistent for Infinite Concentrations}
\label{sec:estim-cons-infin}

When the true value of the estimand is infinite, we see the correct behavior for a consistent estimator. Section \ref{sec:surv-funct-show} explains how the survival functions support this conclusion. Section \ref{sec:finite-estim-shift} explains how the distributions of incorrect finite estimates also supports this conclusion.

\subsubsection{Survival Functions Show Anticipated Behavior}
\label{sec:surv-funct-show}

When the true value of the concentration estimand is infinite, we see for both the density (as in figures \ref{fig:hTPMH_D_survival_plugin}, \ref{fig:hPoDM_100_D_survival_plugin}, \ref{fig:hPoDM_1_D_survival_plugin}, \ref{fig:hExhPoDM_D_survival_plugin} for the plugin estimators and figures \ref{fig:hTPMH_D_survival_MLE}, \ref{fig:hPoDM_100_D_survival_MLE}, \ref{fig:hPoDM_1_D_survival_MLE}, \ref{fig:hExhPoDM_D_survival_MLE} for the ML estimators) and compositional concentration estimators (as in figure \ref{fig:hTPMH_C_survival_plugin} for the plugin estimator and figure \ref{fig:hTPMH_C_survival_MLE} for the ML estimator) that the empirical survival function generally moves to the upper right as the batch size increases. This is the correct behavior for a consistent estimator, since in the case that all estimates were infinite the survival function would be the horizontal line $y=1$.

\begin{figure}[p]
  \centering
  
  \begin{subfigure}{\textwidth}
  \centering \includegraphics[width=\textwidth,height=0.45\textheight,keepaspectratio]{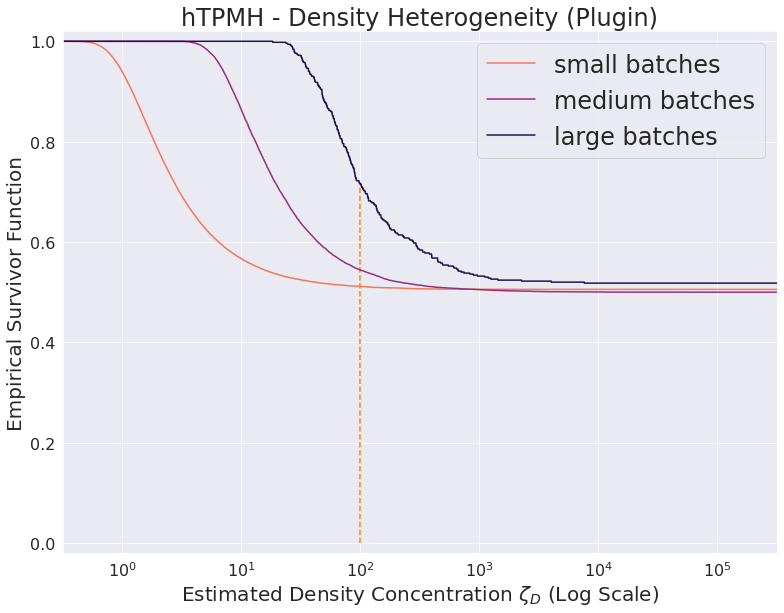}
  \caption[]{}
  \label{fig:hTPMH_D_survival_plugin}
\end{subfigure}

\begin{subfigure}{\textwidth}
  \centering \includegraphics[width=\textwidth,height=0.45\textheight,keepaspectratio]{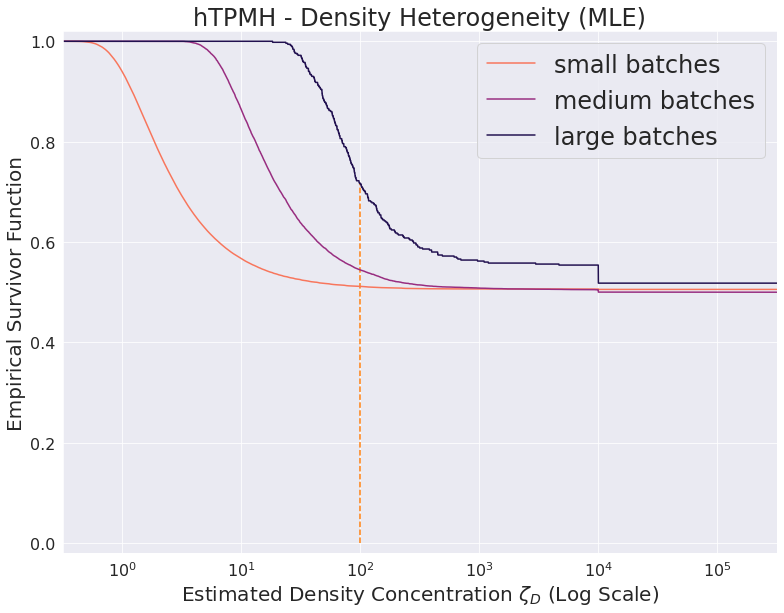}
  \caption[]{}
  \label{fig:hTPMH_D_survival_MLE}
\end{subfigure}

\caption{Survival functions of hTPMH density concentration estimates.}
\label{fig:hTPMH_D_survival}
\end{figure}

\begin{figure}[p]
  \centering
  
  \begin{subfigure}{\textwidth}
  \centering \includegraphics[width=\textwidth,height=0.45\textheight,keepaspectratio]{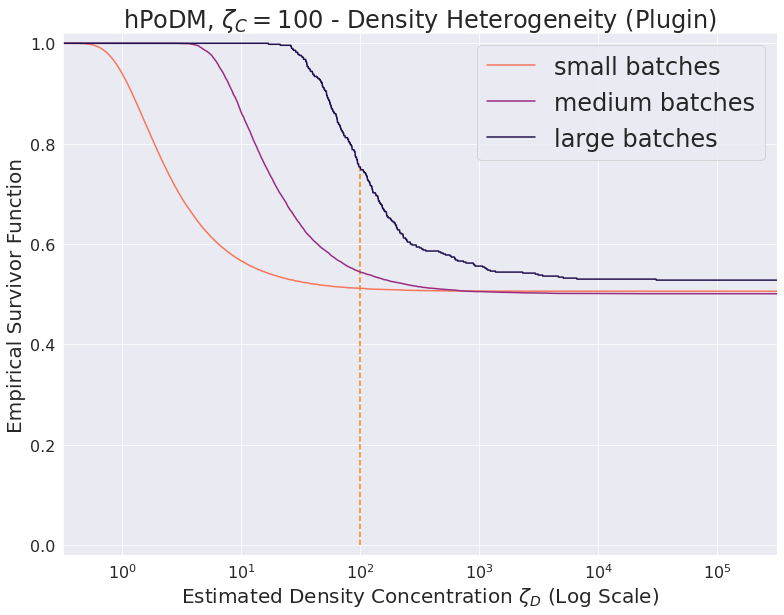}
  \caption[]{}
  \label{fig:hPoDM_100_D_survival_plugin}
\end{subfigure}

\begin{subfigure}{\textwidth}
  \centering \includegraphics[width=\textwidth,height=0.45\textheight,keepaspectratio]{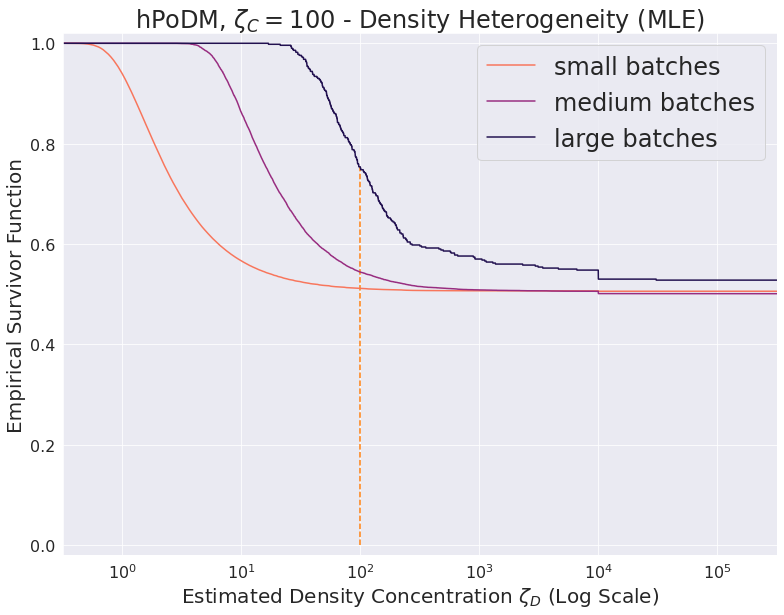}
  \caption[]{}
  \label{fig:hPoDM_100_D_survival_MLE}
\end{subfigure}

\caption{Survival functions of hPoDM ($\cconcentration=100$) density concentration estimates.}
\label{fig:hPoDM_100_D_survival}
\end{figure}

\begin{figure}[p]
  \centering
  \begin{subfigure}{\textwidth}
  \centering
  \includegraphics[width=\textwidth,height=0.45\textheight,keepaspectratio]{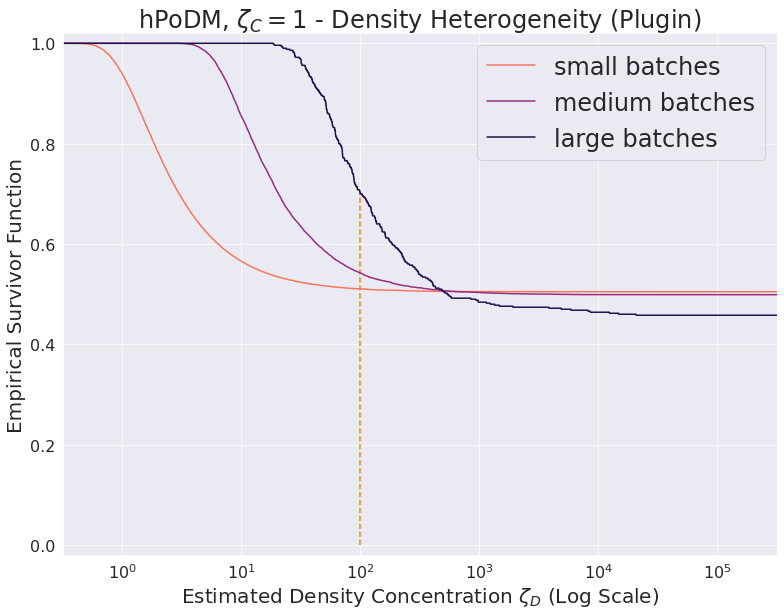}
  \caption[]{}
  \label{fig:hPoDM_1_D_survival_plugin}
\end{subfigure}

\begin{subfigure}{\textwidth}
  \centering
  \includegraphics[width=\textwidth,height=0.45\textheight,keepaspectratio]{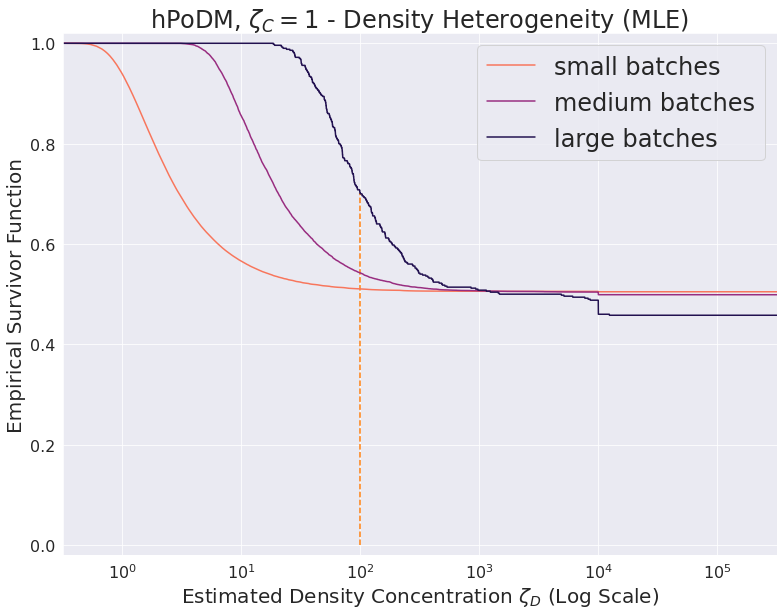}
  \caption[]{}
  \label{fig:hPoDM_1_D_survival_MLE}
\end{subfigure}

\caption{Survival functions of hPoDM ($\cconcentration=1$) density concentration estimates.}
\label{fig:hPoDM_1_D_survival}
\end{figure}

\begin{figure}[p]
  \centering
  \begin{subfigure}{\textwidth}
  \centering
  \includegraphics[width=\textwidth,height=0.45\textheight,keepaspectratio]{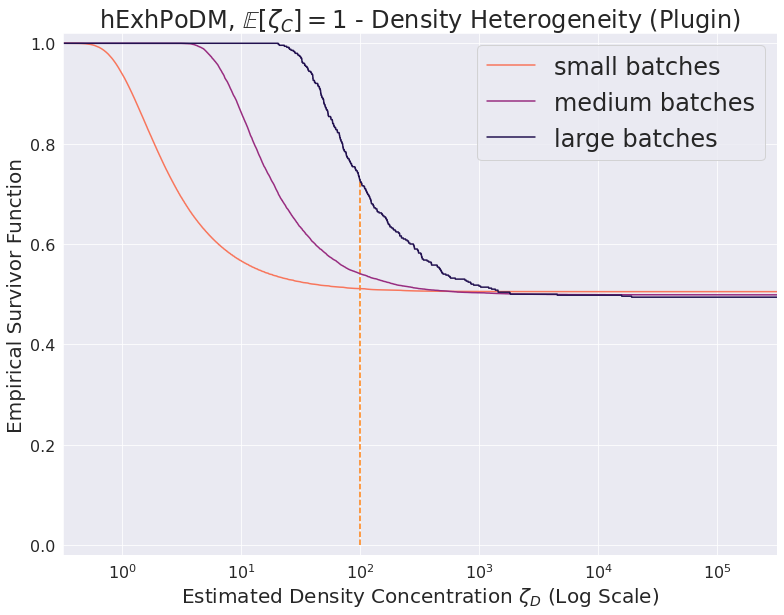}
  \caption[]{}
  \label{fig:hExhPoDM_D_survival_plugin}
\end{subfigure}

\begin{subfigure}{\textwidth}
  \centering
  \includegraphics[width=\textwidth,height=0.45\textheight,keepaspectratio]{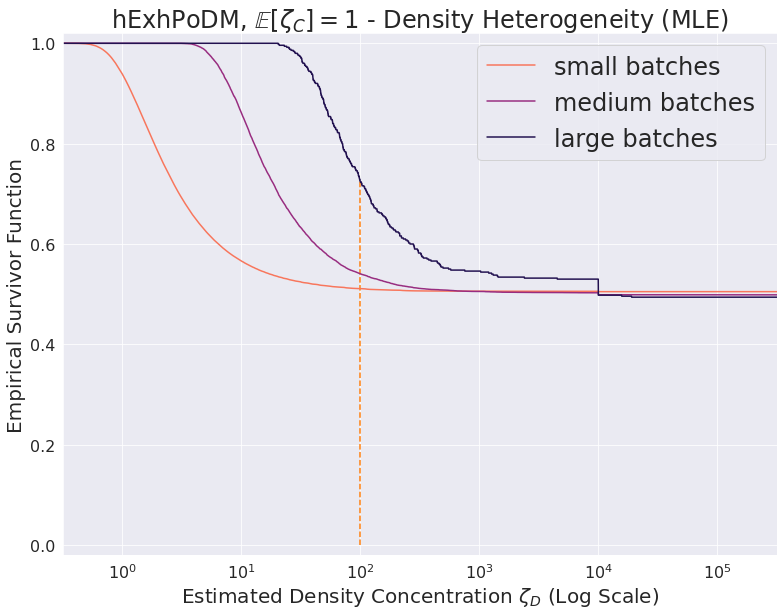}
  \caption[]{}
  \label{fig:hExhPoDM_D_survival_MLE}
\end{subfigure}

\caption{Survival functions of hExhPoDM density concentration estimates.}
\label{fig:hExhPoDM_D_survival}
\end{figure}

\begin{figure}[p]
  \centering
  
  \begin{subfigure}{\textwidth}
  \centering \includegraphics[width=\textwidth,height=0.45\textheight,keepaspectratio]{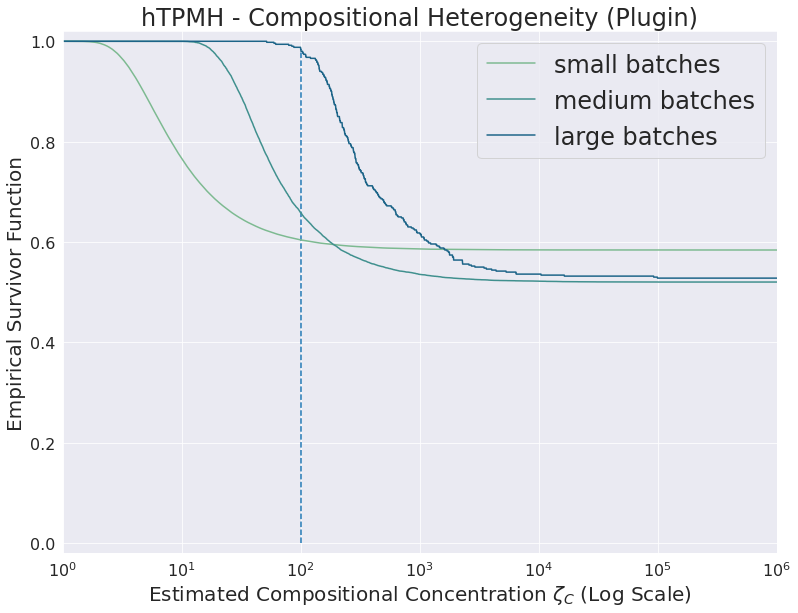}
  \caption[]{}
  \label{fig:hTPMH_C_survival_plugin}
\end{subfigure}

\begin{subfigure}{\textwidth}
  \centering \includegraphics[width=\textwidth,height=0.45\textheight,keepaspectratio]{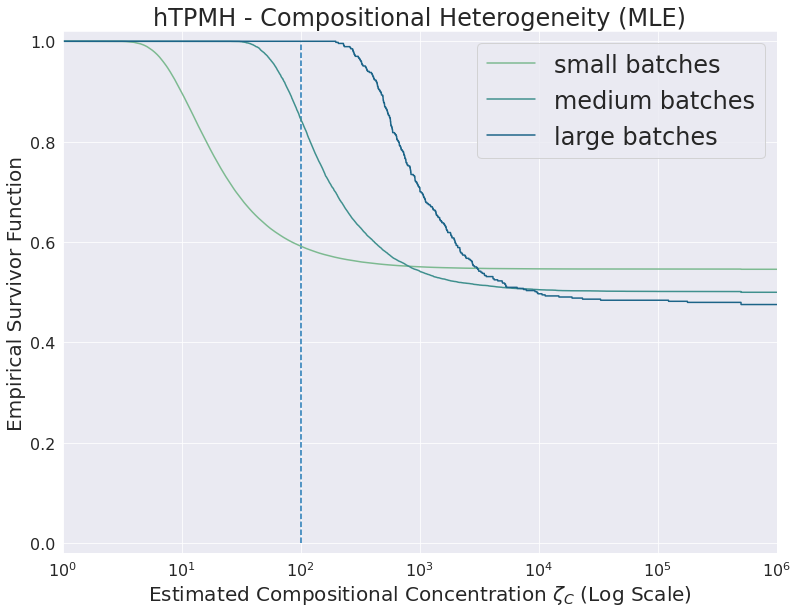}
  \caption[]{}
  \label{fig:hTPMH_C_survival_MLE}
\end{subfigure}

\caption{Survival functions of hTPMH compositional concentration estimates.}
\label{fig:hTPMH_C_survival}
\end{figure}

\subsubsection{Incorrect Finite Estimates Shift Further to the Right as Batch Size Increases}
\label{sec:finite-estim-shift}

Moreover, as the size of the batches increases (cf. figures \ref{fig:hTPMH_D_kde_plugin}, \ref{fig:hPoDM_100_D_kde_plugin}, \ref{fig:hPoDM_1_D_kde_plugin}, \ref{fig:hExhPoDM_D_kde_plugin}, and  \ref{fig:hTPMH_C_kde_plugin} for the plugin estimators and figures \ref{fig:hTPMH_D_kde_MLE}, \ref{fig:hPoDM_100_D_kde_MLE}, \ref{fig:hPoDM_1_D_kde_MLE}, \ref{fig:hExhPoDM_D_kde_MLE}, and \ref{fig:hTPMH_C_kde_MLE} for the ML estimators) the distribution of the remaining estimates shifts further and further to the right. In other words, even when the estimators incorrectly estimate a finite value, the estimated finite value still tends to increase (and thus ``better approximates infinity'') as the data size increases.

\begin{figure}[p]
  \centering
  
  \begin{subfigure}{\textwidth}
  \centering \includegraphics[width=\textwidth,height=0.45\textheight,keepaspectratio]{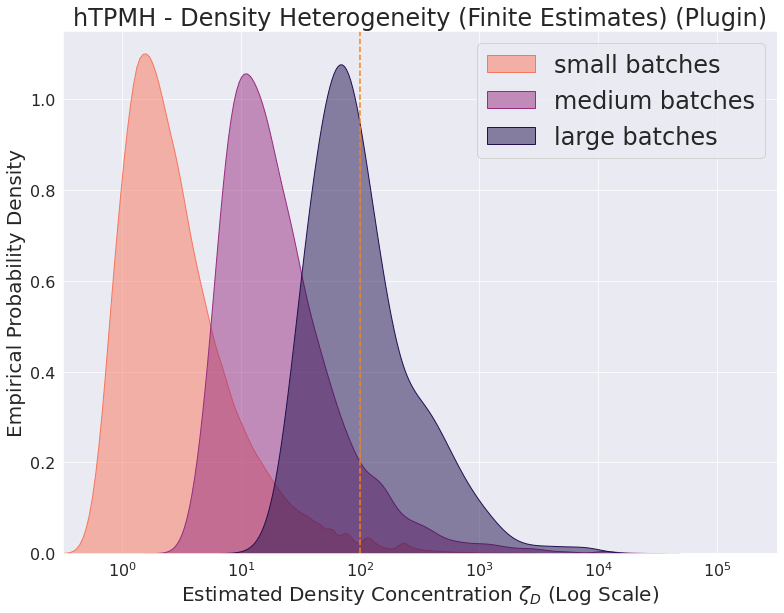}
  \caption[]{}
  \label{fig:hTPMH_D_kde_plugin}
\end{subfigure}

\begin{subfigure}{\textwidth}
  \centering \includegraphics[width=\textwidth,height=0.45\textheight,keepaspectratio]{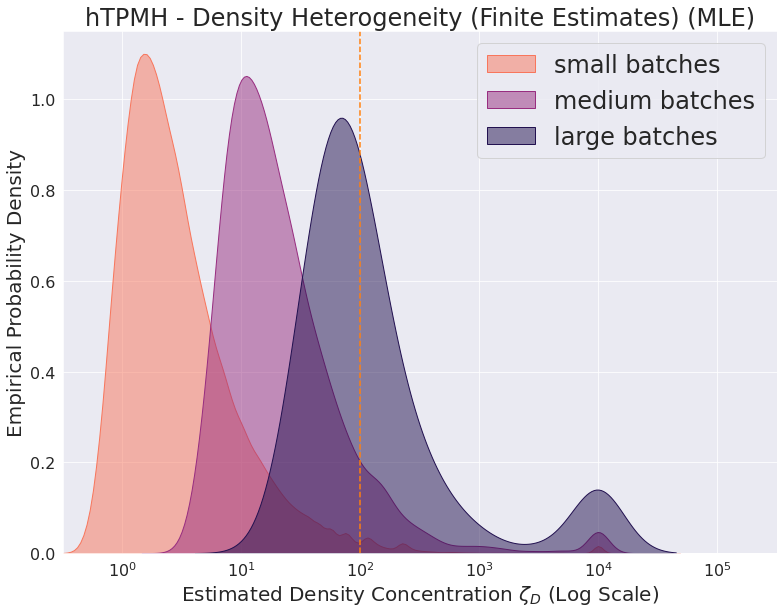}
  \caption[]{}
  \label{fig:hTPMH_D_kde_MLE}
\end{subfigure}

\caption{KDE plots of hTPMH density concentration estimates.}
\label{fig:hTPMH_D_kde}
\end{figure}

\begin{figure}[p]
  \centering
  
  \begin{subfigure}{\textwidth}
  \centering \includegraphics[width=\textwidth,height=0.45\textheight,keepaspectratio]{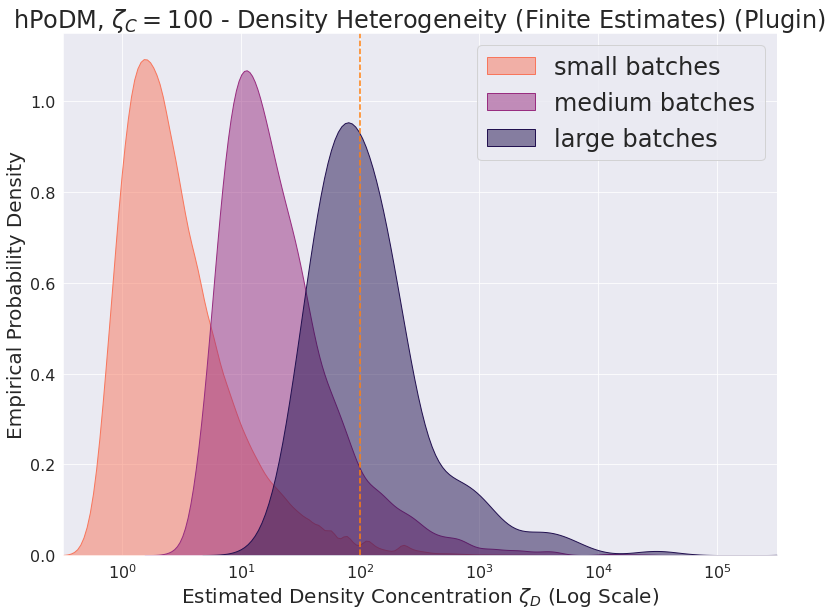}
  \caption[]{}
  \label{fig:hPoDM_100_D_kde_plugin}
\end{subfigure}

\begin{subfigure}{\textwidth}
  \centering \includegraphics[width=\textwidth,height=0.45\textheight,keepaspectratio]{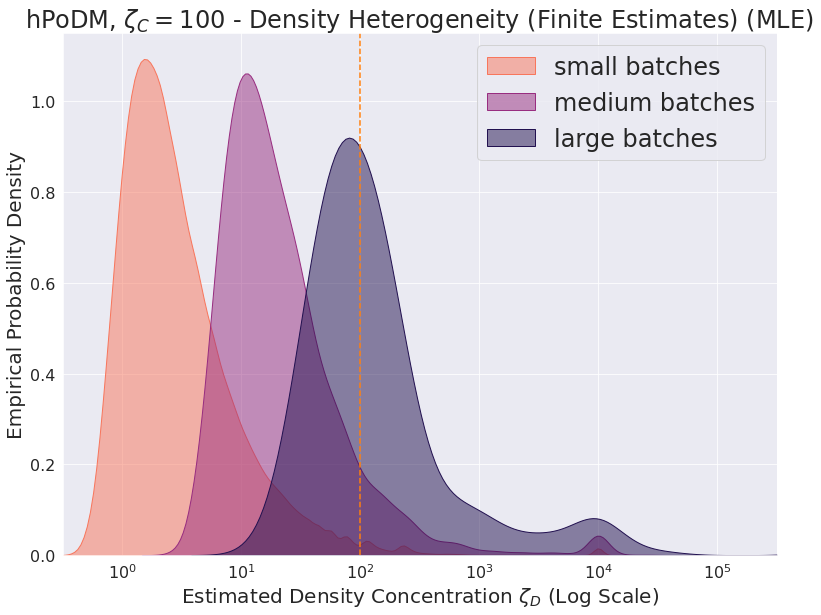}
  \caption[]{}
  \label{fig:hPoDM_100_D_kde_MLE}
\end{subfigure}

\caption{KDE plots of hPoDM ($\cconcentration=100$) density concentration estimates.}
\label{fig:hPoDM_100_D_kde}
\end{figure}

\begin{figure}[p]
  \centering
  
  \begin{subfigure}{\textwidth}
  \centering  \includegraphics[width=\textwidth,height=0.45\textheight,keepaspectratio]{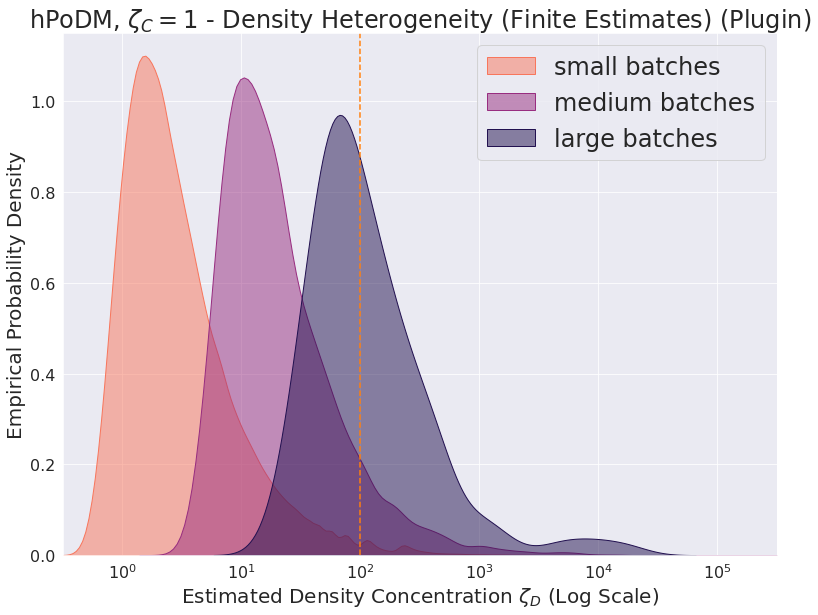}
  \caption[]{}
  \label{fig:hPoDM_1_D_kde_plugin}
\end{subfigure}

\begin{subfigure}{\textwidth}
  \centering
  \includegraphics[width=\textwidth,height=0.45\textheight,keepaspectratio]{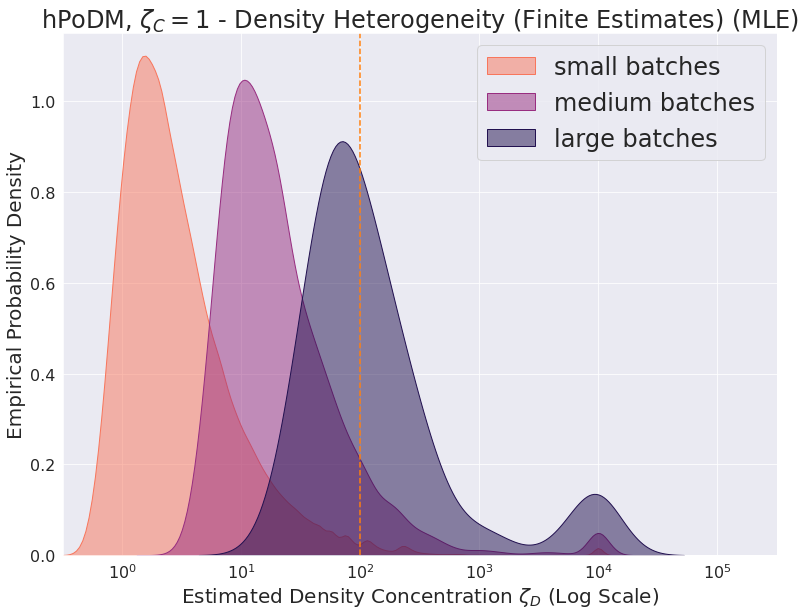}
  \caption[]{}
  \label{fig:hPoDM_1_D_kde_MLE}
\end{subfigure}

\caption{KDE plots of hPoDM ($\cconcentration=1$) density concentration estimates.}
\label{fig:hPoDM_1_D_kde}
\end{figure}

\begin{figure}[p]
  \centering
  
  \begin{subfigure}{\textwidth}
  \centering \includegraphics[width=\textwidth,height=0.45\textheight,keepaspectratio]{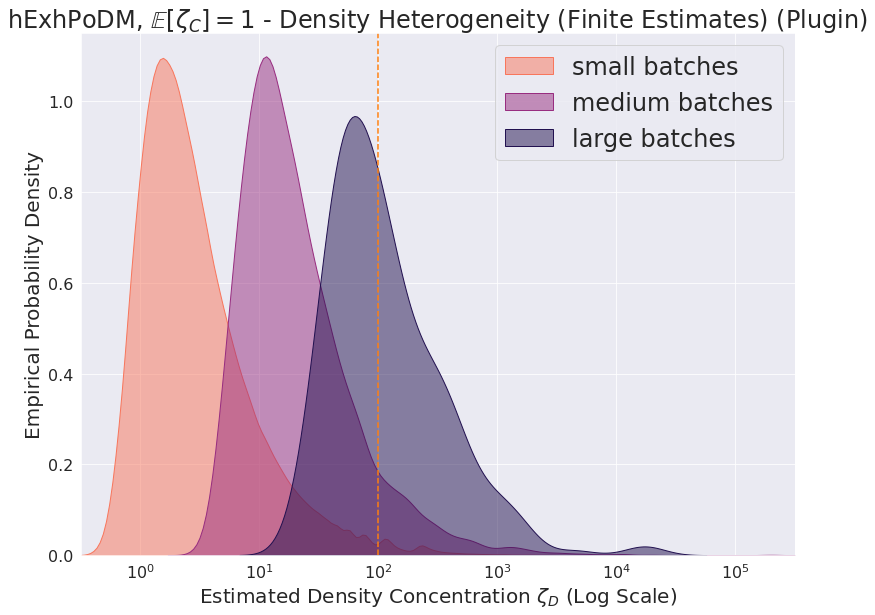}
  \caption[]{}
  \label{fig:hExhPoDM_D_kde_plugin}
\end{subfigure}

\begin{subfigure}{\textwidth}
  \centering
  \includegraphics[width=\textwidth,height=0.45\textheight,keepaspectratio]{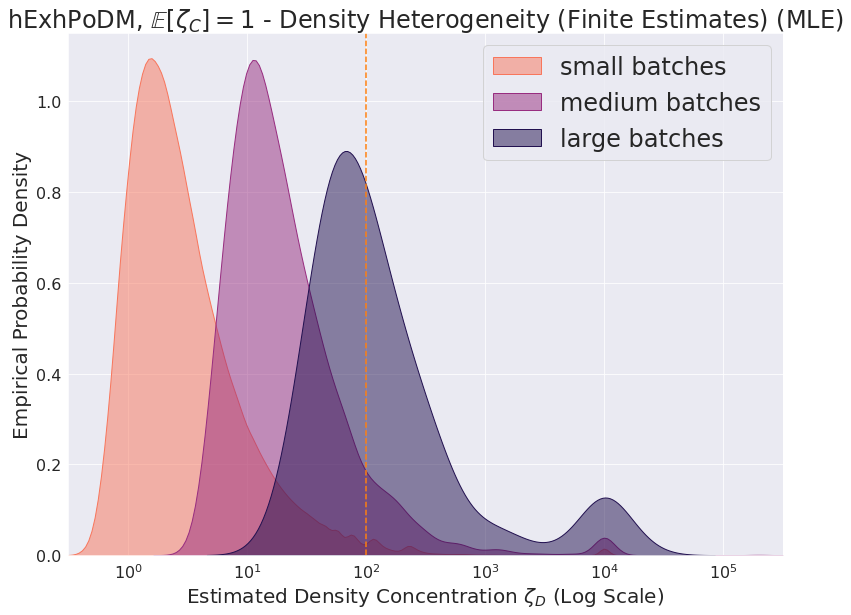}
  \caption[]{}
  \label{fig:hExhPoDM_D_kde_MLE}
\end{subfigure}

\caption{KDE plots of hExhPoDM density concentration estimates.}
\label{fig:hExhPoDM_D_kde}
\end{figure}

\begin{figure}[p]
  \centering
  
  \begin{subfigure}{\textwidth}
  \centering \includegraphics[width=\textwidth,height=0.45\textheight,keepaspectratio]{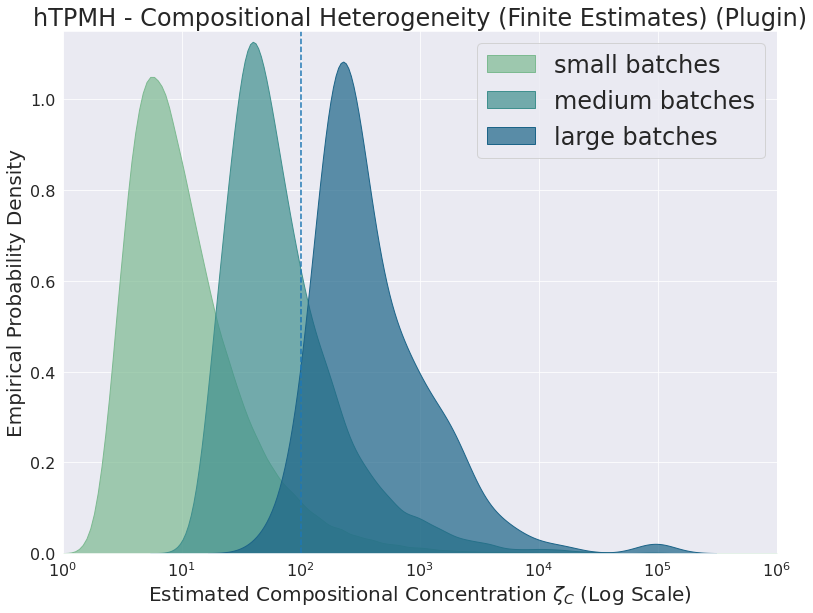}
  \caption[]{}
  \label{fig:hTPMH_C_kde_plugin}
\end{subfigure}

\begin{subfigure}{\textwidth}
  \centering
 \includegraphics[width=\textwidth,height=0.45\textheight,keepaspectratio]{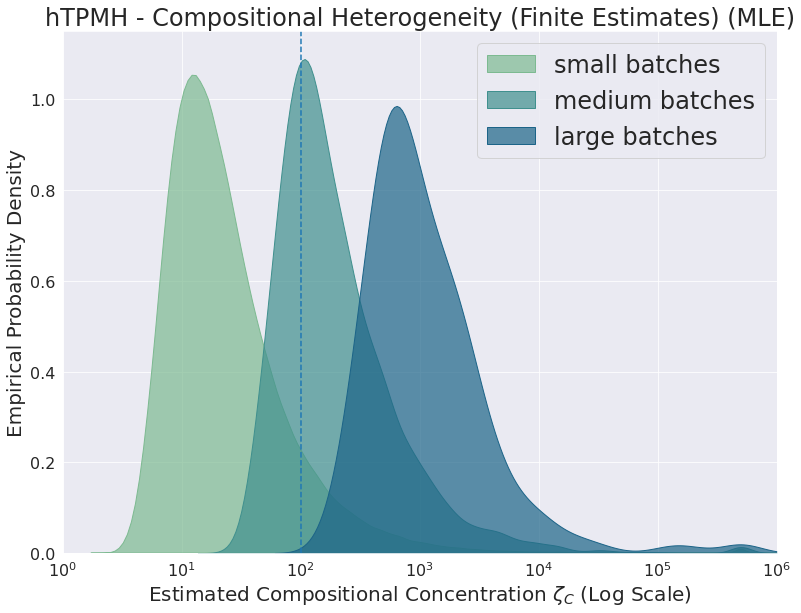}
  \caption[]{}
  \label{fig:hTPMH_C_kde_MLE}
\end{subfigure}

\caption{KDE plots of hTPMH compositional concentration estimates.}
\label{fig:hTPMH_C_kde}
\end{figure}

\subsection{Estimators are Consistent for Finite Concentrations}
\label{sec:estim-cons-finite}

When the true value of the concentration estimand is finite, we also see the correct behavior for a consistent estimator. Section \ref{sec:incorr-infin-estim} explains how the pattern of infinite estimates supports this conclusion. Section \ref{sec:accur-prec-both} explains how the estimators satisfy the two minimal requirements for the behavior of a ``decent'' estimator when the model is correctly specified. Section \ref{sec:accur-prec-both} explains why the estimates are also good even in the case when the model is not correctly specified.

\subsubsection{Incorrect Infinite Estimates Only Occur for Distributions Very Similar to hPoMu}
\label{sec:incorr-infin-estim}

For the distributions which are very similar to the homogeneous case with infinite concentration, the concentration estimates are sometimes infinite (cf. figures \ref{fig:hPoDM_100_C_survival_plugin}, \ref{fig:hNBDM_100_D_survival_plugin}, and \ref{fig:hNBDM_100_C_survival_plugin} for the plugin estimators and figures \ref{fig:hPoDM_100_C_survival_MLE}, \ref{fig:hNBDM_100_D_survival_MLE}, and \ref{fig:hNBDM_100_C_survival_MLE} for the ML estimators). Nevertheless, these figures also show that, as the batch size increases, the proportion of concentration estimates which are infinite decreases. This appears to occur more quickly for the compositional concentration estimators (see figures \ref{fig:hPoDM_100_C_survival_plugin} and \ref{fig:hNBDM_100_C_survival_plugin} for the plugin estimators and figures \ref{fig:hPoDM_100_C_survival_MLE} and \ref{fig:hNBDM_100_C_survival_MLE} for the ML estimators) than for the density concentration estimators (see figure \ref{fig:hNBDM_100_D_survival_plugin} for the plugin estimator and figure \ref{fig:hNBDM_100_D_survival_MLE} for the ML estimator).

\begin{figure}[p]
  \centering
  
  \begin{subfigure}{\textwidth}
  \centering
\includegraphics[width=\textwidth,height=0.45\textheight,keepaspectratio]{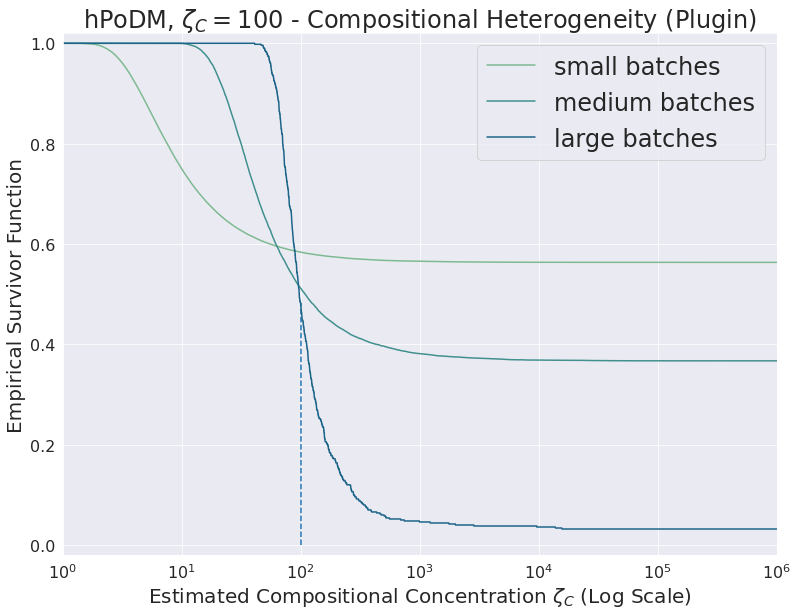}
  \caption[]{}
  \label{fig:hPoDM_100_C_survival_plugin}
\end{subfigure}

\begin{subfigure}{\textwidth}
  \centering
\includegraphics[width=\textwidth,height=0.45\textheight,keepaspectratio]{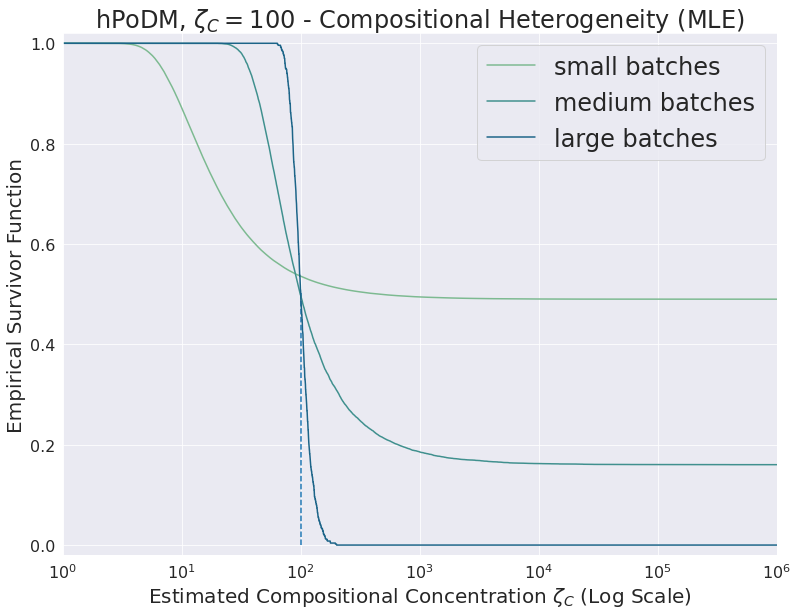}
  \caption[]{}
  \label{fig:hPoDM_100_C_survival_MLE}
\end{subfigure}

\caption{Survival functions of hPoDM ($\cconcentration=100$) compositional concentration estimates.}
\label{fig:hPoDM_100_C_survival}
\end{figure}

\begin{figure}[p]
  \centering
  
  \begin{subfigure}{\textwidth}
  \centering \includegraphics[width=\textwidth,height=0.45\textheight,keepaspectratio]{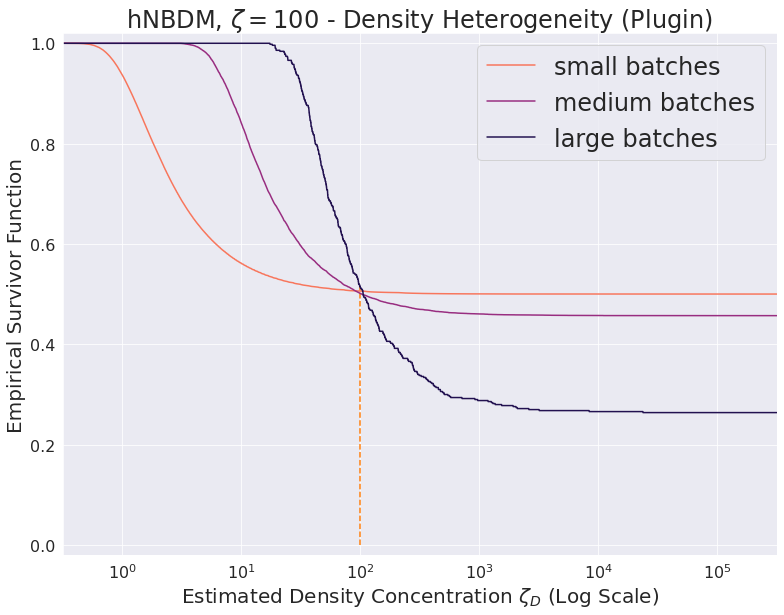}
  \caption[]{}
  \label{fig:hNBDM_100_D_survival_plugin}
\end{subfigure}

\begin{subfigure}{\textwidth}
  \centering
  \includegraphics[width=\textwidth,height=0.45\textheight,keepaspectratio]{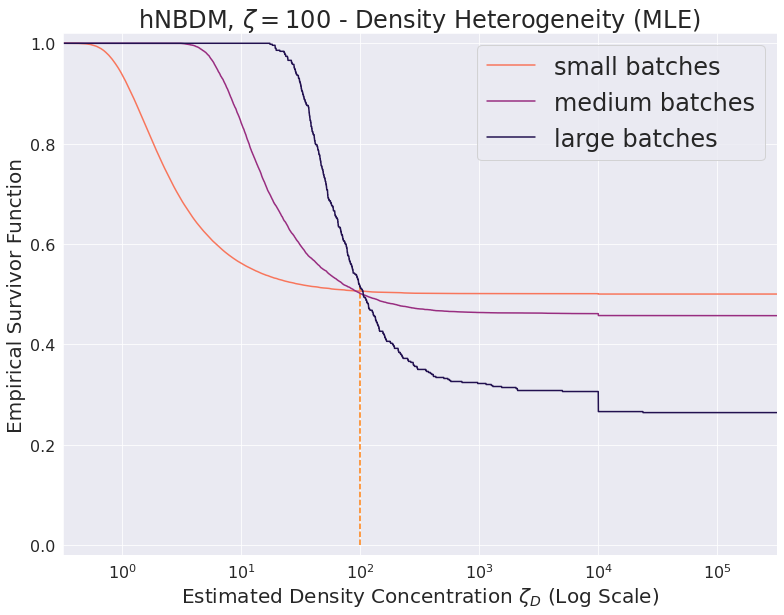}
  \caption[]{}
  \label{fig:hNBDM_100_D_survival_MLE}
\end{subfigure}

\caption{Survival functions of hNBDM ($\concentration=100$) density concentration estimates.}
\label{fig:hNBDM_100_D_survival}
\end{figure}

\begin{figure}[p]
  \centering
  
  \begin{subfigure}{\textwidth}
  \centering \includegraphics[width=\textwidth,height=0.45\textheight,keepaspectratio]{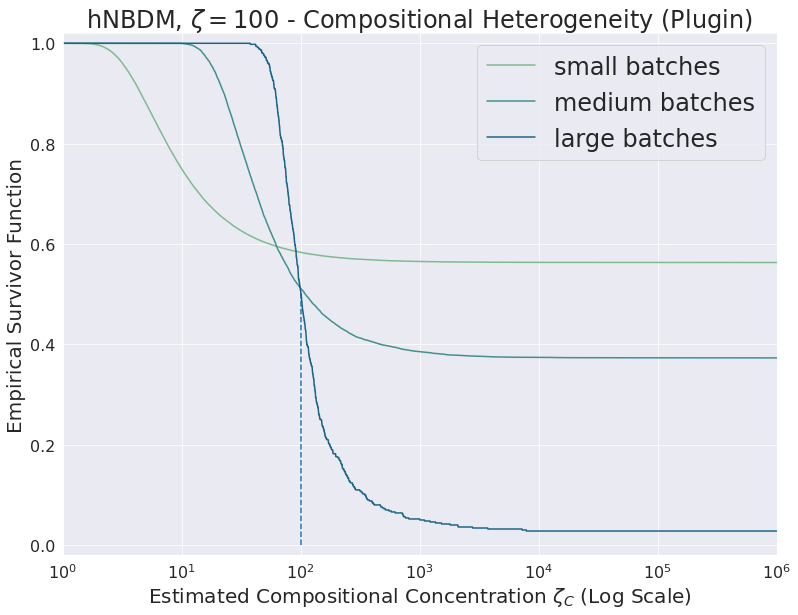}
  \caption[]{}
  \label{fig:hNBDM_100_C_survival_plugin}
\end{subfigure}

\begin{subfigure}{\textwidth}
  \centering
  \includegraphics[width=\textwidth,height=0.45\textheight,keepaspectratio]{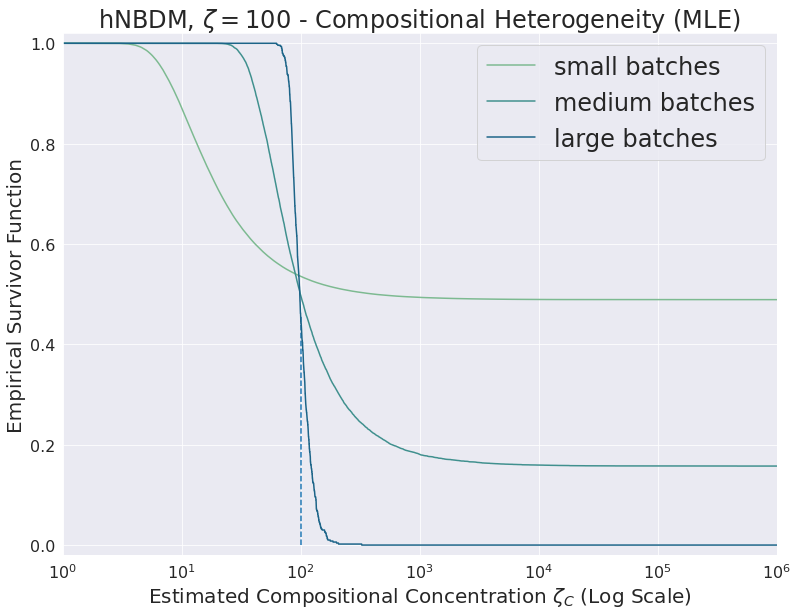}
  \caption[]{}
  \label{fig:hNBDM_100_C_survival_MLE}
\end{subfigure}

\caption{Survival functions of hNBDM ($\concentration=100$) compositional concentration estimates.}
\label{fig:hNBDM_100_C_survival}
\end{figure}

For the distributions which are less similar to hPoMu, when the true value of the concentration estimand is finite, the estimated values are either never or almost never infinite. This is why survival functions are not reported neither for the density concentration estimates of hNBDM with $\zeta=1$ and hExhNBDM with $\mathbb{E}[\zeta] = 1$ nor for the compositional concentration estimates of hPoDM with $\zeta_C=1$, hNBDM with $\zeta =1$, hExhPoDM with $\mathbb{E}[\zeta_C] = 1$, and hExhNBDM with $\mathbb{E}[\zeta] = 1$.

\subsubsection{Accuracy and Precision Both Increase with Batch Size}
\label{sec:accur-prec-both}

When the true value of the concentration estimand is finite, and the model is correctly specified, we see (cf. figures \ref{fig:hPoDM_100_C_kde_plugin}, \ref{fig:hNBDM_100_D_kde_plugin}, \ref{fig:hNBDM_100_C_kde_plugin}, \ref{fig:hPoDM_1_C_kde_plugin}, \ref{fig:hNBDM_1_D_kde_plugin}, and \ref{fig:hNBDM_1_C_kde_plugin} for the plugin estimators and figures \ref{fig:hPoDM_100_C_kde_MLE}, \ref{fig:hNBDM_100_D_kde_MLE}, \ref{fig:hNBDM_100_C_kde_MLE}, \ref{fig:hPoDM_1_C_kde_MLE}, \ref{fig:hNBDM_1_D_kde_MLE}, and \ref{fig:hNBDM_1_C_kde_MLE} for the ML estimators) that the distribution of the (finite) estimates is centered around the true value.

When the limit approached by the estimators is finite, even when the model is \textit{not} correctly specified, the variance of the distribution of estimated concentrations decreases as the batch size increases. Cf. figures \ref{fig:hPoDM_100_C_kde_plugin}, \ref{fig:hNBDM_100_D_kde_plugin}, \ref{fig:hNBDM_100_C_kde_plugin}, \ref{fig:hPoDM_1_C_kde_plugin}, \ref{fig:hNBDM_1_D_kde_plugin}, \ref{fig:hNBDM_1_C_kde_plugin}, \ref{fig:hExhPoDM_C_kde_plugin}, \ref{fig:hExhNBDM_D_kde_plugin}, and \ref{fig:hExhNBDM_C_kde_plugin} for the plugin estimators and figures \ref{fig:hPoDM_100_C_kde_MLE}, \ref{fig:hNBDM_100_D_kde_MLE}, \ref{fig:hNBDM_100_C_kde_MLE}, \ref{fig:hPoDM_1_C_kde_MLE}, \ref{fig:hNBDM_1_D_kde_MLE}, \ref{fig:hNBDM_1_C_kde_MLE}, \ref{fig:hExhPoDM_C_kde_MLE}, \ref{fig:hExhNBDM_D_kde_MLE}, and \ref{fig:hExhNBDM_C_kde_MLE} for the ML estimators.

\begin{figure}[p]
  \centering
  
  \begin{subfigure}{\textwidth}
  \centering \includegraphics[width=\textwidth,height=0.45\textheight,keepaspectratio]{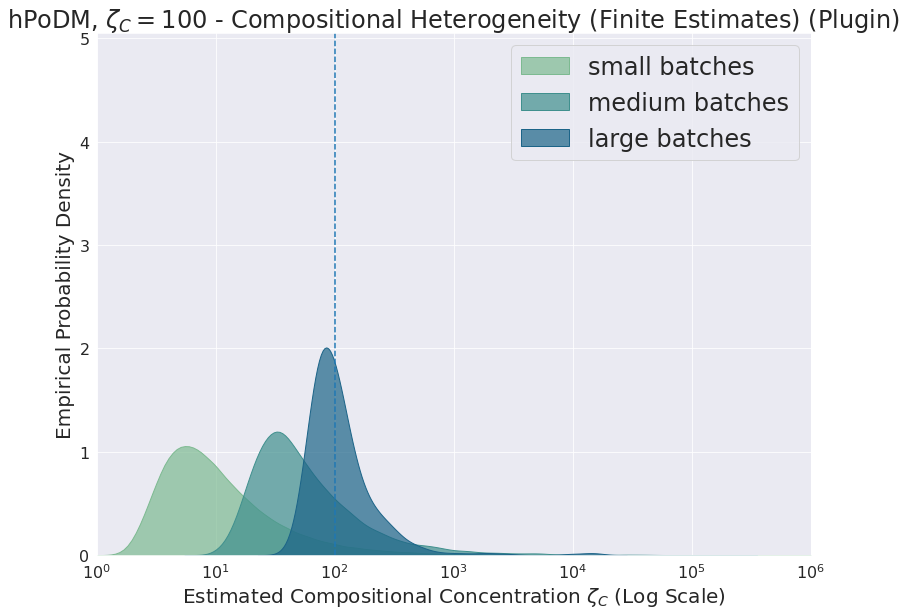}
  \caption[]{}
  \label{fig:hPoDM_100_C_kde_plugin}
\end{subfigure}

\begin{subfigure}{\textwidth}
  \centering
  \includegraphics[width=\textwidth,height=0.45\textheight,keepaspectratio]{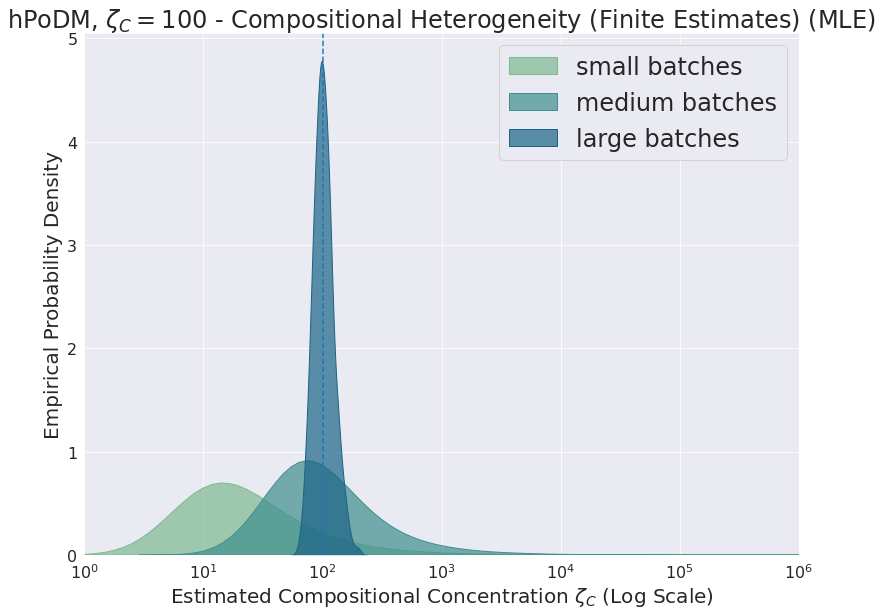}
  \caption[]{}
  \label{fig:hPoDM_100_C_kde_MLE}
\end{subfigure}

\caption{KDE plots of hPoDM ($\cconcentration=100$) compositional concentration estimates.}
\label{fig:hPoDM_100_C_kde}
\end{figure}

\begin{figure}[p]
  \centering
  
  \begin{subfigure}{\textwidth}
  \centering \includegraphics[width=\textwidth,height=0.45\textheight,keepaspectratio]{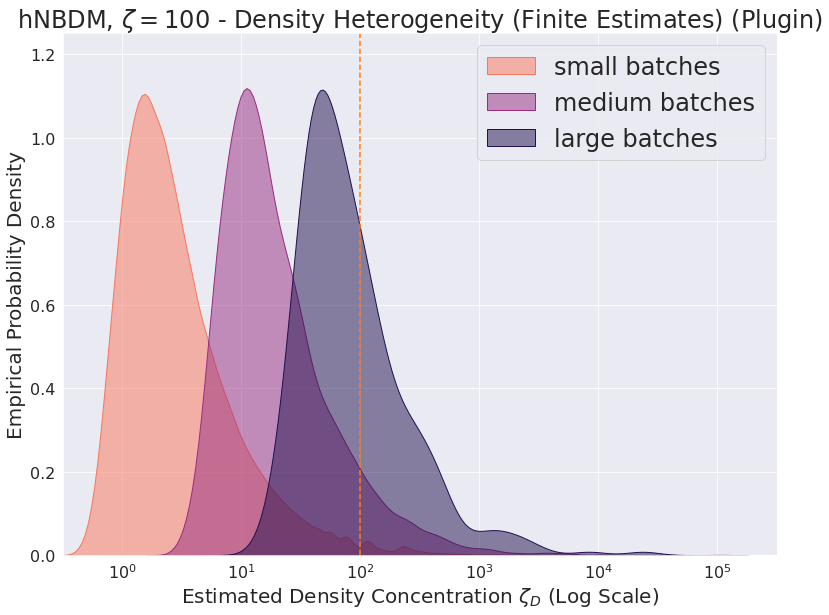}
  \caption[]{}
  \label{fig:hNBDM_100_D_kde_plugin}
\end{subfigure}

\begin{subfigure}{\textwidth}
  \centering
  \includegraphics[width=\textwidth,height=0.45\textheight,keepaspectratio]{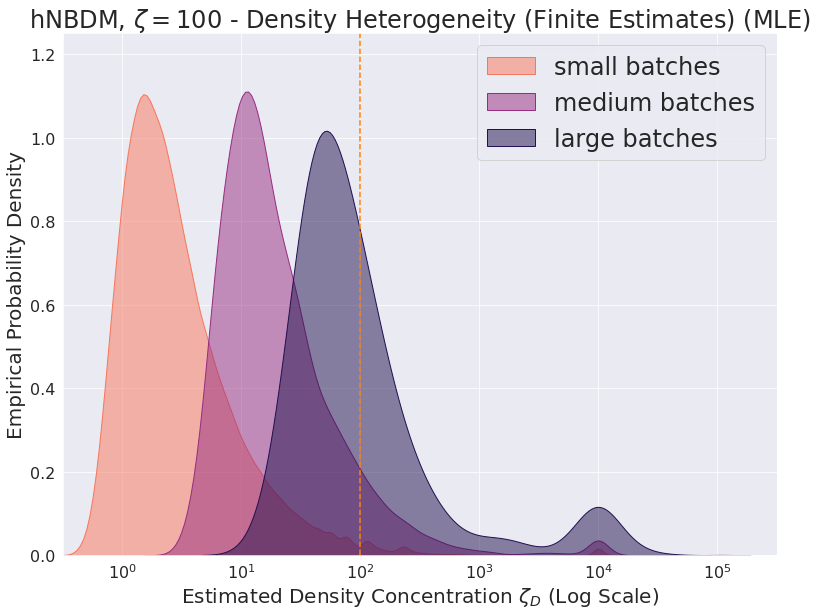}
  \caption[]{}
  \label{fig:hNBDM_100_D_kde_MLE}
\end{subfigure}

\caption{KDE plots of hNBDM ($\concentration=100$) density concentration estimates.}
\label{fig:hNBDM_100_kde}
\end{figure}

\begin{figure}[p]
  \centering
  
  \begin{subfigure}{\textwidth}
  \centering \includegraphics[width=\textwidth,height=0.45\textheight,keepaspectratio]{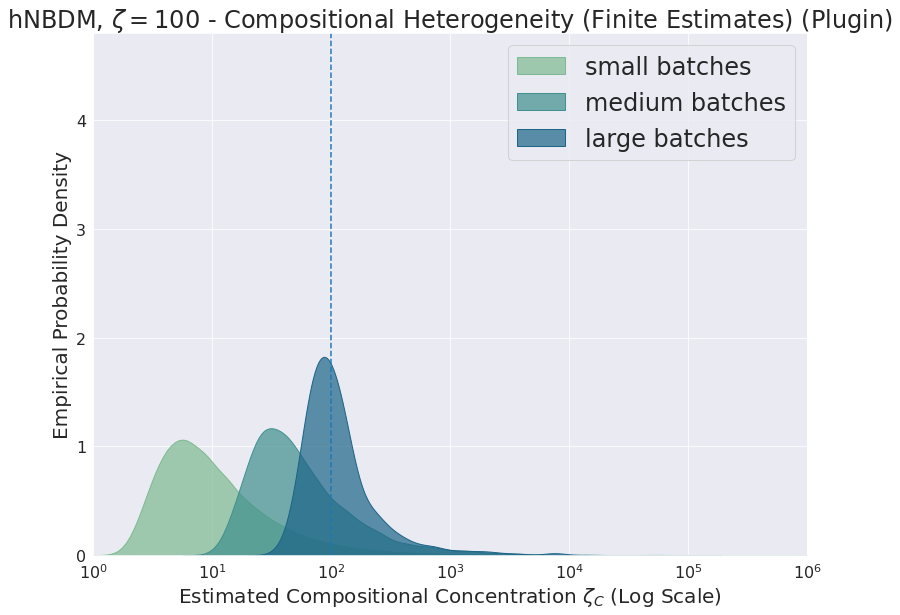}
  \caption[]{}
  \label{fig:hNBDM_100_C_kde_plugin}
\end{subfigure}

\begin{subfigure}{\textwidth}
  \centering
  \includegraphics[width=\textwidth,height=0.45\textheight,keepaspectratio]{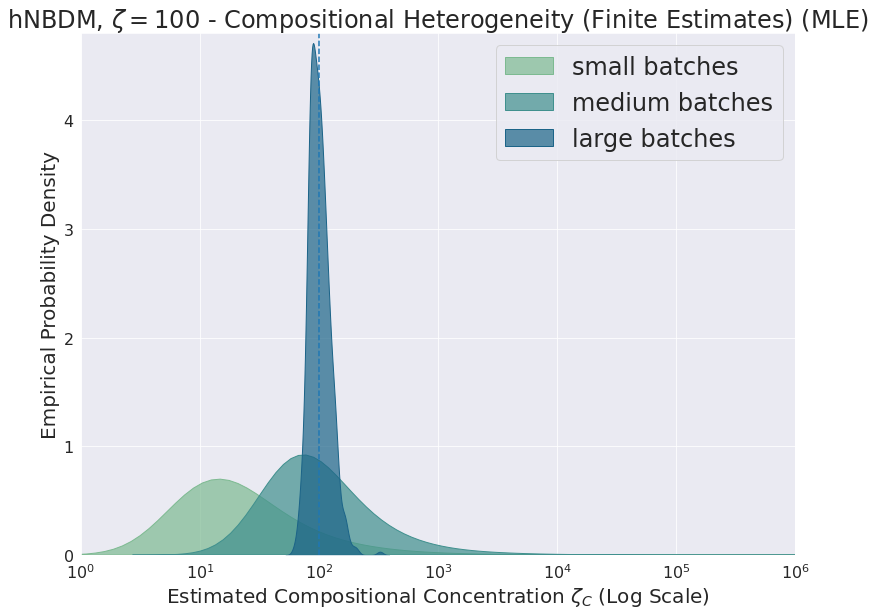}
  \caption[]{}
  \label{fig:hNBDM_100_C_kde_MLE}
\end{subfigure}

\caption{KDE plots of hNBDM ($\concentration=100$) compositional concentration estimates.}
\label{fig:hNBDM_100_C_kde}
\end{figure}

\begin{figure}[p]
  \centering
  
  \begin{subfigure}{\textwidth}
  \centering \includegraphics[width=\textwidth,height=0.45\textheight,keepaspectratio]{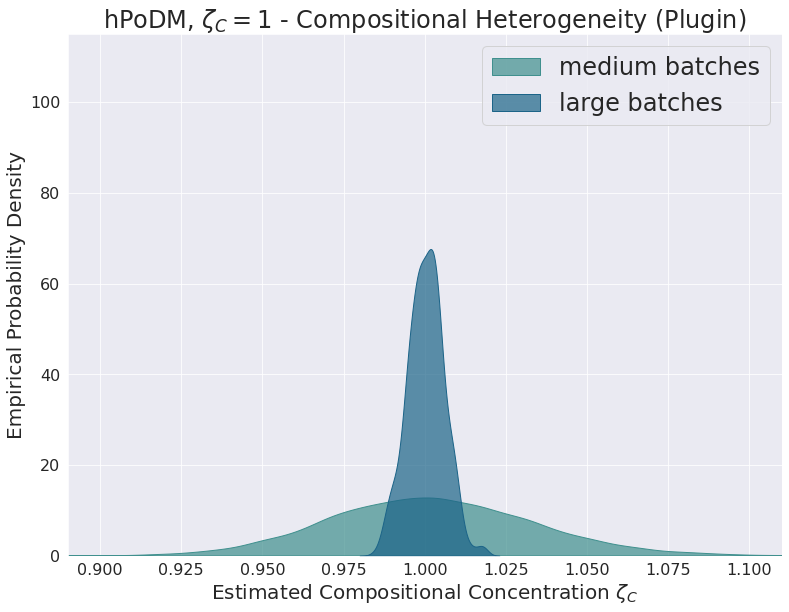}
  \caption[]{}
  \label{fig:hPoDM_1_C_kde_plugin}
\end{subfigure}

\begin{subfigure}{\textwidth}
  \centering
  \includegraphics[width=\textwidth,height=0.45\textheight,keepaspectratio]{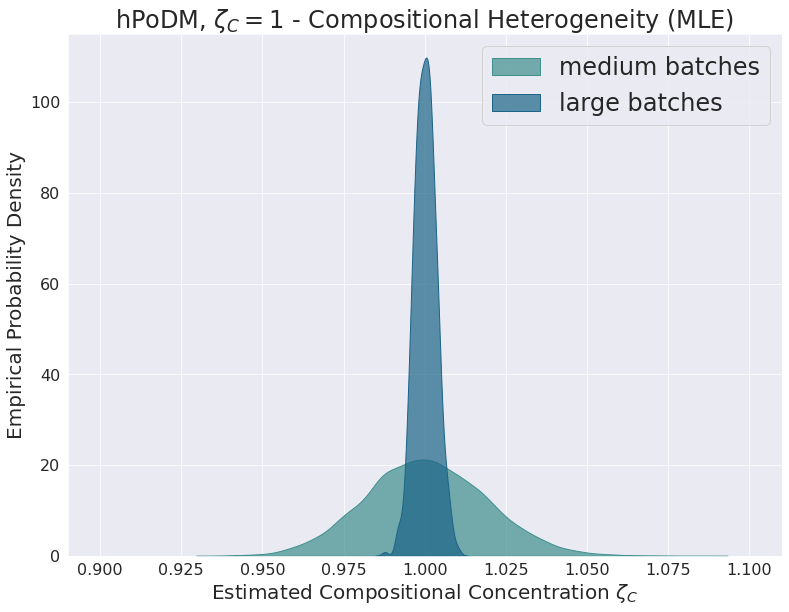}
  \caption[]{}
  \label{fig:hPoDM_1_C_kde_MLE}
\end{subfigure}

\caption{KDE plots of hPoDM ($\cconcentration=1$) compositional concentration estimates.}
\label{fig:hPoDM_1_C_kde}
\end{figure}

\begin{figure}[p]
  \centering
  
  \begin{subfigure}{\textwidth}
  \centering \includegraphics[width=\textwidth,height=0.45\textheight,keepaspectratio]{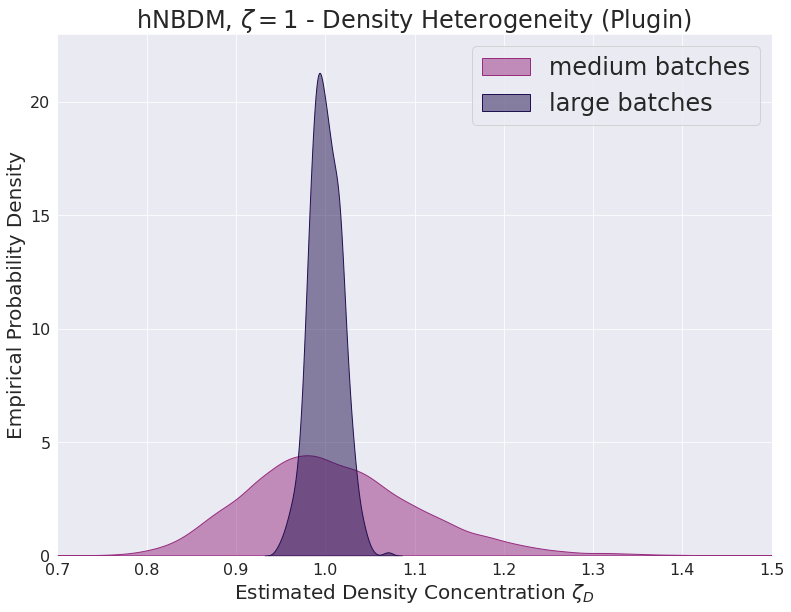}
  \caption[]{}
  \label{fig:hNBDM_1_D_kde_plugin}
\end{subfigure}

\begin{subfigure}{\textwidth}
  \centering
  \includegraphics[width=\textwidth,height=0.45\textheight,keepaspectratio]{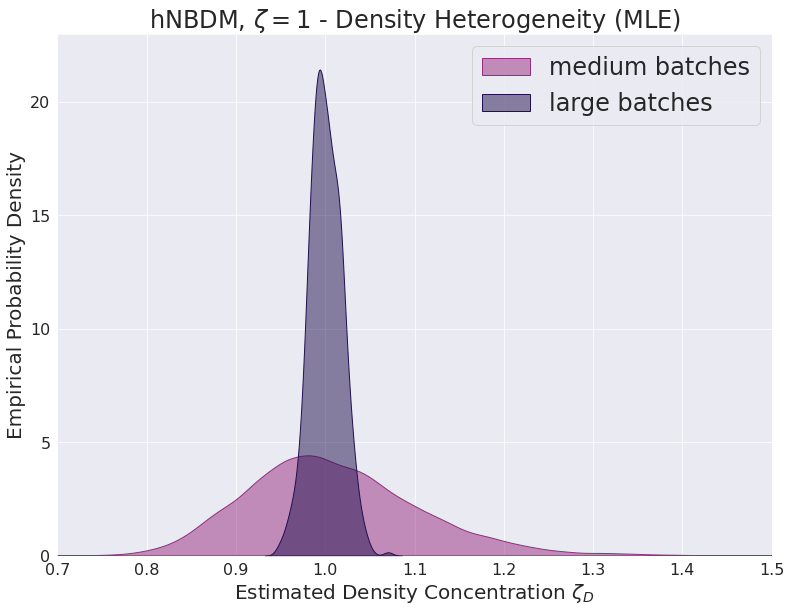}
  \caption[]{}
  \label{fig:hNBDM_1_D_kde_MLE}
\end{subfigure}

\caption{KDE plots of hNBDM ($\concentration=1$) density concentration estimates.}
\label{fig:hNBDM_1_D_kde}
\end{figure}

\begin{figure}[p]
  \centering
  
  \begin{subfigure}{\textwidth}
  \centering \includegraphics[width=\textwidth,height=0.45\textheight,keepaspectratio]{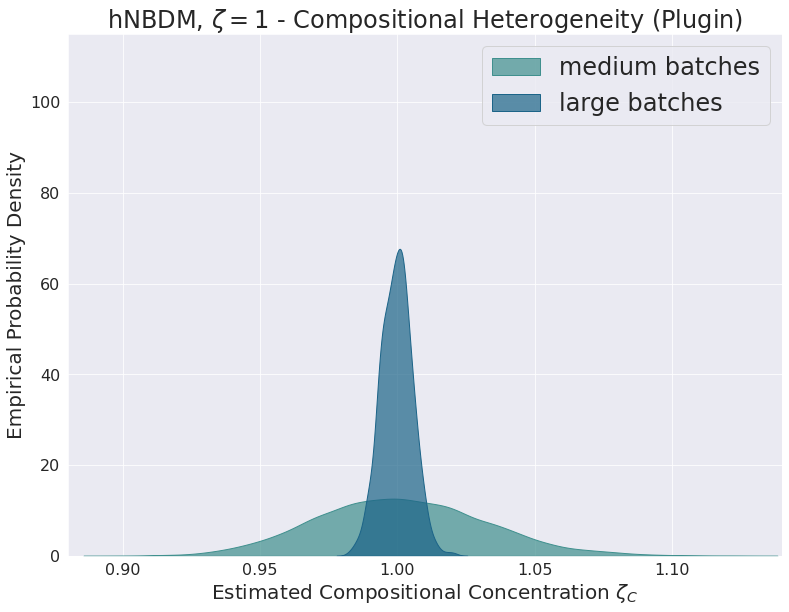}
  \caption[]{}
  \label{fig:hNBDM_1_C_kde_plugin}
\end{subfigure}

\begin{subfigure}{\textwidth}
  \centering
  \includegraphics[width=\textwidth,height=0.45\textheight,keepaspectratio]{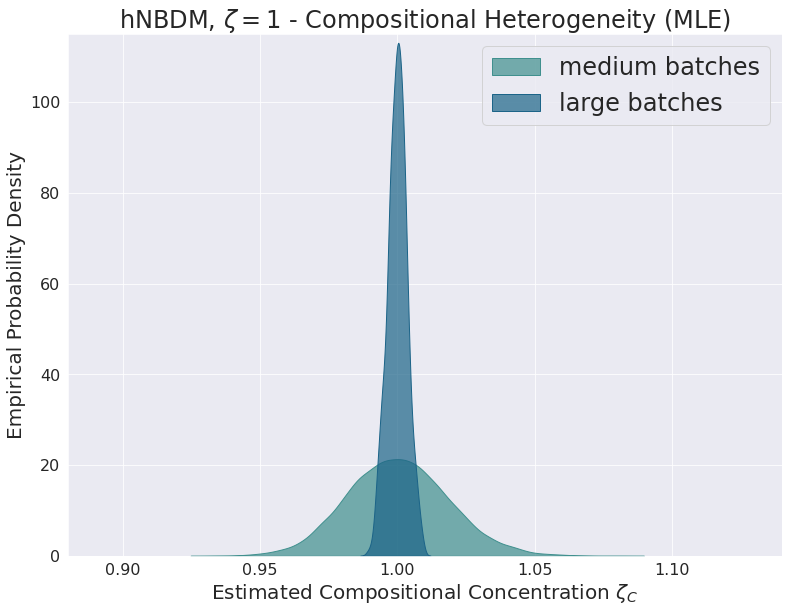}
  \caption[]{}
  \label{fig:hNBDM_1_C_kde_MLE}
\end{subfigure}

\caption{KDE plots of hNBDM ($\concentration=1$) compositional concentration estimates.}
\label{fig:hNBDM_1_C_kde}
\end{figure}

\subsubsection{Results are Qualitatively Correct under Model Misspecification}
\label{sec:results-are-qual}

For hExhPoDM with $\mathbb{E}[\zeta_C] = 1$, and hExhNBDM with $\mathbb{E}[\zeta] = 1$, the model is misspecified\footnote{\label{footnote:misspecified}
Note that technically for hExhPoDM, only the categorical distribution, corresponding to the compositional concentration, is misspecified relative to the chosen estimators. For hExhNBDM both the count distribution, corresponding to the density concentration, and the categorical distribution are misspecified relative to the chosen estimators.
}, since these distributions do not have single values of concentration parameters. Nevertheless, from the results in previous chapters it is also clear that these are the most heterogeneous distributions. These estimators therefore give qualitatively correct answers for these distributions, despite the model being misspecified, because (cf. figures \ref{fig:hExhPoDM_C_kde_plugin}, \ref{fig:hExhNBDM_D_kde_plugin}, and \ref{fig:hExhNBDM_C_kde_plugin} for the plugin estimators and figures \ref{fig:hExhPoDM_C_kde_MLE}, \ref{fig:hExhNBDM_D_kde_MLE}, and \ref{fig:hExhNBDM_C_kde_MLE} for the ML estimators) the lowest concentration estimates occur for these distributions.

\begin{figure}[p]
  \centering
  
  \begin{subfigure}{\textwidth}
  \centering \includegraphics[width=\textwidth,height=0.45\textheight,keepaspectratio]{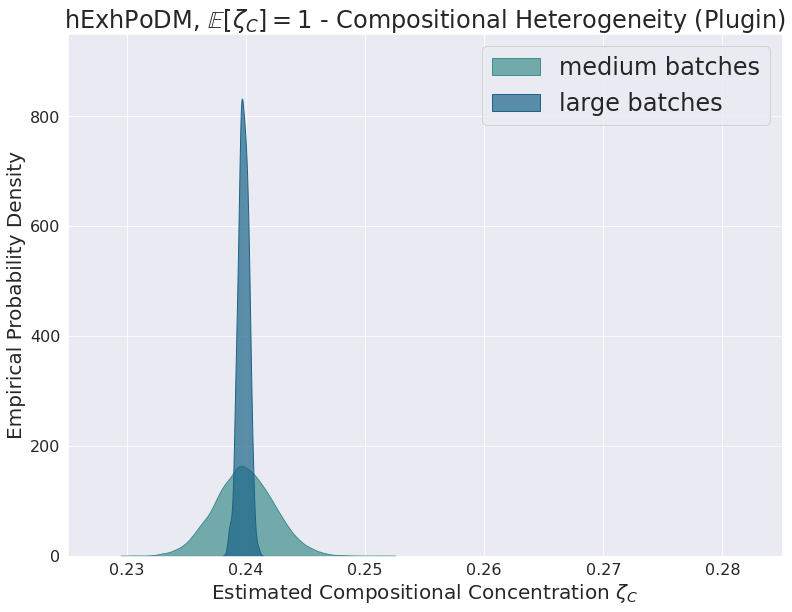}
  \caption[]{}
  \label{fig:hExhPoDM_C_kde_plugin}
\end{subfigure}

\begin{subfigure}{\textwidth}
  \centering
  \includegraphics[width=\textwidth,height=0.45\textheight,keepaspectratio]{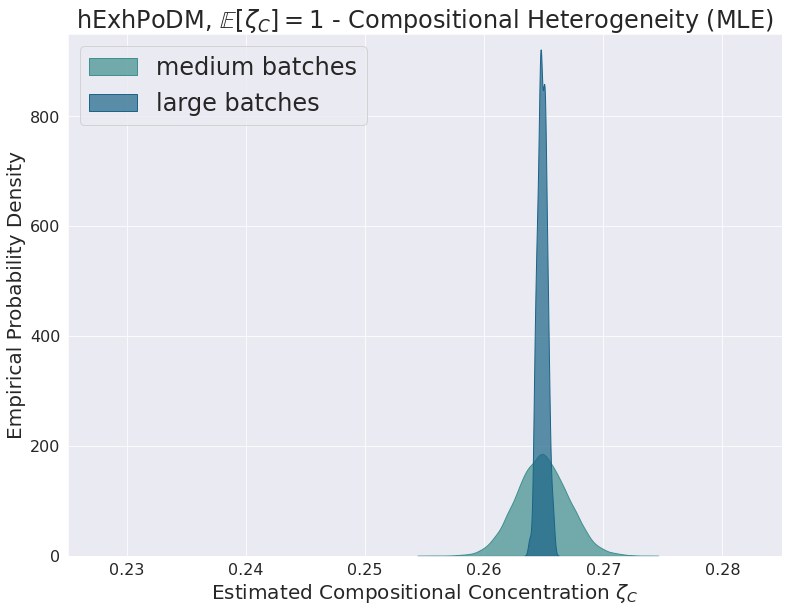}
  \caption[]{}
  \label{fig:hExhPoDM_C_kde_MLE}
\end{subfigure}

\caption{KDE plots of hExhPoDM compositional concentration estimates.}
\label{fig:hExhPoDM_C_kde}
\end{figure}

\begin{figure}[p]
  \centering
  
  \begin{subfigure}{\textwidth}
  \centering \includegraphics[width=\textwidth,height=0.45\textheight,keepaspectratio]{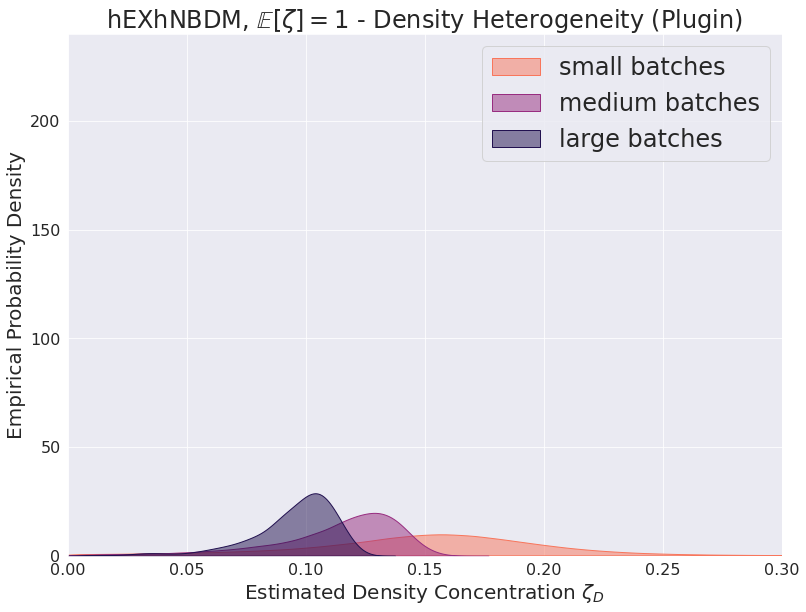}
  \caption[]{}
  \label{fig:hExhNBDM_D_kde_plugin}
\end{subfigure}

\begin{subfigure}{\textwidth}
  \centering
  \includegraphics[width=\textwidth,height=0.45\textheight,keepaspectratio]{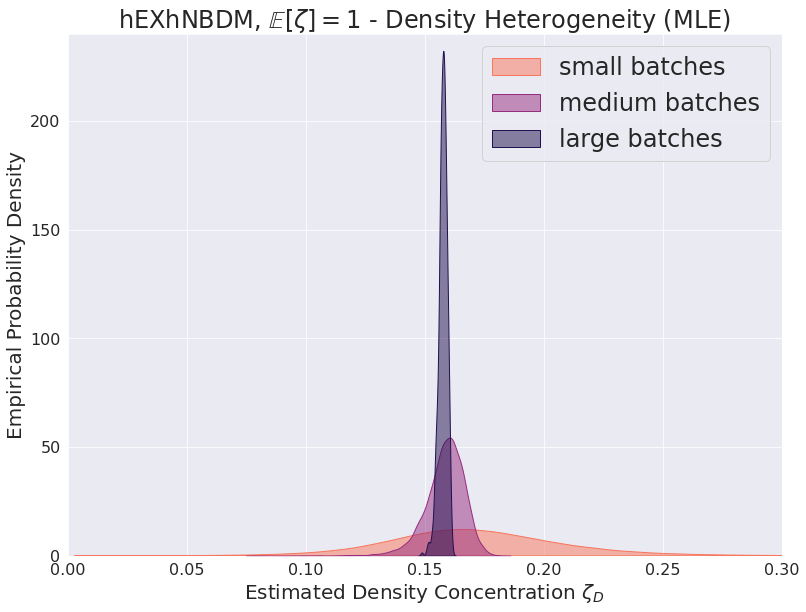}
  \caption[]{}
  \label{fig:hExhNBDM_D_kde_MLE}
\end{subfigure}

\caption{KDE plots of hExhNBDM density concentration estimates.}
\label{fig:hExhNBDM_D_kde}
\end{figure}

\begin{figure}[p]
  \centering
  
  \begin{subfigure}{\textwidth}
  \centering \includegraphics[width=\textwidth,height=0.45\textheight,keepaspectratio]{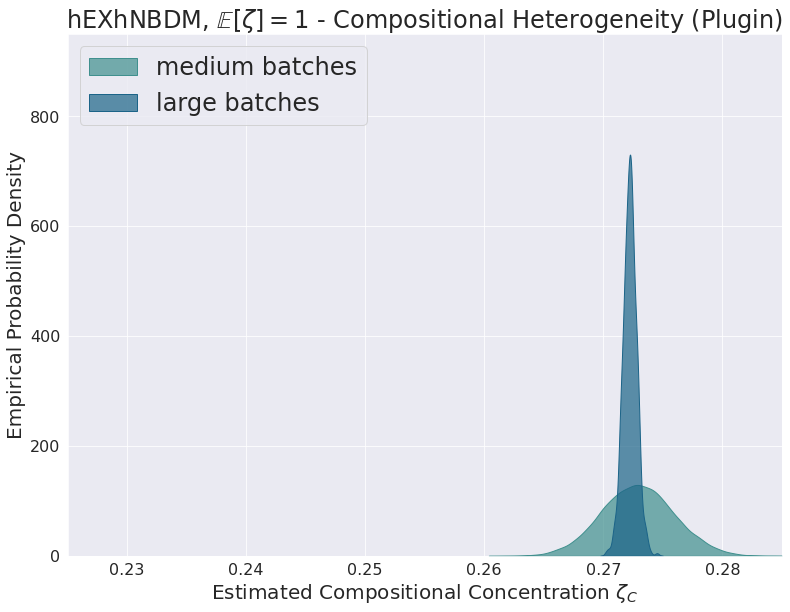}
  \caption[]{}
  \label{fig:hExhNBDM_C_kde_plugin}
\end{subfigure}

\begin{subfigure}{\textwidth}
  \centering
  \includegraphics[width=\textwidth,height=0.45\textheight,keepaspectratio]{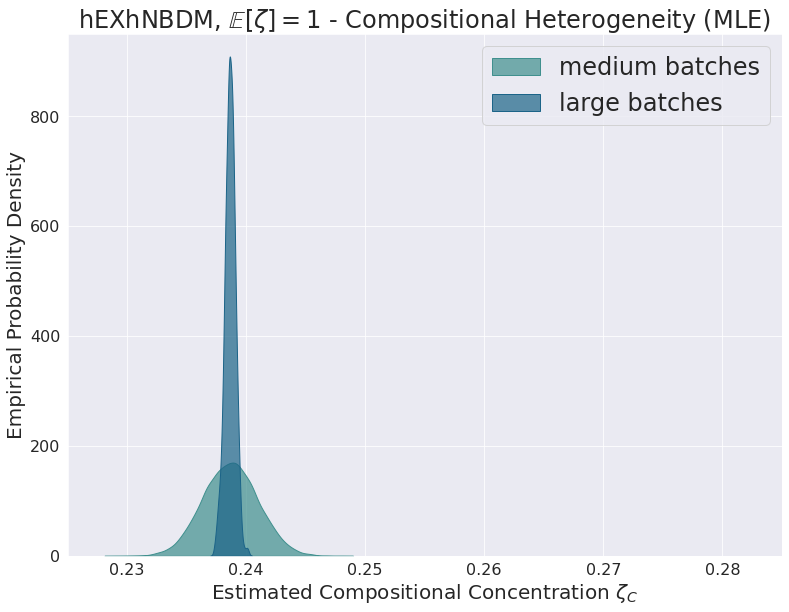}
  \caption[]{}
  \label{fig:hExhNBDM_C_kde_MLE}
\end{subfigure}

\caption{KDE plots of hExhNBDM density concentration estimates.}
\label{fig:hExhNBDM_C_kde}
\end{figure}

\subsection{Comparing ML Estimator and Plugin Estimator Results}
\label{sec:comp-ml-estim}

An important reason for using ML estimators over plugin estimators is increased efficiency. Cf. the discussion from the beginning of section \ref{sec:inference-methods}. Therefore I evaluate whether the precision of the ML estimator appears to increase much more rapidly than that of its plugin counterpart for the density estimators in section \ref{sec:prec-ml-estim} and for the concentration estimators in section \ref{sec:prec-ml-estim-1}. One reason for using the plugin estimators over ML estimators is because the plugin estimators arguably make fewer model assumptions. Again, cf. the discussion at the beginning of section \ref{sec:inference-methods}. Therefore in section \ref{sec:behav-ml-estim} I compare the behaviors of the plugin and ML estimators for the misspecified models.

\subsubsection{Precision of ML Estimator vs. Plugin Estimator for Density Concentration}
\label{sec:prec-ml-estim}

For density concentration, both ML and plugin appear to demonstrate effectively the same efficiency in practice. See figures \ref{fig:hTPMH_D_survival_plugin},  \ref{fig:hTPMH_D_survival_MLE},  \ref{fig:hPoDM_100_D_survival_plugin}, \ref{fig:hPoDM_100_D_survival_MLE},\ref{fig:hPoDM_1_D_survival_plugin}, \ref{fig:hPoDM_1_D_survival_MLE}, \ref{fig:hExhPoDM_D_survival_plugin}, and \ref{fig:hExhPoDM_D_survival_MLE} for survival curves when the true value of the estimand is infinite, figures \ref{fig:hNBDM_100_D_survival_plugin} and \ref{fig:hNBDM_100_D_survival_MLE} for survival curves when the true value of the estimand is finite, figures \ref{fig:hTPMH_D_kde_plugin}, \ref{fig:hTPMH_D_kde_MLE}, \ref{fig:hPoDM_100_D_kde_plugin}, \ref{fig:hPoDM_100_D_kde_MLE}, \ref{fig:hPoDM_1_D_kde_plugin}, \ref{fig:hPoDM_1_D_kde_MLE}, \ref{fig:hExhPoDM_D_kde_plugin}, and \ref{fig:hExhPoDM_D_kde_MLE} for KDE plots when the true value of the estimand is infinite, and figures \ref{fig:hNBDM_100_D_kde_plugin}, \ref{fig:hNBDM_100_D_kde_MLE}, \ref{fig:hNBDM_1_D_kde_plugin}, and \ref{fig:hNBDM_1_D_kde_MLE} for KDE plots when the true value of the estimand is finite.

I double-checked the exact values and the estimates produced by the two estimators really are different, even though these differences are not apparent\footnote{Except for the bumps around $10^4$, see below.} from the aforementioned graphs. This makes sense given that the median difference in the estimated values was never greater than $3 \times 10^{-4}$ (data not shown) and was often smaller (especially for the larger batches). Evidently the difference in the asymptotic efficiency (as computed in \cite{Anscombe1950}) of the negative binomial ML and plugin estimators is negligible for the chosen sample sizes.

Bizarrely, hExhNBDM appeared to be the only distribution for which the ML density estimator was noticeably more efficient than the plugin density estimator, despite (or perhaps because?) of the model misspecification. Compare figures \ref{fig:hExhNBDM_D_kde_plugin} and \ref{fig:hExhNBDM_D_kde_MLE}. I am unsure why this only occurred for hExhNBDM.

The bump in the density concentration estimates which consistently occurs around $10^4$ for the ML estimates appears to be an artifact of the difficulties encountered when relying on Powell's method \cite{Powell} instead of Brent's method \cite{Brent} to find the root of the score, cf. the discussion in section \ref{sec:computing-mles}. See figures \ref{fig:hTPMH_D_survival_MLE}, \ref{fig:hTPMH_D_kde_MLE}, \ref{fig:hPoDM_100_D_survival_MLE}, \ref{fig:hPoDM_100_D_kde_MLE}, \ref{fig:hNBDM_100_D_survival_MLE}, \ref{fig:hNBDM_100_D_kde_MLE}, \ref{fig:hPoDM_1_D_survival_MLE}, \ref{fig:hPoDM_1_D_kde_MLE}, \ref{fig:hExhPoDM_D_survival_MLE}, and \ref{fig:hExhPoDM_D_kde_MLE}. These bumps likely would not occur if the MLE was better implemented.

\subsubsection{Precision of ML Estimator vs. Plugin Estimator for Compositional concentration}
\label{sec:prec-ml-estim-1}

Unlike the situation for the density estimators, the compositional ML estimator is noticeably more efficient than the compositional plugin estimator for all of the distributions, regardless of whether the model was correctly specified. 

See figures \ref{fig:hTPMH_C_survival_plugin} and \ref{fig:hTPMH_C_survival_MLE} for survival curves when the true value of the estimand is infinite. See figures \ref{fig:hPoDM_100_C_survival_plugin}, \ref{fig:hPoDM_100_C_survival_MLE}, \ref{fig:hNBDM_100_C_survival_plugin}, and \ref{fig:hNBDM_100_C_survival_MLE} for survival curves when the true value of the estimand is finite. Despite how difficult it usually is for most of the estimators to distinguish distributions with infinite concentration from those with concentration $\approx 100$, for the large batches the ML estimator remarkably produced almost no infinite estimates. The increased efficiency of the ML estimator is evident from the more rapid decrease compared to the plugin estimator of the survival curves when the true value of the estimand is finite.

See figures \ref{fig:hTPMH_C_kde_plugin} and \ref{fig:hTPMH_C_kde_MLE} for KDE plots when the true value of the estimand is infinite. The greater efficiency of the ML estimator is reflected by the distribution of finite estimates shifting further to the right more quickly. When the true value of the estimand is finite, as well as when the model is misspecified, the greater efficiency of the ML estimator compared to the plugin estimator is reflected by the narrower distribution of estimates. See figures \ref{fig:hPoDM_100_C_kde_plugin}, \ref{fig:hPoDM_100_C_kde_MLE}, \ref{fig:hNBDM_100_C_kde_plugin}, \ref{fig:hNBDM_100_C_kde_MLE}, \ref{fig:hPoDM_1_C_kde_plugin}, \ref{fig:hPoDM_1_C_kde_MLE}, \ref{fig:hNBDM_1_C_kde_plugin}, \ref{fig:hNBDM_1_C_kde_MLE}, \ref{fig:hExhPoDM_C_kde_plugin}, \ref{fig:hExhPoDM_C_kde_MLE}, \ref{fig:hExhNBDM_C_kde_plugin}, and \ref{fig:hExhNBDM_C_kde_MLE}.

\subsubsection{Behavior of ML vs Plugin Estimators Under Misspecification}
\label{sec:behav-ml-estim}

Under the misspecified\footnote{
Cf. footnote \ref{footnote:misspecified} above.
} models hExhPoDM and hExhNBDM, the effect on the limit of switching from plugin to ML was opposite for hExhPoDM compared to hExhNBDM. For the compositional concentration, switching from the plugin estimator to the ML estimator \textit{increased} the limiting value of the estimates for hExhPoDM (cf. figures \ref{fig:hExhPoDM_C_kde_plugin} and \ref{fig:hExhPoDM_C_kde_MLE}), while for hExhNBDM switching from the plugin estimator to the ML estimator \textit{decreased} the limiting value for the estimates (cf. figures \ref{fig:hExhNBDM_C_kde_plugin}, and \ref{fig:hExhNBDM_C_kde_MLE}). For hExhNBDM the effect of switching from the plugin estimator to the ML estimator for the density concentration was opposite the effect for the compositional concentration; the limiting value of the estimates \textit{increased} (cf. figures \ref{fig:hExhNBDM_D_kde_plugin} and \ref{fig:hExhNBDM_D_kde_MLE}). That the plugin and ML estimators converge to different values from each other under these distributions reflects the model misspecification.

\section{Discussion}
\label{sec:hetero_estimator_performance_discussion}

Section \ref{sec:estimators-are-good} discusses why the results from section \ref{sec:hetero_estimator_performance_results} are important for showing that estimates from the estimators will be useful in practice. Section \ref{sec:heter-might-corr} discusses how the results from section \ref{sec:hetero_estimator_performance_results} suggest that future work might be able formulate density and compositional heterogeneities in a nonparametric context. Section \ref{sec:usef-heter-estim} discusses how the results from section \ref{sec:hetero_estimator_performance_results} bring us one step closer to the final goal of this thesis.

\subsection{The Estimators are Good Enough for Use in Practice}
\label{sec:estimators-are-good}

The results show that the proposed estimators both (1) converge towards correct answers as the data size increases and (2) have decreasing variance as the data size increases. Hence they satisfy the minimal requirements one expects of ``adequate'' statistical estimators.

The estimators were already guaranteed to be consistent asymptotically when the statistical model was correctly specified. Nevertheless, that theoretical guarantee does not necessarily translate in practice into estimates usefully approximating the truth for realistic data sizes. (In this case realistic data sizes correspond to the medium or large batches.) Such a failure of ``practical consistency'' occurred, for example, for the version of the plugin compositional concentration estimator that equally weighted the estimates resulting from all pairs of strains. Consistency of an estimator need not imply efficiency of an estimator. The above results show that the proposed estimators are in fact ``practically consistent'', not merely asymptotically consistent. Not only that, but they often still managed to give useful results even for unrealistically small data sizes (the small batches).

\subsection{Heterogeneity Might be Nonparametric}
\label{sec:heter-might-corr}

The plugin estimators also gave sensible results in the cases of model misspecification. This suggests that functions of moments, which are at least similar to the ``effective density concentration'' (\ref{eq:effective_dconcentration}) and ``effective compositional concentration'' (\ref{eq:effective_compositional_concentration_official}) that these estimators converge to, may be useful for describing qualitative notions of ``density heterogeneity'' and ``compositional heterogeneity'' for general\footnote{It probably makes sense only to attempt to define such notions for distributions with finite second moments, or to at least automatically say that any distribution lacking finite second moments is ``infinitely heterogeneous'' without exploring further.} count-categorical distributions. Exploring how to define such notions precisely for more general distributions outside of the ghNBDM family would be a worthwhile goal for future work. This idea resembles the argument from \cite[section 4]{Nakashima1997} that the negative binomial plugin estimator is ``semi-parametric''.

It is worth noting however that, at least for hExhPoDM and hExhNBDM, the ML estimators were also ``unreasonably effective'' at giving sensible results even under model misspecification. Thus we might anticipate either that (1) a ``nonparametric'' or ``semi-parametric'' interpretation of their estimands also exists, or (2) that the ML estimators would not prove equally ``unreasonably effective'' for other misspecified models. \cite{Nakashima1997} appears to argue the latter, at least with respect to the density concentration estimator. Unfortunately this work does not investigate the behavior of these estimators under misspecification enough to distinguish between these two possibilities, cf. footnote \ref{footnote:plugin_robustness}.

\section{Conclusion}
\label{sec:usef-heter-estim}

\paragraph{Practical Implications}
This chapter shows that we have viable statistical estimators for density heterogeneity and compositional heterogeneity. Therefore, neither quantity needs to be treated as an ``unfathomable unknown''. Instead, we are able to actively account for them when simulating or analyzing the data from MOREI. Given real-world data, even of only moderate size, we can get useful estimates for concentration parameters describing realistic levels of heterogeneity. Using these estimates as the parameter values for our simulations, we can make simulations of MOREI more realistic. This in turn enhances our ability to find the methods that best recover true interaction networks from MOREI data.

\paragraph{Next Steps and Open Questions}

An obvious next step would be to clarify the framework for the non-parametric definitions of heterogeneity. From there one could investigate these heterogeneity estimands for a more diverse class of multivariate count distributions than I have considered herein, and then evaluate the performance of various estimators for these within this more general context. For example, giving these estimands a clear definition as pathwise differentiable functionals (or, if this is not possible, then explaining why) would facilitate implementing the targeted learning framework for them.

\begin{coolsubappendices}
\section{Plugin Estimators for Strain Population Frequencies}
\label{sec:append-plug-estim}

When $\expectation{\vabundance(0)|\abundance(0)}$ is Multinomial or Dirichlet-Multinomial distributed, one has that for all $n\ge 1$ and all strains $\strain \in [\Species]$:
\begin{equation}
  \label{eq:frequency_review}
  \freq^{(\strain)} = \frac{\expectation{\abundance[\strain](0) | \abundance(0) = n  }  }{n} \,.
\end{equation}
Averaging over all values of $n \ge 1$ leads to
\begin{equation}
  \label{eq:frequency_review_contd}
  \freq^{(\strain)} = \sum_{n \ge 1} \frac{\expectation{\abundance[\strain](0) | \abundance(0) = n  }  }{n} \probability{\abundance(0) = n | \abundance(0) \ge 1} =: (\freq^{(\strain)})_{eff} \,.
\end{equation}
This leads to the following plugin estimators for the frequencies:
\begin{equation}
  \label{eq:frequency_plugin}
  \efreq^{(\strain)} = \sum_{n \ge 1} \frac{\mean{\abundance[\strain](0) | \abundance(0) = n}}{n} \cdot \frac{ \sum_{\droplet} \indicator{\abundance_{\droplet}(0) = n}  }{ \sum_{\droplet} \indicator{\abundance_{\droplet}(0) \ge 1}  } \,.
\end{equation}

\subsection{Effective Frequencies}
\label{sec:effect-freq}

Even when the left and right hand sides of (\ref{eq:frequency_review_contd}) are not equal, the plugin estimator (\ref{eq:frequency_plugin}) is still consistent for the right hand side of (\ref{eq:frequency_review_contd}), the ``effective frequency'' $(\freq^{(\strain)})_{eff}$. 

Note that for all of the distributions used in the simulations, the true frequencies \textit{do} equal the effective frequencies. Even for hExhNBDM and hExhPoDM this is true, because
\begin{equation}
  \label{eq:hEx_frequencies_match_effective_explanation}
  \freq^{(\strain)} = \int_{\xi > 0} \freq^{(\strain)} \probability{\cconcentration = \xi} \mathrm{d}\xi   = \int_{\xi >0} (\freq^{(\strain)})_{eff} \probability{\cconcentration = \xi} \mathrm{d}\xi   = (\freq^{(\strain)})_{eff}  \,.
\end{equation}
In other words, (\ref{eq:frequency_review_contd}) holds for any individual value of $\cconcentration$ and will therefore still hold after averaging over any choice of ``prior'' for $\cconcentration$.

For distributions such that $\freq^{(\strain)} \not= (\freq^{(\strain)})_{eff}$, the estimators (\ref{eq:frequency_plugin}) will no longer be consistent, and a different choice of estimators $\efreq^{(\strain)}$ which actually are consistent for the true frequencies $\freq^{(\strain)}$ will need to be used inside of (\ref{eq:general_cconcentration_plugin}) to guarantee that (\ref{eq:general_cconcentration_plugin_official}) remains consistent for the effective compositional concentration (\ref{eq:effective_compositional_concentration_official}). 

Alternatively, one could instead change the definition of the effective compositional concentration inside of (\ref{eq:effective_compositional_concentration}) to use the effective frequencies $(\freq^{(\strain)})_{eff}$ in place of the true frequencies $\freq^{(\strain)}$, in which case using (\ref{eq:frequency_plugin}) inside of (\ref{eq:general_cconcentration_plugin}) will continue to ensure that (\ref{eq:general_cconcentration_plugin_official}) is consistent for (\ref{eq:effective_compositional_concentration_official}). Cf. section \ref{sec:effect-comp-conc-1}.

\section{Convergence Sometimes Disappointing for Small Batches}

The variance of the distribution of estimates was sometimes much wider for the small batches than for the medium or large batches. This is unimportant in practice since the small batches represent an unrealistically small problem size.

\pagebreak

\begin{figure}[p]
  \centering
  
  \begin{subfigure}{\textwidth}
  \centering  \includegraphics[width=\textwidth,height=0.45\textheight,keepaspectratio]{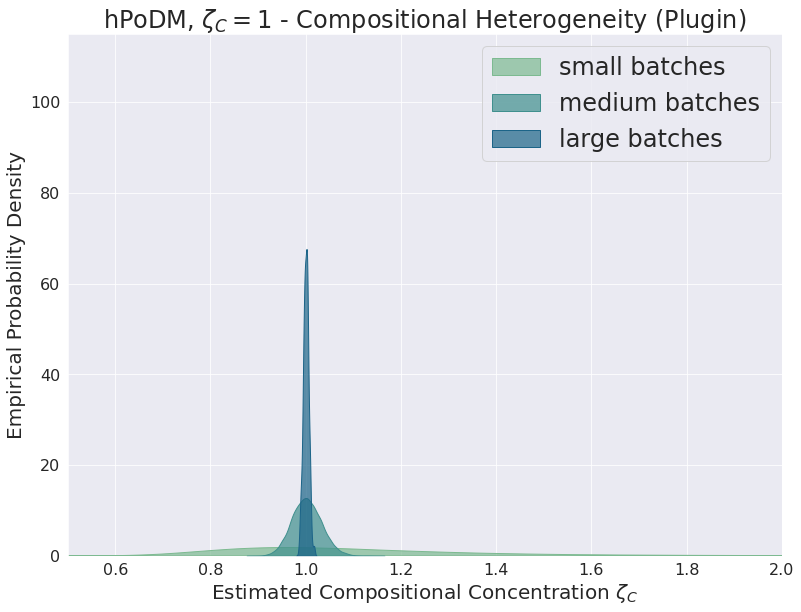}
  \caption[]{}
  \label{fig:hPoDM_1_C_kde_alt_plugin}
\end{subfigure}

\begin{subfigure}{\textwidth}
  \centering
  \includegraphics[width=\textwidth,height=0.45\textheight,keepaspectratio]{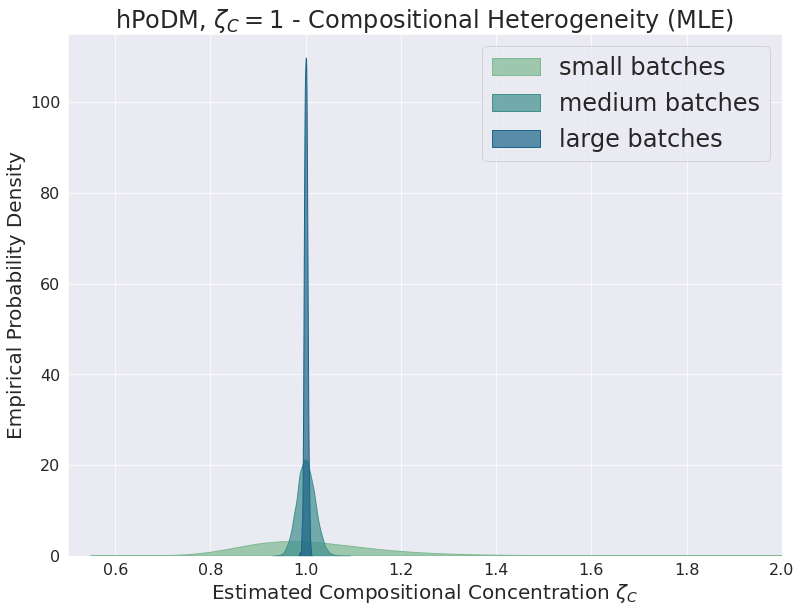}
  \caption[]{}
  \label{fig:hPoDM_1_C_kde_alt_MLE}
\end{subfigure}

\caption%
[~Version of figure \ref{fig:hPoDM_1_C_kde} including small batches.]
{Same as  figure \ref{fig:hPoDM_1_C_kde}, except now including small batches.}
\label{fig:hPoDM_1_C_kde_alt}
\end{figure}

\begin{figure}[p]
  \centering
  
  \begin{subfigure}{\textwidth}
  \centering \includegraphics[width=\textwidth,height=0.45\textheight,keepaspectratio]{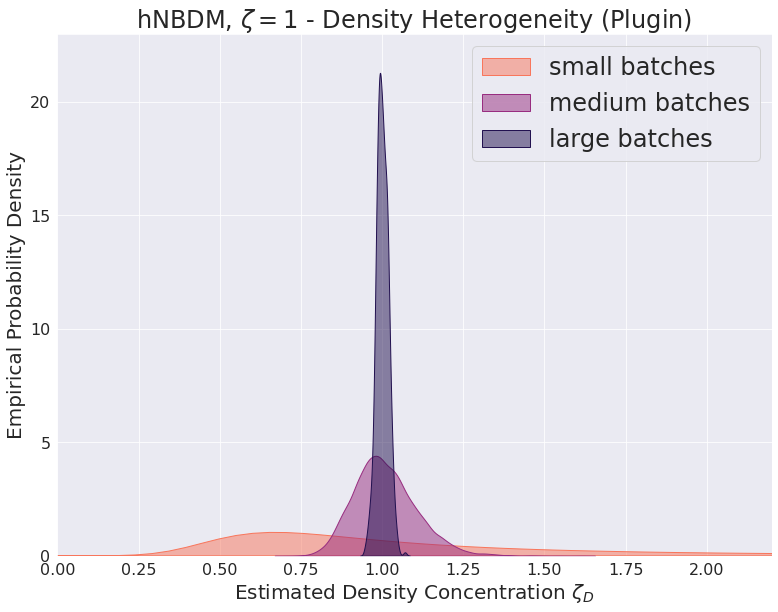}
  \caption[]{}
  \label{fig:hNBDM_1_D_kde_alt_plugin}
\end{subfigure}

\begin{subfigure}{\textwidth}
  \centering
  \includegraphics[width=\textwidth,height=0.45\textheight,keepaspectratio]{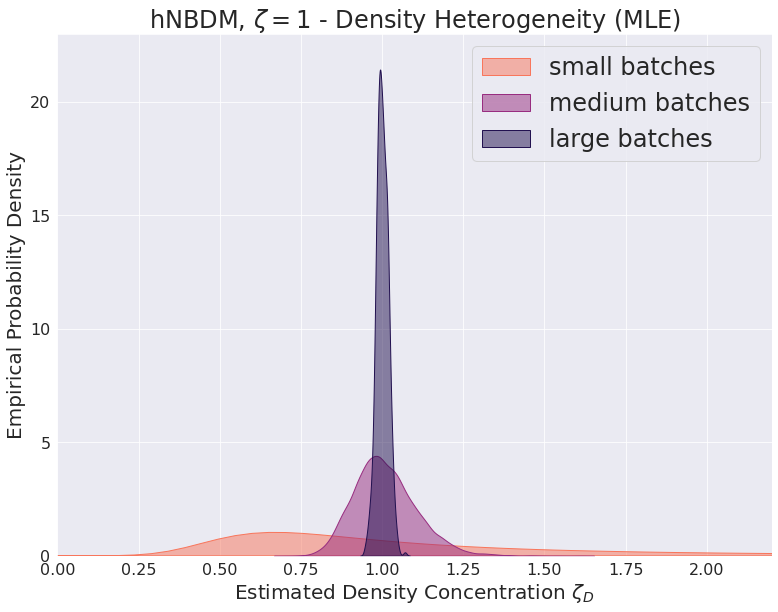}
  \caption[]{}
  \label{fig:hNBDM_1_D_kde_alt_MLE}
\end{subfigure}

\caption%
[~Version of figure \ref{fig:hNBDM_1_D_kde} including small batches.]
{Same as figure \ref{fig:hNBDM_1_D_kde} , except now including small batches.}
\label{fig:hNBDM_1_D_kde_alt}
\end{figure}

\begin{figure}[p]
  \centering
  
  \begin{subfigure}{\textwidth}
  \centering \includegraphics[width=\textwidth,height=0.45\textheight,keepaspectratio]{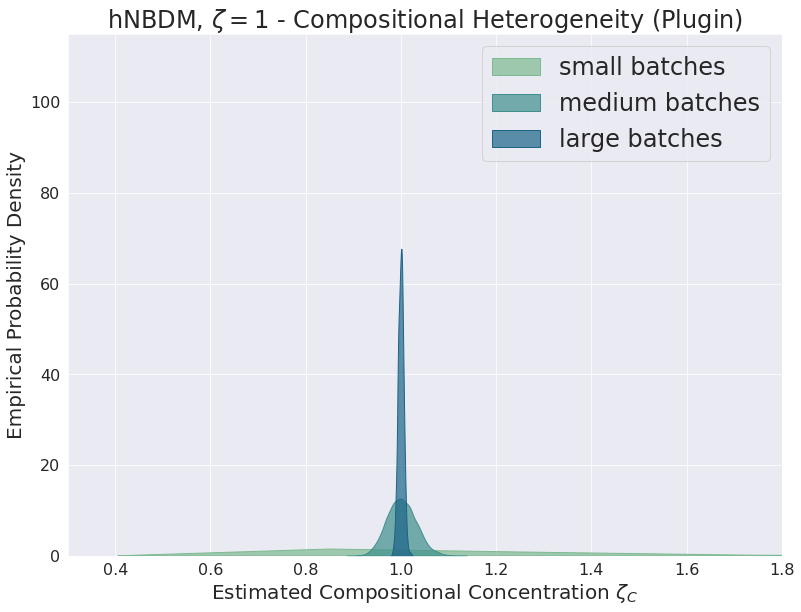}
  \caption[]{}
  \label{fig:hNBDM_1_C_kde_alt_plugin}
\end{subfigure}

\begin{subfigure}{\textwidth}
  \centering
  \includegraphics[width=\textwidth,height=0.45\textheight,keepaspectratio]{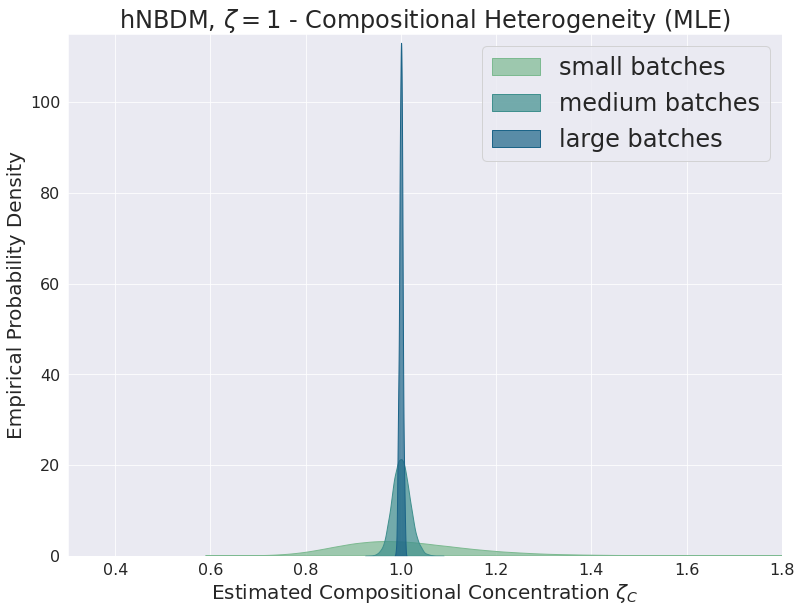}
  \caption[]{}
  \label{fig:hNBDM_1_C_kde_alt_MLE}
\end{subfigure}

\caption%
[~Version of figure \ref{fig:hNBDM_1_C_kde} including small batches.]
{Same as figure \ref{fig:hNBDM_1_C_kde} , except now including small batches.}
\label{fig:hNBDM_1_C_kde_alt}
\end{figure}

\begin{figure}[p]
  \centering
  
  \begin{subfigure}{\textwidth}
  \centering \includegraphics[width=\textwidth,height=0.45\textheight,keepaspectratio]{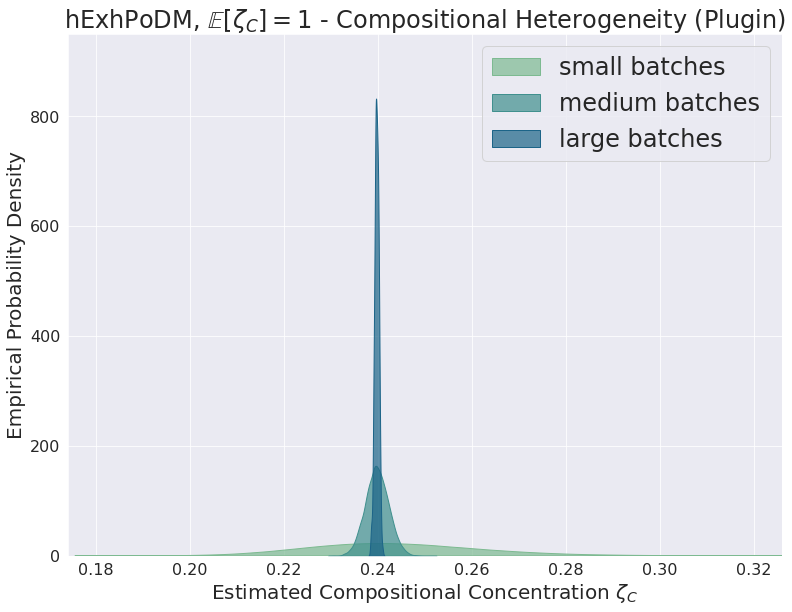}
  \caption[]{}
  \label{fig:hExhPoDM_C_kde_alt_plugin}
\end{subfigure}

\begin{subfigure}{\textwidth}
  \centering
  \includegraphics[width=\textwidth,height=0.45\textheight,keepaspectratio]{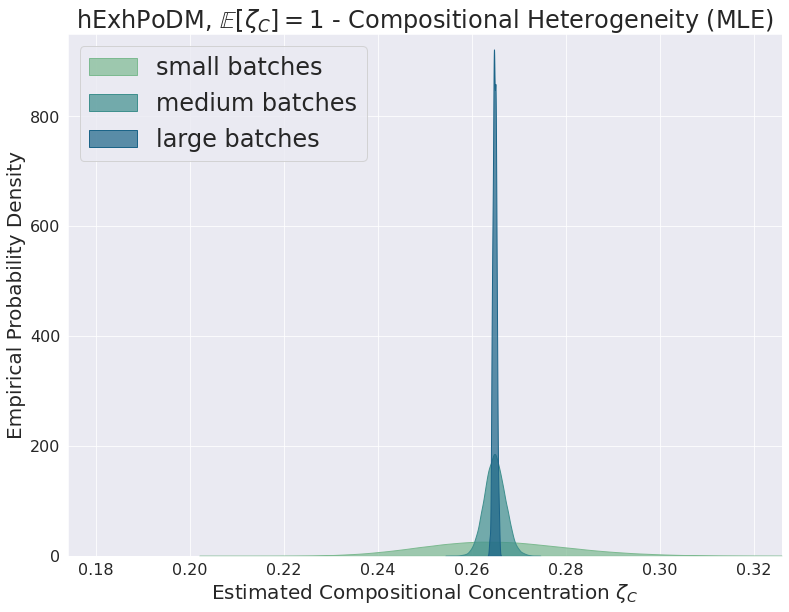}
  \caption[]{}
  \label{fig:hExhPoDM_C_kde_alt_MLE}
\end{subfigure}

\caption%
[~Version of figure \ref{fig:hExhPoDM_C_kde} including small batches.]
{Same as figure \ref{fig:hExhPoDM_C_kde}, except now including small batches.}
\label{fig:hExhPoDM_C_kde_alt}
\end{figure}

\begin{figure}[p]
  \centering
  \begin{subfigure}{\textwidth}
  \centering
  \includegraphics[width=\textwidth,height=0.45\textheight,keepaspectratio]{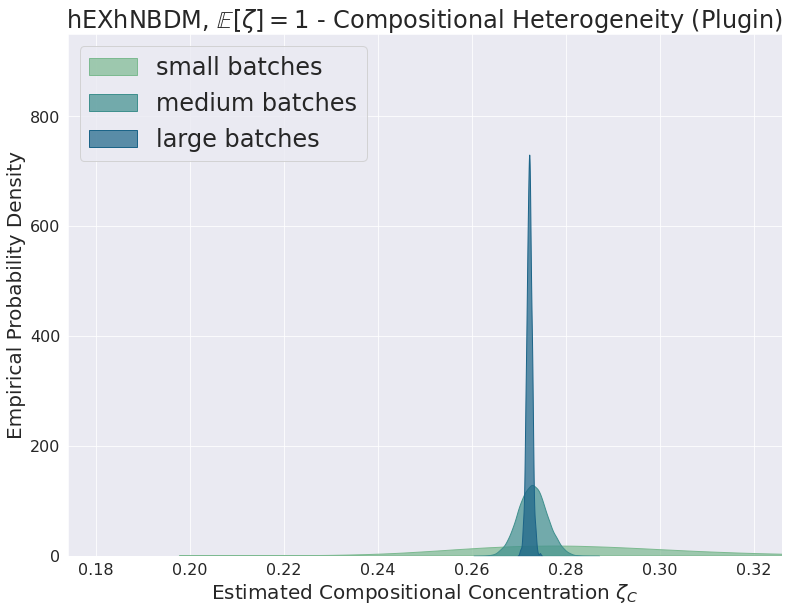}
  \caption[]{}
  \label{fig:hExhNBDM_C_kde_alt_plugin}
\end{subfigure}

\begin{subfigure}{\textwidth}
  \centering
  \includegraphics[width=\textwidth,height=0.45\textheight,keepaspectratio]{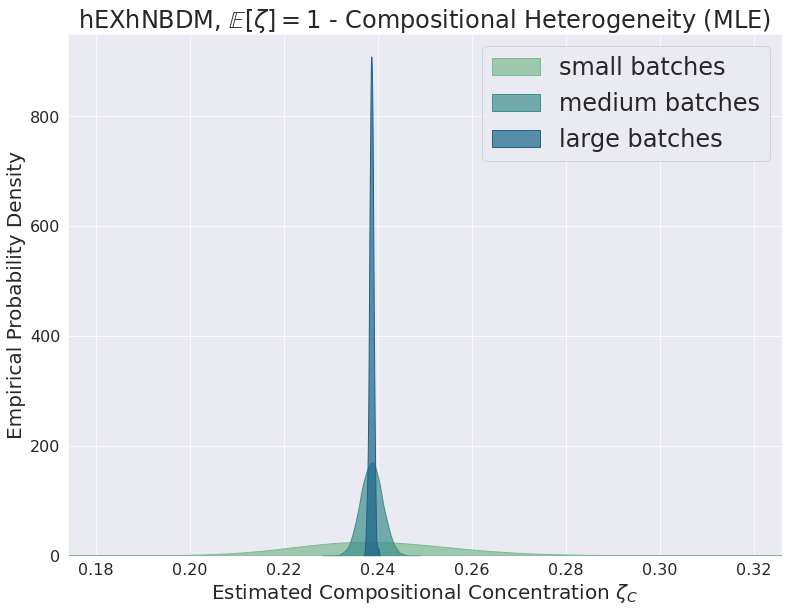}
  \caption[]{}
  \label{fig:hExhNBDM_C_kde_alt_MLE}
\end{subfigure}

\caption%
[~Version of figure \ref{fig:hExhNBDM_C_kde} including small batches.]
{Same as figure \ref{fig:hExhNBDM_C_kde}, except now including small batches.}
\label{fig:hExhNBDM_C_kde_alt}
\end{figure}

\end{coolsubappendices}
\end{coolcontents}

\clearpage
\setcounter{footnote}{0}
\pagestyle{myheadings}
\part{Inference of Ecological Networks}
\label{part:aver-treatm-effects}

The broader field of Part \ref{part:aver-treatm-effects}, and an important problem in microbial ecology, is inference of ecological models of microbial communities \cite{Antwis2017}.
To infer ecological models of microbial communities, we need to understand the interactions between different microbes.
(Cf. again the discussions from before in sections \ref{sec:form-trans-likel}, \ref{sec:form-as-stat}, and \ref{sec:what-why-microbial}.)
It is commonly understood that networks are useful models of ecological interactions\footnote{
Of \textit{pairwise} ecological interactions. Cf. \cite{higher_order} or \cite{Momeni2017} regarding ``higher-order'' interactions.
} \cite{Faust2012} \cite{Angulo2017}.
Likewise, it is also commonly understood that dynamical systems are useful models of ecological interactions \cite{Momeni2017} \cite{Xiao2017} \cite{Angulo2019}.
Finally, at least as early as \cite{Lidicker1979}, it has been commonly understood that directed ecological interactions can be categorized using signs, with ``$+$'' for beneficial effect, ``$0$'' for neutral effect, and ``$-$'' for detrimental effect on the growth of the recipient organism \cite{signs_longitudinal_compositional} \cite{positive_kChip}. By combining all of these observations, it is straightforward to conclude that signed networks \cite{Harary}, networks whose edge weights can be positive or negative, are natural models\footnote{
In the case one starts with a dynamical system model, one can define the edge weights of a corresponding signed network model as coefficients of the dynamical system \cite{Gonze2018}, or as the values of its Jacobian at some point in time \cite{Xiao2017} \cite{Angulo2019}. Cf. the discussion towards the end of section \ref{sec:what-why-microbial}.
} of ecological interactions \cite{Xiao2017}.

Hence Part \ref{part:aver-treatm-effects} investigates the important subfield of how to infer ecological models that can be characterized as signed networks. The signs of edge weights have the same interpretation that signs have above \cite{Lidicker1979} \cite{signs_longitudinal_compositional} \cite{positive_kChip}. Namely, the signs of edge weights categorize the qualitative effects that ecological interactions have on the growth of  recipient organisms (and the magnitudes of the edge weights encode how strong these effects are) \cite{Xiao2017}. (Cf. again section \ref{sec:form-as-stat}.)
Competition assays \cite{competition_assay_defn} and related coculture experiments\footnote{
  Such experiments (especially those involving cell plating or fluorescence imaging) usually produce absolute abundance data, but cf. also e.g. \cite{GG_competition_2} and \cite{signs_longitudinal_compositional} for ideas on how competition assays might still be used to infer microbial interactions even when using relative abundance data. See also sections \ref{sec:how-micr-ecol} and \ref{sec:prep-sequ} for more on the controversies surrounding absolute vs. relative abundance data.
} are ideal for measuring how different microbes affect each other's growth.

Therefore in microbial ecology the problem of inferring signed networks that encode interactions often reduces to the problem of inferring signed networks from the data produced by competition assays and related coculture experiments.
This is why the specific problem that Part \ref{part:aver-treatm-effects} investigates is the inference of signed networks quantifying how microbes affect each other's growth.
This problem corresponds to the ``unpartitioned interaction network'' competition assays from the introduction to Part \ref{part:introduction}, or the equivalently to the ``genes $\times$ genes interaction problem'' from item \ref{item:gene_gene_interactions} of section \ref{broader-field-2}.
The reference \cite{coculture} gives an overview (as of 2014) of several of the technologies used for implementing such experiments.
Even experiments such as \cite{Venturelli} that use fairly straightforward cell plating have made progress in understanding the ecological interactions of microbes.

Yet even if for no other reasons than those of throughput (cf. again section \ref{sec:how-micr-ecol}), the future state-of-the-art experiments for inferring microbial interactions (via their effects on each other's growth) are likely to use droplet-based microfluidics technology.
See the articles \cite{Kintses2010} or \cite{Guo2012} for reviews (as of 2012) of high-throughput biological experiments using such droplet-based microfluidics technology.
Some work using droplet-based microfluidics to study microbial interactions has already been published, cf. e.g. \cite{kChip} \cite{positive_kChip} or \cite{Hsu2019}, but it is expected (cf. again section \ref{sec:using-morei-char}) that in the future the data produced by such experiments for studying microbial interactions will improve even further.

\paragraph{Chapter \ref{chap:log-ratio-coeff}}

The broader field chapter \ref{chap:log-ratio-coeff} belongs to is the inference of ecological models for microbial interactions.
The subfield that chapter \ref{chap:log-ratio-coeff} considers is the use of relative fitness measurements to quantify microbial interactions.
The chapter investigates how a particular measure of relative fitness can be recast into the statistical framework of average treatment effects.
The chapter explains how violations of positivity assumptions are inevitable for this problem, making controlling for confounding difficult.
Finally, chapter \ref{chap:log-ratio-coeff} also gives explicit assumptions under which the estimands are identifiable from the observed data produced by incubated droplets, even though initial states of the droplets are unobserved.

\paragraph{Chapter \ref{chap:network_comparison}}

The broader field of chapter \ref{chap:network_comparison} is the use of signed networks as models for ecological interactions.
In particular, chapter \ref{chap:network_comparison} is motivated by the important issue of how to identify ``best-performing'' estimators of such ecological models.
This chapter investigates how comparisons of ecological interactions can be recast into the statistical framework of loss functions for signed networks.
The chapter explains how avoiding unexpected behavior requires loss functions for signed networks to satisfy what I call herein ``the double penalization principle''.
Starting from examples of loss functions for \textbf{\textit{un}}signed networks, the chapter derives several examples of loss functions for signed networks that satisfy this property.

\clearpage
\pagestyle{headings}

\chapter[Average Treatment Effects (ATEs) for Quantifying Interactions][Average Treatment Effects (ATEs)]{Average Treatment Effects (ATEs) for Quantifying %
  Microbial Interactions}
\label{chap:log-ratio-coeff}

Herein I show how a particular measure of relative fitness can be recast into the statistical framework of average treatment effects.
See sections \ref{sec:poss-choic-resolve} and \ref{sec:estim-simpl-estim}.
I explain how (for this problem) violations of positivity assumptions are inevitable in practice, making controlling for confounding difficult.
See section \ref{sec:conf-adjustm-posit}.
I give explicit assumptions under which the corresponding estimands are identifiable from the observed data produced by incubated droplets, despite that initial states of the droplets are unobserved.
See section \ref{sec:proofs}.

Section \ref{sec:full-data-model} clarifies what the statistical model for the full data is assumed to be. Section \ref{sec:poss-choic-resolve} then clarifies which parameters of the full data statistical model we can choose to target for quantifying the strength and directionality of microbial interactions. Section \ref{sec:observed-data-model} establishes what the statistical model corresponding to the observed data is, highlighting the difference from the full data statistical model that is most important for our choice of target parameters. Section \ref{sec:estim-simpl-estim} expands on this by giving the target parameters for the observed data statistical model that we would like to be able to identify with the target parameters of the full data statistical model. It also briefly states sufficient conditions for such an identifiability result to hold. All in all, the goal of these four sections is to clarify what research question we are choosing to answer, thus firmly positioning our approach to answering our given scientific question (quantifying the strength and directionality of microbial interactions) within the first stage of the targeted learning roadmap \cite{vanderLaan2011}.

Section \ref{sec:estimation} briefly clarifies what the definitions of the relevant plugin estimators for the target parameters of the observed data model would be, and offers an initial explanation of how, starting from the previously established research question, we can proceed further to the estimation stage of the targeted learning roadmap using SuperLearner \cite{vanderLaan2007} and targeted minimum-loss based estimation (TMLE) \cite{vanderLaan2011}. Section \ref{sec:conf-adjustm-posit} clarifies which issues with confounding we might potentially encounter with our observed data, and indicates some possible obstacles and preliminary solutions for addressing those issues. Finally, section \ref{sec:proofs} gives a detailed clarification of the assumptions being used to establish sufficient conditions for identifiability of the full data target parameters from the observed data. It then provides proofs demonstrating why these conditions are indeed sufficient for the intended results. The goal of these last sections is to demonstrate the feasibility of the remaining two stages of the targeted learning roadmap \cite{vanderLaan2011}, estimation and inference, for this research question.

To the extent it is necessary either as a mathematical convenience, or for peace of mind, throughout what follows it can be assumed that the logarithm has had its domain extended from $(0, \infty)$ to $[0, \infty)$ by defining $\log(0) = 0$.

\begin{coolcontents}

\section{Background and Significance}
\label{sec:backgr-sign-7}

\paragraph{Broader field}

The broader field chapter \ref{chap:log-ratio-coeff} belongs to is the inference of ecological models for microbial interactions. The goal is to use data produced by the MOREI experiment to infer these. See \cite{Faust2012} for a review of how networks are often used as such ecological models, and \cite{Xiao2017} (or section \ref{sec:form-as-stat}) for how signed networks in particular can be used to describe microbial interactions. Cf. the discussion earlier from the introduction to Part \ref{part:aver-treatm-effects}.

The subfield that chapter \ref{chap:log-ratio-coeff} considers is the use of relative fitness measurements to quantify microbial interactions. (The ecological model is then a signed network whose weights are the corresponding relative fitness values. Cf. section \ref{sec:how-micr-ecol} or the discussion of ``competition assays'' \cite{competition_assay_defn} from the introduction of Part \ref{part:introduction}.) There are many distinct estimands used as quantitative measures of ``(relative) fitness'' found throughout the literature. Cf. the definitions found in \cite{rb_tnseq}, \cite{competition_assay_defn}, \cite{GG_competition_2}, or \cite{GE_competition_1} for several distinct examples. Definitions are often stated implicitly or imprecisely, contributing to the difficulty of comparing definitions across papers and of analyzing associated statistical problems.

\paragraph{Specific problem}

Herein I investigate how a particular measure of relative fitness can be recast into the statistical framework of average treatment effects. The underlying idea behind this measure is easy to motivate heuristically and very simple (cf. section \ref{sec:heur-deriv-log}). Similar features of other definitions of relative fitness perhaps encourage their primarily informal treatment in the literature. In any case, by making this definition precise (cf. section \ref{sec:poss-choic-resolve}), analysis of the corresponding statistical problems becomes much clearer. In particular, one can readily see that this definition admits an interpretation as an “average treatment effect” estimand. Cf. \cite[chapter 10]{vanderLaan2011} for more about the general theory of average treatment effects, which is substantially well-developed.

At least two papers in the virology literature that study\footnote{
As opposed to the “unpartitioned interactions” that we are interested in, cf. the introduction to Part \ref{part:introduction} for terminology.
} “bipartite interactions”, \cite{GE_competition_2} and  \cite{Holland1991}, appear to use the same (or a sufficiently analogous) definition of relative fitness as that used in this chapter (cf. section \ref{sec:comparison_ate_defns} for the latter.) This is such a straightforward definition that most likely other examples of its use exist in the literature, including those studying non-viral microbes and/or “unpartitioned interactions”. However, I have not yet been able to find additional examples, in part due to how definitions of ``fitness'' in the literature are usually discussed only imprecisely or even implicitly (as in e.g. \cite{Holland1991}).

\paragraph{Particular approach}

Being able to formulate this problem in terms of average treatment effects suggests it should be possible to apply targeted minimum loss based estimation (TMLE) \cite{vanderLaan2011} to this problem.
This chapter demonstrates how one can follow the first two stages of the targeted learning roadmap \cite[section 1.5]{vanderLaan2011} with this choice of estimand: (1) defining the research question, and (2) estimation. Hence this chapter can be thought of as a ``proof of concept'' for applying TMLE, but future work will fill out all of the details of completing the targeted learning roadmap from start to finish.

Because PhenoPath \cite{phenopath} was designed for a problem involving fairly high-throughput data produced using microfluidics, and which is not strictly longitudinal, attempting to modify it to apply to the problem of inferring microbial interactions from MOREI data seems like a natural approach, and more sophisticated than using these simple relative fitness estimates. However, although PhenoPath was suitably modified to apply to this problem, the results of preliminary work suggest that PhenoPath does not work well for this problem.

One of the reasons the method from \cite{phenopath} failed to work well on (simulated) data for this problem in preliminary work is because it was designed for ``gene $\times$ environment'' interactions, not ``gene $\times$ gene'' interactions. (Cf. again item \ref{item:gene_gene_interactions} from section \ref{broader-field-2} for terminology.) Its computational performance for this latter, more computationally demanding, problem proved inadequate. (If $n$ is the number of genes, the latter problem corresponds to $n$ instances of the former problem, cf. footnote \ref{footnote:n_instances} from section \ref{broader-field-2}.)

The method from \cite{phenopath} also seemed unable to handle the sparsity of the data well, much less exploit it. One reason seems to be that the model implemented by \cite{phenopath} inherently treats $0$ values as ``biological zeros'', when in fact for this problem they are almost all ``sampling zeros'' (cf. the terminology from section \ref{sec:positiv-microbiology} and quoted from \cite{Prost2021}). More often than not, when a given strain is not found within a given droplet, the reason is because the strain was never in the droplet in the first place. Yet PhenoPath appears to always interpret this the same way as if the strain had been initially present in the droplet and later killed off by those strains that were found in the droplet.

\section{Full Data Model}
\label{sec:full-data-model}

The full data model $\mathcal{M}^F$ corresponds to the likelihoods from equations (\ref{eq:droplet_full}), (\ref{eq:batch_full}), and (\ref{eq:experiment_full}). This can be interpreted as the marginalization of a continuous time Markov process describing the growth and death of cells over time. The full data model $\mathcal{M}^F$ assumes for any given droplet not only that $\vabundance(\time)$ is known, but also that $\vabundance(0)$ is known as well.

\section{Full Data Average Treatment Effect (ATE) Definitions}
\label{sec:poss-choic-resolve}

For the full data model $\mathcal{M}^F$, denote any average treatment effect (ATE) full data estimand (i.e. assuming that $\vabundance(0)$ is known) corresponding to the effect of strain\footnote{Herein I use ``strains'' to refer equally to strains belonging to the same species(/genus/family/etc.) as well as to strains belonging to different species(/genera/families/etc.), because the distinction is irrelevant for setting up the abstract problem. It may matter for the implementation of a specific experiment.} $\strain_1$ on $\strain_2$ at time $\time$ by $\lrestimandsymbolfull (\time) (\specie_1, \specie_2)$.
In these definitions of ``average treatment effect'', the term corresponding to the treatment group can be interpreted heuristically as the ``numerator'' of a log-ratio, whereas the term corresponding to the control group can be interpreted heuristically as the ``denominator'' of the same log-ratio. Cf. section \ref{sec:heur-deriv-log}. This heuristic interpretation influences the choice of notation used below. Also note that it is implicit that any choice of treatment group and control group do not overlap. Thus the averages corresponding to the two terms of the ATE definition are always taken over two disjoint sets of droplets.

In defining the average treatment effects, note that (as before) we have at least $2$ choices (``picky'' or ``gluttonous'') of treatment groups and at least $2$ choices (``picky'' or ``gluttonous'') of control groups. Cf. again section \ref{sec:defin-thro-intro}, section \ref{sec:how-micr-ecol}, or section \ref{sec:defin-thro}. ATE estimands using a ``picky'' treatment group will be denoted $\lrestimandfull{\relax}{p}{\cdot}(\time)(\specie_1, \specie_2)$ (``p'' for ``\textbf{p}icky''), while ATE estimands using a ``gluttonous'' treatment group will be denoted $\lrestimandfull{\relax}{g}{\cdot} (\time)(\specie_1, \specie_2)$ (``g'' for ``\textbf{g}luttonous''). Likewise, ATE estimands using a ``picky'' control group will be denoted $\lrestimandfull{\relax}{\cdot}{p}(\time)(\specie_1, \specie_2)$ (``p'' for ``\textbf{p}icky''), while ATE estimands using a ``gluttonous'' control group will be denoted  $\lrestimandfull{\relax}{\cdot}{g}(\time)(\specie_1, \specie_2)$ (``g'' for ``\textbf{g}luttonous''). To summarize, within the above framework there are two ways to define the treatment group term of the ATE, and two ways to define the control group term of the ATE, leading to $2 \times 2 = 4$ total possible distinct definitions of ATE (average treatment effect) under consideration. Explicit definitions for all $4$ of these possible choices of full data ATE estimand are given below.

\paragraph{{\small (Full Data)} Picky treatment, picky control:}
\begin{equation}
  \label{eq:lr_estimand_picky_defns_geom_picky_picky}
  \begin{array}{rccl}
\Psi^F_{p/p} (\time)(\specie_1, \specie_2) : & \mathcal{M}^F &\to& \mathbb{R}\\
 \Psi^F_{p/p} (\time)(\specie_1, \specie_2) : & \mathcal{P}_F &\mapsto& \displaystyle \expectation[\mathcal{P}_F]*{\log\left(\frac{\abundance[\specie_2](\time)}{\abundance[\specie_2](0)} \right) |\observed = \{\specie_1, \specie_2\}  }\\
 &&& - \displaystyle \expectation[\mathcal{P}_F]*{\log\left(\frac{\abundance[\specie_2](\time)}{\abundance[\specie_2](0)} \right) | \observed = \{\specie_2\} } \\
  \end{array}
\end{equation}

\paragraph{{\small (Full Data)} Picky treatment, gluttonous control:}
\begin{equation}
  \label{eq:lr_estimand_picky_defns_geom_picky_gluttonous}
  \begin{array}{rccl}
 \Psi^F_{p/g} (\time)(\specie_1, \specie_2) : & \mathcal{M}^F &\to& \mathbb{R} \\
\Psi^F_{p/g} (\time)(\specie_1, \specie_2) : & \mathcal{P}_F &\mapsto& \displaystyle \expectation[\mathcal{P}_F]*{\log\left(\frac{\abundance[\specie_2](\time)}{\abundance[\specie_2](0)} \right) | \observed = \{\specie_1, \specie_2\} } \\
&&& - \displaystyle\expectation[\mathcal{P}_F]*{\log\left(\frac{\abundance[\specie_2](\time)}{\abundance[\specie_2](0)} \right) | \observed \supseteq \{\specie_2\}, \specie_1 \not\in \observed } 
\end{array}
\end{equation}

\paragraph{{\small (Full Data)} Gluttonous treatment, picky control:}
\begin{equation}
  \label{eq:lr_estimand_picky_defns_geom_gluttonous_picky}
  \begin{array}{rccl}
\Psi^F_{g/p} (\time)(\specie_1, \specie_2) : & \mathcal{M}^F &\to& \mathbb{R}    \\
\Psi^F_{g/p} (\time)(\specie_1, \specie_2) : & \mathcal{P}_F &\mapsto& \displaystyle \expectation[\mathcal{P}_F]*{\log\left(\frac{\abundance[\specie_2](\time)}{\abundance[\specie_2](0)} \right) | \observed \supseteq \{\specie_1, \specie_2\}   }\\
&&& - \displaystyle\expectation[\mathcal{P}_F]*{\log\left(\frac{\abundance[\specie_2](\time)}{\abundance[\specie_2](0)} \right) | \observed = \{\specie_2\} } 
  \end{array}
\end{equation}

\paragraph{{\small (Full Data)} Gluttonous treatment, gluttonous control:}
\begin{equation}
    \label{eq:lr_estimand_picky_defns_geom_gluttonous_gluttonous}
\begin{array}{rccl}
\Psi^F_{g/g} (\time)(\specie_1, \specie_2) : & \mathcal{M}^F& \to& \mathbb{R} \\
  \Psi^F_{g/g} (\time)(\specie_1, \specie_2) : & \mathcal{P}_F &\mapsto& \displaystyle
 \expectation[\mathcal{P}_F]*{\log\left(\frac{\abundance[\specie_2](\time)}{\abundance[\specie_2](0)} \right) | \observed \supseteq \{\specie_1, \specie_2\}   }\\
& && - \displaystyle \expectation[\mathcal{P}_F]*{\log\left(\frac{\abundance[\specie_2](\time)}{\abundance[\specie_2](0)} \right) | \observed \supseteq \{\specie_2\}, \specie_1 \not\in \observed  } 
\end{array}
\end{equation}

It turns out (cf. section \ref{sec:proofs}) that these estimands are identifiable from the observed data provided we make the following assumptions:
\begin{itemize}
\item we always have that $\observed = \observed[\time]$,
 \item the marginal distributions of $\vabundance(0)$, i.e. the distributions of $\abundance[1](0)$, $\abundance[2](0)$, \dots, $\abundance[\Strains](0)$, are all mutually independent.
 \end{itemize}
These conditions are sufficient, but they may not be necessary.

\subsection{Stratified Full Data Average Treatment Effect}
\label{sec:strat-full-aver-treatm}

The motivation for using the following definition is explained in section \ref{sec:conf-adjustm-posit}.
\begin{equation}
  \label{eq:stratified_full_data}
  \begin{adjustbox}{max width=\textwidth,keepaspectratio}
 $\displaystyle \begin{array}{rccl}
    \lrestimandsymbolfull (\time)(\specie_1, \specie_2) :  & \mathcal{M}^F& \to & \mathbb{R} \\
    \lrestimandsymbolfull (\time) (\specie_1, \specie_2) : & \mathcal{P}_F & \mapsto & \displaystyle
                                                  \expectationsymbol_{\mathcal{P}_F}  \left[
                                                                                 \expectation[\mathcal{P}_F]*{  \log \left(   \frac{ \abundance[\specie_2](\time)  }{  \abundance[\specie_2](0)  }  \right) |
                                                                                 \straincount[\strain_1](0) = 1 , \vstraincount^{ \{\strain_1, \strain_2\}^c } (0)
                                                                                 }
                                                                                 \right. \\
                                                     &&&  \hphantom{\expectationsymbol_{\mathcal{P}_F} \left[\right]  }
                                                         \left. - \displaystyle
                                                                                 \expectation[\mathcal{P}_F]*{  \log \left(   \frac{ \abundance[\specie_2](\time)  }{  \abundance[\specie_2](0)  }  \right) |
                                                                                 \straincount[\strain_1](0) = 0 , \vstraincount^{ \{\strain_1, \strain_2\}^c } (0)
                                                                                 }
                                                         \right]  \\
                                                           &&=&\displaystyle \sum_{\rvec{b} \in \{ 0,1 \}^{\Strains -2}}  \left[
 \expectation[\mathcal{P}_F]*{  \log \left(  \frac{\abundance[\strain_2](\time)}{\abundance[\strain_2](0)}   \right) |
\straincount[\strain_1](0) = 1, \vstraincount^{\{ \strain_1, \strain_2 \}^c}(0) = \rvec{b}
                                                                }
                                                                \right.   \\
                                                           &&& \displaystyle \left. -
  \expectation[\mathcal{P}_F]*{  \log \left(  \frac{\abundance[\strain_2](\time)}{\abundance[\strain_2](0)}   \right) |
\straincount[\strain_1](0) = 0, \vstraincount^{\{ \strain_1, \strain_2 \}^c}(0) = \rvec{b}        }                                                     \right]
                                                            \!   \cdot \probability[\mathcal{P}_F]*{ \vstraincount^{ \{\strain_1, \strain_2\}^c }(0) = \rvec{b} } \,.
  \end{array}$
\end{adjustbox}
\end{equation}
Observe how this definition implicitly uses a definition of treatment and covariates (cf. section \ref{sec:conf-adjustm-posit}) that ignores the multiple representatives problem (cf. section \ref{sec:append-mult-repr}). The analogous estimand that would correspond to not ignoring the multiple representatives problem appears to not (even in principle) be identifiable from the observed data. By ignoring the multiple representatives problem, we allow ourselves to use the binary strain presence/absence vector $\vstraincount(0)$ (cf. section \ref{sec:dist-distr-numb}), instead of the full strain count vector $\vabundance(0)$, to define the treatment and covariates. Considering that already fairly strong assumptions (cf. section \ref{sec:ident-observed}) are required to conclude that $\vstraincount(0)$ can be determined from the observed data, the assumptions that would be required to determine (even ``approximately'' or ``on average'') $\vabundance(0)$ seem to be hopelessly strong. Considering how the strata of the analogous estimand that does not ignore the multiple representatives problem would be defined in terms of $\vabundance(0)$, this seems to be an insurmountable problem for (directly) estimating such an estimand.

One might try to sidestep this limitation by screening for ``suspicious'' outliers in the identified sample that might correspond to values of $\abundance[\strain_1](0)$ (or $\abundance[{\strain[*]}](0)$ for $\strain[*] \not= \strain_1, \strain_2$) other than $1$. This would be analogous and closely related to the hypothetical strategy discussed at length in the comments at the end of section \ref{sec:no-censoring}, for how to possibly weaken the ``no censoring'' assumption. Thus this strategy would suffer from the same problems, namely requiring a very strong ``signal to noise ratio'', and being very difficult to implement successfully even given a strong ``signal to noise ratio''. Thus it is left to future work to implement and validate such a strategy, if it is even possible.

\section{Observed Data Model}
\label{sec:observed-data-model}

The observed data model $\mathcal{M}^O$ corresponds to the likelihoods from equations (\ref{eq:droplet_observed}), (\ref{eq:batch_observed}), and (\ref{eq:experiment_observed}). The map $\phi^{\mathcal{M}}: \mathcal{M}^F \to \mathcal{M}^O$ sending distributions in the full data model to distributions in the observed data model marginalizes over all of the unobserved time points. Recall from section \ref{sec:using-morei-char} that this means all cell counts at all times are censored, except for the final cell counts. The difference between the full data model and the observed data model that is salient for the definitions of the ATE estimands is that the observed data model assumes that $\vabundance(0)$/$\vstraincount(0)$ is unknown and that only $\vabundance(\time)$ is known.

\section[Observed Data ATE Definitions, Identifiability]{Observed Data ATE Definitions, \\Identifiability}
\label{sec:estim-simpl-estim}

What has so far been glibly ignored is the obvious issue that the droplets are observed at only one time point. In particular, the droplets are not observed upon formation at $\time=0$. There is  no way to observe $\abundance[\specie]_d(0)$ neither for any strain $\strain$ nor for any droplet $\droplet$. Without further assumptions, our targeted estimands, defined in terms of the full data in section \ref{sec:poss-choic-resolve} above, cannot possibly be defined in terms of the observed data alone.

By omitting expressions not belonging to the observed data in the definitions from section \ref{sec:poss-choic-resolve} of the full data ATE estimands, we arrive at the following definitions for observed data ATE estimands. These estimands, rather than ``normalizing'' by the (unobservable) initial counts of cells to get a proper ``growth factor'', instead consider only the final counts of cells. Moreover, instead of defining the treatment and  control groups in terms of the unobservable initial presence/absence of strains, these estimands define the treatment and control groups in terms of the observable presence/absence of strains at time $\time$.

\paragraph{\small{(Observed Data)} Picky treatment, picky control:}
\begin{equation}
  \label{eq:lr_estimand_observed_picky_defns_geom_picky_picky}
  \begin{array}{rccl}
 \Psi_{p/p} (\time)(\specie_1, \specie_2) : & \mathcal{M}^O & \to& \mathbb{R}\\
 \Psi_{p/p} (\time)(\specie_1, \specie_2) : & \mathcal{P}_O & \mapsto &\expectation[\mathcal{P}_O]*{\log\left(\abundance[\specie_2](\time) \right) |\observed[\time] = \{\specie_1, \specie_2\}  }\\
 & &&- \expectation[\mathcal{P}_O]*{\log\left(\abundance[\specie_2](\time) \right) | \observed[\time] = \{\specie_2\} } 
\end{array}
\end{equation}

\paragraph{\small{(Observed Data)} Picky treatment, gluttonous control:}
\begin{equation}
  \label{eq:lr_estimand_observed_picky_defns_geom_picky_gluttonous}
  \begin{array}{rccl}
 \Psi_{p/g} (\time)(\specie_1, \specie_2) : & \mathcal{M}^O& \to& \mathbb{R} \\
\Psi_{p/g} (\time)(\specie_1, \specie_2) : & \mathcal{P}_O &\mapsto &\expectation[\mathcal{P}_O]*{\log\left(\abundance[\specie_2](\time) \right) | \observed[\time] = \{\specie_1, \specie_2\} } \\
&&& - \expectation[\mathcal{P}_O]*{\log\left(\abundance[\specie_2](\time) \right) | \observed[\time] \supseteq \{\specie_2\}, \specie_1 \not\in \observed[\time] } 
  \end{array}
\end{equation}

\paragraph{{\small (Observed Data)} Gluttonous treatment, picky control:}
\begin{equation}
  \label{eq:lr_estimand_observed_picky_defns_geom_gluttonous_picky}
  \begin{array}{rccl}
\Psi_{g/p} (\time)(\specie_1, \specie_2) : & \mathcal{M}^O &\to& \mathbb{R}    \\
\Psi_{g/p} (\time)(\specie_1, \specie_2) : & \mathcal{P}_O &\mapsto& \expectation[\mathcal{P}_O]*{\log\left(\abundance[\specie_2](\time) \right) | \observed[\time] \supseteq \{\specie_1, \specie_2\}   }\\
&&& - \expectation[\mathcal{P}_O]*{\log\left(\abundance[\specie_2](\time) \right) | \observed[\time] = \{\specie_2\} } 
  \end{array}
\end{equation}

\paragraph{{\small (Observed Data)} Gluttonous treatment, gluttonous control:}
\begin{equation}
   \label{eq:lr_estimand_observed_picky_defns_geom_gluttonous_gluttonous}
   \begin{array}{rccl}
  \Psi_{g/g} (\time)(\specie_1, \specie_2) : & \mathcal{M}^O &\to &\mathbb{R} \\
\Psi_{g/g} (\time)(\specie_1, \specie_2) : & \mathcal{P}_O &\mapsto& \expectation[\mathcal{P}_O]*{\log\left(\abundance[\specie_2](\time) \right) | \observed[\time] \supseteq \{\specie_1, \specie_2\}   }\\
&&& - \expectation[\mathcal{P}_O]*{\log\left(\abundance[\specie_2](\time) \right) | \observed[\time] \supseteq \{\specie_2\}, \specie_1 \not\in \observed[\time]  }
   \end{array}
\end{equation}

Under certain assumptions (that are satisfied e.g. by the hPoMu and hNBDM working models), the full data ATE estimands (which assume $\vabundance(0)$ is known) would equal the observed data ATE estimands. This means the full data ATE estimands would be identifiable from the observed data. The assumptions are:
\begin{itemize}
\item we always have that $\observed = \observed[\time]$,
 \item the marginal distributions of $\vabundance(0)$, i.e. the distributions of $\abundance[1](0)$, $\abundance[2](0)$, \dots, $\abundance[\Strains](0)$, are all mutually independent.
 \end{itemize}
Cf. section \ref{sec:proofs} and figure \ref{fig:identifiability} below.

\begin{figure}[H]
  \centering
  \includegraphics[width=\textwidth,height=0.4\textheight,keepaspectratio]{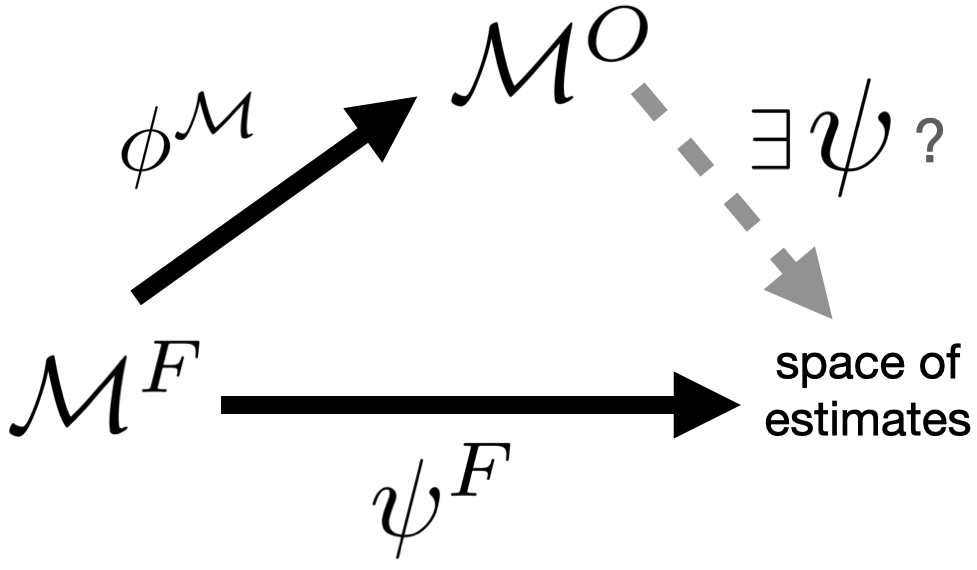}
  \caption[Identifiability of Estimands]{$\mathcal{M}^F$ is the full statistical model, $\mathcal{M}^O$ is the observed statistical model (i.e. corresponding to the observed data and observed likelihoods), $\phi^{\mathcal{M}}$ sends each distribution in the full statistical model to the corresponding distribution of observed data, $\psi^F$ is the estimand mapping defined on the full statistical model answering the scientific question we are interested in. This estimand is called identifiable if there exists a corresponding estimand defined on the observed statistical model such that the above commutative diagram commutes. (For an introduction to commutative diagrams, cf. e.g. \cite{conceptual}.)}
  \label{fig:identifiability}
\end{figure}

Another way to interpret the observed data ATE estimands is as pretending that $\abundance[\specie](0) =1$ always (whenever it is nonzero). This may not be entirely unreasonable. The experiment can be set up so that $1$ is the highest probability nonzero count of cells belonging to any given strain at a droplet's formation. (Cf. the discussion in section \ref{sec:append-mult-repr}.) Hence, even when we do not believe the assumptions made in section \ref{sec:proofs} to be valid, the experiment can still be calibrated to minimize the ``identification gap'' between the full data ATE estimands and the observed data ATE estimands, which makes the estimands useful.

\subsection{Stratified Observed Data Average Treatment Effect}
\label{sec:strat-observ-aver}

The motivation for using the following definition is explained in section \ref{sec:conf-adjustm-posit}.
\begin{equation}
  \label{eq:stratified_observed_estimand}
  \begin{adjustbox}{max width=\textwidth,keepaspectratio}
$\displaystyle  \begin{array}{rccl}
    \lrestimandsymbol(\time) (\strain_1, \strain_2): &  \mathcal{M}^O & \to & \mathbb{R}  \\
    \lrestimandsymbol(\time) (\strain_1, \strain_2): & \mathcal{P}_O & \mapsto & \expectationsymbol_{\mathcal{P}_O} \left[
\expectation[\mathcal{P}_O]*{ \log \left(\abundance[\strain_2](\time) \right) | \straincount[\strain_1](\time) = 1 , \vstraincount^{\{\strain_1, \strain_2\}^c}(\time)  }                                                                                 
 \right.  \\
&&&    \left. \hphantom{ \expectation[\mathcal{P}_O]*{  }  }
    -  \expectation[\mathcal{P}_O]*{ \log\left( \abundance[\strain_2](\time) \right)  | \straincount[\strain_1](\time) = 0 , \vstraincount^{\{ \strain_1, \strain_2 \}^c} (\time) }
    \right] \\
& & = &\displaystyle \sum_{\rvec{b} \in \{ 0,1 \}^{\Strains -2}}  \left[
 \expectation[\mathcal{P}_O]*{  \log \left(  \abundance[\strain_2](\time)   \right) |
\straincount[\strain_1](\time) = 1, \vstraincount^{\{ \strain_1, \strain_2 \}^c}(\time) = \rvec{b}
                                                                }
                                                                \right.   \\
                                                           &&& \displaystyle \left. -
  \expectation[\mathcal{P}_O]*{  \log \left(  \abundance[\strain_2](\time)   \right) |
\straincount[\strain_1](\time) = 0, \vstraincount^{\{ \strain_1, \strain_2 \}^c}(\time) = \rvec{b}        }                                                     \right]
                                                               \cdot \probability[\mathcal{P}_F]*{ \vstraincount^{ \{\strain_1, \strain_2\}^c }(\time) = \rvec{b} } \,.
  \end{array}$
\end{adjustbox}
\end{equation}
Like definition (\ref{eq:stratified_full_data}), this definition also implicitly uses a definition of treatment and covariates (cf. section \ref{sec:conf-adjustm-posit}) that ignores the multiple representatives problem (cf. section \ref{sec:append-mult-repr}). For a discussion of many of the issues that would be entailed by trying to use an estimand analogous to (\ref{eq:stratified_observed_estimand}) that addresses the multiple representatives problem, cf. section \ref{sec:strat-full-aver-treatm}.

\section{Estimation}
\label{sec:estimation}

For a given observed data ATE estimand $\lrestimandsymbol$, let its corresponding (plugin) estimator be denoted $\lrestimatorsymbol$. The latter are defined explicitly below.

\paragraph{{\small (Estimator)} Picky treatment, picky control:}
\begin{equation}
  \label{eq:lr_estimator_picky_defns_geom_picky_picky}
  \begin{split}
\lrestimator{\relax}{p}{p}(\time)(\specie_1, \specie_2) := & \mean*{\log\left(\abundance[\specie_2](\time) \right) |\observed[\time] = \{\specie_1, \specie_2\}  }\\
& - \mean*{\log\left(\abundance[\specie_2](\time) \right) | \observed[\time] = \{\specie_2\} } \\
  \end{split}
\end{equation}

\paragraph{{\small (Estimator)} Picky treatment, gluttonous control:}
\begin{equation}
 \label{eq:lr_estimator_picky_defns_geom_picky_gluttonous}
 \begin{split}
   \lrestimator{\relax}{p}{g}(\time)(\specie_1, \specie_2) := &\mean*{\log\left(\abundance[\specie_2](\time) \right) |\observed[\time] = \{\specie_1, \specie_2\}  }\\
& - \mean*{\log\left(\abundance[\specie_2](\time) \right) | \observed[\time] \supseteq \{\specie_2\}, \specie_1 \not\in \observed[\time] } \\  
 \end{split}
\end{equation}

\paragraph{{\small (Estimator)} Gluttonous treatment, picky control:}
\begin{equation}
  \label{eq:lr_estimator_picky_defns_geom_gluttonous_picky}
  \begin{split}
\lrestimator{\relax}{g}{p}(\time)(\specie_1, \specie_2) := &\mean*{\log\left(\abundance[\specie_2](\time) \right) |\observed[\time] \supseteq \{\specie_1, \specie_2\}  }\\
& - \mean*{\log\left(\abundance[\specie_2](\time) \right) | \observed[\time] = \{\specie_2\} } \\
  \end{split}
\end{equation}

\paragraph{{\small (Estimator)} Gluttonous treatment, gluttonous control:}
\begin{equation}
    \label{eq:lr_estimator_picky_defns_geom_gluttonous_gluttonous}
\begin{split}
  \lrestimator{\relax}{g}{g}(\time)(\specie_1, \specie_2) := &\mean*{\log\left(\abundance[\specie_2](\time) \right) |\observed[\time] \supseteq\{\specie_1, \specie_2\}  }\\
& - \mean*{\log\left(\abundance[\specie_2](\time) \right) | \observed[\time] \supseteq \{\specie_2\}, \specie_1 \not\in \observed[\time] } \\
\end{split}
\end{equation}

\subsection{Stratified Estimator Definition and TMLE}
\label{sec:strat-estim-defin}

The motivation for using the following definition is explained in section \ref{sec:conf-adjustm-posit}.
\begin{equation}
  \label{eq:stratified_estimator}
  \begin{adjustbox}{max width=\textwidth,keepaspectratio}
$\displaystyle  \begin{array}{rcl}
    \lrestimatorsymbol(\time)(\strain_1, \strain_2) &:= & \displaystyle \sum_{\rvec{b} \in \{0,1\}^{\Strains - 2}} \left[
\mean*{ \log\left( \abundance[\strain_2](\time)  \right)  | \straincount[\strain_1](\time) = 1 , \vstraincount^{\{\strain_1, \strain_2\}^c}(\time) = \rvec{b} }     \right.
                                                          \\
 && \displaystyle \left. -   \mean*{ \log\left( \abundance[\strain_2](\time)  \right)  | \straincount[\strain_1](\time) =0 , \vstraincount^{\{\strain_1, \strain_2\}^c}(\time) = \rvec{b} }   \right]
 \cdot \frac{ \displaystyle\sum_{\droplet \in [\Droplets]} \indicator{  \vstraincount_{\droplet}^{\{ \strain_1, \strain_2 \}^c}(\time) = \rvec{b} }  }{\Droplets}   \,.                         
  \end{array} $
\end{adjustbox}
\end{equation}
Note that we can do much better than this na\"{\i}ve plugin estimator when it comes to estimating a stratified ATE estimand of the form (\ref{eq:stratified_full_data}) or (\ref{eq:stratified_observed_estimand}). Even when it is not possible in finite samples to stratify over all possible values of the covariates (cf. sections \ref{sec:outc-treatm-covar} and \ref{sec:why-expect-posit}), for our initial estimate it is still possible to use SuperLearner (see \cite{vanderLaan2007} and \cite[Chapter 3]{vanderLaan2011}) to implement an ensemble machine learning approach that does not require us to make additional assumptions that we do not actually believe. For example, we could use different settings for GLMnet \cite{GLMnet} regression on the binary treatment and covariates as some of the learners in the ensemble considered by SuperLearner.

Afterwards, we can use the targeting step as implemented in e.g. the \textit{R} package '\textit{tmle}' \cite{tmle} for targeted minimum-loss based estimation (TMLE). This adjusts for the high-dimensional covariates in a way that makes the initial estimate produced by SuperLearner unbiased. Targeted learning leads to the most efficient and effective use of the available data possible, because our efforts are focused only on estimating the portion $\mathcal{Q}_O$ of the observed probability distribution $\mathcal{P}_O$ that is relevant for determining our chosen target parameter (\ref{eq:stratified_observed_estimand}). The targeted learning story has already been developed in detail for this kind of estimand, cf. \cite{vanderLaan2011} and \cite{vanderLaan2018}. The first five chapters of \cite{vanderLaan2011} are particularly relevant. In this preliminary exposition I do not go into the full details of implementing the targeted learning roadmap for this class of problem. However that is the goal of future work.

\section{Confounder Adjustment and Positivity}
\label{sec:conf-adjustm-posit}

Although not mentioned explicitly in sections \ref{sec:defin-thro-intro} and \ref{sec:defin-thro}, the gluttonous and picky groupings are two strategies for adjusting for possible confounders. These two strategies lie at opposite ends of a tradeoff. The tradeoff is between, on the one hand, using all droplets that might possibly be informative for inferring the effect of one strain on the other, and, on the other hand, completely controlling for the presence of third strains. To not give up on being able to use all possibly informative droplets, the gluttonous groups give up on being able to completely control for the presence of third strains. In order to not give up on being able to completely control for the presence of third strains, the picky groups give up on being able to use all possibly informative droplets.

As discussed in section \ref{sec:how-micr-ecol}, one reason for choosing to analyze data from MOREI, rather than from e.g. an observational metagenome study, is that the encapsulation in droplets should already control for confounding due to other strains $\strain[*] \not= \strain_1, \strain_2$ as much as possible. In particular, one of the goals is so that the data is as comparable as possible to the ``gold standard'' of manually plating cell cultures (cf. e.g. \cite{Venturelli}), for which\footnote{
  Here $\rvec{0}_{\Strains - 2}$ is used to denote the all-$0$'s vector of length $(\Strains - 2)$.
} $\vstraincount^{\{\strain_1, \strain_2\}^c}(0) = \rvec{0}_{\Strains - 2}$ always. Our goal is that, all other things being equal\footnote{E.g. the experiment is set up reasonably so that the expected number of cells per droplet $\rate$ is not too large, in particular not a number much larger than $2$. Cf. again section \ref{sec:init-dropl-form}.}, in general $\rvec{0}_{\Strains - 2}$ will be the value of $\vstraincount^{\{\strain_1, \strain_2\}^c}(0)$ that occurs with the highest probability. As long as we have enough data to make a reliable, estimate using the picky groups only, that is what we always prefer to do. Whenever we use any droplets for which $\vstraincount^{\{\strain_1, \strain_2\}^c}(0) \not= \rvec{0}_{\Strains - 2}$ while making estimates, it is (almost) \textit{always} due to necessity induced by ``data starvation'' (cf. again the discussion in section \ref{sec:defin-thro-intro}). This last point is important to keep in mind for later.

\subsection{Outcomes, Treatments, Covariates}
\label{sec:outc-treatm-covar}

The problem of confounder adjustment or control can be understood within the framework of splitting the observed data (``$O$'') between (i) outcomes/effects (``$Y$''), (ii) treatments/exposures (``$A$''), and (iii) covariates/potential confounders (``$W$'') as is done e.g. in the references \cite{vanderLaan2011}, \cite{vanderLaan2018}, and \cite{Petersen2010}. Recall that throughout this chapter we are using ATEs to study the effect of strain $\strain_1$ on the growth of strain $\strain_2$.

The definitions in section \ref{sec:estim-simpl-estim} correctly suggest that the outcome $Y$ we are interested in is $\log\left(\abundance[\strain_2](\time) \right)$. When we assume the full data is available and known, then the definitions in section \ref{sec:poss-choic-resolve} correctly suggest that the outcome $Y$ we are interested in becomes $\log\left(\frac{\abundance[\strain_2](\time)}{\abundance[\strain_2](0)}\right)$.

The treatment $A$ we are interested in is the presence/absence $\straincount[\strain_1](0)$ of strain $\strain_1$. Observe how, by using the binary $\straincount[\strain_1](0)$ as the treatment instead of $\abundance[\strain_1](0)$, we are effectively choosing to ignore the multiple representatives problem (cf. section \ref{sec:append-mult-repr}). It is clear that the value of $\straincount[\strain_1](0)$ is the treatment variable, because (given a fixed choice between ``gluttonous'' or ``picky'') a value of $\straincount[\strain_1](0) = 1$, which is equivalent to the condition $\{\strain_1\} \in \observed$, is always what distinguishes the treatment group from the control group, the latter corresponding to $\straincount[\strain_1](0) =0$ or equivalently $\{\strain_1\} \not\in \observed$.

Finally this leaves us to address the covariates, or potential confounders, $W$. These correspond of course to strains $\strain[*] \not= \strain_1, \strain_2$ which are not directly relevant to the effect of strain $\strain_1$ on the growth of $\strain_2$ (although of course they might potentially mediate that effect, cf. e.g. section \ref{sec:form-as-stat}, \cite{Momeni2017}, or \cite{higher_order}). With the length $\Strains$ binary random vector $\vstraincount(0)$ representing the presence/absence of all $\Strains$ strains at time $0$, let $\vstraincount^{\{\strain_1, \strain_2\}^c}(0)$ denote the length $(\Strains-2)$ binary random vector containing the $(\Strains - 2)$ entries of $\vstraincount(0)$ corresponding to the $\Strains - 2$ strains $\strain[*] \not= \strain_1, \strain_2$, i.e. all indices in $\{\strain_1, \strain_2\}^c := [\Strains] \setminus \{\strain_1, \strain_2\}$. Reviewing the definitions of  (\ref{eq:stratified_full_data}), (\ref{eq:stratified_observed_estimand}), and (\ref{eq:stratified_estimator}), and because the definitions of the gluttonous groups differ from those of their picky counterparts only\footnote{
I.e. irrespective of the particular values of the nonzero entries of $\vabundance(0)$.
} in the presence/absence of strains $\strain[*] \not= \strain_1, \strain_2$, it is hopefully clear that the relevant choice of covariates, and thus potential confounders, is $\vstraincount^{\{\strain_1, \strain_2\}^c}(0)$. Observe how considering entries of the vector $\vstraincount(0)$, rather than entries of the vector $\vabundance(0)$, in defining our covariates $W$ again implies that we are choosing to ignore the multiple representatives problem (cf. section \ref{sec:append-mult-repr}). Notice also that the picky definitions amount to requiring that we only consider droplets for which the value of \textit{all} entries of $\vstraincount^{\{\strain_1, \strain_2\}^c}(0)$ is $0$, whereas the gluttonous definitions allow us to consider droplets for which $\vstraincount^{\{\strain_1, \strain_2\}^c}(0)$ has \textit{any} of its $2^{\Strains - 2}$ possible values in $\{0,1\}^{\Strains - 2}$.

\subsection{Stratify to Attempt to Escape Tradeoff}
\label{sec:strat-attempt-escape}

To avoid the tradeoff for which the picky groups and gluttonous groups are positioned at opposite ends, we might like to be able to use all possibly informative droplets while still controlling for the presence of strains $\strain[*] \not= \strain_1, \strain_2$ \textit{in some way}, even if not completely controlling for them by excluding all droplets where they are present. We can achieve such a compromise by stratifying over the effects that occur for all possible values $\rvec{b} \in \{0,1\}^{\Strains - 2}$ of $\vstraincount^{\{\strain_1, \strain_2\}^c}(0)$ and weighting the effects according to the probabilities for $\vstraincount^{\{\strain_1, \strain_2\}^c}(0) = \rvec{b}$. Cf. the definitions (\ref{eq:stratified_full_data}), (\ref{eq:stratified_observed_estimand}), and (\ref{eq:stratified_estimator}). Such an approach in fact corresponds to the standard definition of average treatment effect (cf. chapters $1$ and $2$ of \cite{vanderLaan2011}).

Observe how the conditional expectation estimands for strata $\rvec{b}$ implying the presence of $3$ or more strains in the definitions (\ref{eq:stratified_full_data}) and (\ref{eq:stratified_observed_estimand}) amount to quantifications of the effects of ``higher-order interactions'', cf. section \ref{sec:form-as-stat}. Thus the stratifications in the definitions (\ref{eq:stratified_full_data}), (\ref{eq:stratified_observed_estimand}), and (\ref{eq:stratified_estimator}) can legitimately be interpreted as ``stratifying over all relevant higher-order interactions''. Hence being able to use (\ref{eq:stratified_estimator}) to effectively estimate (\ref{eq:stratified_full_data}) or (\ref{eq:stratified_observed_estimand}) requires being able to effectively estimate all higher-order interactions, which as discussed before in sections \ref{sec:why-empir-distr} and \ref{sec:form-as-stat} is where we begin to encounter serious problems.

In particular, the usefulness of the stratified ATE requires that the experimental treatment assignment (ETA) assumption, also known as the positivity assumption, be satisfied \cite[p. 35]{vanderLaan2011}. Quoting from \cite{Petersen2010}:
\begin{quote}
Positivity violations can arise for two reasons. First, it may be theoretically impossible for individuals with certain covariate values to receive a given exposure of interest... Second, violations or near violations of positivity can arise in finite samples due to chance. 
\end{quote}
The first kind of violation, the theoretical impossibility, corresponds to some of the conditional expectation estimands that are summands in the expression (\ref{eq:stratified_full_data}) not being well-defined. The second kind of violation or near violation, occurring for finite samples, can correspond to some of the plugin estimators for conditional expectations that are summands in the expression (\ref{eq:stratified_estimator}) either not being well-defined in the finite sample\footnote{
 With $0$ observations in the finite sample of at least one of the treatment or control group for that stratum, the corresponding sample conditional mean(s) would not be well-defined.
} or being unreliable/having high variance in the finite sample\footnote{
Due to the number of observations in the finite sample of at least one of the treatment or control group for that stratum being very small, even if the number of observations for both is technically nonzero. The corresponding sample conditional mean(s) would be unreliable.
}. Although the first kind of violation is not relevant here, I argue in the next section that in practice it is impossible for us to avoid the second kind of violation or near violation of the positivity assumption.

\subsection[Why Positivity Violations are Likely Unavoidable]{Why Positivity Violations are Likely Unavoidable in Practice, if not in Principle}
\label{sec:why-expect-posit}

To emphasize, the following bounds are not meant to be precise consequences proven from certain fixed assumptions, they are intended to be entirely heuristic. The basic ideas are that (1) the probability of a given strain appearing in a droplet should be roughly proportional to its relative abundance in the sampling population, and that (2) the probabilities of multiple strains appearing in the same droplet should be (``to $0$th or $1$st order'') roughly multiplicative.

The number of treatment samples for the effect of strain $\strain_1$ on the growth of strain $\strain_2$ can be thought of as roughly\footnote{E.g. assuming all strains have approximately the same relative abundance $\frac{1}{\Strains}$.} proportional to $\frac{1}{\Strains^2}$ or\footnote{
Strictly speaking this should be considered a blatant misuse of Landau Big-Oh notation.
} $O(\frac{1}{\Strains^2})$ in probability, whereas the number of control samples for the effect of strain $\strain_1$ on the growth of $\strain_2$ can be thought of as roughly proportional to $\frac{1}{\Strains}$ or $O(\frac{1}{\Strains})$ in probability. Cf. section \ref{sec:toy-model} for essentially the same idea. 

In particular, the number of samples that contain at least one of each of $I$ given strains $\strain_1$, $\strain_2$, \dots, $\strain_I$ would be roughly proportional to $\frac{1}{\Strains^I}$ or $O(\frac{1}{\Strains^I})$ in probability. Thus, even as the number of strata $\rvec{b} \in \{0,1\}^{\Strains - 2}$ in the definitions (\ref{eq:stratified_full_data}), (\ref{eq:stratified_observed_estimand}), and (\ref{eq:stratified_estimator}) \textit{\textbf{increases}} exponentially with increasing $\Strains$, at the same time the number of samples available for most strata $\rvec{b}$ roughly \textit{\textbf{decreases}} exponentially (in probability) with increasing $\Strains$. Thus, even in the absence of any theoretical positivity violations, i.e. even if $\mathbb{P}( \vstraincount^{\{\strain_1, \strain_2\}^c} (0) = \rvec{b} ) > 0 $ for all strata $\rvec{b}$, in practice we are almost always\footnote{
It should probably go without saying, but this is because it is not physically possible to indefinitely increase the total number of $\Droplets$ exponentially with respect to increasing $\Strains$.
} confronted with massive finite sample positivity violations. This is especially the case in the regime where $\Strains$ is large, which is the regime of greatest scientific novelty and thus interest. Cf. the similar problem motivating \cite{dirichlet_process}.

Cf. also section \ref{sec:why-empir-distr}, which discusses a closely related issue. Note again that conditional expectation estimands for strata $\rvec{b}$ implying the presence of $3$ or more strains effectively correspond to quantifying the effects of ``higher-order interactions''. While not explicitly targeting these higher-order interactions may possibly limit the usefulness of ecological interpretations of any results (cf. e.g. \cite{higher_order} or \cite{Momeni2017} for arguments to this effect), as discussed before already in section \ref{sec:form-as-stat} such a point is irrelevant to the extent that we simply lack the data required to reliably estimate these effects. (All of this also ignores the inevitable computational/algorithmic difficulties associated with attempting to estimate exponentially many parameters, as was mentioned before in section \ref{sec:why-empir-distr}.)

\subsection{Chosen Approaches to Positivity Violations}
\label{sec:how-picky-gluttonous}

Section 6.5 of \cite{Petersen2010} describes any strategy for approaching positivity violations as a ``tradeoff'' between ``altering the target parameter'' and ``improved identifiability''. Earlier I described both the picky and gluttonous approaches as opposite ends of another tradeoff. I explain below how the tradeoff mentioned earlier can be considered a special case of the tradeoff described in \cite{Petersen2010}.

In particular, the picky approach is a special case of the ``trimming'' strategy described in section 6.3 of \cite{Petersen2010}. By restricting the sample to only include the stratum for which $\vstraincount^{\{\strain_1, \strain_2\}^c} (0) = \rvec{0}_{\Strains - 2}$, we improve identifiability by removing the possibility of confounding due to the presence of strains $\strain[*] \not= \strain_1, \strain_2$.

The gluttonous approach does not seem to correspond directly to any of the strategies mentioned explicitly in \cite{Petersen2010} or chapter 8 of \cite{vanderLaan2011}, although it may be analogous to the ``projection'' approach mentioned therein for marginal structural models. The idea of the gluttonous approach is to alter the target parameter by ``collapsing all of the strata into a single stratum''. Because there is now only a single stratum, there is no longer any need to restrict samples according to their stratum membership, which in some sense improves the finite-sample identifiability from observed data. This is, of course, at the cost of substantially altering the research question being asked.

This leads to the question of how one chooses between the two approaches in practice. As implied above, in the introduction to \ref{sec:conf-adjustm-posit}, and sections \ref{sec:defin-thro-intro} and \ref{sec:defin-thro}, the main criterion is ``data starvation''. While it is true in general that the stratum corresponding to $\vstraincount^{\{\strain_1, \strain_2\}^c}(0) = \rvec{0}_{\Strains - 2}$ is usually the stratum with the highest probability of occurrence, especially for smaller finite samples this is often still not enough to guarantee that both sample conditional means in definition (\ref{eq:lr_estimator_picky_defns_geom_picky_picky}) are even well-defined\footnote{Much less calculated from a large enough sample to have manageable variance.}. Thus the ``trimming'' strategy only works when we have sufficient data from our largest stratum. When we lack sufficient data from our largest stratum, then it is perhaps unsurprising that our only remaining choice is to collapse all strata, i.e. adopt the gluttonous approach. Even then that sometimes is still not enough.

Recall from earlier (e.g. sections \ref{sec:toy-model} and \ref{sec:why-expect-posit}) how in general we are much more likely to suffer ``data starvation'' for the treatment groups than for the control groups. With that in mind, preliminary results (not shown) from simulated data appear to support the ``data starvation'' criterion for choosing between the picky and gluttonous approaches. In particular, for combinations involving strains with smaller relative abundances ($\approx 2\%$ or less) and thus having smaller finite-sample sizes, the ``gluttonous treatment and gluttonous control'' approach, cf. equation (\ref{eq:lr_estimator_picky_defns_geom_gluttonous_gluttonous}), appeared to be the most effective. Due to the small sample sizes, ``data starvation'' was evidently occurring for both the treatment and control groups. However, for combinations involving strains with larger relative abundances ($\approx 5-10\%$) the ``gluttonous treatment and picky control'' approach appeared to be the most effective, cf. equation (\ref{eq:lr_estimator_picky_defns_geom_gluttonous_picky}). Due to the slightly larger sample sizes, ``data starvation'' evidently was not a factor for the control groups, allowing the picky approach's stricter controlling for confounding to reap benefits. However, because treatment group sizes are much smaller than control group sizes, even given the larger sample sizes the treatment group sizes were still sufficiently small that the correction for ``data starvation'' implemented by the gluttonous approach remained useful. These preliminary results were not only for average treatment effect estimators, but also for other kinds of estimands (e.g. conditional correlations or rank-based coefficients). That the same pattern occurred for all of these different kinds of estimands strongly suggests that their shared tradeoff, between available sample size and strictness of controlling for confounders, most plausibly explains the pattern.

One might object to the collapsing of all strata implicit in the gluttonous approach as being too odious. However, in choosing to ignore the multiple representatives problem (cf. section \ref{sec:append-mult-repr}) by using binary variables $\vstraincount(0)$ as our exposure and covariates, rather than the count variables $\vabundance(0)$, we were also effectively already choosing to collapse strata. Cf. again section \ref{sec:outc-treatm-covar}. For example, a stratum corresponding to $\straincount[\strain_1](0) = 1$ can be considered collapsing strata corresponding to $\abundance[\strain_1](0)  = 1$, $\abundance[\strain_1](0)  = 2$, $\abundance[\strain_1](0)  = 3$, \dots. Therefore I argue that the gluttonous approach should not be considered any more odious than our previous choice to ignore the multiple representatives problem.

It is true that there is at least one important difference. Collapsing the strata corresponding to $\vstraincount(0)$ is necessitated by a lack of identifiability that in principle may hold for arbitrary, even infinite, sample sizes. In contrast, the further collapse of strata employed by the gluttonous approach is only due to finite-sample constraints. However, as argued before in section \ref{sec:why-expect-posit}, given that the sample sizes necessary to overcome those finite-sample constraints would be exponential, and thus usually not even physically possible, I argue that in practice this difference amounts to an unimportant distinction. Thus even if the gluttonous approach is odious, I argue that in practical terms it is no less of a necessary evil than choosing to ignore the multiple representatives problem.

\section{Sufficient Conditions for Identifiability}
\label{sec:proofs}

Section \ref{sec:ident-observed} in effect addresses the question: why (given that we may be able to ignore the multiple representatives problem, cf. section \ref{sec:append-mult-repr}) is the presence/absence of strains at the final time $\time$ sufficient to completely characterize the presence/absence of strains initially at time $0$? Section \ref{sec:ident-fitness} in effect addresses the question: why can we ignore the multiple representatives problem?

\paragraph{Comment:}
Recall from section \ref{sec:dist-distr-numb} how $\support(\vabundance(0)) = \support(\vstraincount(0))$. Yet at the same time $\observed = \observed[\time]$ if and only if $\vstraincount(0) = \vstraincount(\time)$. Hence claims about the identifiability of $\observed$ can equally be interpreted as claims about the identifiability of $\vstraincount(0)$. In particular, even in the presence of an identification gap between the full data estimand $\lrestimandsymbolfull (\time) (\specie_1, \specie_2)$ and the observed data estimand $\lrestimandsymbol (\time) (\specie_1, \specie_2)$, because the assumptions for identifiability of $\observed$ from the observed data are (arguably) much milder, it is still appealing to define treatment and control groups in terms of $\vstraincount(0)$ instead of $\vabundance(0)$ and thereby neglect the multiple representatives problem (cf. section \ref{sec:append-mult-repr}).

\subsection{Determination of $\observed$ from Observed Data}
\label{sec:ident-observed}

The assumption that
\begin{equation}
  \label{eq:Identifiability_of_support}
  \observed = \observed[\time] \,,
\end{equation}
can be broken down into two assumptions.

\subsubsection{No Spontaneous Generation}
\label{sec:no-spont-gener}

This assumption says that the number of strains within any given droplet can not increase over time, throughout the course of the experiment:
\begin{equation}
  \label{eq:no_spont_gener}
\observed \supseteq \observed[\time]  \,.
\end{equation}
In other words, strains that were not present during the droplet's formation can not later spontaneously appear within the droplet.

This is certainly true if we consider only the incubation of the droplets (cf. section \ref{sec:growth-cells-inside}). As Louis Pasteur already demonstrated in the 19th century using a swan neck flask, the creation of cells of a given strain of microbe requires the prior presence of cells of that strain. The theory of ``spontaneous generation'' is false, so microbes of a given strain can not spontaneously arise from non-living matter, preventing the number of strains from spontaneously increasing. Thus prima facie this assumption appears to be entirely unquestionable.

However, the name of the assumption is somewhat misleading. In practice, the implausible occurrence of spontaneous generation may not actually be necessary for this assumption to be violated. Indeed, experimental error that causes some small percentage of the droplets to be merged incorrectly, or to otherwise incorrectly receive the same barcode (cf. sections \ref{sec:prep-sequ} and \ref{sec:read-counts-vs}) would cause a similar effect to this assumption being violated. Preliminary results suggested that this effect may be negligible for inference. The simulation found little to no deleterious effect on inference caused by the rate of merging error assumed by the simulation. My collaborator also said that MOREI protocol has subsequently been improved to drastically further reduce the rate of merging error. Hence, although this assumption may not be \textit{entirely} unquestionable in practice, it still appears to be a safe assumption to make.

\subsubsection{No Censoring}
\label{sec:no-censoring}

This corresponds to the other direction, namely the assumption that the number of strains cannot decrease over time, throughout the course of the experiment:
\begin{equation}
  \label{eq:no_censoring}
  \observed \subseteq \observed[\time] \,.
\end{equation}
In other words, once a given strain is present in a droplet, it can never disappear. This assumption is actually much less realistic in practice than that from section \ref{sec:no-spont-gener}. In particular, in the case of antagonistic interactions, where one strain harms another, we would expect the antagonized strain to quite often go extinct.

Thus the failure of this identifiability assumption poses particular danger to the inference of antagonistic microbial interactions (negative edge weights in the signed network representation, cf. section \ref{sec:form-as-stat}). However, as discussed in sections \ref{sec:prep-sequ}, \ref{sec:read-counts-vs}, and \ref{sec:modell-pcr-ampl}, given that the data we observe does not actually correspond exactly to the counts of living cells $\vabundance(\time)$ at the time of sequencing $\time$, there are reasons to believe that such censoring concerns could be mitigated by phenomena that occur in practice\footnote{
Preliminary results suggest that, while avoiding censoring this way could increase our ability to infer antagonistic interactions, it could also decrease our ability to infer the magnitude of beneficial interactions (which correspond to positive edges in the signed network representation, cf. again section \ref{sec:form-as-stat}). This is because overcounting the number of living benefit-inducing cells required to induce a certain growth advantage (by including ``relic'' genetic material in the count) causes us to underestimate the strength of the beneficial growth effect. Therefore ``relic'' genetic material is potentially a double-edged sword.
}, e.g. ``relic'' genetic material\footnote{
Cf. section \ref{sec:prep-sequ} and appendix \ref{sec:read-counts-vs} for more discussion of ``relic'' genetic material.
}. In other words, we might expect the sequencing results to be more roughly proportional to the total number of cells that have lived \textit{ever}, rather than necessarily be roughly proportional to the current number of living cells. Cf. section \ref{sec:read-counts-vs} and figure \ref{fig:cell_number_integration} in particular.

Whether censoring occurs depends not only on the living or dying of cells during the second (incubation) phase of the experiment as described in section \ref{sec:using-morei-char}, but also on possible effects of PCR amplification that occurs during the third (sequencing) phase. This presents several possible complications. One is the possible ``swamping out'' in the sequencing results of the signal from strains with small relative abundances. Cf. footnote \ref{footnote:spikein} from section \ref{sec:prep-sequ}. Regardless of whether the genetic material is ``relic'' or from living cells, if the proportion corresponding tor a given strain is small, then its sequence might not show up in the sequencing results after PCR amplification.
Cf. the discussion of ``detection limits'' as a possible cause of ``zero inflation'' in \cite[section 1.1]{McDavid2019}.
Thus a strain might not be registered as having been in the droplet even though the droplet contained genetic material from living and/or dead cells of the strain. (Such effects could be called ``technical zeros'' using the terminology of \cite{Prost2021}.) This is related to ``jackpot effects'', cf. e.g. \cite{Marotz2019}.

In particular, the number of reads after PCR amplification is not necessarily a linear function of the initial number of reads, not even ``on average''. In the case of such ``non-linear PCR amplification'', the relative proportions of reads for each of the strains before PCR amplification may not still be the same after PCR amplification. This is in contrast to the predictions of the simple, na\"ive model mentioned in appendix \ref{sec:modell-pcr-ampl}, for which the number of amplified reads is a constant linear factor times some multiplicative noise. Besides potential non-linearity or multiplicative noise of PCR amplification, a further factor that might exacerbate ``technical zeros'' resulting from ``swamping out'' are sequence-specific PCR amplification biases. See \cite{bias_framework} or appendix \ref{sec:read-counts-vs} for a discussion.

Ultimately though, as stated before at the beginning of chapter \ref{cha:introduction}:
\begin{quote}
The problem of choosing good statistical models of PCR amplification is applicable to an extremely wide range of biological experiments. It should be its own project and fully addressing it is outside of the scope of this work. Previous work that has more substantially considered similar problems includes \cite{PCR1} \cite{PCR2} \cite{PCR3} \cite{PCR4}\cite{bias_framework}\cite{Marotz2019}. 
\end{quote}

Future research directions could include (1) gauging the extent to which experimental artifacts actually do blunt the effects of censoring in practice, (2) developing methods to adjust the analysis to account for the possibility of censoring, (3) gauging how often censoring might occur in practice, or (4) gauging to what extent failure of the no censoring assumption might bias the results (i.e. the size of the identification gap) in practice.

\paragraph{Comments:} This assumption also appears to effectively be necessary to sidestep the issue of whether it makes sense to use the convention that ``$\log(0) = 0$''. Namely because any droplets for which $\abundance[\strain_2](\time) = 0$ are always de facto excluded from the sample of droplets for which we assume that $\abundance[\strain_2](0) \ge 1$. This is because the no censoring assumption implies that such droplets do not exist, and because de facto we have no way to reliably identify them anyway, given that our only observations of the droplets are at time $\time$, and not time $0$.
Any future work which claimed to be able to identify such droplets would then have to put more thought into what convention should be used for ``$\log(0)$''.
Such ``zero inflation'' problems however are not inherently unique to the analysis of MOREI-like data, cf. for example \cite{McDavid2019} or \cite{Prost2021}. (However the problems posed by zeros are in many ways different for the compositional datasets studied in \cite{McDavid2019} or \cite{Prost2021}, whereas the data from MOREI is inherently not compositional, cf. again the discussion of spikein genes from section \ref{sec:prep-sequ}. Cf. e.g. \cite{microbiome_time_series} or \cite{Silverman2017} for subtleties of using compositional data.)

One might try to circumvent the limitations of this assumption as follows, at least for censoring caused by antagonistic interactions. Namely, given an appropriate ``neutral model'' predicting which percentage of droplets should contain certain combinations of strains based on the strains' relative abundances, we could look at whether any combinations have ``suspiciously'' small numbers of observations. From that we might conclude that in fact some of the droplets containing that combination at time $0$ were not counted due to one of the relevant strains becoming extinguished by time $\time$. For example, if strain $\strain_1$ harms strain $\strain_2$, and we see ``suspiciously'' few droplets containing both $\strain_1$ and $\strain_2$, we might conclude that several of the droplets seeming to only contain strain $\strain_1$ most likely originally also contained strain $\strain_2$ but that subsequently strain $\strain_2$ became extinct in those droplets by time $\time$.

Even ignoring the obvious (and potentially insurmountable) difficulty of identifying which droplets seeming to contain only $\strain_1$ actually originally also contained strain $\strain_2$ at time $0$, there are at least two issues with such a proposal.

First, the expected number of droplets containing a given combination of strains might be very small and thus subject to large variance (or at least large ``noise to signal ratio'', cf. the discussion of a similar issue in footnote \ref{footnote:spikein} from section \ref{sec:prep-sequ}). For example, even in a toy neutral model where the observed count of any given combination of strains varies usually by no more than $\pm 2$ from its expected count, if the expected count is $5$ and the observed count is $3$ or $2$, there is very little ``signal'' available with which to distinguish the two possibilities of random variation and censoring due to antagonistic interaction. Cf. the related discussion about practical violations of the positivity assumption in section \ref{sec:conf-adjustm-posit}. This is in contrast to e.g. a situation where the expected count is $50$ and the observed count is $20$ (instead of e.g. $48$ or $47$).

Second, even if we somehow manage to create a large enough total number of droplets $\Droplets$ such that all of the combinations of strains have large enough expected counts to have amenable ``signal to noise ratios'', we are still left with the problem of identifying an appropriate neutral model to compare against in order to flag ``suspicious'' observed counts. Given previous work using the Poisson distribution as the default neutral model for total cell counts in droplets for single-cell sequencing experiments (e.g. \cite{Collins2015}), it is tempting to accept the hPoMu working model as ``the unquestionably appropriate neutral working model''. However, even retaining the assumption that the total number of cells in each droplet is Poisson distributed (e.g. as is also the case for the hPoDM working model), one of the major conclusions of chapter \ref{chap:data_throughput} is that in a neutral model the observed counts can still be ``statistically significantly'' lower than the expected counts predicted by hPoMu. Preliminary work (not shown) also suggested that trying to identify ``suspicious'' low observed counts using hPoMu as a neutral model is most likely doomed to failure, with almost all examples identified as potential instances of censoring due to antagonism being false positives caused by the bias introduced by the misspecified choice of neutral model.

Thus choosing an appropriate neutral model requires either fitting to some relevant data, or establishing a substantial body of research that identifies what is in most scenarios an appropriate neutral model (assuming that such an appropriate ``universal'' choice even exists, which it may not). One reason for recommending the ``$\time = 0$ pre-experiment'' described in section \ref{sec:t=0-experiment} is because, by not incubating the droplets, it allows us to control for the possibility that ``suspicious'' observed counts are due to antagonistic interactions, rather than an incorrectly chosen neutral model. Therefore the ``$\time = 0$ pre-experiment'' is ideally suited for producing data with which to fit an optimal neutral model.

\subsection{Identifiability of $\lrestimandsymbolfull (\time) (\specie_1, \specie_2)$ from Observed Data}
\label{sec:ident-fitness}

With respect to the notation used in the following lemmas below, one has that the picky treatment ATE estimands correspond to the choice:
\begin{equation}
  \label{eq:picky_numerator_powerset_subset}
  \mathscr{O}_{\treatment} :=  \left\{ \Omega \subseteq [\Species] :  \Omega = \{ \specie_1, \specie_2 \}  \right\} \,,
\end{equation}
while the gluttonous treatment ATE estimands correspond to the choice:
\begin{equation}
  \label{eq:gluttonous_numerator_powerset_subset}
  \mathscr{O}_{\treatment} :=  \left\{ \Omega \subseteq [\Species] :  \Omega \supseteq \{ \specie_1, \specie_2 \}  \right\} \,.
\end{equation}
Similarly, the picky control ATE estimands correspond to the choice:
\begin{equation}
  \label{eq:picky_denominator_powerset_subset}
  \mathscr{O}_{\control} :=  \left\{ \Omega \subseteq [\Species] :  \Omega = \{ \specie_2 \}  \right\} \,,
\end{equation}
and the  gluttonous control ATE estimands correspond to the choice:
\begin{equation}
  \label{eq:gluttonous_denominator_powerset_subset}
  \mathscr{O}_{\control} :=  \left\{ \Omega \subseteq [\Species] :  \Omega \supseteq \{ \specie_2 \}, \specie_1 \not\in \Omega  \right\} \,.
\end{equation}

\begin{lemma}
\label{lem:geom_simplify}
  Assuming that
  \begin{itemize}
  \item  the marginal distributions of $\vabundance (0)$ are mutually independent, e.g. as turns out to be true for the hPoMu and hNBDM distributions,
  \item that the distributions of the droplets are independent,
  \item we always have that $\observed[\time] = \observed$ (cf. section \ref{sec:ident-observed}),
  \end{itemize}
then for any two subsets $\mathscr{O}_{\treatment}, \mathscr{O}_{\control} \subseteq \powerset{[\Species]}$ of the power set of $[\Species]$ such that $\specie_2 \in \Omega_{\treatment}$ for all $\Omega_{\treatment} \in \mathscr{O}_{\treatment}$ and $\specie_2 \in \Omega_{\control}$ for all $\Omega_{\control} \in \mathscr{O}_{\control}$, then the estimator
\begin{equation}
  \label{eq:lrestimator_geom}
  \begin{split}
 \lrestimatorsymbol (\time) (\specie_1, \specie_2) := &\mean*{ \log \left(\abundance[\specie_2](\time)\right) | \observed[\time] \in \mathscr{O}_{\treatment} }  \\
&- \mean*{ \log \left(\abundance[\specie_2](\time)\right) | \observed[\time] \in \mathscr{O}_{\control} }
  \end{split}
\end{equation}
is statistically consistent for the full data estimand assuming that $\vabundance(0)$ is known
\begin{equation}
  \label{eq:lrestimand_geom}
  \begin{split}
    \lrestimandsymbolfull (\time) (\specie_1, \specie_2) :=& \expectation*{\log \left( \frac{\abundance[\specie_2](\time)}{\abundance[\specie_2](0)} \right) | \observed \in \mathscr{O}_{\treatment} } \\
& -\expectation*{\log \left( \frac{\abundance[\specie_2](\time)}{\abundance[\specie_2](0)} \right) | \observed \in \mathscr{O}_{\control} } \,.
  \end{split}
\end{equation}
\end{lemma}

\noindent\textit{Proof of Lemma \ref{lem:geom_simplify}:}

\paragraph{Proof of identifiability of $\lrestimandsymbolfull (\time) (\specie_1, \specie_2)$ from the observed data:}~\\
Using the properties of logarithms and the linearity of expectation one has that the expression (\ref{eq:lrestimand_geom}) defining $\lrestimandsymbolfull (\time) (\specie_1, \specie_2)$ above equals
\begin{equation}
\label{eq:geom_simplify}
  \begin{split}
\expectation*{\log(\abundance[\specie_2](\time)) | \observed \in \mathscr{O}_{\treatment}  } 
- \expectation*{\log(\abundance[\specie_2](\time)) | \observed \in \mathscr{O}_{\control} } \\
 -  \left(  \expectation*{\log(\abundance[\specie_2](0)) | \observed \in \mathscr{O}_{\treatment}  }   
- \expectation*{\log(\abundance[\specie_2](0)) | \observed \in \mathscr{O}_{\control}  } \right) \,. 
  \end{split}
\end{equation}
By Lemma \ref{lem:total_expectation} using $\phi(n) = \log(n)$, the expression inside of parentheses in (\ref{eq:geom_simplify}) vanishes. Thus in this case $\lrestimandsymbolfull (\time) (\specie_1, \specie_2) = \lrestimandsymbol (\time) (\specie_1, \specie_2)$:
\begin{equation}
  \label{eq:geom_simplify_part_ii}
  \begin{array}{rcl}
    \lrestimandsymbolfull (\time) (\specie_1, \specie_2) & =&
     \expectation*{\log(\abundance[\specie_2](\time)) | \observed \in \mathscr{O}_{\treatment}  } 
                                                              - \expectation*{\log(\abundance[\specie_2](\time)) | \observed \in \mathscr{O}_{\control} } \\
                                                         & = &
 \expectation*{\log(\abundance[\specie_2](\time)) | \observed[\time] \in \mathscr{O}_{\treatment}  } 
                                                               - \expectation*{\log(\abundance[\specie_2](\time)) | \observed[\time] \in \mathscr{O}_{\control} } \\
    & = & \lrestimandsymbol (\time) (\specie_1, \specie_2) \,,
  \end{array}
\end{equation}
where we used the assumption that $\observed[\time] = \observed$ always.

\paragraph{Proof of consistency of $ \lrestimatorsymbol (\time) (\specie_1, \specie_2) $ for $\lrestimandsymbol (\time) (\specie_1, \specie_2)$:}~\\
Apply the Weak Law of Large Numbers to the first terms of (\ref{eq:lrestimator_geom}) and (\ref{eq:lrestimand_geom}) and the second terms of (\ref{eq:lrestimator_geom}) and (\ref{eq:lrestimand_geom}) respectively, and then use the fact $\mathbf{X}_n \overset{P}{\rightarrow} \mathbf{X}$ and $\mathbf{Y}_n \overset{P}{\rightarrow} \mathbf{Y}$ implies $(\mathbf{X}_n - \mathbf{Y}_n) \overset{P}{\rightarrow} \mathbf{X} - \mathbf{Y}$. $\qed$

\subsubsection{Practical Limitations of Identifiability Result}
\label{sec:pract-limit-ident}

Let $\hat{\boldsymbol{\Psi}}^F$ denote the ``oracle plugin estimator'' for $\lrestimandsymbolfull (\time) (\specie_1, \specie_2)$. (This can not be computed from the observed data, but an ``all-knowing oracle'' knows the observed data and the unobserved data from time $0$, and thus can compute plugin estimators that are functions of $\vabundance(0)$.) By the law of large numbers and the continuous mapping theorem we have that the oracle plugin estimator $\hat{\boldsymbol{\Psi}}^F$ converges in probability to $\lrestimandsymbolfull$, basically for the same reason that the observed data plugin estimator $\lrestimatorsymbol$ converges in probability to $\lrestimandsymbol$.

Now under the assumptions of Lemma \ref{lem:geom_simplify} we have that \textit{both} the oracle plugin estimator $\hat{\boldsymbol{\Psi}}^F$ and the observed data plugin estimator $\lrestimatorsymbol$ converge in probability to the $\lrestimandsymbolfull$ (because $\lrestimandsymbolfull = \lrestimandsymbol$, which again is not true in general). However, this does \textit{not} mean that we also have that the oracle plugin estimator equals the observed data plugin estimator. Indeed, although $(\hat{\boldsymbol{\Psi}}^F - \lrestimatorsymbol)$ approaches $0$ in probability, it is still the case that $(\hat{\boldsymbol{\Psi}}^F - \lrestimatorsymbol) \not= 0$. In particular, for finite samples the difference between the oracle and observed data plugin estimators will usually not be zero\footnote{This is related to the multiple representatives problem that is described in section \ref{sec:append-mult-repr}. Figure \ref{fig:multiple_representatives} illustrates clearly why the oracle and observed data plugin estimators can differ.}, and it may not even converge \textit{quickly} to $0$ in probability.

Thus the usefulness of the result of Lemma \ref{lem:geom_simplify} in practice depends on how quickly the random variable $(\hat{\boldsymbol{\Psi}}^F - \lrestimatorsymbol)$ converges to zero in probability. (We would like a finite-sample ``with high probability'' bound, but an asymptotic convergence rate in probability would not be bad either.)
Because the size of the treatment group $\treatmentsize[\strain_1, \strain_2]$ sample will typically grow much more slowly than the size of the control group $\controlsize[\strain_1, \strain_2]$ sample, the rate at which difference between the oracle and observed data plugin estimators converges to $0$ in probability should probably be bounded by the rate at which the difference\footnote{
Note that the quantity $\mean*{\log \left( \frac{\abundance[\specie_2](\time)}{\abundance[\specie_2](0)} \right) | \observed \in \mathscr{O}_{\treatment}}$ is also an ``oracle estimator'' that cannot be computed from the observed data. In fact it is the first term of $\hat{\boldsymbol{\Psi}}^F$.
} $\mean*{\log \left( \frac{\abundance[\specie_2](\time)}{\abundance[\specie_2](0)} \right) | \observed \in \mathscr{O}_{\treatment}} - \mean*{\log(\abundance[\specie_2](\time)) | \observed[\time] \in \mathscr{O}_{\treatment}}$ converges to $0$ in probability.
The latter difference only approaches zero in probability as the size of the treatment group sample grows $\treatmentsize[\strain_1, \strain_2]$ without bound.
Hence to the extent that $\treatmentsize[\strain_1, \strain_2]$ grows slowly in probability, we should not necessarily expect the difference between the oracle and observed data plugin estimators to converge quickly to $0$ in probability.

\subsubsection{Proofs of Supporting Lemmas}
\label{sec:proofs-supp-lemm}

\begin{lemma}
\label{lem:total_expectation}
  Assuming that
 \begin{itemize}
  \item  the marginal distributions of $\vabundance (0)$ are mutually independent, e.g. as turns out to be true for the hPoMu and hNBDM distributions,
  \item that $\mathscr{O} \subseteq \powerset{[\Species]}$ is such that $\specie_2 \in \Omega$ for all $\Omega \in \mathscr{O}$,
\end{itemize}
then for any measurable $\phi: \mathbb{N} \to \R$ one has
\begin{equation}
  \label{eq:secondly_uber_lemma}
  \expectation*{ \phi \left( \abundance[\specie_2] (0) \right) | \observed \in \mathscr{O} } = \expectation*{ \phi \left( \abundance[\specie_2] (0) \right) | \observed = \{\specie_2\} } \,.
\end{equation}
In particular,  for any two subsets $\mathscr{O}_{\treatment}, \mathscr{O}_{\control} \subseteq \powerset{[\Species]}$ of the power set of $[\Species]$ such that $\specie_2 \in \Omega_{\treatment}$ for all $\Omega_{\treatment} \in \mathscr{O}_{\treatment}$ and $\specie_2 \in \Omega_{\control}$ for all $\Omega_{\control} \in \mathscr{O}_{\control}$ one has that 
\begin{equation}
  \label{eq:secondly_uber_lemma_v2}
  \expectation*{ \phi \left( \abundance[\specie_2] (0) \right) | \observed \in \mathscr{O}_{\treatment} } = \expectation*{ \phi \left( \abundance[\specie_2] (0) \right) | \observed \in \mathscr{O}_{\control} } \,,
\end{equation}
I.e. the value does not depend on the particular subset of the power set, as long as that subset of the power set is such that $\specie_2$ is in every set belonging to it.
\end{lemma}

It may be unnecessary to assume that $\phi$ is non-negative. I want to preclude potential pathological situations where e.g. the infinite summation is convergent but not absolutely convergent, and ensure that tools like the monotone convergence theorem are available.
\\~\\
\noindent \textit{Proof of Lemma \ref{lem:total_expectation}:} Using the law of total expectation, one has that
\begin{equation}
  \label{eq:subset-loq-equiv-proof}
  \begin{split}
    & \expectation*{ \phi\left(\abundance[\specie_2](0) \right) | \observed \in \mathscr{O}  } \\
= & \sum_{\Omega \in \mathscr{O} } \expectation*{ \phi(\abundance[\specie_2](0)) | \observed = \Omega} \probability*{\observed = \Omega | \observed \in \mathscr{O} } \\
= & \sum_{\Omega \in \mathscr{O}  } \expectation*{ \phi(\abundance[\specie_2](0)) | \observed = \{\specie_2\} } \cdot  \probability*{\observed = \Omega | \observed \in \mathscr{O} } \\
= &\expectation*{ \phi(\abundance[\specie_2](0)) | \observed = \{\specie_2\} } \cdot \left(\sum_{\Omega \in \mathscr{O} } \probability*{\observed = \Omega | \observed \in \mathscr{O}} \right) \\
=  &\expectation*{ \phi\left(\abundance[\specie_2](0) \right) | \observed = \{\specie_2\} } \cdot (1) \,,
  \end{split}
\end{equation}
where the second equality follows from Lemma \ref{lem:expectation}, the third equality is the distributive property, and the last equality follows because the events are disjoint and exhaustive, thus their probabilities sum to $1$. $\qed$

\begin{lemma}
\label{lem:expectation}
   Assuming that
 \begin{itemize}
  \item  the marginal distributions of $\vabundance (0)$ are mutually independent, e.g. as turns out to be true for the hPoMu and hNBDM distributions,
\end{itemize}
then for any $\Omega \subseteq [\Species]$ such that $\specie \in \Omega$ one has for measurable $\phi: \mathbb{N} \to \R$ that
\begin{equation}
  \label{eq:uber_lemma}
   \expectation*{ \phi \left( \abundance[\specie] (0) \right) | \observed = \Omega } = \expectation*{ \phi \left( \abundance[\specie] (0) \right) | \observed = \{\specie\} } \,.
\end{equation}
\end{lemma}

\textit{Proof of Lemma \ref{lem:expectation}:} Applying the law of total expectation, one has
\begin{equation}
  \label{eq:30}
  \begin{split}
&    \expectation*{ \phi \left(\abundance[\specie](0) \right) | \observed = \Omega }  \\
= & \sum_{n=1}^{\infty}  \expectation*{\phi \left(\abundance[\specie](0) \right)  | \abundance[\specie](0) = n } \cdot \probability*{\abundance[\specie](0) = n  | \observed = \Omega } \\
= & \sum_{n=1}^{\infty}  \expectation*{\phi \left(\abundance[\specie](0) \right)  |\abundance[\specie](0) =  n } \cdot \probability*{ \abundance[\specie](0) = n  |  \observed = \{\specie\} } \\
= & \expectation*{\phi \left(\abundance[\specie](0) \right)  | \observed = \{\specie\}  } \,,
  \end{split}
\end{equation}
where the second equality was an application of Lemma \ref{lem:probability}, and the last equality another use of the law of total expectation. $\square$

\begin{lemma}
\label{lem:probability}
    Assuming that
 \begin{itemize}
  \item  the marginal distributions of $\vabundance (0)$ are mutually independent, e.g. as turns out to be true for the hPoMu and hNBDM distributions,
\end{itemize}
then for any $\Omega \subseteq [\Species]$ such that $\specie \in \Omega$, and for any $n \in \mathbb{N}$ (i.e. $n>0$) one has
\begin{equation}
  \label{eq:24}
  \probability*{ \abundance[\specie](0) = n | \observed = \Omega } = \probability*{ \abundance[\specie](0) = n | \observed = \{\specie\} } \,.
\end{equation}
In particular the value does not depend on the particular $\Omega$ as long as $\specie \in \Omega$.
\end{lemma}

\textit{Proof of Lemma \ref{lem:probability}:} Noting that the empty product is defined (assumed) to equal $1$, e.g. in the case when $\Omega \setminus \{\specie\} = \emptyset$:
\begin{equation}
  \label{eq:45}
  \begin{split}
    & \probability*{\abundance[\specie](0)  | \observed = \Omega } \\
= & \frac{\displaystyle\probability*{\abundance[\specie](0) = n} \cdot \prod_{\specie[*] \in \Omega \setminus \{\specie\}} \probability*{ \abundance[{\specie[*]}](0) \ge 1 } \cdot \prod_{\specie[*] \not\in \Omega} \probability*{ \abundance[{\specie[*]}](0) = 0 }  }{\probability*{\observed = \Omega}} \\
= &  \frac{\displaystyle\probability*{\abundance[\specie](0) = n} \cdot \prod_{\specie[*] \in \Omega \setminus \{\specie\}} \probability*{ \abundance[{\specie[*]}](0) \ge 1 } \cdot \prod_{\specie[*] \not\in \Omega} \probability*{ \abundance[{\specie[*]}](0) = 0 }  }{ \displaystyle
\probability*{\abundance[\specie](0) \ge 1} \cdot \prod_{\specie[*] \in \Omega \setminus \{\specie\}} \probability*{ \abundance[{\specie[*]}](0) \ge 1 } \cdot \prod_{\specie[*] \not\in \Omega} \probability*{ \abundance[{\specie[*]}](0) = 0 } 
} \\
= & \frac{\probability*{\abundance[\specie](0) =n}}{\probability*{ \abundance[\specie](0) \ge 1 }} \,.
  \end{split}
\end{equation}
The last equality can be seen to clearly not depend on the specific $\Omega$, thus the value is the same as long as $\specie \in \Omega$. $\square$

\section{Conclusion}
\label{sec:conclusion-7}

\paragraph{Findings and Contributions}

I showed how a particular measure of relative fitness can be recast into the statistical framework of average treatment effects.
I explained how (for this problem) violations of positivity assumptions are inevitable in practice, making controlling for confounding difficult.
I gave explicit assumptions under which the corresponding estimands are identifiable from the observed data produced by incubated droplets.

\paragraph{Practical Implications}

This chapter demonstrates that this problem has a difficult balance between controlling for confounding and avoiding ``data starvation''. This makes it particularly important to be able to predict in advance how much data will be available, using the results from part \ref{part:modell-init-form}. By explicitly defining the intended estimands, the specific estimators, and the assumptions required for identifiability from the observed data, this chapter clarifies the goals and limitations of using these estimators.

\paragraph{Next Steps and Open Questions}

This chapter points the way forward to formalizing many other statistical estimators that are informally defined in the microbiological literature, even in the context of other experiments. Doing so will hopefully clarify the properties and performance of such estimators. For the analysis of MOREI data specifically, it would be helpful to identify other possible estimands besides those which were described in this chapter. Simulation studies comparing the performance of different estimators would be particularly useful. The next chapter, chapter \ref{chap:network_comparison}, addresses an important issue that we need to clarify when comparing the performance of different estimators in a simulation study.

\paragraph{Sparsification}
An issue with the estimators proposed in this chapter is their ``lack of sparsity''. The values of the estimators from section \ref{sec:estimation} will almost never be exactly $0$ in finite samples, even if they might approach $0$ asymptotically. Hence the corresponding estimated interaction network will almost always be a complete graph, even if the magnitudes of many, or even most, of the edges might be small.
One approach to address this ``lack of sparsity'' issue would be to impose an arbitrary magnitude threshold, below which any estimate values would automatically be set to $0$. That would be problematic for many reasons, not least of which would include deciding on a choice of threshold value and justifying that choice after it has been made. Another approach would be to throw out the average treatment effects framework altogether and use some modification of the sparse regression and variable selection methods mentioned in e.g. \cite{ruan_interactions} to apply to the ``gene $\times$ gene'' interactions problem we have here. Cf. again item \ref{item:gene_gene_interactions} from section \ref{broader-field-2}. A third approach begins with thinking of sparsity as false discovery control. Then one could attempt to modify the non-parametric method for false discovery control proposed in \cite{Ge2021}, originally designed for the ``gene $\times$ environment'' interactions problem, to apply to the ``gene $\times$ gene'' interactions problem. This modified method could then be used on the ATE estimators. 

\begin{coolsubappendices}
\section{Pre-Existing Notions of Positivity Violations in Microbiology Literature}
\label{sec:positiv-microbiology}
The contrast between the two kinds of positivity violations mentioned in \cite{Petersen2010} appears to be analogous, and perhaps even equivalent, to a similar dichotomy often presented in the existing microbiology literature. When discussing missing data in compositional (i.e. relative abundance) datasets, the authors of \cite{Prost2021} and other sources distinguish between ``structural zeros'' and ``sampling zeros''. It appears that the ``structural zeros'' (also called ``biological zeros'') of \cite{Prost2021} correspond to the situation from \cite{Petersen2010} of when ``it may be theoretically impossible for individuals with certain covariate values to receive a given exposure of interest''. Likewise, it appears that the ``sampling zeros'' (and the related ``technical zeros'') of \cite{Prost2021} correspond to the situation from \cite{Petersen2010} of when ``violations or near violations of positivity can arise in finite samples due to chance''. This suggests that causal inference is relevant to both of these problems previously documented in the microbiology literature. (Although cf. \cite{McDavid2019} for a different interpretation of ``technical zeros''.)

It is also mentioned in \cite{Prost2021} how ``methods that consider the conditional dependency structure of microbial networks (like probabilistic graphical models) should perform better... because [they have] the power to resolve direct from indirect associations (e.g. associations between two species mediated through a third species)''. This appears to correspond to the same motivation underlying confounder adjustment in a causal analysis. The main difference here is that \cite{Prost2021} refers to methods using undirected graphical models and outside of any explicit causal framework, whereas causal inference (more typically) relies on methods using directed graphical models (Bayesian networks). Nevertheless, the strength of the analogies here suggests the potential for future work that more directly applies a causal inference framework to make a major, positive impact on the study of microbial interactions.

\section{Combining Estimates from Different Batches}
\label{sec:combining-log-ratios}

An immediately obvious issue with using the ATE estimands is that they result in different estimates for each batch. Note that, since droplets from different batches were grown for different amounts of time, inherently they are not directly comparable. Especially to the extent that it might be expected that the nature of the interactions could change with time, this could potentially be a useful feature rather than a liability. (See e.g. \cite{umibato} for previous work tackling such issues head-on in the case of time series\footnote{
See \cite{microbiome_time_series} for an introduction to time series geared towards microbial ecologists.
} i.e. longitudinal data.) Collaborators have expressed interest in possibly estimating such functions of time. Readers interested in such approaches should look into the field of functional data analysis, cf. e.g. the monographs \cite{Ramsay2009}, \cite{Kokoszka2021}, \cite{Ramsay2005}, \cite{Horvath2012}, \cite{Hsing2015}, or \cite{Grenander1981} for an introduction. Either of the two reviews \cite{fda_review_1} or \cite{fda_review_2} may also be helpful, if less thorough.

\subsection{Criteria for Choosing Aggregation Method}
\label{sec:choos-aggr-meth}

In this preliminary work the goal will be to choose an aggregate summary of these estimates which
\begin{enumerate}
\item is structurally and conceptually simple,
\item is easy to implement,
\item sensibly incorporates information from all batches.
\end{enumerate}
This is so that any chosen summary method can be and is likely to be used in practice, even by practitioners without extensive quantitative training.

\subsection{Why Incorporate Information from All Batches}
\label{sec:why-incorp-inform}

Given that information from different batches will generally correspond to different microbial growth phases (lag, exponential, stationary), incorporating information from all batches is questionable. After all, arguably from a biological standpoint the growth behavior during exponential phase is what we would most want to quantify. Nevertheless, I still suggest that information from all batches be incorporated, regardless of whether they may be in lag or stationary phases. This is for at least two reasons. 

First, on theoretical or ``philosophical'' grounds, any statistical estimator which is unable to utilize potentially informative data (thus requiring that said data essentially be ``thrown out'') just because said data may be in some sense ``imperfect'' for learning about the problem at hand (e.g. potentially lag or stationary phase batches) is ``inefficient''. In other words, any method that (in general) requires discarding potentially informative data should be avoided on principle if viable alternatives exist. 

Second, on more practical grounds, it could be very inconvenient, frustrating, and time-consuming for the experimenter to run additional statistical analyses on each of the batches to attempt to discern which batches are mostly in what growth phase, as well as to need to repeatedly look up arbitrary rules of thumb for deciding thresholds for e.g. when a batch should be considered ``too far'' into stationary growth phase and have its corresponding data discarded. For the experimenter the simplest and easiest to use possible answer to the question ``Which batches should be used for the data analysis?'' will always be ``All of them''. This is true all the more so when one recalls how there are $O(\Strains^2)$ interactions to consider. Thus, as $\Strains$ becomes large, any ad hoc approach quickly becomes untenable, even if it would have been feasible for a single interaction.

\subsection{Weighting Contributions from Distinct Batches}
\label{sec:weight-contr-from}

Note that the obvious approach of directly averaging over all of the droplets from all of the batches simultaneously does not satisfy the above criteria. Because droplets from different batches are not directly comparable, averaging droplets over all batches to get a single estimate would not be sensible. 

While obviously there cannot be any universal or conclusive answer to the question of which frameworks ``best'' satisfy the aforementioned subjective criteria, another obvious choice would be some convex combination of some fixed function $f$ of the estimands from distinct batches, i.e.
\begin{equation}
  \label{eq:cvx_combo}
  \sum_{\batch=1}^{\Batches} \cvxcoeff{\batch} f \left( \lrestimatormatrix (\time_{\batch}) \right) \,, \quad \quad  \cvxcoeff{\batch} \ge 0 \text{ for all }\batch \in [\Batches] \,, \sum_{\batch=1}^{\Batches} \cvxcoeff{\batch} = 1 \,.
\end{equation}
Because there are legitimate biological reasons to believe that some batches (those in exponential growth phase) would be more informative than other batches in stationary or lag growth phases, in principle the coefficients $\cvxcoeff{\batch}$ in (\ref{eq:cvx_combo}) should be unequally weighted. 

However, without an automated and scientifically well-motivated framework for deciding the values of the weights, from a practical perspective this leads to an even worse generalization of the aforementioned problem with ad hoc thresholds for including or removing batches from the data analysis. Specifically, the problem from section \ref{sec:why-incorp-inform} would corresponding to the special case of deciding a subset of the coefficients $\cvxcoeff{\batch}$ to set to zero and then assuming the remaining non-zero coefficients correspond to equally informative batches and thus be given equal weights. Again, we have $O(\Strains^2)$ interactions to consider, which means one needs to decide the values of the weights $O(\Strains^2)$ times.

Given the constraint of practicality the only viable way forward\footnote{At least in this preliminary work, where deriving such an automated and scientifically well-motivated framework for deciding the values of the weights is outside of scope.} appears to be to brazenly assume that the data from all batches is equally informative for describing the microbial interactions. This means equally weighting all of the coefficients in the convex combination (\ref{eq:cvx_combo}). The result of doing so is simply the arithmetic mean of the $f\left( \lrestimatormatrix (\time_{\batch}) \right)$. 

Given the discussion in section \ref{sec:append-relat-less}, this might prompt the question of whether the geometric mean would be a more suitable summary statistic. Note that the estimators $\lrestimatormatrix$ correspond effectively to log-transformed data. The primary purpose of the logarithm is argabuly to transform multiplicative relationships in $(0, \infty)$ to additive relationships in $(-\infty, \infty)$. Therefore, provided that the ``link function'' $f$ is in some way ``additive'' itself, it seems that an additive summary, as expressed by the arithmetic mean, is more appropriate than the multiplicative summary represented by the geometric mean.

\subsection{Choice of ``Link Function''}
\label{sec:choice-link-function}

This is of course still leaves open the choice of the ``link function'' $f$, for which two alternatives are proposed
\begin{equation}
  \label{eq:link_function_choices}
  f_1 \left(  \lrestimatormatrix (\time_{\batch}) \right) = \lrestimatormatrix (\time_{\batch}) \,, \quad\quad f_2 \left( \lrestimatormatrix (\time_{\batch}) \right) = \frac{\lrestimatormatrix (\time_{\batch})}{||\lrestimatormatrix (\time_{\batch}) ||_1} \,,
\end{equation}
where $||\cdot||_1$ denotes the entrywise $L_1$ norm (i.e. equivalent to vectorizing the matrix and then taking the $L_1$ norm of the resulting vector), and does not denote an operator norm. Cf. section \ref{sec:entryw-l_1-relat}. The choices have been reduced to the (unweighted) average of the unnormalized ATE matrices, or the (unweighted) average of the ATE matrices after they have been normalized.

Starting from the assumption that each batch should be considered equally potentially informative, it follows that the average of the normalized batches is to be preferred. Normalizing the batches beforehand ensures that all of their entries are ``on comparable length scales'' and thus that every batch has ``equal contribution potential''. Taking the average of unnormalized batches, and thus implicitly assuming that the data from all batches are directly comparable, also introduces the question of why different batches are comparable but droplets from different batches are not. (Recall that the assumption that droplets from different batches are not comparable is the motivation for these``hierarchical'' averaging schemes, rather than just averaging over all droplets from all batches.)

As for the choice of norm, the $\lrestimatormatrix (\time_{\batch})$ matrices do not represent linear functions, so operator norms are not appropriate. Instead the matrices are (in physics parlance) entirely ``contravariant''. As for the choice between two common norms for entirely ``contravariant'' data, namely entrywise-$L_1$ and entrywise-$L_2$ (a.k.a. Frobenius), entrywise-$L_1$ seems more suitable. The $L_1$ norm is more ``robust'' to ``extreme'' entries, and it is additive for non-negative data, thus more ``compatible'' with the additive summary represented by the arithmetic mean. The geometric intuition related to the Pythagorean theorem that might be used to justify the use of the $L_2$ norm does not seem applicable here, and in fact could even be used to argue against its use.

\subsection{Dividing Out Time Heuristic Motivation}
\label{sec:dividing-out-time}

Normalization in this context also has the biological interpretation of ``revealing information about growth rates'' by ``dividing out time''. 
For an explanation of this heuristic, first imagine $\Batches$ batches, with each batch $\batch$ grown until time $\time_{\batch}$ and containing $\Species$ droplets, each droplet $\specie$ starting a time zero with exactly one cell of strain $\specie$. 

Under the simple exponential growth model (\ref{eq:exponential_system}), for each batch $\batch$ and strain $\specie$ one will have for the final number of cells:
\begin{equation}
  \label{eq:exp_growth_result_single_droplets}
  \log \left( \abundance[\specie] (\time_{\batch}) \right) = \baserate \time_{\batch} \,,
\end{equation}
where $\baserate$ is the ``intrinsic growth rate'' of strain $\specie$. Without first normalizing the values from each batch, the average log count of cells for strain $\specie$ across all batches is
\begin{equation}
  \label{eq:exp_growth_result_unnormalized_average}
\baserate \cdot  \frac{\time_1 + \cdots + \time_{\Batches}}{\Batches} \,,
\end{equation}
which says as much, or possibly more, about the amount of time each of the batches were grown as it does about the growth rate of strain $\strain$. In particular, rate estimates from batches which were allowed to grow for longer amounts of time will disproportionately influence the final result. However, if one normalized the entries from each batch first, the value recorded for strain $\strain$ from any batch $\batch$ will be
\begin{equation}
  \label{eq: exp_growth_result_normalized_average}
  \frac{\baserate \cdot \time_{\batch}}{\sum_{ {\specie[]}=1}^{\Species} \left(\baserate[{\specie[]}]  \cdot \time_{\batch}  \right) } = \frac{\baserate \cdot \time_{\batch} }{\left(\sum_{ {\specie[]}=1}^{\Species} \baserate[{\specie[]}]\right) \cdot \time_{\batch}   } = \frac{\baserate}{\sum_{{\specie[]}=1}^{\Species} \baserate[{\specie[]}] }  \,,
\end{equation}
which clearly will then also be the average value across all of the batches. 

Thus, under this simple (and blatantly inaccurate) model, first normalizing the batches literally ``divides out'' the amount of time each batch was grown and leaves only information about the intrinsic growth rates from each strain. Moreover, since the resulting value is the same from every batch, the final result corresponds to equal contributions from all batches, rather than the values from batches grown for longer amounts of time contributing disproportionately. The information about the growth rates in (\ref{eq: exp_growth_result_normalized_average}) of course doesn't say anything about their absolute scale, but in this context scientifically what is most interesting is ``qualitative'' information about the growth rates of the strains relative to each other, which clearly is contained in the information from (\ref{eq: exp_growth_result_normalized_average}) above.

Even under the more complicated model of bacterial growth (\ref{eq:simulation_equation}), and with potentially multiple strains in each droplet, the above argument is still true ``to zeroth order''. At time $\time$ the number of cells of strain $\specie$ in a given droplet will be a ``perturbation'' of the amount predicted (\ref{eq:exp_growth_result_single_droplets}) by the exponential growth model (\ref{eq:exponential_system}):
\begin{equation}
  \label{eq:perturbation_of_exp_growth}
\log \left( \abundance[\specie](\time) \right) = \baserate \time + \delta (\time) \,, \quad \delta(\time) := \int_0^{\time} \left[  \sum_{\specie[]=1}^{\Species} \interaction{\specie[]}{\specie} \abundance[\specie](\tau)  - \frac{  \rateterm(\tau)  \limitingfactor(\tau)  }{\carryingcapacity[]} \right] d \tau \,,
\end{equation}
where the notation from equation (\ref{eq:framework}) was used for the purposes of simplification. Arguing using the linearity of expectation and that the ATE estimators and simplified estimands are linear combinations of log counts like those in (\ref{eq:exp_growth_result_single_droplets}) and (\ref{eq:perturbation_of_exp_growth}) above, heuristically speaking it still seems reasonable to expect that the averages of the normalized values should usually give at least roughly better insight into the microbial interactions, as defined by their effects on (relative) growth rates, than the average of the unnormalized values. Thus what seems to have been borne out empirically is also something which could have been anticipated a priori on theoretical grounds.

\subsubsection{Comparison with Previous Work}
\label{sec:comparison_ate_defns}
The argument in this section appears to be very similar to the ideas in \cite{GE_competition_2}, in particular compare equation (4) and the Appendix of that paper. So although derived independently, the ideas from this section have precedent in the literature. I am unsure whether the definition of fitness that they propose as being better is applicable to this problem. In particular, with pseudolongitudinal data it is much more difficult or even impossible to clearly identify timepoints corresponding to the (average) beginning and end of an exponential growth phase. (This was already alluded to in sections \ref{sec:why-incorp-inform} and \ref{sec:weight-contr-from}.) The methods suggested in \cite{GE_competition_2} and its predecessor \cite{Holland1991} both seem to assume that one can clearly identify such timepoints. Valuable future work might investigate the feasibility of identifying such average timepoints from pseudolongitudinal data. That being said, as alluded to already in sections \ref{sec:why-incorp-inform} and \ref{sec:weight-contr-from}, even with such a method for pseudolongitudinal data implementation could still be difficult for the computational/combinatorial reasons that there are $O(\Strains^2)$ interactions for which this method would need to be applied.

\subsection{Treatment of Missing Data}
\label{sec:treatm-miss-data}

The above discussion has so far neglected to consider how in many instances estimates for certain interactions will be missing from certain batches. When normalizing the matrices, the missing entries are treated as if they were found to exactly equal zero. Zero seems to be the most conservative possible choice to impute for missing interaction estimates. In the absence of any evidence for an interaction, the choice is made to prefer the possibility of making the false negative error of assuming no interaction exists when one might exist, rather than possibly making the false positive error of assuming an interaction exists when in fact none exists. Cf. the discussion from appendix \ref{sec:param-estim-inter}.

 Afterwards, when computing the average of the normalized matrices, the ``entrywise average'' is taken, i.e. for each entry the sum of the non-missing values across all batches (equal to the sum of the values over all batches after imputing missing estimates as zero) is divided by the number of batches for which an estimate for the corresponding interaction was not missing (which of course is only strictly less than the total number of batches when an estimate for that interaction is missing from at least one batch). Cf. figure \ref{fig:entrywise_average}.

 \begin{figure}[H]
   \centering
   \includegraphics[width=\textwidth,height=\textheight,keepaspectratio]{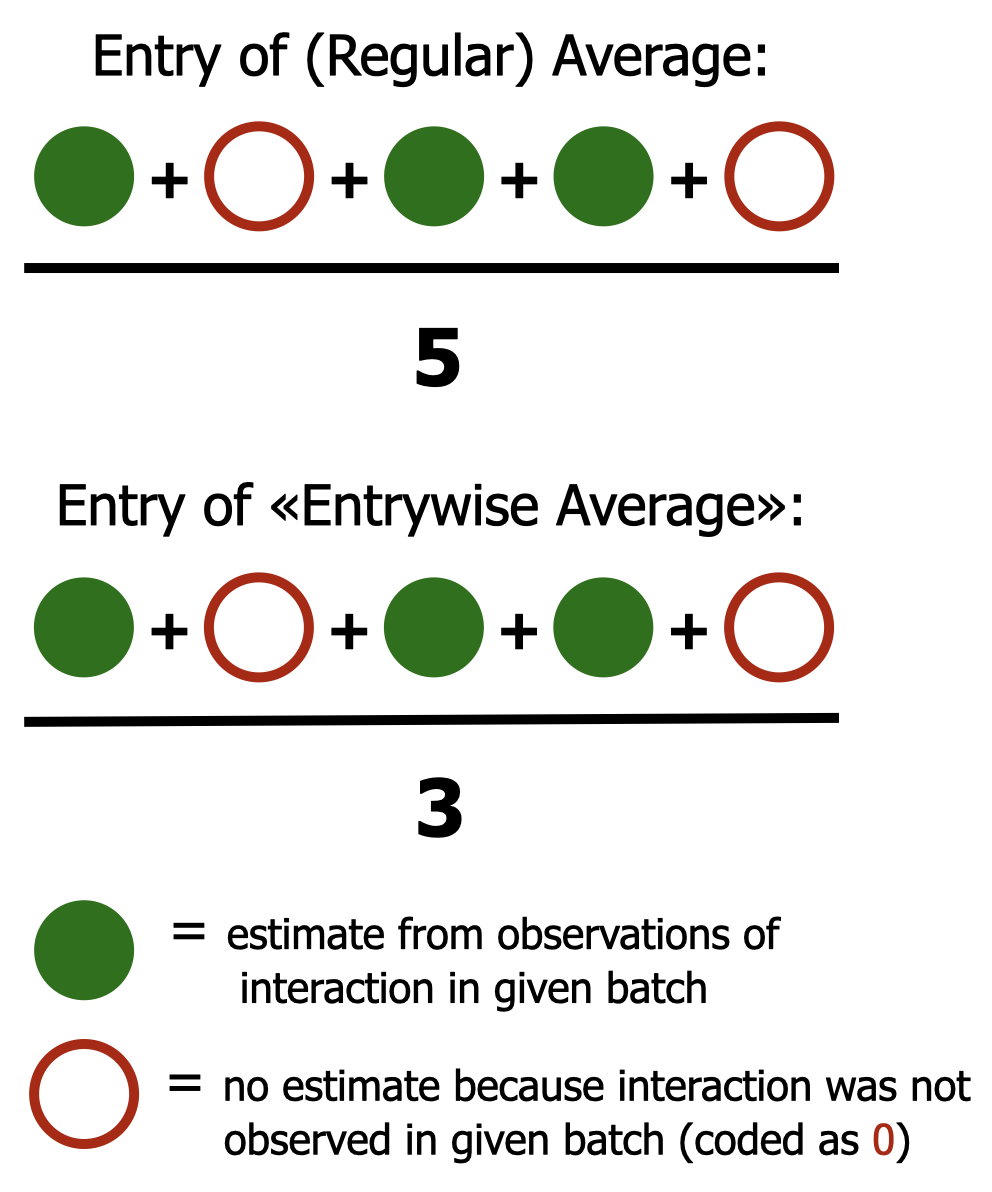}
   \caption[~Two distinct strategies for combining estimates from multiple batches.]{Comparison of ``entrywise average'' with taking the average of the corresponding matrices.}
   \label{fig:entrywise_average} 
 \end{figure}

Compared to dividing the sum of the values (where again the missing values are being treated as the additive identity, zero) by the total number of batches, this should in principle avoid artificially ``deflating'' some entries in the aggregate matrix just because estimates for the corresponding interaction were missing from some batches. (Of course to the extent that such values could be considered more unreliable, one could plausibly also argue that such ``artificial'' deflation/shrinkage of those values would be a feature, not a bug. Without modification however such an argument neglects to consider how the estimates from each batch are themselves averages whose own reliabilities are themselves contingent upon the number of droplets over which the average was formed.) This corresponds again to the assumption that all batches have ``equal potential to be informative''. Thus e.g. in the extreme case that an estimate for a given interaction was available only from one batch, that single estimate should be used ``as is'' in the final aggregate matrix, rather than being deflated or otherwise modified. 

Preliminary investigations of simulations (data not shown) seemed to suggest that this method of handling the missing estimates leads to results which are perhaps slightly more effective (at least in terms of Spearman correlation or relative error) than simply dividing all entries by the total number of batches regardless of the number of missing values. Any difference was difficult to notice however. Because the chosen approach would have because of the above arguments, regardless of empirical performance, any potential empirical performance differences were not examined in any further detail.

\section{Heuristic Derivation}
\label{sec:heur-deriv-log}

Basically the scientific goal is to compare the ``typical growth of strain $\specie_2$ in the presence of strain $\specie_1$'' to the ``typical growth of strain $\specie_2$ (in general)''. The idea is that any effect strain $\strain_1$ has on the growth of strain $\strain_2$ via their interactions will be the only, or at least primary, latent variable explaining any discrepancy between the aforementioned two types of ``typical growth'', and that absence of evidence for any discrepancy can probably be construed as evidence of absence of any underlying interaction causing strain $\specie_1$ to affect strain $\specie_2$. Cf. again sections \ref{sec:defin-thro-intro} and \ref{sec:how-micr-ecol}, or the discussion of ``competition assays'' from the introduction to Part \ref{part:introduction}. As noted before, previous work such as \cite{Momeni2017} or \cite{higher_order} has raised concerns about potential problems with assuming all interactions are pairwise interactions.

One way to define ``growth'' is the multiplicative factor by which the number of cells of a given strain in a droplet has increased relative to its initial population after the predetermined time for its batch has elapsed, with comparison of the growth factors defined via a ratio:
\begin{equation}
  \label{eq:heuristic_ratio}
  \frac{\displaystyle
\left( \begin{matrix} \displaystyle \text{`` typical growth factor }\frac{\abundance[\specie_2](\time_{\batch})}{\abundance[\specie_2](0)} \text{ of cells of strain }\specie_2\\ \text{ in the presence of strain }\specie_1 \text{ for batch }\batch\text{''}   \end{matrix} \right)
}{ \displaystyle
\left( \text{`` typical growth factor }\frac{\abundance[\specie_2](\time_{\batch})}{\abundance[\specie_2](0)} \text{ of cells of strain }\specie_2\text{ in general for batch } \batch \text{''} \right)
} \,.
\end{equation}
However there are several reasons to define ``growth'' in terms of the logarithms of the aforementioned quantities instead. First, doing so accounts for the fact that in the eutrophic regime the number of cells will grow more rapidly the more cells there already are present simply by virtue of the fact that there are more cells present to divide, whereas the logarithm of the number of cells will continue to grow at a constant rate (compare equation \ref{eq:exponential_system} from appendix \ref{chap:relev-diff-equat}). Second, the logarithm corresponds to the natural assumption that e.g. ``causing to grow twice as many cells'' should be considered the opposite of ``causing to grow half as many cells'' as a consequence of the identity $\log\left(\frac{1}{x}\right) = - \log (x)$. Finally, the logarithm transforms multiplicative relationships in $(0, \infty)$, like (\ref{eq:heuristic_ratio}) above, to additive relationships on $(-\infty, \infty)$, the latter being easier to work with in general for numerous reasons. At the same time the logarithm ``preserves order'', in the sense that the logarithm of one number is larger than that of another number if and only if the original number was larger than the other original number, in other words $\log(y) > \log(x)$ if and only if $y > x$, thus log-transforming the data can be said to preserve ``relationships of relative size'' (i.e. which values are largest, which are smallest, etc.). After deciding to log transform, the scientific goal is reduced to the analysis of quantities of the form:
\begin{equation}
  \label{eq:heuristic_log_ratio}
  \begin{split}
    & \left( \begin{matrix}\text{`` typical logarithm growth factor of cells of strain }\specie_2\\ \text{ in the presence of strain }\specie_1 \text{ for batch }\batch\text{''} \end{matrix} \right) \\
- & \left( \text{`` typical logarithm growth factor of cells of strain }\specie_2\text{ in general for batch } \batch \text{''} \right) \,,
  \end{split}
\end{equation}
so-called ``log-ratios'' (because they are logarithms of ratios, while in contrast they are not ratios of logarithms). Note that values of the quantities from (\ref{eq:heuristic_log_ratio}) greater than zero correspond to interactions where the presence of $\specie_1$ promotes the growth of $\specie_2$, values less than zero correspond to interactions where the presence of $\specie_1$ inhibits the growth of $\specie_2$, and values (near) zero correspond to ``null interactions'' where the growth of $\specie_2$ is the same regardless of the presence of $\specie_1$. Cf. the coefficients $\interaction{\strain_1}{\strain_2}$ from equation (\ref{eq:transitions}) of section \ref{sec:form-trans-likel} as well as the related discussion from sections \ref{sec:form-as-stat} and \ref{sec:how-micr-ecol}.

\subsection{Issues with Heuristic Definition}
\label{sec:issu-with-heur}

Even after reducing the scientific goal to the analysis of quantities of the form (\ref{eq:heuristic_log_ratio}), there are still many issues with trying to make this actually precise enough to use as a basis for data analysis. Beginning with definitional issues, i.e. how to rephrase (\ref{eq:heuristic_log_ratio}) so that it may be able to define a concrete statistical estimand, there are at least three such issues:

\begin{enumerate}
\item Does ``typical logarithm'' correspond to the logarithm of the expectation or to the expectation of the logarithm?
\item Which droplets correspond to ``in the presence of $\specie_1$''?
\item Which droplets correspond to ``the typical growth of $\specie_2$ in general''?
\end{enumerate}

\subsection{Discussion}
\label{sec:discussion-1}

Referring to the beginning of section \ref{sec:issu-with-heur}, note that the way the first question ('Does ``typical logarithm'' correspond to the logarithm of the expectation or the expectation of the logarithm?') is phrased alreadly implicitly assumes that one has decided that ``typical'' signifies the use of expected value rather than any other possible distributional summary (e.g. the median or mode). This assumption was made largely, as one might expect, for the sake of simplicity of analysis, i.e. so that it is possible to appeal directly to a law of large numbers (combined with the continuous mapping theorem in the case of the logarithm of the expectation) to guarantee the (theoretical) statistical consistency of the corresponding plug-in estimators (i.e. the values of the estimands with respect to the empirical distribution). Of course in relying on (the standard) laws of large numbers there is also the implicit use of the assumption that all of the droplets in the experiment are mutually statistically independent, an assumption which is most likely to be very questionable at best in practice. In the absence of obviously better alternative models to use (i.e. as opposed to alternative models which would be more complicated, much more difficult to analyze theoretically, and likely even less realistic than assuming the droplets are independent), and since the goal of this work is to introduce this problem for the first time, such simplifying assumptions seemed forgivable if not exactly warranted. Nevertheless worthwhile possible goals for future work would be to examine, from an empirical and/or theoretical perspective, the consequences of either choosing a distributional summary besides the expected value to define ``typical'', or the consequences of relaxing the assumption that all of the individual droplets are mutually statistically independent.

\section{Calculations for hPoMu Working Model}
\label{sec:append-calc-hpomu}

Lemma \ref{lem:probability} allows us to perform some potentially useful calculations in the case of the particularly simple hPoMu working model.

\begin{corollary}
\label{cor:probability}
For any two subsets $\Omega_1, \Omega_2 \subseteq [S]$ such that $s \in \Omega_1 \cap \Omega_2$, for all $n\in \mathbb{N}$ one has under the hPoMu distribution that
\begin{equation}
  \label{eq:prob_corollary}
  \begin{split}
&    \probability*{ \abundance[\specie](0) = n | \observed = \Omega_1  } \\
= & \frac{\displaystyle \frac{(\freq_{\specie} \rate)^n}{n!} }{\displaystyle e^{\freq_{\specie} \rate} - 1} \\ 
= &   \probability*{ \abundance[\specie](0) = n | \observed = \Omega_2 } \,.
  \end{split}
\end{equation}
\end{corollary}

\textit{Proof of Corollary \ref{cor:probability}:} Because ${\abundance[\specie](0) \sim \poisson ( \freq_{\specie} \rate)}$ and from e.g. the law of total probability:
\begin{equation}
  \label{eq:prob_corollary_proof}
  \frac{\probability*{\abundance[\specie](0) = n}}{\probability*{\abundance[\specie](0) \ge 1}} 
= \frac{\displaystyle  e^{-\freq_{\specie} \rate} \left(\frac{(\freq_{\specie}\rate)^n}{n!}\right) }{
\displaystyle  e^{-\freq_{\specie}\rate} \left(\sum_{m=1}^{\infty}\frac{(\freq_{\specie}\rate)^m}{m!} \right)}
= \frac{\displaystyle  \left(\frac{(\freq_{\specie}\rate)^n}{n!}\right) }{\displaystyle e^{\freq_{\specie} \rate} - 1 } \,. \qed
\end{equation}

\begin{corollary}
\label{cor:expectation}
For any two subsets $\Omega_1, \Omega_2 \subseteq [\Species]$ with $\specie \in \Omega_1 \cap \Omega_2$, under the hPoMu distribution one has for any measurable $\phi: \mathbb{N} \to \R$ the following formula:
\begin{equation}
  \label{eq:expectation_corollary}
  \begin{split}
      \expectation*{ \phi\left(\abundance[\specie](0) \right)  |\observed = \Omega_1 } 
=& \frac{\displaystyle\sum_{n=1}^{\infty} \phi(n) \frac{(\freq_{\specie}\rate)^n}{n!} }{\displaystyle e^{\freq_{\specie}\rate} - 1} \\
=&\expectation*{ \phi\left(\abundance[\specie](0) \right)  |\observed = \Omega_2 }  \,.
  \end{split}
\end{equation}
\end{corollary}

\textit{Proof of Corollary \ref{cor:expectation}:} Using the law of total expectation and Corollary \ref{cor:probability}:
\begin{equation}
  \label{eq:31}
  \begin{split}
    & \sum_{n=1}^{\infty} \expectation*{ \phi \left( \abundance[\specie](0) \right) | \abundance[\specie](0) = n } \cdot \probability*{ \abundance[\specie](0) = n | \observed = \{ \specie\} } \\
= &  \sum_{n=1}^{\infty} \phi(n) \cdot \probability*{ \abundance[\specie](0) = n | \observed = \{ \specie\} } \\
= & \frac{\displaystyle \sum_{n=1} \phi(n) \frac{(\freq_{\specie} \rate)^n}{n!} }{\displaystyle e^{\freq_{\specie}\rate} - 1} \,. \qed
  \end{split}
\end{equation}
Some special cases of Corollary \ref{cor:expectation} include the case where $\phi(n) = n$:
\begin{equation}
  \label{eq:34}
   \expectation*{ \abundance[\specie](0) | \observed = \Omega } =  \frac{\freq_{\specie} \lambda e^{\freq_{\specie}\lambda} }{(e^{\freq_{\specie}\lambda} - 1)} \,,
\end{equation}
as well as the case where $\phi(n) = \frac{1}{n}$:
\begin{equation}
  \label{eq:37}
    \expectation*{ \frac{1}{ \abundance[\specie](0) } | \observed = \Omega } = \frac{ \exponint{\freq_{\specie} \lambda}  }{ \left( e^{\freq_{\specie}\lambda} -1\right) }   \,,
\end{equation}
where $\exponint{x}$ refers to a modified version of the exponential integral function, given by the power series:
\begin{equation}
  \label{eq:26}
  \sum_{n=1}^{\infty} \frac{x^n }{n \cdot n!} \,.
\end{equation}
The analytic function (\ref{eq:26}) can be calculated in terms of the usual exponential integral function and the Euler-Mascheroni constant, both of which are implemented for example in SciPy\cite{SciPy}.

\section{Related (Less Useful) Estimands}
\label{sec:append-relat-less}

Regarding the first issue from section \ref{sec:issu-with-heur}, the method of taking expectations first followed by logarithms second will be called ``arithmetic'' estimands and denoted $\lrestimandsymbol^{(a)}(\time)(\specie_1, \specie_2)$ (``$(a)$ '' for ``\textbf{a}rithmetic''). The term ``arithmetic'' refers to how these estimands correspond to the logarithms of the arithmetic means of the relevant growth factors.

The method of taking logarithms first followed by expectations second corresponds to the ATE estimands. For purposes of comparing the ATEs with the ``arithmetic'' estimands, the ATEs can be called ``geometric'' estimands and denoted $\lrestimandsymbol^{(g)}(\time)(\specie_1, \specie_2)$ (``$(g)$'' for ``\textbf{g}eometric''), although this will not be needed below. The term ``geometric'' refers to how these estimands correspond to the logarithms of the geometric means of the relevant growth factors.

The ``arithmetic'' estimands are less useful than the ``geometric'' estimands not only for theoretical reasons, such as how more stringent assumptions (compared to the ``geometric'' estimands) are required for identifiability from the observed data, or their lack of interpretation as an ATE. Preliminary simulation studies also found them to lead to less accurate or reliable estimators of microbial interactions. This was probably due not only to a larger ``identification gap'' but probably also to the general feature of geometric means typically being more robust to outliers than arithmetic means. However, more studies would be required to identify definitive reasons for what appears to be their poorer performance. Nevertheless, for all intents and purposes I would only recommend using the ATE estimands in practice.

If for no other reason than to demonstrate that it is hypothetically possible to use them, the ``arithmetic'' estimands are discussed in detail below. Like the ATEs, the ``arithmetic'' estimands also have versions corresponding to every possible combination of treatment and control groups. Thus, including the possibility of ``arithmetic'' estimands doubles the number of estimands to potentially consider from $4$ to $8$.

\paragraph{{\small (Full Data)} Picky treatment, picky control:}
\begin{equation}
  \label{eq:lr_estimand_picky_defns_arith_picky_picky}
  \begin{array}{rccl}
        \lrestimandfull{a}{p}{p}(\time)(\specie_1, \specie_2) : & \mathcal{M}^F & \to & \mathbb{R} \\
    \lrestimandfull{a}{p}{p}(\time)(\specie_1, \specie_2) : & \mathcal{P}_F & \mapsto & \log \left(\expectation[\mathcal{P}_F]*{ \frac{\abundance[\specie_2](\time)}{\abundance[\specie_2](0)} | \observed = \{\specie_1, \specie_2\} } \right) \\
&&& - \log \left(\expectation[\mathcal{P}_F]*{ \frac{\abundance[\specie_2](\time)}{\abundance[\specie_2](0)} | \observed = \{\specie_2\}  } \right) 
  \end{array}
\end{equation}

\paragraph{{\small (Full Data)} Picky treatment, gluttonous control:}
\begin{equation}
  \label{eq:lr_estimand_picky_defns_arith_picky_gluttonous}
  \begin{array}{rccl}
    \lrestimandfull{a}{p}{g}(\time)(\specie_1, \specie_2) : & \mathcal{M}^F & \to & \mathbb{R}  \\
\lrestimandfull{a}{p}{g}(\time)(\specie_1, \specie_2) : & \mathcal{P}_F & \mapsto &\log \left(\expectation[\mathcal{P}_F]*{ \frac{\abundance[\specie_2](\time)}{\abundance[\specie_2](0)} | \observed = \{\specie_1, \specie_2\} } \right) \\
&&& - \log \left(\expectation[\mathcal{P}_F]*{ \frac{\abundance[\specie_2](\time)}{\abundance[\specie_2](0)} | \observed \supseteq \{\specie_2\}, \specie_1 \not\in \observed } \right) 
  \end{array}
\end{equation}

\paragraph{{\small (Full Data)} Gluttonous treatment, picky control:}
\begin{equation}
  \label{eq:lr_estimand_picky_defns_arith_gluttonous_picky}
  \begin{array}{rccl}
\lrestimandfull{a}{g}{p} (\time)(\specie_1, \specie_2): & \mathcal{M}^F & \to & \mathbb{R}  \\    
\lrestimandfull{a}{g}{p} (\time)(\specie_1, \specie_2): &\mathcal{P}_F & \mapsto & \log \left(\expectation[\mathcal{P}_F]*{ \frac{\abundance[\specie_2](\time)}{\abundance[\specie_2](0)} | \observed \supseteq \{\specie_1, \specie_2\}  } \right)\\
& &&- \log \left(\expectation[\mathcal{P}_F]*{ \frac{\abundance[\specie_2](\time)}{\abundance[\specie_2](0)} | \observed = \{\specie_2\}  } \right) 
  \end{array}
\end{equation}

\paragraph{{\small (Full Data)} Gluttonous treatment, gluttonous control:}
\begin{equation}
  \label{eq:lr_estimand_picky_defns_arith_gluttonous_gluttonous}
  \begin{array}{rccl}
    \lrestimandfull{a}{g}{g} (\time)(\specie_1, \specie_2) : & \mathcal{M}^F & \to & \mathbb{R}  \\
    \lrestimandfull{a}{g}{g} (\time)(\specie_1, \specie_2) : &\mathcal{P}_F & \mapsto & \log \left(\expectation[\mathcal{P}_F]*{ \frac{\abundance[\specie_2](\time)}{\abundance[\specie_2](0)} | \observed \supseteq \{\specie_1, \specie_2\}  } \right)\\
& &&- \log \left(\expectation[\mathcal{P}_F]*{ \frac{\abundance[\specie_2](\time)}{\abundance[\specie_2](0)} | \observed \supseteq \{\specie_2\}, \specie_1 \not\in \observed  } \right) 
  \end{array}
\end{equation}

Note that due to the different signs of the treatment and control terms, Jensen's inequality does not enable any general comparisons or bounds between the ``arithmetic'' estimands and their ``geometric'' counterparts in general. 

\subsection{Observed Data Estimand and Estimator Definitions}
\label{sec:estim-defin}

Just like their ``geometric'' counterparts, the full data ``arithmetic'' estimands also possess observed data counterparts, in the obvious way. See below.

\paragraph{{\small (Observed Data)} Picky treatment, picky control:}
\begin{equation}
  \label{eq:lr_estimand_observed_picky_defns_arith_picky_picky}
  \begin{array}{rccl}
        \lrestimand{a}{p}{p}(\time)(\specie_1, \specie_2) : & \mathcal{M}^O & \to & \mathbb{R} \\
    \lrestimand{a}{p}{p}(\time)(\specie_1, \specie_2) : & \mathcal{P}_O & \mapsto & \log \left(\expectation[\mathcal{P}_O]*{ \abundance[\specie_2](\time) | \observed[\time] = \{\specie_1, \specie_2\} } \right) \\
&&& - \log \left(\expectation[\mathcal{P}_O]*{ \abundance[\specie_2](\time) | \observed[\time] = \{\specie_2\}  } \right) 
  \end{array}
\end{equation}

\paragraph{{\small (Observed Data)} Picky treatment, gluttonous control:}
\begin{equation}
  \label{eq:lr_estimand_observed_picky_defns_arith_picky_gluttonous}
  \begin{array}{rccl}
    \lrestimand{a}{p}{g}(\time)(\specie_1, \specie_2) : & \mathcal{M}^O & \to & \mathbb{R}  \\
\lrestimand{a}{p}{g}(\time)(\specie_1, \specie_2) : & \mathcal{P}_O & \mapsto &\log \left(\expectation[\mathcal{P}_O]*{ \abundance[\specie_2](\time) | \observed[\time] = \{\specie_1, \specie_2\} } \right) \\
&&& - \log \left(\expectation[\mathcal{P}_O]*{ \abundance[\specie_2](\time) | \observed[\time] \supseteq \{\specie_2\}, \specie_1 \not\in \observed[\time] } \right) 
  \end{array}
\end{equation}

\paragraph{{\small (Observed Data)} Gluttonous treatment, picky control:}
\begin{equation}
  \label{eq:lr_estimand_observed_picky_defns_arith_gluttonous_picky}
  \begin{array}{rccl}
\lrestimand{a}{g}{p} (\time)(\specie_1, \specie_2): & \mathcal{M}^O & \to & \mathbb{R}  \\    
\lrestimand{a}{g}{p} (\time)(\specie_1, \specie_2): &\mathcal{P}_O & \mapsto & \log \left(\expectation[\mathcal{P}_O]*{ \abundance[\specie_2](\time) | \observed[\time] \supseteq \{\specie_1, \specie_2\}  } \right)\\
& &&- \log \left(\expectation[\mathcal{P}_O]*{ \abundance[\specie_2](\time) | \observed[\time] = \{\specie_2\}  } \right) 
  \end{array}
\end{equation}

\paragraph{{\small (Observed Data)} Gluttonous treatment, gluttonous control:}
\begin{equation}
  \label{eq:lr_estimand_observed_picky_defns_arith_gluttonous_gluttonous}
  \begin{array}{rccl}
    \lrestimand{a}{g}{g} (\time)(\specie_1, \specie_2) : & \mathcal{M}^O & \to & \mathbb{R}  \\
    \lrestimand{a}{g}{g} (\time)(\specie_1, \specie_2) : &\mathcal{P}_O & \mapsto & \log \left(\expectation[\mathcal{P}_O]*{ \abundance[\specie_2](\time) | \observed[\time] \supseteq \{\specie_1, \specie_2\}  } \right)\\
& &&- \log \left(\expectation[\mathcal{P}_O]*{ \abundance[\specie_2](\time) | \observed[\time] \supseteq \{\specie_2\}, \specie_1 \not\in \observed[\time]  } \right) 
  \end{array}
\end{equation}

For a given observed data ``arithmetic'' estimand $\lrestimandsymbol^{(a)}$, let its corresponding estimator be denoted $\lrestimatorsymbol^{(a)}$. Then the (plugin) estimators corresponding to each ``arithmetic'' estimand are explicitly defined below.

\paragraph{{\small (Estimator)} Picky treatment, picky control:}
\begin{equation}
  \label{eq:lr_estimator_picky_defns_arith_picky_picky}
  \begin{split}
\lrestimator{a}{p}{p}(\time)(\specie_1, \specie_2) := & \log \left(\mean*{ \abundance[\specie_2](\time) | \observed[\time] = \{\specie_1, \specie_2\} } \right) \\
& - \log \left(\mean*{ \abundance[\specie_2](\time) | \observed[\time] = \{\specie_2\}  } \right) 
  \end{split}
\end{equation}

\paragraph{{\small (Estimator)} Picky treatment, gluttonous control:}
\begin{equation}
  \label{eq:lr_estimator_picky_defns_arith_picky_gluttonous}
  \begin{split}
\lrestimator{a}{p}{g}(\time)(\specie_1, \specie_2) := & \log \left(\mean*{ \abundance[\specie_2](\time) | \observed[\time] = \{\specie_1, \specie_2\} } \right) \\
& - \log \left(\mean*{ \abundance[\specie_2](\time) | \observed[\time] \supseteq \{\specie_2\}, \specie_1 \not\in \observed[\time]  } \right)
  \end{split}
\end{equation}

\paragraph{{\small (Estimator)} Gluttonous treatment, picky control:}
\begin{equation}
  \label{eq:lr_estimator_picky_defns_arith_gluttonous_picky}
  \begin{split}
 \lrestimator{a}{g}{p}(\time)(\specie_1, \specie_2) := &  \log \left(\mean*{ \abundance[\specie_2](\time) | \observed[\time] \supseteq \{\specie_1, \specie_2\} } \right) \\
& - \log \left(\mean*{ \abundance[\specie_2](\time) | \observed[\time] = \{\specie_2\}  } \right)
  \end{split}
\end{equation}

\paragraph{{\small (Estimator)} Gluttonous treatment, gluttonous control:}
\begin{equation}
  \label{eq:lr_estimator_picky_defns_arith_gluttonous_gluttonous}
\begin{split}
\lrestimator{a}{g}{g}(\time)(\specie_1, \specie_2) := &\log \left(\mean*{ \abundance[\specie_2](\time) | \observed[\time] \supseteq \{\specie_1, \specie_2\} } \right) \\
& - \log \left(\mean*{ \abundance[\specie_2](\time) | \observed[\time] \supseteq \{\specie_2\}, \specie_1 \not\in \observed[\time]  } \right) 
  \end{split}
\end{equation}
The full data ``arithmetic'' estimands can still be identifiable from the observed data, but only under assumptions which are stronger than those required for the full data ATE (``geometric'') estimands. Cf. Lemma \ref{lem:arith_simplification1} or Lemma \ref{lem:arith_simplification2}.

\subsection{Sufficient Conditions for Identifiability}
\label{sec:suff-cond-ident}

Observe how Lemma \ref{lem:arith_simplify} below corresponds to the ``arithmetic'' estimands in exactly the same way that Lemma \ref{lem:geom_simplify} above corresponds to the ATE estimands.

\begin{lemma}
\label{lem:arith_simplify}
  Assuming that
  \begin{itemize}
  \item  the marginal distributions of $\vabundance (0)$ are mutually independent, e.g. as turns out to be true for the hPoMu and hNBDM distributions,
  \item that the distributions of the droplets are independent,
  \item the assumption (\ref{eq:arith_simplification_assumption1}) from Lemma \ref{lem:arith_simplification1} and/or the assumption (\ref{eq:arith_simplification_assumption2}) from Lemma \ref{lem:arith_simplification2},
   \item we always have that $\observed[\time] = \observed$ (cf. section \ref{sec:ident-observed}),
  \end{itemize}
then for any two subsets $\mathscr{O}_{\treatment}, \mathscr{O}_{\control} \subseteq \powerset{[\Species]}$ of the power set of $[\Species]$ such that $\specie_2 \in \Omega_{\treatment}$ for all $\Omega_{\treatment} \in \mathscr{O}_{\treatment}$ and $\specie_2 \in \Omega_{\control}$ for all $\Omega_{\control} \in \mathscr{O}_{\control}$, then the estimator
\begin{equation}
  \label{eq:lrestimator_arith}
  \begin{split}
    \lrestimatorsymbol^{(a)} (\time) (\specie_1, \specie_2) := & \log \left(  \mean*{ \abundance[\specie_2](\time) | \observed[\time] \in \mathscr{O}_{\treatment} } \right) \\
& - \log \left(  \mean*{ \abundance[\specie_2](\time) | \observed[\time] \in \mathscr{O}_{\control} } \right)
  \end{split}
\end{equation}
is statistically consistent for the estimand
\begin{equation}
  \label{eq:lrestimand_arith}
  \begin{split}
    {\lrestimandsymbolfull}^{(a)} (\time) (\specie_1, \specie_2) := & \log \left(  \expectation*{ \frac{\abundance[\specie_2](\time)}{\abundance[\specie_2](0)} |\observed \in \mathscr{O}_{\treatment} } \right) \\
& - \log \left(  \expectation*{ \frac{\abundance[\specie_2](\time)}{\abundance[\specie_2](0)} |\observed \in \mathscr{O}_{\control} } \right) \,.
  \end{split}
\end{equation}
\end{lemma}

\noindent Note that $\mathscr{O}_{\treatment}$ and $\mathscr{O}_{\control}$ have the same interpretations here as in section \ref{sec:proofs}.
\\~\\
\textit{Proof of Lemma \ref{lem:arith_simplify}:}

\paragraph{Proof of identifiability of ${\lrestimandsymbolfull}^{(a)} (\time) (\specie_1, \specie_2)$ from the observed data:}~\\
Using the result of Lemma \ref{lem:arith_simplification1} and/or Lemma \ref{lem:arith_simplification2} below one immediately has as a consequence that the expression (\ref{eq:lrestimand_arith}) equals
\begin{equation}
  \label{eq:arith_simplification_result}
  \begin{adjustbox}{max width=\textwidth,keepaspectratio}
$\displaystyle  \begin{array}{rcl}
      {\lrestimandsymbolfull}^{(a)} (\time) (\specie_1, \specie_2)  &=&
                                                                        \log \left( \expectation*{ \abundance[\specie_2](\time) | \observed \in \mathscr{O}_{\treatment} } \right) - \log \left( \expectation*{ \abundance[\specie_2](\time) | \observed \in \mathscr{O}_{\control} } \right) \\
& = &
      \log \left( \expectation*{ \abundance[\specie_2](\time) | \observed[\time] \in \mathscr{O}_{\treatment} } \right) - \log \left( \expectation*{ \abundance[\specie_2](\time) | \observed[\time] \in \mathscr{O}_{\control} } \right)                   \\
    & = &  {\lrestimandsymbol}^{(a)} (\time) (\specie_1, \specie_2) \,.
       \end{array}$
     \end{adjustbox}
 \end{equation}
In other words, the full data ``arithmetic'' estimand equals its observed data counterpart, ${\lrestimandsymbolfull}^{(a)} (\time) (\specie_1, \specie_2) = {\lrestimandsymbol}^{(a)} (\time) (\specie_1, \specie_2)$. The penultimate equality used the assumption that we always have $\observed[\time] = \observed$.

\paragraph{Proof of consistency of $ \lrestimatorsymbol^{(a)} (\time) (\specie_1, \specie_2) $ for $\lrestimandsymbol^{(a)} (\time) (\specie_1, \specie_2)$:}~\\
Then using the Weak Law of Large Numbers, the expression inside the logarithm for the first term of (\ref{eq:lrestimator_arith}) converges in probability to the expression inside of the logarithm for the first term of (\ref{eq:lrestimand_arith}), and then applying the Continuous Mapping Theorem with the continuous function $\log$ allows one to conclude that the first term of (\ref{eq:lrestimator_arith}) converges in probability to the first term of (\ref{eq:lrestimand_arith}). Completely analogous reasoning leads one to also conclude that the second term of (\ref{eq:lrestimator_arith}) also converges in probability to the second term of (\ref{eq:lrestimand_arith}). The result then follows using the fact $\mathbf{X}_n \overset{P}{\rightarrow} \mathbf{X}$ and $\mathbf{Y}_n \overset{P}{\rightarrow} \mathbf{Y}$ implies $(\mathbf{X}_n - \mathbf{Y}_n) \overset{P}{\rightarrow} \mathbf{X} - \mathbf{Y}$. $\qed$

\begin{lemma}
\label{lem:arith_simplification1}
  Assuming that
  \begin{itemize}
  \item  the marginal distributions of $\vabundance (0)$ are mutually independent, e.g. as turns out to be true for the hPoMu and hNBDM distributions,
  \item that for any $\mathscr{O} \subseteq \powerset{[\Species]}$ such that $\specie_2 \in \Omega$ for all $\Omega \in \mathscr{O}$, one has
    \begin{equation}
      \label{eq:arith_simplification_assumption1}
      \begin{split}
      & \expectation*{ \frac{\abundance[\specie_2](\time)}{\abundance[\specie_2](0)} | \observed \in \mathscr{O} }   \\
    = &\expectation*{ \abundance[\specie_2](\time) | \observed \in \mathscr{O} } \cdot \expectation*{ \frac{1}{\abundance[\specie_2](0)} | \observed \in \mathscr{O} }
      \end{split}
    \end{equation}
  \end{itemize}
then for any two subsets $\mathscr{O}_{\treatment}, \mathscr{O}_{\control} \subseteq \powerset{[\Species]}$ of the power set of $[\Species]$ such that $\specie_2 \in \Omega_{\treatment}$ for all $\Omega_{\treatment} \in \mathscr{O}_{\treatment}$ and $\specie_2 \in \Omega_{\control}$ for all $\Omega_{\control} \in \mathscr{O}_{\control}$ one has that the expression (\ref{eq:lrestimand_arith}) equals
\begin{equation}
  \label{eq:arith_simplification1_result}
  \log \left( \expectation*{ \abundance[\specie_2](\time) | \observed \in \mathscr{O}_{\treatment} } \right) - \log \left( \expectation*{ \abundance[\specie_2](\time) | \observed \in \mathscr{O}_{\control} } \right) \,.
\end{equation}
\end{lemma}

\textit{Proof of Lemma \ref{lem:arith_simplification1}:} Using assumption (\ref{eq:arith_simplification_assumption1}), as well as the properties of logarithms, one has that (\ref{eq:lrestimand_arith}) above equals
\begin{equation}
  \label{eq:arith_simplification1_proof}
  \begin{split}
    \log \left( \expectation*{\abundance[\specie_2](\time)  | \observed \in \mathscr{O}_{\treatment} }  \right)  - \log \left(  \expectation*{\abundance[\specie_2](\time)  | \observed \in \mathscr{O}_{\control} }   \right) \\
+ \left(  \log \left( \expectation*{  \frac{1}{\abundance[\specie_2](0)} | \observed \in \mathscr{O}_{\treatment} }  \right) - \log \left( \expectation*{ \frac{1}{\abundance[\specie_2](0)}   | \observed \in \mathscr{O}_{\control}  }   \right)   \right) \,.
  \end{split}
\end{equation}
By Lemma \ref{lem:total_expectation} using $\phi(n) = \frac{1}{n}$, the expression inside of parentheses in (\ref{eq:arith_simplification1_proof}) vanishes, since of course $x=y$ also implies that $\log(x)=\log(y)$ (the analogous statement is true for literally any function), and thus the result follows. $\qed$

The condition (\ref{eq:arith_simplification_assumption1}) is equivalent to stating that $\abundance[\specie_2](\time)$ and $\abundance[\specie_2](0)^{-1}$ are conditionally uncorrelated given that $\observed \in \mathscr{O}$. A sufficient (but obviously not necessary) condition for this hypothesis to be true is when $\abundance[\specie_2](\time)$ and $\abundance[\specie_2](0)$ are conditionally independent when given that $\observed \in \mathscr{O}$, since this in turn implies that $\abundance[\specie_2](\time)$ and $\abundance[\specie_2](0)^{-1}$ are conditionally independent given $\observed \in \mathscr{O}$, which in turn implies the conditional uncorrelatedness corresponding to the hypothesis. (Of course, as is probably obvious to the reader, $\abundance[\specie_2](\time)$ and $\abundance[\specie_2](0)$ being at all independent, even if only conditionally, seems like a rather implausible assumption.) 

\begin{lemma}
\label{lem:arith_simplification2}
  Assuming that
  \begin{itemize}
  \item  the marginal distributions of $\vabundance (0)$ are mutually independent, e.g. as turns out to be true for the hPoMu and hNBDM distributions,
  \item that for any $\mathscr{O} \subseteq \powerset{[\Species]}$ such that $\specie_2 \in \Omega$ for all $\Omega \in \mathscr{O}$, one has
    \begin{equation}
      \label{eq:arith_simplification_assumption2}
      \begin{split}
      \expectation*{ \frac{\abundance[\specie_2](\time)}{\abundance[\specie_2](0)} | \observed \in \mathscr{O} }    = \frac{\expectation*{ \abundance[\specie_2](\time) | \observed \in \mathscr{O} } }{ \expectation*{ \abundance[\specie_2](0) | \observed \in \mathscr{O} } }
      \end{split}
    \end{equation}
  \end{itemize}
then for any two subsets $\mathscr{O}_{\treatment}, \mathscr{O}_{\control} \subseteq \powerset{[\Species]}$ of the power set of $[\Species]$ such that $\specie_2 \in \Omega_{\treatment}$ for all $\Omega_{\treatment} \in \mathscr{O}_{\treatment}$ and $\specie_2 \in \Omega_{\control}$ for all $\Omega_{\control} \in \mathscr{O}_{\control}$ one has that the expression (\ref{eq:lrestimand_arith}) equals
\begin{equation}
  \label{eq:arith_simplification2_result}
  \log \left( \expectation*{ \abundance[\specie_2](\time) | \observed \in \mathscr{O}_{\treatment} } \right) - \log \left( \expectation*{ \abundance[\specie_2](\time) | \observed \in \mathscr{O}_{\control} } \right) \,.
\end{equation}
\end{lemma}

\textit{Proof of Lemma \ref{lem:arith_simplification2}:} Using assumption (\ref{eq:arith_simplification_assumption2}) and the properties of logarithms, one has that (\ref{eq:lrestimand_arith}) above equals
\begin{equation}
  \label{eq:arith_simplification2_proof}
  \begin{split}
 \log \left( \expectation*{ \abundance[\specie_2](\time) | \observed\in \mathscr{O}_{\treatment} }  \right) - \log \left( \expectation*{  \abundance[\specie_2](\time) | \observed \in \mathscr{O}_{\control}  } \right) \\
- \left(  \log \left( \expectation*{ \abundance[\specie_2](0) | \observed \in \mathscr{O}_{\treatment} }  \right) - \log \left( \expectation*{  \abundance[\specie_2](0) | \observed \in \mathscr{O}_{\control} }  \right)  \right) \,.
  \end{split}
\end{equation}
Using Lemma \ref{lem:total_expectation} with $\phi(n) = n$, it follows as before that the expression inside of parentheses in (\ref{eq:arith_simplification2_proof}) vanishes, and thus the result follows. $\qed$

As one can see by multiplying both sides of (\ref{eq:arith_simplification_assumption2}) by the denominator of the right side of (\ref{eq:arith_simplification_assumption2}), the condition (\ref{eq:arith_simplification_assumption2}) is equivalent to assuming the conditional uncorrelatedness of the multiplicative growth factor $\frac{\abundance[\specie_2](\time)}{\abundance[\specie_2](0)} $ and the initial cell count $\abundance[\specie_2](0)$ when given that $\observed \in \mathscr{O}$. A sufficient, but again obviously not necessary, condition for this to be true is when the factor by which the population has grown by time $\time$, $\frac{\abundance[\specie_2](\time)}{\abundance[\specie_2](0)} $, and the initial cell count $\abundance[\specie_2](0)$ are conditionally independent when given that $\observed \in \mathscr{O}$. While this still sounds implausible, it is perhaps less so than the above sufficient condition for (\ref{eq:arith_simplification_assumption1}).

\end{coolsubappendices}
\end{coolcontents}

\chapter[Loss Functions for Networks with Mixed-Sign Edge Weights][Loss Functions for Ecological Networks]{Loss Functions for Ecological Networks\\{\large Dissimilarity and Similarity Functions for Quantitatively Comparing Networks with Mixed-Sign Edge Weights}}
\label{chap:network_comparison}

Herein I discuss how comparisons of ecological interactions can be recast into the statistical framework of loss functions for signed networks.
See sections \ref{sec:introduction} and \ref{sec:comparison-networks-mixed}.
I explain how avoiding unexpected behavior requires loss functions for signed networks to satisfy what I call ``the double penalization principle''.
See, for example, section \ref{sec:double-penal-princ}, section \ref{sec:network_comparison_results}, or even section \ref{sec:why-comp-meth}.
Starting from loss functions of unsigned networks, I derive several examples of loss functions for signed networks that satisfy this property.
See section \ref{sec:diss-meas}.

Section \ref{sec:introduction} clarifies the problem and identifies principles to judge whether a method sensibly quantifies the (dis)similarity of two networks with mixed-sign edge weights. Section \ref{sec:general-definitions} provides technical definitions which facilitate the analysis in the rest of the chapter. In section \ref{sec:diss-meas} I propose several examples of methods that satisfy the principles described in section \ref{sec:introduction}. Section \ref{sec:network_comparison_results} demonstrates empirically how methods that satisfy these principles behave compared to methods that do not satisfy these principles. Finally section \ref{sec:network_comparison_discussion} interprets the results and explains which facets of the problem still require further attention.

\begin{coolcontents}
\section{Background and Significance}
\label{sec:backgr-sign-8}

\paragraph{Broader field}

The broader field of chapter \ref{chap:network_comparison} is the use of signed networks \cite{Harary} as models for ecological interactions. See \cite{Faust2012} for a review of network models of interactions in microbial ecology, and \cite{Lidicker1979} or \cite{signs_longitudinal_compositional} for how signs (of network edge weights) can be used to categorize ecological interactions. Positive edges are for interactions that promote the growth of the recipient, and negative edges are for interactions that suppress the growth of the recipient, cf. again section \ref{sec:form-as-stat}. Cf. also the discussion of this earlier from the introduction to Part \ref{part:aver-treatm-effects}. These networks would be inferred using the data from MOREI.

In particular, chapter \ref{chap:network_comparison} is motivated by the important issue of how to identify ``best-performing'' or ``most accurate'' estimators of such ecological models. Our statistical estimation problem has a network with mixed-sign edge weights as its estimand (cf. again section \ref{sec:form-as-stat}), the microbial interaction network \cite{Xiao2017} \cite{Angulo2019}. The issue of what it means for an estimator to be ``best-performing'' is already subtle in the thoroughly studied field of point estimation (cf. e.g. \cite{lehmann_cassella} or \cite{Keener2010}), where the estimand is a much simpler data structure than a signed network. Hence this issue deserves some thought.

\paragraph{Specific problem}

Herein I investigate how comparisons of ecological interactions can be recast into the statistical framework of loss functions for signed networks.
Given multiple choices of estimator, or ``algorithm'', for characterizing signed networks of ecological interactions from MOREI data, we want to know how to decide which choice is best or ``optimal''. Chapter \ref{chap:log-ratio-coeff} established that at least one such choice exists, so the problem is not ``vacuous''.
To ensure that the estimates produced by any ``optimal'' choice are useful, the process for deciding which estimator is ``optimal'' must itself be a ``reasonable'' process.

To decide between estimators, we need to compare their performance. The performance can be defined as how (dis)similar the estimates produced by the estimators are to a reference ``ground truth'' signed network. (Such a ``ground truth'' network is available, for example, when running a simulation and evaluating the performance of the estimators on the simulated data\footnote{
This is another reason why the work in Part \ref{part:modell-init-form} is important. If our simulations do not accurately characterize how much data will be available to infer the edges of the interaction network, whatever conclusions we draw based on that simulation about the effectiveness of various estimators are likely to be distorted.
}. One can also compare with the results inferred from data produced by a more established experiment, such as cell plating, defining those results operationally as the ``ground truth''.)
A function that quantifies the dissimilarity of an estimate to a ``ground truth'' estimand is called a loss function.
(Cf. the discussion of terminology in section \ref{sec:comparison-networks-mixed}.)
An ``optimal'' choice of estimator is then one whose estimates (typically) minimize the value of the loss function with respect to the ``ground truth''.
Therefore, making ``reasonable'' choices of ``optimal'' estimators requires us to identify and use a ``reasonable'' loss function.

Reviews in the literature exist discussing possible choices of such functions in the ``classical'' case of \textit{un}signed networks. The review \cite{Tantardini2019} considers functions that are applicable to directed\footnote{And thus by extension are also applicable to undirected networks, because undirected networks can be formulated as a special case of directed networks in a standard way.} (unsigned) networks, whereas the review \cite{Wills2020} only considers functions that are applicable for undirected (unsigned) networks. However, no previous literature appears to exist discussing loss functions for \textit{signed} networks. If it does, I was unable to find it.

\paragraph{Particular approach}

Because loss functions for signed networks do not seem to be studied in the literature, the approach of this chapter is to provide a baseline for future work in this area. Principles are defined to characterize the behavior of ``reasonable'' loss functions for signed networks, and concrete examples of loss functions that obey these principles are given.

To some extent, how to optimally choose such an algorithm is already obvious: as explained in \cite[chapter 3]{vanderLaan2011}, given multiple possible choices of algorithm, discrete SuperLearner can identify a best possible choice, and SuperLearner \cite{vanderLaan2007} can give us a weighted combination of algorithms that performs even better. However, to make any such choice even SuperLearner needs a specified choice of loss function \cite[section 3.5]{vanderLaan2011}.

Indeed, even regardless of whether one uses SuperLearner, choosing an ``optimal'' estimator requires us to first specify our preferences through the choice of a loss function. If the chosen loss function behaves ``unreasonably'' in some way, then the resulting choice of ``optimal'' estimator will be similarly ``unreasonable''. Quoting from \cite{AndradePacheco2020}:
\begin{quote}
The criteria to define what is optimal depends on what quantity is to be estimated. Hence, it is first necessary to define an objective or utility function, i.e. the measure by which we evaluate the performance of any given design.
\end{quote}
Thus the challenge for this chapter is to do this explicitly for a type of estimand, signed networks, that appears to not have been studied for this purpose before in the literature.

\subsection{Applications of Signed Networks}
\label{sec:previous-work-networks}

Observe how a framework using signed networks has the flexibility to model all possible kinds of ecological interactions, not just trophic interactions\footnote{
At the population level this includes predation and parasitoidism, but in terms of temporally persistent symbioses of individual organisms, this only includes parasitism. 
} where the growth of one organism necessarily comes at the expense of another organism. Previous work \cite{Lopez2019} has discussed comparing such trophic interaction networks\footnote{
As pointed out in \cite{Lopez2019}, (unsigned) directed networks can also describe other types of ecological interactions, not just the trophic interactions mentioned above. However, without using signed edge weights, only one type of ecological interaction can be described by a single network, whereas in real ecology all possible types of interaction can occur in the same ecosystem.
}. In such a network signed edge weights are unnecessary because an edge from $\strain_1$ to $\strain_2$ indicates $\strain_1$ promotes the growth of $\strain_2$ and $\strain_2$ inhibits the growth of $\strain_1$ (necessarily because $\strain_2$ consumes $\strain_1$), while swapping the direction of the edge to be from $\strain_2$ to $\strain_1$ indicates the opposite relationship, that $\strain_2$ promotes the growth of $\strain_1$ and $\strain_1$ inhibits the growth of $\strain_2$ (due to $\strain_1$ consuming $\strain_2$). See \cite{MacKay2020} for further discussion of trophic interaction networks as unsigned, directed networks, analysis of the underlying theory, and comparison with structurally similar problems.

In the signed network framework (cf. \cite{Lidicker1979}, \cite{signs_longitudinal_compositional}, or even \cite{positive_kChip}), the former corresponds to $\strain_1 \overset{+}{\rightarrow} \strain_2$ and $\strain_2 \overset{-}{\rightarrow} \strain_1$, while the latter corresponds to $\strain_2 \overset{+}{\rightarrow} \strain_1$ and $\strain_1 \overset{-}{\rightarrow} \strain_2$. Note also that the signed network framework does not require the assumption that the cause of this asymmetric interaction is one organism consuming the other. For example, humans promote the growth of rats because human waste serves as a food source for rats, while rats can inhibit the growth of humans by serving as a vector for human diseases. The signed network framework also allows modelling mutualistic/mutually supporting interactions (pairs of interactions of the form $\strain_1 \overset{+}{\rightarrow} \strain_2$ and $\strain_2 \overset{+}{\rightarrow} \strain_1$), competitive/mutually antagonistic interactions (pairs of interactions of the form $\strain_1 \overset{-}{\rightarrow} \strain_2$ and $\strain_2 \overset{-}{\rightarrow} \strain_1$), commensalistic interactions (pairs of interactions of the form $\strain_{1/2} \overset{+}{\rightarrow} \strain_{2/1}$ and $\strain_{2/1} \overset{0}{\rightarrow} \strain_{1/2}$), and amensalistic interactions (pairs of interactions of the form $\strain_{1/2} \overset{-}{\rightarrow} \strain_{2/1}$ and $\strain_{2/1} \overset{0}{\rightarrow} \strain_{1/2}$). None of these other kinds of interactions can be described by the unsigned trophic network framework.

Previous work has looked at e.g. comparing ``local'' neighborhoods of nodes in (unweighted)\footnote{
In the framework of \cite{Tantardini2019} what \cite{Zhu2017} addresses is the “unknown node correspondence” comparison problem. In contrast, this work addresses the ``known node correspondence'' comparison problem. Cf. section \ref{sec:known-vs.-unknown}. Other differences are that this work considers ``global'' comparisons of entire networks and allows edge weights to have magnitudes other than $1$.
} signed networks \cite{Zhu2017} or diffusion kernels in signed networks \cite{Qi2008}. However, overall the existing literature on signed networks seems underdeveloped. This is surprising because such networks could presumably model an extremely wide range of phenomena, not just ecological interaction networks. For example, the\footnote{
These correspond to ``gene $\times$ gene'' interaction networks in the terminology of item \ref{item:gene_gene_interactions} of section \ref{broader-field-2}, or equivalently to ``unpartitioned interaction networks'' from the terminology of the introduction to Part \ref{part:introduction}.
} ``gene interaction networks'' defined implicitly in \cite{Mani2008} could be considered (undirected) networks with mixed-sign edge weights if the edge weights were taken to be the differences ($\varepsilon = W_{xy} - E(W_{xy})$ in the notation of the paper) between the observed values of double mutant fitnesses and those expected under a null model of no interactions, with ``positive'' ($\varepsilon > 0$) edges corresponding to ``alleviating'' interactions ($W_{xy} > E(W_{xy})$), and ``negative'' ($\varepsilon < 0$) edges corresponding to ``synergistic'' interactions ($W_{xy} < E(W_{xy})$). More generally, any dynamical system with a “multivariate state space” that has “incomplete connectivity” could potentially be summarized using a network with mixed-sign edge weights. (Cf. section \ref{broader-field-2} for more on this idea.) In particular, it seems that no previous published work discusses what is called, in the framework of \cite{Tantardini2019}, the “known node correspondence” comparison problem for networks with mixed-sign edge weights. That is surprising because the problem seems like an obvious question to ask about a class of objects with extremely broad potential applicability.
If prior literature exists then most likely I was unable to find it due to it using different terminology.

\section{Preliminaries}
\label{sec:introduction}

Section \ref{sec:known-vs.-unknown} clarifies the type of problem considered herein. Section \ref{sec:posit-negat-edges} introduces the main issue with trying to compare signed networks the same way as unsigned networks. Section \ref{sec:definition-networks-mixed} gives some basic definitions, but cf. section \ref{sec:general-definitions} for more details. Sections \ref{sec:monot-princ}, \ref{sec:double-penal-princ}, and \ref{sec:conv-comb-decomp} provide criteria we might like comparison methods to satisfy in order to address these concerns. Section \ref{sec:sparsity-should-not} provides an additional criterion which is also relevant for unsigned networks. Finally section \ref{sec:which-feat-netw} overviews at a very high level the kinds of comparisons of networks which we might want to make herein.

\subsection{Known vs. Unknown Node Correspondence}
\label{sec:known-vs.-unknown}

There are at least two types of network comparison problem which we should distinguish. Following the framework from \cite{Tantardini2019}, we can consider
\begin{itemize}
\item the known node correspondence problem (``fixed nodes, variable edges''), or
\item the unknown node correspondence problem (``variable everything'').
\end{itemize}

For the known node correspondence problem, the identity of the nodes is essential and fixed. The two compared networks must either have the same set of labeled nodes, or a known correspondence between their node sets must be given. The known-node-correspondence problem asks whether the two networks describe similar relationships for the fixed set of nodes. An example of the known node correspondence problem would be to ask how much airline flight routes in Europe changed after the onset of the pandemic. This amounts to comparing two networks (one for before the pandemic and one for after) with the same labeled nodes (the airports), but for which the edges (the flight routes) between those nodes may be different. Cf. section \ref{sec:comparison-networks-mixed}, that attempts to make these ideas more precise.

For the unknown node correspondence problem, the identity of the nodes does not matter. The identity of the nodes may differ between the two networks. The unknown node correspondence asks essentially whether the relationships described by the two networks are ``analogous''. An example of the unknown node correspondence problem would be to ask whether the structure of airports and flight routes is analogous between Europe and China. The set of nodes is obviously different, and a priori there's no reason to believe that e.g. the Shanghai airport should be identified with the London airport rather than the Paris airport or vice versa. Nevertheless one could still sensibly ask e.g. whether it's possible to identify analogous subsets of highly connected ``central hub'' airports in both networks.

The network comparison problem described in section \ref{sec:backgr-sign-8} clearly is known node correspondence. We want to quantitatively assess the ability of statistical estimators to recover the correct set of edges (interactions) for a given set of nodes (microbes). It makes no difference if the estimated network is ``analogous'' to the true network. If the ecological roles of the various microbes are misidentified by ascribing certain relationships to the incorrect pairs of microbes, the estimated network is still a terrible estimate of the truth. Only the known node correspondence problem is relevant for comparing the estimates produced by an estimator with the true value of an estimand.

Unfortunately, the fact that our network comparison problem is known node correspondence also means that most of the methods discussed in \cite{Tantardini2019} would be inapplicable and irrelevant even if we were considering unsigned networks. While comparing distinct microbial ecosystems for analogous network structures is a valid scientific problem and would also require extension of methods for the unknown node correspondence problem to signed networks, it is outside of scope here and left to future work. Herein I try to modify and extend (as necessary) most of the computationally tractable methods\footnote{I.e. not the cut distance, which is not really computationally tractable.} for the known node correspondence problem discussed in \cite{Tantardini2019} to apply also to signed networks.

\subsection{Positive and Negative Edges Qualitatively Distinct}
\label{sec:posit-negat-edges}

In microbial ecology, positive interactions and negative interactions correspond to categorically and qualitatively distinct phenomena. For example, if all growth-promoting (positive) interactions were changed to growth-suppressing (negative) interactions, ecologically speaking the result would be quite different. This is true even if, or in some cases even especially if, the magnitudes of the interactions were left unchanged. Translated to a network representation, this implies that the positive and negative edges of networks with mixed-sign edge weights should be treated as qualitatively distinct. 

Standard network comparison methods assume all edge weights have the same (positive) sign. Therefore applying or extending such methods to networks with mixed-sign edge weights requires caution. We need to avoid situations where, for example, positive and negative weights ``cancel'' in a scientifically meaningless way, or two compared edges with different signs are not penalized due to having similar magnitudes. Section \ref{sec:double-penal-princ} addresses this caution by providing a principle to ``sanity check'' whether a network comparison method may potentially give meaningful results for networks with mixed-sign edge weights.

\subsection{Definition of Networks with Mixed-Sign Edge Weights}
\label{sec:definition-networks-mixed}

These definitions are discussed in more detail in section \ref{sec:general-definitions}.

A ``\textbf{network with mixed-sign edge weights}'' $\graph$ with $\Species \in \mathbb{N}$ nodes is
\begin{enumerate}[label=(\roman*)]
\item a ``\textbf{node set}'' $\nodeset_{\graph}$ (which we can always assume equals $[\Strains] := \{1, \dots, \Strains\}$ without loss of generality, cf. the discussion below),
\item an ``\textbf{edge set}'' $\edgeset_{\graph} \subseteq \nodeset_{\graph} \times \nodeset_{\graph}$ such that $(\strain_1, \strain_2) \in \edgeset_{\graph}$ if and only if there is an edge directed from node $\strain_1$ to node $\strain_2$, and
\item an ``\textbf{edge weight function}'' ${A_{\graph}: \nodeset_{\graph}  \times \nodeset_{\graph} \to \mathbb{R}}$ such that
  \begin{equation}
    \label{eq:signed_network_definition}
    A_{\graph}(\strain_1, \strain_2) = 
    \begin{cases}
      w \in (-\infty, 0) \cup (0, \infty) & (\strain_1, \strain_2) \in \edgeset(\graph) \\
      0 & (\strain_1, \strain_2) \not \in \edgeset(\graph)
    \end{cases} \,.
  \end{equation}
\end{enumerate}
For each edge $(\strain_1, \strain_2) \in \edgeset_{\graph}$, the value $A_{\graph}(\strain_1, \strain_2)$ is the ``\textbf{weight}'' of the edge, the value ${\sign(A_{\graph}(\strain_1, \strain_2)) = \pm 1}$ is the ``\textbf{sign}'' of the edge, and the value ${|A_{\graph}(\strain_1, \strain_2)|}$ is the ``\textbf{magnitude}'' of the edge. Given an explicit identification between $\nodeset_{\graph}$ and $[\Strains]$ (cf. below), there always exists a unique $\Strains \times \Strains$ matrix $\adjacency_{\graph}$, which is called  the ``\textbf{adjacency matrix}'' of the network $\graph$, that corresponds to the edge weight function $A_{\graph}$.

For every pair $\strain_1, \strain_2 \in \nodeset_{\graph}$ such that $\strain_1 \not= \strain_2$, the definition allows for at most one edge directed from $\strain_1$ to $\strain_2$, and for at most one edge directed from $\strain_2$ to $\strain_1$. Likewise, for every $\strain \in \nodeset_{\graph}$, the definition allows for at most one edge (``self-loop'') directed from $\strain$ to $\strain$. I.e. ``multigraphs'' are not considered.

\subsection{Monotonicity Principle}
\label{sec:monot-princ}

Because literature already exists on comparing unsigned networks, a practical paradigm for comparing signed networks focuses on how to extend comparison methods for unsigned networks to also apply to signed networks. Herein we will focus only on extending comparison methods for unsigned networks that satisfy a mild ``well-behavedness'' criterion, termed the ``monotonicity principle''. Cf. figure \ref{fig:monotonicity}.

\begin{figure}[H]
  \centering
  \includegraphics[width=\textwidth]{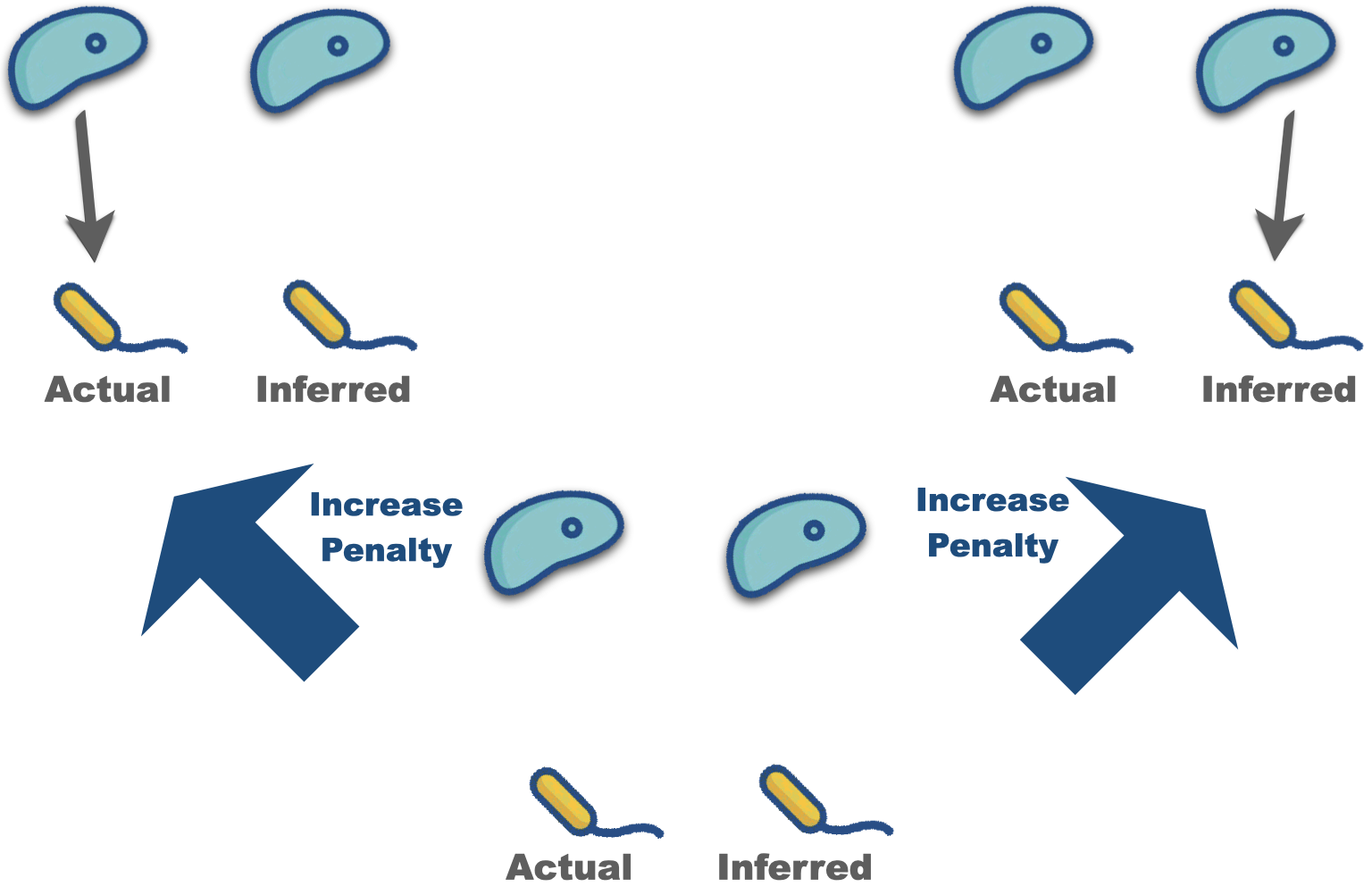}
  \caption[Monotonicity Principle]{Schematic illustration of the monotonicity principle for well-behaved comparison functions of unsigned networks.}
  \label{fig:monotonicity}
\end{figure}

At a high level, the idea of the monotonicity principle is that ``reasonable'' comparison methods for unsigned networks should penalize existence errors no less than they penalize ``true misses'', i.e. situations where an edge is missing in both networks. Cf. section \ref{sec:prec-form-double} for an attempted precise definition.  One can also ask for a ``continuous'' version of the monotonicity principle whereby the size of the penalty for an existence error must never decrease as the magnitude of the weight of the unmatched edge increases. Cf. section \ref{sec:continuity-mag}.

Because the monotonicity principle is a very intuitive property that one would most often tacitly assume to be true by default, we can consider comparison methods for unsigned networks that satisfy the monotonicity principle to be ``reasonable''. Therefore, to implement our paradigm of extending such comparison methods to also apply to signed networks, we need to be able to check whether any proposed extension satisfies the analogous intuitive property for signed networks. This of course first requires us to identify what the analogous intuitive property for signed networks is, which we do below.

\subsection{The Double Penalization Principle}
\label{sec:double-penal-princ}

Confusing two (non-zero) edges with different signs is the only new kind of error that can occur when comparing signed networks that possesses no analogue when comparing unsigned networks.  All other kinds of errors have analogues in the comparison of unsigned networks and hence can be treated using a ``reasonable'' comparison method for unsigned networks. Thus our main challenge, when trying to identify ``reasonable'' ways to extend comparison methods intended for unsigned networks to also apply to signed networks, is ensuring that errors confusing edges with different signs are treated in a ``reasonable'' way.

When interpreting positive and negative edge weights as corresponding to distinct (``equal but opposite'') phenomena, a mistake confusing edges with different signs amounts to the composite of two separate mistakes: both (i) inferring a phenomenon which does not exist, and (ii) failing to infer a phenomenon which does exist. Being worse than either of those two separate mistakes when considered individually, such a composite mistake should therefore receive a penalty that is no smaller than either of the penalties given to the two separate mistakes. This observation leads directly to the double penalization principle.

\textbf{Double penalization principle:} \textit{When the sign of an edge differs between two networks, the resulting penalty should be larger than the maximum of the two penalties that would occur if the edge was missing in either network.} 

In other words, the penalty should equal the maximum penalty that could occur if the edge was missing from either graph, plus an additional (“second”) penalty. Both mistakes, of inferring the wrong sign and of failing to infer the correct sign, should be penalized. The ``composite'' mistake should therefore be ``doubly penalized''. Cf. figure \ref{fig:double_penalization_principle}. The double penalization principle is equivalent to requiring that the monotonicity principle is simultaneously satisfied for both the positive parts and for the negative parts of the signed networks being compared. (Cf. section \ref{sec:posit-negat-parts} for definitions of ``positive part'' and ``negative part'' of a signed network. Both are unsigned networks, with the positive part having weights equal to the magnitudes of edges with positive sign, and analogously for the negative part.)
 
\begin{figure}[H]
  \centering
  \includegraphics[width=\textwidth]{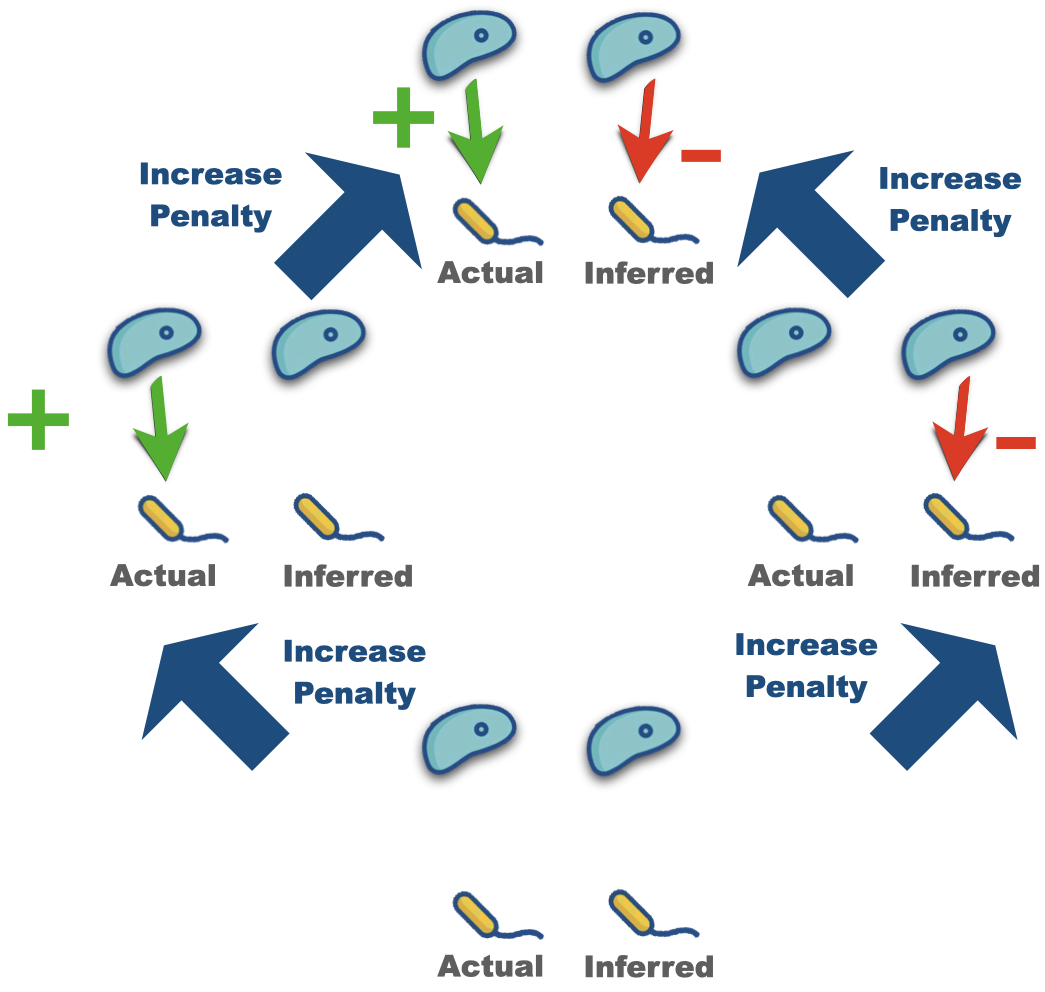}
  \caption{Double Penalization Principle}
  \label{fig:double_penalization_principle}
\end{figure}

Note that the double penalization principle excludes naive comparison methods for signed networks that ``project'' the signed networks being compared onto a space of unsigned networks and then apply a comparison method for unsigned network to the ``projections''. This means that the most straightforward idea for extending comparison methods for unsigned networks to also apply to signed networks fails to create comparison methods for signed networks that either are ``well-behaved'' or ``behave intuitively''. This is true even when the original comparison method for unsigned networks is ``well-behaved'' and ``behaves intuitively'', in the sense that it satisfies the monotonicity principle.

\subsection{Convex Combination Decomposition Property}
\label{sec:conv-comb-decomp}

A sufficient,\textit{ but not necessary}, condition for satisfying the double penalization principle is
\begin{itemize}
\item When the sign of an edge differs between two graphs, the resulting penalty equals the sum of the two penalties that would occur if the edge was set to zero (``removed'') in either graph.
\end{itemize}
An equivalent formulation of the double penalization principle is that the penalty for the edge should equal the maximum of the two penalties that occur when considering either the networks' positive parts only or negative parts only (see section \ref{sec:posit-negat-parts}), plus an additional penalty. This leads to the following equivalent formulation of the above sufficient, \textit{but not necessary}, criterion:
\begin{itemize}
\item When the sign of an edge differs between two graphs, the resulting penalty equals the sum of the penalty from comparing the positive parts of the graphs with the penalty from comparing the negative parts of the graphs.
\end{itemize}
When the penalties are additive across edges, the above criterion means the total penalty is the (weighted) sum of the penalties for the positive parts and negative parts of the graphs. Such (dis)similarity measures can be modified by suitable normalizations to produce (dis)similarity measures whose values equal a convex combination of (i) their value when applied to the positive parts of the networks only and (ii) their value when applied to the negative parts of the networks only. The resulting (dis)similarity measures are said to have the \textbf{convex combination decomposition property}. It follows from the above that the convex combination decomposition property is sufficient, \textit{but not necessary}, to satisfy the double penalization principle.

The convex combination decomposition property is potentially desirable not only because it guarantees that the double penalization principle will be satisfied. (Dis)similarity measures with this property are easy to interpret because understanding how the networks' positive edges vs. their negative edges contribute to the final value is straightforward.

\subsection{Sparsity-Savviness Principle}
\label{sec:sparsity-should-not}

This principle is applicable to general networks, not just signed networks. The fact that an edge is absent from two networks should not lead to an increase in their similarity score. Sparsity should not lead to artifactual similarity. In other words, ``don't reward true misses''. Otherwise, two networks which are very sparse (have a small percentage of all possible edges that \textit{could} exist) could potentially be scored as very similar, even when the structure of the edges they \textit{do} have is very different.

\subsection{Aspects of Networks to Compare}
\label{sec:which-feat-netw}

There are many aspects of networks we might seek to capture in our comparisons. 

``Numerical'' comparisons describe the correspondence of the numerical values of the edge weights. This could refer to only their magnitudes, or to their signed values. Relative error using a given norm (e.g. entrywise $L_1$) is an example. Cf. section \ref{sec:entryw-l_1-relat} for details.

There are also many ways we might seek to compare networks more ``qualitatively''. 

We might seek comparisons that describe the relative ordering of the edge weights (which are largest, which are smallest), either in magnitude, signed value, or both. (Sparsity-adjusted) Spearman correlation is an example. Cf. section \ref{sec:mixed-sign-spearman} for details. 

We might also seek ``qualitative'' comparisons that describe features of network topology. For example, (unweighted) Jaccard similarity describes the presence/absence of edges. Cf. section \ref{sec:mixed-sign-jaccard} for details. Similarly, (unweighted) DeltaCon distance \cite{DeltaCon} describes the correspondences of paths. Cf. section \ref{sec:mixed-sign-deltacon} for details. One might also want to define comparisons to describe the correspondences of other subgraph motifs besides paths.

We can also seek to capture both ``numerical'' and ``qualitative'' aspects of networks in a single comparison. Usually this is done by considering the correspondence of given qualitative features more or less important depending upon the weights of the edges in those features. Examples include weighted Jaccard similarity and weighted DeltaCon distance.

\section{General Definitions}
\label{sec:general-definitions}

This section gives precise definitions which will be used throughout the rest of the chapter for describing or modifying networks. Section \ref{sec:graphs-funct-matr} gives definitions and notation which will be used interchangeably to specify network structure. Section \ref{sec:degree-matrix} defines a related matrix which is important for section \ref{sec:mixed-sign-deltacon} later. Section \ref{sec:skeletons} defines standard ways of ``forgetting'' network structure. Section \ref{sec:posit-negat-parts} defines a standard way of decomposing a network, extending well-known decompositions applicable to the functions and matrices from section \ref{sec:graphs-funct-matr} 
which can be used to represent the network. Finally section \ref{sec:disjoint-union} defines a standard way to join two networks together to create a new (disconnected) network.

\subsection{Representations of Networks as Functions, Sets, and Matrices}
\label{sec:graphs-funct-matr}

 The $(i,j)$th entry of any matrix $\mathbf{M}$ will be denoted $[\mathbf{M}]_{ij}$. $\Species$ denotes the number of strains.

Any function ${\edgefunc:[\Species] \times [\Species] \to \mathbb{R} }$ is equivalent to specifying a (mixed-sign) weighted and directed graph $\graph$ with $\Species$ nodes. The values of $\edgefunc$ are the edges weights. Given any pair $(\specie_1, \specie_2)$, $\edgefunc(\specie_1, \specie_2)$ gives the weight of the edge directed from $\strain_1$ to $\strain_2$, or equals $0$ if no such edge exists. Cf. figure \ref{fig:network_example}. 
Any such \textbf{edge function} ${\edgefunc: [\Species] \times [\Species] \to \mathbb{R} }$ can also be identified with an ${\Species \times \Species}$ \textbf{adjacency matrix} $\adjacency$, where entry $[\adjacency]_{\specie_1 \specie_2}$ of the matrix equals $\edgefunc (\specie_1, \specie_2)$. Thus the $(\strain_1, \strain_2)$'th entry $[\adjacency]_{\specie_1 \specie_2}$ of the adjacency matrix $\adjacency$ gives the weight of the edge from $\strain_1$ to $\strain_2$, or is equal to $0$ if no such edge exists.

Note that here we are only explicitly considering networks that are ``static'' in time, but if we allow the adjacency matrix to vary with time, then the above also applies to time-varying networks (for modelling e.g. time-varying interactions \cite{umibato}), cf. \cite{Angulo2017}.

The set of edges $\edgeset$ of the graph $\graph$ is then defined by the set 
  \begin{equation}
    \label{eq:edge_set_defn}
\edgeset := {\support (\edgefunc) := \{ (\specie_1, \specie_2) \in [\Species] \times [\Species] : \edgefunc(\specie_1, \specie_2) \not=0  \} } \,.     
  \end{equation}

\begin{figure}[H]
  \centering
  \includegraphics[width=\textwidth,height=\textheight,keepaspectratio]{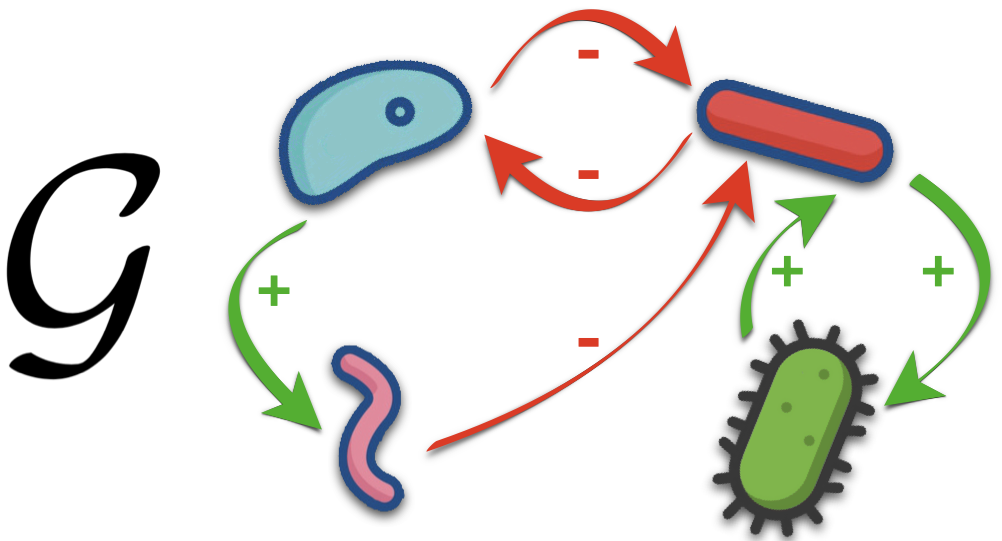}
  \caption[Directed network with signed edges.]{Directed network with signed edges. Bacteria are the nodes and identified with the set ${[\Species] = [4]}$. The arrows are the edges, identified with $\support({\edgefunc})$ where ${\edgefunc:[4]\times[4] \to \mathbb{R}}$. Green arrows are positively weighted edges, identified with the set ${ \{ (\specie_1, \specie_2) \in [4] \times [4] : \edgefunc(\specie_1, \specie_2) > 0  \}}$. Red arrows are negatively weighted edges, identified with the set ${ \{ (\specie_1, \specie_2) \in [4] \times [4] : \edgefunc(\specie_1, \specie_2) < 0  \}}$.}
  \label{fig:network_example}
\end{figure}

\subsection{The Degree Matrix}
\label{sec:degree-matrix}

\textbf{Note:} the material in this section is only used later on in section \ref{sec:mixed-sign-deltacon}.

The degree matrix $\mathbf{D}$ of a network $\graph$ with adjacency matrix $\adjacency$ is defined as
\begin{equation}
  \label{eq:degree_matrix_defn}
   [\textbf{D}]_{\strain_1 \strain_2} :=
  \begin{cases}
    \displaystyle\sum_{\strain[]_2=1}^{\Species} \mathbf{A}_{\strain_1 \strain[]_2}  & \strain_1 = \strain_2 \\
0 & \text{else}
  \end{cases} \,.
\end{equation}
Intuitively speaking, the $\strain$'th entry on the diagonal of $\mathbf{D}$ encodes the ``net influence'' flowing out from node $\strain$. In the case of an unsigned and unweighted network, i.e. an adjacency matrix consisting entirely of $1$'s and $0$'s, this is quantified by the number of the number of edges originating from node $\strain$.

\subsection{Skeletons}
\label{sec:skeletons}

Unsigned skeletons (section \ref{sec:unsigned-skeletons}) preserve the least amount of structure of the original network. Signed skeletons (\ref{sec:signed-skeletons}) and magnitude skeletons (\ref{sec:magnitude-skeletons}) preserve more structure. By definition all notions of ``graph skeleton'' lead to networks with edge sets $\edgeset$ that are the same as that of the original network.

\subsubsection{(Unsigned) Skeletons}
\label{sec:unsigned-skeletons}

Given a network $\graph$ with adjacency matrix $\adjacency$, the \textbf{unsigned skeleton} $\uskel (\graph)$ is defined to be the network with adjacency matrix $\mathbf{B}$ such that 
\begin{equation}
  \label{eq:unsigned_skel_adj}
  [\mathbf{B}]_{ij} :=
  \begin{cases}
    1 & [\adjacency]_{ij} \not = 0 \,, \\
0 & [\adjacency]_{ij} = 0 \,.
  \end{cases}
\end{equation}
The unsigned skeleton $\uskel(\graph)$ preserves only the underlying connectivity/topology of $\graph$, while ``deleting'' all information about edge weights or signs.

\subsubsection{Signed Skeletons}
\label{sec:signed-skeletons}

Given a network $\graph$ with adjacency matrix $\adjacency$, the \textbf{signed skeleton} $\sskel (\graph)$ is defined to be the network with adjacency matrix $\mathbf{B}$ such that 
\begin{equation}
  \label{eq:signed_skel_adj}
  [\mathbf{B}]_{ij} := \sign \left( [\adjacency]_{ij} \right) = 
  \begin{cases}
    1 & [\adjacency]_{ij} > 0 \,, \\
-1 & [\adjacency]_{ij} < 0 \,, \\
0 & [\adjacency]_{ij} = 0 \,.
  \end{cases}
\end{equation}
In addition to the underlying connectivity of the original network $\graph$, the signed skeleton $\sskel(\graph)$ also preserves the signs of the edge weights. The signed skeleton $\sskel(\graph)$ still ``deletes'' all information about edge magnitudes. 

Note that for a network with non-negative edge weights, the unsigned and signed skeletons coincide. Thus the signed skeleton is only a distinct and new notion for the more general setting of networks with mixed-sign edge weights.

\subsubsection{Magnitude Skeletons}
\label{sec:magnitude-skeletons}

Given a network $\graph$ with adjacency matrix $\adjacency$, the \textbf{magnitude skeleton} $\magskel (\graph)$ is defined to be the network with adjacency matrix $\mathbf{B}$ such that 
\begin{equation}
  \label{eq:mag_skel_adj}
  [\mathbf{B}]_{ij} := \abs \left( [\adjacency]_{ij} \right) := \left|[\adjacency]_{ij}\right| =
  \begin{cases}
    [\adjacency]_{ij} & [\adjacency]_{ij} > 0 \,, \\
-[\adjacency]_{ij} & [\adjacency]_{ij} < 0 \,, \\
0 & [\adjacency]_{ij} = 0 \,.
  \end{cases}
\end{equation}
In addition to the underlying connectivity of the original network $\graph$, the magnitude skeleton $\magskel(\graph)$ also preserves the magnitudes of the edge weights. The magnitude skeleton $\magskel(\graph)$ still ``deletes'' all information about edge signs.

Note that for a network with non-negative edge weights, the magnitude skeleton coincides with the original network. Thus the magnitude skeleton is only a distinct and new notion for the more general setting of networks with mixed-sign edge weights. Relatedly, observe how $\uskel(\graph) = \magskel(\sskel(\graph))$.

\subsection{Positive and Negative Parts}
\label{sec:posit-negat-parts}

Any adjacency matrix $\adjacency$ admits the decomposition:
\[
\adjacency = \adjacency^+ - \adjacency^- \,, \quad  [\adjacency^+]_{\specie_1 \specie_2} := \max \{ A_{\specie_1 \specie_2} , 0 \} \,, \quad [\adjacency^-]_{\specie_1 \specie_2} := \max \{ -A_{\specie_1 \specie_2}, 0 \} \,,
\]
which is equivalent to decomposing the edge function $\edgefunc$ in the standard way:
\[
    \edgefunc = \edgefunc^+ - \edgefunc^- \,, \quad  \edgefunc^+ := \max\{\edgefunc, \mathbf{0}\} \,, \quad \edgefunc^- := \max\{-\edgefunc, \mathbf{0}\} \,.
\]
$\adjacency^+$/$\edgefunc^+$ corresponds to its own network, denoted $\graph^+$, the \textbf{positive subnetwork}. Likewise, $\adjacency^-$/$\edgefunc^-$ also corresponds to its own network, denoted $\graph^-$, the \textbf{negative subnetwork}. Cf. figures \ref{fig:positive_network} and \ref{fig:negative_network}.

The edge weights of both $\graph^+$ and $\graph^-$ are by definition all positive, so standard dissimilarity measures for graphs can be applied to each subnetwork separately. This facilitates analysis and corresponds to the point of view that positive and negative edge weights correspond to qualitatively distinct phenomena.

\begin{figure}[H]
  \centering
  \includegraphics[width=\textwidth,height=\textheight,keepaspectratio]{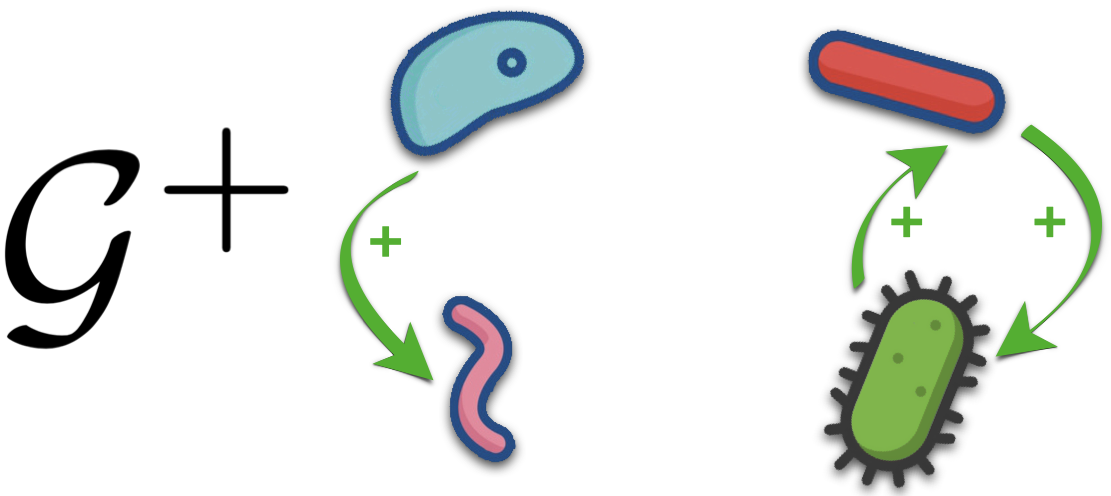}
  \caption{Positive Subnetwork of the Network from Figure \ref{fig:network_example}.}
  \label{fig:positive_network}
\end{figure}

\begin{figure}[H]
  \centering
  \includegraphics[width=\textwidth,height=\textheight,keepaspectratio]{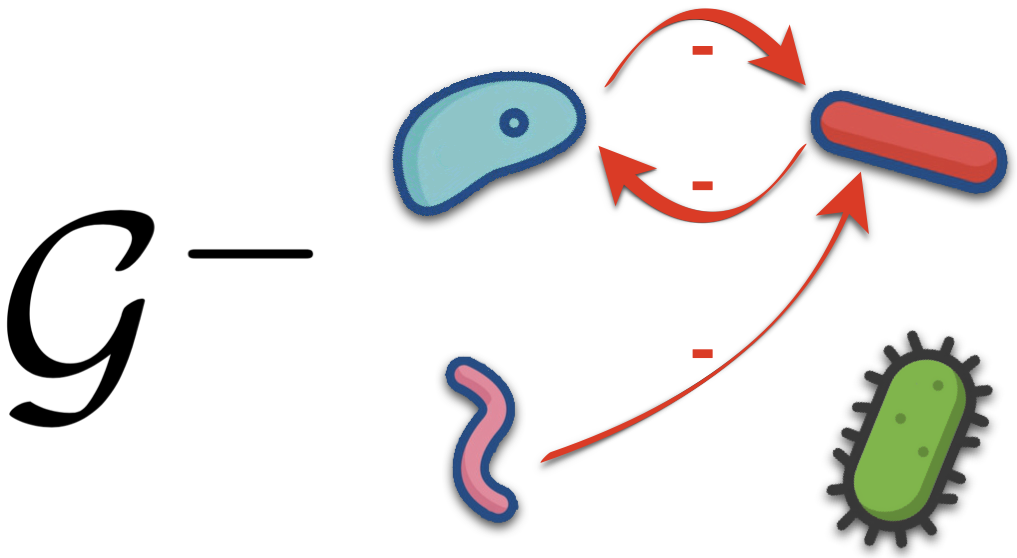}
  \caption{Negative Subnetwork of the Network from Figure \ref{fig:network_example}.}
  \label{fig:negative_network}
\end{figure}

\subsection{Disjoint Union}
\label{sec:disjoint-union}

Given two graphs, $\graph_1$ with adjacency matrix $\adjacency_1$ and $\graph_2$ with adjacency matrix $\adjacency_2$, define their ``\textbf{disjoint union}'' to be the graph $\graph_1 \oplus \graph_2$ corresponding to the adjacency matrix:
\[   \adjacency_1 \oplus \mathbf{A_2} =
  \begin{bmatrix}
    \adjacency_1 & \mathbf{0} \\ \mathbf{0} & \adjacency_2 
  \end{bmatrix} \,,  \]
i.e. the so-called direct sum of the matrices $\adjacency_1$ and $\adjacency_2$. The node set of this graph $\graph_1 \oplus \graph_2$ is the disjoint union of the node sets of $\graph_1$ and $\graph_2$, and likewise the edge set is the disjoint union of their edge sets $\edgeset_1 \sqcup \edgeset_2$. Cf. figure \ref{fig:direct_sum_network}.

\begin{figure}[H]
  \centering
  \includegraphics[width=\textwidth,height=\textheight,keepaspectratio]{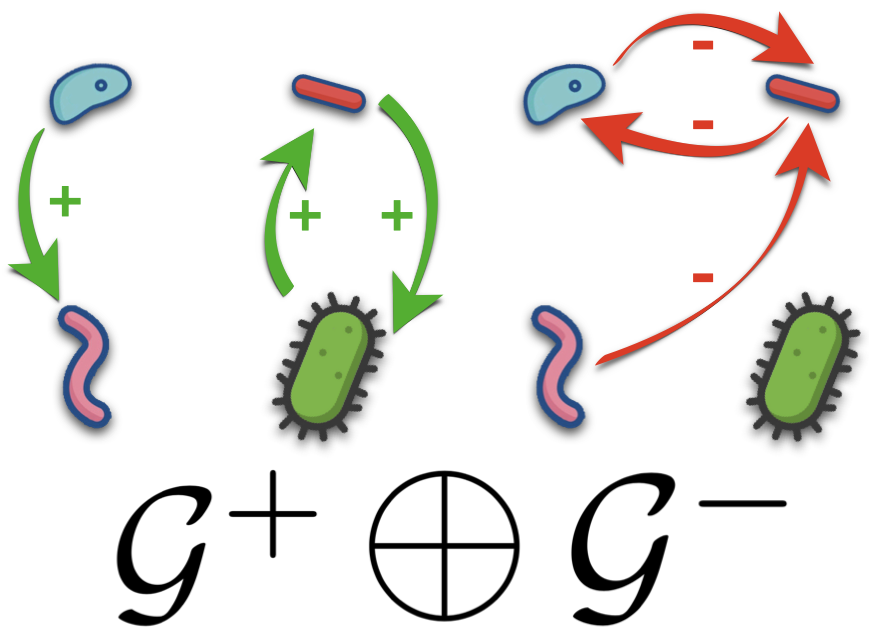}
  \caption[Direct Sum of  the Positive and Negative Subnetworks from Figure \ref{fig:network_example}.]{Direct Sum of the Positive and Negative Subnetworks of the Network from Figure \ref{fig:network_example}. Notice how the nodes are duplicated compared to Figure \ref{fig:network_example}.}
  \label{fig:direct_sum_network}
\end{figure}

\section{(Dis)similarity Measures}
\label{sec:diss-meas}

Below I define and discuss comparison methods applicable to networks with mixed-sign edge weights which span the full range of qualitative and quantitative comparisons explained in section \ref{sec:which-feat-netw}. When relevant, I explain how to modify methods for comparing unsigned networks to create new methods that give sensible results for networks with mixed-sign edge weights. I also explain the extent to which these proposed methods fall in line with the criteria discussed previously in sections \ref{sec:double-penal-princ}, \ref{sec:conv-comb-decomp}, and \ref{sec:sparsity-should-not}. 

Section \ref{sec:entryw-l_1-relat} provides mostly ``quantitative'' comparisons of networks, section \ref{sec:mixed-sign-spearman} provides mostly ``qualitative'' comparisons of relative orderings of edge weights, section \ref{sec:mixed-sign-jaccard} provides ``qualitative'' and ``quantitative'' comparisons of edge presence/absence, and finally section \ref{sec:mixed-sign-deltacon} provides ``qualitative'' and ``quantitative'' comparisons of path correspondence.

\subsection{(Entrywise $L_1$) Relative Error}
\label{sec:entryw-l_1-relat}

Relative error is a ``quantitative'' comparison of networks. If we normalize the edge weights of both networks before computing their relative error, the comparison becomes somewhat more ``qualitative'' in the sense that the input edge-weight-normalized networks better capture the relative sizes of the edge weights. Herein I show that ``out of the box'' the entrywise $L_1$ relative error as defined below automatically considers both positive and negative edges separately and then sensibly recombines the results.

\subsubsection{Entrywise $L_1$ Norm}
\label{sec:entrywise-l_1-norm}

The entrywise $L_1$ norm of an $\Species \times \Species$ matrix $\adjacency$ will herein be defined as
\begin{equation}
  \label{eq:entrywise_L1_defn}
\norm{\adjacency}_{[1]} := \sum_{\strain[]_1 = 1}^{\Species} \sum_{\strain[]_2=1}^{\Species} \left| [\adjacency]_{\strain[]_1 \strain[]_2} \right| \,. 
\end{equation}
In other words, this is the sum of the absolute values of the entries of $\adjacency$, or the $L_1$ norm of a vector resulting from vectorizing $\adjacency$. It is the ``$L_1$ analogue'' of the Frobenius norm.

It is \textit{not} to be confused with $L_1$ \textit{operator} norm of $\adjacency$, which is typically denoted $\norm{\adjacency}_1$. The notation $\norm{\adjacency}_{[1]}$ serves not only to emphasize this distinction, but also to remind of the notation for the entries of a matrix. 

An operator norm would be inappropriate to use here, given that the adjacency matrix $\mathbf{A}$ here is not meant to represent a linear function\footnote{Arguably except during the formation of the Neumann series in the definition of DeltaCon, cf. section \ref{sec:paths-powers-adjac}.} (it is ``entirely contravariant''). The Frobenius norm would also be inappropriate to use here. The Frobenius norm incorrectly ``inflates'' large-magnitude entries and ``deflates'' small-magnitude entries.

\subsubsection{Relative Error Definition}
\label{sec:relative_error_definition}
Given a ``truth'' matrix $\adjacencytruth$ and an ``estimate'' matrix $\adjacencyest$, the error of $\adjacencyest$ relative to $\adjacencytruth$, or just ``relative error'' if clear from context, is
\begin{equation}
  \label{eq:relative_error_definition}
\frac{ \norm{ \adjacencyest - \adjacencytruth   }_{[1]}   }{ \norm{\adjacencytruth}_{[1]} } \,.   
\end{equation}
The fact that relative error normalizes by the norm of one of the networks arguably helps to avoid considering two networks highly similar merely because they are both sparse. Cf. section \ref{sec:sparsity-should-not}. Normalization could help to ensure that a small number of differences is considered important if, due to sparsity, there are only a small number of edges that might be different to begin with.

\subsubsection{Convex Combination Decomposition Property}
\label{sec:conv-comb-decomp-1}

One can directly verify the identity:
\begin{equation}
  \label{eq:pos_neg_decomposition_entrywise_L1_norm}
\norm{ \adjacencyest - \adjacencytruth }_{[1]} =  \norm{ \adjacencyest^+ - \adjacencytruth^+  }_{[1]}  +  \norm{ \adjacencyest^- - \adjacencytruth^- }_{[1]} \,,  
\end{equation}
which also happens to be a special case (for a counting measure) of the measure-theoretic identity $\int |f-g| \mathrm{d}\mu = \int |f^+-g^+| \mathrm{d}\mu + \int |f^--g^-| \mathrm{d}\mu$.

Therefore the overall relative error can be written as a convex combination/weighted average of the ``positive relative error'' and the ``negative relative error''. More specifically, the following is true:
\begin{equation}
  \label{eq:conv_comb_decomposition_relative_error}
\frac{ \norm{ \adjacencyest - \adjacencytruth   }_{[1]}   }{ \norm{\adjacencytruth}_{[1]} } =  c_1 \left( \frac{ \norm{ \adjacencyest^+ - \adjacencytruth^+   }_{[1]}   }{ \norm{\adjacencytruth^+}_{[1]} } \right)  + c_2 \left( \frac{ \norm{ \adjacencyest^- - \adjacencytruth^-   }_{[1]}   }{ \norm{\adjacencytruth^-}_{[1]} }  \right)  \,,  
\end{equation}
where the coefficients above are defined
\begin{equation}
  \label{eq:conv_comb_decomposition_relative_error_coeffs}
c_1 := \frac{ \norm{\adjacencytruth^+}_{[1]}  }{ \norm{\adjacencytruth}_{[1]}  } \,, \quad c_2 := \frac{ \norm{\adjacencytruth^-}_{[1]}  }{  \norm{\adjacencytruth}_{[1]} } \,.
\end{equation}
This also ensures that any positive interactions estimated as negative, or vice versa, will appear twice in the above expression. We can see how the convex combination property implies the double penalization principle is satisfied.

\subsubsection{Normalizing to get More Qualitative Comparisons}
\label{sec:normalizing-get-more}

If a priori we don't believe that the edge weights of the two networks are on the same ``length scale'', e.g. they are measured in different units, but we still want to compare them in some qualitative, ``scale-free'' way, then we can normalize both networks' adjacency matrices before computing their relative error:
\begin{equation}
  \label{eq:normalized_rel_error}
 \frac{ \displaystyle
\norm{ \frac{\adjacencyest}{\norm{\adjacencyest}_{[1]}} - \frac{\adjacencytruth}{\norm{\adjacencytruth}_{[1]}}   }_{[1]}   
}{
\displaystyle
 \norm{\frac{\adjacencytruth}{\norm{\adjacencytruth}_{[1]}} }_{[1]} 
}
=
 \frac{ \displaystyle
\norm{ \frac{\adjacencyest}{\norm{\adjacencyest}_{[1]}} - \frac{\adjacencytruth}{\norm{\adjacencytruth}_{[1]}}   }_{[1]}   
}{1}
=
\norm{ \frac{\adjacencyest}{\norm{\adjacencyest}_{[1]}} - \frac{\adjacencytruth}{\norm{\adjacencytruth}_{[1]}}   }_{[1]}   
 \,.
\end{equation}
If we wanted to compare whether the relatively largest magnitude edge weights in both networks tended to have the same sign, or belong to edges connecting the same nodes, this could be useful. If an edge has the same sign, but three times the magnitude, in $\graph_*$ compared to $\hat{\graph}$, for example, that won't be penalized if the sum of the magnitudes of the edge weights is also three times as large in $\graph_*$ compared to $\hat{\graph}$ (hence why the comparison is ``scale-free'').

Note that the resulting values are necessarily always in the interval $[0,2]$. Also, unlike the relative error in general (\ref{eq:relative_error_definition}), the resulting values are symmetric when interchanging $\adjacencyest$ and $\adjacencytruth$ in the formula (\ref{eq:normalized_rel_error}).

Normalizing both matrices first before computing their relative error arguably also helps to avoid considering two networks highly similar merely because they are both sparse. Cf. again section \ref{sec:sparsity-should-not}.

\subsection{Mixed-Sign Spearman Correlation}
\label{sec:mixed-sign-spearman}

These methods based on Spearman correlation are designed to provide a qualitative comparison of two networks, by comparing the relative ordering of their edge weights. The formation of rank vectors destroys any direct quantitative information about exact edge weight values. The mixed-sign Spearman correlation defined below provides an advantage over ``raw'' Spearman correlation by separately comparing the relative ordering of positive and negative edges and then sensibly recombining the results. By considering this extra information contained in the signs of the edge weights, it is not fooled by drastic changes to the network structure to which the ``raw'' Spearman correlation or Spearman correlation of the weights' magnitudes are completely oblivious.

\subsubsection{Sparsity Adjustment}
\label{sec:sparsity-adjustment}

Given two networks $\graph_1$, $\graph_2$ with adjacency matrices $\adjacency_1$, $\adjacency_2$ respectively, consider the subset of index pairs for which the edge corresponding to each index pair is present in at least one of the two networks:
\begin{equation}
  \label{eq:non_zero_indices}
  \mathcal{I}_{\graph_1,\graph_2} := \left\{ (\strain_1, \strain_2) \in [\Strains] \times [\Strains] : [\adjacency_1]_{\strain_1\strain_2} \not= 0 \text{ or } [\adjacency_2]_{\strain_1\strain_2} \not= 0 \right\}\,.
\end{equation}
Then define a modified vectorization operator $\vectorize_{\graph_1,\graph_2}$ which, given $[\Strains] \times [\Strains]$ matrices, returns vectors of length $|\mathcal{I}_{\graph_1,\graph_2}|$ whose indices correspond to the index pairs belonging to $\mathcal{I}_{\graph_1,\graph_2}$. This is a ``sparsity adjustment'' because it removes values corresponding to index pairs for which no edge exists in either $\graph_1$ or $\graph_2$.

\subsubsection{Rank Vectors}
\label{sec:rank-vectors}

Let $\rank$ denote the ``rank operator'' which replaces each value of a vector with its relative ordering (e.g. the smallest element is assigned $1$, the largest is assigned the length of the vector). Assume $\rank$ is such that ties are replaced with the mean value of the tied positions (as is done by default e.g. in the 'rankdata' function of the SciPy stats library\cite{SciPy}).

Given a network $\graph$ with adjacency matrix $\adjacency$, define the following shorthand:
\begin{equation}
  \label{eq:9}
 \sparrank{\graph} := \rank \left( \vectorize_{\graph_1,\graph_2} \left( \adjacency \right) \right) \,,
\end{equation}
where $\vectorize_{\graph_1,\graph_2}$ is the same sparsity-adjusted vectorization operator that was defined in section \ref{sec:sparsity-adjustment}.

\subsubsection{Positive and Negative Spearman Correlation}
\label{sec:posit-negat-spearm}

The ``\textbf{positive Spearman correlation}'' of $\graph_1$ and $\graph_2$ is defined to be the (sparsity-adjusted) Spearman correlation\footnote{The ``$S$'' in the notation ``$\spearman$'' can be interpreted to refer to either ``Spearman'', ``sparsity-adjusted'', or both. ``$\rho$'' of course is commonly-used notation for a correlation coefficient.} of $\graph_1^+$ and $\graph_2^+$:
\begin{equation}
  \label{eq:positive_spearman_definition}
\spearman^+(\graph_1, \graph_2)  := \frac{
\ecov*{\sparrank{\graph_1^+}, \sparrank{\graph_2^+} }
}{
\sqrt{
\evar*{\sparrank{\graph_1^+}}
}
\cdot
\sqrt{
\evar*{\sparrank{\graph_2^+}}
}
} \,.
\end{equation}
Notice that the sparsity adjustment is in terms of $\graph_1$ and $\graph_2$, not $\graph_1^+$ and $\graph_2^+$. This is noteworthy to the extent that the union of the edge sets of $\graph_1^+$ and $\graph_2^+$ could, in principle, be strictly smaller than the union of the edge sets of $\graph_1$ and $\graph_2$. Therefore one might sensibly argue that the strictest possible sparsity adjustment, that in terms of $\graph_1^+$ and $\graph_2^+$, should be applied when defining a notion of ``positive Spearman correlation''. Anything else might not be compatible with the principle outlined in section \ref{sec:sparsity-should-not}. On the other hand, one could also argue that a correlation which ``preserves the context of the original networks'' admits a more useful interpretation, at least when trying to compare or combine\footnote{
Admittedly my main motivation for defining the sparsity adjustments for the positive and negative Spearman correlations using the edge sets of the original networks was to facilitate the interpretation of the mixed-sign Spearman correlation. If the sparsity adjustment for the positive Spearman correlation was in terms of $\graph_1^+$ and $\graph_2^+$, and that for the negative Spearman correlation in terms of $\graph_1^-$ and $\graph_2^-$, then as far as I can tell there would in general be no straightforward way to write the mixed-sign Spearman correlation as a weighted sum of the positive and negative Spearman correlations. I came up with the notion of mixed-sign Spearman correlation before observing the need for sparsity adjustments, so it perhaps is not altogether surprising that the two notions do not seem to be easily compatible.
} the values of the positive and negative Spearman correlations.

Completely analogously, the ``\textbf{negative Spearman correlation}'' of $\graph_1$ and $\graph_2$ is defined to be the (sparsity-adjusted) Spearman correlation of $\graph_1^-$ and $\graph_2^-$:
\begin{equation}
  \label{eq:negative_spearman_definition}
\spearman^-(\graph_1, \graph_2) :=  \frac{
\ecov*{\sparrank{\graph_1^-}, \sparrank{\graph_2^-}}
}{
\sqrt{
\evar*{\sparrank{\graph_1^-}}
}
\cdot
\sqrt{
\evar*{
\sparrank{\graph_2^-}
}
}
}
\end{equation}
Analogous comments about sparsity adjustments apply here too, of course.

\subsubsection{Vector Concatenation}
\label{sec:vector-concatenation}

The concatenation or ``direct sum'' $\rvec{v}_{1} \oplus \rvec{v}_{2} \in \mathbb{R}^{I_1+I_2}$ of two vectors ${\rvec{v}_1 \in \mathbb{R}^{I_1}}$ and ${\rvec{v}_2\in \mathbb{R}^{I_2}}$ is defined such that
\begin{equation}
  \label{eq:concatenation}
  (\rvec{v}_{1} \oplus \rvec{v}_{2})^{(i)} :=
  \begin{cases}
    v_1^{(i)} & 1 \le i \le I_1 \\
   v_2^{(i-I_1)} & I_1+1 \le i \le I_1 + I_2 \,.
  \end{cases}
\end{equation}
This has the intuitive interpretation of first listing the entries of $\rvec{v}_1$ and then appending to that list the entries of $\rvec{v}_2$.

Some properties are obvious consequences of the definition. For example:
\begin{equation}
  \label{eq:concatenated_vectors_partwise_addition}
  (\rvec{v}_1 + \rvec{w}_1) \oplus (\rvec{v}_2 + \rvec{w}_2) = (\rvec{v}_1 \oplus \rvec{v}_2) + (\rvec{w}_1 \oplus \rvec{w}_2) \,.
\end{equation}
Here is a second example using the standard inner product, $\displaystyle \langle \rvec{x}, \rvec{y} \rangle := \sum_{i=1}^I x^{(i)} y^{(i)}$:
\begin{equation}
  \label{eq: concatenated_vectors_partwise_dot_product}
  \langle \rvec{v}_1 \oplus \rvec{v}_2 , \rvec{w}_1 \oplus \rvec{w}_2 \rangle = \langle \rvec{v}_1, \rvec{w}_1 \rangle + \langle \rvec{v}_2, \rvec{w}_2 \rangle \,.
\end{equation}
As seen above, for many operations on concatenated vectors in $\mathbb{R}^{I_1 + I_2}$ we can treat what happens in $\mathbb{R}^{I_1}$ separately of what happens in $\mathbb{R}^{I_2}$.

\subsubsection{Mixed-Sign Spearman Definition}
\label{sec:definition-1}

Without explaining or motivating anything, the definition may be written as
\begin{equation}
  \label{eq:mixed_sign_spearman}
\spearman^{\pm}(\graph_1, \graph_2)
:=
  \frac{
\ecov*{\sparrank{\graph_1^+} \oplus \sparrank{\graph_1^-}, \sparrank{\graph_2^+} \oplus \sparrank{\graph_2^-}}
}
{
\sqrt{
\evar*{\sparrank{\graph_1^+} \oplus \sparrank{\graph_1^-}}
}
\cdot
\sqrt{
\evar*{\sparrank{\graph_2^+} \oplus \sparrank{\graph_2^-}}
}
} \,.
\end{equation}
Using the sparsity-adjusted rank operator, instead of the typical rank operator $\rank$, prevents two networks from scoring high values merely because both are highly sparse. This is because the sparsity-adjusted rank operator only looks at the subset of edges which are nonzero in at least one of the networks, and does not consider edges missing from both networks. The use of the sparsity-adjusted rank operator follows from the principle outlined in section \ref{sec:sparsity-should-not}.

\subsubsection{Subconvex Combination Decomposition Property}
\label{sec:conv-comb-decomp-2}

The sense in which expression (\ref{eq:mixed_sign_spearman}) considers the positive and negative subnetworks separately and then combines the results is made precise below. 

Because of Lemma \ref{lem:means_of_rank_vectors}, the hypotheses of Lemma \ref{lem:additivity_covariance_concatenated_vectors} are satisfied. Applying Lemma \ref{lem:additivity_covariance_concatenated_vectors} to expressions in both the numerator and denominator of (\ref{eq:mixed_sign_spearman}):
\begin{equation}
  \label{eq:conv_comb_mixed_sign_spearman}
  \begin{split}
\spearman^{\pm} (\graph_1, \graph_2) =
&  \   \frac{
\ecov*{\sparrank{\graph_1^+}, \sparrank{\graph_2^+} }
+
\ecov*{\sparrank{\graph_1^-}, \sparrank{\graph_2^-}}
}{
\sqrt{
\evar*{\sparrank{\graph_1^+}} + \evar*{\sparrank{\graph_1^-}}
}
\cdot
\sqrt{
\evar*{\sparrank{\graph_2^+}} + \evar*{\sparrank{\graph_2^-}}
}
} \\
= & \ 
\cvxcoeff{+}
\frac{
\ecov*{\sparrank{\graph_1^+}, \sparrank{\graph_2^+} }
}{
\sqrt{
\evar*{\sparrank{\graph_1^+}}
}
\cdot
\sqrt{
\evar*{\sparrank{\graph_2^+}}
}
}
+
\cvxcoeff{-}
\frac{
\ecov*{\sparrank{\graph_1^-}, \sparrank{\graph_2^-}}
}{
\sqrt{
\evar*{\sparrank{\graph_1^-}}
}
\cdot
\sqrt{
\evar*{
\sparrank{\graph_2^-}
}
}
} 
\\
= & \ \cvxcoeff{+} \cdot \spearman^{+}(\graph_1, \graph_2) + \cvxcoeff{-} \cdot \spearman^-(\graph_1, \graph_2)
\,,
  \end{split}
\end{equation}
where the coefficients $\cvxcoeff{+}$, $\cvxcoeff{-}$ equal
\begin{equation}
  \label{eq:mixed_sign_spearman_coeffs}
  \begin{split}
    \cvxcoeff{+} = & 
\frac{
\sqrt{
\evar*{\sparrank{\graph_1^+}}
}
\cdot
\sqrt{
\evar*{\sparrank{\graph_2^+}}
}
}{
\sqrt{
\evar*{\sparrank{\graph_1^+}} + \evar*{\sparrank{\graph_1^-}}
}
\cdot
\sqrt{
\evar*{\sparrank{\graph_2^+}} + \evar*{\sparrank{\graph_2^-}}
}
} \,, \\
\cvxcoeff{-} = &
\frac{
\sqrt{
\evar*{\sparrank{\graph_1^-}}
}
\cdot
\sqrt{
\evar*{
\sparrank{\graph_2^-}
}
}
}{
\sqrt{
\evar*{\sparrank{\graph_1^+}} + \evar*{\sparrank{\graph_1^-}}
}
\cdot
\sqrt{
\evar*{\sparrank{\graph_2^+}} + \evar*{\sparrank{\graph_2^-}}
}
} \,.
  \end{split}
\end{equation}

Although in equation (\ref{eq:mixed_sign_spearman_coeffs}) we have $\cvxcoeff{+}, \cvxcoeff{-} \ge 0$, in general we only have that $\cvxcoeff{+} + \cvxcoeff{-} \le 1$, cf. Lemma \ref{lem:subconvex}. In other words, in general the mixed-sign Spearman correlation is only guaranteed to be a \textbf{subconvex combination} of the positive Spearman correlation and the negative Spearman correlation. In special cases it may still also be a convex combination sensu stricto. Because this subconvex property can be derived using the Cauchy-Schwarz inequality, one way to interpret any deficit $ 1 - (\cvxcoeff{+} + \cvxcoeff{-}) > 0$ is as a measurement of how much\footnote{
More precisely, it measures how much $\sparrank{\graph_1^+} \oplus \sparrank{\graph_1^-}$ and $\sparrank{\graph_2^+} \oplus \sparrank{\graph_2^-}$ ``fail to almost align onto the same line''. The Cauchy-Schwarz inequality is an equality only when two vectors are linearly dependent, so in the same (or exactly opposite) directions.
} $\graph_1$ and $\graph_2$ ``fail to align in almost the same direction''. 

Although the subconvex property does still correspond to a weighted sum, and does have some merit for possessing a useful interpretation, it would nevertheless still be fair to argue that the mixed-sign Spearman correlation does not fully live up to the promised easy interpretation from section \ref{sec:conv-comb-decomp}. The mixed-sign Spearman correlation does still satisfy the double penalization principle. Despite in general being a subconvex combination but not necessarily a convex combination, in practice the mixed-sign Spearman correlation also does not seem to necessarily give overly ``conservative'' results ``biased'' towards $0$.

Perhaps one could argue that any such ``shrinkage'' might even be desirable, given that the sparsity adjustments defining the positive and negative Spearman correlations are not as aggressive as possible (cf. section \ref{sec:posit-negat-spearm}). By possibly failing to adhere to the principle of section \ref{sec:sparsity-should-not}, one might be concerned that the positive and negative Spearman correlations artifactually overstate similarities. While this argument might be true to an extent, I would also caution that ``two wrongs don't make a right''. A priori we have no reason to believe that hypothetical ``corrections'' from the subconvex property would (or even could) exactly counterbalance hypothetical artifacts from the chosen sparsity adjustments for the positive and negative Spearman correlations.

\subsection{Mixed-Sign Jaccard Similarity}
\label{sec:mixed-sign-jaccard}

Unweighted Jaccard similarity provides a qualitative comparison of networks by comparing the presence/absence of edges. Weighted Jaccard similarity provides a comparison that is a little quantitative and a little qualitative, by increasing the importance of comparisons for edge pairs where one of the edges has a large edge weight. Mixed-sign versions of both comparison methods provide an advantage over unsigned versions by separately considering presence/absence for positive and negative edges and then sensibly recombining the results. By considering this extra information contained in the signs of the edge weights, the proposed mixed-sign versions are not fooled by drastic changes to the network structure to which the unsigned versions are completely oblivious.

\subsubsection{Binary Classification and Mixed Signs}
\label{sec:binary-class}

Considering positive and negative edges to correspond to distinct phenomena means we effectively have a ternary classification problem (present and positive, present and negative, absent). Our goal is to recast this ternary classification problem as two inter-related binary classification problems, one for the positive edges and one for the negative edges.

To avoid confusion with the edge signs, below I use the following alternative terminology to refer to the four possible binary classification outcomes:
\begin{itemize}
\item ``\textbf{True Hit}'' (TH) := ``True Positive'',
\item ``\textbf{False Hit}'' (FH) := ``False Positive'',
\item ``\textbf{True Miss}'' (TM) := ``True Negative'',
\item ``\textbf{False Miss}'' (FM) := ``False Negative''.
\end{itemize}

\paragraph{Positive Edge Binary Classification Problem}
The positive edge binary classification problem has
\begin{itemize}
\item hits = $\{+\}$,
\item misses = $\{0, -\}$.
\end{itemize}
So for the positive edge binary classification problem:
\begin{itemize}
\item  a ``\textbf{true hit}'' occurs when both the truth and the estimate are positive,
\item a ``\textbf{false hit}'' occurs when the truth is non-positive ($0$ or $-$) but the estimate is positive,
\item a ``\textbf{true miss}'' occurs when the truth is non-positive and the estimate is non-positive, and
\item a ``\textbf{false miss}'' occurs when the truth is positive but the estimate is non-positive.
\end{itemize}

\paragraph{Negative Edge Binary Classification Problem}
Analogously the negative edge binary classification problem has
\begin{itemize}
\item hits = $\{-\}$,
\item misses = $\{0, +\}$.
\end{itemize}
So for the negative edge binary classification problem:
\begin{itemize}
\item a ``\textbf{true hit}'' occurs when both the truth and the estimate are negative,
\item a ``\textbf{false hit}'' occurs when the truth is non-negative ($0$ or $+$) but the estimate is negative,
\item a ``\textbf{true miss}'' occurs when the truth is non-negative and the estimate is non-negative, and
\item a ``\textbf{false miss}'' occurs when the truth is negative but the estimate is non-negative.
\end{itemize}

\paragraph{Origin of Double Penalization}
From this perspective of having two inter-related binary classification problems, confusing a positive interaction with a negative interaction, or vice versa, is arguably both
\begin{itemize}
\item a \textbf{false hit}, inferring a type of interaction which does not exist, and
\item a \textbf{false miss}, failing to infer a type of interaction which does exist.
\end{itemize}
Thus an ideal performance metric should ``doubly penalize'' such errors. This was the original insight which led to the more general double penalization principle from section \ref{sec:double-penal-princ}.

\subsubsection{(Unsigned) Unweighted Definition}
\label{sec:unsign-unwe-defin}

The (unweighted) Jaccard similarity of two graphs $\graph_1$ and $\graph_2$ (with non-negative edge weights) is the Jaccard similarity of their edge sets $\edgeset_1$ and $\edgeset_2$:
\begin{equation}
  \label{eq:unweighted_unsigned_jaccard_defn}
\jaccard (\graph_1, \graph_2 ) := \frac{|\edgeset_1 \cap \edgeset_2|}{|\edgeset_1 \cup \edgeset_2|} = \frac{|\edgeset_1 \cap \edgeset_2|}{|\edgeset_1 \cap \edgeset_2| + |\edgeset_1 \setminus \edgeset_2| + |\edgeset_2 \setminus \edgeset_1|}\,.  
\end{equation}
Recall that 
\[ \edgeset_1 := \support(\edgefunc_1)\,, \edgeset_2 := \support(\edgefunc_2) \]
where $\edgefunc_1 \ge 0$ is the edge function corresponding to $\graph_1$, and likewise for $\edgefunc_2 \ge 0$.

Notice that if $\graph_1$ is an ``estimate'' $\hat{\graph}$, and $\graph_2$ is the ``ground truth'' $\graph_*$ then their (unweighted) Jaccard similarity corresponds to
\begin{equation}
  \label{eq:unsigned_jaccard_binary_classification}
  \jaccard(\hat{\graph}, \graph_*) = \frac{TH}{TH + FH + FM} \,,
\end{equation}
where $TH$ is the number of true hits, $FH$ is the number of false hits, and $FM$ is the number of false misses. Thus Jaccard similarity is no less conservative than the minimum of precision and recall, since it has the same numerator and one additional non-negative term in the denominator. 

It is beneficial that Jaccard similarity does not reward true misses\footnote{$TM$ would denote the number of true misses. Notably this is absent from the numerator of equation (\ref{eq:unsigned_jaccard_binary_classification}). Thus higher values of $TM$ do not lead to higher values of Jaccard similarity.}, because not doing so helps to prevent considering two highly different networks as similar merely because both are highly sparse. Cf. again section \ref{sec:sparsity-should-not}.

As an aside, in meteorology Jaccard similarity (as used for binary classification) is often called the ``critical success index'' or ``threat score''\cite{critical_success_index}. Many binary classification metrics considered improvements over Jaccard similarity (e.g. the ``equitable threat score'') are studied in meteorology\cite{critical_success_index}. Future work examining whether these other binary classification metrics also admit extensions to the comparison of networks of mixed-sign edge weights, and whether such extensions would be useful, would be interesting.

\subsubsection{(Unsigned) Weighted Definition}
\label{sec:unsign-we-defin}

For two graphs $\graph_1$ and $\graph_2$ with non-negative edge weights and edge functions $\edgefunc_1$ and $\edgefunc_2$ as above, their (unsigned) weighted Jaccard similarity is
\begin{equation}
  \label{eq:unsigned_weighted_jaccard_definition}
\jaccard^W (\graph_1, \graph_2) := 
\frac{
\displaystyle
\sum_{i,j} \min \{ \edgefunc_1(i,j), \edgefunc_2(i,j)  \} 
}{
\displaystyle
\sum_{i,j} \max \{ \edgefunc_1(i,j), \edgefunc_2(i,j)  \} 
} \,.  
\end{equation}
By considering the graphs' ``skeletons'', which have edge functions $(\sign \circ \edgefunc_1)$ and $(\sign \circ \edgefunc_2)$, one recovers the regular (unweighted) Jaccard similarity. Specifically
\begin{equation}
  \label{eq:unsigned_unweighted_skels}
\jaccard^W(\uskel(\graph_1), \uskel(\graph_2)) = \jaccard(\graph_1, \graph_2) \,.  
\end{equation}
Recall the definitions of graph (unsigned) skeletons from section \ref{sec:unsigned-skeletons}.

\subsubsection{Positive and Negative Jaccard Similarities}
\label{sec:posit-negat-jacc}

From here on let $\graph_1$, $\graph_2$ denote two general networks with \textit{mixed-sign} edge weights. (In the previous sections \ref{sec:unsign-unwe-defin} and \ref{sec:unsign-we-defin} it was assumed for the sake of simplicity that the networks $\graph_1$, $\graph_2$ had non-negative edge weights.)

One can define the ``positive (unweighted) Jaccard similarity'' of two general networks as:
\begin{equation}
  \label{eq:positive_unweighted_jaccard_similarity_defn}
\jaccard^+ (\graph_1, \graph_2) := \jaccard (\graph_1^+, \graph_2^+) \,,  
\end{equation}
as well as their ``positive weighted Jaccard similarity'', 
\begin{equation}
  \label{eq:positive_weighted_jaccard_similarity_defn}
(\jaccard^W)^+(\graph_1, \graph_2) := \jaccard^W (\graph_1^+, \graph_2^+)\,.  
\end{equation}

Completely analogously, the ``negative (unweighted) Jaccard similarity'' of two general networks is:
\begin{equation}
  \label{eq:negative_unweighted_jaccard_similarity_defn}
\jaccard^- (\graph_1, \graph_2) := \jaccard (\graph_1^-, \graph_2^-) \,,
\end{equation}
and their ``negative weighted Jaccard similarity'', 
\begin{equation}
  \label{eq:negative_weighted_jaccard_similarity_defn}
(\jaccard^W)^-(\graph_1, \graph_2) := \jaccard^W (\graph_1^-, \graph_2^-) \,.
\end{equation}

The above definitions make sense because the positive $\graph^+$ and negative subnetworks $\graph^-$ have non-negative edge weights by definition. Therefore ``standard'' network dissimilarity methods may be applied to them.

Observe how the weighted versions $(\jaccard^W)^+$, $(\jaccard^W)^-$ reduce to their respective unweighted counterparts $\jaccard^+$, $\jaccard^-$ when applied to (signed) skeletons:
\begin{equation}
  \label{eq:signed_jaccard_weighted_reduces_to_unweighted}
  (\jaccard^W)^+ (\sskel(\graph_1), \sskel(\graph_2)) = \jaccard^+ (\graph_1, \graph_2) \,, 
\quad
(\jaccard^W)^- (\sskel(\graph_1), \sskel(\graph_2)) = \jaccard^-(\graph_1, \graph_2) \,.
\end{equation}

\subsubsection{Mixed-Sign Unweighted Jaccard Similarity}
\label{sec:mixed-sign-unwe-2}

 The ``mixed sign (unweighted) Jaccard similarity'' of two networks $\graph_1$, $\graph_2$ with mixed-sign edge weights is defined to equal
 \begin{equation}
   \label{eq:mixed_sign_unweighted_jaccard_similarity_defn}
  \jaccard^{\pm} (\graph_1, \graph_2) := \jaccard( \graph_1^+ \oplus \graph_1^-, \graph_2^+ \oplus \graph_2^-) = 
\frac{|\edgeset_1^+ \cap \edgeset_2^+| +|\edgeset_1^- \cap \edgeset_2^-|}{|\edgeset_1^+ \cup \edgeset_2^+| +|\edgeset_1^- \cup \edgeset_2^-|} \,.   
 \end{equation}
($\edgeset^+$ is the edge set of a graph's positive part, analogously for $\edgeset^-$.)

Notice that if a positive edge is estimated as negative, it will show up once both as a false hit for the negative edges (contributing once to $|\edgeset_1^- \cup \edgeset_2^-|$) as well as a false miss for the positive edges (contributing once to $|\edgeset_1^+ \cup \edgeset_2^+|$). The analogous conclusions hold when a negative edge is estimated as positive.

Therefore, \textit{mixed-sign (unweighted) Jaccard similarity satisfies the desired double penalization property for incorrectly estimated signs}. In fact:
\begin{equation}
  \label{eq:mixed_jaccard_binary_classification}
  \jaccard^{\pm} (\hat{\graph}, \graph_*) = \frac{TH^+ + TH^-}{TH^+ + TH^- + FH^+ + FH^- + FM^+ + FM^-} \,.
\end{equation}
The analogy between (\ref{eq:unsigned_jaccard_binary_classification}) and (\ref{eq:mixed_jaccard_binary_classification}) strongly suggests that mixed-sign Jaccard similarity is ``the correct'' generalization of (unsigned) Jaccard similarity.

Notice how this definition again does not reward true misses\footnote{$TM^+$ would denote the number of true misses for the positive subnetworks, $TM^-$ would denote the number of true misses for the negative subnetworks. Notably these are absent from the numerator of equation (\ref{eq:mixed_jaccard_binary_classification}). Thus neither higher values of $TM^+$, nor higher values of $TM^-$, lead to higher values of mixed-sign Jaccard similarity.}, which could occur merely due to either the positive and/or negative subnetworks being highly sparse. This helps mixed-sign (unweighted) Jaccard similarity to avoid considering two highly different networks as similar merely because both are highly sparse, and/or both have few positive/negative edges. Cf. again section \ref{sec:sparsity-should-not}.

Moreover the mixed-sign (unweighted) Jaccard similarity can be described simply as a convex combination/weighted average of the positive (unweighted) Jaccard similarity and the negative (unweighted) Jaccard similarity:
\begin{equation}
  \label{eq:mixed_sign_unweighted_jaccard_conv_decomposition}
\jaccard^{\pm} (\graph_1, \graph_2) = \cvxcoeff{+}   \jaccard^+ (\graph_1, \graph_2) + \cvxcoeff{-} \jaccard^- (\graph_1, \graph_2) \,,
\end{equation}
with the coefficients of the convex combination above being
\begin{equation}
  \label{eq:mixed_sign_unweighted_jaccard_conv_coefficients}
  \cvxcoeff{+} = \frac{|\edgeset_1^+ \cup \edgeset_2^+| }{|\edgeset_1^+ \cup \edgeset_2^+| +|\edgeset_1^- \cup \edgeset_2^-|} \,, \quad
\cvxcoeff{-} = \frac{|\edgeset_1^- \cup \edgeset_2^-|}{|\edgeset_1^+ \cup \edgeset_2^+| +|\edgeset_1^- \cup \edgeset_2^-|} \,.  
\end{equation}

\subsubsection{Mixed-Sign Weighted Jaccard Similarity}
\label{sec:mixed-sign-weighted-3}

Completely analogously the ``mixed sign weighted Jaccard similarity'' of two networks $\graph_1$, $\graph_2$ with mixed-sign edge weights is defined to equal
\begin{equation}
  \label{eq:mixed_sign_jaccard_weighted_def}
  \begin{split}
   (\jaccard^W)^{\pm}(\graph_1, \graph_2) := & \jaccard^W (\graph_1^+ \oplus   
\graph_1^-, \graph_2^+ \oplus \graph_2^-) \\
=& \frac{
\displaystyle
\sum_{i,j} \min \{ \edgefunc^+_1(i,j), \edgefunc^+_2(i,j)  \} 
+ 
\sum_{i,j} \min \{ \edgefunc^-_1(i,j), \edgefunc^-_2(i,j)  \} 
}{
\displaystyle
\sum_{i,j} \max \{ \edgefunc^+_1(i,j), \edgefunc^+_2(i,j)  \} 
+
\sum_{i,j} \max \{ \edgefunc^-_1(i,j), \edgefunc^-_2(i,j)  \} 
} \,.
\end{split}
\end{equation}
Like the unsigned version (cf. section \ref{sec:unsign-we-defin}), the mixed-sign weighted Jaccard similarity reduces to its unweighted counterpart when applied to skeletons:
\begin{equation}
  \label{eq:weighted_mixed_to_unweighted}
  (\jaccard^W)^{\pm} (\sskel(\graph_1), \sskel(\graph_2)) = \jaccard^{\pm}(\graph_1, \graph_2)\,.
\end{equation}
Like its unweighted counterpart, the mixed-sign weighted Jaccard similarity also enjoys the convex combination decomposition property:
\begin{equation}
  \label{eq:weighted_jaccard_conv_comb}
 (\jaccard^W)^{\pm}(\graph_1, \graph_2) = \cvxcoeff{+}(\jaccard^W)^+ (\graph_1, \graph_2)  + \cvxcoeff{-} (\jaccard^W)^- (\graph_1, \graph_2) \,,
\end{equation}
with the coefficients $\cvxcoeff{+}$, $\cvxcoeff{-}$ of the convex combination being
\begin{equation}
  \label{eq:weighted_jaccard_mixed_coeffs}
  \begin{split}
  \cvxcoeff{+} :=  &
\frac{
\displaystyle
\sum_{i,j} \max \{ \edgefunc^+_1(i,j), \edgefunc^+_2(i,j)  \} 
}{
\displaystyle
\sum_{i,j} \max \{ \edgefunc^+_1(i,j), \edgefunc^+_2(i,j)  \} 
+
\sum_{i,j} \max \{ \edgefunc^-_1(i,j), \edgefunc^-_2(i,j)  \} 
}
\,, 
\\
\cvxcoeff{-} := &
\frac{
\displaystyle
\sum_{i,j} \max \{ \edgefunc^-_1(i,j), \edgefunc^-_2(i,j)  \} 
}{
\displaystyle
\sum_{i,j} \max \{ \edgefunc^+_1(i,j), \edgefunc^+_2(i,j)  \} 
+
\sum_{i,j} \max \{ \edgefunc^-_1(i,j), \edgefunc^-_2(i,j)  \} 
} 
\,. 
  \end{split}
\end{equation}
Therefore, like its unweighted counterpart, mixed-sign weighted Jaccard similarity also satisfies the double penalization principle.

\subsubsection{Mixed-Sign False Discovery Rate and Mixed-Sign False Miss Rate}
\label{sec:mixed-sign-false}

Herein assume again that $\graph_1$ is an ``estimate'' $\hat{\graph}$, and $\graph_2$ is the ``ground truth'', and both have non-negative edge weights. Similar to the ``measure of goodness'' defined by (\ref{eq:unsigned_jaccard_binary_classification}), we can also define two corresponding ``measures of badness'':
\begin{itemize}
\item the \textbf{false discovery rate}:
\begin{equation}
  \label{eq:unsigned_FDR}
  \operatorname{FDR}(\hat{\graph}, \graph_*) := \frac{FH}{TH + FH} \,,
\end{equation}
\item and the \textbf{false miss rate}\footnote{More typically called ``false negative rate'', but that would be confusing in this context.}:
\begin{equation}
  \label{eq:unsigned_false_miss_rate}
  \operatorname{FMR}(\hat{\graph}, \graph_*) := \frac{FM}{TH + FM} \,.
\end{equation}
\end{itemize}
Note that, unlike the definition of Jaccard similarity, these definitions are not symmetric with respect to the roles of $\hat{\graph}$ and $\graph_*$.

Now allow $\hat{\graph}$ and $\graph_*$ to again have mixed-sign edge weights. Analogous to the relationship between definitions (\ref{eq:unsigned_jaccard_binary_classification}) and (\ref{eq:mixed_jaccard_binary_classification}), we can also define mixed-sign versions of the false discovery rate (\ref{eq:unsigned_FDR}) and the false miss rate (\ref{eq:unsigned_false_miss_rate}):
\begin{itemize}
\item the \textbf{mixed-sign false discovery rate}:
\begin{equation}
  \label{eq:mixed_sign_FDR}
  \operatorname{FDR}^{\pm}(\hat{\graph}, \graph_*) :=  \frac{ FH^+ + FH^-}{TH^+ + TH^- + FH^+ + FH^-  } \,,
\end{equation}
\item and the \textbf{mixed-sign false miss rate}:
\begin{equation}
  \label{eq:mixed_sign_false_miss_rate}
  \operatorname{FMR}^{\pm}(\hat{\graph}, \graph_*) := \frac{FM^+ + FM^-}{TH^+ + TH^- + FM^+ + FM^- } \,.
\end{equation}
\end{itemize}
Just like the mixed-sign Jaccard similarity, these can also be written straightforwardly as weighted averages of the corresponding single-sign performance metrics applied to the positive $\hat{\graph}^+, \graph_*^+$ and negative $\hat{\graph}^-, \graph_*^-$ subnetworks.

\subsection{Mixed-Sign DeltaCon Distance}
\label{sec:mixed-sign-deltacon}

Unweighted DeltaCon provides a qualitative comparison of networks by quantifying the correspondence of their paths. Weighted DeltaCon provides a comparison that is a little quantitative and a little qualitative, by increasing the importance of comparisons for paths where some edges have large edge weights. Herein mixed-sign versions of both unweighted and weighted DeltaCon are proposed which are not fooled by drastic changes to the network structure to which the unsigned versions from previous work are completely oblivious.

\subsubsection{Paths and Powers of Adjacency Matrices}
\label{sec:paths-powers-adjac}

Given an adjacency matrix (cf. section \ref{sec:graphs-funct-matr}), for any $n \ge 1$, the matrix $\adjacency^n$ gives information about the paths of length $n$ in the network. (The case where $n=1$ corresponds to paths of length $1$, i.e. the edges.) Specifically, given a pair $(\specie_1, \specie_2)$, the $(\specie_1, \specie_2)$'th entry ${[\adjacency^n]_{\specie_1, \specie_2}}$ of the $n$th power $\adjacency^n$ of the adjacency matrix $\adjacency$ equals the sum over all paths from $\specie_1$ to $\specie_2$ with length exactly $n$ of the product of all of the weights of all of the edges in each path.

For example, if there were two paths of length exactly $3$ from $\specie_1$ to $\specie_2$, and the weights of the edges in the first were $1$, $-4$, $2$, and the weights of the edges in the second were $-2$, $6$, $-5$, then the $(\specie_1, \specie_2)$'th entry $[\adjacency^3]_{\strain_1,\strain_2}$ of $\adjacency^3$ would be $(1)(-4)(2) + (-2)(6)(-5) = -8 + 60 = 52$.

\paragraph{Usefulness of this Formalism}
The usefulness of encoding information about paths this way is uncontroversial in the case of an adjacency matrix with only $1$'s and $0$'s. Then the entries count the number of paths of a given length in between any two nodes. 

In the case of an unsigned weighted graph the usefulness of this notion is perhaps more controversial. One might wonder if it makes more sense to take the sum of the weights along each path rather than the product. It is probably even more controversial in the case of signed and weighted graphs, since taking products over weights with different signs leads to difficult to predict ``fluctuations'' in the sign of the contribution from each path. 

In particular, one might question whether the effect on signs given by multiplying the edges in the path is scientifically meaningful in a given context, or whether (instead of paths potentially cancelling one another out) edges within a path that have opposite signs should be allowed to cancel each other out. 

\paragraph{Usefulness for Microbial Ecology}
I believe there is unlikely to be a universal answer. Instead, whether powers of adjacency matrices encode useful information about a mixed-sign network's paths likely depends on the context established by the specific scientific questions of interest. In this context I believe that powers of adjacency matrices do make sense.

For example, consider a situation where strain\footnote{Herein I use ``strains'' to refer equally to strains belonging to the same species(/genus/family/etc.) as well as to strains belonging to different species(/genera/families/etc.), because the distinction is irrelevant for setting up the abstract problem. It may matter for the implementation of a specific experiment.} $\specie_1$ strongly promotes the growth of strain $\strain_2$, and $\strain_2$ produces compounds toxic to strain $\strain_3$.  Cf. figure \ref{fig:network_example}, where $\strain_1$ could be the blue blob microbe, $\strain_2$ could be the pink spiral microbe, and $\strain_3$ could be the red rod microbe. I think it makes more sense for the resulting path from $\strain_1$ to $\strain_3$ to be given a strong negative weight, which corresponds only to the product of the weights of the individual edges.

If one used the sum of the edge weights instead, one would instead ``cancel'' the effects of $\strain_1$ on $\strain_2$ and of $\strain_2$ on $\strain_3$. That makes no sense, because large numbers of $\strain_1$ would promote large numbers of $\strain_2$, which would in turn produce large amounts of the compound toxic to $\strain_3$. It seems clear that the net indirect effect of $\strain_1$ on $\strain_3$ is strong and negative.

\paragraph{Neumann Series}
One might ask which length of path $n$ should be considered. A possible answer to this is ``all of them'', by using the Neumann series of the adjacency matrix $\adjacency$. For a general matrix $\mathbf{M}$, its Neumann series is
\begin{equation}
  \label{eq:neumann_series_defn}
 \mathbf{M} + \mathbf{M}^2 + \mathbf{M}^3 + \dots + \mathbf{M}^n + \dots = (\mathbf{I} - \mathbf{M})^{-1} \,. 
\end{equation}
This converges\footnote{\label{footnote:neumann_1}This fact is not obvious, but appears to be a ``folk theorem'' such that it is difficult to find a ``canonical'' reference for the proof.} if and only if the spectral radius of $\mathbf{M}$ is less than $1$. Therefore it makes sense to think of the Neumann series as the matrix analogue of the geometric series for (real or complex) numbers.

The spectral radius lower bounds the Frobenius norm and any operator norm\footnote{See footnote \ref{footnote:neumann_1}, because the same comments apply here too.}, so a sufficient (but not necessary) condition for the Neumann series of $\mathbf{M}$ to converge is if the value of one of these norms is less than $1$ for $\mathbf{M}$. Therefore in general convergence of the corresponding Neumann series can be accomplished by rescaling $\mathbf{M}$ by a positive constant.

\subsubsection{DeltaCon Distance Overview}
\label{sec:delt-dist-overv}

For both the unsigned and mixed-sign versions of DeltaCon distance, the general idea can be divided into two steps:
\begin{enumerate}
\item For each of the two networks, define a matrix which encodes the ``path structure'' of the network.
\item Define a distance between the two networks by evaluating a distance between the two matrices produced in the first step.
\end{enumerate}
For the first step we form the Neumann series $\mathbf{S}$ of a suitable matrix $\mathbf{W}$. For the second step, we evaluate (a modified version of) the Matusita (``root Euclidean'') distance of the two matrices $\mathbf{S}_1$, $\mathbf{S}_2$ constructed in the first step.

\subsubsection{DeltaCon Distance Definition}
\label{sec:delt-dist-defin}

The definition of DeltaCon does not use the Neumann series of (a rescaled version of) the adjacency matrix $\adjacency$, but instead uses the Neumann series of a matrix $\mathbf{W}$ based on $\adjacency$. The formula (\ref{eq:W_matrix_definition}) for $\mathbf{W}$ in terms of $\adjacency$ is derived from a linearized version of the belief propagation algorithm called FaBP\cite{FaBP}. The authors of \cite{DeltaCon} state that nevertheless the intuition for using the Neumann series of the matrix $\mathbf{W}$ is still the same as the intuition described above for using the Neumann series of the adjacency matrix $\adjacency$. The scaling constant $\varepsilon$ used below in the definition (\ref{eq:W_matrix_definition}) of this matrix $\mathbf{W}$ has the interpretation of encoding ``the influence between neighboring nodes''\cite{DeltaCon}.

Combining the exact definition\footnote{The definition of $\mathbf{W}$ from \cite{DeltaCon} slightly differs and is not the original definition.} from \cite{FaBP} with the notation from \cite{DeltaCon}, I use the following for the definition of $\mathbf{W}$:
\begin{equation}
  \label{eq:W_matrix_definition}
\mathbf{W} := \frac{1}{1-\varepsilon^2}(\varepsilon \textbf{A} - \varepsilon^2 \textbf{D}   ) \,.  
\end{equation}
(The degree matrix $\mathbf{D}$ is a general notion defined in section \ref{sec:degree-matrix}.) The Neumann series $\mathbf{S}$ of $\mathbf{W}$, which encodes the ``path structure'' of the network, is called the ``similarity matrix'' in \cite{DeltaCon} because each entry can be interpreted as measuring the ``affinity'' of two nodes in terms of the paths between them.
\begin{equation}
  \label{eq:similarity_matrix_definition}
\mathbf{S} :=  (\mathbf{I} - \mathbf{W})^{-1} = \mathbf{I} + \mathbf{W} + \mathbf{W}^2 + \dots \,.  
\end{equation}
To ensure the definition of DeltaCon distance works we need to ensure the convergence of the Neumann series defining the similarity matrix $\mathbf{S}$. To guarantee the convergence of the Neumann series defining $\mathbf{S}$, we choose, for any given input $\adjacency$, a corresponding value for $\varepsilon$ in (\ref{eq:W_matrix_definition}) that guarantees that the spectral radius of the resulting $\mathbf{W}$ defined in terms of $\adjacency$ and $\varepsilon$ will be less than $1$.

The normalization implicit in the definition of $\mathbf{W}$ (\ref{eq:W_matrix_definition}) and of the similarity matrix (\ref{eq:similarity_matrix_definition}) may help DeltaCon distance to avoid judging two highly different networks as being similar merely because both are sparse. Cf. again section \ref{sec:sparsity-should-not}, as well as the comments at the end of section \ref{sec:relative_error_definition}. I am uncertain whether this is really true in practice, but it does seem plausible.

\paragraph{Unsigned Version} The definition given in \cite{DeltaCon} was intended to work for unsigned but weighted networks. Without being modified, it does not work for general networks with mixed-sign edge weights. I describe in detail the proposals made in \cite{DeltaCon} for (1) the value of $\varepsilon$ in the definition (\ref{eq:W_matrix_definition}) of $\mathbf{W}$, and (2) the distance to apply to the similarity matrices $\mathbf{S_1}$ and $\mathbf{S}_2$. Along with using the original definition of $\mathbf{W}$ from \cite{FaBP}, these are the two aspects of the definition from \cite{DeltaCon} that I changed in order to guarantee the new definition would work for all networks with mixed-sign edge weights.

\textbf{Step 1:} The authors of \cite{DeltaCon} propose using the following for $\varepsilon$:
\begin{equation}
  \label{eq:deltacon_paper_epsilon}
  \varepsilon := \frac{1}{1 + \max_i [\mathbf{D}]_{ii}} \,.
\end{equation}
However, they don't include an argument showing why this value (ostensibly) works. This is notable because the previous paper\cite{FaBP} (i) only discussed unweighted (and unsigned) networks, (ii) used a different version of $\mathbf{W}$ (corresponding to (\ref{eq:W_matrix_definition}) above) than the definition used in \cite{DeltaCon}, and (iii) proposed using a slightly different value for $\varepsilon$ (for which a proof was provided).

\textbf{Step 2:} Given two adjacency matrices $\adjacency_1$ and $\adjacency_2$, and defining their corresponding similarity matrices $\mathbf{S}_1$ and $\mathbf{S}_2$ as in equation (\ref{eq:similarity_matrix_definition}), \cite{DeltaCon} defines the (unsigned) DeltaCon distance to be the Matusita distance of $\mathbf{S}_1$ and $\mathbf{S}_2$:
\begin{equation}
  \label{eq:matusita_distance_definition}
\deltacon(\graph_1, \graph_2) := \sqrt{\sum_{i,j} (\sqrt{[\mathbf{S}_1]_{ij}} -\sqrt{[\mathbf{S_2}]_{ij}})^2 } \,.   
\end{equation}
The authors of \cite{DeltaCon} suggest several reasons for using Matusita distance instead of e.g. Frobenius distance, but one important reason they mention is ``boosting'' small entries in $[0,1]$ (given that $\sqrt{x} \ge x$ for all $x \in [0,1]$).

\paragraph{Mixed-Sign Version}
 To modify the definition from \cite{DeltaCon} to apply to networks with mixed-sign edge weights, I made the following two changes.

\textbf{Step 1:} I showed that a sufficient condition for a value of $\varepsilon$ to work with any network with mixed-sign edge weights is for $\varepsilon < \frac{1}{1 + \norm{\adjacency}_{\infty}}$ (see Lemma \ref{lem:deltacon} from section \ref{sec:deltacon}). I also showed that this bound is sharp (see Lemma \ref{lem:deltacon_sharp} from section \ref{sec:deltacon}), in the sense that there exist adjacency matrices $\adjacency$ of networks with mixed-sign edge weights for which choosing $\varepsilon := \frac{1}{1 + \norm{\adjacency}_{\infty}}$ leads to a $\mathbf{W}$ with spectral radius exactly equal to $1$. While such problematic adjacency matrices do not appear to be generic, for safety I chose to use instead:
\begin{equation}
  \label{eq:mixed_sign_deltacon_epsilon}
\varepsilon := \frac{1}{2 + \norm{\adjacency}_{\infty}} \,.  
\end{equation}
In practice, even when $\varepsilon := \frac{1}{1 + \norm{\adjacency}_{\infty}}$ technically works for a given network, if the spectral radius of the resulting $\mathbf{W}$ is still very close to $1$, the DeltaCon distance appears to be poorly behaved, e.g. by producing unreasonably large values (data not shown). Using the more conservative definition (\ref{eq:mixed_sign_deltacon_epsilon}) of $\varepsilon$, the resulting $\mathbf{W}$ matrices had spectral radii bounded further away from $1$, making the resulting values of the DeltaCon distance more stable than they were before.

\textbf{Step 2:} In equation (\ref{eq:matusita_distance_definition}), I replaced $\sqrt{[\mathbf{S}]_{ij}}$ with $\sign \left({[\mathbf{S}]_{ij}} \right) \cdot \sqrt{ | [\mathbf{S}]_{ij} |}$:
\begin{equation}
  \label{eq:signed_matusita_distance_definition}
\deltacon^{\pm} (\graph_1, \graph_2) := \sqrt{\sum_{i,j} \left( \sign([\mathbf{S}_1]_{ij}) \cdot \sqrt{|[\mathbf{S}_1]_{ij}|} -\sign([\mathbf{S}_2]_{ij}) \cdot \sqrt{|[\mathbf{S_2}]_{ij}|} \right)^2 } \,.     
\end{equation}
 The choice to use $\sign \left({[\mathbf{S}]_{ij}} \right) \cdot \sqrt{ | [\mathbf{S}]_{ij} |}$ instead of $\sqrt{[\mathbf{S}]_{ij}}$ appears to be necessary for at least two reasons. First, it ensures that the distance is defined (by not assuming that the entries of $\mathbf{S}$ are necessarily non-negative in the more general case of mixed-sign edge weights). Second, it also guarantees that the double penalization principle (cf. section \ref{sec:double-penal-princ}) is satisfied. At the same time, the new definition (\ref{eq:signed_matusita_distance_definition}) also retains the same ``boosting'' property that helped motivate definition (\ref{eq:matusita_distance_definition}), because $\sqrt{|x|} \ge |x|$ for all $x \in [-1,1]$. So nothing important appears to be lost because of the change, as reflected perhaps in how (\ref{eq:signed_matusita_distance_definition}) reduces to (\ref{eq:matusita_distance_definition}) in the case that the entries of $\mathbf{S}_1$ and $\mathbf{S}_2$ are non-negative.

\section{Methods}
\label{sec:network_comparison_methods}

$1,000$ random networks were generated (section \ref{sec:gener-rand-netw}) and classified according to which sign had the larger number of edges (section \ref{sec:domin-subm-sign}). Three attacks were applied to each of the networks (section \ref{sec:definition-attacks}). The values of the mixed-sign network comparison methods between the original networks and their attacked versions were compared with the corresponding values for variants which do not use the entirety of the mixed-sign network structure or which do not obey the double penalization principle (section \ref{sec:definitions-variants}). Violin plots of the results were generated using Matplotlib \cite{Matplotlib} version 3.4.1 and Seaborn \cite{Seaborn} version 0.11.1. Complete implementation details can be found in the code at \networksgitrepo. See \url{\networksgitrepourl}.

\subsection{Generating Random Networks}
\label{sec:gener-rand-netw}

$1,000$ random networks were independently generated as follows:
\begin{itemize}
\item A hyperparameter $p \in (0,1)$ was selected from the uniform distribution.
\item $100^2$ random entries of an adjacency matrix were independently generated according to a $\operatorname{Beta}(2p, 2(1-p))$ distribution. (Considered as a Dirichlet distribution, this has the same concentration parameter as the uniform distribution.)
\item $0.5$ was subtracted from all entries, and then all entries were multiplied by $2$, with the effect that their support changed from $(0,1)$ to $(-1,1)$.
\item Each of the entries in the adjacency matrix was set to zero independently and with probability $\frac{1}{2}$ (this means the random networks are Erdos-Renyi).
\end{itemize}

\subsection{Predominant and Non-Predominant Sign Edges}
\label{sec:domin-subm-sign}

For each network, if a majority of the nonzero entries were positive, $+$ was the ``predominant sign'' and $-$ the ``non-predominant sign''. If a majority of the nonzero entries were negative, $-$ was the ``predominant sign'' and $+$ the ``non-predominant sign''. Results looking at ``predominant sign edges'' consider only the subnetworks (using the definitions from section \ref{sec:posit-negat-parts}) corresponding to the ``predominant sign'' for any given networks. Similarly for results looking at ``non-predominant sign edges''. Results were split according to the ``predominant sign'' and ``non-predominant sign'' for each network, rather than always grouping the positive subnetworks together and the negative subnetworks together, because the corresponding distributions usually depended only on whether the subnetwork corresponded to the numerically predominant sign for that network, and not to what the particular sign was.

\subsection{Definitions of the Attacks}
\label{sec:definition-attacks}

Three types of attack were applied to each of the random networks. The corresponding values of similiarity/dissimilarity were then computed between each original random network and their attacked counterparts. 

The three types were as follows.

\subsubsection{Shift Attack}
\label{sec:shift-attack}

This attack affected both the signs and magnitudes of the edges of the random network. The effect on the signs was the same as the sign flip attack. First, the predominant sign of the network was identified. If the predominant sign was positive, then twice the maximum value of any edge in the network was subtracted from all nonzero edges, forcing all edges to be negative (the non-predominant sign). If the predominant sign was negative, then twice the absolute value of the minimum value of any edge in the network was added to all nonzero edges, forcing all edges to be positive (the non-predominant sign). In both cases the values of all nonzero edges were shifted by a constant (whence the name), leaving their relative ordering unaffected.

\subsubsection{Magnitude Swap Attack}
\label{sec:magn-swap-attack}

This attack affected the magnitudes of the edges of the random network, but not the signs. The relative ordering of all magnitudes of all nonzero edges was computed. Then the edge with the largest magnitude had its magnitude swapped with the magnitude of the edge with the smallest magnitude, the edge with the second largest magnitude had its magnitude swapped with the magnitude of the edge with the second smallest magnitude, and so on. For all $n$ (less than or equal to the number of nonzero edges), the edge with the $n$th largest magnitude had its magnitude swapped with the magnitude of the edge with the $n$th smallest magnitude.

\subsubsection{Sign Flip Attack}
\label{sec:sign-flip-attack}

This attack affected the signs of the edges of the random network, but not the magnitudes. First, the predominant sign of the network was identified. Then all of the edges corresponding to the predominant sign had their sign flipped (i.e. multiplied by $-1$), while all edges corresponding to the non-predominant sign were unchanged.

\subsection{Definitions of Variants}
\label{sec:definitions-variants}

Mixed-sign network comparison methods were compared against related comparison methods that only considered subsets of the features of the networks. Section \ref{sec:relative-error} explains the definitions of the variants used for (entrywise $L_1$) relative error. Section \ref{sec:spearman-correlation} explains the definitions of the variants used for Spearman correlation. Section \ref{sec:jaccard-similarity} explains the definitions of the variants used for Jaccard similarity. Section \ref{sec:deltacon-distance} explains the definitions of the variants used for DeltaCon distance.

\subsubsection{Relative Error}
\label{sec:relative-error}

First I compared the values of relative error with those values resulting from considering (i) only the magnitudes of the original networks or (ii) only the signs of the original networks. I then compared the distribution of relative error values for the whole network with the corresponding values for the subnetworks corresponding to (i) the predominant sign edges only or (ii) the non-predominant sign edges only.

Whenever computing the relative error of two (sub)networks, the magnitudes of the edges were first normalized by the value of the entrywise $L_1$ norm of the (sub)networks. Cf. section \ref{sec:normalizing-get-more}. This served two purposes. First, it ensured that the distribution of values was confined to $[0,2]$ and thus made the results easier to visualize. Second, it corresponds to what it was done in preliminary work, where only the relative but not absolute sizes of estimates are of interest. The convex combination relationship mentioned in section \ref{sec:conv-comb-decomp-1} is therefore no longer directly relevant. Thus the most accurate interpretation of results is somewhat obscured.

``Magnitudes only'' refers to the relative error between the ``magnitude skeletons'' $\magskel(\graph_1)$, $\magskel(\graph_2)$ of the original networks $\graph_1, \graph_2$. ``Signs only'' refers to the relative error between the ``signed skeletons'' $\sskel(\graph_1), \sskel(\graph_2)$ of the original networks $\graph_1, \graph_2$. (See section \ref{sec:skeletons} for skeleton definitions.)

\subsubsection{Spearman Correlation}
\label{sec:spearman-correlation}

First I compared the values of mixed-sign Spearman correlation with those values resulting from (i) using the raw Spearman correlation, considering (ii) only the magnitudes of the original networks, or (iii) only the signs of the original networks. I then compared the distribution of mixed-sign Spearman correlation values for the whole network with the corresponding (raw) Spearman correlation values for the subnetworks corresponding to (i) the predominant sign edges only or (ii) the non-predominant sign edges only.

``Raw'' Spearman correlation refers to the sparsity-adjusted Spearman correlation of the (unaltered) original networks $\graph_1, \graph_2$. ``Magnitudes only'' refers to the sparsity-adjusted Spearman correlation of the ``magnitude skeletons'' $\magskel(\graph_1), \magskel(\graph_2)$ of the original networks $\graph_1, \graph_2$. Similarly ``signs only'' refers to the sparsity-adjusted Spearman correlation of the ``signed skeletons'' $\sskel(\graph_1), \sskel(\graph_2)$ of the original networks $\graph_1, \graph_2$. (See section \ref{sec:skeletons} for skeleton definitions.) 

As mentioned already in section \ref{sec:definition-1}, only looking at the subset of edges which are nonzero in at least one of the networks (i.e. not considering edges missing from both networks) avoids considering two very different networks as similar due only to them both being highly sparse.

Also note that when one of the compared vectors of edge values was constant (and thus its rank vector had $0$ variance), the Spearman correlation herein is defined by convention to be $0$ (since the ranks from one vector have zero predictive value for predicting the ranks of the other vector).

\subsubsection{Jaccard Similarity}
\label{sec:jaccard-similarity}

First I compared the values of mixed-sign unweighted Jaccard similarity and mixed-sign weighted Jaccard similarity with the values resulting from considering only (i) the presence/absence of edges or (ii) the magnitudes of edges. I then compared the mixed-sign unweighted Jaccard similarity values with the Jaccard similarity values for the subnetworks corresponding to (i) the predominant sign edges only or (ii) the non-predominant sign edges only, and then analogously also for the mixed-sign weighted Jaccard similarity.

``Presence/Absence Only'' refers to the weighted Jaccard similarity of the ``unsigned skeletons'' $\uskel(\graph_1), {\uskel(\graph_2)}$ of the original networks $\graph_1, \graph_2$ (or equivalently the unweighted Jaccard similarity of the ``magnitude skeletons'' $\magskel(\graph_1), {\magskel(\graph_2)}$ of the original networks $\graph_1, \graph_2$). ``Magnitudes only'' refers to the weighted Jaccard similarity of the ``magnitude skeletons'' ${\magskel(\graph_1)}, {\magskel(\graph_2)}$ of the original networks $\graph_1, \graph_2$. (See section \ref{sec:skeletons} for skeleton definitions.) 

Variants called ``Signs Only'' for relative error or Spearman correlation effectively correspond to the mixed-sign unweighted Jaccard similarity (which again is equivalent to the mixed-sign weighted Jaccard similarity of the ``signed skeletons'' of the original networks). More variants are considered than for relative error or Spearman correlation because the unweighted versions of those (their ``Signs only'' variants) are not as interesting as they are for Jaccard similarity.

\subsubsection{DeltaCon Distance}
\label{sec:deltacon-distance}

First I compared the values of mixed-sign unweighted DeltaCon distance and mixed-sign weighted DeltaCon distance with the values resulting from considering only (i) the presence/absence of edges or (ii) the magnitudes of edges. I then compared the mixed-sign unweighted DeltaCon distance values with the DeltaCon distance values for the subnetworks corresponding to (i) the predominant sign edges only or (ii) the non-predominant sign edges only, and then analogously also for the mixed-sign weighted DeltaCon distance.

``Presence/Absence Only'' refers to the DeltaCon distance of the ``unsigned skeletons'' ${\uskel(\graph_1)}, {\uskel(\graph_2)}$ of the original networks $\graph_1, \graph_2$, analogous to Jaccard similarity. ``Magnitudes Only'' refers to the DeltaCon distance of the ``magnitude skeletons'' ${\magskel(\graph_1)}, {\magskel(\graph_2)}$ of the original networks $\graph_1, \graph_2$. Variants called ``Signs Only'' elsewhere effectively correspond to the mixed-sign \textit{unweighted} DeltaCon distance, which by definition equals the mixed-sign weighted DeltaCon distance of the ``signed skeletons'' ${\sskel(\graph_1)}, {\sskel(\graph_2)}$ of the original networks $\graph_1, \graph_2$. (See section \ref{sec:skeletons} for skeleton definitions.)

\section{Results}
\label{sec:network_comparison_results}

Section \ref{sec:relative-error-1} discusses the behavior of variants of relative error for all three attacks. Section \ref{sec:spearman-correlation-1} discusses the behavior of variants of Spearman correlation for all three attacks. Section \ref{sec:jaccard-similarity-1} discusses the behavior of variants of Jaccard similarity for all three attacks. Section \ref{sec:deltacon-distance-1} discusses the behavior of variants of DeltaCon distance for all three attacks.

\subsection{Relative Error}
\label{sec:relative-error-1}

Section \ref{sec:sensitivity-attacks} overviews the results, explaining which variants were sensitive to which attacks. Section \ref{sec:sign-inform-magn} explains the insight gleaned from comparing with the signs only and magnitudes only variants. Section \ref{sec:potent-usef-signs} explains why and when the signs only variant may or may not be useful. Section \ref{sec:effects-conv-comb} explains the indirect evidence for the convex combination property seen in the results. (Note that nevertheless the convex combination property is proven and doesn't actually require empirical evidence to be substantiated.)

\subsubsection{Sensitivity to Attacks}
\label{sec:sensitivity-attacks}

For all three attacks shown in figures \ref{fig:relative_error_shift}, \ref{fig:relative_error_mag}, \ref{fig:relative_error_sign}, we see that the relative error is always sensitive to the major changes that occur. In contrast, the relative error of the signed skeletons of the networks is completely oblivious to the magnitude swap attack in figure \ref{fig:relative_error_mag}, while the relative errors of the magnitude skeletons of the networks is completely oblivious to the sign flip attack in figure \ref{fig:relative_error_sign}. 

{ %

  \newcommand{\figcaption}{Responses of relative error variants to adversarial attacks.}
  \newcommand{\figlabel}{fig:relative_error}
  
\begin{figure}
  \centering
  \begin{subfigure}{\textwidth}
  \centering
  \includegraphics[width=\textwidth,height=0.47\textheight,keepaspectratio]{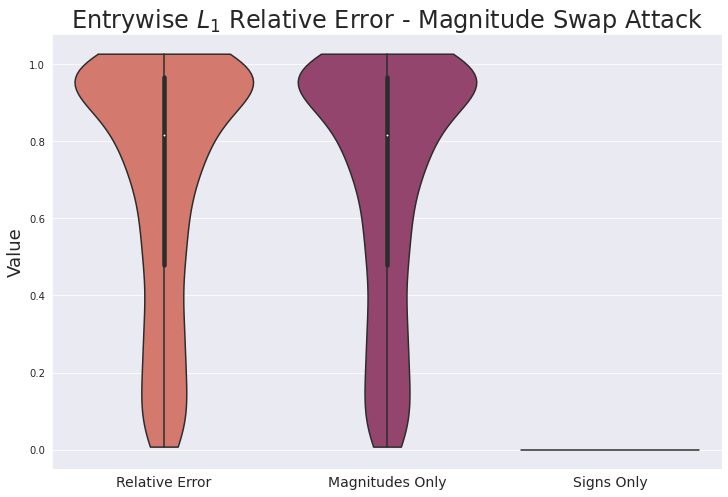}
  \caption[]{}
  \label{fig:relative_error_mag}
\end{subfigure}

\caption[]{\figcaption}
\label{\figlabel}

\end{figure}

\begin{figure}[p]
  \ContinuedFloat
  \centering
  
  \begin{subfigure}{\textwidth}
  \centering
  \includegraphics[width=\textwidth,height=0.45\textheight,keepaspectratio]{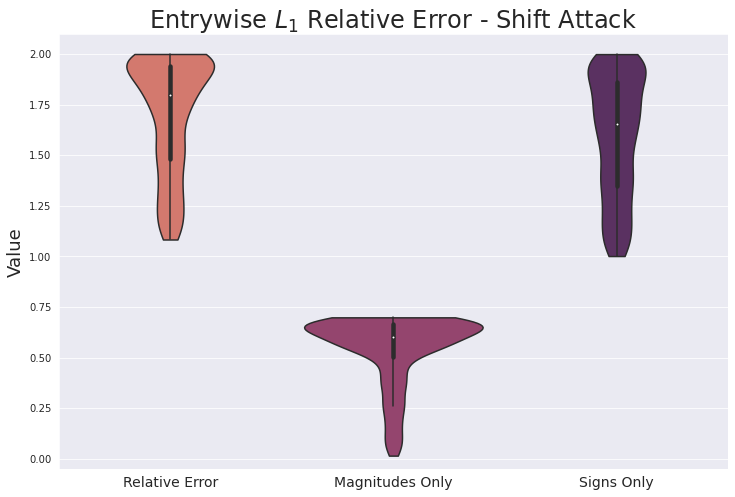}
  \caption[]{}
  \label{fig:relative_error_shift}
\end{subfigure}

\begin{subfigure}{\textwidth}
  \centering
  \includegraphics[width=\textwidth,height=0.45\textheight,keepaspectratio]{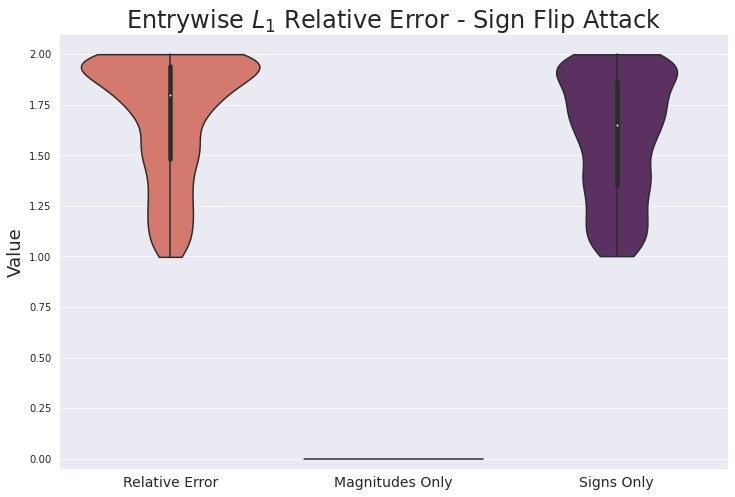}
  \caption[]{}
  \label{fig:relative_error_sign}
\end{subfigure}

\caption{\figcaption}
\label{\figlabel}
\end{figure}

} %

Moreover, even for the sign flip attack in figure \ref{fig:relative_error_sign} when only signs are changed, the wider flare at the top of the distribution for relative error than for ``signs only'' seems to indicate greater sensitivity to the attack. Similarly for the shift attack in figure \ref{fig:relative_error_shift}, the wider flare at the top of the distribution for relative error compared to signs only suggests that relative error is able to combine both sign and magnitude information to be more sensitive to attacks than considering either magnitudes only or signs only. Despite the changes in magnitude the accompany the shift attack, in figure \ref{fig:relative_error_shift} we see that considering magnitudes only still leads to substantially less sensitivity. 

\subsubsection{Sign Information and Magnitude Information}
\label{sec:sign-inform-magn}

Observe also how in figures \ref{fig:relative_error_shift}, \ref{fig:relative_error_mag}, \ref{fig:relative_error_sign}, the results for the relative error are not even weakly intermediate between the relative errors when considering either the magnitudes only or the signs only. In other words, the ``whole'' of using the relative error and considering both the sign and magnitude information in a mixed-sign network is ``greater than the sum of the parts'' of using either the sign information or the magnitude information alone.

\subsubsection{Potential Usefulness for Signs Only Variant}
\label{sec:potent-usef-signs}

Note that, as long as the attack does affect the signs (the shift attack in figure \ref{fig:relative_error_shift} and the sign flip attack in \ref{fig:relative_error_sign}), the signs only variant is actually fairly qualitatively similar to the relative error. This is perhaps to be expected at least inasmuch as the signs only variant does technically satisfy the double penalization principle as well. However this may also be largely an artifact of the magnitudes being bounded to be no greater than $1$ (while the ``magnitudes'' of the signed skeletons, on which the signs only variant acts, are $1$).

\subsubsection{Effects of Convex Combination Decomposition Property}
\label{sec:effects-conv-comb}

Subnetworks were normalized before computing the relative error (cf. sections \ref{sec:relative-error} and \ref{sec:normalizing-get-more}). This means the interpretation as a convex combination as in equation (\ref{eq:conv_comb_decomposition_relative_error}) does not apply. Nevertheless, we still do see in figures \ref{fig:relative_error_subn_shift}, \ref{fig:relative_error_subn_mag}, and \ref{fig:relative_error_subn_sign} results for the relative error which are weakly intermediate between the relative errors for the predominant sign edges and the non-predominant sign edges. Note that for the shift attack in figure \ref{fig:relative_error_subn_shift} and the sign flip attack in figure \ref{fig:relative_error_subn_sign}, the fact that the relative errors are all exactly $1$ is because the corresponding subnetwork has $0$ edges in the attacked versions of the networks.

{ %

  \newcommand{\figcaption}{Relative error combines values for positive and negative parts.}
  \newcommand{\figlabel}{fig:relative_error_subn}
  
\begin{figure}
  \centering
  \begin{subfigure}{\textwidth}
  \centering
  \includegraphics[width=\textwidth,height=0.47\textheight,keepaspectratio]{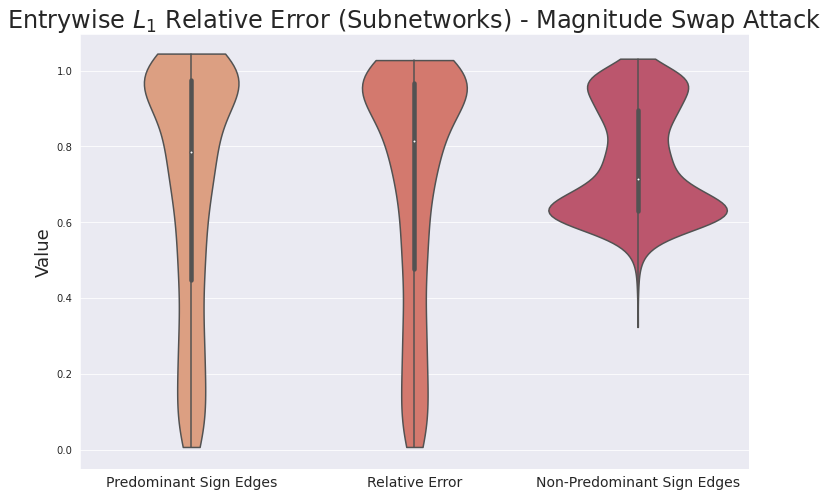}
  \caption[]{}
  \label{fig:relative_error_subn_mag}
\end{subfigure}

\caption[]{\figcaption}
\label{\figlabel}

\end{figure}

\begin{figure}[p]
    \ContinuedFloat
    \centering
    
  \begin{subfigure}{\textwidth}
  \centering
  \includegraphics[width=\textwidth,height=0.45\textheight,keepaspectratio]{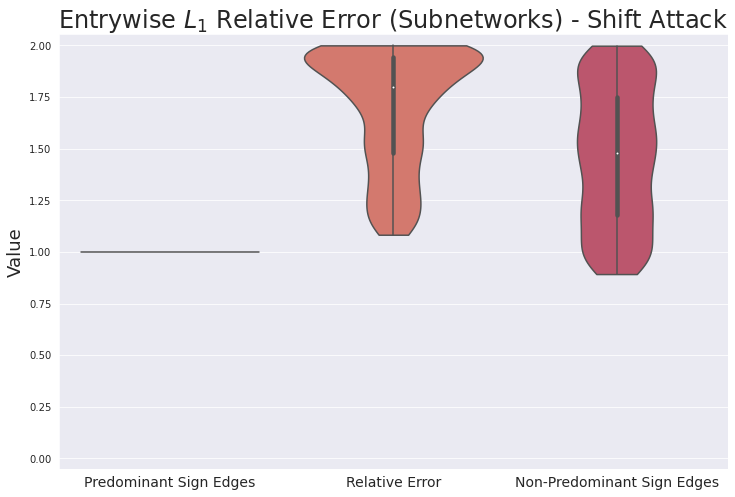}
  \caption[]{}
  \label{fig:relative_error_subn_shift}
\end{subfigure}

\begin{subfigure}{\textwidth}
  \centering
  \includegraphics[width=\textwidth,height=0.45\textheight,keepaspectratio]{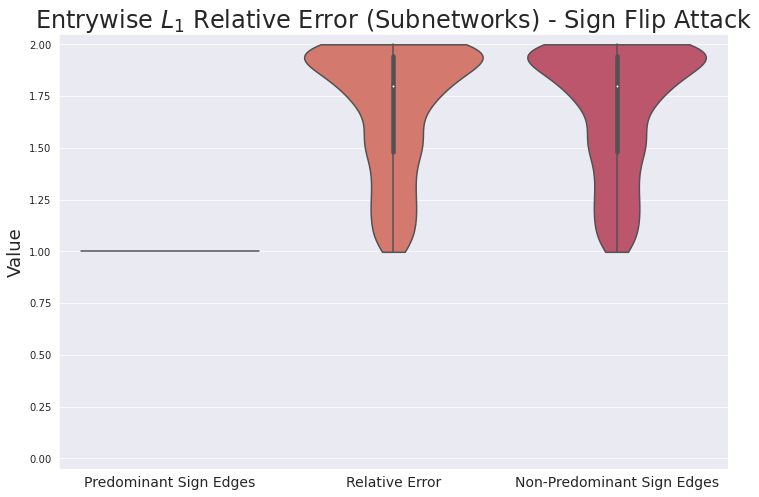}
  \caption[]{}
  \label{fig:relative_error_subn_sign}
\end{subfigure}

\caption{\figcaption}
\label{\figlabel}

\end{figure}

} %

\subsection{Spearman Correlation}
\label{sec:spearman-correlation-1}

Section \ref{sec:sensitivity-attacks-1} overviews which variants were sensitive to which attacks. Section \ref{sec:comb-inform} explains the evidence from the results that the mixed-sign Spearman not only directly combines the information from the positive and negative edges, but also achieves performance superior to that from directly combining the sign and magnitude information.

\subsubsection{Sensitivity to Attacks}
\label{sec:sensitivity-attacks-1}

For all three attacks shown in figures \ref{fig:spearman_shift}, \ref{fig:spearman_mag}, \ref{fig:spearman_sign}, we see that the mixed-sign Spearman correlation is always sensitive to the major changes that occur.

{ %

  \newcommand{\figcaption}{Responses of Spearman correlation variants to adversarial attacks.}
  \newcommand{\figlabel}{fig:spearman}

\begin{figure}
  \centering
  \begin{subfigure}{\textwidth}
  \centering
  \includegraphics[width=\textwidth,height=0.47\textheight,keepaspectratio]{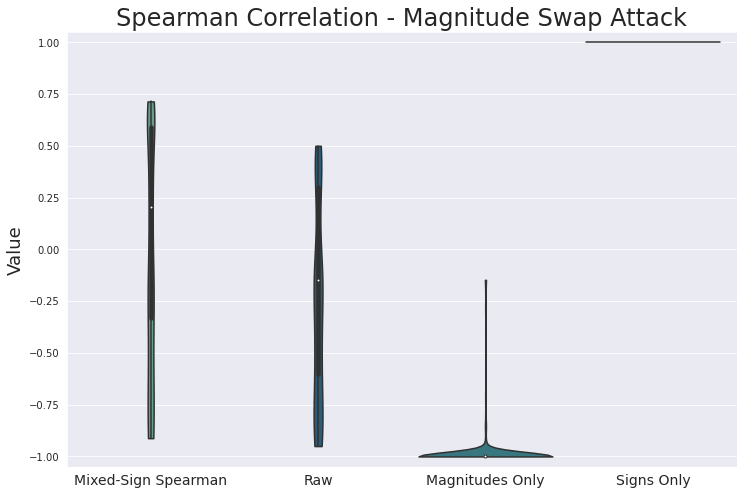}
  \caption[]{}
  \label{fig:spearman_mag}
\end{subfigure}

\caption[]{\figcaption}
\label{\figlabel}

\end{figure}

\begin{figure}[p]
  \ContinuedFloat
  \centering
  
  \begin{subfigure}{\textwidth}
  \centering
  \includegraphics[width=\textwidth,height=0.45\textheight,keepaspectratio]{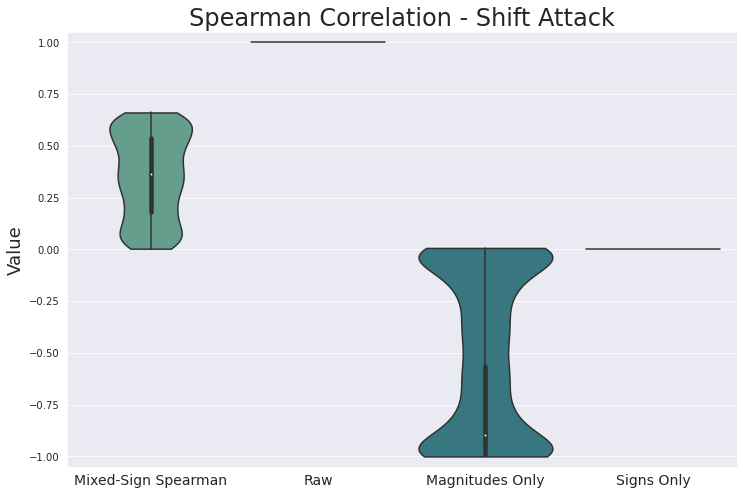}
  \caption[]{}
  \label{fig:spearman_shift}
\end{subfigure}

\begin{subfigure}{\textwidth}
  \centering
  \includegraphics[width=\textwidth,height=0.45\textheight,keepaspectratio]{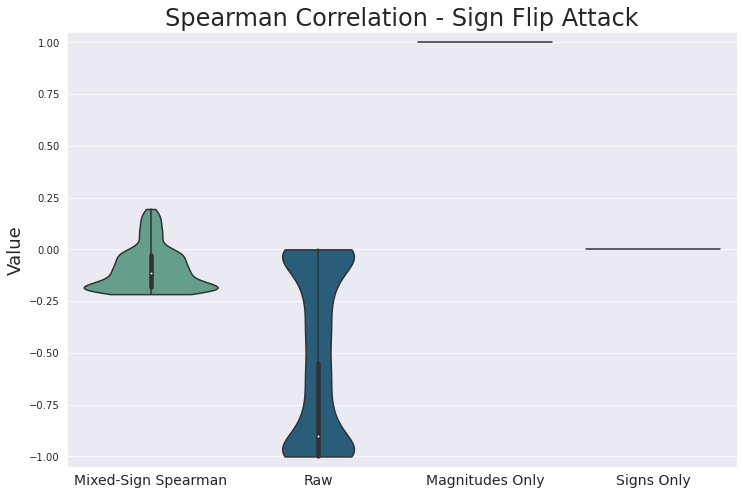}
  \caption[]{}
  
  \label{fig:spearman_sign}
\end{subfigure}

\caption{\figcaption}
\label{\figlabel}

\end{figure}

} %

For the shift attack, we see from figure \ref{fig:spearman_shift} that the raw Spearman correlation is completely oblivious to the attack, while the signs only and magnitudes only variants are unable to detect any similarity to the original underlying network. The mixed-sign Spearman correlation is the only method that is both sensitive to the attack and able to ascertain some similarity to the original underlying network.

 For the magnitude swap attack, we see from figure \ref{fig:spearman_mag} that the signs only variant is (unsurprisingly) completely oblivious, while the magnitudes only variant is (unsurprisingly) completely unable to detect any similarity to the original underlying network. Both the mixed-sign and raw Spearman correlations report an amount of similarity depending on the particular structure of the original network, although the mixed-sign Spearman seems better able to detect similarities and has a wider spread.

 For the sign flip attack, we see from figure \ref{fig:spearman_sign} that the magnitudes only variant is (unsurprisingly) completely oblivious, while the signs only is unable to provide any useful information (because the sign vector for the attacked network is constant). Meanwhile the raw Spearman correlation is never able to detect any similarities to the original underlying network (even though all of the magnitudes and non-predominant sign edges were left unaffected), but the mixed-sign Spearman correlation reports values depending on the particular structure of the original network, correctly reporting some amount of similarity in cases when the non-predominant sign edges constitute a large minority. 

\subsubsection{Combining Information}
\label{sec:comb-inform}

The mixed-sign Spearman correlation is the only variant that is not oblivious to any of the attacks, and which always reports values depending on the particular structure of the underlying network. We also see from figures \ref{fig:spearman_subn_shift}, \ref{fig:spearman_subn_mag}, and \ref{fig:spearman_subn_sign} that the mixed-sign Spearman reports values intermediate between those for the predominant sign and non-predominant sign subnetworks, corresponding to the subconvex combination property mentioned in section \ref{sec:conv-comb-decomp-2} and demonstrating double penalization. The mixed-sign Spearman directly combines information from the positive and negative edges. In contrast, the mixed-sign Spearman can be seen to \textit{not} be intermediate between the magnitudes only and signs only variants in figures \ref{fig:spearman_shift}, \ref{fig:spearman_mag}, \ref{fig:spearman_sign}, demonstrating that the ``whole'' of the magnitude and sign information is greater than the ``sum of the parts'' of the magnitudes only and signs only information.

{ %

    \newcommand{\figcaption}{Spearman correlation combines values for positive and negative parts.}
  \newcommand{\figlabel}{fig:spearman_subn}

\begin{figure}
  \centering
  \begin{subfigure}{\textwidth}
  \centering
  \includegraphics[width=\textwidth,height=0.47\textheight,keepaspectratio]{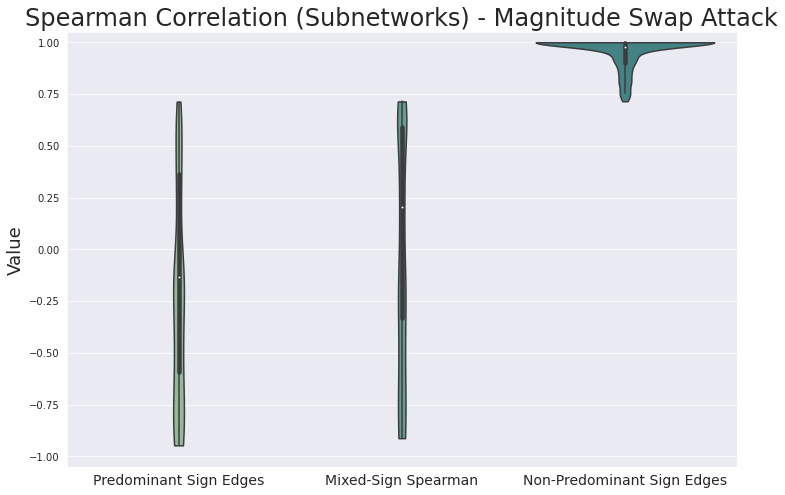}
  \caption[]{}
  \label{fig:spearman_subn_mag}
\end{subfigure}

\caption[]{\figcaption}
\label{\figlabel}

\end{figure}

\begin{figure}[p]
    \ContinuedFloat
    \centering
    
  \begin{subfigure}{\textwidth}
  \centering
  \includegraphics[width=\textwidth,height=0.45\textheight,keepaspectratio]{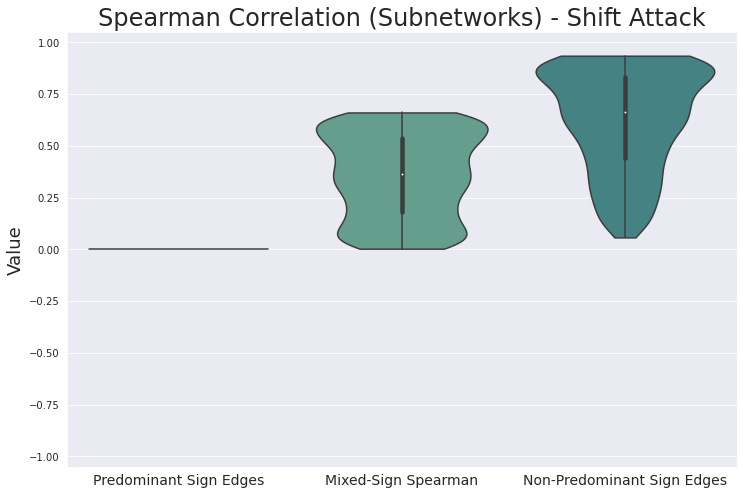}
  \caption[]{}
  \label{fig:spearman_subn_shift}
\end{subfigure}

\begin{subfigure}{\textwidth}
  \centering
  \includegraphics[width=\textwidth,height=0.45\textheight,keepaspectratio]{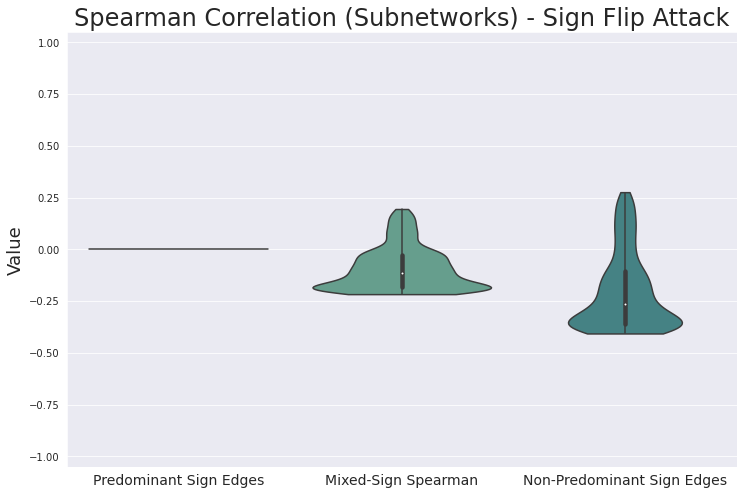}
  \caption[]{}
  \label{fig:spearman_subn_sign}
\end{subfigure}

\caption{\figcaption}
\label{\figlabel}

\end{figure}

} %

\subsection{Jaccard Similarity}
\label{sec:jaccard-similarity-1}

Section \ref{sec:pres-only-always} explains how the standard notion of Jaccard similarity is completely useless for all three attacks. Section \ref{sec:mixed-sign-unwe-1} explains how the results appear to indicate greater importance for sensitivity to sign information than for sensitivity to magnitude information. Section \ref{sec:comb-inform-1} explains how the mixed-sign weighted Jaccard similarity is more sensitive than would follow from directly combining the sign and magnitude information. Section \ref{sec:effects-conv-comb-1} explains how the convex combination property satisfied by both the mixed-sign unweighted and mixed-sign weighted Jaccard similarity manifests itself in the results.

\subsubsection{Presence/Absence Only is Always Completely Oblivious}
\label{sec:pres-only-always}

For all three attacks shown in figures \ref{fig:jaccard_shift}, \ref{fig:jaccard_mag}, and \ref{fig:jaccard_sign}, we see that the presence/absence only variant is completely oblivious (and thus completely useless). This is important, since using Jaccard similarity of presence/absence of edges might have been the default choice of some researchers. We see from this that failing to respect the richer structure of networks with mixed-sign edge weights, and attempting to blindly apply standard methods applicable for less structured networks, can lead to drastically misleading conclusions.

{ %

  \newcommand{\figcaption}{Responses of Jaccard similarity variants to adversarial attacks.}
  \newcommand{\figlabel}{fig:jaccard}

\begin{figure}
  \centering
  
  \begin{subfigure}{\textwidth}
  \centering \includegraphics[width=\textwidth,height=0.47\textheight,keepaspectratio]{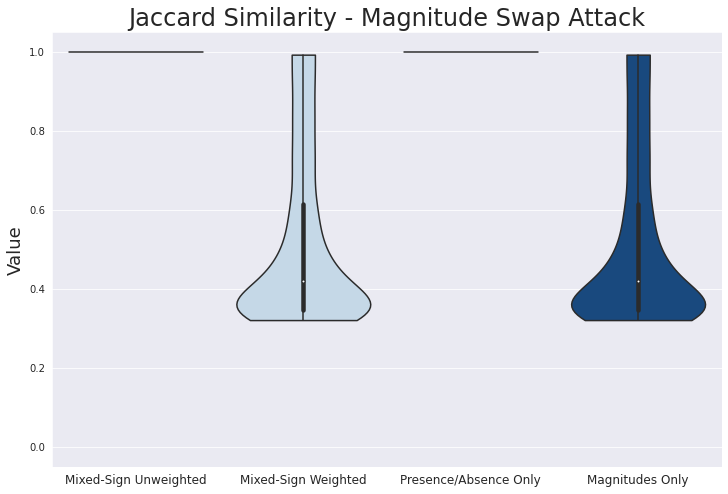}
  \caption[]{}
  \label{fig:jaccard_mag}
\end{subfigure}

\caption[]{\figcaption}
\label{\figlabel}
\end{figure}

\begin{figure}[p]
    \ContinuedFloat
    \centering
    
  \begin{subfigure}{\textwidth}
  \centering
  \includegraphics[width=\textwidth,height=0.45\textheight,keepaspectratio]{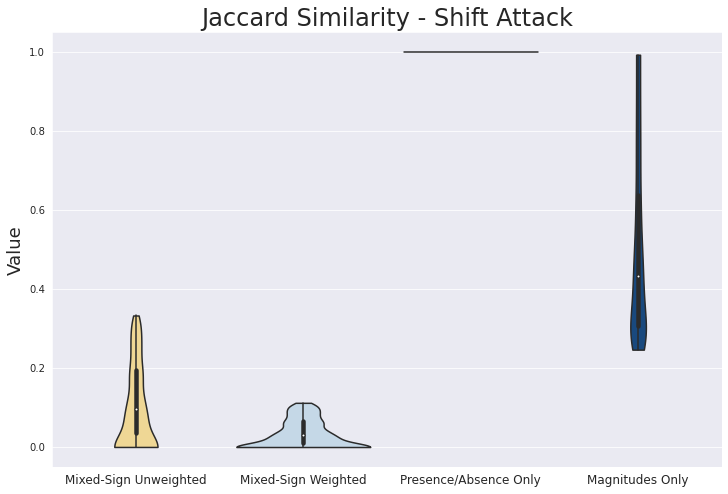}
  \caption[]{}
  \label{fig:jaccard_shift}
\end{subfigure}

\begin{subfigure}{\textwidth}
  \centering
  \includegraphics[width=\textwidth,height=0.45\textheight,keepaspectratio]{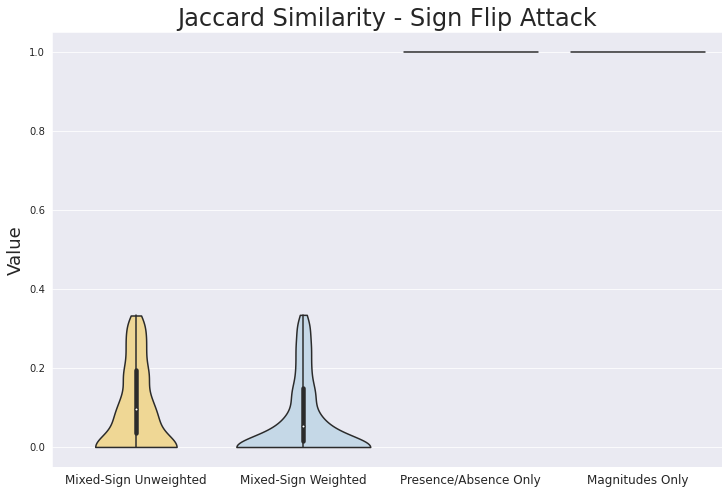}
  \caption[]{}
  \label{fig:jaccard_sign}
\end{subfigure}

\caption{\figcaption}
\label{\figlabel}

\end{figure}
} %

\subsubsection{Sign Information Appears More Important than Magnitude Information}
\label{sec:mixed-sign-unwe-1}

From figure \ref{fig:jaccard_sign}, we see that the magnitudes only variant of Jaccard similarity is completely oblivious to the sign flip attack, as expected since the attack does not change the sign of any edges. For the two attacks shown in figures \ref{fig:jaccard_shift} and \ref{fig:jaccard_mag} which do affect the magnitudes of the edges, we do see some sensitivity on the part of the magnitudes only variant. Moreover, since the magnitudes are the only aspect of the network that changed for the magnitude swap attack, it makes sense that we see in figure \ref{fig:jaccard_mag} that the magnitudes only variant and the mixed-sign weighted Jaccard similarity give the same results, while the mixed-sign unweighted Jaccard similarity is completely oblivious. 

However, the magnitudes only variant is surprisingly insensitive to the shift attack as shown in figure \ref{fig:jaccard_shift}. While it is not necessarily surprising that the mixed-sign weighted Jaccard similarity is more sensitive, it is even the case that the lower range of values of the magnitudes only variant overlaps with the upper range of values of the mixed-sign unweighted variant. This is probably because, even though the mixed-sign unweighted Jaccard similarity technically only looks at the signs of the edges, it nevertheless does obey the double penalization principle in doing so. In contrast the magnitudes only variant does not obey the double penalization principle and largely has ``the wool pulled over its eyes'' when it comes to the shift attack, even though it is not completely oblivious. 

This seems to suggest that, although signs only variants will never be able to detect attacks that only affect magnitudes, they can still be substantially useful in general as long as they obey the double penalization principle. Since the double penalization principle requires sensitivity to signs, but not necessarily to magnitudes, this seems to suggest that sensitivity to signs is more important to sensitivity to magnitudes, \textit{as long as the double penalization principle is satisfied}. This is important to note because blind application of default methods for unsigned networks will generally\footnote{Entrywise $L_1$ relative error is of course a notable exception.} be completely insensitive to signs, much less obey the double penalization principle.
 
\subsubsection{Combining Sign and Magnitude Information}
\label{sec:comb-inform-1}

We see in figure \ref{fig:jaccard_sign} that the distribution of values for the mixed-sign unweighted and mixed-sign weighted Jaccard similarities are most similar for the sign flip attack. This makes sense intuitively because that is the only attack that leaves the magnitudes unaffected. However the wider flare at the bottom of the distribution of the mixed-sign weighted Jaccard similarities in figure \ref{fig:jaccard_sign} suggests that it is even more sensitive to this attack than the mixed-sign unweighted Jaccard similarity, in spite of the attack leaving magnitudes unaffected. This would seem to follow because the distribution of magnitudes is not constant for any of the networks. Thus again we see an instance where the ``whole'' of being sensitive to both sign and magnitude information is greater than the ``sum of the parts'' of the magnitude and sign information, even in instances where seemingly only one of those two types of information would appear to be directly relevant.

\subsubsection{Effects of Convex Combination Decomposition Property}
\label{sec:effects-conv-comb-1}

Figures \ref{fig:u_jaccard_shift}, \ref{fig:u_jaccard_mag}, and \ref{fig:u_jaccard_sign} correctly suggest how the mixed-sign unweighted Jaccard similarity is a convex combination of the unweighted Jaccard similarities for the positive and negative subnetworks. 

{ %

    \newcommand{\figcaption}{Unweighted Jaccard similarity combines values for positive and negative parts.}
  \newcommand{\figlabel}{fig:u_jaccard}

\begin{figure}
  \centering
  \begin{subfigure}{\textwidth}
  \centering
  \includegraphics[width=\textwidth,height=0.47\textheight,keepaspectratio]{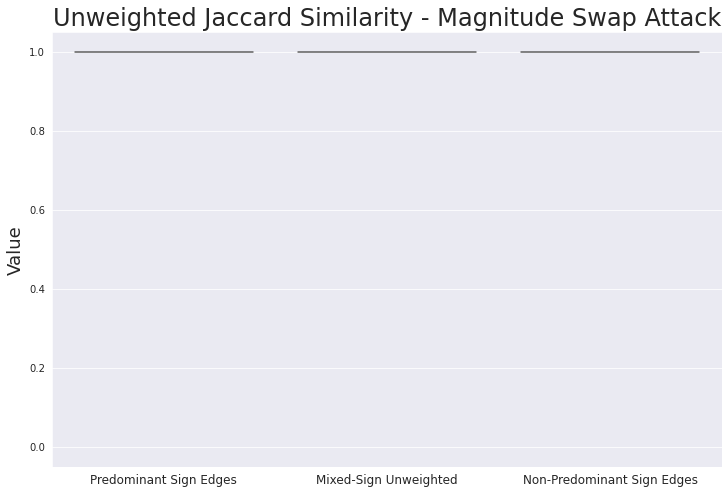}
  \caption[]{}
  \label{fig:u_jaccard_mag}
\end{subfigure}

\caption[]{\figcaption}
\label{\figlabel}

\end{figure}

\begin{figure}[p]
  \ContinuedFloat
  \centering
  
  \begin{subfigure}{\textwidth}
  \centering
  \includegraphics[width=\textwidth,height=0.45\textheight,keepaspectratio]{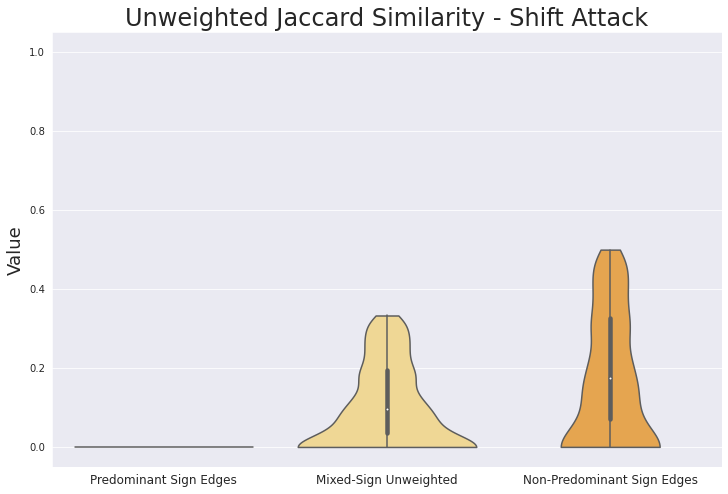}
  \caption[]{}
  \label{fig:u_jaccard_shift}
\end{subfigure}

\begin{subfigure}{\textwidth}
  \centering
  \includegraphics[width=\textwidth,height=0.45\textheight,keepaspectratio]{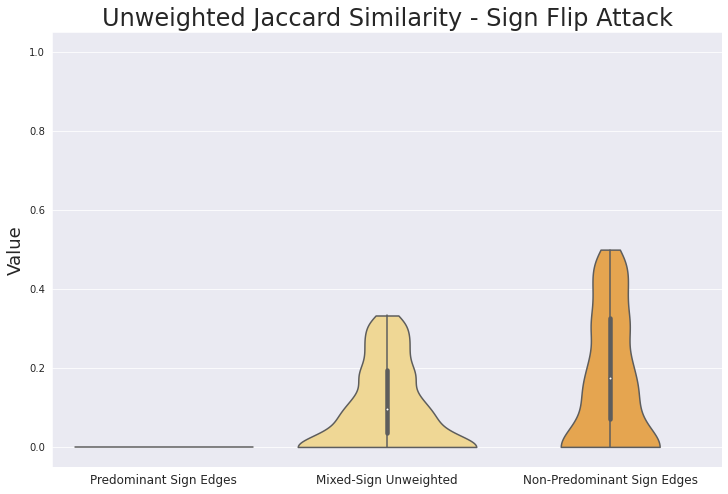}
  \caption[]{}
  \label{fig:u_jaccard_sign}
\end{subfigure}

\caption{\figcaption}
\label{\figlabel}

\end{figure}

} %

Similarly, figures \ref{fig:w_jaccard_shift}, \ref{fig:w_jaccard_mag}, and \ref{fig:w_jaccard_sign} also correctly suggest how the mixed-sign weighted Jaccard similarity is a convex combination of the weighted Jaccard similarities for the positive and negative subnetworks. 

{ %

   \newcommand{\figcaption}{Weighted Jaccard similarity combines values for positive and negative parts.}
  \newcommand{\figlabel}{fig:w_jaccard}

\begin{figure}
  \centering
  \begin{subfigure}{\textwidth}
  \centering
  \includegraphics[width=\textwidth,height=0.47\textheight,keepaspectratio]{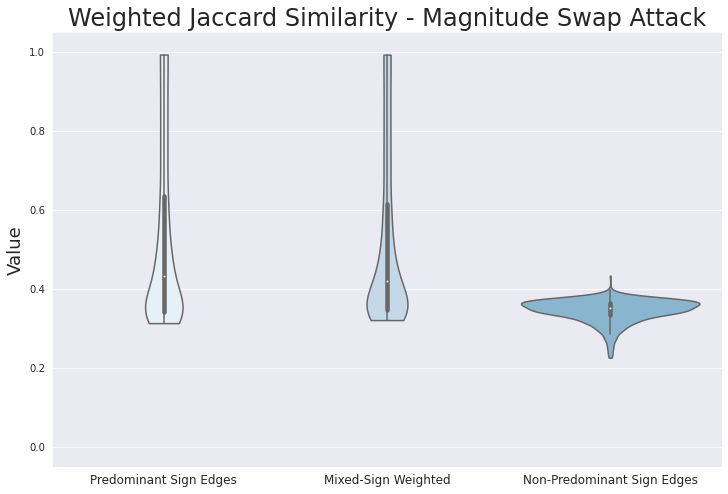}
  \caption[]{}
  \label{fig:w_jaccard_mag}
\end{subfigure}

\caption[]{\figcaption}
\label{\figlabel}

\end{figure}

\begin{figure}[p]
    \ContinuedFloat
    \centering
    
  \begin{subfigure}{\textwidth}
  \centering
  \includegraphics[width=\textwidth,height=0.45\textheight,keepaspectratio]{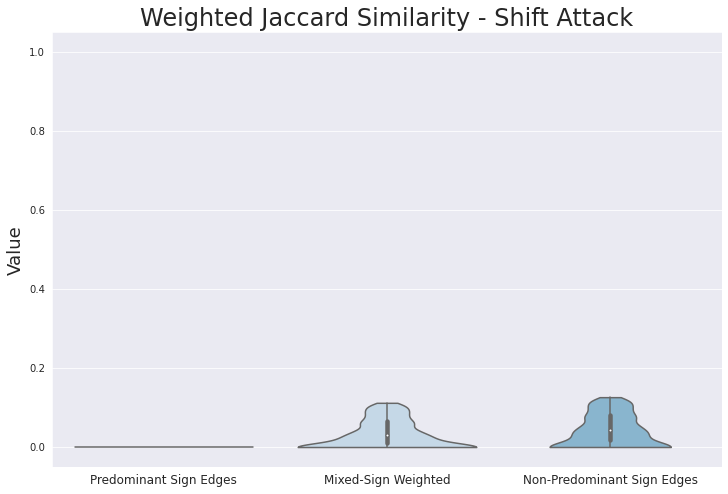}
  \caption[]{}
  \label{fig:w_jaccard_shift}
\end{subfigure}

\begin{subfigure}{\textwidth}
  \centering
  \includegraphics[width=\textwidth,height=0.45\textheight,keepaspectratio]{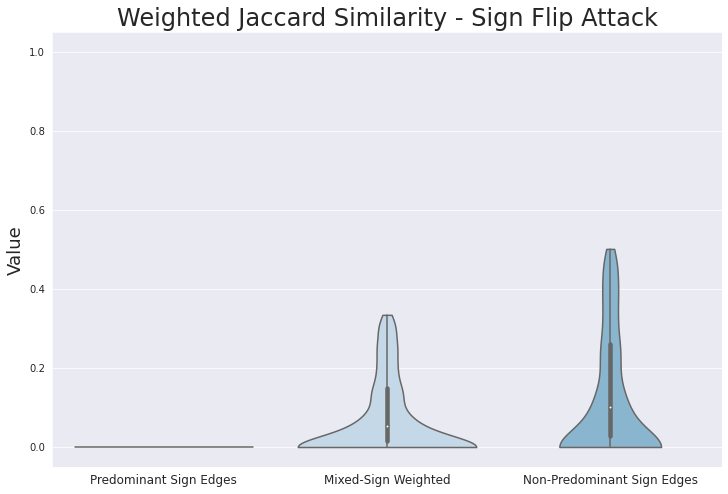}
  \caption[]{}
  \label{fig:w_jaccard_sign}
\end{subfigure}

\caption{\figcaption}
\label{\figlabel}

\end{figure}

} %

Thus both mixed-sign versions of Jaccard similarity incorporate information from both subnetworks in a reasonable fashion.

\subsection{DeltaCon Distance}
\label{sec:deltacon-distance-1}

Only the mixed-sign weighted DeltaCon distance is sensitive to all three attacks, with mixed-sign unweighted and magnitudes only both completely oblivious to at least one attack, and presence/absence only completely oblivious to all three attacks. Mixed-sign weighted is always more sensitive than magnitudes only. Mixed-sign weighted appears slightly less sensitive than mixed-sign unweighted for these attacks, but that may be an artifact.

Section \ref{sec:pres-only-compl} explains the importance of presence/absence only being completely oblivious. Section \ref{sec:mixed-sign-weighted} explains an important caveat when comparing the results for mixed-sign unweighted and mixed-sign weighted DeltaCon distance. Section \ref{sec:mixed-sign-unwe} explains how mixed-sign unweighted may actually still be useful sometimes, despite disregarding information about edge magnitudes. Sections \ref{sec:sens-chang-sign}, \ref{sec:sneis-chang-magn}, and \ref{sec:mixed-sign-weighted-2} give ``sanity checks'' that demonstrate the results are sensible and in line with expectations. Section \ref{sec:mixed-sign-weighted-1} explains a subtle observation that argues strongly for the importance of the double penalization principle. Section \ref{sec:mixed-sign-variants} generalizes the observations from section \ref{sec:mixed-sign-weighted-1}.

\subsubsection{Presence/Absence Only is Always Completely Oblivious}
\label{sec:pres-only-compl}

Just as was the case for Jaccard similarity, the presence/absence only variant of DeltaCon distance can be seen from figures \ref{fig:deltacon_shift}, \ref{fig:deltacon_mag}, and \ref{fig:deltacon_sign} to be completely oblivious to all three attacks. This is again important because it shows that blindly seeking to directly transfer similar methods for less general networks to this context can lead to misleading conclusions. For example the proofs in the FaBP paper \cite{FaBP} seem to assume an unweighted network, as do some code implementations for DeltaCon that can be found. 

{ %

  \newcommand{\figcaption}{Responses of DeltaCon variants to adversarial attacks.}
  \newcommand{\figlabel}{fig:deltacon}
  
\begin{figure}
  \centering
  \begin{subfigure}{\textwidth}
  \centering  \includegraphics[width=\textwidth,height=0.47\textheight,keepaspectratio]{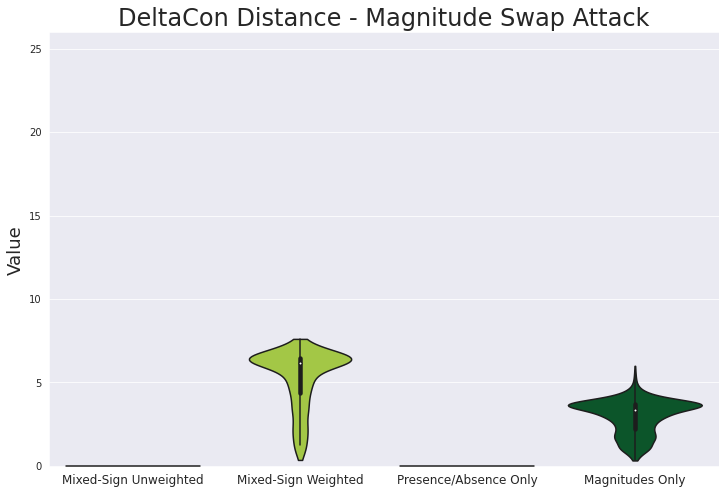}
  \caption[]{}
  \label{fig:deltacon_mag}
\end{subfigure}

\caption[]{\figcaption}
\label{\figlabel}
\end{figure}

\begin{figure}[p]
  \ContinuedFloat
  \centering
  
  \begin{subfigure}{\textwidth}
  \centering
  \includegraphics[width=\textwidth,height=0.45\textheight,keepaspectratio]{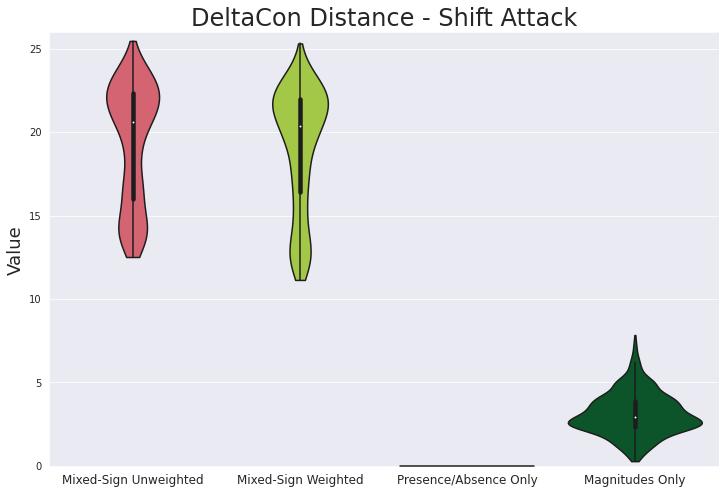}
  \caption[]{}
  \label{fig:deltacon_shift}
\end{subfigure}

\begin{subfigure}{\textwidth}
  \centering
  \includegraphics[width=\textwidth,height=0.45\textheight,keepaspectratio]{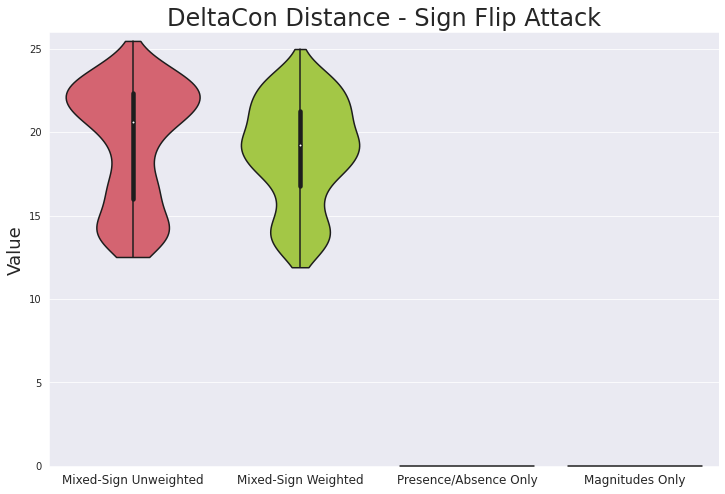}
  \caption[]{}
  \label{fig:deltacon_sign}
\end{subfigure}

\caption{\figcaption}
\label{\figlabel}
\end{figure}

} %

\subsubsection{Mixed-Sign Weighted May Actually Not Be Less Sensitive than Mixed-Sign Unweighted}
\label{sec:mixed-sign-weighted}

That the values of the mixed-sign unweighted DeltaCon distance trend slightly larger than those for the mixed-sign weighted DeltaCon distance in figures \ref{fig:deltacon_shift} and \ref{fig:deltacon_sign} appears to be an artifact of the magnitudes all being less than or equal to $1$. (Thus the ``magnitudes'' of the signed skeleton are in general greater, and mixed-sign unweighted DeltaCon is equivalent to mixed-sign weighted DeltaCon applied to the signed skeletons of the original networks.)

\subsubsection{Mixed-Sign Unweighted Performs Better than Magnitudes Only}
\label{sec:mixed-sign-unwe}

We again see from figure \ref{fig:deltacon_shift} that, as long as the double penalization principle is satisfied by the ``signs only'' variant, sensitivity to signs (mixed-sign unweighted DeltaCon) may be more important than sensitivity to magnitudes for attacks which affect both. The mixed-sign unweighted DeltaCon distance is more sensitive to the shift attack than the magnitudes only DeltaCon distance. One could argue that this is an artifact of the magnitudes being less than or equal to $1$, but that seems unlikely because the distribution of values for the mixed-sign weighted DeltaCon distance for the shift attack is much more similar to that of the mixed-sign unweighted DeltaCon distance than of the magnitudes only variant.

\subsubsection{Sensitivity to Changes in Sign Only}
\label{sec:sens-chang-sign}

Unsurprisingly only the variants that are sensitive to sign, namely the mixed-sign unweighted and mixed-sign weighted DeltaCon distances, can detect that sign flip attack from figure \ref{fig:deltacon_sign} at all. The distribution of values of the mixed-sign weighted DeltaCon distance may be more similar to that of the mixed-sign unweighted DeltaCon distance for the sign flip attack (figure \ref{fig:deltacon_sign}) than for the shift attack (figure \ref{fig:deltacon_shift}) since the latter also involves a change of magnitudes.

\subsubsection{Sensitivity to Changes in Magnitude Only}
\label{sec:sneis-chang-magn}

Unsurprisingly, only the mixed-sign weighted DeltaCon distance and the magnitudes only variant are sensitive to the magnitude swap attack from figure \ref{fig:deltacon_mag} (which only affects magnitude). Not only presence/absence only, but also mixed-sign unweighted DeltaCon, are completely oblivious.

\subsubsection{Mixed-Sign Weighted Picks up Changes in Magnitude}
\label{sec:mixed-sign-weighted-2}

Note that the distributions of values look the same between figures \ref{fig:u_deltacon_shift} and \ref{fig:u_deltacon_sign}, reflecting the fact that the shift and sign flip attacks have the same effect on the signs of the edges, and that the mixed-sign unweighted DeltaCon is only sensitive to changes in sign. 

{ %

    \newcommand{\figcaption}{Unweighted DeltaCon exceeds values for positive and negative parts.}
  \newcommand{\figlabel}{fig:u_deltacon}

\begin{figure}
  \centering
  \begin{subfigure}{\textwidth}
  \centering
  \includegraphics[width=\textwidth,height=0.47\textheight,keepaspectratio]{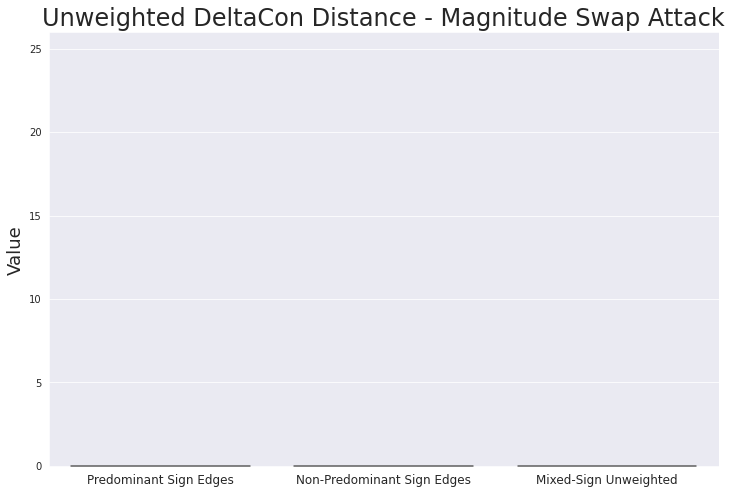}
  \caption[]{}
  \label{fig:u_deltacon_mag}
\end{subfigure}

\caption[]{\figcaption}
\label{\figlabel}

\end{figure}

\begin{figure}[p]
  \ContinuedFloat
  \centering
  
  \begin{subfigure}{\textwidth}
  \centering
  \includegraphics[width=\textwidth,height=0.45\textheight,keepaspectratio]{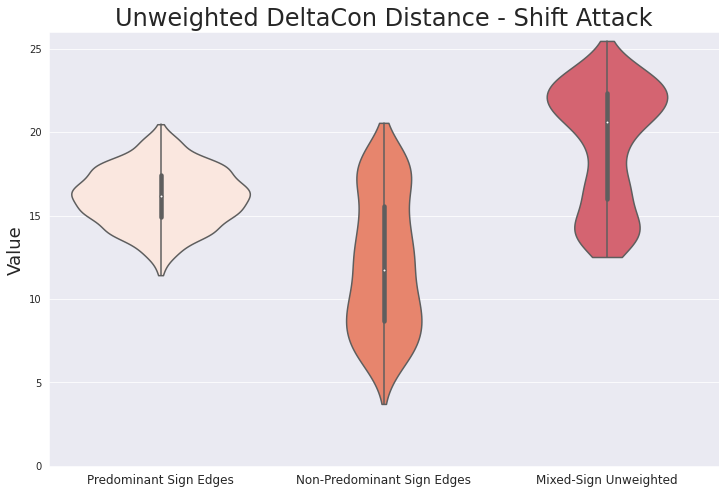}
  \caption[]{}
  \label{fig:u_deltacon_shift}
\end{subfigure}

\begin{subfigure}{\textwidth}
  \centering
  \includegraphics[width=\textwidth,height=0.45\textheight,keepaspectratio]{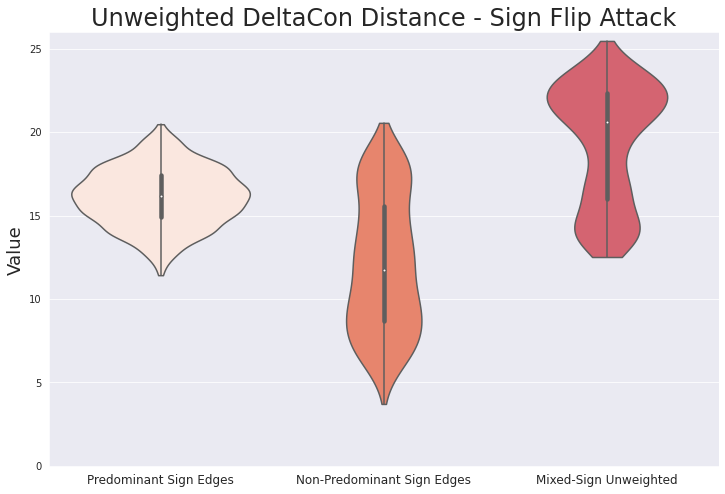}
  \caption[]{}
  \label{fig:u_deltacon_sign}
\end{subfigure}

\caption{\figcaption}
\label{\figlabel}

\end{figure}

} %

In contrast, the distributions of values look different between figures \ref{fig:w_deltacon_shift} and \ref{fig:w_deltacon_sign}, reflecting that the shift attack affects magnitudes (and the sign flip attack does not), and also demonstrating that the mixed-sign weighted version of DeltaCon is truly sensitive to changes in both sign and magnitude.

{ %

    \newcommand{\figcaption}{Weighted DeltaCon exceeds values for positive and negative parts.}
  \newcommand{\figlabel}{fig:w_deltacon}

\begin{figure}
  \centering
  \begin{subfigure}{\textwidth}
  \centering
  \includegraphics[width=\textwidth,height=0.47\textheight,keepaspectratio]{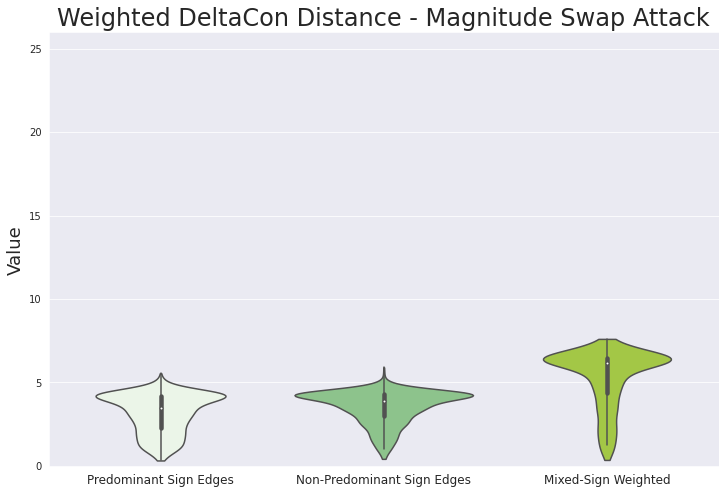}
  \caption[]{}
  \label{fig:w_deltacon_mag}
\end{subfigure}

\caption[]{\figcaption}
\label{\figlabel}

\end{figure}

\begin{figure}[p]
  \ContinuedFloat
  \centering
  
  \begin{subfigure}{\textwidth}
  \centering
  \includegraphics[width=\textwidth,height=0.45\textheight,keepaspectratio]{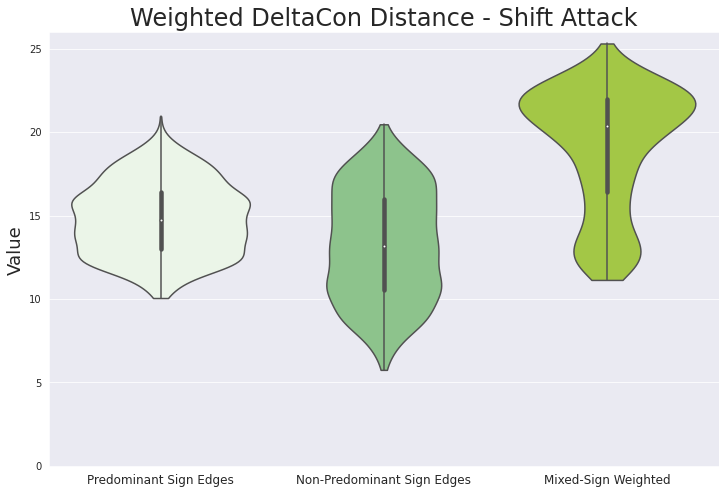}
  \caption[]{}
  \label{fig:w_deltacon_shift}
\end{subfigure}

\begin{subfigure}{\textwidth}
  \centering
  \includegraphics[width=\textwidth,height=0.45\textheight,keepaspectratio]{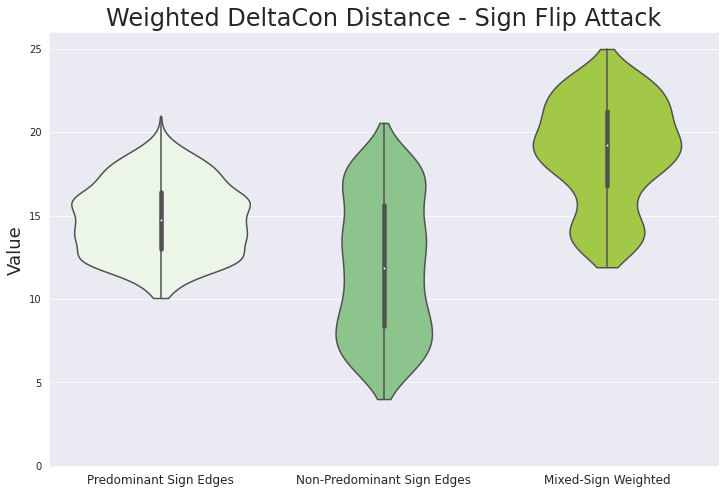}
  \caption[]{}
  \label{fig:w_deltacon_sign}
\end{subfigure}

\caption{\figcaption}
\label{\figlabel}

\end{figure}

} %

\subsubsection{Mixed-Sign Weighted is More Sensitive than Magnitudes Only even when Only Magnitudes Have Changed}
\label{sec:mixed-sign-weighted-1}

In spite of the fact that the attack only directly affects magnitudes, the mixed-sign DeltaCon (which also takes signs into account) appears to be more sensitive to the magnitude swap attack than the magnitudes only version. This may actually be a consequence of the double penalization principle being satisfied, since even though the signs agree for all edges when comparing the original network and its attacked counterpart, the signs may not necessarily agree when comparing the edges of the networks which correspond to interpreting their similarity matrices (\ref{eq:similarity_matrix_definition}) as adjacency matrices of a mixed-sign network. In any case, it is another instance of the ``whole'' of the magnitude and sign information being greater than the ``sum of the parts'' of the magnitude or sign information considered separately. That this occurs is nevertheless still relatively surprising when one notes how the analogous phenomenon does not occur for entrywise $L_1$ relative error (figure \ref{fig:relative_error_mag}) nor for Jaccard similarity (figure \ref{fig:jaccard_mag}). That discrepancy is probably further evidence for the explanation that the phenomenon is a result of DeltaCon's consideration of ``higher-order connectivities'' (i.e. graph motifs consisting of more than one edge).

\subsubsection{Mixed-Sign Variants Always More Sensitive}
\label{sec:mixed-sign-variants}

We see from figures \ref{fig:deltacon_shift}, \ref{fig:deltacon_mag}, \ref{fig:deltacon_sign} that the mixed-sign variants, weighted or unweighted, are more sensitive to all three attacks than considering presence/absence only or magnitudes only. (With the exception of the mixed-sign unweighted variant for the magnitude attack.) This seems to be evidence for double penalization occurring for DeltaCon as well. Figures \ref{fig:u_deltacon_shift} and \ref{fig:u_deltacon_sign} show that the mixed-sign unweighted version is more sensitive than the changes to either the predominant or non-predominant sign subnetworks alone when the attack changes sign, which is further evidence that double penalization is occurring. Figures \ref{fig:w_deltacon_shift}, \ref{fig:w_deltacon_mag}, \ref{fig:w_deltacon_sign} also show that the mixed-sign weighted DeltaCon is more sensitive to all of the attacks than either the predominant or non-predominant sign subnetworks. This is not only because the support of the distributions is higher -- their distributions are also skewed towards higher values, whereas the distributions of values for either subnetwork are more ``evenly'' distributed. This apparent effectiveness of the double penalization principle even for DeltaCon, which is affected by the problem that $(A^+ + A^-)^n \not= (A^+)^n + (A^-)^n$ for ``higher-order connectivities'' ($n \ge 2$), suggests that the double penalization principle is still valuable even when it doesn't make sense to define a mixed-sign (dis)similarity as a convex combination of (dis)similarities for the positive and negative parts of the networks.

\section{Discussion}
\label{sec:network_comparison_discussion}

There are potential limitations to consider when interpreting the general applicability of the results.

Sections \ref{sec:numb-invest-attacks}, \ref{sec:distr-rand-netw}, and \ref{sec:fairness-comparisons} discuss the potential limitations of the chosen simulations. Section \ref{sec:numb-invest-attacks} discusses the number and variety of attacks for which these methods were tested. Section \ref{sec:distr-rand-netw} discusses the statistical properties of the random networks on which these methods were tested. Section \ref{sec:fairness-comparisons} discusses the alternatives that these methods were tested against.

Sections \ref{sec:conv-comb-high}, \ref{sec:node-correspondence}, and \ref{sec:general-methods} discuss the general limitations of the scope of this paper. Section \ref{sec:conv-comb-high} discusses the potential limited applicability of the convex combination decomposition property. Section \ref{sec:node-correspondence} discusses which type of network comparison problems are addressed by this work. Section \ref{sec:general-methods} discusses the limited applicability of this work for extending network comparison methods not discussed in this paper.

\subsection{The Number of Investigated Attacks Was Small}
\label{sec:numb-invest-attacks}

These results only considered a limited number of attacks. The goal of this initial work was to consider attacks which did not affect the overall connectivity of the network and which affected limited aspects of the structure in obvious ways. One can think of many more changes to structure of mixed-sign networks for which we would want (dis)similarity measures to behave well. Future work can investigate whether double penalization leads to good responses for other kinds of attacks.

\subsection{Distribution of Random Networks}
\label{sec:distr-rand-netw}

The goal of this initial work was to consider a large number of networks with varying structure, such that their composite structure might be enough to encompass all ``typical'' networks. For example, always using a uniform distribution for the edge weights, rather than a Beta distribution with a varying hyperparameter, led to results (data not shown) that more reflected the particular nature of the uniform distribution rather than say anything about the network comparison methods themselves. For example, the spread of (dis)similarity values tended to be much smaller, and the effects of the magnitude swap attack were too predictably constrained. The chosen distribution led to mixed-sign networks with a wide range of properties.

Moreover the networks all had an expected sparsity of $50\%$. (Each possible edge was present or absent independently with probability $50\%$.) Future work might examine distributions with less uniform distributions of sparsity. Furthermore future work might want to choose magnitudes and signs more independently, and it might be worthwhile to look at magnitudes sampled from distributions which are not bounded (e.g. the exponential distribution). The Beta distributions have support limited to $(0,1)$ and we made an edge negative if its original magnitude fell in $(0,\frac{1}{2}$) and positive otherwise (see section \ref{sec:gener-rand-netw} for reference).

Also one could argue that I had mostly generated random sparse matrices rather than true ``random networks''. While this concern seems unfounded given the one-to-one correspondence between networks with mixed-sign edge weights and their adjacency matrices (which could correspond to any square matrix), it is nevertheless true that networks with ``realistic'' connectivity have adjacency matrices whose sparsity pattern is non-generic. (In other words, matrices with no $0$ entries are generic under typical random matrix distributions, which correspond to ``complete'' networks with every possible edge present.) I did use a distribution for which networks with many missing edges were generic, but still it could be argued that this did not correspond to ``state of the art'' or ``typical'' models for defining random network topologies. 

Having each edge be present or absent with probability $50\%$ does seem to correspond to an Erdos-Renyi model with parameter $p = \frac{1}{2}$, but that still leaves open the criticism of not having considered Erdos-Renyi for other values of $p$ (or with $p$ as a random hyperparameter), or other models of random network topology. The case where $p=\frac{1}{2}$ corresponds to choosing all possible network topologies (unsigned skeletons) with equal probability, but nevertheless ``most'' such topologies may still be considered unrealistic compared to ``typical real-world networks''. The (dis)similarity methods chosen all adjust implicitly for sparsity somewhat, so it seems unlikely that any bias in the network topologies affects the validity of the overall conclusions. Nevertheless it would still be a valuable issue for future work to investigate.

\subsection{Fairness of Comparisons}
\label{sec:fairness-comparisons}

One could argue that the shown results might make the proposed (dis)similarity measures look better unfairly because they were compared against ``strawmen'' alternatives. Nevertheless, with the exception of the raw spearman and the relative error, it seems most standard (dis)similarity measures that one would use for networks with less structure cannot be applied to networks with mixed-sign edge weights. Therefore to attempt to use such standard methods, it seems conceivable that many people in practice might disregard some of extra structure of networks with mixed-sign edge weights (e.g. by considering only the magnitude skeleton or the unsigned skeleton). So I believe it is fair to compare against such methods. Even to the extent such alternatives are unrealistic, comparing with them illustrates the understanding is lost when specific features of networks with mixed-sign edge weights are ignored, and thus can be justified from a ``pedagogical'' perspective. Future work might try to envision and make comparisons with more ``realistic'' or ``fair'' alternatives that don't obey the double penalization principle.

\subsection{Convex Combinations and Higher-Order Connectivity}
\label{sec:conv-comb-high}

A lot of emphasis was given to (dis)similarity measures that can be decomposed into a convex combination (weighted average) of their values for the positive edge subnetwork and the negative edge subnetwork. This emphasis may not be realistic, because such an approach does not seem to be generally applicable. (Dis)similarity measures which consider graph motifs involving two or more edges most likely require some reference to ``higher-order adjacency matrices'' $\mathbf{A}^n$ describing the number of paths between two nodes with $n$ edges. Although it is true that $\mathbf{A} = \mathbf{A}^+ - \mathbf{A}^-$, in general the ``freshman's dream'' is false, namely $\mathbf{A}^n = (\mathbf{A}^+ - \mathbf{A}^-)^n \not= (\mathbf{A}^+)^n - (\mathbf{A}^-)^n$ for $n \ge 2$. Because the analogue of the ``freshman's dream'' for $n=1$ is most often used to show the convex combination decomposition property (when it exists/is valid), I do not expect analogous properties to be applicable or available for general (dis)similarity measures considering graph motifs involving two or more edges. One could attempt to define reasonable convex combinations in terms of all $2^n$ terms of $(\mathbf{A}^+ - \mathbf{A}^-)^n$ when expanded, but such methods are unlikely to be computationally feasible (unless their computation reduces to something with far fewer terms).

Nevertheless, I do not believe that the failure of the convex combination paradigm to generalize to arbitrary (dis)similarity measures spells trouble for the Double Penalization Principle. Methods with the convex combination decomposition property are only a special case of methods satisfying the double penalization principle. The results for both unweighted and weighted DeltaCon appear to suggest that the double penalization principle remains useful even for (dis)similarity measures that consider ``higher-order connectivities''. Future work considering extensions to networks with mixed-sign edge weights of other (dis)similarity measures considering graph motifs with two or more edges would be worthwhile. For example, this work only considers extending the exact version of DeltaCon, but not also the approximate (and much more scalable) version of DeltaCon defined in \cite{DeltaCon}.

\subsection{Node Correspondence}
\label{sec:node-correspondence}

Using the dichotomy between ``known node correspondence'' and ``unknown node correspondence'' network comparison methods as defined in \cite{Tantardini2019}, as mentioned before this work only considers the known node correspondence problem for networks with mixed-sign edge weights. The extent to which the double penalization principle or other ideas from this work are applicable as well to the unknown node correspondence problem for networks with mixed-sign edge weights is unclear. Future work investigating whether we need to start ``from scratch'' for the unknown node correspondence problem when it comes to principles for extending methods to networks with mixed-sign edge weights would be highly valuable. It seems conceivable that the ideas applicable for the known node correspondence problem might still be applicable at least indirectly, but this still needs to be investigated.

\subsection{General Methods}
\label{sec:general-methods}

This work provides no ``algorithm'' or systematic procedure for extending arbitrary (dis)similarity measures to the case of networks with mixed-sign edge weights. The double penalization principle and the convex composition decomposition properties are merely ``desired properties'' or ``design specifications'' (the former ``essential'', the latter ``optional''). I provide concrete implementations of these principles for only a handful of network comparison methods, and perhaps not even necessarily the most important ones. It seems unlikely that such a general systematic procedure could exist (or at least not any reasonable one). For example, even amongst the three methods with the convex combination decomposition property, the formulas for the convex coefficients were different each time, following from ``just so'' arguments specific to each method. The absence of such a general systematic procedure can be considered to limit the usefulness of this work.

\section{Conclusion}
\label{sec:conclusion-8}

\paragraph{Findings and Contributions}

I discussed how comparisons of ecological interactions can be recast into the statistical framework of loss functions for signed networks. I explained how avoiding unexpected behavior requires loss functions for signed networks to satisfy what I call ``the double penalization principle''. Starting from loss functions of unsigned networks, I derived several specific and concrete examples of loss functions for signed networks that satisfy this property. I also identified a simple property (sufficient but not necessary), the convex combination decomposition property, which implies the double penalization principle.

\paragraph{Practical Implications}

The results show that naive attempts to use pre-existing methods for comparing unsigned networks can lead to useless or even actively misleading results when applied to signed networks. The extra structure of signed networks means that they need to be compared differently. Considering changes in sign alone, or changes in magnitude alone, are usually insufficient. Methods satisfying the double-penalization principle can avoid such misleading results. This work allows us to directly address the difficulties inherent in making meaningful comparisons between signed networks, rather than ignore those difficulties.

\paragraph{Next Steps and Open Questions}

Future work should further clarify the double penalization principle. For example, what are the most general contexts in which it makes sense? In which ways can it be generalized? How does it relate to other criteria we might want a loss function to satisfy? Also worth investigating would be to more directly investigate the statistical consequences of loss functions for signed networks either satisfying or not satisfying certain criteria. For example, if we say that a random sequence of signed networks asymptotically converges to a certain limit network if the value of the loss with respect to that network approaches $0$ in probability, then what properties must such a random sequence of signed networks have? The answer clearly depends on the properties of the chosen loss function. Finally, it should be desirable to connect these ideas to other potentially related areas. For example, to the extent that we can consider a random signed network as a random (sparse) matrix via the adjacency matrix representation, there should be meaningful connections to the already substantially developed field of random matrix theory. Likewise, thinking of networks as functions (e.g. as is done in sections \ref{sec:graphs-funct-matr} or \ref{sec:append-techn-deta}), at least ``morally'' one should expect to be able to apply or modify ideas from functional data analysis \cite{fda_review_1} \cite{Ramsay2005} \cite{Hsing2015} \cite{Kokoszka2021} about what are ``reasonable'' choices of loss functions.

\begin{coolsubappendices}

\section{Lemmas}
\label{sec:appendix:-lemmas}

The symbol $\mathbf{1}_I$ denotes the all ones vector $(1, \dots, 1)$ in $\mathbb{R}^I$, the notation $\mean{\rvec{x}}$ denotes the arithmetic mean (empirical expectation) of $\rvec{x} \in \mathbb{R}^I$, $\mean{\rvec{x}} := \displaystyle \frac{1}{I} \sum_{i=1}^I x^{(i)}$, and $\langle \rvec{x}, \rvec{y} \rangle$ denotes the standard inner product\footnote{Also known as the ``dot product''.} ${\rvec{x}, \rvec{y} \in \mathbb{R}^I}$, ${\displaystyle \langle \rvec{x}, \rvec{y} \rangle := \sum_{i=1}^I x^{(i)} y^{(i)}  }$.

\subsection{Mixed-Sign Spearman Correlation}
\label{sec:mixed-sign-spearman-1}

\begin{lemma}
\label{lem:projection_lemma}
Projection Lemma: when taking the (standard) inner product, ``mean-centering one vector is as good as mean-centering both vectors'':
\begin{equation}
  \label{eq:projection_lemma_statement}
  \begin{split}
\langle \rvec{x} - \mean{\rvec{x}} \cdot \mathbf{1}_I, \rvec{y} - \mean{\rvec{y}} \cdot \mathbf{1}_I \rangle
= &
 \langle \rvec{x} - \mean{\rvec{x}} \cdot \mathbf{1}_I, \rvec{y} \rangle   
\\
= &  \langle \rvec{x} , \rvec{y} - \mean{\rvec{y}} \cdot \mathbf{1}_I \rangle \,.
  \end{split}
\end{equation}
Here we assume $\rvec{x}, \rvec{y} \in \mathbb{R}^I$, and recall that $\mathbf{1}_I \in \mathbb{R}^I$ denotes the all $1$'s vector.
\end{lemma}

\textbf{Proof:} Because inner products are symmetric, 
\begin{equation}
  \label{eq:symmetric_inner_product}
  \langle \rvec{x} - \mean{\rvec{x}} \cdot \mathbf{1}_I, \rvec{y} - \mean{\rvec{y}} \cdot \mathbf{1}_I \rangle 
=
\langle \rvec{y} - \mean{\rvec{y}} \cdot \mathbf{1}_I,\rvec{x} - \mean{\rvec{x}} \cdot \mathbf{1}_I \rangle \,.
\end{equation}
Therefore it suffices to show only the first of the two claimed equalities, namely
\[{\langle \rvec{x} - \mean{\rvec{x}} \cdot \mathbf{1}_I, \rvec{y} - \mean{\rvec{y}} \cdot \mathbf{1}_I \rangle =  \langle \rvec{x} - \mean{\rvec{x}} \cdot \mathbf{1}_I, \rvec{y} \rangle} \,,\]
with the second equality following by applying symmetry (\ref{eq:symmetric_inner_product}), repeating the steps of the proof of the first equality, and then applying symmetry (\ref{eq:symmetric_inner_product2}):
\begin{equation}
  \label{eq:symmetric_inner_product2}
  \langle \rvec{y} - \mean{\rvec{y}} \cdot \mathbf{1}_I, \rvec{x} \rangle = \langle \rvec{x} , \rvec{y} - \mean{\rvec{y}} \cdot \mathbf{1}_I \rangle \,.
\end{equation}
Anyway the conclusion follows readily from the definitions and simple algebra, so providing a proof to this level of detail is probably an over-explanation making the result seem more confusing than it actually is. Nevertheless, at least for the sake of being thorough, below is a computation proving the first equality:
\begin{equation}
  \label{eq:projection_lemma_proof}
  \begin{split}
\left\langle \rvec{x} - \mean{\rvec{x}} \cdot \mathbf{1}_I, \rvec{y} - \mean{\rvec{y}} \cdot \mathbf{1}_I \right\rangle
= &
\langle \rvec{x} - \mean{\rvec{x}} \cdot \mathbf{1}_I, \rvec{y} \rangle
-
\mean{\rvec{y}} \cdot
\langle \rvec{x} - \mean{\rvec{x}} \cdot \mathbf{1}_I, \mathbf{1}_I \rangle \\
= &
\langle \rvec{x} - \mean{\rvec{x}} \cdot \mathbf{1}_I, \rvec{y} \rangle
- \mean{\rvec{y}} \cdot \left(\langle \rvec{x} , \mathbf{1}_I \rangle - \mean{\rvec{x}} \cdot \langle \mathbf{1}_I, \mathbf{1}_I \rangle  \right) \\
= &
\langle \rvec{x} - \mean{\rvec{x}} \cdot \mathbf{1}_I, \rvec{y} \rangle
- \mean{\rvec{y}} \cdot \left( \sum_{i=1}^I x^{(i)} - \mean{\rvec{x}} \cdot I  \right) \\
= &
\langle \rvec{x} - \mean{\rvec{x}} \cdot \mathbf{1}_I, \rvec{y} \rangle
- \mean{\rvec{y}} \cdot \left( \sum_{i=1}^I x^{(i)} - \sum_{i=1}^I x^{(i)}   \right) \\
= & \langle \rvec{x} - \mean{\rvec{x}} \cdot \mathbf{1}_I, \rvec{y} \rangle - 0 \,. \qed
  \end{split}
\end{equation}

For geometric intuition about why we should expect this lemma to be true, consider how $\mean{\rvec{x}}\cdot \mathbf{1}_I$ is the projection\footnote{Herein, whenever I say ``projection'', I mean specifically ``orthogonal projection''.} of $\rvec{x}$ onto the span of $\mathbf{1}_I$. Thus mean-centering $\rvec{x}$ is the same as projecting\footnote{Likewise, herein ``projecting'' always refers to ``orthogonally projecting''.} $\rvec{x}$ onto the orthogonal complement of (the span of) $\mathbf{1}_I$. Hence the projection of $\rvec{y}$ onto the span of $\mathbf{1}_I$, namely $\mean{\rvec{y}} \cdot \mathbf{1}_I$, is orthogonal to the mean-centered version of $\rvec{x}$. So in the expression ${\langle \rvec{x} - \mean{\rvec{x}} \cdot \mathbf{1}_I, \rvec{y} \rangle}$, the contribution from  $\mean{\rvec{y}} \cdot \mathbf{1}_I$ cancels out:
\begin{equation}
  \label{eq:projection_geometric_interpretation}
  \begin{split}
  {\langle \rvec{x} - \mean{\rvec{x}} \cdot \mathbf{1}_I, \rvec{y} \rangle} = &
 {\langle \rvec{x} - \mean{\rvec{x}} \cdot \mathbf{1}_I, (\rvec{y} - \mean{\rvec{y}} \cdot \mathbf{1}_I) + \mean{\rvec{y}} \cdot \mathbf{1}_I \rangle} \\
=& \langle \rvec{x} - \mean{\rvec{x}} \cdot \mathbf{1}_I, \rvec{y} - \mean{\rvec{y}} \cdot \mathbf{1}_I \rangle + \langle  \rvec{x} - \mean{\rvec{x}} \cdot \mathbf{1}_I,  \mean{\rvec{y}} \cdot \mathbf{1}_I \rangle \\
=& \langle \rvec{x} - \mean{\rvec{x}} \cdot \mathbf{1}_I, \rvec{y} - \mean{\rvec{y}} \cdot \mathbf{1}_I \rangle  + 0 \,.    
  \end{split}
\end{equation}
 In general, if $\pi$ denotes an operator projecting onto some subspace, then for analogous reasons it is always true that ${\langle \pi(\rvec{x}) , \pi(\rvec{y}) \rangle} = {\langle \pi(\rvec{x}) , \rvec{y} \rangle} = {\langle \rvec{x} , \pi(\rvec{y}) \rangle}$.

\begin{lemma}
\label{lem:means_of_rank_vectors}
As long as tied values are replaced with the mean of the tied values\footnote{This is the way tied values are replaced by default in SciPy\cite{SciPy}. ``The average of the ranks that would have been assigned to all the tied values is assigned to each value''.}, then any two rank vectors of the same length have the same mean.
\end{lemma}

\textbf{Proof:} In the case that there are no ties, then it follows easily that all rank vectors of length $L$ have the same mean as the vector ${(1, \dots, L)}$. Because addition is commutative, if two vectors are the same up to a permutation of their entries, then they must have the same mean. In the absence of ties, all rank vectors by definition have entries which are permuted from the vector ${(1, \dots, L)}$.

In the case of tied values, we can use the general fact that replacing any subset $\mathcal{S} \subseteq [I]$ of the entries of a vector with the mean value of the entries in the subset does not change the mean of the vector. In the computation below, let $\rvec{v}$ denote the original vector, and let $\rvec{v}'$ denote the vector created by replacing the entries of $\rvec{v}$ corresponding to the subset $\mathcal{S}$ with their mean value:
\begin{equation}
  \label{eq:changing_subset_not_change_mean}
  \begin{split}
    \mean{\rvec{v}'} := & \frac{1}{I} \sum_{i=1}^I (v')^{(i)} \\
= & \frac{1}{I} \left(  \sum_{i \in \mathcal{S}}  (v')^{(i)}   + \sum_{i \not\in \mathcal{S}}  (v')^{(i)}  \right) \\
= &  \frac{1}{I} \left(  \sum_{i \in \mathcal{S}} \left( \frac{1}{|\mathcal{S}|} \sum_{i \in \mathcal{S}} v^{(i)} \right)  + \sum_{i \not\in \mathcal{S}} v^{(i)}  \right) \\
= & \frac{1}{I} \left(  |\mathcal{S}| \cdot \left( \frac{1}{|\mathcal{S}|} \sum_{i \in \mathcal{S}}v^{(i)} \right)
  + \sum_{i \not\in \mathcal{S}} v^{(i)}  \right)   \\
= & \frac{1}{I} \left(   \sum_{i \in \mathcal{S}} v^{(i)}   + \sum_{i \not\in \mathcal{S}} v^{(i)} \right) \\
= & \frac{1}{I} \sum_{i=1}^I v^{(i)} \\
= & \mean{\rvec{v}} \,.
   \end{split}
\end{equation}
Applying this argument for each of the groups of tied elements, we get that the mean of a rank vector with tied values is the same as that of a rank vector without any ties, which in turn is the same as the mean of ${(1,\dots,L)}$. $\square$

\begin{lemma}
\label{lem:additivity_covariance_concatenated_vectors}
\textbf{Conditions for additivity of covariance for concatenated vectors:} Let $\rvec{x}_1, \rvec{x}_2 \in \mathbb{R}^I$, and $\rvec{y}_1, \rvec{y}_2 \in \mathbb{R}^J$. If at least one of 
\begin{itemize}
\item $\mean{\rvec{x}_1} = \mean{\rvec{y}_1}$, or
\item $\mean{\rvec{x}_2} = \mean{\rvec{y}_2}$
\end{itemize}
is true, then covariance is additive for $\rvec{x}_1 \oplus \rvec{y}_1$ and $\rvec{x}_2 \oplus \rvec{y}_2$, i.e.
\begin{equation}
  \label{eq:additive_covariance_concatenation_statement}
  \ecov{\rvec{x}_1 \oplus \rvec{y}_1, \rvec{x}_2 \oplus \rvec{y}_2} = \ecov{\rvec{x}_1, \rvec{x}_2} + \ecov{\rvec{y}_1, \rvec{y}_2}  \,.
\end{equation}
\end{lemma}

\textbf{Proof:} Because covariance is symmetric, following reasoning analogous to that from the proof of Lemma \ref{lem:projection_lemma}, it suffices to give a proof only for the case that $\mean{\rvec{x}_1} = \mean{\rvec{y}_1}$. Note that $\mean{\rvec{x}_1} = \mean{\rvec{y}_1}$ implies ${\mean{\rvec{x}_1 \oplus \rvec{y}_1} = \mean{\rvec{x}_1} = \mean{\rvec{y}_1}}$. 
\begin{equation}
  \label{eq:covariance_concatenation_proof}
  \begin{adjustbox}{max width=\textwidth,keepaspectratio}
$\displaystyle  \begin{split}
      \ecov{\rvec{x}_1 \oplus \rvec{y}_1, \rvec{x}_2 \oplus \rvec{y}_2} = &
\left\langle \rvec{x}_1 \oplus \rvec{y}_1 - \mean{\rvec{x}_1 \oplus \rvec{y}_1} \cdot \mathbf{1}_{I+J} \ ,\  
\rvec{x}_2 \oplus \rvec{y}_2 - \mean{\rvec{x}_2 \oplus \rvec{y}_2} \cdot \mathbf{1}_{I+J}  \right\rangle
 \\
= & 
\left\langle \rvec{x}_1 \oplus \rvec{y}_1 - \mean{\rvec{x}_1 \oplus \rvec{y}_1} \cdot \mathbf{1}_{I+J} \ ,\  
\rvec{x}_2 \oplus \rvec{y}_2  \right\rangle \\
= & 
\left\langle \rvec{x}_1 \oplus \rvec{y}_1 - (\mean{\rvec{x}_1} \cdot \mathbf{1}_I) \oplus (\mean{\rvec{y}_1} \cdot \mathbf{1}_{J})  \  ,\  
\rvec{x}_2 \oplus \rvec{y}_2  \right\rangle \\
= &
\left\langle (\rvec{x}_1 - \mean{\rvec{x}_1} \cdot \mathbf{1}_I)\oplus (\rvec{y}_1 - \mean{\rvec{y}_1} \cdot \mathbf{1}_{J})  \  ,\  
\rvec{x}_2 \oplus \rvec{y}_2  \right\rangle \\
= &
\left\langle \rvec{x}_1 - \mean{\rvec{x}_1} \cdot \mathbf{1}_I \ ,\ \rvec{x}_2 \right\rangle
+
\left\langle \rvec{y}_1 - \mean{\rvec{y}_1} \cdot \mathbf{1}_{J} \ , \ \rvec{y}_2 \right\rangle \\
= &
\left\langle \rvec{x}_1 - \mean{\rvec{x}_1} \cdot \mathbf{1}_I \ ,\ \rvec{x}_2 - \mean{\rvec{x}_2} \cdot \mathbf{1}_I \right\rangle
+
\left\langle \rvec{y}_1 - \mean{\rvec{y}_1} \cdot \mathbf{1}_{J} \ , \ \rvec{y}_2 - \mean{\rvec{y}_2} \cdot \mathbf{1}_J \right\rangle \\
= &
\ecov{\rvec{x}_1, \rvec{x}_2} + \ecov{\rvec{y}_1, \rvec{y}_2} \,.
  \end{split}$
\end{adjustbox}
\end{equation}
The second and second-to-last equalities of (\ref{eq:covariance_concatenation_proof}) use the Projection Lemma, Lemma \ref{lem:projection_lemma}. The third equality of (\ref{eq:covariance_concatenation_proof}) uses ${\mean{\rvec{x}_1 \oplus \rvec{y}_1} = \mean{\rvec{x}_1} = \mean{\rvec{y}_1}}$. The fourth equality of (\ref{eq:covariance_concatenation_proof}) uses property (\ref{eq:concatenated_vectors_partwise_addition}) from section \ref{sec:vector-concatenation}. The fifth equality of (\ref{eq:covariance_concatenation_proof}) uses property (\ref{eq: concatenated_vectors_partwise_dot_product}) from section \ref{sec:vector-concatenation}. The first and last equalities of (\ref{eq:covariance_concatenation_proof}) are just the definition of (empirical) covariance, of course.
$\square$

\begin{lemma}
\label{lem:subconvex}
 \textbf{Subconvex Combination Decomposition Property:} For the coefficients $\cvxcoeff{+}, \cvxcoeff{-}$ from (\ref{eq:mixed_sign_spearman_coeffs}) with the property that \[\spearman^{\pm} (\graph_1, \graph_2) = \cvxcoeff{+} \spearman^+ (\graph_1, \graph_2) + \cvxcoeff{-} \spearman^- (\graph_1, \graph_2) \,, \] we always have that $\cvxcoeff{+} + \cvxcoeff{-} \le 1$. (Trivially, $\cvxcoeff{+}, \cvxcoeff{-} \ge 0$.)
\end{lemma}

\textbf{Proof:} For convenience, define the following variables:
\begin{equation}
  \label{eq:subconvex_proof_shortcuts}
  \begin{split}
     a:= & \evar*{\sparrank{\graph_1^+}} \\
b:= & \evar*{\sparrank{\graph_1^-}}\\
c := & \evar*{\sparrank{\graph_2^+}} \\
d:= & \evar*{
\sparrank{\graph_2^-}}\,.
  \end{split}
\end{equation}
It follows then that the coefficients $\cvxcoeff{+}, \cvxcoeff{-}$ from (\ref{eq:mixed_sign_spearman_coeffs}) equal
\begin{equation}
  \label{eq:subconvex_coeffs_shortcut_form}
  \cvxcoeff{+} = \frac{\sqrt{ac}}{\sqrt{a + b}\cdot \sqrt{c + d}} \,, \quad \cvxcoeff{-} = \frac{\sqrt{bd}}{\sqrt{a + b} \cdot\sqrt{c + d}} \,.
\end{equation}
Thus the inequality $\cvxcoeff{+} + \cvxcoeff{-} \le 1$ is equivalent to
\begin{equation}
  \label{eq:subconvex_equiv_inequality}
  \frac{\sqrt{ac} + \sqrt{bd}}{\sqrt{a+b} \cdot \sqrt{c+d}} \le 1 \quad \iff \quad \sqrt{ac} + \sqrt{bd} \le \sqrt{a+b} \cdot \sqrt{c+d} \,.
\end{equation}
However this follows immediately from the Cauchy-Schwarz inequality, namely
\[{|\langle \rvec{u} , \rvec{v} \rangle| \le \norm{\rvec{u}}_2\norm{\rvec{v}}_2} \,,\]
if we define the vectors $\rvec{u} := (\sqrt{a}, \sqrt{b})$ and $\rvec{v}:= (\sqrt{c}, \sqrt{d})$.

Another, equivalent way to show that the inequality (\ref{eq:subconvex_equiv_inequality}) is true is by using the AM-GM inequality applied to the vector $(ad, bc)$:
\begin{equation}
  \label{eq:AM_GM_strategy}
  \begin{split}
    & \frac{ad + bc}{2} \ge \sqrt{abcd} \\
\iff & ad + bc + (ac + bd) \ge 2 \sqrt{abcd} + (ac + bd) \\
\iff & (a + b)(c+d) \ge (\sqrt{ac} + \sqrt{bd})^2 \\
\iff & \sqrt{a + b} \cdot \sqrt{c + d} \ge \sqrt{ac} + \sqrt{bd} \,.
  \end{split}
\end{equation}
The analysis is somewhat simplified given that we can assume $a,b,c,d \ge 0$. $\square$

\subsection{DeltaCon Distance}
\label{sec:deltacon}

\begin{lemma}
\label{lem:deltacon}
Any choice of  $\varepsilon < \frac{1}{1+ \norm{\adjacency}_{\infty}}$ is sufficient to guarantee that the Neumann series $(\mathbf{I} - \mathbf{W})^{-1}$ converges.
\end{lemma}

\textbf{Proof:} To review, the $\infty$-operator (or ``induced'') matrix norm is the ``maximum absolute row sum'' of a matrix $\mathbf{M}$:
\begin{equation}
  \label{eq:inf_norm_defn}
  \norm{\mathbf{M}}_{\infty} := \max_i \left[ \sum_{j=1}^J  |[\mathbf{M}]_{ij}| \right] \,.
\end{equation}
This operator norm is arguably one of the easiest to use to get a bound on the spectral radius because the definition of the degree matrix $\mathbf{D}$ from equation (\ref{eq:degree_matrix_defn}) of section \ref{sec:degree-matrix} involves the row sums of $\adjacency$.

Recall from section \ref{sec:mixed-sign-deltacon} that the matrix $\mathbf{W}$ used to create the Neumann series $(\mathbf{I} - \mathbf{W})^{-1}$ is defined as
\begin{equation}
  \label{eq:neumann_creating_defn}
  \mathbf{W} := \frac{\varepsilon}{1-\varepsilon^2}\adjacency - \frac{\varepsilon^2}{1 - \varepsilon^2} \mathbf{D} \,.
\end{equation}
Reviewing from section \ref{sec:paths-powers-adjac}, to ensure that the Neumann series $(\mathbf{I}-\mathbf{W})^{-1}$ exists, our goal is to ensure that the spectral radius $\operatorname{sr}(\mathbf{W})$ of $\mathbf{W}$ is less than $1$, $\operatorname{sr}(\mathbf{W}) < 1$, by ensuring the sufficient condition that $\norm{\mathbf{W}}_{\infty} < 1$.

As long as $1 - \varepsilon^2 > 0$, tedious algebra with the definition of $\mathbf{D}$ shows that
\begin{equation}
  \label{eq:deltacon_proof_1}
  \norm{W}_{\infty} = \frac{\varepsilon}{1 - \varepsilon^2} \max_i \left[
\sum_{j \not=i} \left|[\adjacency]_{ij}\right| + \left| [\adjacency]_{ii} - \varepsilon \sum_{j=1}^J [\adjacency]_{ij} \right|
\right] \,.
\end{equation}
Therefore the requirement that $\norm{\mathbf{W}}_{\infty} < 1$ is equivalent to the condition:
\begin{equation}
  \label{eq:deltacon_proof_2}
  \frac{1-\varepsilon^2}{\varepsilon}\norm{\mathbf{W}}_{\infty} = \max_i \left[
\sum_{j \not=i} \left|[\adjacency]_{ij}\right| + \left| [\adjacency]_{ii} - \varepsilon \sum_{j=1}^J [\adjacency]_{ij} \right|
\right] < \frac{1 - \varepsilon^2}{\varepsilon} \,.
\end{equation}
Using the triangle inequality, we can bound $\frac{1 - \varepsilon^2}{\varepsilon} \norm{\mathbf{W}}_{\infty}$ by $(1+\varepsilon)\norm{\adjacency}_{\infty}$:
\begin{equation}
  \label{eq:deltacon_proof_3}
  \begin{split}
  \frac{1-\varepsilon^2}{\varepsilon}\norm{\mathbf{W}}_{\infty} =& \max_i \left[
\sum_{j \not=i} \left|[\adjacency]_{ij}\right| + \left| [\adjacency]_{ii} - \varepsilon \sum_{j=1}^J [\adjacency]_{ij} \right|
\right]  \\
\le &
\max_i \left[
\sum_{j \not= i} \left| [\adjacency]_{ij} \right| + \left| [\adjacency]_{ii} \right| + \varepsilon \sum_{j=1}^J \left| [\adjacency]_{ij}  \right| 
\right] = (1+\varepsilon)\norm{\adjacency}_{\infty}    \,.
  \end{split}
\end{equation}
Therefore it suffices to show that $(1+\varepsilon) \norm{\adjacency}_{\infty} < \frac{1-\varepsilon^2}{\varepsilon} = \frac{(1-\varepsilon)(1+\varepsilon)}{\varepsilon}$, which is equivalent to $\norm{\adjacency}_{\infty} < \frac{1-\varepsilon}{\varepsilon}$, which in turn (as can be shown via tedious algebra) is equivalent to the inequality $\varepsilon < \frac{1}{1 + \norm{\adjacency}_{\infty}}$.

This satisfies $1 - \varepsilon^2 > 0$ as long as $\norm{\adjacency}_{\infty} > 0$, i.e. $\adjacency \not= \mathbf{0}$, where $\mathbf{0}$ denotes the all-zeros matrix (correpsonding to an empty network with no edges). However we can assume that $\adjacency \not= \mathbf{0}$ without loss of generality, because when $\adjacency = \mathbf{0}$, in turn $\mathbf{D} =\mathbf{0}$, implying that $\mathbf{W} = \mathbf{0}$, so that $(\mathbf{I} - \mathbf{W})^{-1} = \mathbf{I}^{-1} = \mathbf{I}$, so the Neumann series exists regardless of our choice of $\varepsilon$.

Therefore we have shown that $\varepsilon < \frac{1}{1+ \norm{\adjacency}_{\infty}}$ is sufficient to guarantee that $\norm{\mathbf{W}}_{\infty} < 1$, which in turn is sufficient to guarantee that $\operatorname{sr}(\mathbf{W}) < 1$, guaranteeing convergence of the Neumann series $(\mathbf{I} - \mathbf{W})^{-1}$.$\quad\square$

By seeking to upper bound the Frobenius norm by $1$, and using very naive bounding techniques (mostly triangle inequality), it is also possible to prove that satisfying the much weaker bound:
\begin{equation}
  \label{eq:frobenius}
  \varepsilon < \frac{1}{1 + \sqrt{\Species}||\mathbf{A}||_{\infty}}
\end{equation}
is sufficient to guarantee convergence of the Neumann series. It is probably possible to refine that argument to get a much better bound from the Frobenius norm, analogous to what was done in a much less general context in \cite{FaBP}.

It follows from the proof of Lemma \ref{lem:deltacon} that if \textit{at least one} of the two \textbf{strict} inequalities $\operatorname{sr}(\mathbf{W}) < \norm{\mathbf{W}}_{\infty}$ or $\frac{1 - \varepsilon^2}{\varepsilon} \norm{\mathbf{W}}_{\infty} < (1+\varepsilon)\norm{\adjacency}_{\infty}$ is true, then we can use $\varepsilon = \frac{1}{1+\norm{\adjacency}_{\infty}}$ without any problems.

However, there do exist matrices $\adjacency$ such that both $\operatorname{sr}(\mathbf{W}) = \norm{\mathbf{W}}_{\infty}$ \textit{and} $\frac{1 - \varepsilon^2}{\varepsilon} \norm{\mathbf{W}}_{\infty} = (1+\varepsilon)\norm{\adjacency}_{\infty}$. These should be the only matrices for which $\varepsilon = \frac{1}{1+\norm{\adjacency}_{\infty}}$ is too large. The general conditions under which $\operatorname{sr}(\mathbf{W}) = \norm{\mathbf{W}}_{\infty}$ is true are obscure to me, but I do know that the second requirement that needs to be satisfied for a counterexample, ${\frac{1 - \varepsilon^2}{\varepsilon} \norm{\mathbf{W}}_{\infty} = (1+\varepsilon)\norm{\adjacency}_{\infty}}$, should be true if and only if all entries in the row of $\adjacency$ with the largest sum of absolute values have the same sign (because then the absence of terms with opposite signs means there are no cancellations and thus that the triangle inequality holds with equality). Through happenstance I have found $\adjacency$ that also satisfy the first requirement, that $\operatorname{sr}(\mathbf{W}) = \norm{\mathbf{W}}_{\infty}$ when $\varepsilon = \frac{1}{1+\norm{\adjacency}_{\infty}}$.

\begin{lemma}
\label{lem:deltacon_sharp}
There exist matrices $\adjacency$ such that defining $\varepsilon = \frac{1}{1 + \norm{\adjacency}_{\infty}}$ leads to $\operatorname{sr}(\mathbf{W}) = 1$ (implying that the Neumann series $(\mathbf{I} - \mathbf{W})^{-1}$ does not converge).
\end{lemma}

\textbf{Proof:} Here is the matrix I found by happenstance that works
\begin{equation}
  \label{eq:counterexample_matrix}
\adjacency = 
  \begin{bmatrix}
        0      &  0 &  0      &  0 &  0 &  -4.5 &  0 &  0 &  0 &  0 \\
        0      &  0 &  0      &  0 &  0 &  0      &  0 &  0 &  0 &  0 \\
        0      &  0 &  0      &  0 &  0 &  0      &  0 &  0 &  0 &  -3.5 \\
        0      &  0 &  0      &  0 &  0 &  0      &  0 &  0 &  0 &  0 \\
        0      &  0 &  0      &  0 &  0 &  0      &  0 &  0 &  0 &  0 \\
        -4.5 &  0 &  0      &  0 &  0 &  0      &  0 &  0 &  0 &  0 \\
        0      &  0 &  0      &  0 &  0 &  0      &  0 &  0 &  0 &  0 \\
        0      &  0 &  0      &  0 &  0 &  0      &  0 &  0 &  0 &  0 \\
        0      &  0 &  0      &  0 &  0 &  0      &  0 &  0 &  0 &  0 \\
        0      &  0 &  -4.5 &  0 &  0 &  0      &  0 &  0 &  0 &  0 
  \end{bmatrix} \,.
\end{equation}
Rescaling a counterexample matrix by a constant does not appear to affect its status as a counterexample, nor does deleting (and/or adding) rows and/or columns consisting entirely of $0$'s. This leads to the simpler counterexample:
\begin{equation}
  \label{eq:counterexample_matrix2}
\adjacency = 
  \begin{bmatrix}
    0 & 0 & 9 & 0 \\
    0 & 0 & 0 & 7 \\
    9 & 0 & 0 & 0 \\
    0 & 9 & 0 & 0
  \end{bmatrix} \,,
\end{equation}
for which $\varepsilon = \frac{1}{1 + \norm{\adjacency}_{\infty}} = \frac{1}{10}$, and thus
\begin{equation}
  \label{eq:counterexample_matrix2b}
\mathbf{W} = 
  \begin{bmatrix}
    -\frac{9}{99} & 0 & \frac{90}{99} & 0 \\
    0 & -\frac{7}{99} & 0 & \frac{70}{99} \\
    \frac{90}{99} & 0 & -\frac{9}{99} & 0 \\
    0 & \frac{90}{99} & 0 & -\frac{9}{99}  
\end{bmatrix} \,.
\end{equation}
Very tedious calculations reveal that indeed the spectral radius of the matrix from (\ref{eq:counterexample_matrix2b}) is $1$. Related counterexamples appear to also include
\begin{equation}
  \label{eq:small_counterexamples}
  \adjacency =
  \begin{bmatrix}
    0 & \pm 9 \\ \pm 9 & 0
  \end{bmatrix} \,, \quad
\adjacency =
\begin{bmatrix}
  0 & \pm 9 & 0 \\ \pm 9 & 0 & 0 \\ 0 & 0 & \pm 9
\end{bmatrix} \,, \quad
\adjacency =
\begin{bmatrix}
  0 & \pm 9 & 0 \\ \pm 9 & 0 & 0 \\ 0 & 0 & \pm x
\end{bmatrix} \ (\forall x \text{ s.t. } |x| < 9) \,.
\end{equation}
Discerning which operations will preserve the property $\operatorname{sr}(\mathbf{W})=\norm{\mathbf{W}}_{\infty}$ is subtle, since e.g. the following matrices appear to \textbf{\textit{not}} be counterexamples:
\begin{equation}
  \label{eq:false_counterexamples}
  \adjacency =
  \begin{bmatrix}
    0 & \pm x \\ \pm 9 & 0
  \end{bmatrix} \,, \quad
  \adjacency =
  \begin{bmatrix}
    0 & \pm 9 \\ \pm x & 0
  \end{bmatrix} \,, \quad
\adjacency =
\begin{bmatrix}
  0 & \pm x & 0 \\ \pm 9 & 0 & 0 \\ 0 & 0 & \pm 9
\end{bmatrix} \ (\forall x \text{ s.t. } |x| < 9) \,.
\end{equation}
In any case, counterexamples clearly exist, but also appear to not be generic. $\square$

\section{Technical Details}
\label{sec:append-techn-deta}

These sections document preliminary steps towards making the underlying ideas discussed in this chapter precise, rigorous, and falsifiable.

\subsection[Node Correspondences and Comparison of Signed Networks]{Node Correspondences and Comparison of Networks with Mixed-Sign Edge Weights}
\label{sec:comparison-networks-mixed}

Let $\graphspace_{[\Strains]}$ denote the space of all networks with mixed-sign edge weights that share the same set of nodes $[\Strains]$. Thus any two networks $\graph_1, \graph_2 \in \graphspace_{[\Strains]}$ differ at most by their edges, in the sense that possibly either $\edgeset_{\graph_1} \not= \edgeset_{\graph_2}$ and/or possibly $A_{\graph_1} \not=A_{\graph_2}$, and never differ by (the ordering or labelling) of their nodes, because by definition of $\graphspace_{[\Strains]}$ we always have that $\nodeset_{\graph_1} = [\Strains] = \nodeset_{\graph_2}$.

Any network $\tilde{\graph}$ with mixed-sign edge weights that has $\Strains$ nodes (not necessarily the integers in $[\Strains]$) can be identified with a network $\graph \in \graphspace_{[\Strains]}$. This identification is not unique and occurs by explicitly choosing a bijection between $[\Strains]$ and the node set $\nodeset_{\tilde{\graph}}$ of $\tilde{\graph}$. Such choice amounts to a fixed, ``correct'' ordering and labelling of the $\Strains$ nodes in $\nodeset_{\tilde{\graph}}$. Cf. Appendix A.3.1 of the Master's thesis \cite{krinsman_2020}.

For ease of reference, given two networks with mixed-sign edge weights $\tilde{\graph}_1$, $\tilde{\graph}_2$ with $\Strains$ nodes and bijective identifications $\nodeset_{\tilde{\graph}_1} \overset{f_1}{\to} [\Strains]$ and $\nodeset_{\tilde{\graph}_2} \overset{f_2}{\to} [\Strains]$, we call the resulting identifications $\nodeset_{\tilde{\graph}_1} \overset{f_2^{-1} \circ f_1}{\longrightarrow} \nodeset_{\tilde{\graph}_2}$ and $\nodeset_{\tilde{\graph}_2} \overset{f_1^{-1} \circ f_2}{\longrightarrow} \nodeset_{\tilde{\graph_1}}$ between the node sets of $\tilde{\graph}_1$ and $\tilde{\graph}_2$ the \textbf{node correspondence} between $\tilde{\graph_1}$ and $\tilde{\graph_2}$. Of course for $\graph_1, \graph_2 \in \graphspace_{[\Strains]}$, the node correspondence is always the identity on $[\Strains]$. Restricting our consideration to $\graphspace_{[\Strains]}$, rather than all networks with mixed-sign edge weights and $\Strains$ nodes, can correctly be thought of as indicating that all node correspondences have already been chosen and fixed.

A function $\graphspace_{[\Strains]} \times \graphspace_{[\Strains]} \to \mathbb{R}$ is called an \textbf{objective function}. An objective function we want to minimize is called a ``penalty'' or ``loss'' function, while an objective function we want to maximize is called a ``reward'' function. A \textbf{dissimilarity function} $\dissimilarity: \graphspace_{[\Strains]} \times \graphspace_{[\Strains]} \to \mathbb{R}$ is an objective function for which higher values are considered ``less similar''. Special classes of dissimilarity functions are often referred to as ``divergences'' or ``distances''. In contrast, a \textbf{similarity function} $\similarity : \graphspace_{[\Strains]} \times \graphspace_{[\Strains]} \to \mathbb{R}$ is an objective function for which higher values are considered ``more similar''.

In \cite{Tantardini2019}, evaluating (analogous) similarity $\similarity$ and/or dissimilarity $\dissimilarity$ functions (for unsigned networks) is called the ``known node correspondence'' comparison problem. Evaluating such objective functions might occur as a subroutine in methods for the ``unknown node correspondence'' comparison problem as defined in \cite{Tantardini2019} or in methods for so-called graph alignment problems. Those problems, including whether or how to modify them for signed networks, are completely outside the scope of discussion of this paper. Herein we deal only with the known node correspondence problem, for which node correspondences are already given and thus require no computational expense.

If a function $\varphi$ is monotone nondecreasing, i.e. $x_1 \le x_2 \implies \varphi(x_1) \le \varphi(x_2)$ for all $x_1, x_2$, then given a dissimilarity function $\dissimilarity: \graphspace_{[\Strains]} \times \graphspace_{[\Strains]} \to \mathbb{R}$ one has that $\varphi(\dissimilarity)$ is always again a dissimilarity function, and likewise given a similarity function $\similarity: \graphspace_{[\Strains]} \times \graphspace_{[\Strains]} \to \mathbb{R}$ one has that $\varphi(\similarity)$ is always again a similarity function. If a function $\psi$ is monotone nonincreasing, i.e. $x_1 \le x_2 \implies \psi(x_1) \ge \psi(x_2)$ for all $x_1, x_2$, then given a dissimilarity function $\dissimilarity: \graphspace_{[\Strains]} \times \graphspace_{[\Strains]} \to \mathbb{R}$ one has that $\psi(\dissimilarity)$ is always a \textit{similarity} function, and likewise given a similarity function $\similarity: \graphspace_{[\Strains]} \times \graphspace_{[\Strains]} \to \mathbb{R}$ one has that $\psi(\similarity)$ is always a \textit{dissimilarity} function. The analysis of dissimilarity functions can always be reduced to the analysis of similarity functions, and vice versa. Thus, when discussing the known node comparison problem, we can always restrict discussion to dissimilarity functions (or to similarity functions) without losing any generality.

This paper asks which qualities or properties make a dissimilarity function (or equivalently, in the sense explained above, a similarity function) for networks with mixed-sign edge weights ``reasonable'' or ``well-behaved''. The Double Penalization Principle (defined below) provides one possible answer.

A dissimilarity minimization problem (equivalently a similarity maximization problem) is of the form:
\begin{equation}
  \label{eq:dissim_minimization} 
  \argmin_{(\graph_1, \graph_2) \in \mathbb{G} \subseteq \graphspace_{[\Strains]}^2} \dissimilarity(\graph_1, \graph_2) \quad \text{or} \quad
  \argmax_{(\graph_1, \graph_2) \in \mathbb{G} \subseteq \graphspace_{[\Strains]}^2} \similarity(\graph_1, \graph_2) \,,
\end{equation}
for some subset $\mathbb{G}$ of $\graphspace_{[\Strains]}^2 \overset{def}{=} \graphspace_{[\Strains]} \times \graphspace_{[\Strains]}$ containing ``admissible'' pairs of networks with mixed-sign edge weights, and for some dissimilarity function $\dissimilarity: \graphspace_{[\Strains]} \times \graphspace_{[\Strains]} \to \mathbb{R}$ or for some similarity function $\similarity: \graphspace_{[\Strains]} \times \graphspace_{[\Strains]} \to \mathbb{R}$. Using the same notation, a dissimilarity maximization problem (equivalently a similarity minimization problem) is of the form:
\begin{equation}
  \label{eq:dissim_maximization}
  \argmax_{(\graph_1, \graph_2) \in \mathbb{G} \subseteq \graphspace_{[\Strains]}^2} \dissimilarity(\graph_1, \graph_2) \quad \text{or} \quad
  \argmin_{(\graph_1, \graph_2) \in \mathbb{G} \subseteq \graphspace_{[\Strains]}^2} \similarity(\graph_1, \graph_2) \,.
\end{equation}
The specific problem motivating this paper is the dissimilarity minimization problem in the case that $\mathbb{G} =  \{ \hat{\graph}_1, \dots, \hat{\graph}_N   \} \times \{ \graph_* \}$, where $\graph_*$ is a ``ground truth'' network and the networks $\hat{\graph}_1, \dots, \hat{\graph}_N$ are ``candidate estimate'' networks. We want to make a \textit{quantitative} statement about which of the $N$ networks\footnote{
$\graph_*$ is in all of the pairs belonging to $\mathbb{G}$, hence the optimization in this case is effectively only over the $\hat{\graph}_1, \dots, \hat{\graph}_N$.
} $\hat{\graph}_1, \dots, \hat{\graph}_N$ is the ``least dissimilar'' to the ``ground truth'' $\graph_*$ and thus the ``best'' estimate of $\graph_*$. Obviously whether the resulting claim, about which estimate is ``best'', is ``reasonable'' depends inherently on whether our choice of dissimilarity function $\dissimilarity$ is itself ``reasonable''. Therefore we must first decide on principles characterizing ``reasonable'' dissimilarity functions in this context before we are able to ``reasonably'' decide which statistical estimators for networks with mixed-sign edge weights perform ``best''.

The computational complexity of solving the motivating optimization problem is $O(C(\Strains) N)$, where $C(\Strains)$ denotes the (maximum) number of operations required to evaluate our chosen dissimilarity function $\dissimilarity$ on a single pair of networks in $\graphspace_{[\Strains]}$. Therefore the computational issues associated with the motivating example are trivial, and thus we will not further consider computational complexity issues herein.

\subsection{Why (Most) Pre-Existing Comparison Methods are Insufficient}
\label{sec:why-comp-meth}

One might try to bypass the comparison problem for signed networks by using a comparison method for unsigned networks. The procedure is the following. Given a pair of signed networks, we could ``project'' both of them into a space of (unsigned) networks with less structure. (Cf. section \ref{sec:skeletons} for examples.) Applying a comparison method for less structured networks to these ``projected'' counterparts then corresponds to defining a ``new'' comparison method on the space of signed networks. Cf. figure \ref{fig:commutative_diagram_projection_procedure}.

\begin{figure}[H]
  \centering
\includegraphics[width=\textwidth,height=\textheight,keepaspectratio]{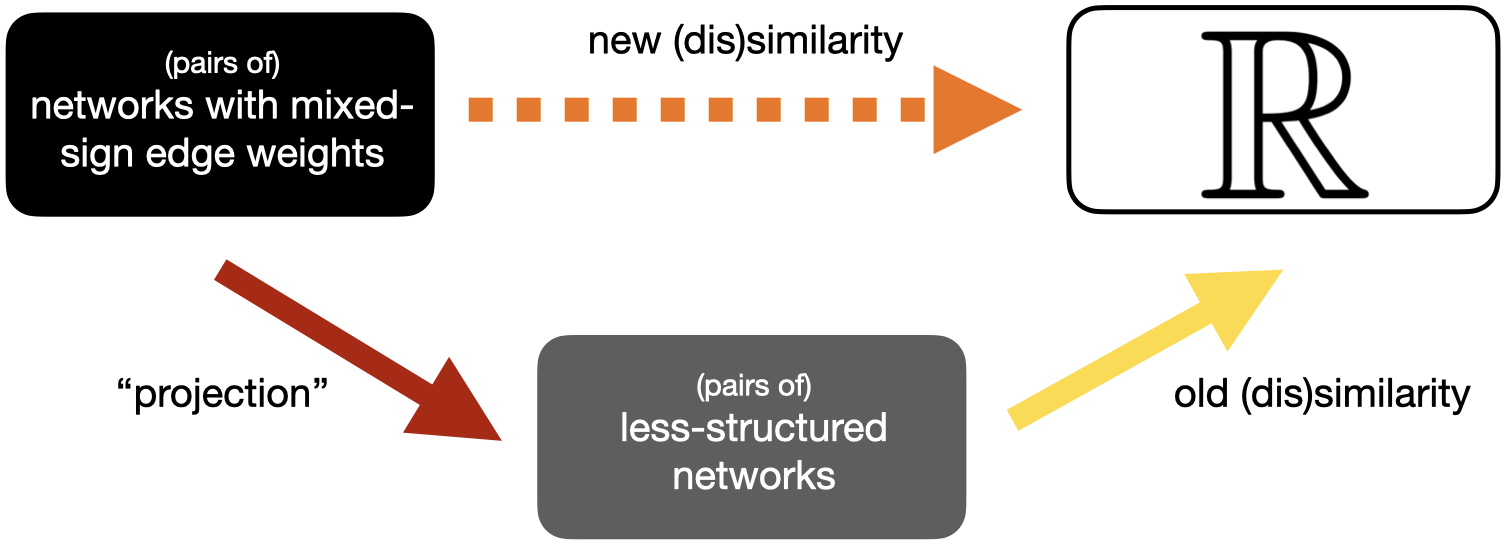}
  \caption[~Schematic of ``projection'' procedure for ``extending'' loss functions.]{``Projection'' procedure for applying comparison methods intended for networks with less structure to networks with mixed-sign edge weights. This is intended to (roughly) correspond to a so-called ``commutative diagram'', i.e. both paths from left to right are equal. Cf. \cite{conceptual} for an accessible introduction to commutative diagrams.}
  \label{fig:commutative_diagram_projection_procedure}
\end{figure}

Using this ``projection'' procedure to apply comparison methods for unsigned networks to signed networks exposes us to potential pitfalls. We need to avoid situations where, for example, positive and negative weights ``cancel'' in a scientifically meaningless way, or two edges with different signs are not considered distinct due to having similar magnitudes. 

For example, positive interactions and negative interactions correspond to categorically and qualitatively distinct phenomena in microbial ecology. Cf. again section \ref{sec:posit-negat-edges}. If all growth-promoting (positive) interactions were changed to growth-suppressing (negative) interactions, ecologically speaking the result would be quite different. This is true even if, or in some cases even especially if, the magnitudes of the interactions were left unchanged, as happens when ``projecting'' onto the space of unsigned networks. This implies that the positive and negative edges of signed networks should always be treated as qualitatively distinct by any well-behaved comparison method, at least for applications to microbial ecology.

The distinction between positive and negative edges is presumably also scientifically important for other applications of signed networks. If the distinction weren't important for a given application, then there would seem to be no good reason to not use an unsigned network for that application instead. Cf. the discussion from the first part of section \ref{sec:previous-work-networks}.

The reason why the ``projection'' procedure is inherently problematic is because two networks that are very distinct as signed and weighted networks may become difficult to distinguish after being ``projected'' into spaces of networks with less structure. This means that comparison methods resulting from the procedure are oblivious to what should be identified as important and obvious differences in the original signed and weighted networks. The ``double penalization principle'' is the most important sanity check for ruling out the possibility of such counter-intuitive behavior. Similar, but less important, principles that can also be used to assess the behavior of a comparison method are also discussed in sections \ref{sec:conv-comb-decomp} and \ref{sec:sparsity-should-not}. The double penalization principle allows us to check whether a given extension of a comparison method for unsigned networks to signed networks implicitly uses the ``projection'' procedure. This allows us to avoid the pitfalls of the ``projection'' procedure and to know when, and even how, to modify comparison methods so as to ensure they don't exhibit counter-intuitive behavior for signed networks.

\subsection{Taxonomy of Single-Edge Comparisons}
\label{sec:taxonomy-single-edge}

Any ``reasonable'' objective function should still be behaving ``reasonably'' when we ``zoom in'' on individual edges. In the following sections we try to make this intuition precise enough so that it can guide our choices about what constitute ``reasonable'' objective functions. First, we propose a way to categorize the kinds of differences that can exist between two networks with mixed-sign edge weights at the level of a single edge\footnote{
Edge here in the sense of an element of $[\Strains] \times [\Strains]$, i.e. a given input $e$ to the ``edge functions'' $A_{\graph_1}$ and $A_{\graph_2}$ (not necessarily lying in either of their supports, i.e. not necessarily in either ``edge set'').
}. Next, we seek to describe how ``reasonable'' objectives should behave in response to perturbations that ``shift'' between these categories of differences. Finally, we aim to propose one way to make precise constraints on the ``global'' behavior of ``reasonable'' objective functions that would result from the earlier considerations about the ``local'' level of single edges.

Let us first analyze the situation without taking magnitude differences into account. This can be interpreted as either (i) focusing on the signed but unweighted case, or (ii) as a first-pass, ``naive'' analysis of the general signed and weighted case. Either way, incorporating discussions of magnitude differences is easier once we have established the initial framework that neglects them. This is because how to correctly interpret any given magnitude difference depends on what the signs of the underlying edges are.

Not taking magnitude differences into account, there are $3^2 = 9$ different possible kinds of single-edge comparisons when taking the order of edges into account:
\begin{equation}
  \label{eq:ordered_edge_comparison_kinds}
  (+,+), (+,0), (+,-), (0,+), (0,0), (0,-), (-, +), (-, 0), (-,-) \,.
\end{equation}
In contrast, when disregarding the order of the edges, there are $\binom{3}{2} + \binom{3}{1} = 6$ different possible kinds of single-edge comparisons:
\begin{equation}
  \label{eq:unordered_edge_comparison_kinds}
  \{ (+,-), (-, +)  \}, \{ (+,0), (0, +) \}, \{ (-,0), (0,-) \}, \{ (+,+) \}, \{ (0,0) \}, \{ (-,-) \} \,.
\end{equation}
Thus by a ``taxonomy'' or ``classification scheme'' of single-edge comparisons, we mean some partition \footnote{
Technically (\ref{eq:unordered_edge_comparison_kinds}) itself is a partition of (\ref{eq:ordered_edge_comparison_kinds}), such that a partition of (\ref{eq:unordered_edge_comparison_kinds}) in practice amounts to nothing more than a coarsening of the corresponding partition of (\ref{eq:ordered_edge_comparison_kinds}), but this is a pedantic, intuitively obvious, and thus not terribly important, point.
} of either (\ref{eq:ordered_edge_comparison_kinds}) or (\ref{eq:unordered_edge_comparison_kinds}). Preferably any such proposed partition has a sensible organizing principles behind it, rather than being arbitrarily chosen. 

We can categorize single-edge comparisons into $2 \times 2 = 4$ categories according to:
\begin{itemize}
\item whether both edge counterparts are non-zero (non-trivial vs. trivial),
\item whether the signs of the edge counterparts coincide (sign agreement vs. sign disagreement).
\end{itemize}

\begin{table}[H]
  \centering
  \begin{tabular}{|c|c|c|}
    \hline
    & sign agreement & sign disagreement \\
    \hline
    trivial &  $\{ (  0, 0 ) \}$    &  $\{  (0, +), (+, 0), (0, -), (-, 0)   \}$   \\
    \hline
    non-trivial & $\{ (+,+), (-,-)    \}$  &  $\{  (+,-), (-,+)  \}$  \\
    \hline
  \end{tabular}
  \caption[~Signs corresponding to each category of single-edge comparison.]{Signs corresponding to each category of single-edge comparison.}
  \label{table:single_edge_comparison}
\end{table}

\subsection{Applying Framework to Constrain Dissimilarity Functions}
\label{sec:apply-fram-constr}

Within this framework let us consider the properties that a ``reasonable'' dissimilarity $\dissimilarity$ function or similarity $\similarity$ function would have. Specifically, we claim that an objective function can be designated ``reasonable'' or not based on its behavior in response to ``perturbations'' that shift single-edge comparisons from one of the above categories to another. This sets aside for now the separate issue of changes in magnitude, i.e. we assume that the magnitude of any non-zero edge is the same constant value, e.g. $1$.

We start from $4 \times 4 = 16$ possible shifts between categories, but I claim we only need to consider $\binom{4}{2} = 6$ of these. Because all non-zero edges are assumed to have the same magnitude (or equivalently for now we neglect to consider changes in magnitude), any shift from the same category to itself can be assumed to be the ``identity'' shift that does nothing. Hence any objective function, reasonable or not, will register no change in values for any of these $4$ ``identity'' shifts. This leaves $16 - 4 =12$ shifts to consider.

\begin{table}[H]
  \centering
  \begin{adjustbox}{max width=\textwidth,keepaspectratio}
  \begin{tabular}{|r |c|c|c|c|c|}
    \cline{3-6}
   \multicolumn{1}{r}{} & & \multicolumn{4}{c|}{After Shift} \\
   \cline{3-6} \multicolumn{1}{r}{}
    & 
    &  {\footnotesize\shortstack{\textbf{trivial}\\\textbf{sign agreement} }} & {\footnotesize\shortstack{\textbf{trivial}\\\textbf{sign disagreement}}}  &  {\footnotesize\shortstack{\textbf{non-trivial}\\\textbf{sign agreement}}} & {\footnotesize\shortstack{\textbf{non-trivial}\\\textbf{sign disagreement}}} \\
    \cline{1-6} 
   \multirow{4}{*}{\shortstack{Before\\Shift}}
   & {\footnotesize\shortstack{\textbf{trivial}\\\textbf{sign agreement}}} & no change & ? & ?  & ? \\
    \cline{2-6}
   & {\footnotesize\shortstack{\textbf{trivial}\\\textbf{sign disagreement}}}  & ? & no change & ? & ? \\
    \cline{2-6}
   & {\footnotesize\shortstack{\textbf{non-trivial}\\\textbf{sign agreement}}} &? &? & no change & ? \\
    \cline{2-6}
    &{\footnotesize\shortstack{\textbf{non-trivial}\\\textbf{sign disagreement}}} &? &? &? & no change \\
    \hline
  \end{tabular}
\end{adjustbox}
  \caption[~Taking ``identity'' shifts into account]{Taking ``identity'' shifts into account}
  \label{tab:beginning_table}
\end{table}

For the remaining $12$ shifts, note that we only need to consider $12 / 2 = 6$ shifts due to the elementary fact that $x_1 \le x_2$ if and only if $x_2 \ge x_1$. For example, how an objective function changes in response to a shift from a trivial sign agreement to a non-trivial sign disagreement completely determines how it changes in response to a shift from a non-trivial sign disagreement to a trivial sign agreement.

\begin{table}[H]
  \centering
  \begin{adjustbox}{max width=\textwidth,keepaspectratio}
  \begin{tabular}{|r |c|c|c|c|c|}
    \cline{3-6}
   \multicolumn{1}{r}{} & & \multicolumn{4}{c|}{After Shift} \\
   \cline{3-6} \multicolumn{1}{r}{}
    & 
    &  {\footnotesize\shortstack{\textbf{trivial}\\\textbf{sign agreement}}} & {\footnotesize\shortstack{\textbf{trivial}\\\textbf{sign disagreement}}}  &  {\footnotesize\shortstack{\textbf{non-trivial}\\\textbf{sign agreement}}} & {\footnotesize\shortstack{\textbf{non-trivial}\\\textbf{sign disagreement}}} \\
    \cline{1-6} 
   \multirow{4}{*}{\shortstack{Before\\Shift}}
   & {\footnotesize\shortstack{\textbf{trivial}\\\textbf{sign agreement}}} & no change & ? & ?  & ? \\
    \cline{2-6}
   & {\footnotesize\shortstack{\textbf{trivial}\\\textbf{sign disagreement}}}  &  {\footnotesize\shortstack{opposite of value\\across diagonal}} & no change & ? & ? \\
    \cline{2-6}
   &{\footnotesize \shortstack{\textbf{non-trivial}\\\textbf{sign agreement}} }&  {\footnotesize\shortstack{opposite of value\\across diagonal}} &  {\footnotesize\shortstack{opposite of value\\across diagonal}} & no change & ? \\
    \cline{2-6}
    &{\footnotesize\shortstack{\textbf{non-trivial}\\\textbf{sign disagreement}} }&{\footnotesize \shortstack{opposite of value\\across diagonal}} & {\footnotesize\shortstack{opposite of value\\across diagonal} }& {\footnotesize\shortstack{opposite of value\\across diagonal}} & no change \\
    \hline
  \end{tabular}
\end{adjustbox}
  \caption[~Pointing out redundant specifications.]{Pointing out redundant specifications.}
  \label{tab:reflection_table}
\end{table}

Elementary considerations already determine how a ``reasonable'' objective should behave for $4$ of the $6$ non-redundant shifts. Namely, it's clear that when shifting from any ``agreement'' to any ``disagreement'' that a ``reasonable'' dissimilarity $\dissimilarity$ function should not decrease and that a ``reasonable'' similarity $\similarity$ function should not increase. I will write ``$\dissimilarity \underline{\uparrow} \,, \similarity \underline{\downarrow}$'' to indicate a non-decreasing change in value for a ``reasonable'' dissimilarity $\dissimilarity$/a non-increasing change in value for a ``reasonable'' similarity $\similarity$.

This is equivalent to the stipulation that, when shifting from any ``disagreement'' to any ``agreement'', a ``reasonable'' dissimilarity $\dissimilarity$ function should not increase and that a ``reasonable'' similarity $\similarity$ function should not decrease. I will write ``$\dissimilarity \underline{\downarrow} \,, \similarity \underline{\uparrow}$'' to indicate a non-increasing change in value for a ``reasonable'' dissimilarity $\dissimilarity$/a non-decreasing change in value for a ``reasonable'' similarity $\similarity$.

\begin{table}[H]
  \centering
  \begin{adjustbox}{max width=\textwidth,keepaspectratio}
  \begin{tabular}{|r |c|c|c|c|c|}
    \cline{3-6}
   \multicolumn{1}{r}{} & & \multicolumn{4}{c|}{After Shift} \\
   \cline{3-6} \multicolumn{1}{r}{}
    & 
    &  {\footnotesize\shortstack{\textbf{trivial}\\\textbf{sign agreement }}} & {\footnotesize\shortstack{\textbf{trivial}\\\textbf{sign disagreement}}}  &  {\footnotesize\shortstack{\textbf{non-trivial}\\\textbf{sign agreement}}} & {\footnotesize\shortstack{\textbf{non-trivial}\\\textbf{sign disagreement}}} \\
    \cline{1-6} 
   \multirow{4}{*}{\shortstack{Before\\Shift}}
   & {\footnotesize\shortstack{\textbf{trivial}\\\textbf{sign agreement}}} &no change & $\dissimilarity \underline{\uparrow} \,, \similarity \underline{\downarrow}$  & ? & $\dissimilarity \underline{\uparrow} \,, \similarity \underline{\downarrow}$  \\
    \cline{2-6}
   & {\footnotesize\shortstack{\textbf{trivial}\\\textbf{sign disagreement}}}  & $\dissimilarity \underline{\downarrow} \,, \similarity \underline{\uparrow}$  & no change & $\dissimilarity \underline{\downarrow} \,, \similarity \underline{\uparrow}$ & ? \\
    \cline{2-6}
   & {\footnotesize\shortstack{\textbf{non-trivial}\\\textbf{sign agreement}}} &{\footnotesize\shortstack{opposite of value\\across diagonal}} &$\dissimilarity \underline{\uparrow} \,, \similarity \underline{\downarrow}$ &no change &$\dissimilarity \underline{\uparrow} \,, \similarity \underline{\downarrow}$ \\
    \cline{2-6}
    &{\footnotesize\shortstack{\textbf{non-trivial}\\\textbf{sign disagreement}}} & $\dissimilarity \underline{\downarrow} \,, \similarity \underline{\uparrow}$ &{\footnotesize\shortstack{opposite of value\\across diagonal}} & $\dissimilarity \underline{\downarrow} \,, \similarity \underline{\uparrow}$ & no change \\
    \hline
  \end{tabular}
\end{adjustbox}
    \caption[~Trivial considerations taken into account]{Trivial considerations taken into account.}
  \label{tab:trivial_specifications}
\end{table}

Note that technically it's redundant to specify the behavior of both a ``reasonable'' dissimilarity $\dissimilarity$ and a ``reasonable'' similarity $\similarity$, because one should transform to the other via a non-increasing transformation\footnote{
If a function $\psi$ is monotone nonincreasing, $x_1 \le x_2 \implies \psi(x_1) \ge \psi(x_2)$ for all $x_1, x_2$, then given a dissimilarity function $\dissimilarity: \graphspace_{[\Strains]} \times \graphspace_{[\Strains]} \to \mathbb{R}$, $\psi(\dissimilarity)$ is always a \textit{similarity} function, and likewise given a similarity function $\similarity: \graphspace_{[\Strains]} \times \graphspace_{[\Strains]} \to \mathbb{R}$ one has that $\psi(\similarity)$ is always a \textit{dissimilarity} function. The analysis of dissimilarity functions can always be reduced to the analysis of similarity functions, and vice versa.
}. I note both behaviors explicitly for the sake of expositional clarity, not mathematical necessity.

\subsection{Choices Available After Imposing ``Trivial'' Constraints}
\label{sec:choic-avail-after}

Comparing with table \ref{tab:trivial_specifications}, we see that this leaves only two kinds of shifts for which the behavior of a ``reasonable'' objective function is not trivially determined. Both remaining shifts are shifts ``within agreement or disagreement'', and ``between non-triviality and triviality''.

\textbf{First}, there is the shift from a trivial sign disagreement to a non-trivial sign disagreement (or equivalently in the sense of table \ref{tab:reflection_table} the shift from a non-trivial sign disagreement to a trivial sign disagreement).

\textbf{Second}, there is the shift from a trivial sign agreement to a non-trivial sign agreement (or equivalently in the sense of table \ref{tab:reflection_table} the shift from a non-trivial sign agreement to a trivial sign agreement). 

Comparing the principles mentioned in section \ref{sec:introduction}, we see that:

\begin{enumerate}
\item the ``double penalization principle'' from section \ref{sec:double-penal-princ} is the choice espoused herein to address shifts from a trivial sign disagreement to a non-trivial sign disagreement, and
  
\item the ``sparsity savviness principle'' from section \ref{sec:sparsity-should-not} is the choice espoused herein to address shifts from a trivial sign agreement to a non-trivial sign agreement.
\end{enumerate}

What the ``correct'' behavior for ``reasonable'' objectives for these two kinds of shifts is perhaps more arguable or subjective than for the others. Certainly the terminology ``trivial'' and ``non-trivial'' belies the opinions espoused herein. Namely, shifting from a ``trivial'' disagreement to a ``non-trivial'' disagreement should ``amplify'' the disagreement, and therefore correspond to a non-decrease for a dissimilarity $\dissimilarity$ function/a non-increase for a similarity $\similarity$ function, i.e. $\dissimilarity \underline{\uparrow} \,, \similarity \underline{\downarrow}$. Likewise, shifting from a ``trivial'' agreement to a ``non-trivial'' agreement\footnote{Again, ceteris paribus, not (yet) taking magnitude differences into account.} should ``amplify'' the agreement, and therefore correspond to a non-increase for a dissimilarity $\dissimilarity$ function/a non-decrease for a similarity $\similarity$ function, i.e. $\dissimilarity \underline{\downarrow} \,, \similarity \underline{\uparrow}$.

\subsection{``Composability'' Constraints for Objective Functions}
\label{sec:comp-constr-object}

One can argue fairly strongly on the basis of preferring ``consistency'', ``composability'', ``continuity'' of behavior of ``rational'' objectives for the double penalization principle.

Sparsity-savviness is compatible with the argument for ``composability'' constraints on ``rational'' objectives, but is also not supported by the argument.

This is analogous to another discrepancy between the double penalization principle and the sparsity-savviness principle. Namely, double penalization is straightforwardly compatible with how to reasonably penalize considering magnitude differences whereas sparsity-savviness is not. See section \ref{sec:taking-magn-diff} below for a further discussion of this.

I believe that both discrepancies in fact share the same underlying cause. Unfortunately however I am not currently sure how to make any of this precise enough to falsify or confirm.

Consider the following undirected network:

\begin{figure}[H]
  \centering
\includegraphics[width=\textwidth,height=\textheight,keepaspectratio]{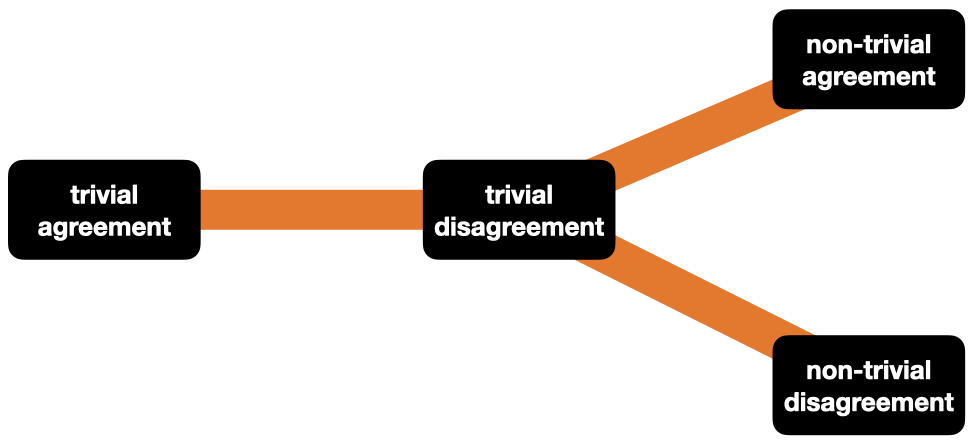}
  \caption[~Spoke graph of ``edit'' relationships of single-edge comparisons.]{Spoke graph for the ``edit'' relationships of the four single-edge comparisons.}
  \label{fig:spoke_graph}
\end{figure}

Each node represents a different category of single-edge comparison, cf. again table \ref{table:single_edge_comparison}. Two nodes are connected if one can be transformed to another by changing, or ``editing'', \textit{one} of the two signs in the comparison to an ``\textbf{adjacent sign}''. By ``adjacent sign'' I mean $\{-, 0\}$ or $\{0, +\}$, corresponding to the ordering of the signs on the number line. For example, starting with a trivial agreement, $(0,0)$, making the second edge positive $(0,+)$ corresponds to a trivial sign disagreement, as does making the first edge negative, $(-,0)$. However making both the second edge positive \textit{and} the first edge negative, $(-,+)$, is considered two changes or ``edits'', and therefore non-trivial sign disagreement is \textit{not} adjacent to trivial sign agreement, but rather separated by a path of length two. This is similar in spirit to the ``edit distance'' discussed in \cite{Lopez2019}, but more abstract and with a different definition of ``edits''.

This spoke graph implies additional consistency conditions for a ``reasonable'' objective function to satisfy, besides the ``trivial'' ones from table \ref{tab:trivial_specifications}. Namely, given a shift between categories corresponding to a path of length two, there are potentially two distinct ``reasonable'' behaviors implied by figure \ref{fig:spoke_graph}: (1) the behavior corresponding to the ``direct jump'' from the beginning of the path to the end of the path, (2) the behavior (if uniquely defined) corresponding to the composition of the behaviors of the two ``edits'' constituting the path. Requiring these two potentially distinct behaviors to coincide is the ``composability'' criterion for ``rational'' objectives alluded to above.

There are six directed paths of length two in the network from figure \ref{fig:spoke_graph}, corresponding to three undirected paths of length two (cf. table \ref{tab:reflection_table} again for why undirected suffices):
\begin{enumerate}
\item A path between ``trivial agreements'' and ``non-trivial agreements'',
\item a path between ``trivial agreements'' and ``non-trivial disagreements'', and
\item a path between ``non-trivial agreements'' and ``non-trivial disagreements''.
\end{enumerate}
The last two belong to the ``trivial'' shifts from table \ref{tab:trivial_specifications} while the first does not.

Looking at all $\binom{3}{2} = 3$ pairs of these three undirected paths, we see that neither of the two pairs that include the relevant path provide unambiguous guidance for the behavior separating ``trivial agreements'' and ``non-trivial agreements'' via the ``composability'' criterion. Cf. figures \ref{fig:sparse_savvy_consistency_1} and \ref{fig:sparse_savvy_consistency_2} below. Hence, as mentioned before, ``composability'' neither provides an argument for, nor an argument against, the sparsity-savviness principle. Therefore one must look to other reasons to justify the Sparsity-Savviness Principle as a desideratum.

\begin{figure}[p]
  \centering
  
  \begin{subfigure}{\textwidth}
  \centering
  \includegraphics[width=\textwidth,height=0.47\textheight,keepaspectratio]{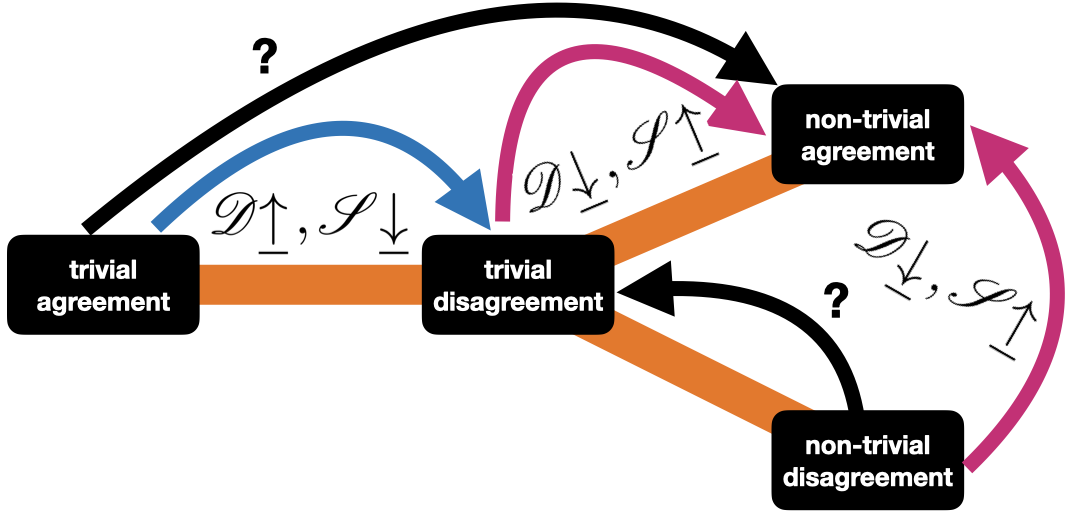}
  \caption[]{These two paths fail to usefully constrain any unknown shift, even if we determined the shift between trivial and non-trivial disagreements.}
  \label{fig:sparse_savvy_consistency_1}
\end{subfigure}

\begin{subfigure}{\textwidth}
  \centering
  \includegraphics[width=\textwidth,height=0.47\textheight,keepaspectratio]{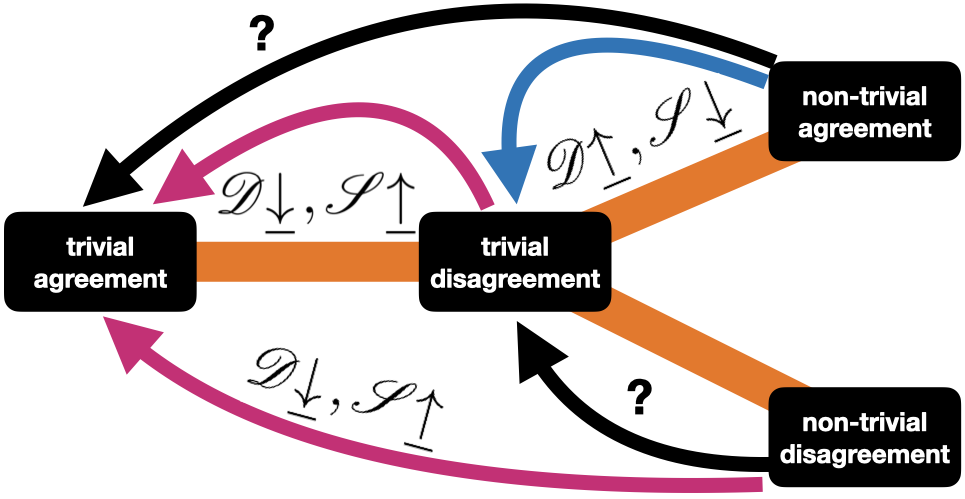}
  \caption[]{These two paths \textit{also} fail to usefully constrain any unknown shift, even if we determined the shift between trivial and non-trivial disagreements.}
  \label{fig:sparse_savvy_consistency_2}
\end{subfigure}

\caption[~Sparsity-savviness is not motivated by consistency with other basic criteria.]{Using only the ``trivial'' behavior and the ``composability'' criterion, no useful constraints emerge for the shift between trivial agreement and non-trivial agreement.}
\label{fig:sparse_savvy_consistency}
\end{figure}

However, the third pair of undirected paths provides an unambiguous choice of behavior for one of the two shifts not decided by ``trivial'' considerations alone. Cf. figure \ref{fig:double_penalization_consistency} below. Thus ``composability'' argues strongly in favor of requiring the double penalization principle.
 
\begin{figure}[H]
  \centering
  \includegraphics[width=\textwidth,height=\textheight,keepaspectratio]{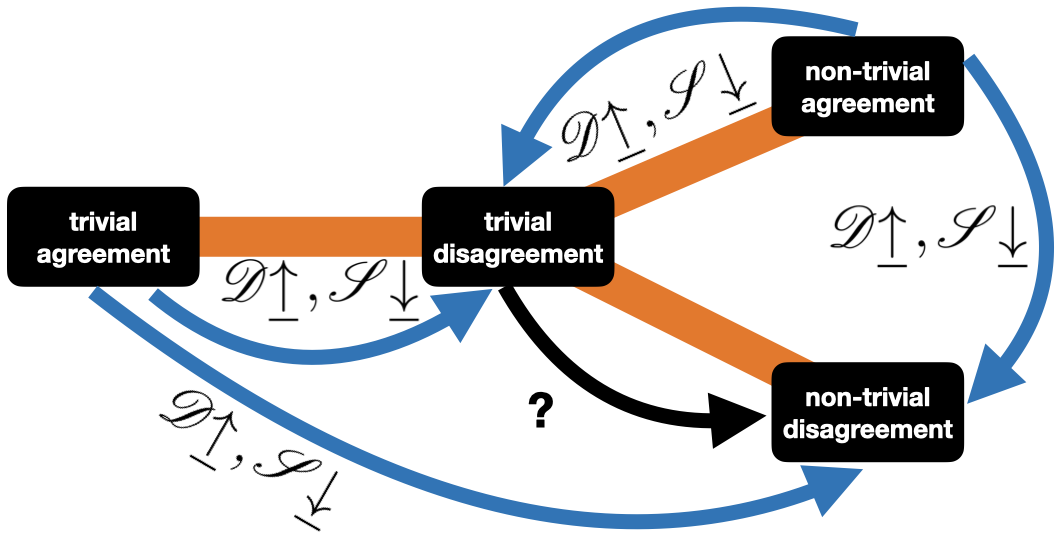}
  \caption[~Consistency of double penalization with ``trivial'' behavior and ``composability''.]{The ``trivial'' behavior combined with the ``composability'' criterion suggest only one possible option for the behavior of the unknown shift. Thus the double penalization principle is highly motivated.}
  \label{fig:double_penalization_consistency}
\end{figure}

Technically the constraint from figure \ref{fig:double_penalization_consistency}, that going from trivial agreements to non-trivial disagreements never increases similarity, can be satisfied without imposing the double penalization principle. Doing so would require extra work however. Namely, for a given objective, one would have to demonstrate that the first step (trivial agreements to trivial disagreements) of the two part path from trivial agreements to non-trivial disagreements would always non-increase similarity more than any increase in similarity that might possibly occur during the second step (trivial disagreements to non-trivial disagreements), i.e. that the change corresponding to the first step would always be large enough to cancel out that from the second step and thus able to determine the overall trend by itself.

On the other hand, as figure \ref{fig:double_penalization_consistency} demonstrates, if we only consider objective functions to be ``rational'' when shifts from trivial disagreements to non-trivial disagreements non-decrease dissimilarity $\dissimilarity \underline{\uparrow}$ (i.e. non-increase similarity $\similarity \underline{\downarrow}$), no extra work is ever required to satisfy the ``composability'' constraint. Restricting to such objective functions of course corresponds to requiring that the double penalization principle be satisfied.

\subsection{Taking Magnitude Differences into Account}
\label{sec:taking-magn-diff}

The above subsections limited the discussion to unweighted (weights $\pm 1$) networks, both to simplify the discussion and to serve as a starting point for understanding the general weighted case. Preliminary steps towards the latter goal are taken in this section.

I conjecture (cf. section \ref{sec:comp-constr-object}) that the greater compatibility of the double penalization principle, compared to the sparsity-savviness principle, with both (i) the ``composability'' constraint and (ii) consideration of magnitude differences, has one underlying cause.

(Assume in what follows that $x,y,z_1, z_2 > 0$.) In the general case where non-zero magnitudes can have values besides $1$, a non-trivial sign agreement (between the corresponding edges of the compared networks) of the form $(+z_1, +z_2)$ could very well deserve a penalty, depending on the size of the magnitude difference $|z_1 - z_2|$. This is in explicit contrast to the unweighted case, where in the analogous situation $|z_1 - z_2| = 0$ always.

Of course, in both the unweighted and weighted cases, a non-trivial sign disagreement of the form $(+x, -y)$ always deserves a penalty. To summarize:
\begin{itemize}
\item In the unweighted case from before, there are never penalties arising from magnitude differences of non-trivial agreements, so there is no need to consider how to balance those with the penalties arising from sign differences of non-trivial disagreements.
  
\item In the weighted case, we now must consider both kinds of penalties. 
\end{itemize}
We need to consider whether, and to what extent, the ability to impose penalties for both of these kinds of differences might come at the expense of one another. This occurs especially when considering changes in the objective functions for shifts from non-trivial agreements $(+z_1, +z_2)$ to non-trivial disagreements $(+x, -y)$ (and vice versa). To answer that, we need to know when the penalty for $(+z_1, +z_2)$ smaller or larger than that for $(+x, -y)$.

According to the ``composability'' constraint, for ``rational'' objective functions the change in value when shifting from $(+z_1, +z_2)$ to $(+x, -y)$ should correspond to ``composing''\footnote{I.e. not necessarily additively, but at least monotonically.} the changes in value associated with any ``path'' from $(+z_1, +z_2)$ to $(+x, -y)$.

\begin{figure}[p]
  \centering

  \begin{subfigure}{\textwidth}
    \centering
\includegraphics[width=\textwidth,height=0.62\textheight,keepaspectratio]{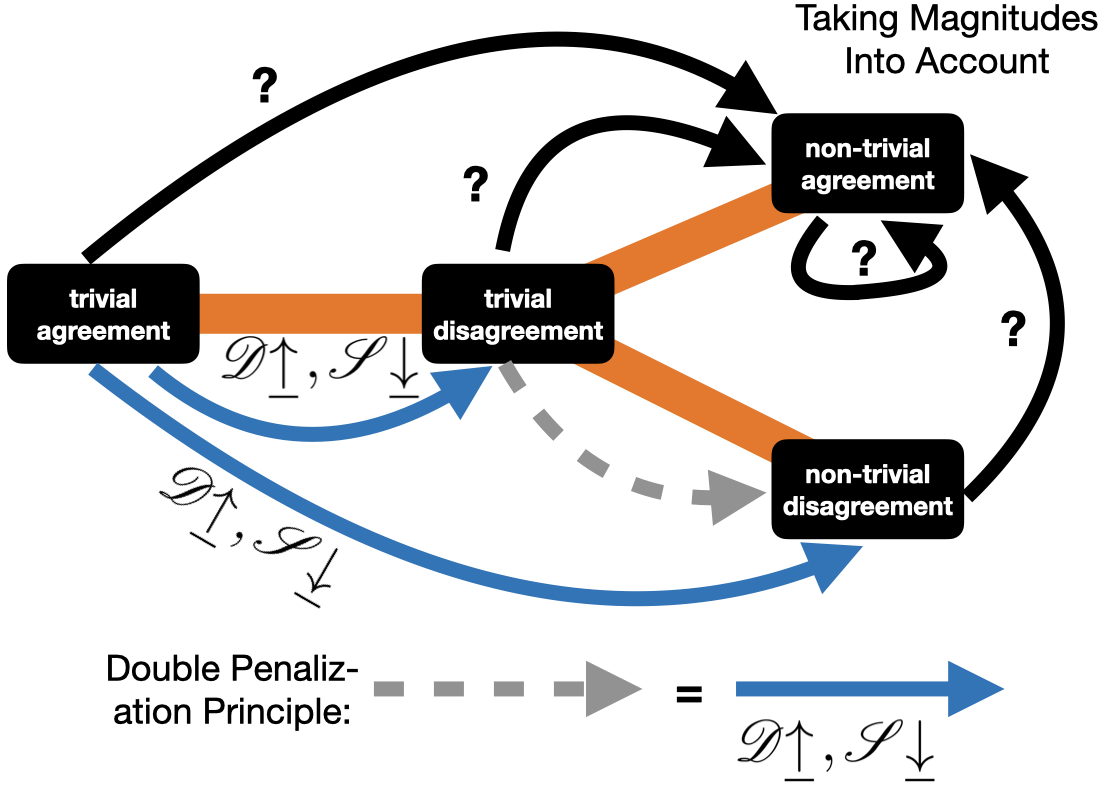}
    \caption[]{Generalized ``composability'' criteria for weighted and signed networks.}
    \label{fig:composability_weighted_signed}
  \end{subfigure}

  \begin{subfigure}{\textwidth}
    \centering
\includegraphics[width=\textwidth,height=0.33\textheight,keepaspectratio]{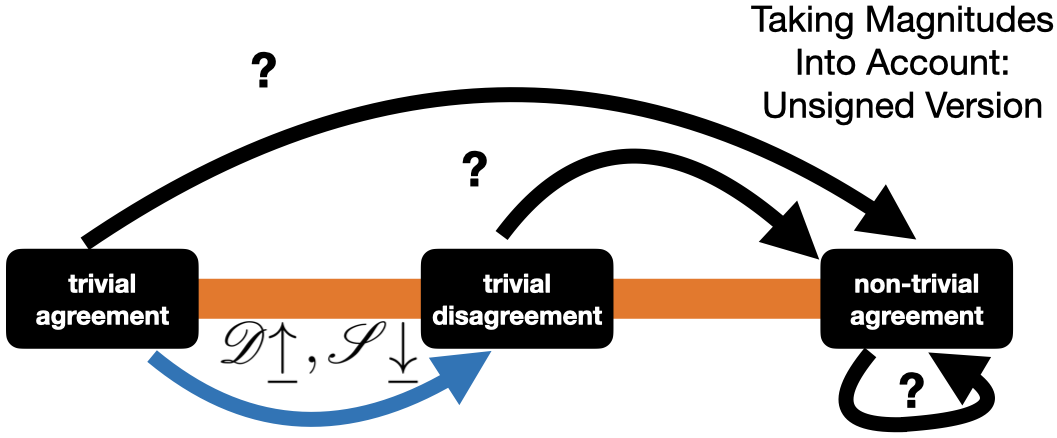}
    \caption[]{Generalized ``composability'' criteria that already apply for weighted and \textit{un}signed networks.}
    \label{fig:composability_weighted_unsigned}
  \end{subfigure}
  
  \caption[~Generalized ``composability'' criteria for weighted networks.]{Generalized ``composability'' criteria for weighted networks.}
  \label{fig:composability_weighted}
\end{figure}

Figure \ref{fig:composability_weighted_signed} depicts what those possible paths are. Here are two explicitly:

\begin{equation}
  \label{eq:magnitude_path_from_nta_to_ntda_1}
  \begin{array}{rcccccl}
(+z_1, +z_2) &\to& (+x, +z_2)& \to &(+x, 0)& \to& (+x, -y) \\
 \operatorname{NTA} & \to & \operatorname{NTA} & \to & \operatorname{TDA} & \to & \operatorname{NTDA}
  \end{array}
\end{equation}

\begin{equation}
  \label{eq:magnitude_path_from_nta_to_ntda_2}
  \begin{array}{rcccccccl}
(+z_1, +z_2) &\to & (0, +z_2) & \to & (0, 0) & \to & (+x, 0) & \to & (+x, -y) \\
 \operatorname{NTA} & \to & \operatorname{TDA} & \to & \operatorname{TA} & \to & \operatorname{TDA} & \to & \operatorname{NTDA}    
  \end{array}
\end{equation}
(In the above, ``NTA'' stands for ``non-trivial agreement'', ``NTDA'' stands for ``non-trivial disagreement'', ``TA'' stands for ``trivial agreement'', and ``TDA'' stands for ``trivial disagreement''.) We can get the remaining paths by switching e.g. the roles of $z_1$ and $z_2$, or of $x$ and $y$. I don't believe that doing so alters any of the following conclusions.

In figure \ref{fig:composability_weighted_signed}, all of the unknown values (``?'') that depend on how the magnitudes of $z_1$, $z_2$, $x$, and $y$ compare correspond to the following shifts (zero-step and one-step paths):

\begin{equation}
  \label{eq:unknown_mag_arrow_1}
\begin{array}{rcl}
(+z_1, +z_2) & \to & (+x, +z_2\\
 \operatorname{NTA} & \to & \operatorname{NTA}  
\end{array}
\end{equation}

\begin{equation}
  \label{eq:unknown_mag_arrow_2}
\begin{array}{rcl}
(+z_1, +z_2) & \to & (0, +z_2)\\
 \operatorname{NTA} & \to & \operatorname{TDA}  
\end{array}
\end{equation}

\begin{equation}
  \label{eq:unknown_mag_arrow_3}
\begin{array}{rcl}
(+x, +z_2) &  \to & (+x, 0)\\
 \operatorname{NTA} & \to & \operatorname{TDA}  
\end{array}
\end{equation}

How to weight the changes in the objective function value corresponding to each shift, and whether the changes are increases or decreases in penalty, depends on what the specific magnitudes are, and which functions we choose for computing penalties as a function of magnitude differences. These are what determine the ``?'' arrows in figure \ref{fig:composability_weighted_signed}.

None of these choices have anything to do with the double penalization principle. The double penalization principle only dictates the behavior of the arrows on the bottom of figure \ref{fig:composability_weighted_signed}, namely that they should not decrease dissimilarity. This amounts to making the gray dashed arrow, which corresponds to both $(+x, 0) \to (+x,-y)$ and $(0, -y) \to (+x, -y)$, another blue, similarity non-increasing arrow, in addition to the arrows corresponding to $(0,0) \to (+x, -y)$, $(0,0) \to (+x, 0)$, and $(0,0) \to (0, y)$.

In fact all of the uncertain arrows correspond to choices that already have to be made for purely unsigned networks. Cf. figure \ref{fig:composability_weighted_unsigned}. Thus, if we are already starting with a choice of objective function for weighted unsigned networks, we are already mostly done as long as we impose the ``composability'' constraint. In other words, in order to determine the ``?'' arrow from non-trivial disagreements to non-trivial agreements (and reverse) for weighted and signed networks, i.e. the shifts $(+z_1, +z_2) \leftrightarrow (+x, -y)$, all we need to do is:
\begin{itemize}
\item settle on a choice of comparison function for unsigned weighted networks to start with -- this settles the other three ``?'' arrows (again cf. figure \ref{fig:composability_weighted_unsigned}),
  
\item ensure that the gray dashed arrow at the bottom of figure \ref{fig:composability_weighted_signed} is determined (most likely by insisting that the double penalization principle be satisfied),
  
\item and require the ``composability'' constraint from section \ref{sec:comp-constr-object} above be satisfied.
\end{itemize}
At \textit{no} point do we ever have to explicitly or directly compare penalties for non-trivial agreements and non-trivial disagreements, because their implicit relative weighting follows from the three ingredients above, all of which we would have already demanded anyway.

Hence, whether the double penalization principle is satisfied is not directly relevant for determining implicit relative weighting of penalties for NTAs and and NTDAs.

What instead determines that implicit relative weighting is the original comparison function for unsigned and unweighted networks that we start with, as well as how we ensure that monotonic ``composability'' is satisfied. We should not need to \textit{directly} compare NTA and NTDA penalties if we are starting from and extending an objective function for unsigned networks. More concretely:
\begin{itemize}
\item the weighting of the NTA penalties relative to trivial agreements and disagreements should be determined by the original objective function for weighted, unsigned networks that is being extended, and
  
\item the weighting of the NTDA penalties relative to trivial agreements and disagreements should be determined by the means of satisfying the double penalization principle.
\end{itemize}
Then whatever ensures ``composability'' will then step in to use the above two points to implicitly determine the relative weighting of the NTA and NTDA penalties.

Of course, the devil is in the details. One would usually hope that monotonic ``composability'' can be ensured for the signed extension in the same way it was ensured for the unsigned version, thus requiring no extra work. However, given how vaguely ``composability'' is ``defined'' herein, this guidance is currently at best of limited use to a practitioner. To be more unambiguous, in the case of entrywise-$L_1$ relative error and Jaccard similarity, this monotone ``composability'' follows from the additivity of the expressions in the numerators. I am not even entirely certain it is satisfied for the other objectives, and may not be. Still, I hope the overall guiding principle is still at least heuristically useful.

It seems that the compatibility of the double penalization principle with the ``composability'' constraint implies that it can be satisfied without making any explicit choice of ``relative weighting'' between (i) penalties for sign differences and (ii) penalties for magnitude differences. In contrast, the fact that the sparsity-savviness principle does require such a choice of ``relative weighting'' seems to then imply its lack of inherent compatibility with the ``composability'' constraint (roughly the contrapositive statement). Again, as mentioned before, I am currently not sure how to make these conjectures precise enough to falsify.

\subsection{Precise Formulation of Double Penalization Principle}
\label{sec:prec-form-double}

Given a network with mixed-sign edge weights $\graph$, as well as a subset $E \subseteq \edgeset(\graph)$, define the ``\textbf{erased edges} network'' or ``zeroed out edges network'' $\mathcal{O}_E(\graph)$ as the network with mixed-sign edge weights such that
\begin{equation}
  \label{eq:erased_edges_single_graph_defn}
  \nodeset(\mathcal{O}_E(\graph)) = \nodeset(\graph) \,, \quad \edgeset(\mathcal{O}_{E}(\graph)) = \edgeset(\graph) \setminus E \,,
  \quad
  A_{\mathcal{O}_E(\graph)} :=
  \begin{cases}
    A_{\graph}(e) & e \not\in E \,, \\ 0 & e \in E
  \end{cases} \,.
\end{equation}
Sometimes it helps to think of $\mathcal{O}_E(\graph)$ as a ``sub-network'' of the original network $\graph$. It can also be useful to think of $\mathcal{O}_E(\graph)$ as a ``sparsifying perturbation'' of the original network $\graph$.

A ``\textbf{non-trivial sign disagreement}'' between two networks with mixed-sign edge weights $\graph_1, \graph_2$ occurs for a given edge $e \in \edgeset(\graph_1) \cap \edgeset(\graph_2)$ if and only if ${\sign(A_{\graph_1}(e)) \not= \sign(A_{\graph_2}(e)) }$. Recall how, \textit{by definition}, that an edge $e$ is in the edge set $\edgeset(\graph)$ of a given network $\graph$, $e \in \edgeset(\graph)$, if and only if $A_{\graph}(e) \not=0$. In particular, a ``\textbf{trivial sign disagreement}'' between two networks with mixed-sign edge weights $\graph_1, \graph_2$ occurs for a given edge $e$ if and only if $\sign(A_{\graph_1}(e)) \not= \sign(A_{\graph_2}(e))$ and either $\sign(A_{\graph_1}(e)) = 0$ or $\sign(A_{\graph_2}(e)) =0$, \textit{but not both}, i.e. either $e \not\in \edgeset(\graph_1)$ or $e \not\in \edgeset(\graph_2)$ (but not both). Similarly, a ``\textbf{trivial sign agreement}'' between two networks with mixed-sign edge weights $\graph_1, \graph_2$ occurs for a given edge $e$ if and only if both $e \not\in \edgeset(\graph_1)$ \textit{and} $e \not\in \edgeset(\graph_2)$, i.e. ${\sign(A_{\graph_1}(e)) = 0 = \sign(A_{\graph_2}(e))}$. Finally, a ``\textbf{non-trivial sign agreement}'' between two networks with mixed-sign edge weights $\graph_1, \graph_2$ occurs for a given edge $e$ if and only if $e \in \edgeset(\graph_1) \cap \edgeset(\graph_2)$ and $\sign(A_{\graph_1}(e))  = \sign(A_{\graph_2}(e))$.

Observe how, given a non-trivial sign disagreement for an edge $e \in \edgeset(\graph_1) \cap \edgeset(\graph_2)$, there are \textit{two} erasures of $e$ that would replace the non-trivial sign disagreement with a trivial sign disagreement, namely (1) erasing $e$ in $\graph_1$ but not in $\graph_2$, and (2) erasing $e$ in $\graph_2$ but not in $\graph_1$. The motivating idea behind the ``Double Penalization Principle'' is that we should never ``reward'' non-trivial sign disagreements over trivial sign disagreements. The penalty associated with a non-trivial sign disagreement should be no less severe than either of the two penalties that are associated with the two trivial sign disagreements resulting from erasing the corresponding edge. This idea is illustrated in figure \ref{fig:double_penalization_principle} above.

For two networks $\graph_1$ and $\graph_2$ with mixed-sign edge weights, their ``\textbf{set of non-trivial sign disagreements}'' $\mathscr{E}_{\pm}(\graph_1, \graph_2) \subseteq \edgeset(\graph_1) \cap \edgeset(\graph_2)$ is the set of ``shared'' edges such that
\begin{equation}
  \label{eq:nontrivial_sign_disagreements_defn}
  \mathscr{E}_{\pm}(\graph_1, \graph_2) := \{ e \subseteq \edgeset(\graph_1) \cap \edgeset(\graph_2) : \sign(A_{\graph_1}(e)) \not= \sign(A_{\graph_2}(e))   \} \,.
\end{equation}
 
A dissimilarity function $\dissimilarity$ satisfies the ``weak double penalization principle'' (figure \ref{fig:weak_double_penalization_principle}, contrast figure \ref{fig:double_penalization_principle} above) if and only if, given two networks with mixed-sign edge weights $\graph_1$ and $\graph_2$, for every pair of subsets $E_1, E_2 \subseteq \mathscr{E}_{\pm}(\graph_1, \graph_2)$ with $E_1 \cap E_2 = \emptyset$, one has that
\begin{equation}
  \label{eq:weak_double_penalization}
\dissimilarity(\graph_1, \graph_2) \ge \dissimilarity(\mathcal{O}_{E_1}(\graph_1), \mathcal{O}_{E_2}(\graph_2)) \,,
\end{equation}
i.e. replacing non-trivial sign disagreements with trivial sign disagreements can only decrease the dissimilarity, but \textit{never} increase it. Observe that the condition that $E_1 \cap E_2 = \emptyset$ is what enforces that we are always replacing non-trivial sign disagreements with trivial sign disagreements. If we removed the condition that $E_1 \cap E_2 = \emptyset$, we would also be including the replacement of non-trivial sign disagreements by trivial sign agreements.

\begin{figure}
  \centering \includegraphics[width=\textwidth,height=\textheight,keepaspectratio]{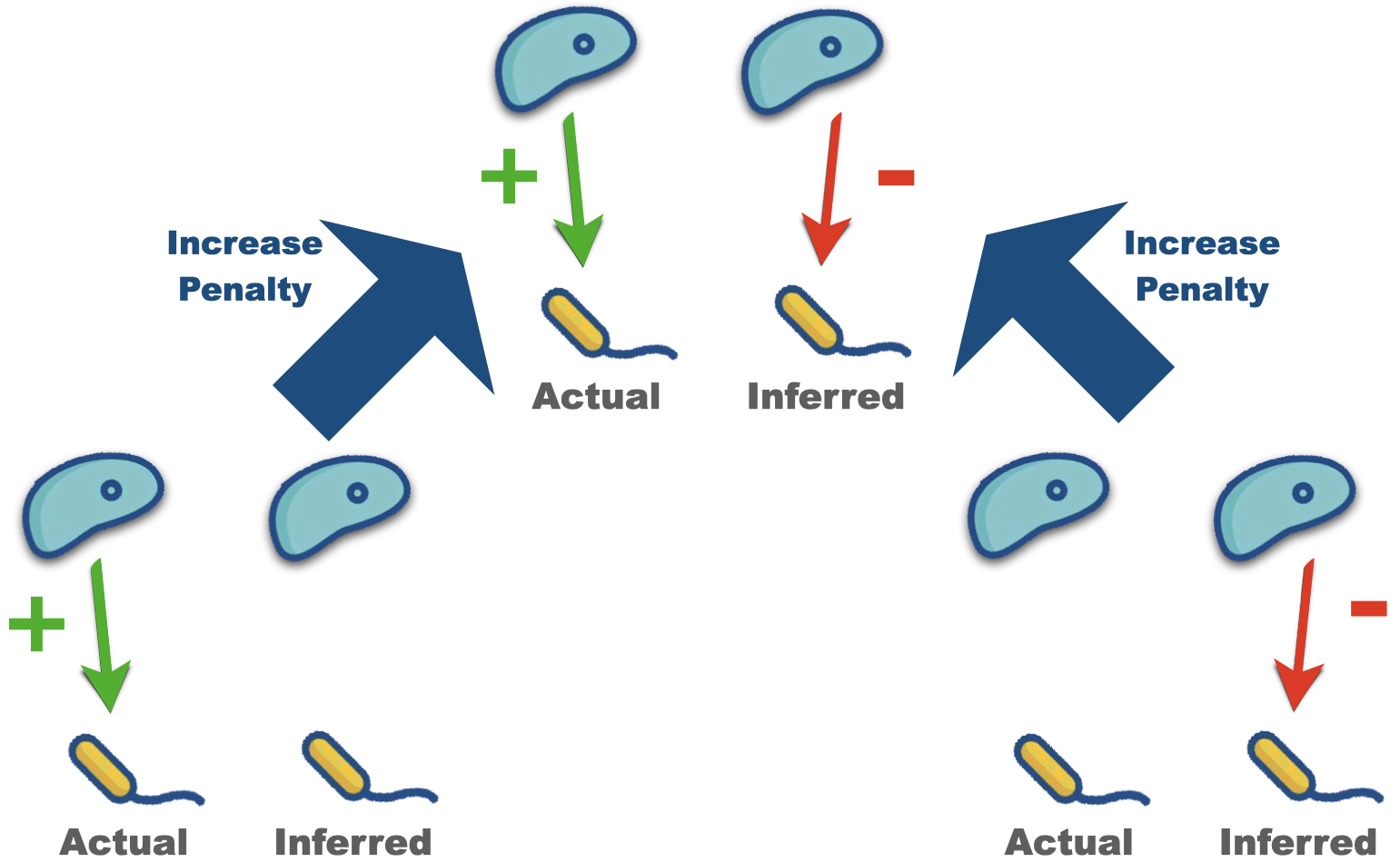}
  \caption[~Schematic for weak double penalization principle.]{Schematic for weak double penalization principle.} 
  \label{fig:weak_double_penalization_principle}
\end{figure}

Note that based on the weak double penalization property being satisfied alone it is a priori unclear for which $E_1, E_2 \subseteq \mathscr{E}_{\pm}(\graph_1, \graph_2)$, $E_1 \cap E_2 = \emptyset$ the minimum
  \begin{equation}
    \label{eq:weak_double_penalization_min}
   \min_{\substack{E_1, E_2 \subseteq \mathscr{E}_{\pm}(\graph_1, \graph_2) \\ E_1 \cap E_2 = \emptyset   }}  \dissimilarity(\mathcal{O}_{E_1}(\graph_1), \mathcal{O}_{E_2}(\graph_2))
  \end{equation}
 is satisfied, although a minimizer must exist due to the finiteness of the optimization space.

Given two networks $\graph_1, \graph_2$ with mixed-sign edge weights, their ``set of trivial sign disagreements'' is $\edgeset(\graph_1) \Delta \edgeset(\graph_2)$, i.e. $(\edgeset(\graph_1) \setminus \edgeset(\graph_2)) \sqcup (\edgeset(\graph_2) \setminus \edgeset(\graph_1))$. Then a dissimilarity function $\dissimilarity$ is called ``\textbf{monotonic}'' if and only if, for any two networks with mixed-sign edge weights $\graph_1$, $\graph_2$, and every $E_1 \subseteq \edgeset(\graph_1) \setminus \edgeset(\graph_2)$ and every $E_2 \subseteq \edgeset(\graph_2) \setminus \edgeset(\graph_1)$, one has
  \begin{equation}
    \label{eq:monotonic_dissimilarity_definition}
    \dissimilarity(\graph_1, \graph_2) \ge \dissimilarity(\mathcal{O}_{E_1} (\graph_1), \mathcal{O}_{E_2}(\graph_2)) \,,
  \end{equation}
i.e. replacing trivial sign disagreements with trivial sign agreements can only decrease the dissimilarity, but \textit{never} increase it.

Observe how, when considering unsigned (weighted) networks as special cases of networks with mixed-sign edge weights such that all nonzero edges are positive, all sign disagreements are trivial sign disagreements. Moreover, the monotonicity property is obviously desirable in that case. The monotonicity property says that replacing true misses with either false hits or false misses (cf. terminology from section \ref{sec:binary-class}) should always increase the dissimilarity. This interpretation remains valid in the case of general networks with mixed-sign edge weights, because trivial sign agreements always correspond to true misses, and trivial sign disagreements always correspond to false hits or false misses.

The monotonicity property implies (is it also equivalent to?) the relationship
\begin{equation}
  \label{eq:monotonic_dissimilarity_max}
  \dissimilarity(\graph_1, \graph_2) = \max_{\substack{E_1 \subseteq \edgeset(\graph_1) \setminus \edgeset(\graph_2) \\  E_2 \subseteq \edgeset(\graph_2) \setminus \edgeset(\graph_1) }} \dissimilarity(\mathcal{O}_{E_1} (\graph_1), \mathcal{O}_{E_2}(\graph_2)) \,.
\end{equation}

\textbf{Claim:} The monotonicity property implies (is it also equivalent to?)
\begin{equation}
  \label{eq:monotonic_dissimilarity_min}
  \dissimilarity(\mathcal{O}_{\edgeset(\graph_1) \setminus \edgeset(\graph_2)} (\graph_1),  \mathcal{O}_{\edgeset(\graph_2) \setminus \edgeset(\graph_1)} (\graph_2)   ) = \min_{\substack{E_1 \subseteq \edgeset(\graph_1) \setminus \edgeset(\graph_2) \\  E_2 \subseteq \edgeset(\graph_2) \setminus \edgeset(\graph_1) }} \dissimilarity(\mathcal{O}_{E_1} (\graph_1), \mathcal{O}_{E_2}(\graph_2)) \,.
\end{equation}

A dissimilarity function $\dissimilarity$ is said to satisfy the ``(strong) \textbf{double penalization principle}'' if and only if, (1) $\dissimilarity$ is monotonic, and (2) given any two networks with mixed-sign edge weights $\graph_1$ and $\graph_2$, for \textit{every} pair of subsets $E_1, E_2 \subseteq \mathscr{E}_{\pm}(\graph_1, \graph_2)$ (i.e. possibly but not necessarily $E_1 \cap E_2 \not= \emptyset$) one has that
\begin{equation} 
  \label{eq:strong_double_penalization_definition}
  \dissimilarity(\graph_1, \graph_2) \ge \dissimilarity(\mathcal{O}_{E_1}(\graph_1) , \mathcal{O}_{E_2}(\graph_2)) \,,
\end{equation}
i.e. replacing non-trivial sign disagreements with either trivial sign disagreements \textit{or} trivial sign agreements can only decrease the similarity, but \textit{never} increase it\footnote{
Without the possible loophole of replacing trivial sign disagreements with trivial sign agreements nevertheless somehow being able to increase the dissimilarity, which I think might technically be permitted if we only required (\ref{eq:strong_double_penalization_definition}) without also explicitly requiring $\dissimilarity$ to be monotonic, albeit I'm not certain.
}.

\textbf{Claim:} A dissimilarity function $\dissimilarity$ satisfies the (strong) double penalization principle if and only if $\dissimilarity$ both is monotonic and satisfies the weak double penalization principle.

More important is the following:

\begin{quote}
\textbf{Claim:} A dissimilarity function $\dissimilarity$ satisfies the (strong) double penalization principle \textit{if and only if} $\dissimilarity$ satisfies the monotonicity property on both the positive and negative parts simultaneously.
\end{quote}

To the extent that the monotonicity property is highly motivated, perhaps even ``trivially'' so, for unsigned networks, the truth of the above claim would then imply that the (strong) double penalization principle is correspondingly strongly motivated for signed networks.

Observe how the double penalization principle is agnostic about whether (i) [the dissimilarity increase for replacing a trivial sign agreement with a trivial sign disagreement] is larger or smaller than (ii) [the dissimilarity increase for replacing a trivial sign disagreement with a non-trivial sign disagreement], i.e. it is agnostic about the ``relative weighting'' of the dissimilarity increases associated with each kind of replacement. The double penalization principle only requires that both kinds of replacement do not decrease the dissimilarity. Combined together, both requirements then imply that replacing a trivial sign agreement with a non-trivial sign disagreement also not decrease the dissimilarity.

Non-trivial sign disagreements correspond to a false hit or false miss for both the positive parts and the negative parts of the networks (cf. the sections \ref{sec:posit-negat-parts} and \ref{sec:binary-class} for relevant terminology). Thus the double penalization principle requires that a false hit/miss for the positive parts of the networks can not be combined with a false miss/hit for the negative parts of the network to decrease the dissimilarity.

The double penalization property implies (is it also equivalent to?) the relationship
  \begin{equation}
    \label{eq:double_penalization_max}
    \dissimilarity(\graph_1, \graph_2) = \max_{\substack{E_1, E_2 \subseteq \mathscr{E}_{\pm}(\graph_1, \graph_2)   }} \dissimilarity(\mathcal{O}_{E_1}(\graph_1) , \mathcal{O}_{E_2}(\graph_2))  \,.
  \end{equation}

\textbf{Claim:} the double penalization property implies (is it also equivalent to?)
  \begin{equation}
    \label{eq:double_penalization_min}
    \dissimilarity(\mathcal{O}_{\mathscr{E}_{\pm}(\graph_1, \graph_2)}(\graph_1), \mathcal{O}_{\mathscr{E}_{\pm}(\graph_1, \graph_2)}(\graph_2)) = \min_{\substack{E_1, E_2 \subseteq \mathscr{E}_{\pm}(\graph_1, \graph_2)   }} \dissimilarity(\mathcal{O}_{E_1}(\graph_1) , \mathcal{O}_{E_2}(\graph_2))  \,.
  \end{equation}

For two networks $\graph_1$ and $\graph_2$ with mixed-sign edge weights, their ``\textbf{set of non-trivial sign agreements}'' $\mathscr{E}_{=}(\graph_1, \graph_2) \subseteq \edgeset(\graph_1) \cap \edgeset(\graph_2)$ is the set of ``shared'' edges such that
\begin{equation}
  \label{eq:nontrivial_sign_agreements_defn}
  \mathscr{E}_{=}(\graph_1, \graph_2) := \{ e \subseteq \edgeset(\graph_1) \cap \edgeset(\graph_2) : \sign(A_{\graph_1}(e)) = \sign(A_{\graph_2}(e))   \} \,.
\end{equation}

A \textit{similarity} function $\similarity$ is ``\textbf{sparsity-savvy}'' if and only if, for any given networks with mixed-sign edge weights $\graph_1$ and $\graph_2$, for every subset $E \subseteq \mathscr{E}_{=}(\graph_1, \graph_2)$ one has that
  \begin{equation}
    \label{eq:sparsity_savvy_definition}
    \similarity(\graph_1, \graph_2) \ge \similarity(\mathcal{O}_E(\graph_1), \mathcal{O}_E(\graph_2)) \,,
  \end{equation}
i.e. that replacing a trivial sign agreement with a non-trivial sign agreement at most only increase the similarity, but \textit{never} decrease it. Cf. section \ref{sec:sparsity-should-not}. In particular we can't ``game'' the similarity by considering only sparse networks, or removing common edges, or ``embedding networks'' into spaces with higher numbers of nodes (i.e. adding completely disconnected nodes, or adding blocks of all $0$'s to the adjacency matrices). Keep in mind that this property is not actually entirely trivial, at least for weighted networks. This is because the difference in magnitudes for a trivial sign agreement is always $0$, whereas the difference in magnitudes for a non-trivial sign agreement can be arbitrarily large. So the requirement is actually substantial/not to be taken for granted.

When considering unsigned (weighted) networks as special cases of signed networks whose nonzero edges are positive, there are no non-trivial sign disagreements, i.e. all sign disagreements are trivial sign disagreements. The sparsity-savviness principle says that replacing true hits with true misses should never increase the similarity. Such an interpretation remains valid in the case of general signed (weighted) networks. Even then non-trivial sign agreements always correspond to true hits for either the positive subnetworks or for the negative subnetworks (cf. the definitions in sections \ref{sec:posit-negat-parts} and \ref{sec:binary-class}).

\begin{table}[H]
  \centering
  \begin{adjustbox}{max width=\textwidth,keepaspectratio}
  \begin{tabular}{|c|c|c|}
    \hline
    & sign agreement & sign disagreement \\
    \hline
    trivial & {\small $\{ (  TM_+, TM_-  ) \}$ }   &  {\small$\{  (FH_+, TM_-), (TM_+, FH_- ), (FM_+, TM_-), (TM_+, FM_-)    \}$}   \\
    \hline
    non-trivial & {\small$\{ (TH_+, TM_-), (TM_+, TH_-)    \}$}  &  {\small$\{  (FH_+, FM_-), (FM_+, FH_-)   \}$}  \\
    \hline
  \end{tabular}
\end{adjustbox}
\end{table}

For the sake of clarity and comparison, let's consider the analogous binary classification notions for unsigned networks:

\begin{table}[H]
  \centering
  \begin{tabular}{|c|c|c|}
    \hline
    & sign agreement & sign disagreement \\
    \hline
    trivial &  $\{ (TM)  \}$    &  $\{  (FH), (FM)  \}$   \\
    \hline
    non-trivial & $\{ (TH)   \}$  &  \text{do not exist} \\
    \hline
  \end{tabular}
\end{table}

Note that the characteristics for unsigned networks should emerge as a special case of those for general signed networks, and indeed they do.

\textbf{Definition:} Given $\Strains \times \Strains$ matrices, a \textbf{distance metric} $d: \mathbb{R}^{[\Strains] \times [\Strains]} \times \mathbb{R}^{[\Strains] \times [\Strains]} \to [0, \infty)$ is a function satisfy the axioms (1) for any matrices $\mathbf{M}_1$ and $\mathbf{M}_2$ one has $d(\mathbf{M}_1, \mathbf{M}_2) = 0$ if and only if $\mathbf{M}_1 = \mathbf{M}_2$, (2) $d(\mathbf{M}_1, \mathbf{M}_2) = d(\mathbf{M}_2, \mathbf{M}_1)$ for any matrices $\mathbf{M}_1$ and $\mathbf{M}_2$, and (3) $d(\mathbf{M}_1, \mathbf{M}_3) \le d(\mathbf{M}_1, \mathbf{M}_2) + d(\mathbf{M}_2, \mathbf{M}_3)$ for any matrices $\mathbf{M}_1$, $\mathbf{M}_2$, and $\mathbf{M}_3$. The distance metric is said to turn $\mathbb{R}^{[\Strains] \times [\Strains]}$ into a ``metric space''.

\textbf{Claim:} Given a dissimilarity function $\dissimilarity$ such that for any $\graph_1, \graph_2 \in \graphspace_{[\Strains]}$ one has that
\[{\dissimilarity(\graph_1, \graph_2) = d(\adjacency_{\graph_1}, \adjacency_{\graph_2})} \]
for some distance metric $d$ on the space of matrices, then $\dissimilarity$ (a) is monotone, and (b) satisfies the (strong) double penalization property.

In particular, when using dissimilarity functions directly derived from distance metrics on arbitrary matrices, we don't have to worry much. This isn't necessarily true if the distance metric is only defined for non-negative matrices (corresponding to unsigned networks).

\textbf{Definition:} A \textbf{distance metric} $d$ on the space of unsigned networks with known node correspondences $\graphspace_{[\Strains]}^{\ge 0}$ satisfies the axioms (1) for any unsigned networks $\mathcal{U}_1$ and $\mathcal{U}_2$ one has $d(\mathcal{U}_1, \mathcal{U}_2) = 0$ if and only if $\mathcal{U}_1 = \mathcal{U}_2$ , (2) $d(\mathcal{U}_1, \mathcal{U}_2) = d(\mathcal{U}_2, \mathcal{U}_1)$ for any unsigned networks $\mathcal{U}_1$ and $\mathcal{U}_2$, and (3) $d(\mathcal{U}_1 , \mathcal{U}_3) \le d(\mathcal{U}_1, \mathcal{U}_2) + d(\mathcal{U}_2, \mathcal{U}_3)$ for any unsigned networks $\mathcal{U}_1$, $\mathcal{U}_2$, and $\mathcal{U}_3$.

\textbf{Claim:} Given a dissimilarity function $\dissimilarity$ such that for any $\graph_1, \graph_2$ one has that
\[{\dissimilarity(\graph_1, \graph_2)  =  d(\magskel(\graph_1), \magskel(\graph_2)  )   } \]
for some distance metric $d$ on unsigned networks, where $\magskel(\graph)$ denotes the magnitude skeleton of $\graph$, then in general $\dissimilarity$ does \textbf{\textit{not}} satisfy\footnote{
It may still satisfy monotonicity -- honestly I'm not sure either way.
} the double penalization principle.

Even starting with the most ``well-behaved'' dissimilarity functions on unsigned networks, dissimilarity functions for general networks with mixed-sign edge weights derived from the ``projection procedure'' will not satisfy this basic ``well-behavedness'' property.

However, we should still have the following

\textbf{Claim:} Given a dissimilarity function $\dissimilarity$ such that for any $\graph_1, \graph_2$ one has that
\[{ \dissimilarity(\graph_1, \graph_2) = c_1 d(\graph_1^+, \graph_2^+) + c_2 d(\graph_1^-, \graph_2^-) }\]
for $c_1, c_2 \ge 0$, $c_1$ and $c_2$ are allowed to depend on $\graph_1$ and $\graph_2$, and $d$ is a distance metric on unsigned networks, then $\dissimilarity$ \textit{will} satisfy the double penalization principle.

The above claim only applies to conic (i.e. non-negative) combinations of dissimilarity functions. However I believe that the analogous claim for similarity functions is true as well (or at least have been implicitly assuming as much):

\textbf{Claim:} Given a $\similarity$ function such that for any $\graph_1, \graph_2$ one has that
\[{ \similarity(\graph_1, \graph_2) = c_1 \mathcal{S}(\graph_1^+, \graph_2^+) + c_2 \mathcal{S}(\graph_1^-, \graph_2^-) }\]
for $c_1, c_2 \ge 0$, $c_1$ and $c_2$ are allowed to depend on $\graph_1$ and $\graph_2$, and $\mathcal{S}$ is a similarity function on unsigned networks satisfying the (similarity function version of) monotonicity, then $\similarity$ \textit{will} satisfy the (strong) double penalization property.

I think the above claim is true, or at the very least it seems I have been implicitly assuming as much. Similarly (pun unintended) it seems I have also been implicitly assuming

\textbf{Claim:} Given a $\similarity$ function such that for any $\graph_1, \graph_2$ one has that
\[{ \similarity(\graph_1, \graph_2) = c_1 \mathcal{S}(\graph_1^+, \graph_2^+) + c_2 \mathcal{S}(\graph_1^-, \graph_2^-) } \]
for $c_1, c_2 \ge 0$, $c_1$ and $c_2$ are allowed to depend on $\graph_1$ and $\graph_2$, and $\mathcal{S}$ is a similarity function on unsigned networks satisfying the sparsity-savviness property, then $\similarity$ \textit{will} satisfy the sparsity-savviness property.

\subsection{Continuity with respect to Magnitude}
\label{sec:continuity-mag}

Consider the plane ($\mathbb{R}^2$) with coordinates denoted $(a, i)$, ``$a$'' for ``actual'', ``$i$'' for ``inferred''. The first coordinate corresponds to the weight of the edge of the actual network in a single-edge comparison, whereas the second coordinate corresponds to the weight of the edge of the inferred network in a single-edge comparison.

Then we can divide/partition the plane into $4$ regions:
\begin{enumerate}
\item
  \begin{equation}
    \label{eq:non-trivial-agreements-region}
    \{ (a,i) : (a > 0 \text{ and } i > 0) \text{ or }(a <0 \text{ and } i < 0)  \} \,.
  \end{equation}
  These are the non-trivial sign agreements.
Geometrically on the plane, these correspond to the interiors of the upper right and lower left quadrants.
\item 
  \begin{equation}
    \label{eq:non-trivial-disagreements-region}
    \{ (a,i): (a>0 \text{ and } i < 0) \text{ or }(a < 0 \text{ and } i > 0)  \} \,.
  \end{equation}
  These are the non-trivial sign disagreements.
  Geometrically on the plane these correspond to the interiors of the upper left and lower right quadrants.
\item
  \begin{equation}
    \label{eq:trivial-disagreements-region}
    \{  (a,i): (a = 0 \text{ and } i \not= 0) \text{ or }(a \not=0 \text{ and } i =0) \} \,.
  \end{equation}
  These are the trivial sign disagreements.
  Geometrically on the plane these correspond to the union of the $a$ and $i$ axes with the origin removed.
\item
  \begin{equation}
    \label{eq:trivial-agreements-region}
    \{ (a,i): a = 0 \text{ and } i = 0  \} = \{  (0,0)\} \,.
  \end{equation}
  These are the trivial sign agreements.
  Geometrically on the plane these correspond to the origin.
\end{enumerate}

Figure \ref{fig:regions} illustrates this heuristically:
\begin{figure}[H]
  \centering
  \includegraphics[width=\textwidth,height=\textheight,keepaspectratio]{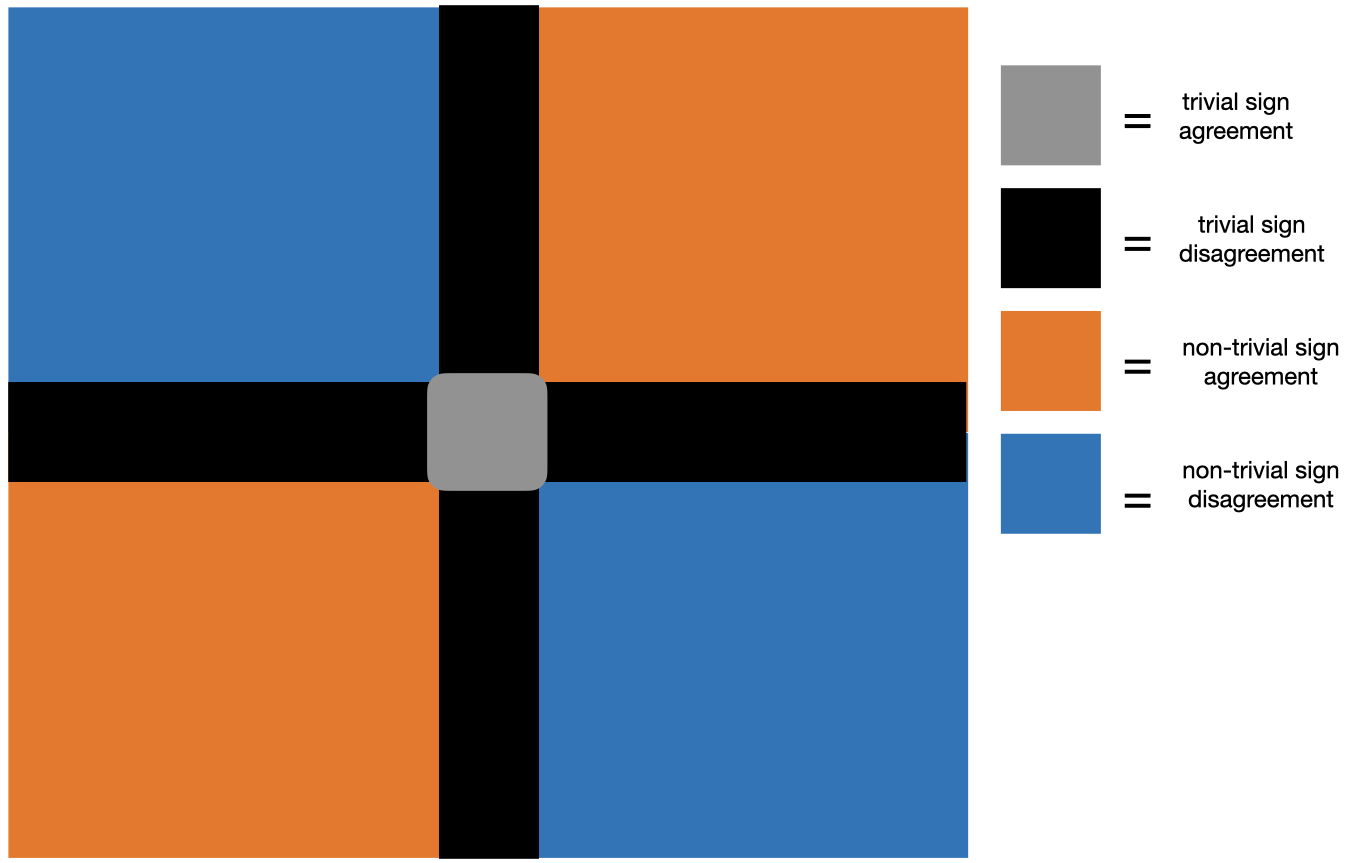}
  \caption[~The ``single-edge comparison plane''.]{The ``single-edge comparison plane'' described in the main text.}
  \label{fig:regions}
\end{figure}

Now consider any continuous function $[0,1] \to \mathbb{R}^2$, $t \mapsto (a(t), i(t))$, a so-called \textit{path}. For any such path, we can look at the sequence of above regions that it passes through. Close inspection reveals that the set of all such allowable sequences corresponds to the set of all possible paths (in the graph-theoretic sense) in the undirected network from figure \ref{fig:spoke_graph}. (At least when prohibiting diagonal movements in the immediate vicinity of the origin, something which is admittedly better motivated in the unweighted case.)

The continuity/composability property, at least intuitively speaking or heuristically or roughly, corresponds to requiring that for any such path in the ``single-edge comparison plane'', and any given objective function, its values are continuous as a function of the $t$-parametrized family of graph comparisons corresponding to any given edge in the graph-comparison, and the start and end points.

Specifically, say we are comparing the actual graph $\graph_*$ and the inferred graph $\hat{\graph}$. We focus on some particular edge $e \in [\Strains] \times [\Strains]$. This corresponds to a tuple $(a,i) = (e_*, \hat{e})$, where $e_* := A_{\graph_*}(e)$ and $\hat{e} := A_{\hat{\graph}}(e)$. Then we can choose some path $[0,1] \to \mathbb{R}^2$ with $ (e_*, \hat{e})$ as the starting point, i.e. $(a(0), i(0)) =  (e_*, \hat{e})$. Then this gives us a parameterized family of tuples of networks, where each $t \in [0,1]$ gets sent to $(\graph_*(t), \hat{\graph}(t))$, again with $(\graph_*(0), \hat{\graph}(0)) = (\graph_*, \hat{\graph})$, and where $A_{\graph_*(t)}(e) = a(t)$ and $A_{\hat{\graph}(t)}(e) = i(t)$ (the values of the adjacency function are assumed unchanged for all other edges). Composing this parameterized family of tuples of networks with our objective function, we are asking for the composite function $[0,1] \to \mathbb{R}$ to be continuous, irrespective of which edge we choose, and which path $[0,1] \to \mathbb{R}^2$ was chosen.

Call such a path ``coordinate-wise non-decreasing'' if, for all $t_1 < t_2$ in $[0,1]$, we have both $a(t_1) \le a(t_2)$ \textit{and} $i(t_1) \le i(t_2)$.

Of course, when we consider unweighted network, there is no longer any ``continuous plane'' of pairs of edge weight values to send paths into. But we still would like to consider ``discretized paths'' where the only allowed ``discretized paths'' correspond again to paths (in the graph-theoretic sense) in the undirected network from figure \ref{fig:spoke_graph}.

Also of course in the case of weighted networks, non-trivial sign agreements are no longer straightforward. In other words, even when the signs (non-trivially) agree, we can now still have differences in magnitude which we would usually want to penalize as well. The only places in the region of non-trivial sign agreements where we don't have magnitude disagreements are along the intersection with the line $a = i$. This corresponds to a line with slope $1$. In general, any line with slope bounded in between $0$ and $+\infty$ passes through the two quadrants defining the region of non-trivial sign agreements and thus corresponds to a certain departure (possibly null when the slope is $1$) from absence of magnitude of disagreement. (Hypothetically we could choose to associate this with the non-negative projective line, but that's probably a topic for another time.) In the limits as the slopes approach either $0$ or $+\infty$, we have that the non-trivial sign agreements actually approach trivial sign disagreements. Therefore, continuity combined with the desire to penalize trivial sign disagreements \textit{require us to penalize at least some non-trivial sign agreements}. (Again this is restricted only to the case of weighted networks, which is the only context where non-zero magnitude differences are possible for non-trivial sign agreements.)

While all of this certainly poses some complications for the sparsity-savviness principle, in particular probably requiring that it can only be satisfied in some asymptotic sense which we won't bother to attempt to precisely define, it actually does not pose any challenges or complications for satisfying a (continuous version of) the double penalization principle. To discuss this, it probably makes sense to begin by discussing a continuous version of monotonicity (i.e. comparing trivial sign agreements with trivial sign disagreements in the case of weighted networks). The main idea however is that (the continuous version of\footnote{But then presumably also as a consequence the discretized version}) the double penalization principle does not require any constraints on the behavior of paths passing through the quadrants corresponding to non-trivial sign agreements. It only regulates paths which pass from trivial sign-disagreements to non-trivial sign disagreements, to ensure they display analogous monotonicity behavior to paths from trivial sign agreements to trivial sign disagreements. (I.e. how the objective function behaves in the upper right and lower left quadrants is irrelevant.)

Consider any path starting from $(0,0)$ and such that either $a(t) = 0$ or $i(t) = 0$ for all $t \in [0,1]$ (but not both). Then for the coordinate of the path that is not constant and identically $0$, assume that it is monotone (non-decreasing or non-increasing, it does not matter which). Then any dissimilarity $\dissimilarity$ satisfying the continuous monotonicity property will monotonically \textit{non-decrease} along any such path. I.e. making one of the edge counterparts being compared increasingly nonzero should only worsen the score.

Now the double penalization principle is motivated by the fact that, when looking at any point in the non-trivial sign disagreement quadrants (lower right or upper left), projecting the point onto the \textit{closed} quadrant $\{ a \ge 0 \text{ and } i \ge 0 \}$ represents the point as a trivial disagreement for the positive part of the network, whereas projecting the point onto the \textit{closed} quadrant $\{ a \le 0 \text{ and } i \le 0 \}$ portrays it as a trivial disagreement for the negative part of the network. So intuitively speaking, we would like, at least for paths which are suitably coordinate-wise monotone restricted within one of the non-trivial sign disagreement quadrants, for the objective function to ``obey monotonicity for both the positive and negative parts'', i.e. for a dissimilarity function to monotonically non-decrease. Again, this is not at all a restriction on the behavior in the non-trivial sign agreement quadrants and thus whether the objective satisfies the double penalization principle is completely independent of e.g. the relative weighting of sign errors versus magnitude errors. Cf. the figure \ref{fig:dpp_regions}, where the lighter shaded areas indicate the regions that the (strong and continuous) double penalization principle implies should correspond to greater dissimilarity values than the corresponding large white dots.

\begin{figure}[p]
  \centering
  
  \begin{subfigure}{\textwidth}
  \centering
  \includegraphics[width=\textwidth,height=0.41\textheight,keepaspectratio]{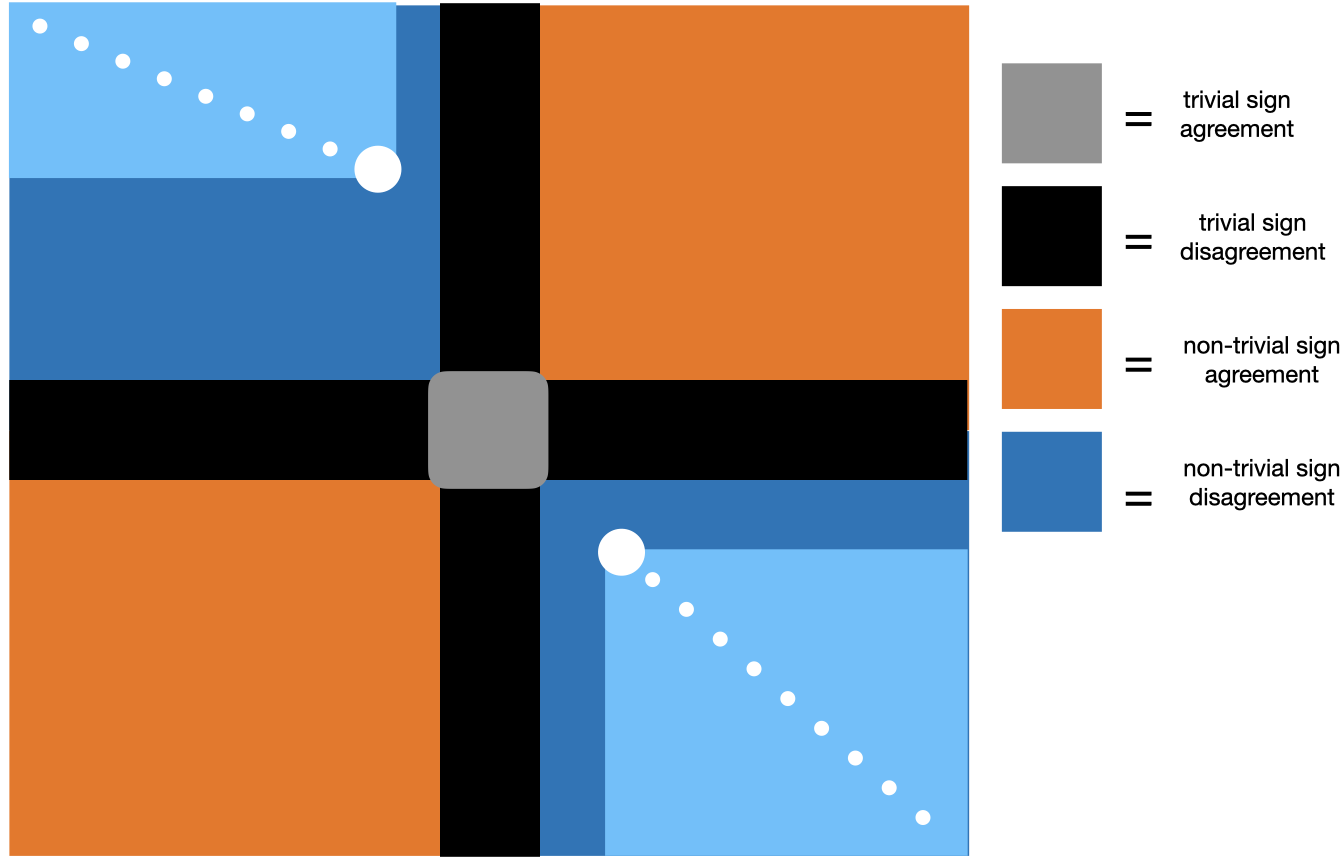}
  \caption[]{Large white dot has smallest dissimilarity compared to everything in the corresponding lightened regions. Additionally white dots indicate the direction of ``increasing isoclines of dissimilarity''.}
  \label{fig:dpp_regions}
\end{subfigure}

\begin{subfigure}{\textwidth}
  \centering
  \includegraphics[width=\textwidth,height=0.41\textheight,keepaspectratio]{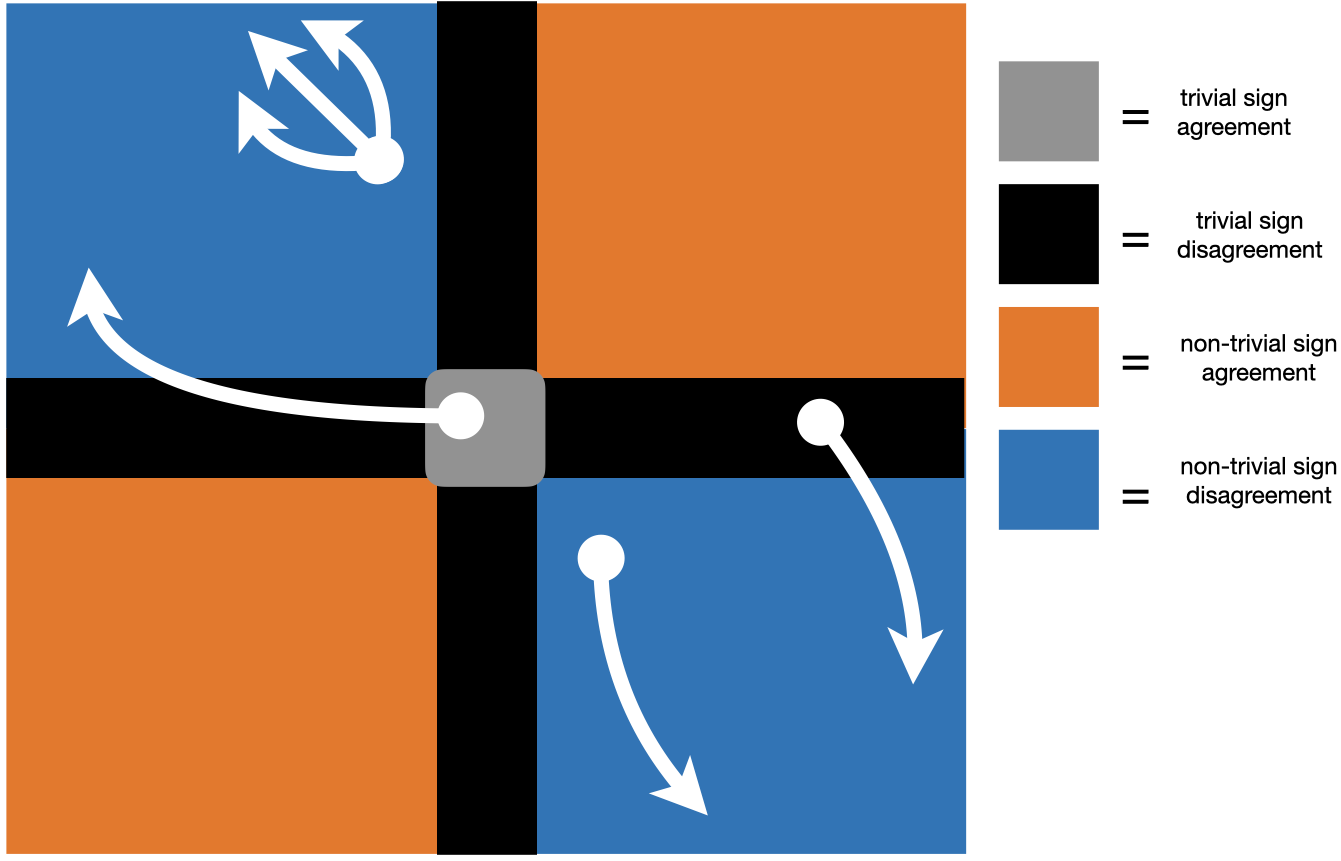}
  \caption[]{Examples of paths to which the continuous double penalization principle applies. Large white dots are the start points of the paths, arrows indicate the end points and directions of the paths.}
  \label{fig:dpp_trajectories}
\end{subfigure}

\caption[~Schematic of continuous version of double penalization principle.]{Schematic depiction of continuous version of double penalization principle.}
\label{fig:dpp_comparison_plane}
\end{figure}

In the lower right quadrant, a suitable path to which the continuous version of the double penalization principle applies is one for which $a(0) \ge 0$ and $i(0) \le 0$, and then for all $t_1, t_2 \in [0,1]$ with $t_1 < t_2$, $a(t_1) \le a(t_2)$ and $i(t_1) \ge i(t_2)$. (If the path coordinate functions are differentiable, then basically we are requiring that $a'(t) \ge 0$ whereas $i'(t) \le 0$.) Likewise, in the upper left quadrant, suitable paths are those for which $a(0) \le 0$ and $i(0) \ge 0$ and such that for all $t_1, t_2 \in [0,1]$ with $t_1 < t_2$ we have $a(t_1) \ge a(t_2)$ and $i(t_1) \le i(t_2)$. (If the path coordinate functions are differentiable, then basically we are requiring that $a'(t) \le 0$ and $i'(t) \ge 0$.) Then for all such paths, the (continuous version of) the double penalization principle requires that a dissimilarity function monotonically non-decrease with $t$. (I feel like requiring the objective to be continuous along with requiring the discretized definition above should be equivalent to this continuous version of the double penalization principle, but if it's true the formal argument would be very technical and boring so I at least currently will not bother.) Notice in particular that the continuous version of the double penalization principle does \textit{not} require the dissimilarity to increase (non-decrease) fastest or slowest along a specific choice of this class of paths, only that it be monotonically non-decreasing along all of them. (If we assume the start and end points of all of the paths under consideration are the same, then only the ``rate of change'' along the paths can differ, because obviously then the net difference between the values of the dissimilarity function at the beginning and end of the path must be the same.)

The figure \ref{fig:dpp_trajectories} indicates examples of such allowable trajectories.

\subsection{Considerations for Generalized Networks}
\label{sec:cons-gener-netw}

The arguments and notions of e.g. the monotonicity principle, double penalization principle, or sparsity-savviness principle, do not inherently need the comparison functions be defined for ``discrete'', finite networks. To clarify the ``true'' underlying ideas, what follows is a preliminary attempt to demonstrate their generalizability to other comparison functions. Hopefully this will be a preliminary step towards future work that makes all of the concepts and principles expressed in this chapter precise enough to be falsifiable.

  \subsubsection{Generalized Edge Spaces}
  Let a locally compact Polish space\footnote{
Equivalent to a locally compact second-countable Hausdorff space (a.k.a. ``LCCB'' space), among other characterizations.
} $E$ correspond to a ``generalized edge space''.

\textbf{Examples:}
\begin{itemize}
\item For a (finite) network, $E = [\Strains] \times [\Strains]$ (equipped with the usual discrete\footnote{
Like the discrete topology for $\mathbb{N} \times \mathbb{N}$, this is also the same as the Euclidean subspace topology inherited from $\mathbb{R}^2$.
  } topology).
  
\item For an infinite network, $E = \mathbb{N} \times \mathbb{N}$ (equipped with the usual discrete topology).
  
\item For a graphon, $E = [0,1] \times [0,1]$ (equipped with the usual Euclidean subspace topology).
\end{itemize}

  \subsubsection{Generalized Edge/Adjacency Functions}
  Given a ``generalized edge space $E$'', i.e. a locally compact Polish space, a ``generalized edge (adjacency) function'' $A: E \to \mathbb{R}$ will be interpreted as a \textit{continuous} function $E \to \mathbb{R}$.

  Note that in the case that the topology of $E$ is discrete, then all functions $E \to \mathbb{R}$ are continuous. However, this is a meaningful restriction e.g. in the graphon case. Analysis extending these notions to general Borel-measurable functions (or showing that such an extension is impossible) is left to future work.

  Note that because $E$ is locally compact Polish, and $\mathbb{R}$ is Polish, the space of all continuous functions $C(E, \mathbb{R})$ is itself a Polish space when equipped with the standard compact-open topology and, moreover, is an exponential object in the category of topological spaces (in particular evaluation of functions is continuous).

  \subsubsection{Indicators as Pointwise Limits}

  \begin{lemma}
    \label{lem:1}
  Let $(X, \tau)$ be a topological space, $Y \subseteq X$. Assume that the indicator function $\indicator{Y} : X \to [0,1]$ can be written as the monotone non-decreasing pointwise limit of some sequence of continuous functions $f_n: X \to [0,1]$, i.e. for all $x \in X$ one has $\lim\limits_{n \to \infty} f_n(x) = \indicator{Y}(x)$ and for all $n_1, n_2 \in \mathbb{N}$, $n_1 \le n_2$, $f_{n_1}(x) \le f_{n_2}(x)$. Then $Y$ must be (1) open and (2) $F_{\sigma}$.
  \end{lemma}

  \paragraph{Proof of Lemma \ref{lem:1}}
  First, note that $\{1\}$ and $\{0\}$ are closed subsets of $[0,1]$, whereas as $[0,1)$ and $(0,1]$ are open subsets $[0,1]$.

  By definition, the inverse image of $\{1\}$ under $\indicator{Y}$ is exactly the set $Y$, so the sets of values where the $f_n$ equal $1$ must approach $Y$. Because the $f_n$ are monotone increasing and bounded above by $1$, for $n_1 \le n_2$ we must have $f_{n_1}^{-1}(\{1\}) \subseteq f_{n_2}^{-1}(\{1\})$, i.e. the inverse images of $\{1\}$ under the $f_n$ must be non-decreasing as well and thus ``approach $Y$ from the inside''. More concretely:
  \begin{equation}
    \label{eq:Y_is_f_sigma}
    Y = \bigcup_{n=1}^{\infty} f_n^{-1}(\{1\}) = \left( \bigcap_{n=1}^{\infty} f_n^{-1}([0,1)) \right)^c \,,
  \end{equation}
with the last equality following from DeMorgan's Law. Because all of the $f_n$ are continuous, the $f_n^{-1}(\{1\})$ are all closed and the $f_n^{-1}([0,1))$ are all open. Hence $Y$ is the countable union of closed sets, thus $F_{\sigma}$. (Equivalently $Y$ is the complement of a countable intersection of open sets, i.e. the complement of a $G_{\delta}$ set.)

  At the same time, the inverse image of $\{0\}$ under $\indicator{Y}$ is the set $Y^c$, so the sets of values where the $f_n$ equal $0$ must approach $Y^c$. Because the $f_n$ are monotone increasing and bounded below by $0$, for $n_1 \le n_2$ we must have $f_{n_1}^{-1}(\{0\}) \supseteq f_{n_2}^{-1}(\{0\})$, i.e. the inverse images of $\{0\}$ under the $f_n$ must be non-increasing and thus ``approach $Y^c$ from the outside''. More concretely:
  \begin{equation}
    \label{eq:Y_is_open}
    Y = (Y^c)^c = \left( \bigcap_{n=1}^{\infty} f_n^{-1}(\{0\})  \right)^c = \bigcup_{n=1}^{\infty} f_n^{-1}((0,1]) \,,
  \end{equation}
  with the last equality again following from DeMorgan's Law. Again because all of the $f_n$ are continuous, the $f_n^{-1}(\{0\})$ are all closed and the $f_n^{-1}((0,1])$ are all open. The arbitrary intersection of closed sets is closed, and the arbitrary union of open sets is open, so (\ref{eq:Y_is_open}) means that $Y$ is the complement of a closed set, or equivalently an open set. $\square$

  \begin{corollary}
    \label{cor:2}
Let $(X, \tau)$ be a topological space, $Z \subseteq X$. Assume that the indicator function $\indicator{Z} : X \to [0,1]$ can be written as the monotone non-\textit{in}creasing pointwise limit of some sequence of continuous functions $g_n: X \to [0,1]$, i.e. for all $x \in X$ one has $\lim\limits_{n \to \infty} g_n(x) = \indicator{Z}(x)$ and for all $n_1, n_2 \in \mathbb{N}$, $n_1 \le n_2$, $g_{n_1}(x) \ge g_{n_2}(x)$. Then $Z$ must be (1) closed and (2) $G_{\delta}$.    
\end{corollary}

\textbf{Proof of Corollary \ref{cor:2}:} Take $Y:= Z^c$ and $f_n := 1 - g_n$. These satisfy the hypotheses of Lemma \ref{lem:1}. Because $Y=Z^c$ is open and $F_{\sigma}$, it follows that its complement $Z$ must be closed and $G_{\delta}$. $\square$

  \subsubsection{$\sigma$-compact subsets}
  \begin{lemma}
    \label{lem:sigma-compact}
    Let $X$ be a $\sigma$-compact and Hausdorff\footnote{
Hausdorffness is technically somewhat overkill. Because a closed subset of a compact set is always compact, all we need is for all compact sets to be closed (a so-called ``KC space''), and even then only for one direction. That being said, Hausdorffness is a much weaker assumption than the assumption that $E$ be a locally compact Polish space.
} space. Then a subset $Y$ is $\sigma$-compact\footnote{
Because compactness is an ``intrinsic'' topological property, i.e. $Z \subseteq X$ is compact in its subspace topology if and only if it is compact in the ``ambient'' topology of $X$, the same is also true of $\sigma$-compactness. For other properties we would need to be more careful and specify whether we meant in the subspace topology of $Y$ or in the ``ambient'' topology of $X$.
} if and only if $Y$ is $F_{\sigma}$.
  \end{lemma}

  \paragraph{Proof of Lemma \ref{lem:sigma-compact}}
  $(\implies)$ Assume $Y$ is $\sigma$-compact, i.e. $Y = \bigcup_{n=1}^{\infty} \tilde{K}_n$ with all $\tilde{K}_n$ compact. Because the $\tilde{K}_n$ are compact, they are closed. So $Y$ is the countable union of closed sets, i.e. $F_{\sigma}$.

  $(\impliedby)$ Assume $Y$ is $F_{\sigma}$, i.e. $Y = \bigcup_{n=1}^{\infty} F_n$ with all $F_n$ closed. Because $X$ is $\sigma$-compact, $X = \bigcup_{m=1}^{\infty} K_m$ with all $K_m$ compact. Then because $X \cap Y = Y$, we have that
  \begin{equation*}
Y = \bigcup_{m=1}^{\infty} \bigcup_{n=1}^{\infty} (K_m \cap F_n) \,.
\end{equation*}
Because the $K_m$ are compact and the $F_n$ are closed, for all $m$ and $n$ we have that $K_m \cap F_n$ is compact. Therefore $Y$ is the countable union of compact sets, i.e. $\sigma$-compact. $\square$

  \subsubsection{Relevant Facts}
  \begin{itemize}
  \item Locally compact Polish spaces are (by definition) metrizable, and metrizable spaces are perfectly normal (because they are normal and closed sets are $G_{\delta}$).
    
  \item Locally compact and second-countable implies $\sigma$-compact, all Polish spaces are second-countable, so locally compact Polish spaces are $\sigma$-compact.

  \item In a metrizable space open sets are\footnote{This is dual to the aforementioned fact that closed sets are $G_{\delta}$.} $F_{\sigma}$, thus by Lemma \ref{lem:sigma-compact} and the above facts we know that every open subset of a locally compact Polish space is $\sigma$-compact.
  \end{itemize}

  \subsubsection{Approximating Open Indicators from Below}
  \begin{lemma}
    \label{lem:monotone_limit}
    Let $U$ be an open subset of a locally compact Polish space $E$. Then there exists a monotone non-decreasing sequence of continuous functions $f_n: E \to [0,1]$ such that the indicator $\indicator{U}: E \to [0,1]$ is the pointwise limit,  i.e. for all $e \in E$ one has $\lim\limits_{n \to \infty} f_n(e) = \indicator{U}(e)$ and for all $n_1, n_2 \in \mathbb{N}$, $n_1 \le n_2$, $f_{n_1}(e) \le f_{n_2}(e)$. (Using Lemma \ref{lem:1} it follows that the existence of such a sequence exactly characterizes open subsets of $E$.)
  \end{lemma}

  \paragraph{Proof of Lemma \ref{lem:monotone_limit}}
  This seems to only\footnote{
Specifically that $E$ be perfectly normal. However, $E$ is perfectly normal if and only if it is normal and every closed set is $G_{\delta}$, every closed set is $G_{\delta}$ if and only if every open set is $F_{\sigma}$, so $E$ is perfectly normal if and only if it is normal and all open subsets are $F_{\sigma}$. In particular metrizability is not directly relevant here, even though all metrizable spaces are perfectly normal. That being said, known counterexamples of perfectly normal spaces that are not metrizable, e.g. the Arens-Fort space or the Sorgenfrey line, are all fairly ``weird'' and thus not likely to be relevant to the practice of data science.
} require that $E$ be normal and that all open subsets be $F_{\sigma}$. First, using the fact that $U$ open is $F_{\sigma}$, write $U=\bigcup\limits_{n=1}^{\infty} F_n$ with all $F_n$ closed\footnote{We can assume without loss of generality that the sequence of closed sets is increasing, by taking $\tilde{F}_n := \bigcup\limits_{m \le n} F_m$ if necessary, noting the fact that the union of finitely many closed sets is again always closed.}. Next follows an induction argument.

For the base case where $n=1$, note that the fact that $E$ is perfectly normal means Urysohn's lemma can be applied to the closed sets $F_1$ and $U^c$ in such a way to give a continuous function $f_1 := g_1: E \to [0,1]$ such that $g_1(e) = 0$ for all $e \in U^c$, $g_1(e) = 1$ for all $e \in F_1$, and (this is the part requiring $E$ to be perfectly normal) $0 < g_1(e) < 1$ for all $e \in U \setminus F_1$. Because there are no $n$ smaller than $1$, the monotonicity requirement is vacuously satisfied.

Now assume for $n\ge 1$ the induction hypothesis holds that for all $m \le n$ there exists a continuous $f_m: E \to [0,1]$ such that $f_m(e) = 0$ for all $e \in U^c$, $f_m(e) = 1$ for all $e\in F_m$, $0 < f_m(e) < 1$ for all $e \in U \setminus F_m$, and that for all $n_1, n_2 \le n$  with $n_1 \le n_2$ that one has for all $e \in E$ that $f_{n_1}(e) \le f_{n_2}(e)$. To prove the claim, we now need to show that there exists a continuous $f_{n+1}: E \to [0,1]$ such that $f_{m+1} = 0$ for all $e \in U^c$, $f_{n+1}(e)=1$ for all $e \in F_{n+1}$, $0 < f_{n+1}(e) <1$ for all $e \in U \setminus F_{n+1}$, and that for all $m \le n$ that for all $e \in E$ one has $f_m(e) \le f_{n+1}(e)$.

Because $E$ is perfectly normal, again Urysohn's lemma can be applied to the closed sets $F_{n+1}$ and $U^c$ in such a way to give a continuous function $g_{n+1}: E \to [0,1]$ such that $g_{n+1}(e)=0$ for all $ e \in U^c$, $g_{n+1}(e) = 1$ for all $e \in F_{n+1}$, and $0 < g_{n+1}(e) < 1$ for all $ e \in U \setminus F_{n+1}$. Now define $f_{n+1} := \max\{ g_{n+1} , f_n \}$. As the maximum of two continuous functions, $f_{n+1}$ is itself continuous. Likewise, because for all $e \in U^c$ we have that $f_n(e) = g_{n+1}(e) = 0$, we also have that $f_{n+1}(e) = 0$ for all $ e \in U^c$. Because $f_n(e) \le 1$ for all $e \in E$, and $g_{n+1}(e) = 1$ for all $e \in F_{n+1}$, it follows that for all $e \in F_{n+1}$ one has
\begin{equation*}
f_{n+1}(e) = \max\{ f_n(e), g_{n+1}(e) \} = \max \{ f_n(e), 1 \} = 1 \,.  
\end{equation*}
Finally, because $F_n \subseteq F_{n+1}$, we have that $U \setminus F_{n+1} \subseteq U \setminus F_n$, so $0 < f_n(e) < 1$ for all $ e \in U \setminus F_n$ in turn implies that $0 < f_n(e) < 1$ for all $e \in U \setminus F_{n+1}$. Therefore
\begin{equation*}
  0 < f_{n+1}(e) = \max\{ f_n(e), g_{n+1}(e)  \} < 1
\end{equation*}
 for all $e \in U \setminus F_{n+1}$. Therefore we have proven the induction step.

The only claim that remains to be verified is that for all $e \in E$ one has  $\lim\limits_{n \to \infty} f_n(e) = \indicator{U}(e)$. This of course splits into two cases: first verifying that, for all $e \in U^c$, that $\lim\limits_{n \to \infty} f_n(e) = \indicator{U}(e) = 0$, and second verifying that, for all $e \in U$, that $\lim\limits_{n \to \infty} f_n(e) = \indicator{U}(e) = 1$.

For the first verification, note that by definition of the $f_n$ we have that for all $ e \in U^c$ one has $f_n(e) = 0$. Therefore for every $ e \in U^c$ the corresponding sequence $\{f_n(e)\}$ is the constant sequence $0$, which obviously converges and with $0$ as its limit.

For the second verification observe that, because the $F_n$ are assumed to be increasing, i.e. $F_{n+1} \supseteq F_n$ for all $n$, \textit{and} that $\bigcup\limits_{n=1}^{\infty} F_n = U$ (i.e. we can't have the $F_n$ ``stabilize'' at some proper subset of $U$) that for every $e \in U$ there must exist some $n_e$ such that for all $n \ge n_e$, $ e \in F_n$. Thus for all $n \ge n_e$, by definition of the $f_n$ we have that $f_n(e) = 1$. Thus for every $e \in U$ the sequence $\{f_n(e)\}$ equals $1$ for all but finitely many terms, again obviously forcing the sequence to converge and to the limit $1$.

  \paragraph{Comments about Approximating Open Indicators from Below}
  In the case that $E$ is locally compact Polish, as explained above it follows that $E$ is both locally compact and $\sigma$-compact. Because the finite union of compact sets is again compact, being $\sigma$-compact is equivalent to equaling an \textit{increasing} countable union of compact sets. Because any open subset of a locally compact space is again locally compact, any open subset of $E$ is locally compact. Likewise, in a locally compact Polish space any open set is $F_{\sigma}$ and $F_{\sigma}$ subsets of all $\sigma$-compact spaces are themselves $\sigma$-compact, thus any open subset of $E$ is also $\sigma$-compact. Therefore we can assume without loss of generality in the case that $E$ is a locally compact Polish space that the increasing sequence of closed sets whose union equals the open subset $U$ consists exclusively of compact sets, and that for each $n$ we have $K_{n+1} \supseteq Int(K_{n+1}) \supseteq K_n$, where $Int(K_{n+1})$ denotes the topological interior (with respect to either $E$ or $U$, both notions coincide because $U$ is open). Such ``(strong) exhaustions by compact sets'' can be technically useful.

  \subsubsection{Use for Censoring Subsets}
  Let $U$ be an open subset of a ``generalized edge space'' $E$, i.e. a locally compact Polish space $E$. Given a monotonically non-decreasing sequence of continuous functions $f_n: E \to [0,1]$ which converge pointwise to the indicator function of $U$, it follows that the sequence of functions $1 - f_n$ are a non-increasing sequence of continuous functions which converge pointwise to the indicator function of $U^c$. If we replace/augment such a convergent sequence to get a convergent net indexed by the directed set $[0,1]$, then we get non-increasing (as $t$ in creases) functions that converge pointwise to the indicator of $U^c$ as $ t \to 1$. Re-indexing by $1 -t$, we get non-decreasing (as $t$ increases) functions that converge pointwise to the indicator of $U^c$ as $t \to 0$. The details of this will be explained below, but this is to be interpreted as progressively removing the ``censoring'' of the subset $U$ as $t$ increases, approaching no censoring as $t \to 1$, and full censoring as $t \to 0$. While the important aspect for definitions is that the ``easing of censoring'' is done monotonically, it is also nice to know that this can be constructed in a way that is continuous.

  Specifically, first consider the sequence $t_n$ defined on $[0,1]$, and converging towards $0$, defined for all $n \in \mathbb{N}$ such that\footnote{
This is probably the simplest example, but any monotonically decreasing sequence bounded within $[0,1]$ with $t_0=1$ and whose limit is $0$ will work.
} $t_n = 2^{-n}$. Define ``censoring functions'' $\xi_t: E \to [0,1]$ for all $t \in [0,1]$ as follows. First, $\xi_0$ is the constant function identically equal to $1$ for all $e \in E$. Then for all $n \ge 1$, define $\xi_{t_n}$ to be $\xi_{t_n}:= 1 - f_n$. Then for any other value of $t \in [0,1]$, first identify the $m \in \mathbb{N}$ such that $t_m = \sup_{n \in \mathbb{N}} \{ t_n: t_n < t \}$. Such an $m$ must exist because the $t_n$ are monotonically decreasing and thus there can only be finitely many of them greater than $t$. Then basically define $\xi_t$ as the corresponding ``interpolation'' between $\xi_{t_{m-1}}$ and $\xi_{t_m}$, i.e.
\begin{equation*}
  \xi_t = \frac{ (t_m - t)}{(t_m - t_{m-1})} \xi_{t_{m-1}} + \frac{(t_m - t_{m-1}) - (t_m -t)}{(t_m - t_{m-1})} \xi_{t_m} \,.
\end{equation*}
From these definitions it follows that:
\begin{itemize}
\item the $\xi_t$ are all continuous for $t > 0$ (being the convex combination of continuous functions),
  
\item as $t \to 1$ the $\xi_t$ approach the constant function identically equal to $1$,
  
\item as $t \to 0$ the $\xi_t$ approach the (discontinuous) indicator function of $U^c$ (which equals $0$ on $U$ and $1$ everywhere else),
  
\item the $\xi_t$ are monotonically non-decreasing as $t$ increases, i.e. for all $t_1 < t_2$ we have for all $e \in E$ that $\xi_{t_1}(e) \le \xi_{t_2}(e)$.
\end{itemize}
Thus the $\xi_t$ define a continuous ``homotopy'' from complete censoring of $U$ as $t \to 0$ to no ``censoring'' at all as $t \to 1$. For $t > 0$, the $\xi_t \in C(E, \mathbb{R})$ (i.e. are continuous), so their graph-loci are closed subsets of $E \times \mathbb{R}$. Thus I expect, i.e. conjecture and have not rigorously verified, that we would have a homotopy, in the rigorous sense of a continuous function $[0,1] \to (E \times \mathbb{R}$) connecting the graph-loci of the $\xi_t$ for $t > 0$ with (as $t \to 0$) the closed subset of $E \times \mathbb{R}$ that equals $(U^c \times \{1\}) \cup (\partial(U) \times [0,1] ) \cup (U \times \{0\})$ (which contains, but is a strict superset of, the graph-locus of the discontinuous indicator function of $U$). Here $\partial(U)$ equals the topological boundary of $U$, i.e. the closure of $U$ minus the interior of $U$, or equivalently $U^c$ (which is closed and thus equals its own closure) minus its interior (because the topological boundary of any subset equals that of its complement).

  \subsubsection{Monotone Uncensoring of $U \subseteq E$}
  \begin{definition}
    A family of functions $(\xi_t: E \to \mathbb{R} )_{t \in [0,1]}$ such that
    \begin{itemize}
    \item for all $t > 0$ we have $\xi_t \in C(E, \mathbb{R})$ (i.e. $\xi_t$ is continuous),
      
    \item the graph-loci of the $\xi_t$ ``vary continously with $t$,
      
    \item as $ t \to 1$ the $\xi_t$ approach the constant function $1: E \to \mathbb{R}$ pointwise,
      
    \item as $t \to 0$ the $\xi_t$ approach the indicator function of $U^c$ pointwise,
      
    \item the $\xi_t$ are pointwise monotonically non-decreasing as $t$ increases.
    \end{itemize}
  \end{definition}
I claim to have demonstrated previously that such a monotone uncensoring exists for any open subset $U \subseteq E$, and that satisfying the first condition (that the $\xi_t$ be continuous for $t > 0$) requires that $U \subseteq E$ be open.

  \subsubsection{Comments about Monotone Uncensorings}
  Future work should probably seek to determine whether there is a sensible way to require the $\xi_t$ for $t > 0$ to only be Borel-measurable, rather than necessarily continuous. Again, we showed above that ``continuous'' monotone uncensorings should only exist for open subsets of $E$, so one question such future work would have to address is which kinds of subsets of $E$ would admit ``measurable'' monotone uncensorings.

  The definition of (continuous) monotone uncensoring is used to define the (continuous versions of the) monotonicity property and double penalization property for comparison functions of (generalized) networks. Thus it is important to verify that such families of functions actually do exist.

  When $E$ has the discrete topology (and thus every subset is open), the condition that $U$ be open is obviously not at all restrictive.

\end{coolsubappendices}
\end{coolcontents}

\clearpage

\bibliographystyle{amsalpha}
\bibliography{report.bib}

\clearpage

\begin{appendices}
  
\setcounter{chapter}{0}
\renewcommand{\theHchapter}{A.\arabic{chapter}}
\renewcommand{\thechapter}{A.\arabic{chapter}}

\chapter{Relevant Background on Differential Equations}
\label{chap:relev-diff-equat}

The goals of the simulations were to follow the precedent of prior work \cite{Venturelli} \cite{Fisher2014} \cite{Marino2014} \cite{eco_model_time_series} \cite{Mounier2008} in postulating a likelihood model similar to the generalized Lotka-Volterra equations (cf. again section \ref{sec:form-trans-likel}), while also being more biologically realistic in both (i) not having any restrictions on interaction coefficients (which would seem to be arbitrary from a biological point of view), and (ii) addressing the capacity for nutrient depletion and waste product accumulation to moderate changes in population size, something which seems especially relevant, and even inappropriate to ignore, inside of the highly resource-limited (oligotrophic) environment of a microfluidic droplet.

\section{Exponential Growth: Unlimited Resources}
\label{sec:expon-growth:-unlim}

The exponential differential equation describes the idealized situation where the rate of change in the population is directly proportional to the current number of cells:

\begin{equation}
  \label{eq:exponential}
\frac{\mathrm{d}
\counts }{ \mathrm{d}t} = \baserate[] \counts \,,
  \end{equation}
which is equivalent (for populations with more than zero cells) to the even simpler equation:

\begin{equation}
  \label{eq:log_form_exponential}
\frac{\mathrm{d} (\log
\counts)}{ \mathrm{d}t} = \baserate[] \,.
  \end{equation}
The second form emphasizes how this idealized situation corresponds to what is often referred to as a constant growth (or decay) rate.

\section{Nutrient Limitation and Indirect Interactions}
\label{sec:nutr-limit-indir}

The exponential model is often considered adequate for describing the growth under copiotrophic conditions immediately following the initial lag phase. However, since the effects of nutrient depletion and waste product accumulation increase with the size of the population, for large enough populations it becomes unrealistic to model the population size's effect on the rate of change as being constant. Instead an additional multiplicative factor can be included to describe how growth is expected to slow down as nutrients are depleted and waste products accumulate:

\begin{equation}
  \label{eq:17}
\frac{\mathrm{d} (\log
x)}{ \mathrm{d}t} = \baserate[] \left( 1 - \frac{x}{\carryingcapacity[]}  \right) \,.
  \end{equation}
This is called the logistic differential equation. As the population size continues to approach the constant $\carryingcapacity[]$, growth slows down (asymptotically) towards nothing, causing $\carryingcapacity[]$ to represent an upper bound on the total possible population size. This upper bound is often referred to as the \textit{carrying capacity} of the environment which supports the population.

Assuming that the population sizes of any given strain do not affect the population sizes of the other strains, it is straightforward to model the simultaneous exponential growth of multiple strains via a system of differential equations:

\begin{equation}
  \label{eq:exponential_system}
\frac{\mathrm{d} (\log
\counts[\specie]  )}{ \mathrm{d}t} = \baserate \,.
  \end{equation}

When still assuming that the population sizes of different strains do not affect one another, it is also straightforward in exactly the same manner to model the simultaneous logistic growth of multiple strains using a system of differential equations:

\begin{equation}
  \label{eq:logistic_system}
\frac{\mathrm{d} (\log \counts[\specie] )}{ \mathrm{d}t} = \baserate \left( 1 - \frac{\counts[\specie]}{ \carryingcapacity  }  \right) \,.
  \end{equation}

  The above system of equations (\ref{eq:logistic_system}) is unrealistic, however, inasmuch as it assumes not only that the population sizes of different strains do not directly affect one another, but also because it assumes that the environment has a separate carrying capacity for each strain, independent of that for all other strains.

Especially in an environment like the interior of a single droplet in a microfluidics experiment, the same nutrient supply will often be common/shared among all strains, and the accumulation of waste products will affect all strains simultaneously. In other words, in a limited resource environment strains will almost always interact with each other at least \textit{indirectly}, with the \textit{indirect} interactions being mediated by the nutrient supply and concentration of waste products. Therefore it is often more realistic to assume a single carrying capacity for the total aggregate population size of all strains in the environment. The following system of equations better models the aforementioned situation:

\begin{equation}
\label{eq:common_carrying_capacity}
\frac{\mathrm{d} (\log \counts[\specie] )}{ \mathrm{d}t} = \baserate  \left(1 - \frac{\sum_{{\specie[] } } \counts[{\specie[]} ]}{ \carryingcapacity[]  } \right) \,,
  \end{equation}
since as the total aggregate population size of all strains (asymptotically) approaches the common and shared carrying capacity $\carryingcapacity[]$ of the environment, the growth of all strains will simultaneously slow (asymptotically) towards zero. It is worth noting though that in practice different strains can have wildly varying metabolisms, such that what constitutes a nutrient or a harmful waste product could differ substantially between strains, and making the above description overly simplistic. For examples of models accounting for the potentially different metabolisms of different strains, see e.g. \cite{Estrela2022}, \cite{Goyal2021}, \cite{Estrela2021}, \cite{Erez2020}, \cite{Goldford2018}, or \cite{Momeni2017}. (For an overview, see \cite{Gonze2018}.) However, fully addressing this issue falls outside of the scope of the current manuscript and is left for future work.

\subsection{Comparison with Previous Work}
\label{sec:comp-with-prev-indirect}

The paper \cite{Marino2014} uses equations analogous to (\ref{eq:logistic_system}), with $\carryingcapacity = \carryingcapacity[]$ for all $\strain$ for a common $\carryingcapacity[]$, and claims that this alone is sufficient alone to enforce the constraint that $\sum_{{\specie[] } } \counts[{\specie[]} ] \le \carryingcapacity[]$ always. However, that does not seem to be true. As far as I can tell, with an equation of that form alone, there is no reason at all why the (logarithmic) derivatives of all strains should approach $0$ as $\sum_{{\specie[] } } \counts[{\specie[]} ] \to \carryingcapacity[]$. If the growth rates do not asymptotically approach $0$ as the total population sizes $\sum_{{\specie[] } } \counts[{\specie[]} ] $ of all strains approaches the supposedly common carrying capacity $\carryingcapacity[]$, then there is no reason to expect in general that $\sum_{{\specie[] } } \counts[{\specie[]} ] \le \carryingcapacity[]$.

For example, based on (\ref{eq:logistic_system}), as $\counts[\strain_1] \to \carryingcapacity[]/2$ and $\counts[\strain_2] \to \carryingcapacity[]/2$ the growth rates of $\strain_1$ and $\strain_2$ would not seem to be approach $0$. Yet under those circumstances $\sum_{{\specie[] } } \counts[{\specie[]} ] \le \carryingcapacity[]$. Hence, absent a flaw in the above reasoning, it seems clear that (\ref{eq:logistic_system}) is incapable of enforcing a common carrying capacity constraint shared by all strains simultaneously.

In contrast, it is obviously the case for the proposed equation (\ref{eq:common_carrying_capacity}) that the growth rates of all strains will asymptotically approach $0$ as $\sum_{{\specie[] } } \counts[{\specie[]} ] \to \carryingcapacity[]$.

\section{Direct Interactions}
\label{sec:direct-interactions}
 
Even focusing on the initial regime where diminishing nutrient supplies and increasing waste product concentrations have not yet become significant factors affecting growth, the assumption in (\ref{eq:exponential_system}) that the growth of any given strain's population is unaffected by the population sizes of other strains can still be unrealistic. This is of course because of \textit{direct} interactions between different strains of microbes, which can either support or inhibit their growth, and which are the primary focus of interest of the current manuscript. The generalized Lotka-Volterra equations seek to account for direct interactions by adding to (\ref{eq:exponential_system}) a linear interaction term:

\begin{equation}
\label{eq:glv_equations}
\frac{\mathrm{d} (\log \counts[\specie] )}{ \mathrm{d}t} = \baserate + \sum_{\specie[]=1}^{\Species} \interaction{\specie[]}{ \specie} \counts[{\specie[]}] \,.
\end{equation}
The model describes interactions between different strains through the magnitude and sign of the coefficients in the portion of the affine expressions above which are a linear combination of the population sizes of all of the strains. Positive coefficients correspond to mutualistic interactions between strains whereas negative coefficients correspond to antagonistic interactions between strains.

\section{Combining Indirect and Direct Interactions}
\label{sec:comb-indir-direct}

The simulations used in this paper seek to capture both indirect and direct interactions between microbial strains by combining models (\ref{eq:common_carrying_capacity}) and (\ref{eq:glv_equations}) above in a straightforward way:

\begin{equation}
\label{eq:simulation_equation}
  \frac{\mathrm{d} (\log
\counts[\specie])}{ \mathrm{d}t} =\left(  \baserate + \sum_{\specie[]=1}^{\Species} \interaction{\specie[]}{ \specie} \counts[{\specie[]}] \right) \left(1 - \frac{\sum_{{\specie[]}} \counts[{\specie[]}]}{ \carryingcapacity[]  } \right) \,.
\end{equation}
This preserves the property that the change in population size will slow towards zero as the total aggregate population size approaches the environment's carrying capacity, while still retaining ability for the population sizes of various strains to directly affect one another. This makes sense even in the case of antagonistic interactions, since as the growth of the antagonizing strain slows towards zero (along with the growth of all growing strains), one would expect that the magnitude of the antagonizing interaction would also slow towards zero, making it sensible for the decay in the population size in the antagonized strain to slow towards zero as well.

Also, unlike the regular generalized Lotka-Volterra equations, it is intended that, for any possible choice of direct interaction coefficients, the population sizes of any solutions will remain bounded for any initial condition for which the carrying capacity of the aggregate population is not exceeded. (Cf. for example \cite{Venturelli} which excludes certain combinations of coefficient values to prevent the possibility of blow-up. To me that seems arbitrary from a purely biological point of view.) This is an additional source of realism for (\ref{eq:simulation_equation}), since there does not seem to be any a priori biological reason for why certain combinations of values of interaction coefficients should be prohibited. This quote from \cite{Marino2014} is precedent for the above viewpoint on parameter constraints:
\begin{quote}
  One approach to improving parameter estimation is to superimpose constraints on the parameters; however, unless these constraints can be justified both biologically and mathematically, they should not be enforced because the results can be greatly affected.
\end{quote}

\subsection{Comparison with Previous Work}
\label{sec:comp-with-prev-combined}

Note that the derivation of (\ref{eq:simulation_equation}) given above is not the first attempt to combine logistic growth constraints with gLV-like direct interactions. Notably \cite{Marino2014} proposes an equation of the form (modulo logarithms and differences in notation):
\begin{equation}
\label{eq:murine_model_equation}
  \frac{\mathrm{d} (\log
\counts[\specie])}{ \mathrm{d}t} =  \baserate  \left(1 - \frac{ \counts[{\specie[]}]}{ \carryingcapacity[]  } \right) + \sum_{\specie[]=1}^{\Species} \interaction{\specie[]}{ \specie} \counts[{\specie[]}]  \,.
\end{equation}
In section \ref{sec:comp-with-prev-indirect} above I already explained why the proposed implementation of a common carrying capacity in (\ref{eq:murine_model_equation}) does not seem to enforce the intended constraint $\sum_{{\specie[] } } \counts[{\specie[]} ] \le \carryingcapacity[]$. Let us apply the fix proposed in \ref{sec:comp-with-prev-indirect}, namely switching the attempted carrying capacity implementation from a form similar to (\ref{eq:logistic_system}) to a form similar instead to (\ref{eq:common_carrying_capacity}):
\begin{equation}
\label{eq:murine_model_equation_improved}
  \frac{\mathrm{d} (\log
\counts[\specie])}{ \mathrm{d}t} = \baserate \left(1 - \frac{\sum_{{\specie[]}} \counts[{\specie[]}]}{ \carryingcapacity[]  } \right) + \sum_{\specie[]=1}^{\Species} \interaction{\specie[]}{ \specie} \counts[{\specie[]}]  \,.
\end{equation}
However (\ref{eq:murine_model_equation_improved}) also seems to fail to enforce the intended constraint $\sum_{{\specie[] } } \counts[{\specie[]} ] \le \carryingcapacity[]$, for the same reason that (\ref{eq:logistic_system}) does. Namely, without distributing the ``logistic factor'' across the entire expression, which \textit{is} done in (\ref{eq:simulation_equation}), again there does not seem to be any reason why we should have for (\ref{eq:murine_model_equation_improved}) that all growth rates simultaneously approach $0$ as $\sum_{{\specie[] } } \counts[{\specie[]} ] \to \carryingcapacity[]$. In contrast, (\ref{eq:simulation_equation}) was explicitly designed so that all growth rates simultaneously approach $0$ as $\sum_{{\specie[] } } \counts[{\specie[]} ] \to \carryingcapacity[]$. Hence it seems much more plausible to me that (\ref{eq:simulation_equation}) enforces the intended constraint $\sum_{{\specie[] } } \counts[{\specie[]} ] \le \carryingcapacity[]$, whereas I do not see why either (\ref{eq:murine_model_equation}) or (\ref{eq:murine_model_equation_improved}) would. Thus (\ref{eq:simulation_equation}) seems to be an improvement on previous work.

\section{Common Framework}
\label{sec:common-framework}

All of the above systems of equations fit into the following framework:

\begin{equation}
  \label{eq:framework}
    \frac{\mathrm{d} (\log
\counts[\specie])}{ \mathrm{d}t} = \rateterm (\counts[1], \dots, \counts[\Species]) \left(1 - \frac{\limitingfactor(\counts[1], \dots, \counts[\Species])}{ \carryingcapacity  } \right) \,,
\end{equation}
where $\rateterm$ corresponds to the growth of the given strain under idealized conditions, and $\limitingfactor$ is a quantity for which the external environment imposes some finite carrying capacity $\carryingcapacity$, such that $\limitingfactor \rightarrow  \carryingcapacity$ causes all changes in the population size of strain $\specie$ to slow towards zero. For equations (\ref{eq:exponential_system}), (\ref{eq:logistic_system}), and (\ref{eq:common_carrying_capacity}), one has that $\rateterm = \baserate$, whereas for equations (\ref{eq:glv_equations}) and (\ref{eq:simulation_equation}) one has instead that $\rateterm = \baserate + \sum_{\specie[]=1}^{\Species} \interaction{\specie[]}{ \specie} \counts[{\specie[]}]$. For equations (\ref{eq:exponential_system}) and (\ref{eq:glv_equations}) one has that $\limitingfactor = 0$, for equation (\ref{eq:logistic_system}) one has that $\limitingfactor = \counts[\specie]$, and for equations (\ref{eq:common_carrying_capacity}) and (\ref{eq:simulation_equation}) one has instead that $\limitingfactor = \sum_{{\specie[]}} \counts[{\specie[]}]$ as well as that $\carryingcapacity = \carryingcapacity[]$ for all $\specie \in [\Species]$ for a given fixed constant $\carryingcapacity[]$.

Of course in order for its effect on changes in the population size of strain $\specie$ to have any clear interpretation as an external limitation on the magnitude of changes in the system, one needs that $0 \le \limitingfactor \le \carryingcapacity$ so that the factor multiplying $\rateterm$ is always a value between $0$ and $1$. As mentioned already before, $\limitingfactor = \sum_{{\specie[]}} \counts[{\specie[]}]$ satisfies this property in equations (\ref{eq:common_carrying_capacity}) and (\ref{eq:simulation_equation}), or at least for the region of phase space between the two hyperplanes $\sum_{{\specie[]}} \counts[{\specie[]}] =0$ and $\sum_{{\specie[]}} \counts[{\specie[]}] = \carryingcapacity[]$ (recall that $\carryingcapacity[1]=\dots=\carryingcapacity[\Species]=\carryingcapacity[]$ for these equations). This is fine since the aformentioned region is the only biologically realistic/sensible/meaningful region of phase space anyway, and any solution of (\ref{eq:simulation_equation}) beginning in that region will remain inside that region for all time.

\section{``Competitive Lotka-Volterra Equations''}
\label{sec:comp-lotka-volt}

There is at least one other similar system of differential equations intended to model both the direct interactions corresponding to the system (\ref{eq:glv_equations}) and external limitations due to finiteness of nutrient supply and the accumulation of waste products, which is often called the ``Competitive Lotka-Volterra equations''. For the competitive Lotka-Volterra equations, using the general framework of (\ref{eq:framework}) one has 

\begin{equation}
  \label{eq:competitive-lv}
  \rateterm = \baserate \,, \quad\quad\quad \limitingfactor = \sum_{{\specie[]}=1}^{\Species} \interaction{\specie[]}{\specie} \counts[{\specie[]}]\,.
\end{equation}
Apart from the issue that $\carryingcapacity$ may have different values for different strains (which can easily be rectified), a bigger issue with these equations, and why they were not used for these simulations, is that there is no way to guarantee that $0 \le \limitingfactor \le 1$, at least not without restrictions on the interaction coefficients which from a biological point of view would be arbitrary. Moreover $ \limitingfactor = \sum_{{\specie[]}=1}^{\Species} \interaction{\specie[]}{\specie} \counts[{\specie[]}]$ does not correspond to any scientifically meaningful or experimentally observable quantity, making the meaning of the restrictions imposed on the growth of strain $s$ by the $(1 - \limitingfactor/\carryingcapacity )$ factor difficult, if not impossible, to give any scientifically meaningful interpretation. In contrast, $\limitingfactor = \sum_{{\specie[]}} \counts[{\specie[]}]$ is not only possible to measure experimentally, but is probably easier to measure than any of the $\counts[\specie]$ individually (since the accurate assignment of microbes' taxonomic identifications can be challenging to do in a high-throughput manner).

This last reason for not using the competitive Lotka-Volterra equations is more subjective, but it seems that the purpose of the linear term in the generalized Lotka-Volterra equations (\ref{eq:glv_equations}) is to generalize the constant expression used for $\rateterm$ in (\ref{eq:exponential_system}) with an affine expression instead, while still leaving the absence of constraints on growth unchanged when compared to (\ref{eq:exponential_system}). Therefore, the competitive Lotka-Volterra equations (\ref{eq:competitive-lv}), in using the linear expression for the growth-constraining term $\limitingfactor$, and leaving the growth describing factor $\rateterm$ as a constant and thus ungeneralized, seems to not follow the ``spirit'' of the generalized Lotka-Volterra equations (\ref{eq:glv_equations}), and thus did not seem adequately comparable to prior work \cite{Venturelli} \cite{Fisher2014} \cite{eco_model_time_series} \cite{Mounier2008} using those equations.

\chapter{Simulating Growth of Microbes in Droplets}
\label{chap:simul-impl}

\begin{figure}[H]
  \centering
  \includegraphics[width=\textwidth,keepaspectratio]{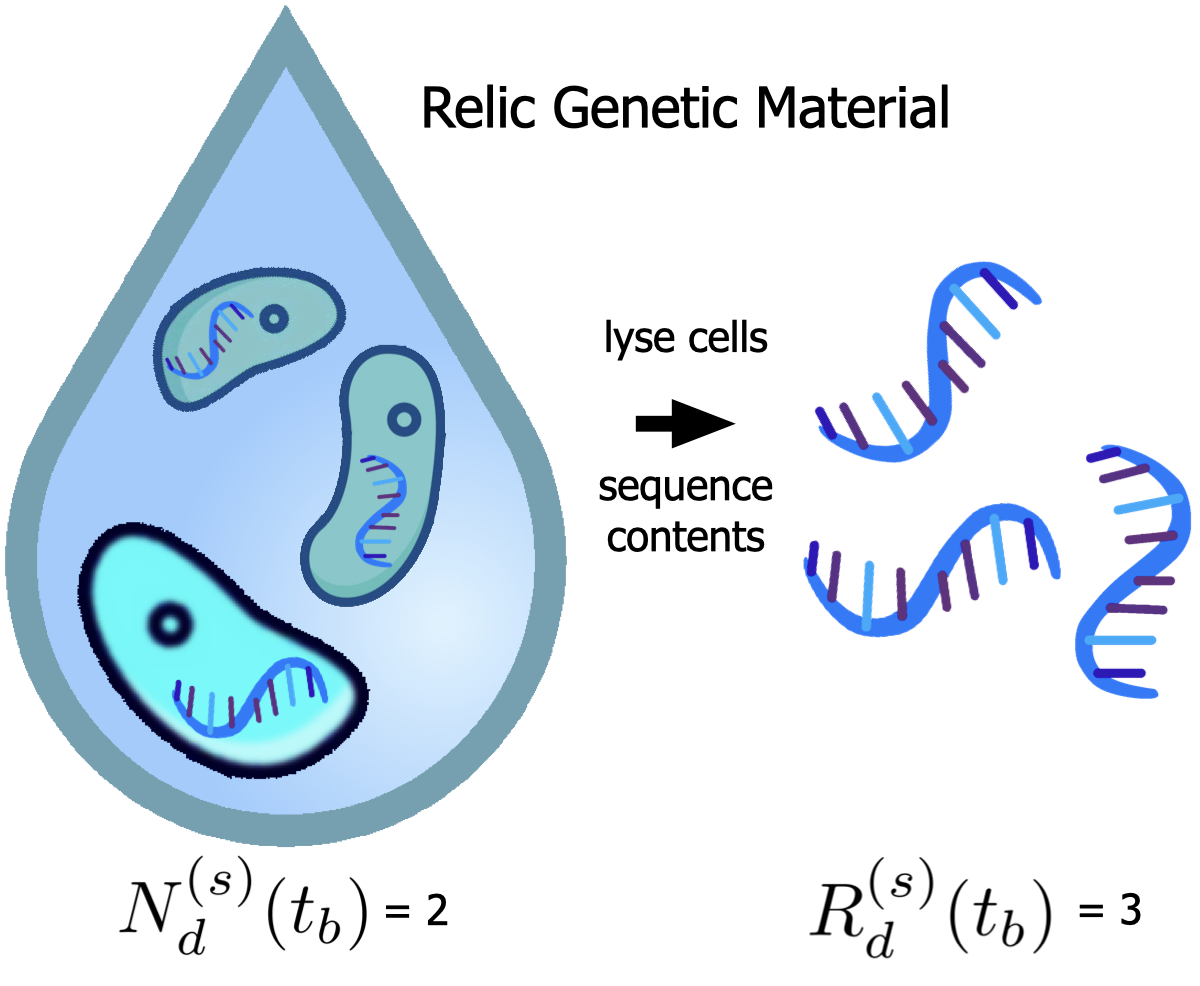}
  \caption[]{At the end of the growth period, two cells are still alive in the droplet, but one is dead. After lysing the cells and sequencing their contents, the reported number of reads is three. The dead cell is effectively counted as if it were alive. Cf. figure \ref{fig:relic_dna_intro}.}
\label{fig:relic_dna}
\end{figure}

In practice, observed measurements of the numbers of reads $\vreads_{\droplet}(\time_{\batch})$ for each strain in a droplet $\droplet$ from batch $\batch$, meant to correspond to the numbers of cells from each strain currently alive $\vabundance_{\droplet}(\time_{\batch})$ in droplet $\droplet$ at time $\time_{\batch}$, will instead better correspond to the numbers of cells which had been alive at any time before time $\time_{\batch}$ (and which have not yet had their genetic material scavenged by other microbes). Compare with Figure \ref{fig:relic_dna}. This discrepancy is because all cells, living or dead, must be lysed in order to be able to sequence the contents of the (previously) living cells. This leads to the contents of any cells which had died before being lysed to most likely also have their genetic material sequenced. One of the goals of the simulations was to explicitly account for this discrepancy between theory and measurements. 

Another goal of these simulations was to explicitly account for the reality of resource limitations in these small droplets, which clearly will not allow for indefinite exponential growth. We required that any solution achieving that goal also not impose any restrictions on the allowed interaction coefficients, since there would not seem to be any biological justification for any such restrictions. In particular, such restrictions seem to most often be made for no other reason than preventing unbounded growth. Thus, in principle, a solution correctly modelling resource limitations should work for any interaction coefficients. Resource limitations inherently limit growth and should be sufficient to do so without artificial constraints on the interaction coefficients. (Cf. the argument in section \ref{sec:comb-indir-direct}.)

A further goal was to have a simulation model that was comparable to prior work \cite{Venturelli} \cite{eco_model_time_series} \cite{Mounier2008} \cite{Fisher2014} \cite{Marino2014}, while also not sacrificing accuracy or relevance for this problem in terms of failing to accomplish the first and second goals. For example, although it was found in \cite{Venturelli} that the generalized Lotka-Volterra equations provided a good approximation to the observed dynamics, this was for manually grown plate cultures which were continuously replated as necessary in order to ensure that the measured populations never encountered resource deprivations. Chapter \ref{chap:relev-diff-equat} describes in detail how the third and second goals were reconciled. 

One benefit of satisfying the last goal above is that it allows looking at the performance of the parametric estimator described in appendix \ref{sec:param-estim-inter} in a context which is both ``fair'' and ``realistic''. ``Fair'' in the sense of being reasonably similar to the models used implicitly in prior work \cite{Venturelli} \cite{eco_model_time_series} \cite{Marino2014} \cite{Mounier2008} \cite{Fisher2014} from which it was derived. ``Realistic'' in the sense that the original models are still significantly misspecified\footnote{
I assume that parametric models are almost always misspecified in practice. Cf. \cite{vanderLaan2011} for discussion.
}.

\section{Modelling Cell Counts under Nutrient Limitation}
\label{sec:modell-cell-counts}

Starting from equation (\ref{eq:simulation_equation}), a discretization is applied to define an Euler step update, and then a stochastic noise term $\noise$ is added to yield
\begin{equation}
\label{eq:simulation_update}
\begin{split}
  &\log \left( \abundance[\specie](\time\!+\!\Delta\time) \right) - \log \left( \abundance[\specie](\time)\right) \\
=&
\begin{cases}
  \left( \Delta \time \left(\displaystyle  \baserate + \sum_{\specie[]=1}^{\Species} \interaction{\specie[]}{ \specie} \abundance[{\specie[]}](\time) \right)  + \noise \right) \frac{\left(\displaystyle \carryingcapacity[] - \sum_{{\specie[]}=1}^{\Species} \abundance[{\specie[]}](\time) \right) }{ \displaystyle \carryingcapacity[]} & \text{when }\abundance[\specie](\time) \not = 0 \\
0 & \text{when }\abundance[\specie](\time)  = 0
\end{cases}
\end{split}
\end{equation}
where all of the $\noise$ are i.i.d. $\gaussian (0, \noisescale^2  \Delta \time)$ for some ``noise scale'' $\noisescale > 0$ (the second parameter is the variance, so e.g. the standard deviation is $\noisescale \sqrt{\Delta \time}$).

The equation (\ref{eq:simulation_update}) has a relationship with (\ref{eq:simulation_equation}) that is analogous to the relationship that the assumed updates (\ref{eq:transitions}) from previous work \cite{Venturelli} \cite{Mounier2008} \cite{Fisher2014} \cite{eco_model_time_series} have with the generalized Lotka-Volterra equations (\ref{eq:glv_equations}). 

This may be equivalent (cf. \cite[section 6.2 or part IV]{Kloeden} to check my work, because I doubt this is entirely correct) to a discretization of the following SDE model:
\begin{equation}
\label{eq:simulation_update_SDE_version}
\begin{split}
  & \mathrm{d}\abundance[\specie](\time) \\
=&
\begin{cases}
 \abundance[\specie](\time) \left( \mathrm{d} \time \left(\displaystyle  \baserate + \sum_{\specie[]=1}^{\Species} \interaction{\specie[]}{ \specie} \abundance[{\specie[]}](\time) \right)  + \noisescale \mathrm{d}W(\time) \right) \frac{\left(\displaystyle \carryingcapacity[] - \sum_{{\specie[]}=1}^{\Species} \abundance[{\specie[]}](\time) \right) }{ \displaystyle \carryingcapacity[]} & \text{when }\abundance[\specie](\time) \not = 0 \\
0 & \text{when }\abundance[\specie](\time)  = 0
\end{cases}
\end{split}
\end{equation}
where $\mathrm{d}W(\time)$ denotes a standard Wiener process/Brownian motion.

Using the notation of equation (\ref{eq:framework}) from chapter \ref{chap:relev-diff-equat}, the update (\ref{eq:simulation_update}) can be written as
\begin{equation}
\label{eq:simulation_update_framework}
\log \left( \abundance[\specie](\time\!+\!\Delta\time) \right) - \log \left( \abundance[\specie](\time)\right) = 
\begin{cases}
  \left( (\Delta \time)\rateterm  + \noise \right) \frac{ \carryingcapacity[] - \limitingfactor}{ \carryingcapacity[]} & \text{when }\abundance[\specie](\time) \not = 0 \\
0 & \text{when }\abundance[\specie](\time)  = 0
\end{cases}
\end{equation}
using the definitions
\begin{equation}
  \label{eq:simulation_update_framework_defns}
  \begin{split}
    \rateterm = & \baserate + \sum_{\specie[]=1}^{\Species} \interaction{\specie[]}{ \specie} \abundance[{\specie[]}](\time) \\
\limitingfactor = &  \sum_{{\specie[]}=1}^{\Species} \abundance[{\specie[]}](\time)
  \end{split} \,.
\end{equation}
It is possible, due to the errors introduced by both discretization and stochasticization, that the population of a droplet in the simulation might temporarily exceed the carrying capacity, which corresponds to a special case of $\sign(\carryingcapacity[] \!-\! \limitingfactor ) = -1$. Without modification to equation (\ref{eq:simulation_update}), this would lead to a biologically extremely unrealistic indefinite explosion in population growth for any strain $\specie$ whose growth due to microbial interactions $\rateterm$ was negative (i.e. decreasing growth). Defining
\begin{equation}
  \label{eq:fix_signs}
  \fixsigns(x,y) :=
  \begin{cases}
    - \sign(x) \sign(y) & \sign(x) = \sign(y) = - 1 \\
1 & \text{else}
  \end{cases} \,,
\end{equation}
the aforementioned problem is fixed by using the update equation
\begin{equation}
\label{eq:simulation_update_framework_sign_fix}
\log \left( \abundance[\specie](\time\!+\!\Delta\time) \right) - \log \left( \abundance[\specie](\time)\right) = 
\begin{cases}
  \fixsigns(\rateterm, (\carryingcapacity[]\!-\!\limitingfactor))\left( (\Delta \time)\rateterm  + \noise \right) \frac{ \carryingcapacity[] \!-\! \limitingfactor}{ \carryingcapacity[]} & \abundance[\specie](\time) \not = 0 \\
0 & \abundance[\specie](\time)  = 0
\end{cases}
\end{equation}
whose corresponding system of differential equations
\begin{equation}
  \label{eq:sign_fix_framework}
    \frac{\mathrm{d} (\log
\counts[\specie])  }{ \mathrm{d}t} = \fixsigns(\rateterm, (\carryingcapacity[]-\limitingfactor)) \rateterm (\counts[1], \dots, \counts[\Species]) \left(1 - \frac{\limitingfactor(\counts[1], \dots, \counts[\Species])}{ \carryingcapacity[] }  \right) \,,
\end{equation}
has solutions which vary smoothly in phase space except for a region of measure zero, where they still vary continuously. Thus any concerns about the realism of equation (\ref{eq:simulation_update_framework}) due to non-smoothness seem likely to be minor in comparison to e.g. the known potential issues for correctly modelling the dynamics of even the unmodified generalized Lotka-Volterra equations (\ref{eq:glv_equations}) in low dimensions \citeAppendix{discrete_glv} (i.e. few strains, e.g. ~2-3) introduced by discretization (especially via Euler steps, cf. e.g. \citeAppendix{Ushiki1982}) or stochasticization (discussed e.g. in the case of continuous time SDE models in \citeAppendix{glv_sde}). Empirically, it seems that with $\Delta \time$ chosen to be small enough, the dynamics should (almost) always be constrained to those of (\ref{eq:simulation_update_framework}), so the dynamics in (\ref{eq:simulation_update_framework_sign_fix}) can be considered a fallback or  last-resort failsafe that usually is irrelevant but nevertheless guarantees no unbounded growth.

\section{Read Counts vs Cell Counts}
\label{sec:read-counts-vs}

In the simulation used for the subsequent analysis, the updates used in (\ref{eq:simulation_update_framework}) are used to define what is assumed to be the true (pseudo)count of the numbers of cells of each strain in each droplet. (Fractional values of the pseudocounts can, for the sake of argument, be assumed to represent incomplete cell division.) We can make further changes so that the simulation keeps track of both (1) the actual number of cells $\vabundance_{\droplet}(\time)$  (until $\time = \time_{\batch}$) that would be predicted to be observed as a consequence of the updates (\ref{eq:simulation_update_framework}) using a perfect experimental apparatus, as well as (2) the number of (16S V4 region) reads $\vreads_{\droplet}(\time)$ that are likely to be observed in practice.

Three modifications are used to model the number of reads for each strain which the researcher is likely to measure. 

One modification accounts for the accumulation of relic genetic material (RNA/DNA) in the droplets, a second modification accounts for the fact that different strains are likely to have different copy numbers of the 16S rRNA gene, and a final modification accounts for the possibility of nonzero droplet merging error at the end of the experiment. Note that no attempts were made to directly account for stochasticities in the amount of marker introduced in each droplet (which is assumed to have a low signal-to-noise ratio ${\frac{\sigma}{\mu}}$ due to the efforts of the experimenter), or for potential PCR amplification biases. However the modification employed to account for 16S rRNA copy number variations resembles (at least superficially) the modelling of PCR amplification bias proposed in previous work \cite{bias_framework}, and thus can likely without loss of generality be assumed to account for both.

\begin{figure}[H]
  \centering
  \includegraphics[width=\textwidth,height=0.93\textheight,keepaspectratio]{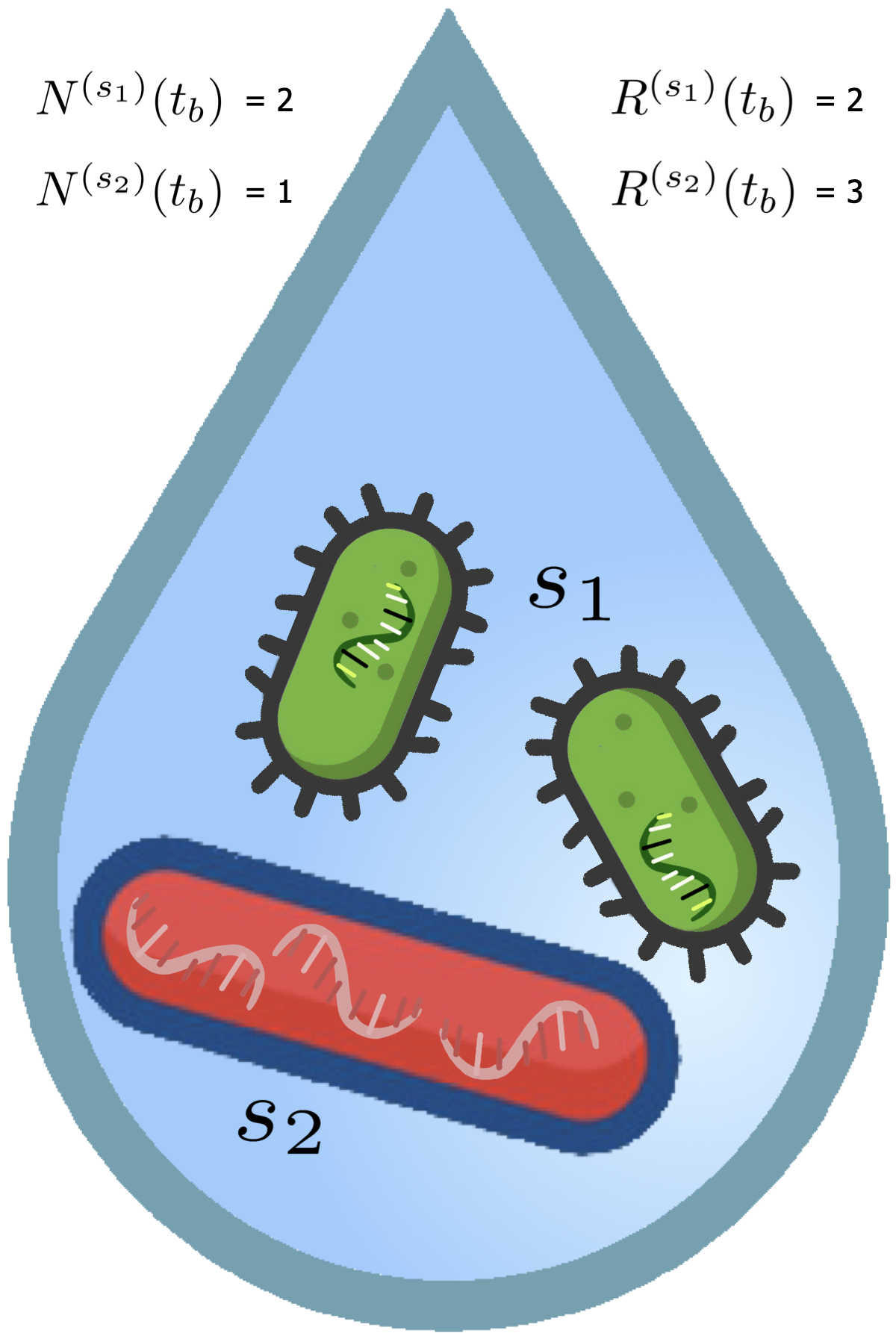}
  \caption[~Copy number problem]{One strain outnumbers the other strain 2:1. Yet because the latter strain has a larger 16S gene copy number, the observed ratio of reads is 2:3.}
  \label{fig:copy_number}
\end{figure}

\begin{figure}[H]
  \centering
  \includegraphics[width=\textwidth,height=\textheight,keepaspectratio]{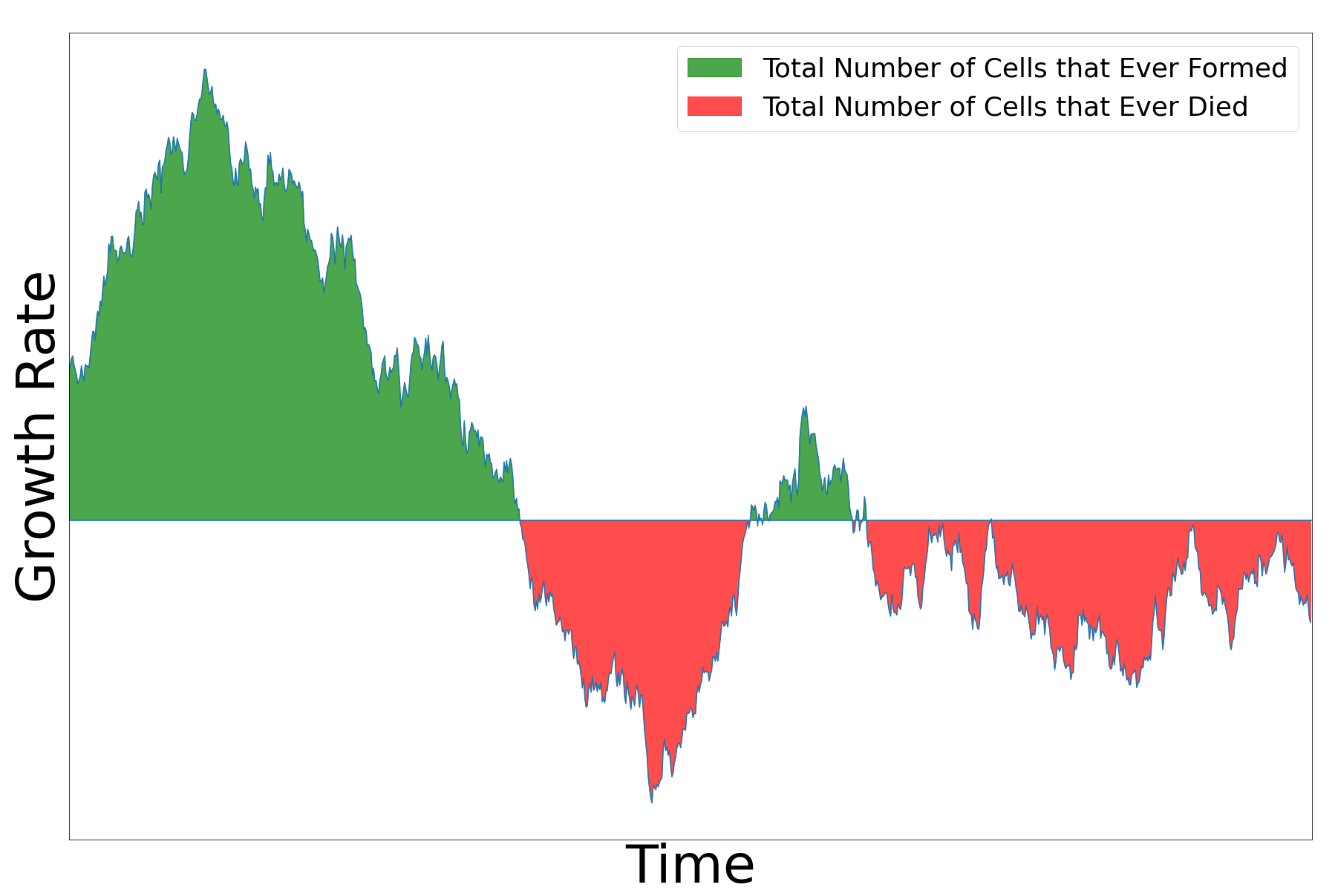}
  \caption[~Counting all cells that have ever lived.]{The true number of cells observed at the end of the simulation is the green area minus the red area, whereas the number of reads actually observed at the end of the simulation is proportional to the green area only. \textit{(NB: This is a heuristic plot illustrating the idea and not data from an actual simulation.})}
  \label{fig:cell_number_integration}
\end{figure}

For each time step, the change in the observed number of reads $\reads[\specie]_{\droplet}(\time)$ for each strain $\specie$ in a given droplet $\droplet$ is assumed to follow from the true (pseudo)counts $\abundance[\specie]_{\droplet}(\time)$ of the numbers of cells for each strain using the update steps:
\begin{equation}
  \label{eq:reads_definitions}
 \vreads_{\droplet}(\time_{\batch[]\!+\! 1}) - \vreads_{\droplet}(\time) =  \left( \copynumbersbias \ast \left( \vabundance_{\droplet}(\time\!+\!\Delta\time) - \vabundance_{\droplet}(\time)  \right)  \right)_+ \,,
\end{equation}
where $\ast$ denotes the entrywise or Hadamard product, $(x)_+$ denotes the positive part of $x$, i.e. $\max\{x,0\}$, and $\copynumbersbias$ denotes the vector whose entries contain the 16S rRNA copy numbers (and/or PCR amplification biases) for each strain. The idea is to take the total number of cells of each strain which have ever been alive in the droplet and then multiply those numbers by the corresponding 16S copy numbers/PCR amplification biases. This is since the total number of currently living cells $\counts[\specie](\time) = \int_0^{\time} \frac{\mathrm{d}\counts[\specie](\tau)}{\mathrm{d}\tau} \mathrm{d}\tau$ is equal to the total number of cells which have ever lived $\int_0^{\time} \left(\frac{\mathrm{d}\counts[\specie](\tau)}{\mathrm{d}\tau} \right)_+ \mathrm{d}\tau$ minus the total number of cells which have ever died $\int_0^{\time} \left(\frac{\mathrm{d}\counts[\specie](\tau)}{\mathrm{d}\tau} \right)_- \mathrm{d}\tau$, $(x)_- := \max\{-x,0\}$. Compare with Figure \ref{fig:cell_number_integration}.

Thus the difference between the total number of cells which have ever lived and the total number of currently living cells, after being multiplied by the copy numbers and/or PCR amplification biases $\copynumbersbias$, can be assumed to approximately correspond to the amount of relic genetic material (RNA/DNA) in the droplet. (At least assuming little to no cell lysis and/or nucleoside scavenging has occurred before the time $\batch$ when the droplet's contents are finally measured.) Thus vector $\copynumbersbias$ (entrywise) multiplied by the total number of cells which have ever lived can be assumed to correspond to the total number of reads observed from both living cells and from relic genetic material, thus (\ref{eq:reads_definitions}) encompasses both of the first two aforementioned modifications.

Regarding the last of the three aforementioned modifications, at the end of the simulation, if a non-zero merging error rate $\merging$ is assumed, then $\lfloor \Droplets \cdot (\merging/2) \rfloor$ droplets are randomly selected without replacement to be merged into other droplets, and another $\lfloor \Droplets \cdot (\merging/2) \rfloor$ droplets are randomly selected without replacement to be the droplets which are randomly merged into. Then for each of the $\lfloor \Droplets \cdot (\merging/2) \rfloor$ pairs of droplets, a new droplet consisting of the sum of the observed contents from each droplet of the pair is created, with the values from the two older droplets discarded (as they are assumed to be unobserved). (The indices of which droplets were merged with which other droplets are recorded, and the original droplets for the cell pseudocounts are left unchanged.)

\section{Modelling PCR Amplification}
\label{sec:modell-pcr-ampl}

A (naively) simple model assumes that the PCR amplification factor for each kind $k$ of genetic material (either the spikein gene or the 16S rRNA gene of one of the $\Strains$ strains) equals a constant $a$ multiplied by a kind-specific instance of a log-normal random variable $\xi^{(k)}$. (This assumes that some sort of multiplicative central limit theorem is applicable.) Thus after amplification the (amplified) reads $\rvec{A}_{\droplet}(\time_{\batch})$ corresponding to droplet $\droplet$ would equal
 \begin{equation}
   \label{eq:PCR_amplification}
  \rvec{A}_{\droplet}(\time_{\batch}) := \vreads_{\droplet}(\time_{\batch}) \ast (a_{\droplet}\rvec{\boldsymbol{\xi}}) 
= \left(
a_{\droplet}\xi^{(1)}_{\droplet} \reads[1]_{\droplet}(\time_{\batch}), \dots, a_{\droplet} \xi^{(\strain)}_{\droplet} \reads[\strain]_{\droplet}(\time_{\batch}), \dots, a_{\droplet} \xi^{(\Strains)}_{\droplet} \reads[\Strains]_{\droplet}(\time_{\batch})
\right) \,.
 \end{equation}
Letting $M$ denote the copy number for the spikein gene (which recall is Poisson distributed, with a high expectation, cf. footnote \ref{footnote:spikein} from section \ref{sec:prep-sequ}), it follows that after normalizing by the observed number of spikein gene amplicons, the observed number of reads $\rvec{O}_{\droplet}(\time_{\batch})$ corresponding to droplet $\droplet$ is
\begin{equation}
\label{eq:normalized_and_amplified}
\rvec{O}_{\droplet}(\time_{\batch}) :=
\left( \frac{1}{a_{\droplet} \xi^{M}_{\droplet} M_{\droplet}} \right) \cdot \rvec{A}_{\droplet}(\time_{\batch})
= 
\left(  
\frac{\xi^{(1)}_{\droplet}}{\xi^{M}_{\droplet} M_{\droplet}} \reads[1]_{\droplet} (\time_{\batch}) ,
\dots
\frac{\xi^{(\strain)}_{\droplet}}{\xi^{M}_{\droplet} M_{\droplet}} \reads[\strain]_{\droplet} (\time_{\batch}),
\dots,
\frac{\xi^{(\Strains)}_{\droplet}}{\xi^{M}_{\droplet} M_{\droplet}} \reads[\Strains]_{\droplet} (\time_{\batch})
 \right) \,.
\end{equation}
Then the hope is that $\expectation{M} \cdot \rvec{O}_{\droplet}(\time_{\batch}) \approx \vabundance_{\droplet}(\time_{\batch})$, or at least ${\expectation{M}\cdot \rvec{O}_{\droplet}(\time_{\batch}) \approx \vreads_{\droplet}(\time_{\batch})}$, ``on average'' (i.e. the variance of the $\log(\xi_{\droplet})$'s is not too large).

\section{Example Cell and Read Count Trajectories}
\label{sec:example-trajectories}

For all simulations, the carrying capacity $\carryingcapacity[]$ was set to $\carryingcapacity[] = 268$ (based on what my collaborator said was realistic). Likewise for all simulations, the timestep $\Delta \time$ was set to $0.0001$, the ``noise scale'' was set to $\noisescale = 8$, and the cells were assumed to be grown ``in the fifth batch'' with $200$ timesteps separating each batch (so the simulation was ran for $1,000$ time steps total). All simulations were initialized with one cell each of the two strains.

The copy numbers for both strains were always set to be $1$.

\subsection{Cell Count Trajectories}
\label{sec:cell-count-traj}

\begin{figure}[H]
  \centering
 \includegraphics[width=\textwidth,height=\textheight,keepaspectratio]{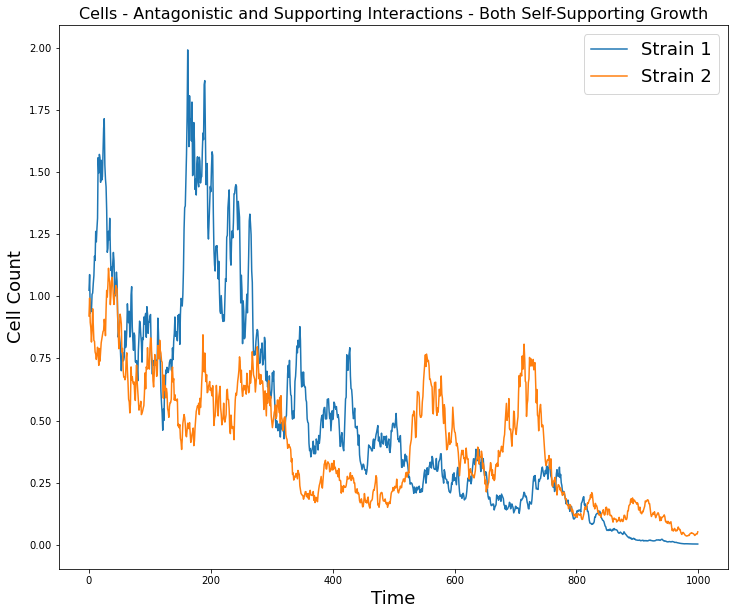}
  \caption[]{
  \begin{equation*}
  \begin{bmatrix}
    \baserate[1] \\
\baserate[2]
  \end{bmatrix} 
=
\begin{bmatrix}
  0.09417735 \\ 0.97562235
\end{bmatrix}
\,, \quad
\begin{bmatrix}
  \interaction{1}{1} & \interaction{1}{2} \\
\interaction{2}{1} & \interaction{2}{2}
\end{bmatrix}
=
\begin{bmatrix}
  0.5479121 &  -0.12224312 \\
  0.71719584 & 0.39473606
\end{bmatrix}
\end{equation*}
}
\label{fig:cell_trajectory_1}
\end{figure}

\begin{figure}[H]
  \centering
  \includegraphics[width=\textwidth,height=\textheight,keepaspectratio]{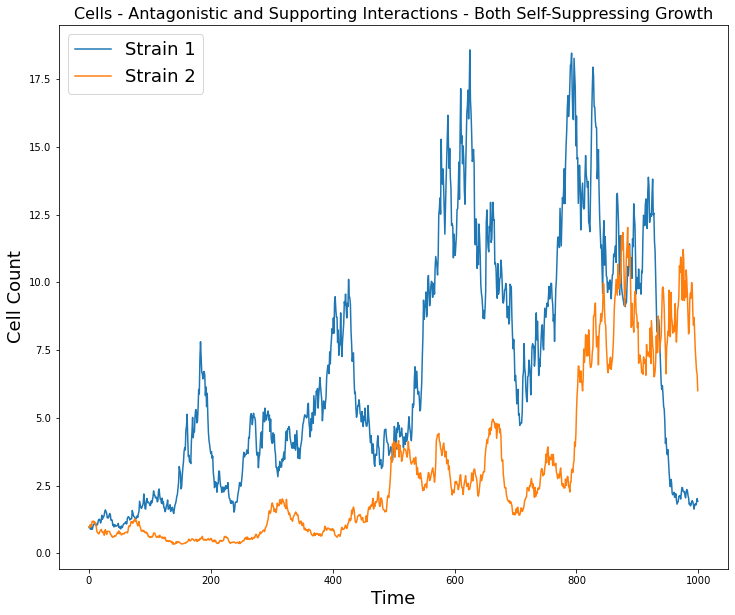}
\caption[]{
  \begin{equation*}
  \begin{bmatrix}
    \baserate[1] \\
\baserate[2]
  \end{bmatrix} 
=
\begin{bmatrix}
0.4797985 \\  0.97804575
\end{bmatrix}
\,, \quad
\begin{bmatrix}
  \interaction{1}{1} & \interaction{1}{2} \\
\interaction{2}{1} & \interaction{2}{2}
\end{bmatrix}
=
\begin{bmatrix}
-0.96753909 & -0.09498414 \\
 0.68530924 & -0.35219747
\end{bmatrix}
\end{equation*}
}
\label{fig:cell_trajectory_2}
\end{figure}

\begin{figure}[H]
  \centering
  \includegraphics[width=\textwidth,height=\textheight,keepaspectratio]{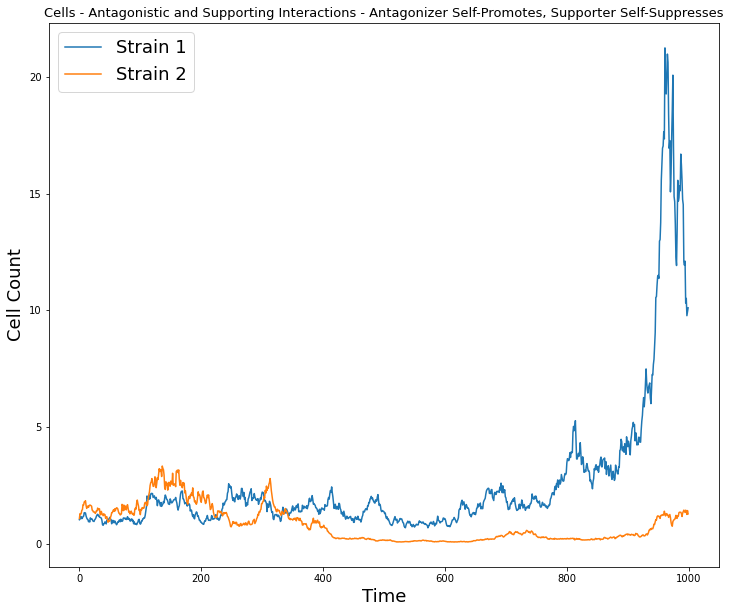}
  \caption[]{  \begin{equation*}
  \begin{bmatrix}
    \baserate[1] \\
\baserate[2]
  \end{bmatrix} 
=
\begin{bmatrix}
0.92088119 \\ 0.26796687
\end{bmatrix}
\,, \quad
\begin{bmatrix}
  \interaction{1}{1} & \interaction{1}{2} \\
\interaction{2}{1} & \interaction{2}{2}
\end{bmatrix}
=
\begin{bmatrix}
0.62393464 & -0.37883923 \\
 0.56723451 & -0.34407233
\end{bmatrix}
\end{equation*}}
\label{fig:cell_trajectory_3}
\end{figure}

\begin{figure}[H]
  \centering
  \includegraphics[width=\textwidth,height=\textheight,keepaspectratio]{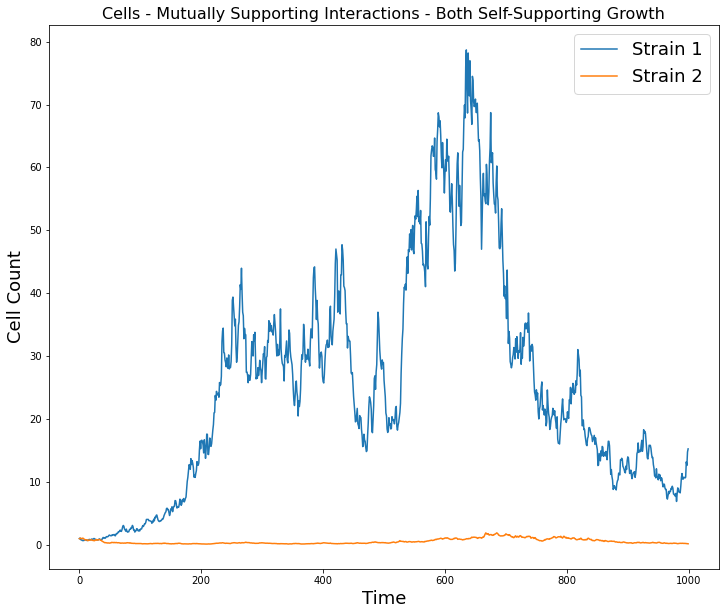}
  \caption[]{  \begin{equation*}
  \begin{bmatrix}
    \baserate[1] \\
\baserate[2]
  \end{bmatrix} 
=
\begin{bmatrix}
0.32162234 \\ 0.82774067
\end{bmatrix}
\,, \quad
\begin{bmatrix}
  \interaction{1}{1} & \interaction{1}{2} \\
\interaction{2}{1} & \interaction{2}{2}
\end{bmatrix}
=
\begin{bmatrix}
0.42400205 & 0.45574743 \\
0.83065306 & 0.03716118
\end{bmatrix}
\end{equation*}}
\label{fig:cell_trajectory_4}
\end{figure}

\begin{figure}[H]
  \centering
  \includegraphics[width=\textwidth,height=\textheight,keepaspectratio]{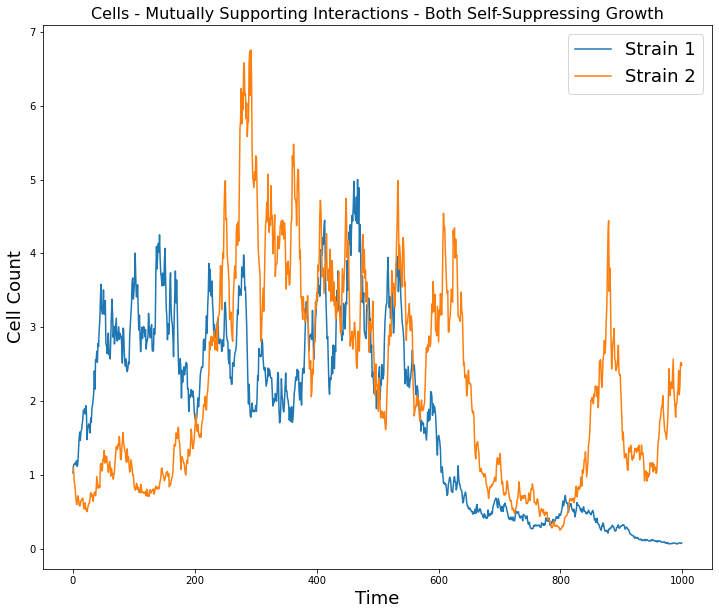}
  \caption[]{  \begin{equation*}
  \begin{bmatrix}
    \baserate[1] \\
\baserate[2]
  \end{bmatrix} 
=
\begin{bmatrix}
0.4602045 \\  0.34569703
\end{bmatrix}
\,, \quad
\begin{bmatrix}
  \interaction{1}{1} & \interaction{1}{2} \\
\interaction{2}{1} & \interaction{2}{2}
\end{bmatrix}
=
\begin{bmatrix}
-0.14107603 &  0.78315155 \\
 0.30476027 & -0.204964 
\end{bmatrix}
\end{equation*}}
\label{fig:cell_trajectory_5}
\end{figure}

\begin{figure}[H]
  \centering
  \includegraphics[width=\textwidth,height=\textheight,keepaspectratio]{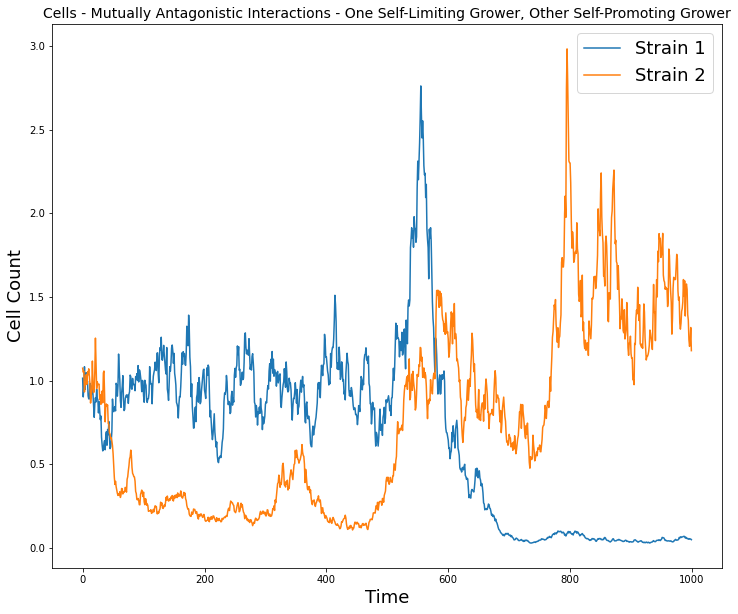}
  \caption[]{  \begin{equation*}
  \begin{bmatrix}
    \baserate[1] \\
\baserate[2]
  \end{bmatrix} 
=
\begin{bmatrix}
0.22324414 \\ 0.69651731
\end{bmatrix}
\,, \quad
\begin{bmatrix}
  \interaction{1}{1} & \interaction{1}{2} \\
\interaction{2}{1} & \interaction{2}{2}
\end{bmatrix}
=
\begin{bmatrix}
-0.15129588 & -0.05373991 \\
-0.41705514 &  0.58090085
\end{bmatrix}
\end{equation*}}
\label{fig:cell_trajectory_6}
\end{figure}

\begin{figure}[H]
  \centering
  \includegraphics[width=\textwidth,height=\textheight,keepaspectratio]{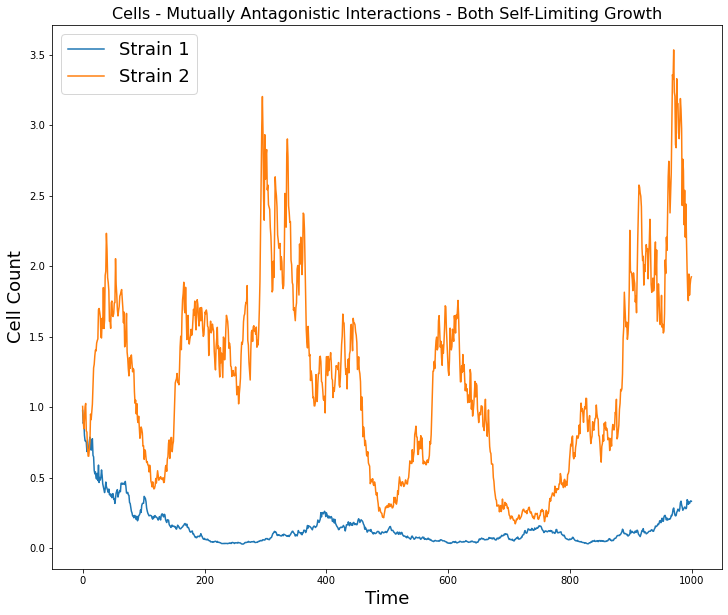}
  \caption[]{  \begin{equation*}
  \begin{bmatrix}
    \baserate[1] \\
\baserate[2]
  \end{bmatrix} 
=
\begin{bmatrix}
0.3970483 \\  0.18062232
\end{bmatrix}
\,, \quad
\begin{bmatrix}
  \interaction{1}{1} & \interaction{1}{2} \\
\interaction{2}{1} & \interaction{2}{2}
\end{bmatrix}
=
\begin{bmatrix}
-0.07416439 & -0.17117972 \\
-0.79256721 & -0.27501624
\end{bmatrix}
\end{equation*}}
\label{fig:cell_trajectory_7}
\end{figure}

\subsection{Read Count Trajectories}
\label{sec:read-count-trajectories}

\begin{figure}[H]
  \centering
  \includegraphics[width=\textwidth,height=\textheight,keepaspectratio]{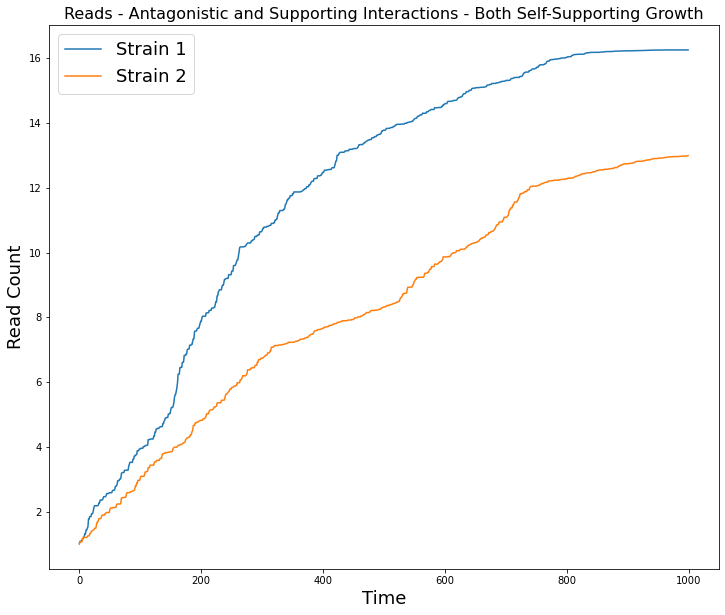}
  \caption[]{  \begin{equation*}
  \begin{bmatrix}
    \baserate[1] \\
\baserate[2]
  \end{bmatrix} 
=
\begin{bmatrix}
  0.09417735 \\ 0.97562235
\end{bmatrix}
\,, \quad
\begin{bmatrix}
  \interaction{1}{1} & \interaction{1}{2} \\
\interaction{2}{1} & \interaction{2}{2}
\end{bmatrix}
=
\begin{bmatrix}
  0.5479121 &  -0.12224312 \\
  0.71719584 & 0.39473606
\end{bmatrix}
\end{equation*}}
\label{fig:read_trajectory_1}
\end{figure}

\begin{figure}[H]
  \centering
  \includegraphics[width=\textwidth,height=\textheight,keepaspectratio]{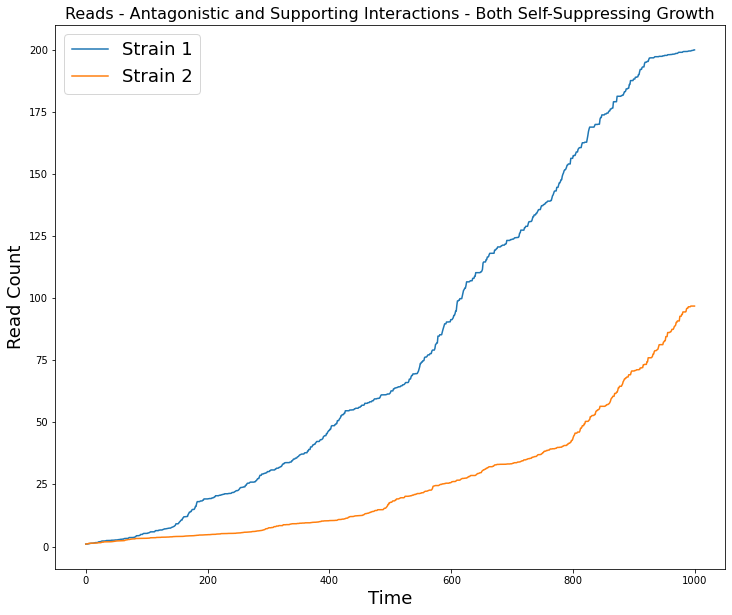}
  \caption[]{  \begin{equation*}
  \begin{bmatrix}
    \baserate[1] \\
\baserate[2]
  \end{bmatrix} 
=
\begin{bmatrix}
0.4797985 \\  0.97804575
\end{bmatrix}
\,, \quad
\begin{bmatrix}
  \interaction{1}{1} & \interaction{1}{2} \\
\interaction{2}{1} & \interaction{2}{2}
\end{bmatrix}
=
\begin{bmatrix}
-0.96753909 & -0.09498414 \\
  0.68530924 & -0.35219747
\end{bmatrix}
\end{equation*}}
\label{fig:read_trajectory_2}
\end{figure}

\begin{figure}[H]
  \centering
  \includegraphics[width=\textwidth,height=\textheight,keepaspectratio]{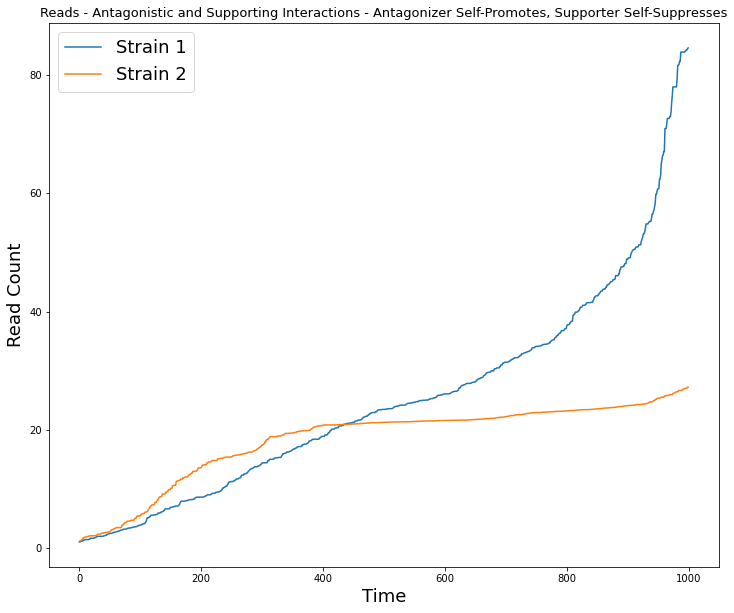}
  \caption[]{  \begin{equation*}
  \begin{bmatrix}
    \baserate[1] \\
\baserate[2]
  \end{bmatrix} 
=
\begin{bmatrix}
0.92088119 \\ 0.26796687
\end{bmatrix}
\,, \quad
\begin{bmatrix}
  \interaction{1}{1} & \interaction{1}{2} \\
\interaction{2}{1} & \interaction{2}{2}
\end{bmatrix}
=
\begin{bmatrix}
0.62393464 & -0.37883923 \\
 0.56723451 & -0.34407233
\end{bmatrix}
\end{equation*}}
\label{fig:read_trajectory_3}
\end{figure}

\begin{figure}[H]
  \centering
  \includegraphics[width=\textwidth,height=\textheight,keepaspectratio]{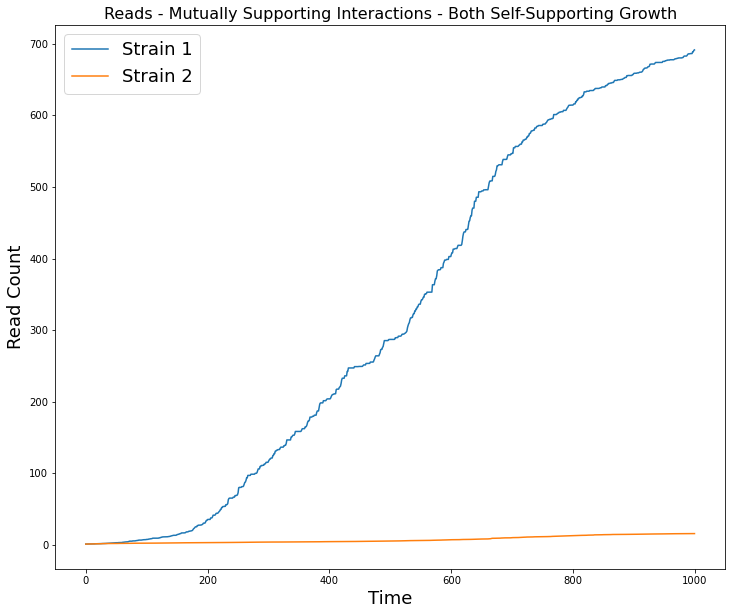}
  \caption[]{  \begin{equation*}
  \begin{bmatrix}
    \baserate[1] \\
\baserate[2]
  \end{bmatrix} 
=
\begin{bmatrix}
0.32162234 \\ 0.82774067
\end{bmatrix}
\,, \quad
\begin{bmatrix}
  \interaction{1}{1} & \interaction{1}{2} \\
\interaction{2}{1} & \interaction{2}{2}
\end{bmatrix}
=
\begin{bmatrix}
0.42400205 & 0.45574743 \\
0.83065306 & 0.03716118
\end{bmatrix}
\end{equation*}}
\label{fig:read_trajectory_4}
\end{figure}

\begin{figure}[H]
  \centering
  \includegraphics[width=\textwidth,height=\textheight,keepaspectratio]{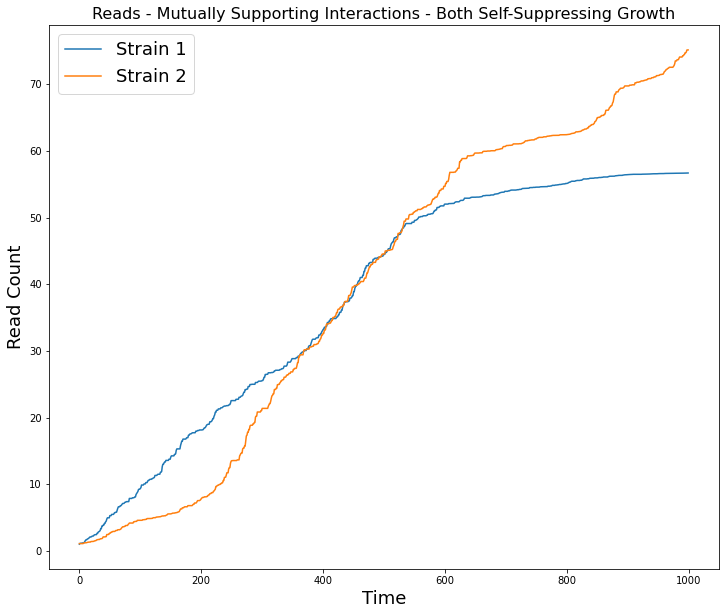}
  \caption[]{  \begin{equation*}
  \begin{bmatrix}
    \baserate[1] \\
\baserate[2]
  \end{bmatrix} 
=
\begin{bmatrix}
0.4602045 \\  0.34569703
\end{bmatrix}
\,, \quad
\begin{bmatrix}
  \interaction{1}{1} & \interaction{1}{2} \\
\interaction{2}{1} & \interaction{2}{2}
\end{bmatrix}
=
\begin{bmatrix}
-0.14107603 &  0.78315155 \\
 0.30476027 & -0.204964 
\end{bmatrix}
\end{equation*}}
\label{fig:read_trajectory_5}
\end{figure}

\begin{figure}[H]
  \centering
  \includegraphics[width=\textwidth,height=\textheight,keepaspectratio]{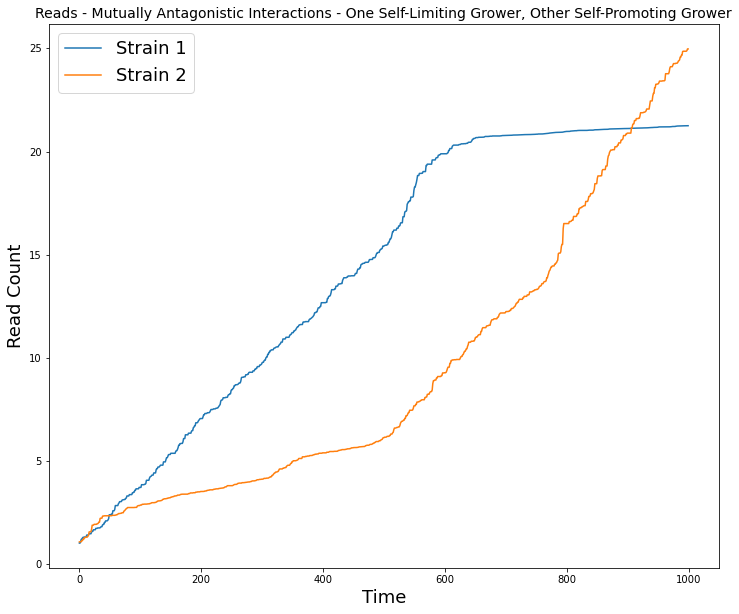}
  \caption[]{  \begin{equation*}
  \begin{bmatrix}
    \baserate[1] \\
\baserate[2]
  \end{bmatrix} 
=
\begin{bmatrix}
0.22324414 \\ 0.69651731
\end{bmatrix}
\,, \quad
\begin{bmatrix}
  \interaction{1}{1} & \interaction{1}{2} \\
\interaction{2}{1} & \interaction{2}{2}
\end{bmatrix}
=
\begin{bmatrix}
-0.15129588 & -0.05373991 \\
-0.41705514 &  0.58090085
\end{bmatrix}
\end{equation*}}
\label{fig:read_trajectory_6}
\end{figure}

\begin{figure}[H]
  \centering
  \includegraphics[width=\textwidth,height=\textheight,keepaspectratio]{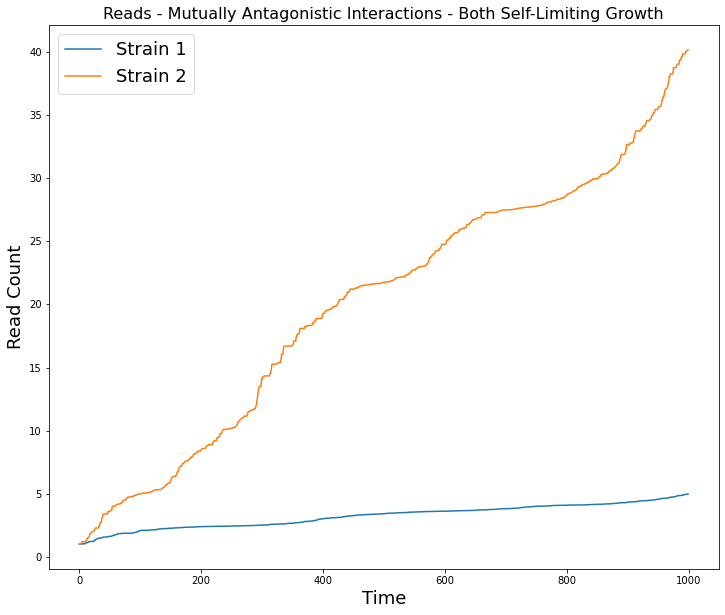}
  \caption[]{  \begin{equation*}
  \begin{bmatrix}
    \baserate[1] \\
\baserate[2]
  \end{bmatrix} 
=
\begin{bmatrix}
0.3970483 \\  0.18062232
\end{bmatrix}
\,, \quad
\begin{bmatrix}
  \interaction{1}{1} & \interaction{1}{2} \\
\interaction{2}{1} & \interaction{2}{2}
\end{bmatrix}
=
\begin{bmatrix}
-0.07416439 & -0.17117972 \\
-0.79256721 & -0.27501624
\end{bmatrix}
\end{equation*}}
\label{fig:read_trajectory_7}
\end{figure}

\chapter{Parametric Estimation of Interaction Coefficients}
\label{sec:param-estim-inter}

The goal of this section is to modify the regression-based methodology used in \cite{Venturelli} \cite{eco_model_time_series} \cite{Marino2014} \cite{Fisher2014} \cite{Mounier2008} to not require longitudinal data, and to be applicable to pseudo-longitudinal data. (Unfortunately preliminary data suggests that the resulting methodology does not work well in practice.) Future work should compare the ideas herein with the ideas in \cite{Sikorski2017}, which proposes modifying the aforementioned methodology to be applicable to a different\footnote{I.e. as opposed to the kind of cross-sectional data that pseudo-longitudinal data is.} kind of cross-sectional data. Cf. section \ref{sec:specific-problem-2}.

\section{Details}
\label{sec:details}

Given any estimand which is a conditional expectation, i.e. of the form $  \expectation{ X | A}$ for some random variable $X$ and some event $A$ in the underlying sigma-algebra, in what follows the corresponding empirical (arithmetic) mean will be denoted
\begin{equation}
  \label{eq:conditional_empircal_mean}
  \mean{X | A} :=  \frac{\sum_{\droplet \in [\Droplets]} X_{\droplet} \indicator{A} }{  \sum_{\droplet \in [\Droplets]} \indicator{A} } \,.
\end{equation}
In particular the unconditional empirical mean is
\begin{equation}
  \label{eq:empirical_mean_appendix}
  \mean{X} := \frac{1}{\Droplets} \sum_{\droplet \in [\Droplets]} X_{\droplet} \,.
\end{equation}

If one assumes the transition likelihoods are implicitly specified by the relationships:
\begin{align}
  \tag{(\ref{eq:transitions}) revisited}
  \begin{cases}
    \log(\abundance[\specie](\time_{\batch[]})) - \log(\abundance[\specie](\time_{\batch[]\!-\!1})) =
 \baserate + \displaystyle\sum_{\specie[*]=1}^{\Species} \interaction{\specie[*]}{\specie} \abundance[{\specie[]}] (\time_{\batch[]\! -\! 1}) + \noise & \abundance[\specie](0) > 0 \\
\abundance[\specie](\time_{\batch[]}) = 0 & \abundance[\specie](0) = 0
  \end{cases} \,,
\end{align}
for all $\batch[] \in [\batch]$ and all $\specie \in [ \Species]$ (as discussed in further detail in section \ref{sec:form-trans-likel}), then as a result of the linearity of (conditional) expectation it follows that for any $\batch \in [\Batches]$
\begin{align}
  \label{eq:cond_exp_transitions}
  &\expectation*{\log(\abundance[\specie](\time_{\batch})) | \abundance[\specie](0) > 0  } - \expectation*{ \log(\abundance[\specie](\time_{\batch\!-\!1}))  | \abundance[\specie](0) > 0   }\\
 =& \baserate + \displaystyle\sum_{\specie[*]=1}^{\Species} \interaction{\specie[*]}{\specie} \expectation*{ \abundance[{\specie[]}] (\time_{\batch\! -\! 1}) | \abundance[\specie](0) > 0 } \,.
\end{align}
Thus one would then hope that for any $b \in [\Batches]$ as the number of observations increases 
\begin{align}
  \label{eq:cond_mean_transitions}
  &\mean*{\log(\abundance[\specie](\time_{\batch})) | \abundance[\specie](0) > 0  }  - \mean*{ \log(\abundance[\specie](\time_{\batch\!-\!1}))  | \abundance[\specie](0) > 0   }\\
 \approx & \baserate + \displaystyle\sum_{\specie[*]=1}^{\Species} \interaction{\specie[*]}{\specie} \mean*{ \abundance[{\specie[]}] (\time_{\batch\! -\! 1}) | \abundance[\specie](0) > 0 } \,.
\end{align}
On this basis, below, for all $\specie \in [\Species]$ the least squares estimators
\begin{align}
  \label{eq:least_sq_cond_mean_transitions}
 (\hat{\baserate}, \hat{\interaction{1}{\specie}}, \dots, \hat{\interaction{\Species}{\specie}} ) := &\argmin_{(\tilde{\baserate}, \tilde{\interaction{1}{\specie}}, \dots, \tilde{\interaction{\Species}{\specie}}) } 
\sum_{\batch=2}^{\Batches} \left| \left| \vphantom{\displaystyle\sum_{\specie[*]=1}^{\Species}}
\left(\mean*{\log(\abundance[\specie](\time_{\batch})) | \abundance[\specie](0) > 0  } \right. \right. \right. \\
& \left. \phantom{\argmin_{(\tilde{\baserate}, \tilde{\interaction{1}{\specie}}, \dots, \tilde{\interaction{\Species}{\specie}}) }} 
- \mean*{ \log(\abundance[\specie](\time_{\batch\!-\!1}))  | \abundance[\specie](0) > 0   } \right) \\
&\left.\left. - \left(\tilde{\baserate} + \displaystyle\sum_{\specie[*]=1}^{\Species} \tilde{\interaction{\specie[*]}{\specie}} \mean*{ \abundance[{\specie[]}] (\time_{\batch\! -\! 1}) | \abundance[\specie](0) > 0 } \right)  \right| \right|_2^2\,.
\end{align}
are considered. 

Technically speaking, it is actually necessary to consider the set:
\begin{align}
  \label{eq:glv_indices_def}
  \glvindices := &  \left\{ \batch \in [\Batches] \setminus \{1\} : \text{ the events }\{ \abundance[\specie](\time_{\batch}) > 0 \}\text{ and }\{ \abundance[\specie](\time_{\batch\! -\!1}) > 0 \} \right. \\
&  \left. \vphantom{\left\{ \batch \in [\Batches] \setminus \{1\} : \text{ the events }\{ \abundance[\specie](\time_{\batch}) > 0 \} \right.}\phantom{\batch \in [\Batches] \setminus \{1\} :}
\text{ are both observed at least once}  \right\} \,,
\end{align}
such that the computed estimators actually are
\begin{align}
  \label{eq:least_sq_cond_mean_transitions_fancy}
 (\hat{\baserate}, \hat{\interaction{1}{\specie}}, \dots, \hat{\interaction{\Species}{\specie}} ) := &\argmin_{(\tilde{\baserate}, \tilde{\interaction{1}{\specie}}, \dots, \tilde{\interaction{\Species}{\specie}}) } 
\sum_{\batch \in \glvindices} \left| \left| \vphantom{\displaystyle\sum_{\specie[*]=1}^{\Species}}
\left(\mean*{\log(\abundance[\specie](\time_{\batch})) | \abundance[\specie](0) > 0  } \right. \right. \right. \\
& \left. \phantom{\argmin_{(\tilde{\baserate}, \tilde{\interaction{1}{\specie}}, \dots, \tilde{\interaction{\Species}{\specie}}) }} 
- \mean*{ \log(\abundance[\specie](\time_{\batch\!-\!1}))  | \abundance[\specie](0) > 0   } \right) \\
&\left.\left. - \left(\tilde{\baserate} + \displaystyle\sum_{\specie[*]=1}^{\Species} \tilde{\interaction{\specie[*]}{\specie}} \mean*{ \abundance[{\specie[]}] (\time_{\batch\! -\! 1}) | \abundance[\specie](0) > 0 } \right)  \right| \right|_2^2\,.
\end{align}
When $\glvindices = \emptyset$, the estimates are marked as missing. Just like for the estimands from chapter \ref{chap:log-ratio-coeff}, in analyses of simulation data, when estimates of the interaction coefficients are missing, they are conservatively set to $0$. (I.e. in the absence of any evidence for an interaction, the choice is made to prefer the possibility of making the fase negative error of assuming no interaction exists when one might exist, rather than possibly making the false positive error of assuming an interaction exists when in fact none exists.)

Note that starting from (\ref{eq:transitions}) and considering instead the relationship:
\begin{align}
  \label{eq:cond_exp_transitions_glv}
  &\expectation*{\log(\abundance[\specie](\time_{\batch})) | \abundance[\specie](\time_{\batch\!-\!1})=n, \abundance[\specie](0) > 0  }\\
& - \expectation*{ \log(\abundance[\specie](\time_{\batch\!-\!1}))  |\abundance[\specie](\time_{\batch\!-\!1})=n,  \abundance[\specie](0) > 0   }\\
 =& \baserate + \displaystyle\sum_{\specie[*]=1}^{\Species} \interaction{\specie[*]}{\specie} \expectation*{ \abundance[{\specie[]}] (\time_{\batch\! -\! 1}) | \abundance[\specie](\time_{\batch\!-\!1})=n,  \abundance[\specie](0) > 0 } \,,
\end{align}
following reasoning analogous to (\ref{eq:least_sq_cond_mean_transitions_fancy}) would lead directly to the linear regression used in e.g. \cite{Venturelli} \cite{Marino2014} \cite{Fisher2014} \cite{eco_model_time_series} \cite{Mounier2008} to estimate interaction coefficients. However, estimating the quantities of the form:
\begin{equation}
  \label{eq:regression_conditional_exp_term}
  \expectation*{\log(\abundance[\specie](\time_{\batch})) | \abundance[\specie](\time_{\batch\!-\!1})=n, \abundance[\specie](0) > 0  } 
\end{equation}
requires being able to observe the value of $\abundance[\specie]$ at more than one time point for a single droplet, which is impossible for the data relevant to this study. As a reminder, this data is not actually longitudinal. Thus it seems one is limited to using (\ref{eq:least_sq_cond_mean_transitions_fancy}), or something conceptually similar, when attempting to modify the approach from previous work \cite{Venturelli} \cite{Fisher2014} \cite{Marino2014} \cite{eco_model_time_series} \cite{Mounier2008} to be applicable to this data.

\section{Comments}
\label{sec:comments}

The only reason for including the parameters $\baserate$ in the model (\ref{eq:transitions}) used in this work is simply to ensure that ``baseline'' or ``constant'' growth effects are not misattributed as being part of the effects of the microbial interactions. Indeed, unlike the papers \cite{Venturelli}, \cite{Marino2014}, and \cite{Mounier2008}, the previous work \cite{Fisher2014} did not include (analogues of) the parameters $\baserate$ in their model, so there would be precedent for omitting them.

The assumption that the temporal dynamics can be adequately described by the equations (\ref{eq:transitions}) also implies at least two important assumptions:\\

\begin{enumerate}
\item Because the equations (\ref{eq:transitions}) allow for indefinite population growth, using them implicitly assumes that all of our observations correspond only to the regime before resource scarcity and other factors begin to constrain the growth of microbial cells. Cf. the discussion from section \ref{sec:growth-cells-inside}.
\item The equations (\ref{eq:transitions}) implicitly assume that only pairwise interactions between different microbial species are relevant for determining their rates of growth, and that there is no non-negligible effect from higher-order interactions involving three or more distinct species. This is reflected in the absence of any terms that are a function of the abundances of more than two species. This has been recognized as a major limitation of these equations for accurately describing microbial interactions in previous work \cite{Momeni2017}. Cf. also \cite{higher_order} or the discussion from section \ref{sec:form-as-stat}.
\end{enumerate}
The second assumption was also reflected in the experimental design used in \cite{Venturelli}, which nevertheless found the equations (\ref{eq:transitions}) adequate for describing the microbial interactions which they observed. In previous work the first assumption has been enforced via constraints on $\beta_{\specie}$ and $\alpha_{\specie_1 \specie_2}$ which are not applied here, since such constraints seem to lack a biological justification, and they are unnecessary for the specific simulation model (\ref{eq:simulation_update_framework}) that I used.

This quote from \cite{Marino2014} is precedent for the above viewpoint on parameter constraints:
\begin{quote}
  One approach to improving parameter estimation is to superimpose constraints on the parameters; however, unless these constraints can be justified both biologically and mathematically, they should not be enforced because the results can be greatly affected.
\end{quote}

\end{appendices}

\clearpage

\let\etalchar\undefined
\bibliographystyleAppendix{amsalpha}
\bibliographyAppendix{appendices/newcites}

\end{document}